\documentclass[a4paper,11pt,twoside,notitlepage]{report}
\usepackage[T1]{fontenc}

\usepackage[encapsulated]{CJK}
\usepackage{ucs}
\usepackage[utf8x]{inputenc} 
\usepackage{lmodern}

\usepackage{fancyhdr}
\usepackage[a4paper]{geometry}
\usepackage[toc,page]{appendix}
\pdfoutput=1 
\usepackage{cite}
\usepackage{caption}

\usepackage{slashed}
\usepackage{feynmf}
\usepackage{amsmath}
\usepackage{amssymb}

\usepackage{placeins}
\usepackage{default}
\usepackage{textcomp}

\usepackage{graphicx}

\newcommand{\sectionc}[2]{%
\section{#1}%
\label{#2}%
\markboth{\thesection \hspace*{1mm} #1}{}%
\noindent \rule{\linewidth}{0.5pt}%
\vspace{5mm}}

\usepackage{titlesec}

\titleformat
{\chapter} 
[display] 
{\bfseries\LARGE\slshape} 
{} 
{0.0ex} 
{
\rule{\textwidth}{1pt}%
\vspace{1ex}
\centering
} 
[
\vspace{-0.5ex}%
\rule{\textwidth}{0.3pt}
] 

\newcommand{\japanesetext}[1]{\begin{CJK*}{UTF8}{min}#1\end{CJK*}}

\title{{\center\LARGE Properties of minimally doubled fermions}}
\author{Johannes Heinrich Weber}

\begin{document}
\begin{titlepage}
\pagestyle{empty}
\begin{center}
\begin{minipage}{0.75\linewidth}
    \centering
    \rule{1.0\linewidth}{0.01\linewidth}\par
    \vspace{1cm}
    {\uppercase{\Large Properties of minimally doubled fermions\par}}
    \vspace{1cm}
    \rule{1.0\linewidth}{0.01\linewidth}\par
    \vspace{2cm}
    {\Large Dissertation zur Erlangung des Grades\par}
    {\Large ``Doktor der Naturwissenschaften''\par}
    \vspace{1cm}
    {\Large vorgelegt am Fachbereich\par}
    {\Large Physik, Mathematik und Informatik\par}
    {\Large der Johannes Gutenberg-Universität in Mainz\par}
        \includegraphics[bb=0 0 180 120,width=1.0\linewidth]{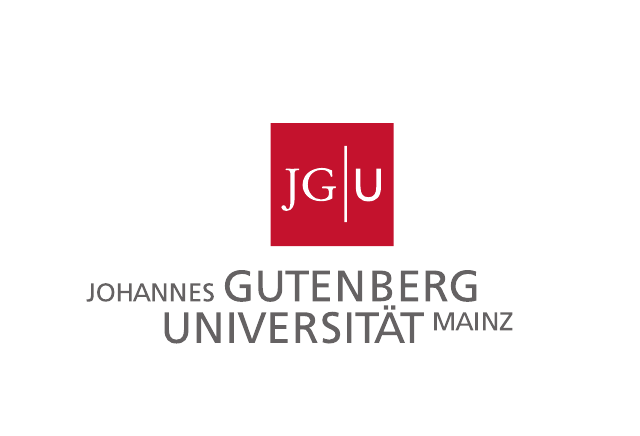}\par
    \rule{1.0\linewidth}{0.01\linewidth}\par
    \vspace{1cm}
    {\Large Johannes Heinrich Weber\par}
    {\Large geboren in Gro\ss-Gerau\par}
    \vspace{2cm}
    {\Large Mainz, den 13. M\"{a}rz 2015}\par
    \vspace{1cm}
    \rule{1.0\linewidth}{0.01\linewidth}\par
\end{minipage}
\end{center}
\clearpage
\end{titlepage}
{
\pagestyle{empty}
{
~\vfill
\begin{tabular}{lr}
  Johannes Heinrich Weber: \textsl{Properties of Minimally doubled fermions} & \hspace{3cm} \\
   Einreichung der Promotion: 13. M\"{a}rz 2015 \\
   Mündliches Kolloquium: 30. Oktober      2015  \\
   Datum der Promotion: 14. Dezember      2015  & D77 
\end{tabular}
}
\clearpage
}

{
\pagestyle{empty}
{
\pagestyle{empty}
{\centering \textbf{Abstract}\par}\vskip1em

\noindent
The majority of quark actions in lattice QCD encounter difficulties with chiral symmetry and its spontaneous symmetry breaking, which is realised only in compliance with the Nielsen-Ninomiya No-Go theorem. 
Minimally doubled fermions are a category of discretisations of strictly local chiral fermions which reproduce two degenerate quark flavours in the continuum limit. 
The Dirac operator complies with the No-Go theorem by having two poles at different points in the Brillouin zone. 
Their alignment implies that hypercubic symmetry is explicitly broken at finite lattice spacing.\newline

\noindent
Bori\c{c}i-Creutz and Karsten-Wilczek (KW) fermions are two different variants of minimally doubled fermions, which are studied in perturbation theory. Renormalisation properties are studied and three counterterms, which arise due to explicit loss of hypercubic symmetry, are determined perturbatively. Inclusion of counterterms removes the anisotropy from self-energy and vacuum polarisation at one-loop level. Corrections to local bilinears are calculated and chiral symmetry is tested. Vector and axial symmetry currents are derived and their conservation is verified.\newline
\noindent
Numerical simulations of KW fermions in the quenched approximation are used to determine non-perturbative renormalisation criteria. The relevant fermionic counterterm is tuned using two complementary approaches. First, the action's aniso\-tropy is reflected in an anisotropic pseudoscalar mass. The minimum of the mass anisotropy defines the tuned action. Second, some hadronic correlation functions have oscillating contributions. Restoration of the tree-level frequency spectrum is used as complementary condition, which agrees within errors. Due to lack of a numerically robust non-perturbative condition, the marginal fermionic counterterm is fixed to the estimate from perturbation theory.\newline
\noindent
The hadron spectrum is studied with a properly tuned KW action. As the spontaneously broken chiral symmetry SU(2)$ _A $ is reduced to U(1)$ _A $ due to the fermion action, only one pseudoscalar meson is a \mbox{(Pseudo-)}~Goldstone boson at finite cutoff. The pseudoscalar channel is studied and Goldstone boson-like behaviour including quenched chiral logarithms is observed, which agrees with phenomenological predictions. A second channel using $ \gamma^0 $ instead of $ \gamma^5 $ is studied. Its ground state scales with the bare quark mass like a \mbox{(Pseudo-)}~Goldstone boson, but retains a residual mass in the chiral limit, which vanishes as $ \mathcal{O}(a^2) $ in the continuum limit. This strongly indicates that the $ \gamma^0 $~channel contains a pseudoscalar meson which is affected by lattice artifacts. These simulations in the chiral regime indicate that KW fermions are not affected by exceptional configurations.\newline

\noindent
The first numerical study of KW fermions is a cornerstone for future applications. Simulations using dynamical KW fermions and including missing quark-disconnected contributions will overcome present deficiencies. Vector mesons and nucleons are the next milestones in a study of the hadron spectrum. 
}

\pagestyle{empty}
{
\pagebreak
\pagestyle{empty}
\begin{center}
\textbf{Zusammenfassung}
\end{center}

\noindent
Die Mehrheit der Quarkwirkungen in Gitter QCD leidet an Problemen mit chiraler Symmetrie und deren spontaner Brechung, die nur gem\"{a}\ss\ des Nielsen-Ninomiya Theorems realisiert wird.
Minimal verdoppelte Fermionen sind eine Kategorie von Diskretisierungen strikt lokaler, chiraler Fermionen, die im Kontinuums\-limes zwei entartete Quark Flavours reproduzieren.
Der Dirac Operator erf\"{u}llt das Theorem, da er zwei Polstellen an ungleichen Punkten der Brillouin-Zone hat.
Diese Anordnung zieht explizite Brechung der hyperkubische Symmetrie bei endlichem Gitterabstand nach sich. \newline

\noindent
Bori\c{c}i-Creutz und Karsten-Wilczek (KW) Fermionen sind zwei Varianten, die in St\"{o}\-rungs\-theorie untersucht werden. Aufgrund expliziter hyper\-kubischer Symmetriebrechung erfordert Renormierung drei Counterterme, die perturbativ bestimmt werden.
Mittels der Counterterme wird die An\-iso\-tropie aus Selbst-Energie und Vakuum-Polarisation auf Ein-Schleifen Niveau entfernt.
Korrekturen zu bilinearen Operatoren werden berechnet und chirale Symmetrie \"{u}berpr\"{u}ft. 
Vektorielle und axiale Symmetriestr\"{o}me werden hergeleitet und deren Erhaltung verifiziert.\newline
\noindent
Numerische Simulation von KW Fermionen in Quenched Approximation dienen der Bestimmung nicht-perturbativer Renormierungsbedingungen.
Der relevante fermionische Counterterm wird mit zwei komplement\"{a}ren Ans\"{a}tzen justiert.
Erstens spiegelt sich die An\-iso\-tropie der Wirkung in einer Anisotropie der pseudoskalaren Masse wider.
Das Minimum der Massenanisotropie definiert die kalibrierte Wirkung.
Zweitens haben manche hadronische Korrelatoren oszillierende Anteile.
Wiederherstellung des Tree-level Frequenzspektrums ist eine komplement\"{a}re Bedingung, die innerhalb der Fehler \"{u}ber\-ein\-stimmt.
Mangels einer numerisch robusten, nicht-perturbativen Bedingung wird der marginale fermionische Counterterm auf die Vorhersage der Störungstheorie fixiert.\newline
\noindent
Das Hadronenspektrum wird mit einer korrekt eingestellten KW Wirkung untersucht.
Da die spontan gebrochene chirale Symmetrie SU(2)$ _A $ durch die Fermionwirkung zu U(1)$ _A $ reduziert ist, ist nur eines der pseudoskalaren Mesonen bei endlichem Cutoff ein \mbox{(Pseudo-)}~Goldstone Boson.
In der Untersuchung des pseudoskalaren Kanals wird Goldstone Boson-artiges Verhalten einschließlich Quenched Chiral Logarithms im Einklang mit ph\"{a}nomenologischen Vorhersagen beobachtet.
Ein zweiter Kanal mit $ \gamma^0 $ anstelle von $ \gamma^5 $ wird untersucht.
Dessen Grundzustand skaliert mit der nackten Quarkmasse wie ein \mbox{(Pseudo-)}~Goldstone Boson, aber beh\"{a}lt eine residuale Masse im chiralen Limes, welche wie $ \mathcal{O}(a^2) $ im Kontinuumslimes verschwindet.
Dies ist ein starkes Indiz, dass der $ \gamma^0 $~Kanal ein von Gitterartefakten betroffenes pseudoskalares Meson enth\"{a}lt. Diese Simulationen im chiralen Regime zeigen, dass KW Fermionen nicht von Exceptional Configurations betroffen sind. \newline

\noindent
Die erste numerische Studie von KW Fermionen ist Grundlage zuk\"{u}ftiger Anwendungen.
\"{U}berwindung derzeitiger Unzul\"{a}nglichkeiten 
erfordert Simulationen mit dynamischen KW Fermionen unter Ber\"{u}cksichtigung fehlender Quark-disconnected Beitr\"{a}ge.
Vektormesonen und Nukleonen sind die n\"{a}chsten Meilensteine der Untersuchung des Hadronenspektrums.
}

\pagestyle{empty}
{
\newpage
\enlargethispage*{\baselineskip}
\pagestyle{empty}
{\centering\textbf{\japanesetext{概略}}\par}\vskip1em
\pagestyle{empty}
\noindent
\japanesetext{格子上における量子色力学のクォーク作用の多くはカイラル対称性を保持することが困難であり、Nielsen-二宮の定理にある通り, 制限が加わることが知られている。}
\japanesetext{Minimally doubled fermions(MDF)は, 局在するカイラル・フェルミ粒子を離散化することによって、連続極限における縮退のフレーバー数が2となるように定式化されている。}
\japanesetext{ブリユアン領域において二つのゼロ点が存在するならば、そのようなディラック演算子は上記の定理を満たす。}
\japanesetext{有限の格子間隔の場合には, 四次元の格子対称性の破れが上記のゼロ点の方向性の結果となる。}
\newline

\noindent
\japanesetext{Bori\c{c}i-Creutz, 及び, Karsten-Wilczek(KW)フェルミオンは, 二種類の異なるMDFであり, 本研究では, 
これらのMDFを摂動論を用いて検証する。}
\japanesetext{四次元の格子対称性の破れにより, 繰り込みは３つの, 異方性を持ったcounter term (CT) によって特徴付けられる。また, これらのCTは摂動論によって求まる。}
\japanesetext{このCTを取り込むことによって, 自己エネルギーと真空偏極の異方性は１-ループのオーダーで取り除かれる。}
\japanesetext{また, 双一次の演算子の１-ループ補正を計算することによってカイラル対称性を検討する。}
\japanesetext{さらに, ベクトル対称性, 及び, 軸性ベクトル対称性のカレントを導出し, これらを通して保存則を確認する。}
\newline
\noindent
\japanesetext{クェンチ近似を用いた数値シミュレーションによって, KWフェルミオンの非摂動的な繰り込み条件の設定を行う。}
\japanesetext{フェルミ作用の三次元のCTは, 以下の２つの要素から構成される手法によって計算される。}
\japanesetext{第一に, 遷移行列の異方性が擬スカラーの質量の異方性に映し出されることから, }
\japanesetext{質量における異方性の最小値によって, 作用を定義することができる。}
\japanesetext{次に, 異なる極を含むハドロン相関関数が振動する成分を持つにことに注目する。}
\japanesetext{この振動成分の周波数スペクトルを再現するための条件を求めることができ, この条件によって再現されたスペクトルは, 元の振動に誤算の範囲で一致することが確認できる。}
\japanesetext{また, 四次元格子上のCTを数値的に評価するための必要条件を特定するために, 前述の摂動論による予測を使用する。}
\newline
\noindent
\japanesetext{上記の繰り込み条件によって改良されたKW作用を用いて, ハドロン・スペクトルの研究を行う。}
\japanesetext{自発的に破れるカイラル対称性は, 格子上におけるフェルミオンの定式化によりSU(2)$ _A $ が破れ,U(1)$ _A $だけが残る。このことから, 有限の格子間隔の場合には,擬スカラー中間子のうちの一つのみが南部ボソンの役割を担う。}
\japanesetext{擬スカラー・チャンネルには, このような現象論的な予測に従う, クェンチ・カイラル対数を含んだ南部ボソンとの対応が見られる。}
\japanesetext{次に, $ \gamma^5 $ の代わり $ \gamma^0 $ を用いるチャンネルの研究を行う。}
\japanesetext{本チャネルにおける基底状態の質量スペクトラムは南部ボソンと類似しているが, 有限の格子間隔による寄与がカイラル極限に残ったとしても, 連続極限においては $ \mathcal{O}(a^2) $ で消滅することが示される。}
\japanesetext{このことから, $ \gamma^0 $ チャンネルの基底状態は,量子エラーの影響を受けている擬スカラーであるということが結論付けられる。}
\newline

\noindent
\japanesetext{本研究におけるMDFの数値的な検証は, 今後のMDF研究の礎になるであろう。}
\japanesetext{一方で, 現在までの数値計算には不十分な点が残る。今後の進展として, MDFによって定式化された動的フェルミオンの導入, 及び, quark-disconnected 図の計算は欠かせないと考える。}
\japanesetext{また, ベクトル中間子と核子は, 更なるハドロン・スペクトル研究の目標である。}
\pagestyle{empty}
}

\pagestyle{empty}
\newpage
}

\pagenumbering{Roman}
\tableofcontents

\newpage
\pagenumbering{arabic}
\addcontentsline{toc}{chapter}{Introduction}  
\chapter*{Introduction}

\noindent
The recent discovery~\cite{Aad:2012tfa} of a Higgs boson at CERN, which is at the level of present knowledge not dissimilar to the Higgs boson of the Standard Model (SM)~\cite{Glashow:1961tr, PhysRevLett.19.1264, Salam:1968rm}, strongly supports the SM as the appropriate theory of particle physics at the energy scales, which are nowadays accessible in experimental observations. The Nobel prize of physics 2013 was awarded for the theoretical suggestion of the Higgs mechanism~\cite{1963PhRv..130..439A, 1964PhRvL..13..321E, 1964PhRvL..13..508H, 1964PhRvL..13..585G} of spontaneous breakdown of the electroweak SU(2) symmetry as the origin of a non-vanishing vacuum expectation value (VEV) of a Higgs field. It requires the existence of at least one scalar Higgs boson as a particle, contributes longitudinal polarisation degrees of freedom to the weak gauge bosons and generates the mass of all elementary particles through Yukawa couplings between massless, bare fields and a Higgs field with its non-vanishing VEV. 
\noindent
Whereas the non-trivial electroweak vacuum structure of the Higgs field sets the scale of elementary particle masses in the SM through its VEV, it is not the dominant contribution to the mass of the visible universe. Instead the internal structure of baryons, bound states of strongly interacting elementary fields, is responsible for the overwhelming majority of this mass. The constituent quark masses, which are due to the Higgs mechanism, contribute only about $ 2\% $ to the proton's total mass. The mass of hydrogen is completely dominated by the proton mass, which exceeds the electron mass (due to the Higgs mechanism) by nearly a factor $ 2000 $. Thus, the mass of the visible universe, which mainly consists of hydrogen and helium atoms, is almost entirely due to the strong interactions. Hence, the mass and structure of the visible universe cannot be understood without a thorough comprehension of the strong interactions.\newline

\noindent
Since the advent of particle accelerators in the late 1940's and early 1950's, experimental observations of a multitude of previously unknown particles necessitated a rethinking of subatomic physics in general and of the strong interactions in particular. The previously known particle spectrum contained protons, electrons and photons, which were readily accessible in many atomic systems already in the 1920's, as well as neutrons~\cite{Chadwick:1932ma}, positrons~\cite{1931RSPSA.133...60D}, muons~\cite{1933PhRv...43..491A, PhysRev.50.263} and, as postulated particles, Pauli's neutrino~\cite{Pauli:1930pc} and Yukawa's meson~\cite{Yukawa:1935xg}. The muon was considered as a candidate for the meson, since its mass agreed with the postulated meson mass within a factor of two. Electric charge and isospin~\cite{Heisenberg:1932dw} were considered as the charges of elementary quanta.
\noindent
With the experimental discovery of pions and kaons in 1947~\cite{Lattes:1947mw, Rochester:1947mi}, the pion fit into the isospin picture as Yukawa's meson. However, kaon decays indicated new physics as their time scales greatly exceeded those of pion-nucleon reactions. A new quantum number called strangeness~\cite{PhysRev.86.663} was introduced, which is violated in weak decays. As more and more additional hadrons (the `particle zoo') were discovered in new particle accelerator facilities, theoretical models of hadrons could not keep up with the speed of discovery. In the meantime, the development of Quantum Electrodynamics (QED)~\cite{Tomonaga:1946zz, PhysRev.73.416, PhysRev.74.1439, PhysRev.75.486, PhysRev.75.1736, PhysRev.76.749, PhysRev.76.769, PhysRev.80.440} as the quantised U(1)~gauge theory of the photon and the electron advanced very quickly. With high precision calculations such as the Lamb shift in the spectrum of hydrogen~\cite{PhysRev.72.241, PhysRev.72.339}, QED was already an established theory, while the model of the strong interactions was still in its infancy.\newline

\noindent
Experimental pion-nucleon scattering data did not match the prediciton of a renormalisable, pseudoscalar pion-nucleon coupling, which predicted large $ s $-wave contributions~\cite{PhysRev.86.793.2, PhysRev.91.155, PhysRev.88.1053}. Instead, the energy dependence of pion-nucleon interactions was in good agreement with gradient-coupling models in which low-energy pions decouple from the nucleons and from other pions, since their coupling is proportional to their $ 4 $-momenta. The shift symmetry of these soft pions led to an interpretation of the pion as an almost massless \mbox{(Pseudo-)}~Goldstone boson of a spontaneously broken chiral symmetry of the strong interactions ~\cite{PhysRev.122.345, PhysRev.127.965}. An explicit, soft breaking of the chiral symmetry was considered the source of the finite pion mass. However, the nature of the fundamental chiral fermions was still obscure.
\noindent
Detailed studies of weak pion decays~\cite{Goldberger:1958tr, Goldberger:1958vp} revealed a relation between the axial charge of the neutron in beta decay and the pion-nucleon coupling constant from the gradient-coupling model, which was coined the Goldberger-Treiman relation. It is understood in terms of the \textit{Partially Conserved Axial Current} (PCAC) hypothesis~\cite{GellMann:1964tf}. Axial currents do not preserve the vacuum of the strong interactions, but instead serve as interpolating operators for pseudoscalar meson fields and create one-particle states. The same axial currents participate in the \textit{V-A-coupling} of the weak interactions and mediate the weak decay of the neutron. Application of axial current operators on external particle wave-functions successfully procured amplitudes involving soft pions from amplitudes without them through the use of \textit{chiral Ward identities}~\cite{PhysRevLett.18.188}. This current algebra was successfully applied to a variety of strong processes, even though the fundamental carriers of chiral charge were still missing.\newline

\noindent
Gell-Mann and Ne'eman introduced an approximate global SU(3) symmetry of the hadron spectrum~\cite{GellMann:1961ky,Ne'eman:1961cd} as \textit{the eightfold way}. Hadrons with equal space-time quantum numbers $ J^{PC} $ and similar masses were considered as an SU(3) multiplet. Isospin and strangeness quantum numbers were united in a group-theoretical sense. The eightfold way described the known particle spectrum with great success, but predicted new states~\cite{PhysRev.125.1067}. Though the existence of Delta baryons with spin-{3}/{2} had been known since 1952~\cite{Anderson:1952nw, PhysRev.86.106}, a spin-{3}/{2} baryon decuplet implied the existence of a new baryon with strangeness $ S=-3 $, which would have to be a long-lived state with a unique decay signature via three weak decays. This Omega baryon $\Omega^{-} $ was discovered only in 1964~\cite{PhysRevLett.12.204} after its earlier prediction in the eightfold way. 
\noindent
Hypothetical fermionic constituents were suggested by Gell-Mann~\cite{GellMann:1964nj, GellMann:1964xy} as the origin of the \textit{flavour} SU(3) symmetry. As part of a spin-{3}/{2} baryon decuplet, three states at the decuplet's corners ($ \Delta^{++} $, $ \Delta^- $, $ \Omega^- $) must have fully-symmetric spin and flavour wave functions of all constituents. However, as fermions, they must have a totally antisymmetric wave function. This implies antisymmetry in a new quantum number of the constituents, which was called \textit{colour}.\newline

\enlargethispage*{\baselineskip}
\noindent
These hypothetised fundamental carriers of flavour SU(3) and colour SU(3) quantum numbers were referred to as \textit{quarks}~\cite{GellMann:1964nj}. Whereas deep inelastic electron scattering on nucleons indeed revealed point-like constituents~\cite{Bjorken:1968dy} inside of the nucleons, quarks could not be isolated in a detector. Fritzsch, Gell-Mann and Leutwyler suggested~\cite{Fritzsch:1973pi} that the quarks' colour charge is coupled to an octet of \textit{gluons}, fictitious neutral gauge bosons in the adjoint representation of the colour gauge group with dynamics of a Yang-Mills field~\cite{Yang:1954ek}. Their theory of the strong interactions had already reached its modern form except for the number of flavours. However, a mechanism of colour confinement for quarks and gluons, which simultaneously allowed for Bjorken scaling~\cite{Bjorken:1968dy}, was still lacking.
\noindent
With the proof of asymptotic freedom by Gross, Politzer and Wilczek~\cite{Gross:1973ju, Gross:1973id, Gross:1974cs, Politzer:1974fr} using perturbation theory and Wilson's demonstration of a colour confinement mechanism on a space-time lattice in the strong coupling limit~\cite{Wilson:1974sk}, the theory of quarks and gluons quickly started to gain acceptance. It is nowadays included in the Standard Model as Quantum Chromodynamics (QCD), a quantised SU(3) gauge theory, and covers the interactions of eight gauge fields, massless vector bosons called gluons, amongst each other and with six species of quarks, which are massive spin-{1}/{2} fields. The six species are labelled as flavours \textit{up}, \textit{down}, \textit{strange}, \textit{charm}, \textit{bottom} and \textit{top}. They are identical copies which differ only in their masses and their electroweak couplings\footnote{Whereas flavour eigenstates of the strong interactions have definite mass and electric charge, weak interactions couple to quarks in a different basis. This is the origin of the CKM matrix~\cite{Cabibbo:1963yz,Kobayashi:1973fv}.}. \newline

\noindent
QCD can be studied successfully in the framework of perturbation theory only in the high energy regime. The running coupling is small at high four-momentum transfer, since the colour anti-screening effect of the self-interacting gluon field is lessened at short distances, and quarks and gluons are asymptotically free. In stark contrast, quarks and gluons or coloured composita thereof cannot exist as almost free particles at lower energies due to a linearly rising colour potential, which is the root of confinenment. Potentials in general and the quark potential in particular are non-perturbative effects. Any perturbation series with a finite number of terms inevitably fails to bring forth such a potential.
\noindent
Since gluons are fields in the adjoint representation and quarks are fields in the fundamental representation of the colour gauge group SU(3), only composita which belong to the trivial representation can exist as stable \textit{hadrons} at the hadronic scale. Mesons with one valence quark-antiquark pair ($ |q\bar q\rangle $) and baryons with three totally colour-antisymmetric valence quarks ($ |qqq\rangle $) are the only hadrons for which experimental evidence exists. Even though they are allowed by the gauge symmetry of QCD, glueballs without any valence quarks (e.g. $ |gg\rangle $ or $ |ggg\rangle $) as well as exotic meson-like states (e.g. $ |q\bar q g\rangle $ or $ |q\bar qq\bar q \rangle $) and pentaquarks ($ |qqqq\bar q\rangle $) still lack experimental evidence\footnote{The charmonium spectrum contains X, Y and Z~states~\cite{Choi:2003ue,Aubert:2005rm} that cannot be explained as $ |q\bar q\rangle $ states, but their detailed nature is still unclear. They are candidates for $ |q\bar q g\rangle $- or $ |q\bar qq\bar q \rangle $-states.}.\newline

\enlargethispage*{\baselineskip}
\begin{figure}[htb]
 \begin{picture}(360,120)
  \put(090.0, 0.0){\includegraphics[bb=0 0 180 120, scale=0.20]{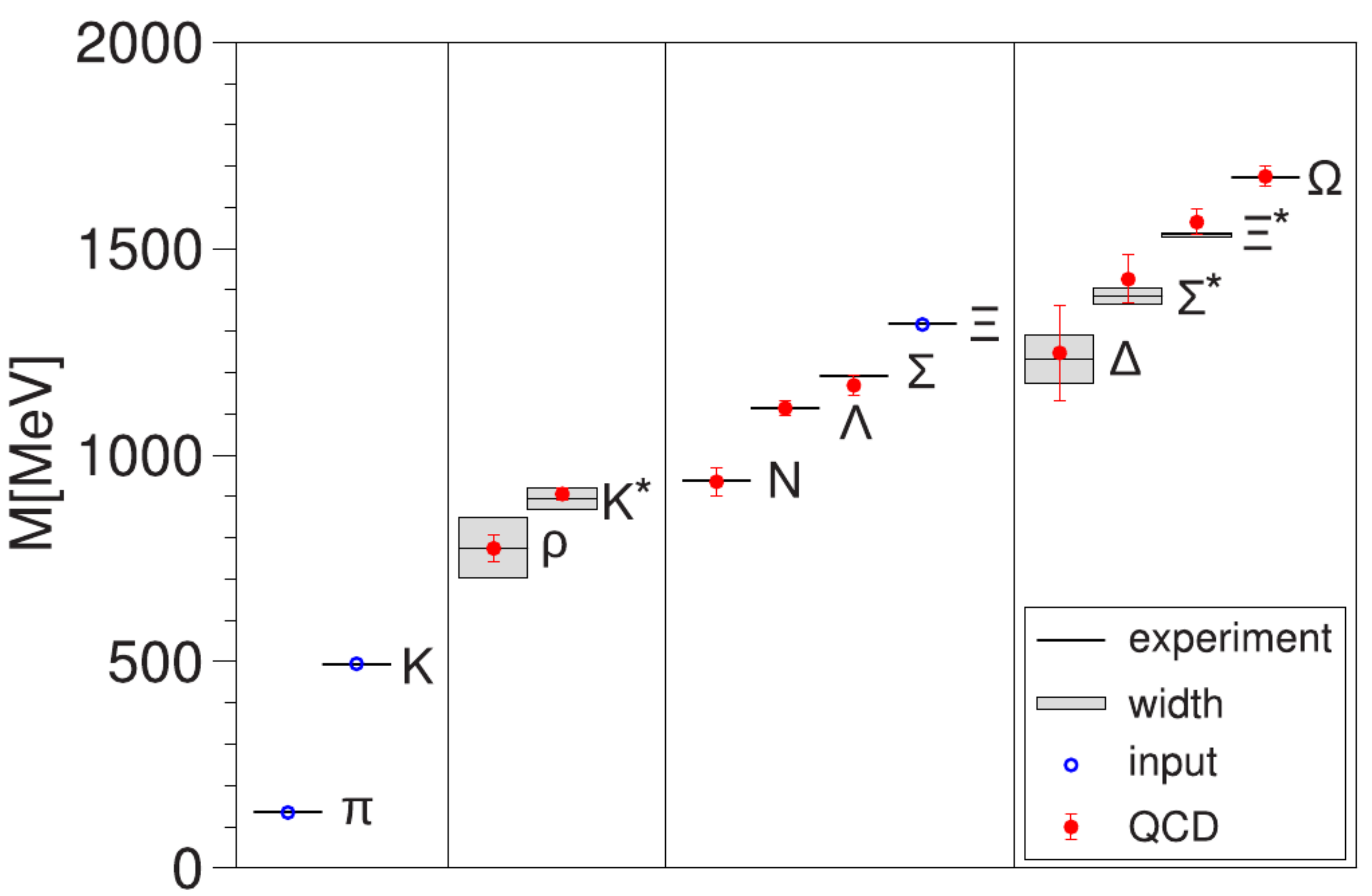}}
 \end{picture}
 \caption{
 Ab initio calculations with lattice QCD reproduce the experimental light hadron spectrum within errors. The figure is taken from~\protect\cite{Durr:2008zz}. 
  }
 \label{fig: BMW spectrum}
\end{figure}

\noindent
Any approach aimed at understanding the spectrum of QCD with ab initio calculations must necessarily involve non-perturbative methods. Observables are calculated without the need for a perturbation series with the path integral (\mbox{cf.} section \ref{sec: Quantisation of QCD with the path integral approach}), which can be evaluated numerically in computer simulations. An estimate of the path integral based on a representative subset of configuration space with controlled statistical errors is obtained using importance sampling, where configurations are weighted with the classical action on a discretised Euclidean space-time. The lattice spacing serves as a non-perturbative cutoff, which removes ultraviolet divergences. This is the strategy of lattice QCD, which has attained remarkable success in the description of the strong interactions that is evident in figure \ref{fig: BMW spectrum}. 
\noindent
Nevertheless, a shortcoming which is intrinsically connected to the lattice regularisation is the intricacy of the implementation of chiral symmetry~\cite{Nielsen:1981hk,Niedermayer:1998bi}. Wilson fermions~\cite{Wilson:1974sk}, which had been used in the calculations of~\cite{Durr:2008zz}, explicitly break chiral symmetry at finite cutoff. Though Ginsparg-Wilson fermions~\cite{Ginsparg:1981bj} retain a chiral symmetry for one single quark species at finite lattice spacing, their Dirac operators lack ultralocality and turn out as numerically costly. Staggered fermions~\cite{Susskind:1976jm}, which retain an ultralocal, non-singlet chiral symmetry, include four mass-degenerate species of quarks. Simulations with dynamical staggered fermions make use of a rooting procedure, which reduces the number of mass-degenerate sea quark species to one or two. Though the root of a determinant is mathematically well-defined, rooting a quantum field theory with non-singlet properties may lead to pathologies and fail to reproduce QCD in the continuum limit. This rooting procedure is subject of ongoing and controversial discussions (\mbox{cf.}~\cite{Creutz:2008nk}). 
\noindent
Minimally doubled fermions~\cite{Karsten:1981gd} are a type of lattice fermions with two mass-degenerate species and an ultralocal, non-singlet chiral symmetry. Since the Dirac operator is ultracolal, numerical application of these lattice fermions is supposedly similarly simple and fast as Wilson fermions. If the species are interpreted as an isospin doublet of light quarks, such actions may provide a compromise between the theoretical cleanliness of Ginsparg-Wilson fermions and the numerical efficiency of Wilson fermions. \newline

\enlargethispage*{\baselineskip}
\noindent
The study of \textit{properties of minimally doubled fermions} is the subject of this thesis. Among the different kinds of minimally doubled fermions know at present, Karsten-Wilczek~\cite{Karsten:1981gd, Wilczek:1987kw} and Bori\c{c}i-Creutz fermions~\cite{Creutz:2008sr, Borici:2008ym} are studied within this thesis using various different approaches. Since these actions explicitly break the hypercubic lattice symmetry at finite lattice spacing, there is a risk that these theories fail to reproduce an isotropic continuum theory due to operator mixing if these anisotropic effects are not properly accounted for by renormalisation. Therefore, the main focus of this thesis lies on renormalisation of minimally doubled fermion actions. \newline

Chapter \ref{sec: Foundations} outlines foundations of continuum and lattice QCD. Wilson and Ginsparg-Wilson fermions are juxtaposed with minimally doubled fermions. Results from other research groups concerning minimally doubled fermions are touched. 
\noindent
Chapter \ref{sec: Perturbative studies} covers analytical investigations of minimally doubled fermions using lattice perturbation theory~\cite{Capitani:2009yn, Capitani:2009ty,Capitani:2010nn, Capitani:2010ht}. The fermionic self-energy and the fermionic contribution to the vacuum polarisation are calculated and their anisotropies are removed using the counterterms of the theory. Renormalisation factors of fermionic bilinears are calculated and the symmetry currents of the theory are derived. 
\noindent
Chapter \ref{sec: Structure and symmetry} contains analytical studies of the structure of minimally doubled fermions in the na\"{i}ve continuum limit. A formal decomposition of the spinors into a pair of fields suggests the existence of oscillating contributions  in some correlation functions. Later on, it is concluded from $ CP\Theta $~symmetry of Karsten-Wilczek fermions in the quenched approximation that correlation functions with the same charge conjugation eigenstates at source and sink are invariant under time reflection. 
\noindent
Chapter \ref{sec: Numerical studies} encompasses numerical studies of Karsten-Wilczek fermions in the quenched approximation~\cite{Weber:2013tfa}. Foreseeable difficulties for numerical simulations are discussed and an overview of the setup of simulations is given. Methods for dealing with a lack of symmetry under time reflection are discussed. No signature for this broken reflection symmetry is observed in data. Next, two independent approaches to non-perturbative tuning of the relevant counterterm's coefficient are discussed in detail. The first approach, which minimises the anisotropy that is observed in the mass of the pseudoscalar ground state, is scrutinised with regard to its various systematical uncertainties, which are dominated by finite size effects and the chiral extrapolation. The second approach restores the frequency spectrum of an oscillating ratio of correlation functions to its tree-level form. These oscillations are related to fermion doubling in a comparison with na\"{i}ve and Wilson fermions and systematical uncertainties of the approach are examined. The numerical part is concluded with an investigation of the scaling behaviour of the ground states of two different channels using the non-perturbatively tuned theory. The ground state of the $ \gamma^5 $~channel scales like a (Pseudo-) Goldstone boson including quenched chiral logarithms at finite cutoff. Since the ground state of the $ \gamma^0 $~channel scales like a (Pseudo-) Goldstone boson up to $ \mathcal{O}(a^2) $-corrections, it is identified as a pseudoscalar that is massive in the chiral limit due to lattice artefacts. The study with pseudoscalar masses below $ 300\,\mathrm{MeV} $ shows that Karsten-Wilczek fermions are not affected by exceptional configurations. Finally, chapter \ref{sec: Conclusions} summarises the results and provides an outlook to future applications of minimally doubled fermions.

\chapter{Foundations}\label{sec: Foundations}

\sectionc{QCD in the continuum}{sec: QCD in the continuum}

\noindent
The topic of this section is a brief review of QCD as a quantised field theory of the strong interactions. The first focus lies on the fields of QCD and the structure of its action. The symmetries of QCD are discussed in the absence of electroweak interactions. The second focus lies on the quantisation of the QCD action within the path integral formalism as it had been conceived by Feynman~\cite{RevModPhys.20.367} and on renormalisation. In general, this part follows mainly~\cite{Peskin:1995ev}. The treatment of chiral symmetry follows~\cite{Scherer:2002tk}. Notational conventions are guided by~\cite{Gattringer:2010zz}.

\subsection{The QCD action and its symmetries}

\subsubsection{The fields and the action of QCD}

QCD is a quantised SU(3)~gauge theory which includes massive matter fields called quarks and massless gauge fields called gluons. The quarks are described by Dirac four-spinors in the fundamental representation of the gauge group,
\begin{equation}
  \psi_{\alpha}^{a,(f)}(x),\quad \bar\psi_{\alpha}^{a,(f)}(x),
  \label{eq: quark fields}
\end{equation}\normalsize
which carry spinor indices $ \alpha $, colour indices $ a $ and flavour indices $ (f) $ as defined in the appendix (\ref{app: Indices}). The spinor field $ \psi^{(f)}(x) $ has twelve independent components at each point $ x $, which are represented by anti-commuting Grassmann numbers due to their compliance with Fermi statistics. The $ \psi^{(f)}(x) $ can be interpreted as a component of a six-dimensional vector $ \psi(x) $ in flavour space. The adjoint spinor field is defined by $ \bar \psi = \psi^\dagger \gamma^0 $, where $ \gamma^0 $ is the Dirac matrix associated with time (\mbox{cf.} appendix \ref{app: Minkowski and Euclidean space-time}). In the path integral approach to quantisation, $ \bar\psi $ and $ \psi $ are treated as independent degrees of freedom.

\noindent
The gluons are described by four-vector fields in the gauge group's adjoint representation,
\begin{equation}
  A_\mu^{ab}(x) \equiv A_\mu^{c}(x) T_{c}^{ab},
  \label{eq: gluon fields}
\end{equation}\normalsize
which carry space-time indices $ \mu $ and colour indices $ a,b $ as defined in the appendix (\ref{app: Indices}).  The gluon fields $ A_\mu^{ab}(x) $, which are $ 3\times3 $-matrices can be expressed through eight hermitian fields $ A_\mu^{c}(x) $ using the eight generators $ T_c $ of the gauge group SU(3) as in  \mbox{eq.}~(\ref{eq: gluon fields}). In total, the gluon field $ A $ has 32 components at each space-time point $ x $. Since it is a massless spin-1 vector field, only sixteen components are independent. \newline

\noindent
The action of QCD is a functional of the fields $ \psi $, $ \bar\psi $ and $ A $. The action functional has a piece which depends only on the gluon field, which is called gluon action  $ S^{g}[A] $, pure gauge action or Yang-Mills action, which reads
\small
\begin{equation}
  S^\mathrm{g}[A] = \int\limits_{\mathbb{M}^4} \frac{1}{4 g_0^2} F^{\mu\nu,ab}(x) F_{\nu\mu}^{ba}(x) \ d^4x,
\label{eq: gluon action}
\end{equation}\normalsize
and a piece which involves quark and gauge fields, which is called quark action or fermion action $ S^{f}[\psi,\bar\psi,A] $,
\small
\begin{equation}
  S^\mathrm{f}[\psi, \bar\psi, A] = \sum\limits_{(f)} \int\limits_{\mathbb{M}^4} \bar\psi_{\alpha}^{a,(f)}(x) \left( i\gamma^\mu_{\alpha\beta} D^{ab}_{\mu} - m_0^{(f)}\delta^{ab}\delta{\alpha\beta} \right) \psi_{\beta}^{b,(f)}(x) \ d^4x.
\label{eq: quark action}
\end{equation}\normalsize
Among the seven parameters of the QCD action, there is one bare coupling constant $ g_0 $ and six bare quark masses $ m_0^{(f)} $.  
The coupling $ g_0 $ of QCD cannot be defined for on-shell quantities like the coupling $ e $ of QED, since QCD is a confining theory and its charge carriers, quarks and gluons, cannot be observed in an asymptotic state under on-shell conditions. This is why the coupling of QCD must be defined with methods of the renormalisation group (\mbox{cf.} section \ref{sec: Quantisation of QCD with the path integral approach}).
\noindent
The six masses fall apart into a set of light masses and another set of heavy masses (\mbox{cf.} table \ref{tab: quark masses}). QCD has a natural scale, $ \Lambda^{QCD} $, which is a function of its seven parameters. Typical four-momenta of states of the QCD spectrum are of the order of $ \Lambda^{QCD} $, which is about $ 200\,\mathrm{MeV} $. The strange quark mass $ m_s $ is considered as light in some applications and as heavy in others, since it compares with $ \Lambda^{QCD} $ within a factor of two. The up and down quark masses, $ m_u $ and $ m_d $, however, are always considered as light.
\begin{table}[hbt]
\center\footnotesize
 \begin{tabular}{|c|c|c|c|}
  \hline
  \multicolumn{4}{|c|}{Light quarks} \\
  \hline
  $ f $ & $ m^{(f)} $ (exp.) $ [\mathrm{MeV}] $ & $ m^{(f)} $ (th.) $ [\mathrm{MeV}] $ & e.m. charge $ [e] $ \\
  \hline
  $ u $ & $ 2.3^{+0.7}_{-0.5} $ & $ 2.16\pm0.11 $ & $ +2/3 $\\
  \hline
  $ d $ & $ 4.8^{+0.5}_{-0.3} $ & $ 4.68\pm0.14\pm0.07 $ & $ -1/3 $ \\
  \hline
  $ s $ & $ 95\pm5 $ & $ 93.8\pm1.5\pm1.9 $ & $ -1/3 $ \\
  \hline
  \hline
  \multicolumn{4}{|c|}{Heavy quarks} \\
  \hline
  $ f $ & \multicolumn{2}{|c|}{$ m^{(f)} $ (exp.) $ [\mathrm{GeV}] $} & e.m. charge $ [e] $  \\
  \hline
  $ c $ & \multicolumn{2}{|c|}{$ 1.275\pm0.025 $} & $ +2/3 $ \\
  \hline
  $ b $ & \multicolumn{2}{|c|}{$ 4.18\pm0.03 $} & $ -1/3 $ \\
  \hline
  $ t $ & \multicolumn{2}{|c|}{$ 173.07\pm0.52\pm0.72 $} & $ +2/3 $ \\
  \hline
 \end{tabular}
 \caption{Quarks are separated by $ \Lambda^{QCD} $ into light flavours ($ m^{(f)} \lesssim \Lambda^{QCD} $) and heavy flavours ($ m^{(f)} \gtrsim \Lambda^{QCD} $). Quark masses are taken from ~\cite{PDG:2012} and ~\cite{Aoki:2013ldr}.}
 \label{tab: quark masses}
\end{table}
Since the quark action is explicitly written as a sum of different flavours, the following discussion is exactly equal for all flavours. A concise notation for one flavour reads
\begin{equation}
  S^\mathrm{QCD}[\psi, \bar\psi, A] = \int\limits_{\mathbb{M}^4} \bar\psi(x) \left( i\gamma^\mu D_{\mu} - m_0 \right) \psi(x) + \frac{1}{2g_0^2} \mathrm{Tr} \left(F^{\mu\nu}(x) F_{\mu\nu}(x)\right) \ d^4x
\label{eq: concise QCD action}
\end{equation}\normalsize
and has to be interpreted in the sense of \mbox{eqs.}~(\ref{eq: gluon action}) and (\ref{eq: quark action}). The kinetic term of the quark action contains Dirac matrices $ \gamma^\mu $ (\mbox{cf.} appendix \ref{app: Dirac matrices}) and the covariant derivative,
\begin{equation}
  D^{ab}_{\mu} \psi_{\alpha}^{b}(x) \equiv
  \left(\delta^{ab}\partial_\mu + i A_\mu^{ab}(x)\right)\psi_{\alpha}^{b}(x),
\label{eq: covariant derivative}
\end{equation}\normalsize
which couples quark and gluon fields. The covariant derivative implicitly depends on the space-time point $ x $ through the gauge field $ A_\mu(x) $. The commutator of two covariant derivatives at the same space-time point $ x $ gives rise to the field strength tensor $ F_{\mu\nu}(x) $,
\begin{equation}
  F^{ab}_{\mu\nu}(x) \equiv -i\left[D^{ac}_{\mu},D^{cb}_{\nu} \right] \equiv \partial_\mu A_\nu^{ab}(x)- \partial_\nu A_\mu^{ab}(x) +i \left[ A_\mu^{ac}(x),A_\nu^{cb}(x) \right].
\label{eq: field strength tensor}
\end{equation}\normalsize
The field strength tensor $ F_{\mu\nu} $, which is a $ 3\times3 $-matrix just like the gauge field $ A_\mu $ can be rewritten in a vector notation with hermitian fields $ F^{c}_{\mu\nu} $ multiplied by the generators of the gauge group SU(3).

\subsubsection{Gauge symmetry in QCD}

The QCD action is invariant under local gauge transformations $ \Omega(x) = e^{i\omega^c(x)T_c} $, which are fully parameterised by eight real parameters $ \omega^c(x) $, since the gauge group SU(3) is an eight-dimensional compact Lie group. Gauge transformations are simultaneously applied to all fields which lie in non-trivial representations of SU(3),
\begin{equation}
  \left.\begin{array}{rcl}
  \psi_{\alpha}^{a}(x) &\stackrel{\Omega(x)}{\to}& \Omega^{ab}(x) \psi_{\alpha}^{b}(x), \\
  \bar\psi_{\alpha}^{a}(x) &\stackrel{\Omega(x)}{\to}& \bar\psi_{\alpha}^{b}(x)(\Omega^{ab})^\dagger(x), \\
  F^{ab}_{\mu\nu}(x) &\stackrel{\Omega(x)}{\to}& \Omega^{ac}(x) F^{cd}_{\mu\nu}(x) (\Omega^{bd})^{\dagger}(x).
  \end{array}\right.
  \label{eq: gauge transformation}
\end{equation}\normalsize
The invariance of the quark action in \mbox{eq.}~(\ref{eq: quark action}) is achieved by the covariant derivative of the fermion field, which transforms like the fermion field itself in \mbox{eq.}~(\ref{eq: covariant derivative}). This is due to the nontrivial behaviour of the gauge field $ A_\mu(x) $ under gauge transformations:
\begin{align}
  D^{ab}_{\mu} \psi_{\alpha}^{b}(x) \stackrel{\Omega(x)}{\to}&\ 
  \Omega^{ab}(x) D^{bc}_{\mu}\psi_{\alpha}^{c}(x), \\
  A_\mu^{ab}(x) \stackrel{\Omega(x)}{\to}&\ 
  \Omega^{ac}(x) A_\mu^{cd}(x)(\Omega^{bd})^\dagger(x)  +i\left(\partial_\mu \Omega^{ac}(x)\right)(\Omega^{bc})^\dagger(x)  
  = A_\mu^{ab}(x) +D^{ab}_{\mu}\omega^{bc}(x).
  \label{eq: gauge transformations with covariant derivatives}
\end{align}\normalsize
The covariant derivative is related to the concept of parallel transport in the framework of differential geometry. Space-time $ \mathbb{M}^4 $ and the gauge group SU(3) form a vector bundle in the sense that each space-time point $ x $ is vested with its own copy of SU(3). Therefore, fields which belong to non-trivial representations of the gauge group at different points transform differently under gauge transformations. A gauge transporter $ U(x,x+a\hat{e}_\mu) $ is introduced, which compensates for different copies of SU(3) at base points $ x $ and $ x+a\hat{e}_\mu $. A sensible derivative of a field $ \psi(x) $ is thus defined in the limit of vanishing distance $ a $,
\small
\begin{align}
  D_\mu^{bc}\psi^c(x) \stackrel{!}{=}&\
  \lim_{a\to0} \frac{1}{a}\left\{U^{bc}(x,x+a\hat{e}_\mu)\psi^c(x+a\hat{e}_\mu)-\psi^b(x)\right\} \nonumber \\ 
  =&\ \lim_{a\to0} U^{bc}(x,x+a\hat{e}_\mu)\frac{1}{a}\Big\{\psi^c(x+a\hat{e}_\mu)-\psi^c(x)\Big\} 
  +\lim_{a\to0} \frac{1}{a}\Big\{U^{bc}(x,x+a\hat{e}_\mu)-\delta^{bc}\Big\}\psi^c(x) \nonumber \\
  =&\ \left(\delta^{bc}\partial_\mu +[\partial_\mu U(x^\prime,x)]^{bc}_{x^\prime=x}\right)\psi^c(x) ,
\end{align}\normalsize
where $ U^{bc}(x,x)=\delta^{bc} $ was used. A comparison with \mbox{eq.}~(\ref{eq: covariant derivative}) reveals
\begin{equation}
  [\partial_\mu U(x^\prime,x)]^{bc}_{x^\prime=x} = iA_\mu^{bc}(x),\quad U(x,x+a\hat{e}_\mu)=1+ia\hat{e}_\mu A_\mu^{c}(x)T_c +\mathcal{O}(a^2) \equiv U_\mu(x).
  \label{eq: infinitesimal gauge transporter}
\end{equation}\normalsize
The gauge field $ A_\mu $ is the connection of the covariant derivative. The gauge transporter is generalised to arbitrary differentiable paths $ \mathcal{C} $ and the parallel transporter reads
\begin{equation}
  U(x,y)=\exp{\left(i \int\limits_x^y T_c \sum\limits_{\mu=0}^3 A_\mu^c ds_\mu  \right)},
\end{equation}\normalsize
which is a Wilson line between the points $ x $ and $ y $.

\subsubsection{Discrete symmetries of QCD}

Besides its gauge symmetry, the QCD action retains its structure under three individual discrete symmetry transformations, which are charge conjugation (often referred to as $ C $~symmetry), parity transformation ($ P $) and time reversal (often referred to as $ \Theta $~symmetry). Parity and time reversal are global symmetries, since the transformations connect fields with different space-time arguments. Parity is defined with a unitary operator acting on fields and space-time points,
\begin{equation}
  \left.\begin{array}{rcl}
  (x_0,\mathbf{x}) &\stackrel{P}{\to}& (x_0,-\mathbf{x}), \\
  \psi(x_0,\mathbf{x}) &\stackrel{P}{\to}& \gamma_0 \, \psi(x_0,-\mathbf{x}), \\
  \bar\psi(x_0,\mathbf{x}) &\stackrel{P}{\to}& \bar\psi(x_0,-\mathbf{x}) \, \gamma_0, \\
  \left(A_0(x_0,\mathbf{x}),\mathbf{A}(x_0,\mathbf{x})\right) &\stackrel{P}{\to}& \left(A_0(x_0,-\mathbf{x}),-\mathbf{A}(x_0,\mathbf{x})\right),
  \end{array}\right.
  \label{eq: parity transformation}
\end{equation} \normalsize
whereas time reversal must be defined with an anti-unitary operator, which enacts complex conjugation on all c-numbers and operators alongside its operation on fields and space-time points,
\begin{equation}
  \left.\begin{array}{rcl}
  (x_0,\mathbf{x}) &\stackrel{\Theta}{\to}& (-x_0,\mathbf{x}), \\
  \psi(x_0,\mathbf{x}) &\stackrel{\Theta}{\to}& \gamma_1\gamma_3 \, \psi(-x_0,\mathbf{x}) , \\
  \bar\psi(x_0,\mathbf{x}) &\stackrel{\Theta}{\to}& -\bar\psi(x_0,-\mathbf{x}) \,  \gamma_1\gamma_3 , \\
  \left(A_0(x_0,\mathbf{x}),\mathbf{A}(x_0,\mathbf{x})\right) &\stackrel{\Theta}{\to}& -\left(-A_0(-x_0,\mathbf{x}),\mathbf{A}(-x_0,\mathbf{x})\right).
  \end{array}\right.
  \label{eq: time reflection for Minkowski space-time}
\end{equation} \normalsize
Lastly, charge conjugation replaces particles by antiparticles and vice versa. Since space-time points are unaltered, charge conjugation can be locally defined with the charge conjugation matrix $ C $ of \mbox{eq.}~(\ref{eq: charge conjugation matrix}),
\begin{equation}
  \left.\begin{array}{rcl}
  \psi(x_0,\mathbf{x}) &\stackrel{C}{\to}& C \, \bar\psi^T(x_0,\mathbf{x}) , \\
  \bar\psi(x_0,\mathbf{x}) &\stackrel{C}{\to}& -\psi^T(x_0,\mathbf{x}) \, C , \\
  \left(A_0(x_0,\mathbf{x}),\mathbf{A}(x_0,\mathbf{x})\right) &\stackrel{C}{\to}& -\left(A_0(x_0,\mathbf{x}),\mathbf{A}(x_0,\mathbf{x})\right).
  \end{array}\right.
  \label{eq: charge conjugation}
\end{equation} \normalsize
The product $ CP\Theta $ leaves the structure of any field theory invariant, since the action is scalar and must necessarily be invariant under space-time transformations.

\subsubsection{Approximate chiral symmetry in QCD}

\noindent
Nevertheless, the QCD action has additional symmetries, some of which are only approximately realised. These approximate symmetries can be identified, when the quarks are sufficiently lighter and external four-momenta are sufficiently smaller than $ \Lambda^{QCD} $. On the one hand, small external four-momenta in QCD processes as well as the light quark masses can be treated as a perturbation to a theory with vanishing external four-momenta and quark masses. On the other hand, heavy quark states are strongly suppressed as they are produced only far off shell in processes with low external four-momenta and hence do not contribute to asymptotic states.
\noindent
As a result, QCD processes with small external four-momenta can be described by an effective theory of interactions between hadronic states consisting of the light flavours only, whereas the heavy degrees of freedom are integrated out. The a priori unknown effective coupling constants and the symmetry of interactions of such an effective theory are determined by QCD.\newline

\noindent
Right-handed and left-handed components of the $ N_f=3 $ light quark fields $ \psi^{(f)} $ are defined using the chirality projectors of \mbox{eq.}~(\ref{eq: chirality projectors}),
\begin{equation}
  \psi_{R,L}^{(f)}(x)=P_{R,L}\psi^{(f)}(x),  \quad \bar\psi_{R,L}^{(f)}(x)=\bar\psi^{(f)}(x)P_{L,R}.
\end{equation}\normalsize
In the chiral limit, where light quark masses are set to zero, left- and right-handed quarks decouple in the chiral quark action,
\begin{equation}
  S^{\mathrm{ch.}}[\psi, \bar\psi, A] = \sum\limits_{f=1}^{N_f}\ \int\limits_{\mathbb{R}^4} \bar\psi_R^{(f)}(x) \left( i\gamma^\mu D_{\mu} \right) \psi_R^{(f)}(x)+\bar\psi_L^{(f)}(x) \left( i\gamma^\mu D_{\mu} \right) \psi_L^{(f)}(x) \ d^4x.
\label{eq: chiral quark action}
\end{equation}\normalsize
The action is invariant under two individual sets of global symmetry transformations for right-handed or left-handed quarks,
\begin{equation}
  \left.\begin{array}{rcl}
  \psi_{R,L}^{\prime(g)}(x)&=& \exp{\left(-i\sum\limits_{j=1}^8 \Theta_{R,L}^j T_{j}^{gf}\right)} e^{-i\Theta_{R,L}} \psi_{R,L}^{(f)}(x), \\
  \bar\psi_{R,L}^{\prime(g)}(x)&=&\bar\psi_{R,L}^{(f)}(x) \exp{\left(i\sum\limits_{j=1}^8 \Theta_{R,L}^j T_{j}^{fg}\right)} e^{i\Theta_{R,L}},
  \end{array}\right.
\end{equation}\normalsize
where the $ T_j $ are the generators of a flavour symmetry group SU(3) (\mbox{cf.} appendix \ref{app: SU(N)-Matrices}). The independent transformations of right-handed and left-handed fields can be equivalently represented by vector and axial transformations, which simultaneously transform right-handed and left-handed fields with $ \Theta_V^j=(\Theta_R^j+\Theta_L^j)/2 $, $ \Theta_A^j=(\Theta_R^j-\Theta_L^j)/2 $ and $ \Theta_V=(\Theta_R+\Theta_L)/2 $. The action of the axial transformation on the flavour group's three-vectors includes the chirality matrix $ \gamma_5 $,
\begin{equation}
  \left.\begin{array}{rcl}
  \psi^{\prime(g)}(x)&=& \exp{\left(-i\sum\limits_{j=1}^8 \Theta_A^j T_{j}^{gf}\gamma_5\right)} \psi^{(f)}(x), \\
  \bar\psi^{\prime(g)}(x)&=& \bar\psi^{(f)}(x) \exp{\left(-i\sum\limits_{j=1}^8 \Theta_A^j T_{j}^{fg}\gamma_5\right)},
  \end{array}\right.
\end{equation}\normalsize
and leaves the chiral quark action invariant. This flavoured axial symmetry $ \mathrm{SU(3)}_A $ is spontaneously broken in QCD and the axial charge operators do not annihilate the vacuum. Instead, when applied to the vacuum state, they create massless pseudoscalar bosons, which have the same quantum numbers as the axial charge operators. These are the \mbox{(Pseudo-)}~Goldstone bosons of QCD. In this idealised case, pions, kaons and the eta meson are massless and fully degenerate in all interactions.\newline

\noindent
These axial transformations cease to be symmetry transformations in the presence of light quark masses. Therefore, the spontaneously broken axial symmetry is explicitly broken by the light quark masses and the \mbox{(Pseudo-)}~Goldstone bosons acquire small masses, while their degeneracies are partially lifted. In the spectrum of QCD, they can be identified as the light pseudoscalar meson octet. In a limiting case with degenerate up and down quark masses, $ m_u=m_d=m_{ud}<m_s $, isospin is realised as a symmetry and pions are fully degenerate and considerably lighter than the kaons. In the real world, isospin symmetry is broken by the finite mass difference of up and down quarks as well as their different electroweak charges.
\noindent
The presence of light quark masses reduces the flavoured vector symmetries as well. The singlet vector transformation is preserved as a symmetry transformation for arbitrary quark masses and is connected to baryon number conservation. For degenerate light quarks, all flavoured vector symmetries are preserved and the perfect flavour symmetry of the eightfold way of Gell-Mann and Ne'eman is realised. The divergences of the axial symmetry currents are proportional to products of pseudoscalar quark bilinears and combinations of quark masses in the form which was realised in the PCAC relation.

\subsection{Quantisation of QCD with the path integral approach}\label{sec: Quantisation of QCD with the path integral approach}

\subsubsection{The path integral}
The path integral approach to quantisation as it had been suggested by Feynman~\cite{RevModPhys.20.367} expresses quantum mechanical amplitudes, such as the time evolution operator $ U(x_i,t_i;x_f,t_f) $ between states $ x_i $ and $ x_f $, as a functional integral of the observables weighted with the classical action over all possible paths between initial and final points:
\begin{equation}
  \langle U(x_i,t_i;x_f,t_f) \rangle= \int\limits_{x(t_i)=x_i}^{x(t_f)=x_f} \mathcal{D}x(t) e^{iS[x(t)]}.
\end{equation}\normalsize
Paths in the neighbourhood of the classical solution $ x_{cl}(t) $ dominate the path integral and produce small quantum corrections, whereas outlying paths are suppressed by strong cancellations.
\noindent
The path integral is generalised to quantum field theories (QFTs), where observables are functionals of elementary field variables. The path integral of QCD reads
\begin{equation}
  \langle \mathcal{O} \rangle= \frac{1}{Z} \int \mathcal{D}\bar\psi \mathcal{D}\psi \mathcal{D}A\ \mathcal{O}[\psi,\bar\psi,A]\ e^{iS^{QCD}[\psi,\bar\psi,A]},
  \label{eq: QCD path integral, Minkowski space-time}
\end{equation}\normalsize
where $ Z = \langle 1 \rangle $ is a constant and $ S^{QCD}[\psi,\bar\psi,A] $ is defined in \mbox{eq.}~(\ref{eq: concise QCD action}). However, there are two obstacles which must be overcome before the path integral of QCD can be evaluated.\newline

\noindent
Firstly, due to gauge invariance, the integration overcounts gauge field configurations, which are connected by gauge transformations. The na\"{\i}ve gluon propagator,
\begin{equation}
  \left(\partial^2 g_{\mu\nu}-\partial_\mu\partial_\nu\right)D_g^{\nu\rho,\,ab}(x-y)\stackrel{!}{=}i\delta_\mu^\rho\delta^{(4)}(x-y)\delta^{ab},
\end{equation}\normalsize
has no solution for gauge fields, which are connected to vanishing $ A $ by the gauge transformations of \mbox{eq.}~(\ref{eq: gauge transformations with covariant derivatives}) and thus consist only of a derivative of a scalar function.
These physically equivalent configurations must be culled by an ingenious gauge-fixing procedure, which has been developed by Fadeev and Popov ~\cite{Faddeev:1967fc}. An arbitrary gauge-fixing parameter $ \xi $ is introduced, which modifies the gluon propagator,
\begin{equation}
  \left(\partial^2 g_{\mu\nu}-\left(1-\frac{1}{\xi}\right)\partial_\mu\partial_\nu\right)D_g^{\nu\rho,\,ab}(x-y)\stackrel{!}{=}i\delta_\mu^\rho\delta^{(4)}(x-y)\delta^{ab},
  \label{eq: gauge-fixed gluon propagator}
\end{equation}\normalsize
and unphysical degrees of freedom, which are called Faddeev-Popov ghosts, are coupled to the gluon field and cancel the contributions from unphysical polarisations of gauge bosons. Since physical observables are naturally gauge invariant, any effects due to particular choices of the gauge-fixing parameter $ \xi $ or contributions of unphysical gauge field or ghost degrees of freedom cancel completely in any observables.
\noindent
Secondly, non-perturbative evaluation of the path integral of QCD must be restricted to an evaluation of its kernel on a relatively small number of representative gauge field configurations due to technical reasons. Following the idea of importance sampling, a small but representative ensemble of configurations which yields a good estimate of the full path integral is required. Hence, the majority of configuration must belong to the neighbourhood of the classical path. However, an explicit bias which generally rejects outlying configurations is not acceptable. Gauge field configurations can be produced with Monte-Carlo methods and their vicinity to the classical path is determined from the numerical value of the action. Yet, because a complex exponential $ e^{iS^{QCD}} $ is not a valid weight factor that would have to be real, importance sampling cannot be applied straightforwardly.\newline

\noindent
Both of these obstacles are lifted simultaneously by the approach of Lattice QCD, which uses the analytic continuation of the QCD action of \mbox{eq.}~(\ref{eq: concise QCD action}) on a discretised Euclidean space-time (\mbox{cf.} appendix \ref{app: Minkowski and Euclidean space-time}). The discretised space-time requires another representation of the gauge fields in terms of the gauge transporters of \mbox{eq.}~(\ref{eq: infinitesimal gauge transporter}), which are integrated with the gauge-invariant Haar measure. Thus, the Haar measure dispenses with gauge-fixing. The exponential factor changes into $ e^{-S_E^{QCD}} $, which is a real, suitable weight factor for importance sampling. Lattice QCD is addressed in section \ref{sec: Lattice QCD}.

\subsubsection{Renormalisation}
The seven parameters of the QCD action, the bare coupling $ g_0 $ and the six bare masses $ m_0^{(f)} $, as well as the field variables $ \psi $, $ \bar\psi $ and $ A $ are subject to renormalisation in the quantised theory. The quantities in the action are defined in the absence of interactions, whereas the physical fields naturally include any interaction effects in full. Quantum corrections, which are diagrammatically represented by loops in a perturbative approach, lead to finite and divergent contributions beyond the bare quantities. These corrections are absorbed into coefficients of counterterms at any given renormalisation scale $ M $, where the physical values of the parameters of the theory are known.
\noindent
In a so-called renormalisable quantum field theory, a finite number of counterterms is sufficient for the complete absorption of all quantum corrections into their coefficients. The same coefficients are repeatedly modified at each order in a perturbative expansion. In an effective field theory, absorption requires an infinite number of counterterms. However, if the quantum corrections are computed in a well-defined perturbative expansion, the number of counterterms which are required for absorbing divergences at each order of the perturbation theory is still finite and divergences can be removed order by order.\newline

\noindent
Due to the requirement of renormalisation, the coupling constant $ g $ of the interactions which are the origin of the quantum corrections receives quantum corrections itself. These quantum corrections to the coupling constant $ g $ are determined by the Callan-Symanzik $ \beta $ function of QCD,
\begin{equation}
  \beta(g) \equiv M\frac{\partial g}{\partial M},
  \label{eq: beta function}
\end{equation}\normalsize
which describes the dependence of the coupling constant $ g $ on changes of the renormalisation scale $ M $. In that sense the coupling constant is actually a running coupling parameter. In the case of QED, a renormalisation scale $ M $ is chosen as the physical scale of on shell electrons, which can propagate as asymptotic states. Therefore, the classical electron mass and electric charge can be used in the renormalisation procedure.
\newline

\noindent
On the contrary, due to the colour confinement property of QCD, neither the quarks in the fundamental representation nor the gluons in the adjoint representation of the gauge group can propagate as asymptotic states into a detector. However, they can behave as almost free particles when they carry very high four-momenta. The beta function of QCD (in the chiral limit) reads
\begin{align}
  \beta(g)=&-\frac{g^3}{(4\pi)^2}\left(\beta_0+\frac{g^2}{(4\pi)^2}\beta_1\right) + \mathcal{O}(g^7), 
  \label{eq: QCD beta function}\\
  \beta_0=&\left(\frac{11}{3}C(G)-\frac{4}{3}N_f C(R)\right), \\
  \beta_1=&\left(\frac{34}{3}C^2(G)-N_f C(R)\left(\frac{20}{3} C(G)+4C_2(R)\right)\right),
\end{align}\normalsize
where the Casimir operators of the adjoint and fundamental representations, $ C(G) $, $ C(R) $ and $ C_2(R) $, are defined in appendix~\ref{app: SU(N)-Matrices}. $ N_f $ is the number of massless quark flavours in the theory. If $ N_f<(11/4)\, C(G)/C(R)=16.5 $, the coefficient $ \beta_0 $ is positive and the $ \beta $-function has negative curvature in $ g $. Therefore, there is a high four-momentum scale, where the coupling strength vanishes and quarks and gluons behave as free particles. This property is aymptotic freedom of QCD, which was discovered by Gross, Politzer and Wilczek~\cite{Gross:1973ju, Gross:1973id, Gross:1974cs, Politzer:1974fr}. Asymptotic freedom is the reason why the coupling constant of QCD can be measured at the electroweak scale. The world average of measurements of the coupling is given in the particle data book~\cite{PDG:2012} as
\begin{equation}
  \alpha_s(M_Z^2)\equiv \frac{g^2(M_Z^2)}{4\pi} = 0.1184\pm0.0007,
\end{equation}\normalsize
which must be evolved with renormalisation group methods to the scale of interest.

\sectionc{Lattice QCD}{sec: Lattice QCD}

\noindent
This part follows mostly~\cite{Gattringer:2010zz}. It includes a brief introduction to Lattice QCD (LQCD) as a discretisation of an analytic continuation of the continuum action of \mbox{eq.}~(\ref{eq: concise QCD action}) to Euclidean space-time as it had been conceived for the first time by Wilson~\cite{Wilson:1974sk}. The gluon action is discretised as the Wilson plaquette action and Wilson fermions are introduced as the standard case of lattice fermions, which is compared in section~\ref{sec: Minimally doubled fermions} with minimally doubled fermion actions. A lattice realisation of chiral symmetry according to the Ginsparg-Wilson equation~\cite{Ginsparg:1981bj} employing Ginsparg-Wilson fermions with an Overlap operator~\cite{Neuberger:1997fp} is presented for contrasting juxtaposition.

\subsection{Discretisation of QCD on a Euclidean space-time}

\noindent
The analytic continuation of the QCD action from Minkowski space-time to Euclidean space-time is rather straightforward. The Osterwalder-Schrader approach~\cite{Osterwalder:1973kn, Osterwalder:1973dx} is used for the Wick rotation\footnote{The terminology is taken over from~\cite{vanNieuwenhuizen:1996tv,Waldron:1997re}.}~(\mbox{cf.} appendix~\ref{app: Minkowski and Euclidean space-time}). Wick rotation is applied to \mbox{eq.}~(\ref{eq: concise QCD action}) and produces an overall factor $ i $,
\begin{equation}
  S_M^{QCD}[\psi^M,\bar\psi^M,A^M] = i S_E^{QCD}[\psi^E,\bar\psi^E,A^E],
\end{equation}\normalsize
which multiplies the Euclidean QCD action (without Euclidean indexing),
\begin{equation}
  S^{QCD}[\psi,\bar\psi,A]  = \int\limits_{\mathbb{R}^4} \bar\psi(x) \left(\gamma^\mu D_\mu + m_0 \right) \psi(x) - \frac{1}{4} \mathrm{Tr} \left(F_{\mu\nu}(x) F_{\mu\nu}(x)\right) \ d^4x.
\end{equation}\normalsize
Euclidean space-time is discretised according to appendix~\ref{app: Continuous space-time and discrete space-time lattices} using \mbox{eqs.}~(\ref{eq: Euclidean space-time lattice}) and (\ref{eq: site vectors}). The lattice spacing $ a $, the intrinsic length scale of the discretised theory, is treated as a small quantity and its inverse serves as a cutoff removing UV divergences. The limit $ a \to 0 $ of tree-level quantities is called the \textit{na\"{i}ve continuum limit}. However, since interactions mix terms of different powers in the lattice spacing $ a $, an appropriate renormalisation prescription is required in the determination of the continuum limit.\newline

\noindent
Fermion fields are defined on the lattice sites labelled by $ n $. Partial derivatives cannot be defined on a lattice and have to be discretised as finite differences. Two na\"{i}ve versions,
\begin{align}
  \nabla_\mu  \psi_n=&\ \frac{1}{a}\left(\psi_{n+\hat{e}_\mu}-\psi_n\right) = \partial_\mu \psi(x) -\frac{a}{2}\psi(x) +\mathcal{O}(a^2), \\
  \nabla_\mu^*\psi_n=&\ \frac{1}{a}\left(\psi_n-\psi_{n-\hat{e}_\mu}\right) = \partial_\mu \psi(x) +\frac{a}{2}\psi(x) +\mathcal{O}(a^2),
\end{align}\normalsize
are combined to a symmetrised form\footnote{Only the symmetrised form has the desired hermiticity properties.} with reduced discretisation errors,
\begin{equation}
  \frac{1}{2}(\nabla_\mu+\nabla_\mu^*)\psi_n = \frac{1}{2a}\left(\psi_{n+\hat{e}_\mu}-\psi_{n-\hat{e}_\mu}\right) = \partial_\mu \psi(x) +\mathcal{O}(a^2).
\end{equation}\normalsize
As fermion fields in the kinetic term sit on neighbouring sites, they transform differently under gauge transformations. A gauge-invariant kinetic term uses gauge links $ U^\mu_n $,
\begin{equation}
  U^\mu_n \equiv U^\mu(an,a(n+\hat{e}_\mu)) = e^{ia A^\mu(a(n+\hat{e}_\mu/2))}=e^{ia A^\mu_{n+\hat{e}_\mu/2}},
  \label{eq: gauge links}
\end{equation}\normalsize
which are the Euclidean analogue of the parallel transporters of \mbox{eq.}~(\ref{eq: infinitesimal gauge transporter}). Since gauge transporters $ U^\mu_n $ instead of the connection $ A^\mu_{n+\hat{e}_\mu/2} $ are the field variables in LQCD, \mbox{eq.}~(\ref{eq: gauge links}) defines the connection in LQCD. Gauge links $ U^\mu_n $ are defined by the end sites $ n $ and $ n+\hat{e}_\mu $ which they connect. The particular definition of the site label of the connection, which is exactly in the middle between the ends of the gauge link, simplifies calculations in lattice perturbation theory in section~\ref{sec: Perturbative studies}. Thus, the symmetrically discretised covariant derivative reads
\small
\begin{equation}
  D^\mu[U] \psi_n = \frac{1}{2a}\left(U^\mu_n\psi_{n+\hat{e}_\mu}-U^{\mu\dagger}_{n-\hat{e}_\mu}\psi_{n-\hat{e}_\mu}\right) =\left(\partial^\mu+iA^\mu(x)\right) \psi(x) +\mathcal{O}(a^2) = D^\mu \psi(x) +\mathcal{O}(a^2)
  \label{eq: continuum expansion of the lattice covariant derivative}
\end{equation}\normalsize
and is conveniently expressed as a matrix with two site indices,
\begin{equation}
  D^\mu_{n,m}[U] = \frac{1}{2a}\left(U^\mu_n\delta_{n+\hat{e}_\mu,m}-U^{\mu\dagger}_{n-\hat{e}_\mu}\delta_{n-\hat{e}_\mu,m}\right).
  \label{eq: lattice covariant derivative}
\end{equation}\normalsize
The contraction of the discretised covariant derivative $ D^\mu_{n,m}[U] $ with Dirac matrices $ \gamma^\mu $,
\small
\begin{equation}
  D^N_{n,m} = \sum\limits_{\mu=0}^3 \gamma^\mu D^\mu_{n,m}[U],
  \label{eq: naive Dirac operator}
\end{equation}\normalsize
is referred to as the na\"{i}ve Dirac operator and implicitly depends on the local gauge fields $ U $. It is a remarkable feature of the na\"{i}ve Dirac operator that it connects only fields on even with fields on odd sites as defined in \mbox{eq.} (\ref{eq: even-odd site number}), which is inherited by many other discretised Dirac operators. It is employed in the na\"{i}ve fermion action with mass $ m_0 $,
\begin{equation}
  S^{N}[\psi,\bar\psi,U]  = a^4\sum\limits_{n,m\in\Lambda} \bar\psi_n \left(D^N_{n,m} + m_0\delta_{n,m}\right) \psi_m.
  \label{eq: naive fermion action}
\end{equation}\normalsize
However, the action of \mbox{eq.} (\ref{eq: naive fermion action}) is not well-suited for simulations of QCD due to the so-called doubling problem, which is covered in section \ref{sec: Doubling problem and No-Go theorem}.\newline

\noindent
Since the field strength tensor $ F_{\mu\nu}(x) $ is defined through explicit use of derivatives, discretisation of the gauge action of \mbox{eq.}~(\ref{eq: gluon action}) requires a lattice representation of the field strength tensor that uses only finite differences. Any chain of neighbouring gauge links is a discretised expression for a Wilson line. If the Wilson line returns to its origin in a so-called Wilson loop, it becomes a gauge-invariant object. The smallest possible Wilson loop in the $ \mu $-$ \nu $ plane is the so-called plaquette $ U_n^{\mu\nu} $, which can be expanded in the gauge fields $ A $ using \mbox{eq.}~(\ref{eq: gauge links}),
\begin{align}
   U_n^{\mu\nu}=&U^\mu_n U^\nu_{n+\hat{e}_\mu}U^{\mu\dagger}_{n+\hat{e}_\nu}U^{\nu\dagger}_n 
   \label{eq: plaquette} \\
   =& e^{iaA^\mu_{n+\hat{e}_\mu/2}}\ e^{iaA^\nu_{n+\hat{e}_\mu+\hat{e}_\nu/2}}\ e^{-iaA^\mu_{n+\hat{e}_\nu+\hat{e}_\mu/2}}\ e^{-iaA^\nu_{n+\hat{e}_\nu/2}} \nonumber \\
   =&e^{ia^2(\nabla^\mu A^\nu_{n+\hat{e}_\nu/2}-\nabla^\nu A^\mu_{n+\hat{e}_\mu/2}+i\left[A^\mu_{n+\hat{e}_\nu+\hat{e}_\mu/2},A^\nu_{n+\hat{e}_\nu/2}\right])}+\mathcal{O}(a^3) \nonumber \\
   =&e^{ia^2F^{\mu\nu}(x)}+\mathcal{O}(a^3),
\end{align}\normalsize
until it yields the exponentiated field strength tensor up to higher orders in the lattice spacing. Hence, the plaquette as an exponentiated multiple of a discretised version of the field strength tensor is used in the so-called plaquette action,
\begin{equation}
  S^{p}[U]  = \sum\limits_{n\in\Lambda} \sum\limits_{\mu<\nu} \frac{2}{g_0^2}\mathrm{Re}\,\mathrm{Tr}(1-U^{\mu\nu}_n),
  \label{eq: plaquette action}
\end{equation}
which was originally conceived by Wilson~\cite{Wilson:1974sk}. The inverse coupling constant is usually abbreviated as
\begin{equation}
  \beta=\frac{6}{g_0^2},
\end{equation}\normalsize
and is non-analytically related the lattice spacing $ a $. The product of lattice spacing and length of the time direction plays the role of an inverse temperature $ aN_0=1/(k_BT) $.

\subsection{Symmetries of Lattice QCD}\label{sec: symmetries of Lattice QCD}
Most symmetries of the Minkowski space-time QCD action are carried over to the Euclidean QCD action, though some symmetries are replaced by a counterpart. The symmetry under transformations with elements of the proper orthochronous Lorentz group $ \mathcal{L}_+^\uparrow $ together with the spatial rotation group SO(3) are substituted by an SO(4) symmetry, since Euclidean time is fully equivalent to any spatial direction. The na\"{i}ve discretisation procedure breaks the rotation group SO(4) down to the hypercubic group $ W_4 $. The symmetry transformations of \mbox{eqs.}~(\ref{eq: parity transformation}) and~(\ref{eq: charge conjugation}), parity $ P $ and charge conjugation $ C $, fully carry over to Euclidean space-time. Due to the full equivalence of the four directions of Euclidean space-time, the parity transformation can be generalised to reflections $ P_\mu $, which leave only the $ \hat{e}_\mu $~direction invariant,
\begin{equation}
  \left.\begin{array}{rcl}
  x_\nu        &\stackrel{P_\mu}{\to}& P_\mu x_\nu \equiv (-1)^{1-\delta_{\mu\nu}} x_\nu, \\
  \psi(x)      &\stackrel{P_\mu}{\to}& \gamma^\mu \, \psi(P_\mu x), \\
  \bar\psi(x)  &\stackrel{P_\mu}{\to}& \bar\psi(P_\mu x) \, \gamma^\mu, \\
  A_\lambda(x) &\stackrel{P_\mu}{\to}& (-1)^{1-\delta_{\mu\lambda}} A_\lambda(P_\mu x).
  \end{array}\right.
  \label{eq: generalised parity transformation}
\end{equation} \normalsize
Hermiticity of the Minkowski space-time quark action is not carried over to Euclidean space-time in the OS approach for Euclidean field theories~\cite{Osterwalder:1973kn, Osterwalder:1973dx}. Nevertheless, $ \gamma^5 $~hermiticity of the Dirac operator,
\begin{equation}
  \gamma^5 D^\dagger \gamma^5 = D,
  \label{eq: gamma^5 hermiticity}
\end{equation}\normalsize
is a property of Euclidean Dirac operators that are analytic continuations of hermitian Dirac operators on Minkowski space-time. The product $ Q=\gamma^5 D $ is a hermitian Dirac operator if $ D $ is $ \gamma^5 $~hermitian.
\newline

\noindent
Time reflection $ \Theta $ of the Euclidean space-time QCD action is realised with three successive generalised parity transformations with cyclical spatial indices, e.g. $ P_1 P_2 P_3 $. The reflection is generalised to reflection of the single $ \hat{e}_\mu $~direction in Euclidean space-time,
\begin{equation}
  \left.\begin{array}{rcl}
  x_\nu       &\stackrel{\Theta_\mu}{\to}& \Theta_\mu x_\nu \equiv (-1)^{\delta^{\mu\nu}} x_\nu, \\
  \psi(x)     &\stackrel{\Theta_\mu}{\to}& \gamma^\mu\gamma^5 \, \psi(\Theta_\mu x), \\
  \bar\psi(x) &\stackrel{\Theta_\mu}{\to}& \bar\psi(\Theta_\mu x) \, \gamma^5 \gamma^\mu, \\
  A^\nu(x)    &\stackrel{\Theta_\mu}{\to}& \delta^{\mu\nu}A^\nu(\Theta_\mu x).
  \end{array}\right.
  \label{eq: time reflection for Euclidean space-time}
\end{equation} \normalsize
In a discretised theory, time-reflection can be performed with two different methods. Either one temporal site index is fixed and the lattice is reflected at this hyperplane, which is called site-reflection. Or the invariant hyperplane is chosen in the middle of a link between two neighbouring temporal hyperplanes, which is called link-reflection. If either site- or link-reflection is a symmetry of the discretised theory, the continuum limit satisfies $ \Theta $~symmetry.
\noindent
The condition, whether the Minkowski space-time QFT, which is obtained from any Euclidean space-time QFT by the analytic continuation back to Minkowski space-time with an inverse Wick rotation, satisfies $ \Theta $~symmetry or not has been covered by Osterwalder \mbox{et al.}~\cite{Osterwalder:1973dx, Osterwalder:1977pc}. The associated Minkowski space-time QFT satisifies $ \Theta $~symmetry, if any correlation function, which depends only on fields at positive times (defined with either site- or link reflection) is necessarily positive.

\subsection{Doubling problem and No-Go theorem}\label{sec: Doubling problem and No-Go theorem}
The na\"{i}vely discretised fermion action of \mbox{eq.}~(\ref{eq: naive fermion action}) contains spurious fermionic degrees of freedom, which are called doublers. This property is apparent in the momentum space representation of the na\"{i}ve fermion action in the free theory with trivial gauge fields,
\begin{equation}
    S^{N}[\psi,\bar\psi]  = \frac{1}{|\Lambda|}\sum\limits_{k\in\widetilde{\Lambda}} \bar\psi(k) \left(\frac{i}{a}\sum\limits_{\mu=0}^3 \gamma^\mu \sin{(ak_\mu)} + m_0 \right)\psi(k),
  \label{eq: naive fermion action in momentum space}
\end{equation}
where $ |\Lambda| $ is defined in appendix~\ref{app: Continuous space-time and discrete space-time lattices}. On the one hand, the na\"{i}ve Dirac operator's chiral limit $ D^N(k)=\tfrac{i}{a} \gamma^\mu \sin{(ak_\mu)} $ vanishes, if all four-momentum components satisfy
\begin{equation}
  k_\mu=n_\mu \frac{\pi}{a},\ n_\mu=\{0,1\}.
  \label{eq: naive doublers}
\end{equation}\normalsize
In a four-dimensional lattice, each dimension contributes a factor two for a total of sixteen poles of $ D^N(k) $. On the other hand, the continuum Dirac operator $ D(k)=i \gamma^\mu k_\mu $ has only one pole, if all momentum components vanish. Because each pole represents a real and degenerate fermion species\footnote{Since these poles differ by local spin factors~\cite{Karsten:1980wd}, lattice fermions are called species in the following.}, the na\"{i}vely discretised Dirac operator has fifteen spurious species compared to the Dirac operator of QCD on a space-time continuum. The degeneracy is due to invariance of the action of \mbox{eq.}~(\ref{eq: naive fermion action}) under any products (\mbox{$ \mu \in \{0,1,2,3\} $}) of the following unitary symmetry transformations:
\begin{equation}
  \psi_n \to \chi_n = T_n^\mu \psi_n, \quad \bar{\psi}_n \to \bar{\chi}_n = \bar\psi_n (T_n^\mu)^{-1}, \quad T_n^\mu=(-1)^{n_\mu} Q^\mu, \quad Q^\mu=\gamma^\mu\gamma^5.
  \label{eq: naive species symmetry}
\end{equation}\normalsize
These are the spin factors (\mbox{cf. eq.}~($ 2.10 $) in~\cite{Karsten:1980wd}) that reflect the symmetry between the sixteen species. Since the number of zero modes of the na\"{i}ve Dirac operator is doubled for every space-time direction, these spurious degrees of freedom are called doublers and their presence is known as the doubling problem. Appropriate modification of the na\"{i}ve Dirac action solves the doubling problem.\newline

\noindent
However, Nielsen and Ninomiya presented a No-Go theorem~\cite{Nielsen:1981hk}, which identified limitations to such modifications. According to Niedermayer~\cite{Niedermayer:1998bi}, there is a limitation that no Dirac operator can simultaneously satisfy
\begin{center}
 \begin{tabular}{cll}
  (a) & $ D_{n,m} $ is local & (bounded by $ Ce^{-\gamma |n-m|} $) \\
  (b) & $ D(p) = i\gamma^\mu p_\mu + \mathcal{O}(ap^2) $ for $ p \ll \pi/a $   \\
  (c) & $ D(p) $ is invertible for $ p\neq 0 $ & (has no massless doublers) \\
  (d) & $ \gamma^5 D + D \gamma^5 = 0 $ & (chiral symmetry)
 \end{tabular}
\end{center}
In brief, ultralocal lattice discretisations of the Dirac operator cannot have an odd number of chiral fermions without failing to reproduce the continuum Dirac operator in the na\"{i}ve continuum limit. Therefore, any modification of the na\"{i}ve Dirac operator must lack at least one of these properties. First, Wilson's discretisation of the quark action, which explicitly sacrifices the chiral symmetry is discussed. Second, the Ginsparg-Wilson relation~\cite{Ginsparg:1981bj}, as a possible non-ultralocal realisation of chiral symmetry on a space-time lattice, and Neuberger's overlap operator~\cite{Neuberger:1997fp} as an example of Ginsparg-Wilson fermions are covered. Finally\footnote{This is by no means exhaustive, e.g. staggered fermions are a third type of lattice fermions with a large number of different subtypes.}, minimally doubled fermions of the types suggested by Karsten~\cite{Karsten:1981gd} and Wilczek~\cite{Wilczek:1987kw} as well as the types suggested by Creutz~\cite{Creutz:2007af} and Borici~\cite{Borici:2007kz} are introduced. The presence of exactly two chiral zero modes inevitably breaks the hypercubic symmetry of the space-time lattice.

\subsection{Wilson fermions}\label{sec: Wilson fermions}
Wilson's solution~\cite{Wilson:1974sk} to the doubling problem adds to the na\"{i}ve Dirac operator of \mbox{eq.}~(\ref{eq: naive Dirac operator}) an addititional operator, that constitutes the Wilson term,
\begin{equation}
  D^W_{n,m}=\mathbf{1}\sum\limits_{\mu=0}^3 \frac{r}{2a}\left(2\delta_{n,m}-\left(U^\mu_n\delta_{n+\hat{e}_\mu,m}+U^{\mu\dagger}_{n-\hat{e}_\mu}\delta_{n-\hat{e}_\mu,m}\right)\right),
  \label{eq: Wilson operator}
\end{equation}\normalsize
which trivially vanishes in the na\"{i}ve continuum limit. In the free theory with trivial gauge fields, it can be expressed through
\begin{equation}
  D^W \psi_n = -\frac{ar}{2} \mathbf{1}\sum\limits_{\mu=0}^3\nabla_\mu\nabla^*_\mu \psi_n
  = -\frac{r}{2} \mathbf{1}\sum\limits_{\mu=0}^3\left(\nabla_\mu-\nabla^*_\mu\right) \psi_n,
\end{equation}\normalsize
which shows that it is a discretised form of a d'Alembertian operator. The momentum space representation of the Wilson term in the free theory,
\begin{equation}
  D^W(k)=\mathbf{1}\sum\limits_{\mu=0}^3 \frac{r}{a}\left(1-\cos{(ak_\mu)}\right),
  \label{eq: Wilson operator in momentum space}
\end{equation}\normalsize
formally resembles a four-momentum dependent mass term. It vanishes at \mbox{$ k=(0,0,0,0) $}, which corresponds to the continuum Dirac operator's pole. All spurious fermion modes acquire \textit{mass terms} that diverge in the continuum limit. Due to this infinite mass, the doublers decouple from the dynamics of the continuum fermion. The Wilson term, however, explicitly breaks chiral symmetry:
\begin{equation}
  (D^N+D^W)\gamma^5 + \gamma^5(D^N+D^W) = 2\gamma^5 D^W.
\end{equation}
If the lattice spacing $ a $ is sufficiently small, cutoff effects due to the Wilson term break chiral symmetry softly like the quark masses. Then cutoff effects can be treated as a perturbation to chiral symmetry and the spectrum of QCD can be recovered in the continuum limit. Otherwise, the spontaneous breakdown of chiral symmetry due to QCD cannot be studied numerically. Subtle properties of QCD due to the broken chiral symmetry might be obscured in studies with Wilson fermions.

\subsection{Ginsparg-Wilson fermions}\label{sec: Ginsparg-Wilson fermions}
In 1981, Ginsparg and Wilson~\cite{Ginsparg:1981bj} suggested a discretised form of chiral symmetry as
\begin{equation}
  D^{GW} \gamma^5+\gamma^5 D^{GW} = aD^{GW} \gamma^5 D^{GW},
  \label{eq: Ginsparg-Wilson equation}
\end{equation}\normalsize
which is known as the Ginsparg-Wilson equation. \mbox{Eq.}~(\ref{eq: Ginsparg-Wilson equation}) requires that any Ginsparg-Wilson Dirac operator $ D^{GW} $ has non-vanishing contributions for arbitrarily separated lattice sites. Locality in the usual sense (ultralocality) is therefore sacrificed in these Dirac operators. However, an infinitesimal chiral transformation of the quark fields with
\begin{equation}
  \psi^\prime =e^{i\alpha\gamma^5(\mathbf{1}-\tfrac{a}{2}D^{GW})}\psi, \quad \bar\psi^\prime =\bar\psi e^{i\alpha(\mathbf{1}-\tfrac{a}{2}D^{GW})\gamma^5}
\end{equation}\normalsize
leaves a Ginsparg-Wilson quark action invariant,
\begin{equation}
  L[\psi^\prime,\bar\psi^\prime] = \bar\psi^\prime D^{GW} \psi^\prime = \bar\psi D^{GW} \psi = L[\psi,\bar\psi].
\end{equation}\normalsize
Lattice chirality projectors are defined with a non-local Ginsparg-Wilson chirality matrix,
\begin{equation}
  \hat{\gamma}^5 = \gamma^5(\mathbf{1}-aD^{GW}), \quad \hat{P}_R =\frac{\mathbf{1}+\hat{\gamma}^5}{2}, \quad \hat{P}_L =\frac{\mathbf{1}-\hat{\gamma}^5}{2}
  \label{eq: Ginsparg-Wilson chirality matrices}
\end{equation}\normalsize
and Ginsparg-Wilson fermions with a finite quark mass are obtained with
\begin{equation}
  D^{GW,m} = \omega D^{GW}+m\mathbf{1}, \quad \omega= 1-\frac{am}{2}.
\end{equation}\normalsize
For a long time, no Dirac operator that qualified as a Ginsparg-Wilson Dirac operator $ D^{GW} $ was discovered until the 1990s. Kaplan~\cite{Kaplan:1992bt} suggested domain wall fermions, Hasenfratz \mbox{et al.}~\cite{Hasenfratz:1993sp} suggested perfect actions and Neuberger~\cite{Neuberger:1997fp} suggested the overlap operator, which all qualify as Ginsparg-Wilson fermions. The overlap operator reads
\begin{equation}
  D^{ov} = \frac{1}{a}\left(\mathbf{1}+\gamma^5\, \mathrm{sign}(Q)\right) = \frac{1}{a}\left(\mathbf{1}+\frac{\gamma^5 Q}{\sqrt{Q^2}}\right),
  \label{eq: overlap Dirac operator}
\end{equation}\normalsize
where $ Q $ is the overlap kernel, which can be any undoubled hermitian Dirac operator with real eigenvalues such as $ Q=\gamma^5 (D^N+D^W) $. Due to the presence of the sign function, the overlap Dirac operator is not a sparse matrix in space-time even with a sparse kernel matrix $ Q $. It must be approximated with considerable numerical effort and is computationally quite expensive. Nevertheless, it is shown by Hernandez~\mbox{et al.}~\cite{Hernandez:1998et} that the magnitude of components of the Ginsparg-Wilson Dirac operator decreases at least exponentially with the distance between sites,
\begin{equation}
  \left|(D^{GW}_{n,m})^{\bar{a}\bar{b}}_{\alpha\beta}\right|\leq C \exp{(-\gamma \| n-m\|)}.
\end{equation}\normalsize
As the coefficients $ C $ and $ \gamma $ are independent of the gauge configuration and the lattice spacing $ a $, the exponential fall-off shrinks to zero and a local field theory is recovered in the continuum limit. In that sense, the non-locality of the overlap operator can be considered as only a technical problem of calculations.

\sectionc{Minimally doubled fermions}{sec: Minimally doubled fermions}

\noindent
Minimally doubled fermions are another type of discretised fermion actions, which have explicit ultralocal chiral symmetry. An additional ultralocal operator $ D^{md} $ that anti\-commutes with the standard chirality matrix $ \gamma^5 $ is added to the na\"{i}ve fermion action. $ D^{md} $ reduces the number of poles of the lattice Dirac operator to two. Since these poles are located at different points on a line through the Brillouin zone, their displacement explicitly necessitates a reduction of the hypercubic symmetry of the na\"{i}ve Dirac operator. Their different momentum support implies that they acquire different spin factors. Chiral fermions in the continuum limit can be defined as linear combinations of fields on neighbouring sites~\cite{Wilczek:1987kw}. These two fermions show sensitivity to fluctuations of local gauge fields which may lift the degeneracy between both at finite lattice spacing. This is the analogue of taste-splitting and taste-breaking for staggered fermions~\cite{KlubergStern:1983dg, Blum:1996uf}.
Two different types of minimally doubled fermions, Karsten-Wilczek~\cite{Karsten:1981gd, Wilczek:1987kw} and Bori\c{c}i-Creutz fermions~\cite{Creutz:2007af, Borici:2007kz, Creutz:2008sr, Borici:2008ym} are covered in more detail. Other types of minimally doubled fermions such as twisted ordering fermions~\cite{Misumi:2010ea} and generalisations of Karsten-Wilcek and Bori\c{c}i-Creutz types~\cite{Capitani:2013zta, Capitani:2013fda, Capitani:2013iha} are not covered here.

\subsection{Karsten-Wilczek fermions}\label{sec: KW fermions}

\noindent
Karsten-Wilczek fermions were suggested in response to the No-Go theorem of Nielsen and Ninomiya~\cite{Nielsen:1981hk} by Karsten~\cite{Karsten:1981gd} as a strictly local, chiral Dirac operator with the minimal number of chiral fermion species. The additional term vanishes only at the two temporal doublers and hence lessens the degeneracy and removes fourteen spurious fermions. Wilczek generalised the action further through the Wilczek parameter (called $ \zeta $ throughout this thesis) and suggested a gluonic counterterm~\cite{Wilczek:1987kw}. Pernici studied the action with the Karsten-Wilczek term in the $ \hat{e}_1 $ direction~\cite{Pernici:1994yj}. He suggested to consider the two fermion species as two flavours of QCD, constructed the transfer matrix and discussed fermionic counterterms. Moreover, he showed that the free action has a different species symmetry than the na\"{i}ve action. 
\noindent
In the wake of the minimally doubled fermion revival by Creutz~\cite{Creutz:2007af} and Bori\c{c}i~\cite{Borici:2007kz}, Bedaque \mbox{et al.}~\cite{Bedaque:2008xs} revisited the symmetries of the Karsten-Wilczek action. A study of the perturbative renormalisation of Karsten-Wilczek fermions at one-loop level~\cite{Capitani:2009ty, Capitani:2010nn, Capitani:2010ht} independently derived its counterterm structure in perturbation theory (\mbox{cf.} section~\ref{sec: Perturbative studies}). 
Tiburzi~\cite{Tiburzi:2010bm} studied the spin-flavour structure and the axial anomaly with Karsten-Wilczek fermions and Creutz suggested an approach to flavours of minimally doubled fermions using point-split fields~\cite{Creutz:2010qm}. Later, Misumi \mbox{et al.}  studied the index theorem~\cite{Creutz:2010bm} with Karsten-Wilczek fermions, their spin-flavour representation~\cite{Kimura:2011ik} and an interpretation as fermions with flavoured chemical potential~\cite{Misumi:2012uu,Misumi:2012eh}.
The additional operator of the Karsten-Wilczek term reads
\begin{equation}
  D^{KW}_{n,m} =\sum\limits_{\mu\neq\underline{\alpha}} \frac{i\zeta}{2a}\gamma^{\underline{\alpha}}\left(2\delta_{n,m}-\left(U^\mu_n\delta_{n+\hat{e}_\mu,m}+U^{\mu\dagger}_{n-\hat{e}_\mu}\delta_{n-\hat{e}_\mu,m}\right)\right),
  \label{eq: KW operator}
\end{equation}\normalsize
and is added to the na\"{i}ve Dirac operator of \mbox{eq.} (\ref{eq: naive Dirac operator}) to form the full Karsten-Wilczek operator. $ \hat{e}_{\underline{\alpha}} $ is the unit vector along the alignment of the Karsten-Wilczek term and can be any one of the four Euclidean directions. The additional operator breaks the hypercubic group $ W_4 $ down to the cubic subgroup $ W_3 $, which excludes the $ \hat{e}_{\underline{\alpha}} $~direction. Any combination of permutations of axes and reflections $ P_\mu $ within the orthogonal complement of the $ \hat{e}_{\underline{\alpha}} $~direction ($ \mu\neq\underline{\alpha} $) is a symmetry of the Karsten-Wilczek action. However, neither the operators $ \Theta_{\underline{\alpha}} $ for $ n_{\underline{\alpha}} $~reflection of the $ \hat{e}_{\underline{\alpha}} $~direction nor $ C $ for charge conjugation are symmetry operators of the Karsten-Wilczek action, since neither leaves the Karsten-Wilczek operator form-invariant. Nevertheless, the combination $ \Theta_{\underline{\alpha}}C=C\Theta_{\underline{\alpha}} $ leaves the Karsten-Wilczek action form-invariant. Therefore, the product $ CP\Theta $ is a symmetry of the Karsten-Wilczek action for any choice of $ \underline{\alpha} $. The two doublers are located on the $ \underline{\alpha} $~axis,
\begin{equation}
  k_{\underline{\alpha}}=n_{\underline{\alpha}}\frac{\pi}{a},\ n_{\underline{\alpha}}=\{0,1\};\ k_\mu=0,\ \mu\neq\underline{\alpha}.
  \label{eq: KW doublers}
\end{equation}\normalsize
The mechanism which removes the spurious doublers is evident in the free momentum space Karsten-Wilczek action,
\small
\begin{equation}
  S^{KW}[\psi,\bar\psi]  = \frac{1}{|\Lambda|}\sum\limits_{k\in\widetilde{\Lambda}} \bar\psi(k) \left(\frac{i}{a}\sum\limits_{\mu=0}^3 \left\{\gamma^\mu \sin{(ak_\mu)} + \zeta(1-\delta_{\mu\underline{\alpha}}) \gamma^{\underline{\alpha}}(1-\cos{(ak_\mu)})\right\} + m_0 \right)\psi(k).
  \label{eq: free KW action in momentum space}
\end{equation}\normalsize
The magnitude of the Karsten-Wilczek term exceeds the magnitude of the $ \gamma^{\underline{\alpha}} $-component of the na\"{i}ve kinetic term, if the Wilczek parameter satisfies $ |\zeta|>1/2 $ at any of the spurious doublers $ k_\mu = \pi/a,\ \mu\neq\underline{\alpha}$,
\begin{equation}
  |\zeta|(1-\cos{(ak_\mu)}) = 2|\zeta| > 1 \geq |\sin{(ak_{\underline{\alpha}})}|.
\end{equation}\normalsize
Thus, the massless Dirac equation $ (D^{N}(k)+D^{KW}(k))\psi(k)=0 $ has no solution, if any of the four-momentum components is non-zero except $ k_{\underline{\alpha}} $. This leads to Pernici's symmetry transformation between mirror fermions~\cite{Pernici:1994yj}, 
\begin{equation}
  \psi_n \to (-1)^{n_{\underline{\alpha}}} \psi_{R_{\underline{\alpha}} n}, \quad \bar{\psi}_n \to \bar{\psi}_{R_{\underline{\alpha}} n} (-1)^{n_{\underline{\alpha}}}, \quad R_\mu n_\nu = (1-2\delta_{\mu\nu}) n_\nu,
  \label{eq: KW species symmetry}
\end{equation}
which can be understood as a combination of $ n_{\underline{\alpha}} $~reflection $ \Theta_{\underline{\alpha}} $ of \mbox{eq.}~(\ref{eq: time reflection for Euclidean space-time}) and the unitary operator $ T_n^{\underline{\alpha}} $ of \mbox{eq.}~(\ref{eq: naive species symmetry}). Thus, the combination of any two of the three operators $ C $, $ \Theta_{\underline{\alpha}} $ and $ T_n^{\underline{\alpha}} $ leaves the Karsten-Wilczek action form-invariant.
\noindent
The Karsten-Wilczek action requires three counterterms~\cite{Pernici:1994yj, Bedaque:2008xs, Capitani:2009ty, Capitani:2010nn}, which are the only further relevant and marginal operators in the operator product expansion, which comply with the Karsten-Wilczek term's symmetry. The two fermionic counterterm operators read
\begin{align}
  D^{3}_{n,m} =& c^{KW}(g_0)\ \frac{i}{a}\gamma^{\underline{\alpha}}\delta_{n,m},
  \label{eq: ferm dim 3 KW counterterm operator} \\
  D^{4}_{n,m} =& d^{KW}(g_0)\ \gamma^{\underline{\alpha}} D^{\underline{\alpha}}_{n,m}[U],
  \label{eq: ferm dim 4 KW counterterm operator}
\end{align}
where $ D^{\underline{\alpha}}_{n,m}[U] $ is the $ \underline{\alpha} $~component of the usual lattice covariant derivative of \mbox{eq.} (\ref{eq: lattice covariant derivative}). It amounts to a renormalisation of the fermionic speed of light in the $ \hat{e}_{\underline{\alpha}} $~direction. The gluonic counterterm is due to a carryover of the anisotropy to the gauge fields and reads
\begin{equation}
  S^{4p}[U]  = \sum\limits_{n\in\Lambda} \sum\limits_{\mu\neq\underline{\alpha}} d_P^{KW}(g_0)\frac{\beta}{3}\mathrm{Re}\,\mathrm{Tr}(1-U^{\mu\underline{\alpha}}_n).
  \label{eq: gluonic KW counterterm}
\end{equation}\normalsize
It amounts to a renormalisation of the gluonic speed of light in the $ \hat{e}_{\underline{\alpha}} $~direction. The dimension three counterterm has the same symmetry ($ C\Theta_{\underline{\alpha}} $ instead of $ C $ and $ \Theta_{\underline{\alpha}} $ individually) as the Karsten-Wilczek term. However, both dimension four counterterms are form-invariant under charge conjugation, any $ n_{\mu} $~reflection ($ P_\mu $ or $ \Theta_\mu $) and the unitary operators ($ T_n^{\underline{\alpha}} $).
Thus, the full Karsten-Wilczek action has three counter\-terms,
\begin{align}
  S^{f}[\psi,\bar\psi,U]  =&\ a^4\sum\limits_{n,m\in\Lambda} \bar\psi_n \left(D^N_{n,m}+D^{KW}_{n,m}+D^{3}_{n,m}+D^{4}_{n,m} + m_0\delta_{n,m}\right) \psi_m, \label{eq: KW fermion action} \\
  S^{g}[U]  =&\ \sum\limits_{n\in\Lambda} \sum\limits_{\mu<\nu} \frac{\beta}{3}\mathrm{Re}\,\mathrm{Tr}(1-U^{\mu\nu}_n)\left(1+d_P^{KW}(g_0)\delta_{\mu\underline{\alpha}}\right), \label{eq: KW gluon action}
\end{align}\normalsize
with a priori unknown coefficients $ c^{KW}(g_0)$, $d^{KW}(g_0) $ and $ d_P^{KW}(g_0) $, which depend on the parameters of the theory (in the chiral limit only $ g_0 $ and $ \zeta $). These counterterm coefficients\footnote{The label $ ^{KW} $ of the coefficients and the coupling dependence are usually omitted in the following.} must be tuned in order to restore isotropy to the continuum limit.

\subsection{Bori\c{c}i-Creutz fermions}\label{sec: BC fermions}

\noindent
Bori\c{c}i-Creutz fermions were suggested by Creutz~\cite{Creutz:2007af} as a four-dimensional generalisation of graphene. Graphene layers of single atom thickness had been extracted for the first time by Geim and Novoselov in 2004~\cite{Geim:2004} and they demonstrated that electrons in graphene propagate as a two-dimensional gas of massless Dirac fermions~\cite{Novoselov:2005kj}. Bori\c{c}i placed these four-dimensional graphene actions on orthogonal lattices~\cite{Borici:2007kz}. Bedaque \mbox{et al.} studied symmetries of these actions~\cite{Bedaque:2008xs}, introduced an iso-doublet notation and pointed out the non-singlet structure of the Bori\c{c}i-Creutz operator. They demonstrated that Bori\c{c}i-Creutz actions on totally symmetric hyperdiamond lattices do not reproduce the correct continuum limit~\cite{Bedaque:2008jm}. Cichy \mbox{et al.}~\cite{Cichy:2008gk} studied cutoff effects of Bori\c{c}i-Creutz fermions at tree level and Bori\c{c}i found scaling behaviour of pseudoscalar masses with Bori\c{c}i-Creutz fermions consistent with predictions of chiral perturbation theory~\cite{Borici:2008ym}. Misumi and Kimura~\cite{Kimura:2009di, Kimura:2009qe} studied hyperdiamond lattices of different dimensionalities and demonstrated that minimal doubling and hyperdiamond structure are compatible only in two dimensions. One-loop renormalisation properties were determined in perturbative studies of Bori\c{c}i-Creutz fermions~\cite{Capitani:2009yn, Capitani:2009ty, Capitani:2010nn, Capitani:2010ht}. The spin-flavour representation of Bori\c{c}i-Creutz fermions was covered in~\cite{Kimura:2011ik}.
The additional operator is
\begin{equation}
  D^{BC}_{n,m} =\sum\limits_{\mu} \frac{i\zeta}{2a}\left(\Gamma-\gamma^\mu\right)\left(\left(U^\mu_n\delta_{n+\hat{e}_\mu,m}+U^{\mu\dagger}_{n-\hat{e}_\mu}\delta_{n-\hat{e}_\mu,m}\right)-2\delta_{n,m}\right),
  \label{eq: BC operator}
\end{equation}\normalsize
and its fermionic bilinear is the Bori\c{c}i-Creutz term with parameter\footnote{Some authors (e.g.~\cite{Kimura:2011ik}) use a generalised Bori\c{c}i-Creutz action with arbitrary values of $ \zeta $. A brief discussion why these actions are not considered here is presented at the end of this section.} $ \zeta=\pm1 $ and where
\begin{equation}
  \Gamma\equiv \frac{1}{2}\sum\limits_{\nu=0}^{3}\gamma^\nu.
  \label{eq: BC Gamma}
\end{equation}\normalsize
The additional operator is added to the na\"{i}ve Dirac operator of \mbox{eq.}~(\ref{eq: naive Dirac operator}) to form the full Bori\c{c}i-Creutz operator. A transformed set of Dirac matrices,
\begin{equation}
  \gamma^{\mu\prime}\equiv\Gamma \gamma^\mu \Gamma = \Gamma-\gamma^\mu,
\end{equation}\normalsize
allows for a convenient notation of the additional operator,
\begin{equation}
  D^{BC}_{n,m} =\sum\limits_{\mu} \frac{i\zeta}{2a}\gamma^{\mu\prime}\left(U^\mu_n\delta_{n+\hat{e}_\mu,m}+U^{\mu\dagger}_{n-\hat{e}_\mu}\delta_{n-\hat{e}_\mu,m}\right)-\frac{i\zeta}{a}2\Gamma\delta_{n,m}.
  \label{eq: convenient BC operator}
\end{equation}\normalsize
The symmetry of the Bori\c{c}i-Creutz action has been studied by Creutz~\cite{Creutz:2008sr}. The residual $ W_3 $ cubic symmetry of the Bori\c{c}i-Creutz action can be made transparent most easily using a real, self-inverse and symmetric matrix,
\begin{equation}
  A=\frac{\zeta}{2}\left(\begin{array}{rlrl}
    +1&+1&+1&+1\\+1&-1&+1&-1\\+1&+1&-1&-1\\+1&-1&-1&+1
  \end{array}\right) = A^{-1}=A^T=A^*,
\end{equation}\normalsize
which transforms the site basis $ \hat{e}_\mu $ to a new orthonormal basis $ \hat{f}_{\nu} $ ($ \hat{f} $ basis),
\begin{equation}
  \hat{f}_{\nu} = A_{\nu\mu} \hat{e}_\mu, \quad  \hat{e}_\mu = A_{\mu\nu} \hat{f}_{\nu}.
  \label{eq: BC f-basis}
\end{equation}\normalsize
The Dirac matrices in the $ \hat{f} $ basis are linear combinations of the old Dirac matrices,
\begin{equation}
  \Gamma^{\nu} =  A^{\nu\mu} \gamma^\mu, \quad  \gamma^\mu = A^{\mu\nu} \Gamma^{\nu}, \quad \text{in particular}\quad \zeta\Gamma \equiv \Gamma^{0}.
\end{equation}\normalsize
The na\"{i}ve Dirac operator is form-invariant under the coordinate transformation to the $ \hat{f} $ basis and the additional operator written in the $ \hat{f} $ basis reads
\begin{align}
  D^{BC}_{n,m} =&-\sum\limits_{\nu,\mu} \frac{i}{2a}\Gamma^{\nu} A^{\nu\mu}\left(U^\mu_n\delta_{n+\hat{e}_\mu,m}+U^{\mu\dagger}_{n-\hat{e}_\mu}\delta_{n-\hat{e}_\mu,m}\right) 
  \label{eq: BC operator W4 symmetric piece}\\
  &+\sum\limits_{\mu} \frac{i}{a}\Gamma^{0} A^{0\mu}\left(U^\mu_n\delta_{n+\hat{e}_\mu,m}+U^{\mu\dagger}_{n-\hat{e}_\mu}\delta_{n-\hat{e}_\mu,m}\right)+ \frac{i}{a}2\Gamma^{0}\delta_{n,m}
  \label{eq: BC operator symmetry breaking piece},
\end{align}\normalsize
where the residual cubic symmetry of the linear span
\begin{equation}
  \mathrm{span}(V)=\left\{\sum\limits_{i=1}^{3}\lambda_i\hat{f}_{i}\ |\ \hat{f}_{1},\hat{f}_{2},\hat{f}_{3}\right\}
\end{equation}\normalsize
is evident in \mbox{eq.} (\ref{eq: BC operator W4 symmetric piece}).
\noindent
However, any reflection of the $ \hat{f}_{0} $ direction changes the sign of the last term in \mbox{eq.}~(\ref{eq: BC operator symmetry breaking piece}). Moreover, a simultaneous reflection of all directions ($ P\Theta = P_f\Theta_f $) changes the overall sign of the additional operator of \mbox{eq.}~(\ref{eq: BC operator}). Because charge conjugation induces a sign change in the additional operator as well, the Bori\c{c}i-Creutz action satisfies $ CP\Theta $~symmetry. The two doublers are aligned on the $ \hat{f}_{0} $-axis,
\begin{equation}
  k_{0}=n_{0}\frac{\pi}{a},\ n_{0}=\{0,1\};\ k_{\nu}=0,\ \nu\neq0.
  \label{eq: BC doublers, f-basis}
\end{equation}\normalsize
Whereas this looks very much alike \mbox{eq.}~(\ref{eq: KW doublers}), the length of one period on the axes in the $ \hat{f} $ basis is $ 4\pi/a $ instead of $ 2\pi/a $. Hence, the Bori\c{c}i-Creutz term in \mbox{eq.}~(\ref{eq: BC operator}) and the Karsten-Wilczek term in \mbox{eq.}~(\ref{eq: KW operator}) are not exactly rotated analogues. Only even sites in the site basis $ \hat{e}_\mu $ are mapped on points with integer labels in the $ \hat{f} $ basis $ \hat{f}_{\nu} $. Odd sites in the site basis are mapped on points with half-integer labels in the $ \hat{f} $ basis and vice versa. In terms of the site basis $ \hat{e}_\mu $, the doublers are located at
\begin{equation}
  k_\mu=n \zeta \frac{\pi}{2a}\ \forall \mu,\ n=\{0,1\},
  \label{eq: BC doublers, site basis}
\end{equation}\normalsize
where all components are equal. The mechanism, which removes the spurious doublers is covered in great detail by Creutz~\cite{Creutz:2008sr}. Among the two pieces of the additional operator, the first is unitarily equivalent to the na\"{i}ve Dirac operator with shifted four-momentum,
\begin{equation}
  \frac{i\zeta}{a}\gamma^{\mu\prime} \cos{(ak_\mu)} = \Gamma \left(\frac{i}{a} \gamma^{\mu} \sin{\left(ak_\mu+\zeta\frac{\pi}{2}\right)}\right) \Gamma,
\end{equation}\normalsize
and vanishes at four-momenta that maximize the na\"{i}ve Dirac operator. The second piece is a constant
\begin{equation}
  -\frac{i\zeta}{a}2\Gamma = -\frac{i}{a}\sum\limits_{\mu}\gamma^\mu\zeta = -\frac{i\zeta}{a}\sum\limits_{\mu}\gamma^{\mu\prime}
\end{equation}\normalsize
that cancels the with the first piece exactly at the continuum pole of the na\"{i}ve Dirac operator ($ k_\mu=0\ \forall\ \mu$), whereas it cancels with the na\"{i}ve Dirac operator exactly a single pole of the first piece of the additional operator of the Bori\c{c}i-Creutz term \mbox{($ k_\mu=\zeta\pi/(2a)\ \forall\ \mu$)}. Any other pole of either operator is only incompletely cancelled by the second piece and does not persist as a pole of the full Bori\c{c}i-Creutz operator. Therefore, minimal doubling is achieved and the free momentum space Bori\c{c}i-Creutz action reads
\small
\begin{equation}
  S^{BC}[\psi,\bar\psi]  = \frac{1}{|\Lambda|}\sum\limits_{k\in\widetilde{\Lambda}} \bar\psi(k) \left(\frac{i}{a}\sum\limits_{\mu=0}^3 \left\{\gamma^\mu \sin{(ak_\mu)} + \zeta\gamma^{\mu\prime}(\cos{(ak_\mu)})\right\} + \left\{m_0-\frac{i\zeta}{a}2\Gamma\right\} \right)\psi(k).
  \label{eq: free BC action in momentum space}
\end{equation}\normalsize
The Bori\c{c}i-Creutz action requires three counterterms~\cite{Bedaque:2008xs, Capitani:2009yn, Capitani:2009ty, Capitani:2010nn, Capitani:2010ht}, which are the only further relevant and marginal operators in the operator product expansion that comply with the Bori\c{c}i-Creutz term's symmetry. The two fermionic counterterm operators read
\begin{align}
  D^{3}_{n,m} =& c^{BC}(g_0)\ \frac{i}{a}\Gamma\delta_{n,m},
  \label{eq: ferm dim 3 BC counterterm operator} \\
  D^{4}_{n,m} =& d^{BC}(g_0)\  \frac{1}{2}\Gamma\sum\limits_{\mu=0}^{3} D^{\mu}_{n,m}[U],
  \label{eq: ferm dim 4 BC counterterm operator}
\end{align}
where $ D^{\mu}_{n,m}[U] $ are the $ \mu $-components of the usual lattice covariant derivative of \mbox{eq.}~(\ref{eq: lattice covariant derivative}). The dimension four counterterm can be understood as the $ \Gamma^{0} D^{0}_{n,m}[U] $ component of the na\"{i}ve Dirac operator in the $ \hat{f} $ basis. The gluonic counterterm is due to a carryover of the anisotropy to the gauge fields and reads
\begin{equation}
  S^{4p}[U] = \sum\limits_{n\in\Lambda} \sum\limits_{\mu,\nu,\rho=0}^{3} d_P^{BC}(g_0) \left(\widehat{F}^{\mu\rho}_n\widehat{F}^{\rho\nu}_n\right),
  \label{eq: gluonic BC counterterm}
\end{equation}\normalsize
where $ \widehat{F}^{\mu\rho}_n $ is some lattice discretisation of the gluon field strength tensor. In the $ \hat{f} $-basis, the term takes a form which clearly underscores the analogy to \mbox{eq.}~(\ref{eq: gluonic KW counterterm}),
\begin{equation}
  S^{4p}[U] = \sum\limits_{n\in\Lambda} \sum\limits_{\rho=1}^{3} d_P^{BC}(g_0) \frac{1}{4}\left(\widehat{F}^{0\rho}_n\widehat{F}^{\rho0}_n\right).
\end{equation}\normalsize
The dimension four counterterms amount to different renormalisation of the speed of light for fermions and gluons  in the $ \hat{f}_{0} $-direction. The dimension three counterterm has the same $ CP\Theta $~symmetry as the additional term of \mbox{eq.}~(\ref{eq: BC operator}). However, both dimension four counterterms are additionally form-invariant under charge conjugation as well as any $ n_\mu $~reflection ($ P_\mu $ or $ \Theta_\mu $). The full Bori\c{c}i-Creutz action contains its three counterterms,
\begin{align}
  S^{f}[\psi,\bar\psi,U]  =& a^4\sum\limits_{n,m\in\Lambda} \bar\psi_n \left(D^N_{n,m}+D^{BC}_{n,m}+D^{3}_{n,m}+D^{4}_{n,m} + m_0\delta_{n,m}\right) \psi_m, \label{eq: BC fermion action} \\
  S^{g}[U]  =& \sum\limits_{n\in\Lambda} \sum\limits_{\mu<\nu} \frac{\beta}{3}\mathrm{Re}\,\mathrm{Tr}(1-U^{\mu\nu}_n)+
  \sum\limits_{\rho=0}^{3} 2 d_P^{BC}(g_0) \left(\widehat{F}^{\mu\rho}_n\widehat{F}^{\rho\nu}_n\right), \label{eq: BC gluon action}
\end{align}\normalsize
with a priori unknown coefficients $ c^{BC}(g_0)$, $d^{BC}(g_0) $ and $ d_P^{BC}(g_0) $, which depend on the parameters of the theory $ g_0 $. These counterterm coefficients\footnote{The label $ ^{BC} $ of the coefficients and the coupling dependence are usually omitted in the following.} must be tuned in order to restore isotropy to the continuum limit.
The additional operator could be generalised~\cite{Kimura:2011ik} with $ \zeta \neq \pm1 $, if the dimension-four counterterm were included with a changed coefficient,
\begin{equation}
  D^{4,mod}_{n,m} = \left(d^{BC}(g_0)+2(|\zeta|-1)\right)\ \frac{1}{2}\Gamma\sum\limits_{\mu=0}^{3} D^{\mu}_{n,m}[U].
  \label{eq: modified fermionic dim 4 BC counterterm operator}
\end{equation}
The action would maintain minimal doubling with the same zero modes as in \mbox{eq.} (\ref{eq: BC doublers, site basis}). However, \mbox{eq.}~(\ref{eq: modified fermionic dim 4 BC counterterm operator}) changes the na\"{i}ve continuum limit of the Dirac operator, which is
\begin{equation}
  \lim_{a \to 0} D^{mod}(p) = \lim_{a\to 0} D^{N}(p)+D^{4,mod}(p) = 
  i\sum\limits_{\mu}\left(\gamma^\mu+(|\zeta|-1)\Gamma\right) p_\mu.
\end{equation}\normalsize
Hence, the na\"{i}ve continuum limit of the modified Bori\c{c}i-Creutz actions fails to agree with the correct continuum action $ i\sum_\mu \gamma^\mu p_\mu $ unless $ |\zeta|=1 $. This is why these modified actions are not given any further consideration here.

\chapter{Perturbative studies}\label{sec: Perturbative studies}

\noindent
Though the objective of LQCD is to study QCD in the non-perturbative regime,  Lattice Perturbation Theory (LPT) is an important tool. Here, LPT studies are applied to determine renormalisation properties of minimally doubled fermions. Their counter\-terms are a manifestiation of the anisotropy and their coefficients are calculated at one-loop level. It is shown how isotropy is restored to the continuum limit in perturbation theory up to $ \mathcal{O}(g_0^4) $ corrections through application of counterterms. 
\noindent
However, only coefficients of counterterms in the continuum limit are known at one-loop level. If there are different lattice operators with the same continuum limit, a one-loop LPT calculation cannot necessarily discern which counterterm operator is favourable at finite cutoff. This is important if counterterm operators include lattice derivatives such as \mbox{eqs.}~(\ref{eq: ferm dim 4 KW counterterm operator}) and~(\ref{eq: ferm dim 4 BC counterterm operator}), as one-loop calculations are ignorant of their discretisation\footnote{Two-loop coefficients explicitly discriminate between these choices, since counterterms have to be included in loop calculations up to $ \mathcal{O}(g_0^4) $.}. Nevertheless, the dependence of one-loop coefficients on tree-level parameters (such as $ \zeta $) restricts possible choices of lattice counterterm operators. 
\noindent
LPT calculations demonstrate the renormalisability of minimally doubled fermions (\mbox{cf.} sections~\ref{sec: Fermionic self-energy} and~\ref{sec: Fermionic contribution to the vacuum polarisation}). 
LPT substantiates the statements~\cite{Bedaque:2008xs,Tiburzi:2010bm} that the axial symmetry current of minimally doubled fermions is conserved in the perturbative regime for arbitrary gauge coupling. Vector and axial symmetry currents, which are derived with chiral Ward-Takahashi identities have renormalisation factors equal to one, since contributions from proper vertex renormalisation and legs cancel exactly (\mbox{cf.} section~\ref{sec: Local bilinears and symmetry currents}).

\noindent
The chapter starts with a discussion of general technical aspects of LPT is section~\ref{sec: Technical aspects of lattice perturbation theory}, which are taken from~\cite{Capitani:2002mp}. The next section~\ref{sec: Propagators and vertices} covers propagators and vertices of minimally doubled fermions, which were already presented in~\cite{Capitani:2009yn, Capitani:2009ty, Capitani:2010nn, Capitani:2010ht}. The calculation of one-loop corrections is then covered in more detail in section~\ref{sec: One-loop corrections}. Finally, the concept of boosted perturbation theory (BPT)~\cite{Lepage:1992xa} is introduced and the perturbative estimates from BPT for non-perturbative coefficients are presented in section~\ref{sec: Boosted perturbation theory}. Lastly, results of perturbative studies are summarised as interim findings (I) in section~\ref{sec: Interim findings (I)}.

\sectionc{Technical aspects of lattice perturbation theory}{sec: Technical aspects of lattice perturbation theory}
LPT has been covered in great detail by Capitani in~\cite{Capitani:2002mp}. This section covers only those concepts from chapters~15 and~18 of~\cite{Capitani:2002mp}, which are necessary in the following. These are the power counting theorem of Reisz~\cite{Reisz:1987da, Reisz:1987pw, Reisz:1987px, Reisz:1987hx}, the subtraction scheme for divergent integrals from Kawai \mbox{et al.}~\cite{Kawai:1980ja} and basic bosonic integrals, which were first studied by Caracciolo \mbox{et al.}~\cite{Caracciolo:1991cp}.

\subsection{The power counting theorem of Reisz}

Any LPT integral $ I $ with $ L $ loops has a generic structure (\cite{Capitani:2002mp}, p.~208, \mbox{eq.}~($ 15.1 $))
\begin{equation}
  I = \int\limits_{-\pi/a}^{+\pi/a} \frac{d^4k_1}{(2\pi)^4} \ldots  \int\limits_{-\pi/a}^{+\pi/a} \frac{d^4k_L}{(2\pi)^4}
  \frac{V(k,q;m,a)}{C(k,q;m,a)},
  \label{eq: generic LPT integral}
\end{equation}\normalsize
where $ k_j $ are internal loop four-momenta, $ q_j $ are external four-momenta, $ m $ represents masses of all particles on internal lines and $ a $ is the lattice spacing. The denominator is written as a product of denominators of individual propagators (\cite{Capitani:2002mp}, p.~208, \mbox{eq.}~($ 15.2 $)),
\begin{equation}
  C(k,q;m,a) = \prod\limits_{i=1}^{L} C_i(l_i;m,a),
\end{equation}\normalsize
with line momenta $ l_i(k,q) $, which can be expressed as linear combinations of internal four-momenta $ k_j $ and external four-momenta $ q_j $. The numerator and denominator have to satisfy a set of conditions (V1,V2, C1,C2,C3 and L1,L2 in~\cite{Capitani:2002mp}). C3 is quoted here verbatim (\cite{Capitani:2002mp}, p.~209, \mbox{eq.}~($ 15.7 $)), because it is not satisfied in LPT calculations with minimally doubled fermions without additional precautions:
\begin{quote}
  ``(C3) There exist positive constants $ a_0 $ and $ A $ such that
\begin{equation}
 |C_i(l_i;m,a)| \geq A (\hat{l}_i^2+m_i^2)
 \label{eq: C3 condition}
\end{equation}\normalsize
 for all $ a\leq a_0 $ and all $ l_i $'s.''
\end{quote}
The variable $ \hat{l} $ on the right hand side of \mbox{eq.}~(\ref{eq: C3 condition}) is defined as
\begin{equation}
  \hat{l}_\mu \equiv \frac{2}{a}\sin{\left(\frac{al_\mu}{2}\right)}.
  \label{eq: hat momentum}
\end{equation}\normalsize
Lattice integrals at finite $ a $ are necessarily finite in the ultraviolet (UV) region. Instead, their divergence is entirely in the infrared (IR). Evaluation of any LPT integral requires knowledge of its superficial degree of divergence ( \cite{Capitani:2002mp}, p.~$ 209 $, \mbox{eq.}~($ 15.12 $)). The superficial degree of divergence $ \mathrm{deg}\, V $ of a numerator $ V $ is defined as
\begin{equation}
  V(\lambda k,q;m,\lambda a) \stackrel{\lambda \to \infty}{=} K \lambda^{\mathrm{deg}\, V} +\mathcal{O}(\lambda^{\mathrm{deg}\ V-1}),
\end{equation}\normalsize
where $ K\neq0 $. With an analogous definition of $ \mathrm{deg}\, C $ for the denominator, the full superficial degree of divergence is given by
\begin{equation}
  \mathrm{deg}\ I = 4L +\mathrm{deg}\, V - \mathrm{deg}\, C.
  \label{eq: superficial degree of divergence}
\end{equation}\normalsize
In the one-loop calculations of section \ref{sec: One-loop corrections}, chiral fermions are assumed. A common mass parameter is introduced in all denominators as a mass regularisation in the infrared.

\subsection{Subtraction scheme for lattice integrals}

One-loop integrals $ I $, which are calculated in section~\ref{sec: One-loop corrections} depend on one or two external four-momenta $ p $ and $ q $, where the latter case applies within this thesis only to quadratically divergent contributions to the vacuum polarisation with $ p=q $. The subtraction scheme is demonstrated here for integrals with only one external four-momentum $ p $, 
\begin{equation}
  I=\int\limits_{-\pi/a}^{\pi/a}\frac{d^4k}{(2\pi)^2} \mathcal{I}(k,p),
\end{equation}\normalsize
which are representative of logarithmically or linearly divergent contributions to the fermionic self-energy. Following the scheme introduced by Kawai \mbox{et al.}~\cite{Kawai:1980ja}, the divergent integral is artificially split into two pieces $ J $ and $ I-J $, where
\begin{equation}
  J=\int\limits_{-\pi/a}^{\pi/a}\frac{d^4k}{(2\pi)^2} \mathcal{I}(k,0) + 
  \sum\limits_{\mu=0}^3 {\left[ p_\mu \int\limits_{-\pi/a}^{\pi/a}\frac{d^4k}{(2\pi)^2} \left.\frac{\partial\, \mathcal{I}(k,p)}{\partial p_\mu}\right|_{p=0} \right]}
  \label{eq: split integral J}
\end{equation}\normalsize
is obtained as a Taylor expansion of $ I $ up to first order in the external four-momentum $ p $. Thus, external particles are assumed as belonging to the neighbourhood of the standard pole $ k_\mu=0\ \forall\ \mu $. 
All integrands in $ J $ are evaluated for vanishing external four-momenta and in the chiral limit, which greatly simplifies the calculation. Renormalisation of the quark mass is calculated from a Taylor expansion in the bare mass, where integrals are evaluated in the chiral limit. 
\noindent
However, even for IR-finite cases of the original integral $ I $, both integrals $ J $ and $ I-J $ are IR-divergent and require an intermediate IR-regularisation scheme. The second piece, $ I-J $, is UV-finite and is evaluated in the limit $ a \to 0 $. The first piece $ J $ can be simplified further with subtractions of well-known basic integrals $ \widetilde{J} $, which share the same IR divergence. Hence, the difference $ J-\widetilde{J} $ is IR-finite. The lattice spacing $ a $ is absorbed by rescaling the integration momentum,
\begin{equation}
  ak \to k^\prime,
  \label{eq: rescaling of integration momentum}
\end{equation}
and the integral itself is transformed into a pure number multiplying a power of the lattice spacing $ a $. The subtracted IR divergence is obtained from basic bosonic integrals $ \widetilde{J} $, which are known at very high precision. The basic integrals that are used in the presented calculations are expressed with the notation of \mbox{eq.}~(\ref{eq: hat momentum}) as
\begin{align}
 \mathcal{B}(1)     \equiv& \int\limits_{-\pi/a}^{\pi/a} \frac{d^4k}{(2\pi)^4} \frac{1}{\sum_\mu{\hat{k}_\mu^2}} = \frac{1}{a^2}Z_0, 
  \label{eq: basic integral B(1)}\\
 \mathcal{B}(1;1,1) \equiv& \int\limits_{-\pi/a}^{\pi/a} \frac{d^4k}{(2\pi)^4} \frac{\hat{k}_{\underline{\mu}}^2\hat{k}_{\underline{\nu}}^2}{\sum_\lambda{\hat{k}_\lambda^2}} = 4 a^2 Z_1, 
  \label{eq: basic integral B(1;1,1)}\\
  \mathcal{B}(2)    \equiv&  \int\limits_{-\pi/a}^{\pi/a} \frac{d^4k}{(2\pi)^4} \frac{1}{\left(\sum_\mu{\hat{k}_\mu^2}+M^2\right)^2} =
  \frac{1}{16\pi^2}\left(-\log{(aM)^2)} -\gamma_E+F_0\right)
  \label{eq: basic integral B(2)}.
\end{align}\normalsize
The Euler-Mascheroni constant $ \gamma_E $ also appears in continuum loop integrals, 
\begin{equation}
  \gamma_E=0.57721566490153286.
  \label{eq: Euler's constant}
\end{equation}\normalsize
These basic integrals are taken from~\cite{Capitani:2002mp}, p.~254, \mbox{eqs.}~($ 18.23 $),~($ 18.24 $) and~($ 18.26 $). Other bosonic integrals are defined in~\cite{Capitani:2002mp}, p.~254, \mbox{eq.}~($ 18.15 $) using \mbox{eq.} (\ref{eq: hat momentum}) as
\begin{equation}
  \mathcal{B}(p;n_0,n_1,n_2,n_3) = \int\limits_{-\pi/a}^{\pi/a} \frac{d^4k}{(2\pi)^4} \frac{\hat{k}_0^{2n_0}\hat{k}_1^{2n_1}\hat{k}_2^{2n_2}\hat{k}_3^{2n_3}}{\left(\sum_\mu{\hat{k}_\mu^2}+M^2\right)^p}
  \label{eq: general bosonic integral B}
\end{equation}\normalsize
and satisfy recursion relations, which are summarised in appendix~\ref{sec: Recursion relations for bosonic integrals}.

\begin{table}[hbt]
\center
 \begin{tabular}{|c|l|}
  \hline
  \multicolumn{2}{|c|}{Basic bosonic constants} \\
  \hline
  $ Z_0 $ & $ 0.154933390231060214084837208 $ \\
  $ Z_1 $ & $ 0.107781313539874001343391550 $ \\
  $ F_0 $ & $ 4.369225233874758 $ \\
  \hline  
 \end{tabular}
 \caption{Numerical values of basic bosonic constants in LPT are taken from~\cite{Capitani:2002mp}, p.~$ 254 $., Table~$ 2 $. These constants are defined in \mbox{eqs.} (\ref{eq: basic integral B(1)}), (\ref{eq: basic integral B(1;1,1)}) and (\ref{eq: basic integral B(2)}).}
 \label{tab: basic bosonic constants}
\end{table}

\subsection{Numerical integration and finite volume effects}

\noindent
The remaining lattice integrals $ J-\widetilde{J} $ are evaluated with discrete summation methods for periodic analytic functions taken from~\cite{Luscher:1985wf}. Integrals $ I $ of analytic periodic functions $ f(k) $ are estimated as
\begin{equation}
  I = \int \frac{d^4k}{(2\pi)^4} f(k) = \frac{1}{N^4}\sum_{\mu=0}^3\sum_{n_\mu=1}^N f(2\pi\frac{ n}{N}) + \mathcal{O}(e^{-\epsilon N}) \equiv I(N)+ \mathcal{O}(e^{-\epsilon N}),
\end{equation}\normalsize
where $ \epsilon $ is given by the absolute value of the singularity of the integrand which is closest to the real axis (\mbox{eqs.}~($ 5.42 $)-($ 5.44 $) in chapter~5.3 of~\cite{Luscher:1985wf}). The convergence of the integrals is accelerated by changing the integration variable according to (\mbox{eqs.}~($ 5.46 $)-($ 5.51 $) in chapter~5.3 of~\cite{Luscher:1985wf}) as
\begin{equation}
  k=k^\prime -\alpha \sin{(k^\prime)},\quad 0\leq \alpha < 1,
\end{equation}\normalsize
which moves the singularity away from the real axis. 
The integrand takes the new form
\begin{equation}
  \hat{f}(k^\prime) = (1-\alpha\cos{k^\prime})f(k(k^\prime))
\end{equation}\normalsize
and the summation is replaced by
\begin{equation}
  I = \frac{1}{N^4}\sum_{\mu=0}^3\sum_{n_\mu=1}^N \hat{f}(2\pi\frac{ n}{N}) + \mathcal{O}(e^{-\hat{\epsilon} N}) \equiv \hat{I}(N) + \mathcal{O}(e^{-\hat{\epsilon} N}),
\end{equation}\normalsize
where $ \hat{\epsilon}=\mathcal{O}(1) $. The summation is performed for various discretisations of the Brillouin zone with different $ N $. According to \mbox{eq.}~($ 6.5 $) in chapter~6 of~\cite{Luscher:1985wf}, the finite volume effects for lattice one-loop diagrams with engineering dimension $ \delta $ in the absence of external momenta and masses take the form
\begin{equation}
  D(f)\sim a^{\delta-4}[A+B \log(1/N)]+(aN)^{\delta-4}\sum_{m=0}^\infty a_m (1/N)^m.
\end{equation}\normalsize
Therefore, multiple $ \hat{I}(N) $ with large, but similar $ N $ are extrapolated to $ N\to \infty $ with
\begin{equation}
  \hat{I}(N) = I_\infty+\frac{dI}{N^2}.
\end{equation}\normalsize

\sectionc{Propagators and vertices}{sec: Propagators and vertices}

\noindent
Propagators and vertices for Karsten-Wilczek and Bori\c{c}i-Creutz fermions have been already presented in~\cite{Capitani:2009yn, Capitani:2009ty, Capitani:2010nn, Capitani:2010ht}. Their properties are covered in detail in the next two sections. Propagators and vertices are derived from the exponentiated action in the path integral with variational methods in the limit of weak coupling $ g_0 $. 
\noindent
Since perturbative expressions are lengthy, the following abbreviations are used:
\begin{equation}
  \left.\begin{array}{rcl}
  s^\mu_p &\equiv& \sin{(ap_\mu)} \\
  c^\mu_p &\equiv& \cos{(ap_\mu)} \\
  \hat{s}^\mu_p &\equiv& 2\,\sin{(ap_\mu/2)} \\
  \hat{c}^\mu_p &\equiv& 2\,\cos{(ap_\mu/2)} 
  \end{array}\right.
  \label{eq: notation for LPT}
\end{equation}\normalsize
When the Euclidean index is omitted in terms within parentheses that include any of the expressions defined in \mbox{eq.}~(\ref{eq: notation for LPT}), their index is to be understood as being summed over (\mbox{e.g.}~$ (\hat{s}_p)\equiv \sum_\mu (\hat{s}^\mu_p)$, $ (\hat{s}_p\hat{s}_p\hat{c}_p)\equiv \sum_\mu (\hat{s}^\mu_p\hat{s}^\mu_p\hat{c}^\mu_p)$). Summations, which are restricted to all but the $ \hat{e}_{\underline{\alpha}} $~direction (\mbox{e.g.}~$ (\hat{s}_p\hat{c}_p)_{\perp} \equiv \sum_{\mu\neq\underline{\alpha}} (\hat{s}^\mu_p\hat{c}^\mu_p)$) are abbreviated with an additional lower  index ``$ _\perp $''. For a similar purpose, the abbreviation $ \varrho^{\mu\nu} $ is introduced:
\begin{equation}
  \varrho^{\mu\nu} \equiv 1-\delta^{\mu\nu}.
  \label{eq: definition of varrho^munu}
\end{equation}\normalsize

\noindent
The propagator is obtained as the inverse of the free momentum space Dirac operator,
\begin{equation}
  S(p;m,a) = \frac{D^\dagger(p;m,a)}{D(p;m,a)D^\dagger(p;m,a)}.
\end{equation}\normalsize
The vertices must be derived from a variation of the connection $ A^\mu $ instead of the gauge links $ U^\mu $, which are related by \mbox{eq.}~(\ref{eq: gauge links}). Moreover, the coupling constant must be extracted from the definition of the connection as $ A^\mu=g_0\mathcal{A}^\mu $. Since one-loop corrections are of order $ \mathcal{O}(g_0^2) $, the gauge links have to be expanded up to order $ \mathcal{O}(g_0^2) $,
\begin{equation}
  U^\mu_n = 1+ iag_0 \mathcal{A}^\mu_{n+\hat{e}_\mu/2} - \frac{a^2g_0^2}{2!} \mathcal{A}^\mu_{n+\hat{e}_\mu/2} \mathcal{A}^\mu_{n+\hat{e}_\mu/2} + \mathcal{O}(g_0^3)
\end{equation}\normalsize
which generates vertices of fermions with two gluons that produce tadpole diagrams of LPT, which do not have counterparts in continuum QCD. Each gluon field $ \mathcal{A}^\mu_{n+\hat{e}_\mu/2} $ is expressed through its Fourier transform $ \mathcal{A}^\mu(k) $.
\noindent
The gluon propagator is taken from~\cite{Capitani:2002mp}, p.~144, \mbox{eq.}~($ 5.62 $). It is obtained from an expansion of the plaquette action and requires use of the Fadeev-Popov procedure~\cite{Baaquie:1977hz}. Once a gauge has been fixed using a gauge fixing parameter $ \xi $, the gluon propagator reads
\small
\begin{equation}
  G_{\mu\nu}^{ab}(p;M,a) \equiv \delta^{ab}G_{\mu\nu}(p;M,a),\quad G_{\mu\nu}(p;M,a) \equiv \frac{\delta^{\mu\nu} - (1-\xi)\frac{\hat{s}^\mu_p\hat{s}^\nu_p}{\left(\hat{s}_p\right)^2+(aM)^2}}{\left(\hat{s}_p\right)^2+(aM)^2},
  \label{eq: gluon propagator}
\end{equation}\normalsize
where a mass parameter $ M $ is included only for the purpose of mass regularisation. The overall denominator is referred to as the gluon denominator $ D^g(p;M,a) = \left(\hat{s}_p\right)^2+(aM)^2 $.

\subsection{Karsten-Wilczek fermions}
The fermion propagator for Karsten-Wilczek fermions is obtained by inverting the free Dirac operator in \mbox{eq.} (\ref{eq: free KW action in momentum space}). It reads
\small
\begin{equation}
  S(p;\zeta,m_0,a)=a \frac{-i(\gamma \cdot s_p) 
  -i\frac{\zeta}{2} \gamma^{\underline{\alpha}} (\hat{s}_p)_\perp^2 +am_0 }
  {(s_p)^2 +\frac{\zeta^2}{4} \left((\hat{s}_p)_\perp^2\right)^2+(am_0)^2
  +\zeta s^{\underline{\alpha}}_p(\hat{s}_p)_\perp^2}
  \label{eq: KW propagator in momentum space}
\end{equation}\normalsize
and is multiplied by a Kronecker symbol $ \delta^{bc} $ in SU(3) space. Its denominator, which is later referred to as $ D(p;\zeta,m_0,a) $, vanishes at both doublers of \mbox{eq.} (\ref{eq: KW doublers}), which clearly violates condition C3 of \mbox{eq.} (\ref{eq: C3 condition}). 
Whether or not the second doubler generates an IR divergence depends on properties of the other contributions to the loop integral.
\noindent
The propagator has the correct continuum limit $ S(p;\zeta,m_0,0) = (\sum_\mu i\gamma^\mu p_\mu + m_0)^{-1} $ in the neighbourhood of its first doubler and can be interpreted as a quark propagator in the  limit $ a \to 0 $. In the neighbourhood of the second doubler, the four-momentum $ p $ must be expanded around the pole in line with~\cite{Karsten:1980wd,Tiburzi:2010bm} as
\begin{equation}
  p_\mu=\frac{\pi}{a} \delta^{\mu\underline{\alpha}}+q_\mu.
\end{equation}\normalsize
Expansion of the propagator yields
\begin{equation}
  S(q+\frac{\pi}{a}\hat{e}_{\underline{\alpha}};\zeta,m_0,a)=a \frac{-i\sum_\mu{\left(\gamma^\mu s^\mu_q (1-2 \delta^{\mu\underline{\alpha}})\right)} 
  -i\zeta \gamma^{\underline{\alpha}} (\hat{s}_q)_\perp^2 +am_0 }
  {(s_q)^2 +\frac{\zeta^2}{4} \left((\hat{s}_q)_\perp^2\right)^2+(am_0)^2
  -\zeta s^{\underline{\alpha}}_q(\hat{s}_q)_\perp^2}
\end{equation}\normalsize
which does not match a free quark in the limit $ a \to 0 $: 
\begin{equation}
  \lim_{a \to 0} S(q+\frac{\pi}{a}\hat{e}_{\underline{\alpha}};\zeta,m_0,a) = \left(\sum_\mu i\gamma^\mu q_\mu (1-2 \delta^{\mu\underline{\alpha}}) + m_0\right)^{-1}.
\end{equation}\normalsize
If the poles correspond to degenerate quarks in the limit $ a \to 0 $, they must be related by a unitary transformation like \mbox{eq.}~(\ref{eq: naive species symmetry} for na\"{i}ve fermions). It is defined as
\begin{equation}
  \lim_{q \to 0} \chi(q) \equiv \mathcal{Q} \psi(q+\frac{\pi}{a}\hat{e}_{\underline{\alpha}}), \quad 
  \lim_{q \to 0} \bar\chi(q) \equiv \bar\psi(q+\frac{\pi}{a}\hat{e}_{\underline{\alpha}}) \mathcal{Q}^\dagger.
\end{equation}\normalsize
with matrices $ \mathcal{Q} $ and $ \mathcal{Q}^\dagger $.
Thus, the propagator is transformed by left and right multiplication with their inverse matrices. $ \mathcal{Q} $ and $ \mathcal{Q}^\dagger $ equal $ Q^{\underline{\alpha}} $ of \mbox{eq.}~(\ref{eq: naive species symmetry}) up to phase factors,
\begin{equation}
  \mathcal{Q}= e^{i\vartheta} i \gamma^{\underline{\alpha}}\gamma^5, \quad
  \mathcal{Q}^\dagger = e^{-i\vartheta} i \gamma^{\underline{\alpha}}\gamma^5 = \mathcal{Q}^{-1},
  \label{eq: KW species rotation matrices}
\end{equation}\normalsize
and bring the continuum limit of the shifted propagator in the neighbourhood of the second pole to the original form in the neighbourhood of the first pole. The phase $ \vartheta $ of the matrix $ \mathcal{Q} $ is unrestricted and taken as zero for convenience. The propagator is called $ S_\chi(q;\zeta,m_0,a) $ after four-momentum shift and unitary transformation reads
\begin{align}
  S_\chi(q;\zeta,m_0,a) &\equiv \mathcal{Q}^\dagger S(q+\frac{\pi}{a}\hat{e}_{\underline{\alpha}};\zeta,m_0,a)\mathcal{Q} 
  = a\frac{-i \left(\gamma \cdot s_q \right) 
  +i\zeta \gamma^{\underline{\alpha}} (\hat{s}_q)_\perp^2 +am_0 }
  {(s_q)^2 +\frac{\zeta^2}{4} \left((\hat{s}_q)_\perp^2\right)^2+(am_0)^2
  -\zeta s^{\underline{\alpha}}_q(\hat{s}_q)_\perp^2},
  \label{eq: KW species conjugated propagator}
\end{align}\normalsize
which satisfies
\begin{equation}
  S_\chi(p;\zeta,m_0,a) = S(p;-\zeta,m_0,a).
  \label{eq: KW relation between two species}
\end{equation}\normalsize
Since the Karsten-Wilczek term of \mbox{eq.}~(\ref{eq: KW operator}) changes its sign under charge conjugation $ C $, transforming the fermions with $ (-1)^{n_{\underline{\alpha}}} \mathcal{Q} C $ (with arbitrary phase $ \vartheta $) and gauge fields with $ C $ is a local symmetry. The propagator always (with the exception of the limit $ a \to 0 $) includes both doublers, 
though $ S(p;\zeta,m_0,a) $ and $ S\chi(p;\zeta,m_0,a) $ differ by $ \mathcal{O}(a) $~effects. Since left and right multiplication of the chirality matrix $ \gamma^5 $ with matrices $ \mathcal{Q} $ and $ \mathcal{Q}^\dagger $ changes its sign as $ \mathcal{Q}^\dagger\gamma^5\mathcal{Q} = -\gamma^5  $, the two doublers always correspond to fermion modes with opposite chirality in compliance with the No-Go theorem of Nielsen and Ninomiya~\cite{Nielsen:1981hk}. 
\noindent
The fermion-fermion-gluon vertex reads
\begin{equation}
  V_1^\mu(p,q;\zeta,a)= -i\frac{g_0}{2}\left(\gamma^\mu \hat{c}^\mu_{p+q} +\zeta \gamma^{\underline{\alpha}}\varrho^{\underline{\alpha}\mu}\hat{s}^\mu_{p+q}\right)
  \label{eq: KW vertex V1}
\end{equation}\normalsize
and is multiplied by an SU(3) generator $ (T^d)^{bc} $, where $ d $ is the gluon's colour index and $ b $ and $ c $ are the fermions' colour indices. The fermion-fermion-gluon-gluon vertex reads
\begin{equation}
  V_2^{\mu\nu}(p,q;\zeta,a)= ia\frac{g_0^2}{4}\delta^{\mu\nu}\left(\gamma^\mu \hat{s}^\mu_{p+q} -\zeta \gamma^{\underline{\alpha}}\varrho^{\underline{\alpha}\mu}\hat{c}^\mu_{p+q}\right)
  \label{eq: KW vertex V2}
\end{equation}\normalsize
and is multiplied by an anticommutator of two SU(3) generators (\mbox{cf.} \mbox{eq.}~(\ref{eq: product of SU(3) generators})),
\begin{equation}
  \left\{T^d,T^e\right\}^{bc}=\left(\frac{1}{3}\delta^{de}\delta^{bc}+d^{def}(T^f)^{bc}\right),
\end{equation}\normalsize
where $ d $ and $ e $ are colour indices of the two emitted gluons. $ b $ and $ c $ are colour indices of the fermion fields. The four-momentum $ p $ is attributed to an incoming fermion field and $ q $ is attributed to an outgoing fermion field in both vertices.

\subsection{Bori\c{c}i-Creutz fermions}

\noindent
Propagators and vertices for Bori\c{c}i-Creutz fermions are obtained like those for Karsten-Wilczek fermions. In particular, the SU(3) structure is identical to the previous case and is therefore not repeated here. The fermion propagator is derived by inverting the free momentum space Bori\c{c}i-Creutz Dirac operator of \mbox{eq.}~(\ref{eq: free BC action in momentum space}) and reads
\begin{equation}
  S(p;\zeta,m_0,a) = a \frac{N(p;\zeta,m_0,a)}{D(p;\zeta,m_0,a)},
  \label{eq: BC propagator in momentum space}
\end{equation}\normalsize
with
\small
\begin{align}
  N(p;\zeta,m_0,a) &= -i\sum_\mu{\left(\gamma^\mu s^\mu_p 
  -\frac{\zeta}{2} \gamma^{\mu\prime} (\hat{s}_p^\mu)^2\right)} +am_0, 
  \label{eq: BC propagator N} \\
  D(p;\zeta,m_0,a) &= (s_p)^2 \!+\! \zeta^2\left((c_p)^2 - 2(c_p)+4\right)
  - \zeta \left(2(s_p)+2 (s_p c_p) - (s_p)(c_p)\right)+(am_0)^2.
  \label{eq: BC propagator D}
\end{align}\normalsize
Its denominator~$ D(p;\zeta,m_0,a) $ vanishes at both doublers of \mbox{eq.}~(\ref{eq: BC doublers, site basis}) and thus violates condition~C3 of \mbox{eq.}~(\ref{eq: C3 condition}). 
Whether or not the second doubler generates an IR divergence depends on properties of the other contributions to the loop integral.
\noindent
The propagator has the correct continuum limit $ S(p;\zeta,m_0,0) = (\sum_\mu i\gamma^\mu p_\mu + m_0)^{-1} $ in the neighbourhood of its first doubler and can be interpreted as a quark propagator in the limit $ a \to 0 $. In line with the considerations of\cite{Karsten:1980wd,Bedaque:2008xs}, the four-momentum $ p $ must be expanded around the pole in the neighbourhood of the second doubler as
\begin{equation}
  p_\mu=\zeta \frac{\pi}{2a}+q_\mu.
\end{equation}\normalsize
Expansion of the propagator yields
\begin{equation}
  S(q+\frac{\pi}{2a};\zeta,m_0,a) =a \frac{N^\prime(q;\zeta,m_0,a)}{D^\prime(q;\zeta,m_0,a)}
  \label{eq: BC unrotated propagator at second doubler}
\end{equation}\normalsize
with
\small
\begin{align}
  N^\prime(q;\zeta,m_0,a) =&\ -i\sum_\mu{\left(\gamma^\mu \zeta c^\mu_q
  -\zeta \gamma^{\mu\prime} (1+\zeta s^\mu_q)\right)} +am_0 \nonumber \\
  D^\prime(q;\zeta,m_0,a) =&\
  (\zeta c_q)^2 +\zeta^2\left((\zeta s_q)^2+2 \zeta (s_q)+4\right)
  -\zeta^2 \left(2(c_q)-2(c_q \zeta s_q)+(c_q)(\zeta s_q)\right)+(am_0)^2 \nonumber \\
  =&\ \zeta^4(s_q)^2 +\zeta^2 \left((c_q)^2-2 (c_q)+4\right)
  +\zeta^3 \left(2(s_q)+2 (s_q c_q)- (s_q)(c_q)\right)+(am_0)^2.
\end{align}\normalsize
\mbox{Eq.}~(\ref{eq: BC unrotated propagator at second doubler}) does not match a free quark in the limit $ a \to 0 $: 
\begin{equation}
  \lim_{a \to 0} S(q+\zeta \frac{\pi}{2a};\zeta,m_0,a) = \frac{i \zeta^2\sum_\mu \gamma^{\mu\prime} q_\mu + m_0}{\zeta^4\sum_\mu q_\mu^2 + m_0^2}.
  \label{eq: BC propagator cl in the neighbourhood of the second doubler}
\end{equation}\normalsize
If the poles correspond to degenerate quarks in the limit $ a \to 0 $, a unitary transformation similar to \mbox{eq.}~(\ref{eq: naive species symmetry} for na\"{i}ve fermions) must relate the fields. With a similar notation but different matrices $ \mathcal{Q} $ and $ \mathcal{Q}^\dagger $ that are fixed in \mbox{eq.}~(\ref{eq: BC species rotation matrices}), it is defined as
\begin{equation}
  \lim_{q \to 0} \chi(q) \equiv \lim_{q \to 0} \mathcal{Q} \psi(q+\zeta \frac{\pi}{2a}), \quad 
  \lim_{q \to 0} \bar\chi(q) \equiv \lim_{q \to 0} \bar\psi(q+\zeta \frac{\pi}{2a}) \mathcal{Q}^\dagger.
\end{equation}\normalsize
Therefore, the propagator is transformed by left and right multiplication with the inverse matrices. The shifted propagator of \mbox{eq.}~(\ref{eq: BC propagator cl in the neighbourhood of the second doubler}) can be transformed into a free quark propagator only if $ \zeta^2=1 $. Left and right multiplication of the propagator with
\begin{equation}
  \mathcal{Q}= e^{i\vartheta} i \Gamma\gamma^5, \quad
  \mathcal{Q}^\dagger = e^{-i\vartheta} i \Gamma\gamma^5 = \mathcal{Q}^{-1}
  \label{eq: BC species rotation matrices}
\end{equation}\normalsize
bring the continuum limit of the shifted propagator in the neighbourhood of the second pole to the original form in the neighbourhood of the first pole. The phase $ \vartheta $ of the matrix $ \mathcal{Q} $ is unrestricted and taken as zero for convenience. The propagator is called $ S_\chi(q;\zeta,m_0,a) $ after four-momentum shift and unitary transformation reads
\small
\begin{align}
  S_2(q;\zeta,m_0,a) &\equiv \mathcal{Q}^\dagger S(q+\frac{\pi}{2a};\zeta,m_0,a)\mathcal{Q} \\
  &= \frac{a\left( -i\sum_\mu{\left(\zeta^2 \gamma^\mu s^\mu_q
  -\frac{\zeta}{2} \gamma^{\mu\prime} (\hat{s}^\mu_q)^2\right)} +am_0 \right)}
  { \zeta^4(s_q)^2 +\zeta^2 \left((c_q)^2-2 (c_q)+4\right)
  +\zeta^3 \left(2(s_q)+2 (s_q c_q) - (s_q)(c_q)\right)+(am_0)^2} \nonumber,
\end{align}\normalsize
which satisfies
\begin{equation}
  S_\chi(p;\pm 1,m_0,a) = S(p;\mp 1,m_0,a),
  \label{eq: BC relation between two species}
\end{equation}\normalsize
which strongly resembles the corresponding relation for Karsten-Wilczek fermions of \mbox{eq.}~(\ref{eq: KW relation between two species}). Moreover, as the Bori\c{c}i-Creutz term changes its sign under charge conjugation $ C $, transforming fermion fields with $ i^{-n_\Sigma} Q C $ and $ i^{n_\Sigma} QC $ (with arbitrary phase $ \vartheta $) and gauge fields with $ C $ is a local symmetry, where $ n_\Sigma $ is defined in \mbox{eq.}~(\ref{eq: even-odd site number}).  The propagator always (with the exception of the limit $ a \to 0 $) includes both doublers, 
though $ S(p;\zeta,m_0,a) $ and $ S\chi(p;\zeta,m_0,a) $ differ by $ \mathcal{O}(a) $~effects, which has been already pointed out by Bedaque \mbox{et al.}~\cite{Bedaque:2008xs}. Since left and right multiplication of the chirality matrix $ \gamma^5 $ with matrices $ \mathcal{Q} $ and $ \mathcal{Q}^\dagger $ changes its sign as $ \mathcal{Q}^\dagger\gamma^5\mathcal{Q} = -\gamma^5  $, the two doublers always describe fermion modes with opposite chirality in compliance with the No-Go theorem of Nielsen and Ninomiya~\cite{Nielsen:1981hk}.
\noindent
The fermion-fermion-gluon vertex reads
\begin{equation}
  V_1^\mu(p,q;\zeta,a)= -i\frac{g_0}{2}\left(\gamma^\mu \hat{c}^\mu_{p+q} -\zeta \gamma^{\mu\,\prime}\hat{s}^\mu_{p+q}\right)
  \label{eq: BC vertex V1}
\end{equation}\normalsize
and the fermion-fermion-gluon-gluon vertex reads
\begin{equation}
  V_2^{\mu\nu}(p,q;\zeta,a)= ia\frac{g_0^2}{4}\delta^{\mu\nu}\left(\gamma^\mu \hat{s}^\mu_{p+q} +\zeta \gamma^{\mu\,\prime}\hat{c}^\mu_{p+q}\right),
  \label{eq: BC vertex V2}
\end{equation}\normalsize
where fermionic four-momenta represent incoming fermions and the colour structure is the same as in the previous case (\mbox{cf.} text below \mbox{eqs.}~(\ref{eq: KW vertex V1} and~\ref{eq: KW vertex V2})).

\sectionc{One-loop corrections}{sec: One-loop corrections}
After relevant technical aspects are summarised in section~\ref{sec: Technical aspects of lattice perturbation theory} and necessary propagators and vertices are derived in section~\ref{sec: Propagators and vertices}, the calculation of one-loop corrections for minimally doubled fermions is covered. Fermionic self-energy and the fermionic contribution to the vacuum polarisation are discussed in detail.

\subsection{Fermionic self-energy}\label{sec: Fermionic self-energy}

\begin{figure}[htb]
 \begin{center}
 \begin{fmffile}{se_diagrams}
  \unitlength=1mm
  \parbox{40mm}{
  \begin{fmfgraph*}(40,30)
   \fmfpen{thick}
   \fmfleft{i1} \fmfright{o1}
   \fmf{fermion,label=$ p $}{i1,v1}
   \fmf{fermion,label=$ p $}{v2,o1}
   \fmf{fermion,label=$ k $,straight}{v1,v2}
   \fmfdot{v1,v2}
   \fmffreeze
   \fmf{gluon,label=$ p-k $,left,l.dist=50,tension=0.0}{v1,v2}
  \end{fmfgraph*}
  }
  \hspace{20mm}
  \parbox{40mm}{
  \begin{fmfgraph*}(40,30)
   \fmfpen{thick}
   \fmfleft{i1} \fmfright{o1}
   \fmf{fermion,label=$ p $}{i1,v1}
   \fmf{fermion,label=$ p $}{v1,o1}
   \fmfdot{v1}
   \fmffreeze
   \fmf{gluon,label=$ k $,right,label.side=right,l.dist=100,tension=1.1}{v1,v1}
  \end{fmfgraph*}
  }
 \end{fmffile}
 \unitlength=1pt
 \vspace{-22pt}
 \caption{The $ 1 $-loop contribution to the fermionic self-energy consists of a sunset (left) and a tadpole (right) diagram.}
 \label{fig: s-e diagrams}
 \end{center}
\end{figure}
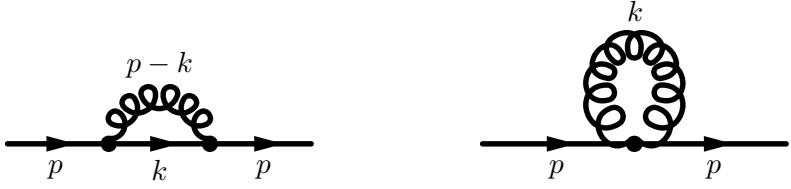

\noindent
The fermionic self-energy is calculated from the sum of the sunset diagram (left side of figure~\ref{fig: s-e diagrams}) and the tadpole diagram (right side of figure~\ref{fig: s-e diagrams}),
\begin{equation}
  I^{ad}(p;\zeta,m_0,a) = I_{s}^0(p;\zeta,m_0,a)\cdot C_s^{ad} + I_t^{0}(p;\zeta,m_0,a) \cdot C_t^{ad}.
  \label{eq: fermionic self-energy}
\end{equation}\normalsize
The sunset diagram\footnote{Since the gluon propagator of \mbox{eq.}~(\ref{eq: gluon propagator}) is isotropic and relatively simple compared to fermion propagators of \mbox{eqs.}~(\ref{eq: KW propagator in momentum space}) and~(\ref{eq: BC propagator in momentum space}), the momentum difference $ p-k $ is assigned to the gluon propagator.} involves colour indices $ a,b,c,d \in \{1,2,3\}$ and $ e,f \in \{1,\ldots,8\} $ as
\begin{align}
  I_{s}^0(p;\zeta,m_0,a)\cdot C_s^{ad} &=& \int\limits_{-\pi/a}^{+\pi/a} \frac{d^4k}{(2\pi)^4} \sum\limits_{\mu,\nu=0}^{3} V_1^\mu(p,k) S(k;\zeta,m_0,a) V_1^\nu(k,p) 
  \nonumber\\ 
  && \times 
  \sum\limits_{b,c;e,f}(T^e)^{ab} G_{\mu\nu}^{ef}(p-k;0,a) \delta^{bc} (T^f)^{cd}
\end{align}\normalsize
The tadpole diagram includes a symmetry factor $ 1/2 $,
\small
\begin{equation}
  I_t^0(p;\zeta,m_0,a) \cdot C_t^{ad} =  \frac{1}{2} \int\limits_{-\pi/a}^{+\pi/a} \frac{d^4k}{(2\pi)^4} \sum\limits_{\mu,\nu=0}^{3} V_2^{\mu\nu}(p,p) 
  \sum\limits_{b,c}
   \left(\frac{1}{3}\delta^{bc}\delta^{ad}+\sum\limits_{e} d^{bce}(T^e)^{ad}\right) G_{\mu\nu}^{bc}(k;0,a),
\end{equation}\normalsize
and involves colour indices $ a,d \in \{1,2,3\}$ and $ b,c,e \in \{1,\ldots,8\} $. The SU(3)~structure is independent of the discretisation and collapses to $ C_F\,\delta^{ad} $ after using \mbox{eq.}~(\ref{eq: Casimir operator C_F}),
\begin{align}
  C_{s}^{ad} &= \sum\limits_{b,c=1}^{3}\sum\limits_{e,f=1}^{8}(T^e)^{ab} \delta^{ef} \delta^{bc} (T^f)^{cd} = \sum\limits_{e=1}^{8}(T^eT^e)^{ad} & = C_F \delta^{ad}, \\
  C_{t}^{ad} &= \sum\limits_{b,c}
   \frac{1}{2}\left(\frac{1}{3}\delta^{bc}\delta^{ad}+\sum\limits_{b,c}d^{bce}(T^e)^{ad}\right)
 = \frac{1}{6}\sum\limits_{e=1}^{8}\delta^{ad} & = C_F \delta^{ad},
\end{align}\normalsize
which is included into the integrals $ I_{s,t}(p;\zeta,m_0,a) =C_F I_{s,t}^0(p;\zeta,m_0,a) $ as
\begin{align}
  I_s(p;\zeta,m_0,a)&= C_F \int \frac{d^4k}{(2\pi)^4} \sum\limits_{\mu,\nu} V_1^\mu(p,-k) S(k;\zeta,m_0,a) V_1^\nu(k,-p) G_{\mu\nu}(p-k;0,a),
  \label{eq: colourless fermionic self-energy sunset diagram} \\
  I_t(p;\zeta,0,a) &= C_F \int \frac{d^4k}{(2\pi)^4} \sum\limits_{\mu,\nu} V_2^{\mu\nu}(p,p) G_{\mu\nu}(k;0,a) \equiv I_t(p;\zeta,a).
  \label{eq: colourless fermionic self-energy tadpole diagram}
\end{align}\normalsize

\subsubsection{Tadpole diagram}

\noindent
Without internal fermions in the tadpole diagram, there is no problem with condition~C3 of \mbox{eq.}~(\ref{eq: C3 condition}). Mass regularisation is not required, since the superficial degree of divergence is negative, $ \mathrm{deg}\ I_t = -1$. Any $ \xi $~dependence of a power divergence must cancel exactly with the sunset diagram.
\noindent
For Karsten-Wilczek fermions, the tadpole diagram reads
\begin{align}
  I_t(p;\zeta,a) 
  =& ia^3\frac{g_0^2 C_F}{4} \int \frac{d^4k}{(2\pi)^4} \sum\limits_{\mu,\nu} \delta^{\mu\nu} \Bigg( \frac{ \left(\gamma^\mu \hat{s}^\mu_{2p} -\zeta \gamma^{\underline{\alpha}}\varrho^{\underline{\alpha}\mu}\hat{c}^\mu_{2p}\right)}{\left(\hat{s}_k\right)^2}
  \nonumber \\
  &- (1-\xi) (\hat{s}^\mu_k\hat{s}^\nu_k) 
  \frac{\left(\gamma^\mu \hat{s}^\mu_{2p} -\zeta \gamma^{\underline{\alpha}}\varrho^{\underline{\alpha}\mu}\hat{c}^\mu_{2p}\right)}{\left(\hat{s}_k\right)^2\left(\hat{s}_k\right)^2}\Bigg),
\end{align}\normalsize
which is simplified algebraically to (terms that vanish for $ a\to 0 $ have been dropped)
\small
\begin{equation}
  I_t(p;\zeta,a) 
  = ia\frac{g_0^2 C_F}{2}  \left( a\gamma\cdot p- 3\zeta \gamma^{\underline{\alpha}}\right) \left( 1- \frac{(1-\xi)}{4}\right) \int \frac{d^4k}{(2\pi)^4} \frac{1}{\left(\hat{s}_k\right)^2}.
\end{equation}\normalsize
The loop integral is identical to \mbox{eq.}~(\ref{eq: basic integral B(1)}) and the tadpole contribution reads
\begin{equation}
  I_t(p;\zeta,a) 
  = \frac{g_0^2 C_F Z_0}{2} \left( 1- \frac{(1-\xi)}{4}\right) \left( i\gamma\cdot p- i\frac{3\zeta}{a} \gamma^{\underline{\alpha}}\right).
  \label{eq: KW s-e tadpole}
\end{equation}\normalsize
\noindent
For Bori\c{c}i-Creutz fermions, the tadpole diagram reads
\begin{align}
  I_t(p;\zeta,a) 
  =& ia^3\frac{g_0^2 C_F}{4} \int \frac{d^4k}{(2\pi)^4} \sum\limits_{\mu,\nu} (\delta^{\mu\nu}) \Bigg(
  \frac{\left(\gamma^\mu \hat{s}^\mu_{2p} +\zeta \gamma^{\mu\,\prime}\hat{c}^\mu_{2p}\right)}{\left(\hat{s}_k\right)^2}
  \nonumber \\&
  - (1-\xi) 
  (\hat{s}^\mu_k\hat{s}^\nu_k) 
  \frac{\left(\gamma^\mu \hat{s}^\mu_{2p} +\zeta \gamma^{\mu\,\prime}\hat{c}^\mu_{2p}\right)}{\left(\hat{s}_k\right)^2\left(\hat{s}_k\right)^2}\Bigg),
\end{align}\normalsize
which is simplified algebraically to (terms that vanish for $ a\to 0 $ have been dropped)
\begin{equation}
  I_t(p;\zeta,a) 
  = ia\frac{g_0^2 C_F}{2} \left(a \gamma \cdot p + 2\zeta\Gamma\right) \left( 1-\frac{(1-\xi)}{4}\right)
  \int \frac{d^4k}{(2\pi)^4} \frac{1}{\left(\hat{s}_k\right)^2}.
\end{equation}\normalsize
Again, the loop integral is identical to \mbox{eq.}~(\ref{eq: basic integral B(1)}) and the tadpole contribution reads
\begin{equation}
  I_t(p;\zeta,a) 
  = \frac{g_0^2 C_F Z_0}{2} \left( 1- \frac{(1-\xi)}{4}\right) \left(i \gamma \cdot p + i\frac{2\zeta}{a}\Gamma \right).
  \label{eq: BC s-e tadpole}
\end{equation}\normalsize

\subsubsection{Sunset diagram}

\noindent
The integral $ I_s(p;\zeta,m_0,a) $ of the sunset diagram is split into two pieces, the lattice integral $ J(p,m_0;\zeta,M,a) $, and the continuum integral $ I_s(p;\zeta,m_0,a)-J(p,m_0;\zeta,M,a) $. The external momenta $ p $ and the mass $ m_0 $ are treated as small and the integral $ J(p,m_0;\zeta,M,a) $ is subjected to a Taylor expansion in $ p $ and $ m_0 $,
\small
\begin{align}
  J(p,m_0;\zeta,M,a)&= \sum\limits_{\chi} i\gamma^\chi J_3^\chi(\zeta,a) +\sum\limits_{\chi,\theta}
  i\gamma^\chi p_\theta J_4^{\theta\chi}(\zeta,M,a) + m_0 J_{m}(\zeta,M,a)
  \label{eq: s-e sunset int J} \\
  J_3^\chi(\zeta,0,a) &= 
  \!\!\int\!\!\sum\!\!\mathrm{tr}
\left\{\gamma^\chi V_1^\mu(0,k) S(k;\zeta,0,a) V_1^\nu(k,0)  G_{\mu\nu}(-k;0,a)\right\}
  \label{eq: s-e sunset int J3}, \\
  J_4^{\theta\chi}(\zeta,M,a) &=
  \!\!\int\!\!\sum\!\!\mathrm{tr}
  \left\{\gamma^\chi \frac{\partial \left\{V_1^\mu(p,k) S(k;\zeta,0,a) V_1^\nu(k,p) G_{\mu\nu}(p-k;0,a)\right\}}{\partial p_\theta}\right\}_{p=0}
  \label{eq: s-e sunset int J4}, \\
  J_{m} (\zeta,M,a) &=
   m_0\!\!\int\!\!\sum\!\!\mathrm{tr}
  \left\{\frac{\partial \left\{V_1^\mu(0,k) S(k;\zeta,m_0,a) V_1^\nu(k,0) G_{\mu\nu}(-k;0,a)\right\}}{\partial m_0}\right\}_{m_0=0}
  \label{eq: s-e sunset int Jm}.
\end{align}\normalsize
Each self-energy integral in \mbox{eqs.}~(\ref{eq: s-e sunset int J3}),~(\ref{eq: s-e sunset int J4}) and~(\ref{eq: s-e sunset int Jm}) is integrated and summed as
\begin{equation}
  \int\!\!\sum\!\!\mathrm{tr} \left( \ldots \right) \equiv \int\limits_{-\pi/a}^{+\pi/a} \frac{d^4k}{(2\pi)^4} \sum\limits_{\mu,\nu} \frac{1}{4}\mathrm{tr} \left( \ldots \right).
\end{equation}\normalsize
$ J_4^{\theta\chi}(\zeta,M,a) $ and $ J_{m} (\zeta,M,a) $ are regularised in the IR with a small mass term $ M^2 $ in each denominator. The indices $ \theta $ of the Taylor expansion in $ p $ and $ \chi $ of the projection to Dirac matrices may differ, since the anisotropy of the fermion action allows for the persistence of complicated combinations of indices in the continuum limit.

\subsubsection{Karsten-Wilczek fermions}

\noindent
For Karsten-Wilczek fermions, the power divergent integral $ J_3(\zeta,a) $ reads
\small
\begin{align}
  J_3(\zeta,a) =& \sum_\chi i\gamma^\chi J_3^\chi(\zeta,a), 
  \nonumber \\
  J_3^\chi(\zeta,a) =& a^3\frac{g_0^2 C_F}{4}
  \!\!\int\!\!\sum\!\!\mathrm{tr}\ \Big\{
  \gamma^\chi \left(\gamma^\mu \hat{c}^\mu_{k} +\zeta \gamma^{\underline{\alpha}}\varrho^{\underline{\alpha}\mu}\hat{s}^\mu_{k}\right) 
  \left( (\gamma \cdot s_k)
  +\frac{\zeta}{2} \gamma_{\underline{\alpha}} (\hat{s}_k)_\perp^2 \right)
  \nonumber \\
  &\times
  \left(\gamma^\nu \hat{c}^\nu_{k} +\zeta \gamma^{\underline{\alpha}}\varrho^{\underline{\alpha}\nu}\hat{s}^\nu_{k}\right) \Big\} \frac{\delta^{\mu\nu}-(1-\xi)\frac{\hat{s}^\mu_{-k}\hat{s}^\nu_{-k}}{D^{g}(-k;0,a)}}{D^{KW}(k;\zeta,0,a)D^{g}(-k;0,a)}.
  \label{eq: KW s-e int J3}
\end{align}\normalsize
The fermion propagator has two separate divergences at the poles of \mbox{eq.}~(\ref{eq: KW doublers}). However, the gluon denominator vanishes only at the standard pole. Thus, the superficial degree of divergence is $ \mathrm{deg}\ I_1 = -1 $ at the first pole. Since $ \mathrm{deg}\ I_2 = 1 $ at the second pole, it does not contribute to the divergence of the integral. Even though condition~C3 of \mbox{eq.}~(\ref{eq: C3 condition}) is not met by all propagators, there is no extra divergence and the power counting theorem of Reisz is applicable. Due to the negative degree of divergence, IR~regularisation is not needed. 
\noindent
The integral of \mbox{eq.}~(\ref{eq: KW s-e int J3}) is split into a Feynman gauge part ($ \mathcal{J}_0=J_3^\chi|_{\xi=1} $) and a gauge fixing part ($ \mathcal{J}_1=(\xi-1) \left(\partial J_3^\chi/\partial \xi\right) $), which is the rest of \mbox{eq.}~(\ref{eq: KW s-e int J3}). Numerators $ N_0 $ of $ \mathcal{J}_0 $ and $ N_1 $ of $ \mathcal{J}_1 $ are simplified algebraically to \mbox{eqs.}~(\ref{eq: N_0 algebraically simplified}) and (\ref{eq: N_1 algebraically simplified}). Since only $ k_{\underline{\alpha}} $ contributes to the denominator as an odd power, odd powers of other Euclidean components of $ k $ in the numerators $ N_0 $ and $ N_1 $ are integrated to zero. Hence, symmetry restricts the power divergence to the $ \gamma^{\underline{\alpha}}\, $-component. Numerators collapse to
\begin{align}
  N_0 &=\ \delta^{\chi\underline{\alpha}} \Bigg\{ 
  \left(s^{\underline{\alpha}}_k +\frac{\zeta}{2}(\hat{s}_k)_\perp^2\right) 
  \left((\hat{c}^{\underline{\alpha}}_k)^2-(\hat{c}_k)_\perp^2 +\zeta^2 \left(\hat{s}_k\right)_\perp^2\right) 
  + 4\zeta (s_k)_\perp^2\Bigg\},
  \label{eq: KW s-e sunset p div N0}
  \\
  N_1 &=\ \delta^{\chi\underline{\alpha}}\Bigg\{
  s^{\underline{\alpha}}_k \left(4 (s_k)^2
  +3\zeta^2 \left(\left(\hat{s}_k\right)_\perp^2\right)^2\right) 
  +
  \zeta \left(\hat{s}_k\right)_\perp^2\left(4 \left(s^{\underline{\alpha}}_k\right)^2 +
   2 (s_k)^2+\frac{\zeta^2}{2}\left(\left(\hat{s}_k\right)_\perp^2\right)^2 \right)\Bigg\}.
  \label{eq: KW s-e sunset p div N1}
\end{align}\normalsize
It is noteworthy that the numerators contain only terms which are even in $ k^{\underline{\alpha}} $ and odd in $ \zeta $ and vice versa, whereas the denominators add even powers of $ k^{\underline{\alpha}} $ and $ \zeta $ to odd powers of both. Thus, the overall contribution from $ J_3^\chi $ to $ c_{1L} $ is necessarily an odd function\footnote{
The contribution to $ c_{1L} $ from the tadpole in \mbox{eq.}~(\ref{eq: KW s-e tadpole}) is directly proportional to $ \zeta $.
} 
of $ \zeta $. The final form of the power divergent integral $ J_3^\chi(\zeta,a) $ reads
\begin{align}
  J_3^\chi(\zeta,a) =&\ ia^3\frac{g_0^2 C_F}{4}\delta^{\chi\underline{\alpha}}
  \int \frac{d^4k}{(2\pi)^4} \frac{N_0-(1-\xi)\frac{N_1}{D^{g}(-k;0,1)}}{D^{KW}(k;\zeta,0,1)D^{g}(-k;0,1)}.
\end{align}\normalsize
It is evaluated numerically for $ \zeta=+1 $ after rescaling (\mbox{cf.}~\mbox{eq.}~(\ref{eq: rescaling of integration momentum})) and yields
\begin{equation}
  J_3(+1,a) = \frac{i}{a}\gamma^{\underline{\alpha}}\frac{g_0^2 C_F}{16\pi^2} \left(7.166866 - 9.17479(1-\xi) \right). 
  \label{eq: KW s-e sunset c}
\end{equation}\normalsize

\noindent
Mass renormalisation is due to the integral $ J_{m}(\zeta,M,a) $ of \mbox{eq.}~(\ref{eq: s-e sunset int Jm}), which reads
\begin{align}
  J_{m_0}(m_0;\zeta,M,a) =&\ m_0 J_m(\zeta,M,a), 
  \nonumber \\
  J_{m}(\zeta,M,a) =&\ a^4 \frac{g_0^2 C_F}{4}
  \!\!\int\!\!\sum\!\!\mathrm{tr}\ \Big\{
  \left(\gamma^\mu \hat{c}^\mu_{k} +\zeta \gamma^{\underline{\alpha}}\varrho^{\underline{\alpha}\mu}\hat{s}^\mu_{k}\right) 
  \left(\gamma^\nu \hat{c}^\nu_{k} +\zeta \gamma^{\underline{\alpha}}\varrho^{\underline{\alpha}\nu}\hat{s}^\nu_{k}\right) \Big\} 
  \nonumber \\
  &\times\
  \frac{\delta^{\mu\nu}-(1-\xi)\frac{\hat{s}^\mu_{-k}\hat{s}^\nu_{-k}}{D^{g}(-k;M,a)}}{D^{KW}(k;\zeta,M,a)D^{g}(-k;M,a)}
  \label{eq: KW s-e int Jm}
\end{align}\normalsize
for Karsten-Wilczek fermions. The fermion propagator has two separate divergences at the poles of \mbox{eq.}~(\ref{eq: KW doublers}). However, the gluon denominator vanishes only at the standard pole. Thus, the superficial degree of divergence is $ \mathrm{deg}\ I_1 = 0 $ at the first pole. Since $ \mathrm{deg}\ I_2 = 2 $ at the second pole, it does not contribute to the logarithmic divergence of the integral. Though the condition~C3 of \mbox{eq.} (\ref{eq: C3 condition}) is not met by all propagators, there is no extra divergence and the power counting theorem of Reisz is applicable. 
\noindent
\mbox{Eq.}~(\ref{eq: KW s-e int Jm}) is split into Feynman gauge ($ \mathcal{J}_2 =J_m|_{\xi=1} $) and gauge fixing ($ \mathcal{J}_3 =(\xi-1)\left(\partial J_m/\partial \xi\right) $) parts. Numerators $ N_2 $ of $ \mathcal{J}_2 $ and $ N_3 $ of $ \mathcal{J}_3 $ are simplified algebraically to
\begin{align}
  N_{2} &=\ (\hat{c}_k)^2 + \zeta^2 (\hat{s}_k)_\perp^2 \\
  N_{3} &=\ 4(s_k)^2 + \zeta^2 \left((\hat{s}_k)_\perp^2\right)^2 + 4\zeta s^{\underline{\alpha}}_k (\hat{s}_k)_\perp^2.
\end{align}\normalsize
Since $ N_2 $ and $ N_3 $ contain only terms which are even in $ k $ and $ \zeta $ or odd in $ k^{\underline{\alpha}} $ and $ \zeta $, the one-loop coefficient $ \Sigma_2 $ of the mass renormalisation is an even function of $ \zeta $. Thus, the lattice contribution to mass renormalisation reads
\begin{align}
  J_m(\zeta,M,a) =&\ a^4 \frac{g_0^2 C_F}{4} 
  \int \frac{d^4k}{(2\pi)^4} \frac{N_2-(1-\xi)\frac{N_3}{D^{g}(-k;M,a)}}{D^{KW}(k;\zeta,M,a)D^{g}(-k;M,a)}
\end{align}\normalsize
and would still be IR-divergent without $ M^2 $ in the denominator. The IR~regulator is removed from $ J_m(\zeta,M,a) $ by subtracting base integrals $ \widetilde{J}_m(M,a) $ such as \mbox{eq.}~(\ref{eq: basic integral B(2)}) with appropriate prefactors. The difference $ J_m(\zeta,M,a)-\widetilde{J}_m(M,a) $ is IR-finite for any $ M^2 $ by construction. It is evaluated for $ \zeta=+1 $ after rescaling (\mbox{cf.}~\mbox{eq.}~(\ref{eq: rescaling of integration momentum})) and removing the IR~regulator,
\begin{equation}
  J_m(+1,0,a) - \widetilde{J}_m(0,a) = \frac{g_0^2 C_F}{16\pi^2} \left(-1.200712\right).
\end{equation}
Next, the IR~regularised base integrals with the necessary prefactors,
\begin{align}
  \widetilde{J}_m(M,a) \equiv&\ g_0^2 C_F \cdot 4\left(\mathcal{B}(2)-(1-\xi)\mathcal{B}(3;1)\right) 
  \nonumber \\
  =&\ g_0^2 C_F \left(
  -15.168038 +4 \log{(aM)^2}+(1-\xi)\left(3.292010 - \log{(aM)^2}\right)\right),
  \label{eq: tilde Jm}
\end{align}\normalsize
are added to the finite integral to obtain the complete IR~regularised lattice integral,
\begin{align}
  J_{m_0}(m_0;+1,M,a) =&\ m_0 \frac{g_0^2 C_F}{16\pi^2} \Big\{4\log{(aM)^2}-16.36875 
  \nonumber \\
  &\ + (1-\xi)\left(-\log{(aM)^2}+3.292010\right)\Big\}.
  \label{eq: KW s-e sunset Sigma_2}
\end{align}\normalsize
\noindent
Lastly, wavefunction renormalisation is the most laborious part of the self-energy's calculation. It is due to the integral $ J_4(p;\zeta,M,a) $ of \mbox{eq.}~(\ref{eq: s-e sunset int J4}), which reads
\begin{align}
  J_4 (p;\zeta,M,a) =& \sum_{\theta,\chi} i\gamma^\chi p_\theta J_4^{\theta\chi}(\zeta,M,a), 
  \nonumber \\
  J_4^{\theta\chi}(\zeta,M,a) =& a^4 \frac{g_0^2 C_F}{4}
  \int \frac{d^4k}{(2\pi)^4} \Big\{ \left(\frac{N_4}{D_4} + \frac{N_5}{D_5}\right) - (1 - \xi)\left(\frac{N_6 + N_7}{D_5} +\frac{N_8}{D_6}\right) \Big\}
  \label{eq: KW s-e int J4}
\end{align}\normalsize
for Karsten-Wilczek fermions. Denominators $ D_4 $ - $ D_6 $ and numerators $ N_4 $ - $ N_8 $ are shown in appendix \ref{app: Sunset diagram for Karsten-Wilczek fermions} in \mbox{eqs.}~(\ref{eq: KW s-e sunset wf ren den D4}) -~(\ref{eq: KW s-e sunset wf ren full N8}). Since any pieces of the algebraically simplified numerators of \mbox{eqs.} (\ref{eq: KW s-e sunset wf ren N4 algebraically simplified}) - (\ref{eq: KW s-e sunset wf ren N8 algebraically simplified}) contain only terms which are either even in $ k $ and $ \zeta $ or odd in both, one-loop coefficients $ \Sigma_{1} $ and $ d_{1L} $ that are obtained from $ J_4^{\theta\chi}(\zeta,M,a) $ are even functions\footnote{The $ 1 $-loop contribution to $ \Sigma_{1} $ from the tadpole diagram is independent of $ \zeta $. Since $ \Sigma_{1} $ receives a contribution from a continuum integral without any $ \zeta $ dependence, it cannot be an odd function of $ \zeta $.} 
of $ \zeta $. Since odd momenta in the denominator are restricted to $ k^{\underline{\alpha}} $, any odd powers of momenta $ k^\theta $  and $ k^\chi $ contribute only to the integral if the indices $ \theta $ and $ \chi $ are matched with each other or with $ \underline{\alpha} $. This last stage of algebraic simplification of the numerators requires equality of the indices $ \theta $ and $ \chi $:
\small
\begin{align}
  N_{4} = -\delta^{\theta\chi}&
  \left\{(s^\theta_k)^2 \big[2 - \zeta^2 \left(\delta^{\theta\underline{\alpha}}-\varrho^{\theta\underline{\alpha}}(1 + (\hat{s_k})_\perp^2) \right)\big]
  - \zeta s^{\underline{\alpha}}_k\big[
  \varrho^{\theta\underline{\alpha}}(\hat{c}^\theta_k)^2 - \delta^{\theta\underline{\alpha}}(\hat{s}_k)_\perp^2\big]\right\}, 
  \label{eq: KW s-e sunset wf ren N4} \\
  N_{5} =
   -\delta^{\theta\chi}& \Big\{ 2(s^\theta_k)^2 \big[
  (\hat{c}_k)^2-2(\hat{c}^\theta_k)^2
  -\zeta^2\delta^{\theta\underline{\alpha}}(\hat{s}_k)_\perp^2\big]
  \nonumber \\
  &+ \zeta s^{\underline{\alpha}}_k \big[
  8  (s^\theta_k)^2 + \delta^{\theta\underline{\alpha}}
  \left(8 \left((s_k)_\perp^2 - (s^{\underline{\alpha}}_k)^2\right)
  +(\hat{s}_k)_\perp^2 \left((\hat{c}^{\underline{\alpha}}_k)^2-(\hat{c}_k)_\perp^2
  +\zeta^2 ((\hat{s})_\perp^2) \right)
  \right)\big]\Big\}, 
  \label{eq: KW s-e sunset wf ren N5} \\
  N_{6} = 
  -\delta^{\theta\chi}& \Big\{2(\hat{s}^\theta_k)^2 \big[ (s_k)^2 - \zeta^2 (\hat{s}_k)_\perp^2 \left(\varrho^{\theta\underline{\alpha}} - \frac{(\hat{s}_k)_\perp^2}{2}\right)\big]
  + \zeta s^{\underline{\alpha}}_k (\hat{s}_k)_\perp^2 \big[
  3(\hat{s}^\theta_k)^2 
  - \zeta^2 \delta^{\theta\underline{\alpha}} (\hat{s}_k)_\perp^2
  \big]\Big\}, 
  \label{eq: KW s-e sunset wf ren N6} \\
  N_{7} = 
  -\delta^{\theta\chi}& \Big\{2\big[ (\hat{c}^\theta_k)^2 \left((s_k)^2 + \zeta^2 ((\hat{s}_k)_\perp^2)^2\right)
  + \left(2 - \delta^{\theta\underline{\alpha}}\right) \zeta^2 (s^\theta_k)^2 (\hat{s})_\perp^2\big]
  + 2\zeta s^{\underline{\alpha}}_k (\hat{c}^\theta_k)^2 (\hat{s}_k)_\perp^2 \Big\}, 
  \label{eq: KW s-e sunset wf ren N7} \\
  N_{8} = 
  -\delta^{\theta\chi}& 4\Big\{
  (s^\theta_k)^2 \big[ 4(s)^2 
  + 
  \zeta^2\left(2\delta^{\theta\underline{\alpha}} 
  + 1\right)((\hat{s}_k)_\perp^2)^2 \big]
  + 
  \zeta s^{\underline{\alpha}}_k (\hat{s}_k)_\perp^2 \big[
  4 (s^\theta_k)^2 
  + 
  \delta^{\theta\underline{\alpha}} \frac{8-\zeta^2((\hat{s}_k)_\perp^2)^2}{4}
  \big] \Big\}.
  \label{eq: KW s-e sunset wf ren N8}
\end{align}\normalsize
After the simplified numerators are obtained, the superficial degree of divergence is calculated as $ \mathrm{deg}\ I_1 = 0 $ at the first pole. Since $ \mathrm{deg}\ I_2 = 2 $ at the second pole, it does not contribute to the divergence of the integral. Though condition~C3 of \mbox{eq.} (\ref{eq: C3 condition}) is not fulfilled by all propagators, there is no extra divergence and the power counting theorem of Reisz is applicable. Since ratios $ N_4/D_4 $ and $ N_6/D_5 $ are due to discretisation effects, they do not contribute to the superficial degree of divergence. 
Moreover, any structure proportional to $ \delta^{\theta\underline{\alpha}} $ or $ \varrho^{\theta\underline{\alpha}} $ is either multiplied by $ \zeta^2 $ or by $ \zeta s^{\underline{\alpha}}_k $. The latter must combine with $ \zeta s^{\underline{\alpha}}_k $ in the fermionic denominator. Hence, the coefficient $ d_{1L} $ of the anisotropic dimension-four counterterm is entirely due to even powers of $ \zeta $. 
The integral $ J_4^{\theta\chi}(\zeta,M,a) $ has three divergent pieces, which require IR~regularisation with a mass $ M^2 $ in all denominators. The IR~regulator is removed from $ J_4^{\chi\theta}(\zeta,M,a) $ by subtracting base integrals $ \widetilde{J}_4^{\chi\theta}(M,a) $ of the form of \mbox{eq.}~(\ref{eq: general bosonic integral B}) with appropriate prefactors. The difference $ J_4^{\chi\theta}(\zeta,M,a)-\widetilde{J}_4^{\chi\theta}(M,a) $ is IR-finite  for any $ M^2 $ by construction. It is evaluated for $ \zeta=+1 $  after rescaling (\mbox{cf.}~\mbox{eq.}~(\ref{eq: rescaling of integration momentum})) and removing the IR~regulator,
\begin{equation}
  J_4^{\theta\chi}(+1,0,a) -\widetilde{J}_4^{\theta\chi}(0,a)
  = \frac{g_0^2 C_F}{16\pi^2}\delta^{\theta\chi}\left(2.29985 -0.12554\,\delta^{\theta\underline{\alpha}} +(1-\xi)\left(2.558262\right)\right).
\end{equation}\normalsize
Next, the IR~regularised base integrals with the necessary prefactors,
\begin{align}
  \widetilde{J}_4^{\theta\chi}(M,a)
  &=\ g_0^2 C_F \delta^{\theta\chi} \cdot 4\left(\mathcal{B}(3,1) +(1-\xi)\cdot 2 \left(\mathcal{B}(3,1)-\mathcal{B}(4,2)-3\mathcal{B}(3,1,1)\right)\right) 
  \nonumber \\
  &=\ \frac{g_0^2 C_F}{16\pi^2} \left(\log{(aM)^2}- 3.292010 +(1-\xi)\left(-\log{(aM)^2}+3.62534\right)\right),
  \label{eq: tilde J4}
\end{align}\normalsize
are added to the finite integral to obtain the complete IR~regularised lattice integral,
\begin{align}
  J_4(p;+1,M,a) =& \sum_\theta i\gamma^\theta p_\theta\frac{g_0^2 C_F}{16\pi^2} 
  \Big\{\log{(aM)^2}-0.99216-0.12554\,
  \delta^{\theta\underline{\alpha}} 
  \nonumber \\
  &+ (1-\xi)\left(-\log{(aM)^2}+6.350272\right)\Big\}.
  \label{eq: KW s-e sunset Sigma_1}
\end{align}\normalsize

\subsubsection{Bori\c{c}i-Creutz fermions}

\noindent
The preceding procedure for Karsten-Wilczek fermions is closely mirrored for Bori\c{c}i-Creutz fermions. Since technical details are very similar, individual steps are not repeated here. The superficial degree of divergence is the same for both discretisations. The second divergence of the fermion propagator violates condition~C3 of \mbox{eq.}~(\ref{eq: C3 condition}) without violating the power counting theorem of Reisz due to the presence of the gluon propagator. Mass renormalisation is applied and the subtracted integrals $ \widetilde{J}_m(M,a) $ of \mbox{eq.}~(\ref{eq: tilde Jm}) and $ \widetilde{J}_4^{\theta\chi}(M,a) $ of \mbox{eq.}~(\ref{eq: tilde J4}) are identical.
\noindent
For Bori\c{c}i-Creutz fermions, the power divergent integral $ J_3^\chi(\zeta,a) $ reads
\begin{align}
  J_3(\zeta,a) =& \sum_\chi i\gamma^\chi J_3^\chi(\zeta,a), 
  \nonumber \\
  J_3^\chi(\zeta,a) =& ia^3\frac{g_0^2 C_F}{4}
  \!\!\int\!\!\sum\!\!\mathrm{tr}\ \Big\{
  \gamma^\chi 
  \left(\gamma^\mu \hat{c}^\mu_{k} -\zeta \gamma^{\mu\,\prime}\hat{s}^\mu_{k}\right) 
  \sum\limits_{\lambda}\left(\gamma_\lambda s^\lambda_p 
  -\zeta \gamma_\lambda^\prime (1-c^\lambda_p)\right)
  \nonumber \\
  &\times
  \left(\gamma^\nu \hat{c}^\nu_{k} -\zeta \gamma^{\nu\,\prime}\hat{s}^\nu_{k}\right)
  \Big\} \frac{\delta^{\mu\nu}-(1-\xi)\frac{\hat{s}^\mu_{-k}\hat{s}^\nu_{-k}}{D^{g}(-k;0,a)}}{D^{BC}(k;\zeta,0,a)D^{g}(-k;0,a)}.
  \label{eq: BC s-e int J3}
\end{align}\normalsize
The algebra is presented in appendix \ref{app: Sunset diagram for Bori\c{c}i-Creutz fermions}. $ \zeta=+1 $ is fixed and $ J_3^\chi(+1,a) $ reads
\begin{equation}
  J_3(+1,a) = \frac{i}{a} \sum_\chi\left(\frac{1}{2}\gamma^\chi\right) \frac{g_0^2 C_F}{16\pi^2} \left(5.07558 +6.11653(1-\xi) \right).
  \label{eq: BC s-e sunset c}
\end{equation}\normalsize
Due to the different anisotropic Dirac structure, there is no Kronecker symbol for Euclidean indices here. The combination $ \sum_\chi\left(\tfrac{1}{2}\gamma^\chi\right)=\Gamma $ is the matrix of \mbox{eq.}~(\ref{eq: BC Gamma}). 
\noindent
Mass renormalisation is calculated from $ J_{m_0}(m_0;\zeta,M,a) $ of \mbox{eq.} (\ref{eq: s-e sunset int Jm}) in the same way as for Karsten-Wilczek fermions. For Bori\c{c}i-Creutz fermions, the integral reads
\small
\begin{equation}
  J_m(\zeta,M,a) = a^4 \frac{g_0^2 C_F}{4}
  \!\!\int\!\!\sum\!\!\mathrm{tr}\ \Big\{\!
  \left(\gamma^\mu \hat{c}^\mu_{k} -\zeta \gamma^{\mu\,\prime}\hat{s}^\mu_{k}\right)\! 
  \left(\gamma^\nu \hat{c}^\nu_{k} -\zeta \gamma^{\nu\,\prime}\hat{s}^\nu_{k}\right)\! 
  \Big\} 
  \frac{\delta^{\mu\nu}-(1-\xi)\frac{\hat{s}^\mu_{-k}\hat{s}^\nu_{-k}}{D^{g}(-k;M,a)}}{D^{BC}(k;\zeta,M,a)D^{g}(-k;M,a)}.
  \label{eq: BC s-e int Jm}
\end{equation}\normalsize
The algebra is shown in appendix~\ref{app: Sunset diagram for Bori\c{c}i-Creutz fermions}. $ \zeta=+1 $ is fixed and $ J_{m_0}(m_0;+1,M,a) $ reads
\begin{align}
  J_{m_0}(m_0;+1,M,a) =& m_0 \frac{g_0^2 C_F}{16\pi^2} \Big\{4\log{(aM)^2} -21.48729 
  \nonumber \\&
  + (1-\xi)\left(-\log{(aM)^2}+3.292010\right)\Big\}.
  \label{eq: BC s-e sunset Sigma_2}
\end{align}\normalsize
\noindent
Wavefunction renormalisation is due to the integral $ J_4 (p;\zeta,M,a) $ of \mbox{eq.}~(\ref{eq: s-e sunset int J4}), which is again the hardest part of the self-energy's calculation. The integral $ J_4 (p;\zeta,M,a) $ reads
\begin{align}
  J_4 (p;\zeta,M,a) =& \sum_{\theta,\chi} i\gamma^\chi p_\theta J_4^{\theta\chi}(\zeta,M,a), 
  \nonumber \\
  J_4^{\theta\chi}(\zeta,M,a) =& i a^4 \frac{g_0^2 C_F}{4}
  \int \frac{d^4k}{(2\pi)^4} \Big\{ \left(\frac{N_4}{D_4}+\frac{N_5}{D_5}\right)-(1-\xi)\left(\frac{N_6+N_7}{D_5}+\frac{N_8}{D_6}\right) \Big\}
  \label{eq: BC s-e int J4}
\end{align}\normalsize
for Bori\c{c}i-Creutz fermions, where denominators $ D_4 $ - $ D_6 $ and numerators $ N_4 $ - $ N_8 $ are presented in appendix \ref{app: Sunset diagram for Bori\c{c}i-Creutz fermions} in \mbox{eqs.}~(\ref{eq: BC s-e sunset wf ren den D4}) -~(\ref{eq: BC s-e sunset wf ren full N8}). The algebra, which is even more tedious than for Karsten-Wilczek fermions is also shown in appendix~\ref{app: Sunset diagram for Bori\c{c}i-Creutz fermions}. $ \zeta=+1 $ is fixed and $ J_4(p;+1,M,a) $ reads
\begin{align}
  J_4(p;+1,M,a) =& 
  \sum_{\theta\chi} i\gamma^\chi p_\theta \frac{g_0^2 C_F}{16\pi^2} \left(\delta^{\theta\chi} 
  \Big\{\log{(aM)^2} - 3.42642\right) + \frac{1}{2}\cdot 1.52766
  \nonumber \\
  &+ \delta^{\theta\chi} (1 - \xi)\left(-\log{(aM)^2} + 6.350272\right)\Big\}.
  \label{eq: BC s-e sunset Sigma_1}
\end{align}\normalsize

\subsubsection{Continuum integral}\label{sec: continuum integral for fermionic self-energy}

\noindent
The continuum integral $ I(p;\zeta,m_0,0)-J(p,m_0;\zeta,M,0) $ is independent of discretisations:
\begin{equation}
 K(p,m_0;M) = I(p;0,m_0,0)- \lim_{M^2\to0} J(p,m_0;0,M,0).
\end{equation}\normalsize
Even though $ K(p,m_0;M) $ is already UV~finite and IR~regularised, it is easily evaluated using dimensional regularisation. The associated additional scale $ \mathcal{M} $ cancels between $ I $ and $ J $. The external four-momentum $ p $ contributes only in the gluon propagator,
\begin{equation}
G_{\mu\nu}(p-k;0)
 = \frac{1}{(p-k)^2+M^2}\left( \delta^{\mu\nu} - (1-\xi) \left(\frac{(p-k)^\mu(p-k)^\nu}{(p-k)^2}\right)\right),
\end{equation}\normalsize
and the continuum integral reads
\small
\begin{align}
  K(p,m_0;M)
  =&\ -g_0^2 C_F \Bigg(\int\limits_{-\infty}^{+\infty} \frac{d^dk}{(2\pi)^d} \mathcal{M}^{4-d}
  \sum\limits_{\mu,\nu} \left( 
  \gamma^\mu \frac{\sum_\lambda -i\gamma^\lambda k_\lambda + m_0}{k^2}
  \gamma^\nu \right) G_{\mu\nu}(p-k) 
  \nonumber \\
  &\ + \int\limits_{-\infty}^{+\infty} \frac{d^dk}{(2\pi)^d} \mathcal{M}^{4-d} 
  \sum\limits_{\mu,\nu} \left(
  \gamma^\mu \frac{\sum_\lambda -i\gamma^\lambda k_\lambda}{k^2+M^2} 
  \gamma^\nu \right) \sum_\theta p_\theta 
  \left(\frac{\partial G_{\mu\nu}(p-k;M)}{\partial p_\theta}\right)_{p=0} 
  \nonumber\\
  &\ + \int\limits_{-\infty}^{+\infty} \frac{d^dk}{(2\pi)^d} \mathcal{M}^{4-d} 
  \sum\limits_{\mu,\nu} \left( 
  \gamma^\mu \frac{m_0}{k^2+M^2} \gamma^\nu \right) G_{\mu\nu}(k;M) \Bigg).
\end{align}\normalsize
$ K(p,m_0;M) $ is evaluated to
\begin{align}
  K(p,m_0;M) =&\
  \frac{g_0^2 C_F}{16\pi^2}\Bigg\{ i\gamma\cdot p \left(\log{\frac{p^2}{M^2}}-2 -(1-\xi)\left(\log{\frac{p^2}{M^2}}-\frac{5}{3}\right)\right)
  \nonumber \\
  &+
  m_0\left(4\left(\log{(\frac{p^2}{M^2})-2}\right)-(1-\xi)\left(\log{(\frac{p^2}{M^2})-\frac{5}{2}}\right)\right) \Bigg\}.
  \label{eq: s-e sunset continuum integral}
\end{align}\normalsize

\subsubsection{Full self-energy}

\noindent
With the calculation of the continuum integral, all pieces of the LPT self-energy calculation are completed. The full fermionic self-energy of Karsten-Wilczek fermions is the sum of \mbox{eqs.}~(\ref{eq: KW s-e tadpole}),~(\ref{eq: KW s-e sunset c}),~(\ref{eq: KW s-e sunset Sigma_2}),~(\ref{eq: KW s-e sunset Sigma_1}) and~(\ref{eq: s-e sunset continuum integral}) and reads
\begin{equation}
  \Sigma(p,m_0) = i\gamma\cdot p \Sigma_1(p) +m_0 \Sigma_2(p) +  \frac{i}{a}\gamma^{\underline{\alpha}}c_{1L}(g_0) + i \gamma^{\underline{\alpha}}p_{\underline{\alpha}} d_{1L}(g_0),
  \label{eq: KW full s-e}
\end{equation}\normalsize
where the one-loop coefficients are given by
\begin{align}
  \Sigma_1(p) =& \frac{g_0^2 C_F}{16\pi^2}\left\{\log{(ap)^2} + 9.24089 + (1 - \xi)\left(-\log{(ap)^2} + 4.792010 \right)\right\}, 
  \label{eq: KW s-e full Sigma_1} \\
  \Sigma_2(p) =& \frac{g_0^2 C_F}{16\pi^2}\left\{4\log{(ap)^2} - 24.36875  + (1 - \xi)\left(-\log{(ap)^2} + 5.792010 \right)\right\}, 
  \label{eq: KW s-e full Sigma_2} \\
  c_{1L}(g_0) =& \frac{g_0^2 C_F}{16\pi^2} \cdot \left(-29.53228\right) , 
  \label{eq: KW s-e full c} \\
  d_{1L}(g_0) =& \frac{g_0^2 C_F}{16\pi^2} \cdot \left(-0.12554\right) . 
  \label{eq: KW s-e full d}
\end{align}\normalsize
\noindent
With the inclusion of fermionic counterterms of \mbox{eqs.}~(\ref{eq: ferm dim 3 KW counterterm operator}) and~(\ref{eq: ferm dim 4 KW counterterm operator}), the one-loop propagator for Karsten-Wilczek fermions reads
\begin{align}
  \frac{1}{i\slashed{p} +m_0} +\ & 
  \frac{1}{i\slashed{p} +m_0}
  \Big\{i\slashed{p}\Sigma_1 +m_0 \Sigma_2 
  +\frac{i}{a}\gamma^{\underline{\alpha}}(c_{1L} - c)
  + i\gamma^{\underline{\alpha}}p_{\underline{\alpha}}(d_{1L} - d) \Big\}
  \frac{1}{i\slashed{p} +m_0} 
  \nonumber \\ =&\
  \frac{1}{i\slashed{p} (1-\Sigma_1) + m_0 (1-\Sigma_2) + \frac{i}{a}\gamma^{\underline{\alpha}} (c - c_{1L}) + i\gamma^{\underline{\alpha}}p_{\underline{\alpha}}(d - d_{1L}) }.
\end{align}\normalsize
Once the coefficients of the counterterms are set to
\begin{align}
  c(g_0) =&\ c_{1L}(g_0) + \mathcal{O}(g_0^4), \\
  d(g_0) =&\ d_{1L}(g_0) + \mathcal{O}(g_0^4),
\end{align}\normalsize
the anisotropy is completely removed from the renormalised propagator at one-loop level,
\begin{equation}
  \Sigma(p,m_0) = \frac{Z_2}{i\slashed{p} +Z_m m_0},
  \label{eq: 1-loop propagator}
\end{equation}
which has standard form. Wavefunction ($ Z_2 $) and mass ($ Z_m $) renormalisation factors are
\begin{align}
  Z_2 =&\ (1-\Sigma_1)^{-1},
  \label{eq: Z_2} \\
  Z_m =&\ 1-(\Sigma_2-\Sigma_1).
  \label{eq: Z_m}
\end{align}\normalsize

\noindent
The full fermionic self-energy of Bori\c{c}i-Creutz fermions has different anisotropic terms but an analogous structure,
\begin{equation}
  \Sigma(p,m_0) = i\gamma\cdot p \Sigma_1(p) +m_0 \Sigma_2(p) +  \frac{i}{a}\Gamma c_{1L}(g_0) + i \Gamma \sum\limits_\mu p_{\mu} d_{1L}(g_0),
  \label{eq: BC full s-e}
\end{equation}\normalsize
where the one-loop coefficients are given by
\begin{align}
  \Sigma_1(p) =& \frac{g_0^2 C_F}{16\pi^2}\left\{\log{(ap)^2} + 6.80663 + (1 - \xi)\left(-\log{(ap)^2} + 4.792010 \right)\right\}, 
  \label{eq: BC s-e full Sigma_1} \\
  \Sigma_2(p) =& \frac{g_0^2 C_F}{16\pi^2}\left\{4\log{(ap)^2} - 29.48729  + (1 - \xi)\left(-\log{(ap)^2} + 5.792010 \right)\right\}, 
  \label{eq: BC s-e full Sigma_2} \\
  c_{1L}(g_0) =& \frac{g_0^2 C_F}{16\pi^2} \cdot \left(+29.54170\right) , 
  \label{eq: BC s-e full c} \\
  d_{1L}(g_0) =& \frac{g_0^2 C_F}{16\pi^2} \cdot \left(+1.52766\right) . 
  \label{eq: BC s-e full d}
\end{align}\normalsize
Once fermionic counterterms of \mbox{eqs.}~(\ref{eq: ferm dim 3 BC counterterm operator}) and~(\ref{eq: ferm dim 4 BC counterterm operator}) with coefficients
\begin{align}
  c(g_0) =& c_{1L}(g_0) + \mathcal{O}(g_0^4), \\
  d(g_0) =& d_{1L}(g_0) + \mathcal{O}(g_0^4)
\end{align}\normalsize
are included, the Bori\c{c}i-Creutz fermion propagator recovers its isotropy and takes the standard form of \mbox{eq.}~(\ref{eq: 1-loop propagator}) with $ Z_2 $ and $ Z_m $ given by \mbox{eqs.}~(\ref{eq: Z_2}) and~(\ref{eq: Z_m}) using coefficients $ \Sigma_1 $ and $ \Sigma_2 $ of \mbox{eqs.}~(\ref{eq: BC s-e full Sigma_1}) and (\ref{eq: BC s-e full Sigma_2}).\newline

\noindent
Both fermionic self-energies share the same gauge-fixing contributions proportional to $ (1-\xi) $. This is necessarily true for all lattice fermions due to gauge invariance. As a consequence, the anisotropic terms proportional to $ (1-\xi) $ must invariably cancel between tadpole and sunset diagrams.
\noindent
Coefficients of dimension-three counterterms are very similar but have different sign. This seems natural since the on-site pieces of Karsten-Wilczek term in \mbox{eq.} (\ref{eq: KW operator}) and of Bori\c{c}i-Creutz term in \mbox{eq.} (\ref{eq: BC operator}) have opposite sign. In contrast, the smallness of the coefficient of the dimension four counterterm of the Karsten-Wilczek action is not thoroughly understood at present.

\subsection{Local bilinears and symmetry currents}
\label{sec: Local bilinears and symmetry currents}

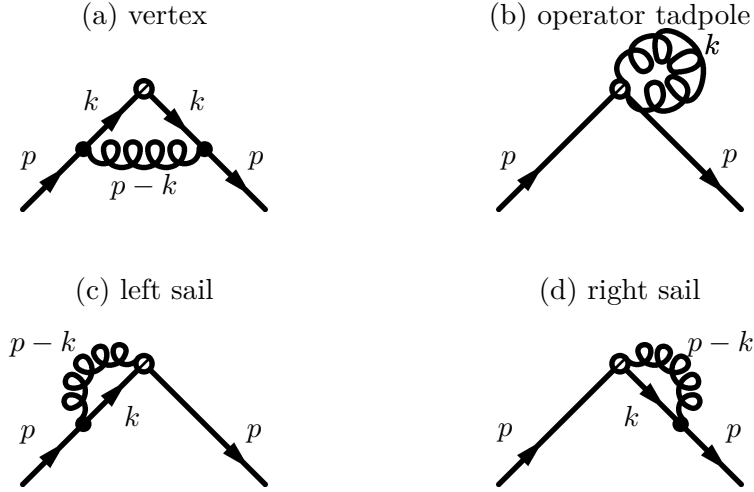
\begin{figure}[htb]
 \begin{center}
 \begin{fmffile}{fbvtx_diagrams}
  \unitlength=1mm
  \vskip1em\vspace{-33pt}
  \noindent\hspace{10mm}
  \parbox{40mm}{
  \begin{fmfgraph*}(40,40)
   \fmfpen{thick}
   \fmfleft{i1,i2} \fmfright{o1,o2}
   \fmf{fermion,label=$ p $,l.dist=-130,straight}{i1,v1}
   \fmf{fermion,label=$ k $,l.dist=-130,straight}{v1,v2}
   \fmf{fermion,label=$ k $,l.dist=-130,straight}{v2,v3}
   \fmf{fermion,label=$ p $,l.dist=+ 50,straight}{v3,o1}
   \fmf{phantom}{i2,v4,o2}
   \fmf{phantom}{v2,v4}
   \fmfdot{v1,v3}
   \fmfblob{3thick}{v2}
   \fmfv{label=\text{(a) vertex},l.dist=-200}{v4}
   \fmffreeze
   \fmf{gluon,label=$ p-k $,right,l.dist=+80,straight, weight=0.0}{v1,v3}
  \end{fmfgraph*}
  }
  \hspace{20mm}
  \parbox{40mm}{
  \begin{fmfgraph*}(40,40)
   \fmfpen{thick}
   \fmfleft{i1,i2} \fmfright{o1,o2}
   \fmf{fermion,label=$ p $,l.dist=-120,straight}{i1,v1}
   \fmf{plain,straight}{v1,v2}
   \fmf{plain,straight}{v2,v3}
   \fmf{fermion,label=$ p $,l.dist=+ 50,straight}{v3,o1}
   \fmfblob{3thick}{v2}
   \fmf{phantom}{i2,v4,o2}
   \fmf{phantom}{v2,v4}
   \fmfv{label=\text{(b) operator tadpole},l.dist=-200}{v4}
   \fmffreeze
   \fmf{gluon,rubout,right,label=$ k $,label.side=right,l.dist=100,tension=1.1}{v2,v2}
   \end{fmfgraph*}
  }
  \vskip1em\vspace{-22pt}
  \hspace{10mm}
  \parbox{40mm}{
  \begin{fmfgraph*}(40,40)
   \fmfpen{thick}
   \fmfleft{i1,i2} \fmfright{o1,o2}
   \fmf{fermion,label=$ p $,l.dist=-135,straight}{i1,v1}
   \fmf{fermion,label=$ k $,l.dist=+50,straight}{v1,v2}
   \fmf{plain,straight}{v2,v3}
   \fmf{fermion,label=$ p $,l.dist=+ 50,straight}{v3,o1}
   \fmfdot{v1}
   \fmfblob{3thick}{v2}
   \fmf{phantom}{i2,v4,o2}
   \fmf{phantom}{v2,v4}
   \fmfv{label=\text{(c) left sail},l.dist=-200}{v4}
   \fmffreeze
   \fmf{gluon,label=$ p-k $,left,l.dist=+ 30,tension=0.0}{v1,v2}
   \end{fmfgraph*}
  }
  \hspace{20mm}
  \parbox{40mm}{
  \begin{fmfgraph*}(40,40)
   \fmfpen{thick}
   \fmfleft{i1,i2} \fmfright{o1,o2}
   \fmf{fermion,label=$ p $,l.dist=-135,straight}{i1,v1}
   \fmf{plain,straight}{v1,v2}
   \fmf{fermion,label=$ k $,l.dist=+ 50,straight}{v2,v3}
   \fmf{fermion,label=$ p $,l.dist=+ 60,straight}{v3,o1}
   \fmfdot{v3}
   \fmfblob{3thick}{v2}
   \fmf{phantom}{i2,v4,o2}
   \fmf{phantom}{v2,v4}
   \fmfv{label=\text{(d) right sail},l.dist=-200}{v4}
   \fmffreeze
   \fmf{gluon,label=$ p-k $,left,l.dist=+ 30,tension=0.0}{v2,v3}
   \end{fmfgraph*}
  }
  \vskip1em\vspace{-11pt}
 \end{fmffile}
 \unitlength=1pt
 \caption{Four diagrams are required for the proper vertex renormalisation of the symmetry currents. Merely the vertex diagram (a) is sufficent for local bilinears.}
 \label{fig: vertex diagrams for renormalisation of bilinears}
 \end{center}
\end{figure}

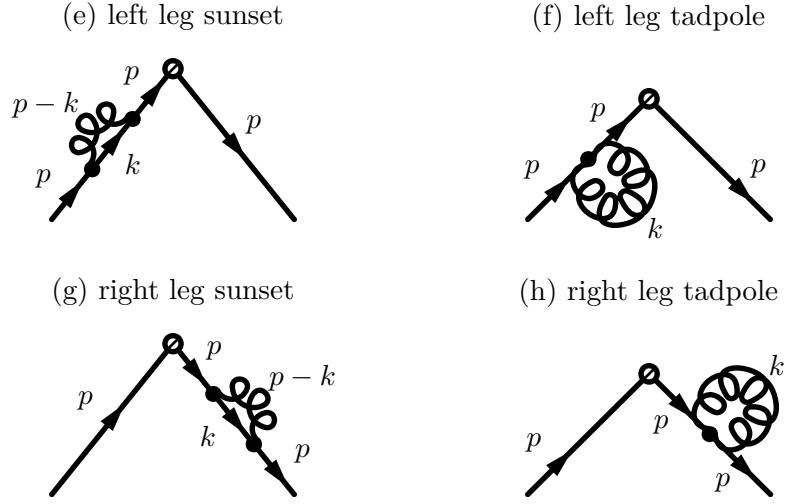
\begin{figure}[htb]
 \begin{center}
 \begin{fmffile}{fbleg_diagrams}
  \unitlength=1mm
  \vskip1em\vspace{-33pt}
  \noindent\hspace{10mm}
    \parbox{40mm}{
  \begin{fmfgraph*}(40,40)
   \fmfpen{thick}
   \fmfleft{i1,i2} \fmfright{o1,o2}
   \fmf{fermion,label=$ p $,l.dist=-135,straight}{i1,v1}
   \fmf{fermion,label=$ k $,l.dist=+ 50,straight}{v1,v2}
   \fmf{fermion,label=$ p $,l.dist=+ 50,straight}{v2,v3}
   \fmf{plain}{v3,v4}
   \fmf{fermion,label=$ p $,l.dist=+ 50,straight}{v4,v5}
   \fmf{plain}{v5,o1}
   \fmfdot{v1,v2}
   \fmfblob{3thick}{v3}
   \fmf{phantom}{i2,v6,o2}
   \fmf{phantom}{v3,v6}
   \fmfv{label=\text{(e) left leg sunset},l.dist=-200}{v6}
   \fmffreeze
   \fmf{gluon,label=$ p-k $,left,l.dist=+ 30,tension=0.0}{v1,v2}
   \end{fmfgraph*}
  }
  \hspace{20mm}
  \parbox{40mm}{
  \begin{fmfgraph*}(40,40)
   \fmfpen{thick}
   \fmfleft{i1,i2} \fmfright{o1,o2}
   \fmf{fermion,label=$ p $,l.dist=-135,straight}{i1,v1}
   \fmf{fermion,label=$ p $,l.dist=-115,straight}{v1,v2}
   \fmf{plain}{v2,v3}
   \fmf{fermion,label=$ p $,l.dist=+ 50,straight}{v3,o1}
   \fmfdot{v1}
   \fmfblob{3thick}{v2}
   \fmf{phantom}{i2,v4,o2}
   \fmf{phantom}{v2,v4}
   \fmfv{label=\text{(f) left leg tadpole},l.dist=-170}{v4}
   \fmffreeze
   \fmf{gluon,label=$ k $,label.side=right,l.dist=100,tension=1.1}{v1,v1}
   \end{fmfgraph*}
  }
  \vskip1em\vspace{-22pt}
  \hspace{10mm}
  \parbox{40mm}{
  \begin{fmfgraph*}(40,40)
   \fmfpen{thick}
   \fmfleft{i1,i2} \fmfright{o1,o2}
   \fmf{plain}{i1,v1}
   \fmf{fermion,label=$ p $,l.dist=-135,straight}{v1,v2}
   \fmf{plain}{v2,v3}
   \fmf{fermion,label=$ p $,l.dist=+ 50,straight}{v3,v4}
   \fmf{fermion,label=$ k $,l.dist=-135,straight}{v4,v5}
   \fmf{fermion,label=$ p $,l.dist=-135,straight}{v5,o1}
   \fmfdot{v4,v5}
   \fmfblob{3thick}{v3}
   \fmf{phantom}{i2,v6,o2}
   \fmf{phantom}{v3,v6}
   \fmfv{label=\text{(g) right leg sunset},l.dist=-200}{v6}
   \fmffreeze
   \fmf{gluon,label=$ p-k $,left,l.dist=+ 40,tension=0.0}{v4,v5}
   \end{fmfgraph*}
  }
  \hspace{20mm}
  \parbox{40mm}{
  \begin{fmfgraph*}(40,40)
   \fmfpen{thick}
   \fmfleft{i1,i2} \fmfright{o1,o2}
   \fmf{fermion,label=$ p $,l.dist=-135,straight}{i1,v1}
   \fmf{plain}{v1,v2}
   \fmf{fermion,label=$ p $,l.dist=+ 50,straight}{v2,v3}
   \fmf{fermion,label=$ p $,l.dist=-115,straight}{v3,o1}
   \fmfdot{v3}
   \fmfblob{3thick}{v2}
   \fmf{phantom}{i2,v4,o2}
   \fmf{phantom}{v2,v4}
   \fmfv{label=\text{(h) right leg tadpole},l.dist=-170}{v4}
   \fmffreeze
   \fmf{gluon,label=$ k $,label.side=right,l.dist=100,tension=1.1}{v3,v3}
   \end{fmfgraph*}
  }
  \vskip1em\vspace{-11pt}
 \end{fmffile}
 \unitlength=1pt
 \caption{Renormalisation of the legs contributes four further diagrams to the renormalisation of local bilinears and symmetry currents.}
 \label{fig: leg diagrams for renormalisation of bilinears}
 \end{center}
\end{figure}

Renormalisation factors for local fermionic bilinears are required for many applications. They are computed by evaluating the vertex diagram (a) of figure~\ref{fig: vertex diagrams for renormalisation of bilinears} and adding the self-energy contribution for the legs of figure~\ref{fig: leg diagrams for renormalisation of bilinears}. Because both fermion actions preserve chiral symmetry, they each have identical scalar and pseudoscalar renormalisation factors. Moreover, vector and axial-vector renormalisation factors are identical due to chiral symmetry for each action. These local currents undergo anisotropic mixing, which is not absorbed by the counterterms of their actions. The tensor current, however, does not have anisotropic contributions. One-loop results for local currents are presented as a summary.
\noindent
Symmetry currents are covered in more detail. They are constructed using infinitesimal vector and axial transformations, which yield point-split current operators. In the chiral limit, the point-split axial-vector current is conserved as well. The symmetry currents further require contributions from two sails ((c) and (d)) and from the operator tadpole (b) of figure~\ref{fig: vertex diagrams for renormalisation of bilinears}. The sum of these proper contributions exactly cancels the self-energy contribution from the legs of figure~\ref{fig: leg diagrams for renormalisation of bilinears}. Therefore, the renormalisation constant is exactly one before any counterterms are included.

\subsubsection{Karsten-Wilczek fermions}

\noindent
Vertex corrections for insertion of scalar or pseudoscalar densities are equal: $ \Lambda_P=\Lambda_S \gamma^5 $. The vertex correction of the scalar density equals $ -\Sigma_2(p) $ of \mbox{eq.}~(\ref{eq: KW s-e full Sigma_2}) and reads
\begin{equation}
  \Lambda_S = \frac{g_0^2 C_F}{16\pi^2} \left(-4\log{(ap)^2} + 24.36875  + (1 - \xi)\left(\log{(ap)^2} - 5.792010 \right)\right)
\end{equation}\normalsize
Thus, the renormalisation factor of the pseudoscalar density,
\begin{equation}
  Z_P = 1-(\Lambda_P+\Sigma_1),
\end{equation}\normalsize
is the inverse of the mass renormalisation factor $ Z_m $ of \mbox{eq.}~(\ref{eq: Z_m}). 
\noindent
Local vector and axial currents undergo anisotropic renormalisation. The vertex correction to the local vector current reads
\begin{equation}
  \Lambda_V^\mu = \frac{g_0^2 C_F}{16\pi^2} \cdot \gamma^\mu \left(-\log{(ap)^2} \!+\! 10.44610 \!-\! 2.88914 \!\cdot\! \delta^{\mu\underline{\alpha}} \!+\! (1 \!-\! \xi)\left(\log{(ap)^2} \!-\! 4.792010\right)\right)
\end{equation}\normalsize
and the vertex correction to the axial current is obtained as $ \Lambda_A^\mu =\Lambda_V^\mu \gamma^5 $. Finally, the tensor current does not undergo anisotropic renormalisation and reads
\begin{equation}
  \Lambda_T^{\mu\nu} = \frac{g_0^2 C_F}{16\pi^2} \sigma^{\mu\nu} \left(4.17551+(1-\xi)\left(L-3.792010\right)\right).
\end{equation}\normalsize

\noindent
Symmetry currents are constructed by applying infinitesimal vector and axial transformations to the fermion fields in the action,
\begin{equation}
  \left.\begin{array}{lrclrc}
  \delta_V \psi_n &=& i \alpha^V_n \psi_n, & \delta_V \bar\psi_n &=& -i \alpha^V_n \bar\psi_n,  \\
  \delta_A \psi_n &=& i \alpha^A_n \gamma^5 \psi_n, & \delta_A \bar\psi_n &=& +i \alpha^A_n \bar\psi_n \gamma^5,
  \end{array}\right.
  \label{eq: vector and axial transformations}
\end{equation}\normalsize
which yield point-split symmetry currents
\begin{align}
  V^\mu_n =& 
  \bar\psi_n
  \frac{
  \gamma^\mu(1 + d\delta^{\mu\underline{\alpha}})-i \zeta \gamma^{\underline{\alpha}}\varrho^{\underline{\alpha}\mu}
  }{2}
  U^\mu_n \psi_{n+\hat{e}_\mu}
  +\bar\psi_{n+\hat{e}_\mu}
  \frac{
  \gamma^\mu(1 + d\delta^{\mu\underline{\alpha}})+i \zeta \gamma^{\underline{\alpha}}\varrho^{\underline{\alpha}\mu}
  }{2}
  U^{\mu\dagger}_n \psi_{n}
  , 
  \label{eq: KW cvc}
  \\
  A^\mu_n =& 
  \bar\psi_n
    \frac{
  \gamma^\mu(1 + d\delta^{\mu\underline{\alpha}})-i \zeta \gamma^{\underline{\alpha}}\varrho^{\underline{\alpha}\mu}
  }{2}
  \gamma^5 U^\mu_n \psi_{n+\hat{e}_\mu}
  +\bar\psi_{n+\hat{e}_\mu}
  \frac{
  \gamma^\mu(1 + d\delta^{\mu\underline{\alpha}})+i \zeta \gamma^{\underline{\alpha}}\varrho^{\underline{\alpha}\mu}
  }{2}
  \gamma^5 U^{\mu\dagger}_n \psi_{n}
  .
  \label{eq: KW cac}
\end{align}\normalsize
Whereas the vector transformation $ \delta_V \psi_n $ commutes with the matrices $ \mathcal{Q} $ and $ \mathcal{Q}^\dagger $ of \mbox{eq.}~(\ref{eq: KW species rotation matrices}), the axial transformation anticommutes with $ \mathcal{Q} $ and $ \mathcal{Q}^\dagger $,
\begin{equation}
  \gamma^5 \mathcal{Q} = -\mathcal{Q} \gamma^5, \quad 
  \mathcal{Q}^\dagger \gamma^5 = -\gamma^5 \mathcal{Q}^\dagger.
  \label{eq: non-singlet gamma5}
\end{equation}\normalsize
The axial transformation produces a current which has opposite sign for both doublers~\footnote{The axial current is interpreted as an isovector current unrelated to the axial anomaly in~\cite{Tiburzi:2010bm}. There, an explicit expression for the isosinglet axial current is presented and the Abelian axial anomaly for Karsten-Wilczek fermions in two dimensions is derived from it perturbatively.}.
\noindent
Only the vector symmetry current of \mbox{eq.}~(\ref{eq: KW cvc}) is covered in detail here, since the axial current of \mbox{eq.}~(\ref{eq: KW cac}), which is a symmetry current in the chiral limit, is treated identically. Evaluation of the vertex correction, the two sails and the operator tadpole of figure~\ref{fig: vertex diagrams for renormalisation of bilinears} yields the proper one-loop contribution to the vector symmetry current
\begin{equation}
  \frac{g_0^2 C_F}{16\pi^2} \gamma^\mu \left(-\log{(ap)^2} -9.24089 + 0.12554 \cdot \delta^{\mu\underline{\alpha}} + (1 - \xi)\left(\log{(ap)^2}-4.79201 \right)  \right),
  \label{eq: KW cvc 1-loop proper contribution}
\end{equation}\normalsize
if the counterterm's contribution in \mbox{eq.}~(\ref{eq: KW cvc}) is excluded. Using $ \Sigma_1 $ and $ d_{1L} $ of \mbox{eqs.}~(\ref{eq: KW s-e full Sigma_1}) and (\ref{eq: KW s-e full d}), this proper contribution is expressed as
\begin{equation}
  \gamma^\mu (-\Sigma_1(p)-d_{1L}(g_0)\delta^{\mu\underline{\alpha}}).
\end{equation}\normalsize
Since the self-energy contribution from the legs not only includes the wavefunction renormalisation $ Z_2 = (1-\Sigma_1)^{-1} $, but also the anisotropic contribution from $ d_{1L} $, the full renormalisation factor amounts to
\begin{equation}
  \gamma^\mu (-\Sigma_1-d_{1L}\delta^{\mu\underline{\alpha}}+(1-\Sigma_1)^{-1}+d_{1L}\delta^{\mu\underline{\alpha}}) =\gamma^\mu \cdot 1
\end{equation}\normalsize
and supports the claim of current conservation. In particular, the vector symmetry current is conserved even before inclusion of counterterms. Hence it cannot be used as a device for tuning the coefficient $ d $ in contradiction to previous conclusions~\cite{Capitani:2010nn}. In particular, the Ward identity is satisfied only if the point-split symmetry current is used instead of the local axial current.

\subsubsection{Bori\c{c}i-Creutz fermions}

\noindent
Vertex corrections for insertion of scalar or pseudoscalar densities are equal: $ \Lambda_P=\Lambda_S \gamma^5 $. The vertex correction of the scalar density equals $ -\Sigma_2(p) $ of \mbox{eq.}~(\ref{eq: BC s-e full Sigma_2}) and reads
\begin{equation}
  \Lambda_S = \frac{g_0^2 C_F}{16\pi^2} \left(-4\log{(ap)^2} + 29.48729  + (1 - \xi)\left(\log{(ap)^2} - 5.792010 \right)\right).
\end{equation}\normalsize
Local vector and axial currents undergo anisotropic renormalisation. The vertex correction to the local vector current reads
\begin{eqnarray}
  \Lambda_V^\mu &=& \frac{g_0^2 C_F}{16\pi^2} \cdot \sum\limits_\nu \gamma^\nu \Bigg(\delta^{\mu\nu}\left(-\log{(ap)^2} + 9.54612\right) - \frac{1}{2} \cdot 0.10037  
  \nonumber \\ &+& 
  \delta^{\mu\nu}(1 - \xi)\left(\log{(ap)^2} - 4.792010\right)\Bigg)
\end{eqnarray}\normalsize
and the vertex correction to the axial current is $ \Lambda_A^\mu =\Lambda_V^\mu \gamma^5 $. Lastly, the tensor current does not undergo anisotropic renormalisation and reads
\begin{equation}
  \Lambda_T^{\mu\nu} = \frac{g_0^2 C_F}{16\pi^2} \sigma^{\mu\nu} \left(2.16548+(1-\xi)\left(L-3.792010\right)\right).
\end{equation}\normalsize
\noindent
Symmetry currents which are constructed using transformations of \mbox{eq.}~(\ref{eq: vector and axial transformations}) read
\begin{align}
  V^\mu_n =& 
  \bar\psi_n
  \frac{
  \gamma^\mu + d^{BC} \Gamma+i\gamma^{\mu \prime}
  }{2}
  U^\mu_n \psi_{n+\hat{e}_\mu}
  +
  \bar\psi_{n+\hat{e}_\mu}
  \frac{
  \gamma^\mu + d^{BC} \Gamma-i\gamma^{\mu \prime}
  }{2}
  U^{\mu\dagger}_n \psi_{n}
  , 
  \label{eq: BC cvc}\\
  A^\mu_n =& 
  \bar\psi_n
  \frac{
  \gamma^\mu + d^{BC} \Gamma+i\gamma^{\mu \prime}
  }{2}
  \gamma^5 U^\mu_n \psi_{n+\hat{e}_\mu}
  +
  \bar\psi_{n+\hat{e}_\mu}
  \frac{
  \gamma^\mu + d^{BC} \Gamma-i\gamma^{\mu \prime}
  }{2}
  \gamma^5 U^{\mu\dagger}_n \psi_{n}
  .
  \label{eq: BC cac}
\end{align}\normalsize
Analogously to the case for Karsten-Wilczek fermions in\mbox{eq.}~(\ref{eq: non-singlet gamma5}), the axial transformation anticommutes with the matrices of \mbox{eq.}~(\ref{eq: BC species rotation matrices}). 
\noindent 
Again, only the vector symmetry current is covered in detail. The axial current, which is conserved in the chiral limit, is treated equally. Evaluation of the vertex correction, the sails and the operator tadpole of figure~\ref{fig: vertex diagrams for renormalisation of bilinears} yields the proper one-loop contribution to the vector symmetry current
\small
\begin{equation}
  \frac{g_0^2 C_F}{16\pi^2} \sum\limits_\nu \gamma^\nu \Big\{-\delta^{\mu\nu}\left(\log{(ap)^2} +6.80664\right)
  -\frac{1}{2} \cdot 1.52766
  +\delta^{\mu\nu}(1 - \xi)\left(\log{(ap)^2}-4.79201 \right)  \Big\},
  \label{eq: BC cvc 1-loop proper contribution}
\end{equation}\normalsize
if the counterterm's contribution in \mbox{eq.}~(\ref{eq: BC cvc}) is excluded. This proper contribution is expressed using $ \Sigma_1 $ and $ d_{1L} $ of \mbox{eqs.}~(\ref{eq: BC s-e full Sigma_1}) and~(\ref{eq: BC s-e full d}) and yields
\begin{equation}
  \sum\limits_\nu \gamma^\nu (-\delta^{\mu\nu}\Sigma_1(p)-\frac{1}{2} \cdot d_{1L}(g_0)),
\end{equation}\normalsize
which combines with the self-energy contribution from the legs to
\begin{equation}
  \sum\limits_\nu \gamma^\nu (-\delta^{\mu\nu}\Sigma_1-\frac{1}{2} \cdot d_{1L} + \delta^{\mu\nu} (1-\Sigma_1)^{-1}+\frac{1}{2} \cdot d_{1L}\delta^{\mu\underline{\alpha}}) =\gamma^\mu \cdot 1.
\end{equation}\normalsize
Hence, the vector symmetry current is conserved even before inclusion of counterterms and the Ward identity is satisfied only if the point-split symmetry current is used instead of the local axial current.

\subsection{Fermionic contribution to the vacuum polarisation}\label{sec: Fermionic contribution to the vacuum polarisation}

\begin{figure}[htb]
 \begin{center}
 \begin{fmffile}{vp_diagrams}
  \unitlength=1mm
  \parbox{40mm}{
  \begin{fmfgraph*}(40,30)
   \fmfpen{thick}
   \fmfleft{i1} \fmfright{o1}
   \fmf{gluon,label=$ p $,l.dist=100}{i1,v1}
   \fmf{gluon,label=$ p $,l.dist=100}{v2,o1}
   \fmf{fermion,label=$ p+k $,left,tension=1.0}{v1,v2}
   \fmfdot{v1,v2}
   \fmffreeze
   \fmf{fermion,label=$ k $,left,tension=1.0}{v2,v1}
  \end{fmfgraph*}
  }
  \hspace{20mm}
  \parbox{40mm}{
  \begin{fmfgraph*}(40,30)
   \fmfpen{thick}
   \fmfleft{i1} \fmfright{o1}
   \fmf{gluon,label=$ p $,l.dist=100}{i1,v1}
   \fmf{gluon,label=$ p $,l.dist=100}{v1,o1}
   \fmfdot{v1}
   \fmffreeze
   \fmf{fermion,label=$ k $,right,label.side=right,l.dist=50,tension=0.9}{v1,v1}
  \end{fmfgraph*}
  }
 \end{fmffile}
 \unitlength=1pt
 \vspace{-11pt}
 \caption{The one-loop fermionic contribution to the vacuum polarisation consists of a bubble (left) and a tadpole (right) diagram, the latter of which cancels the quadratic power divergence of the former.}
 \label{fig: vp diagrams}
 \end{center}
\end{figure}
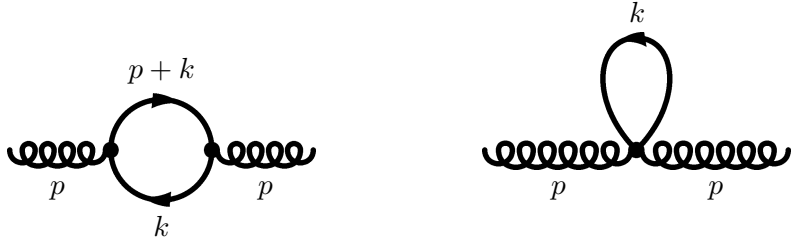
\noindent
Since Karsten-Wilczek and Bori\c{c}i-Creutz fermions break hypercubic symmetry, it must be expected that fermion loops communicate this anisotropy to quantities that are isotropic at tree level. First and foremost, the validity of the Ward-Takahashi identity has to be confirmed, because the tensor structure of the vacuum polarisation tensor is modified in an anisotropic theory. The fermionic contribution to the vacuum polarisation, which is simply referred to as ``the vacuum polarisation'' here for brevity's sake, reveals the extent of induced anisotropy. The gluonic dimension four counterterm is used to restore isotropy to its continuum limit.
\noindent
The bubble diagram (left side of figure~\ref{fig: vp diagrams}) of the vacuum polarisation introduces a new aspect, which has not been encountered in the preceding LPT calculations. Due to the fact that all internal lines are fermion propagators, both poles of the fermion propagator contribute on equal footing. Hence, a na\"{i}ve approach to the bubble diagram does not satisfy condition C3 of \mbox{eq.}~(\ref{eq: C3 condition}) for the applicability of the power counting theorem of Reisz.
 \newline
 
\noindent
The vacuum polarisation is calculated from the sum of the bubble diagram and the tadpole diagram (right side of figure \ref{fig: vp diagrams}),
\begin{equation}
  I^{ad\,\theta\chi}(p;\zeta,a) = I_b^{0\,\theta\chi}(p;\zeta,a) \cdot C_b^{ad} + I_t^{0\,\theta\chi}(p;\zeta,a) \cdot C_t^{ad},
  \label{eq: vacuum polarisation, fermionic contribution}
\end{equation}\normalsize
where space-time parts of bubble and tadpole diagrams are given by
\begin{align}
  I_b^{0\,\theta\chi}(p;\zeta,a) =& -\int\limits_{-\pi/a}^{+\pi/a} \frac{d^4k}{(2\pi)^4} \mathrm{tr} \left\{ V_1^\theta(k,p+k) S(p+k;\zeta,0,a) V_1^\chi(p+k,k) S(k;\zeta,0,a) \right\}, 
  \\
  I_t^{0\,\theta\chi}(p;\zeta,a) =&  - \int\limits_{-\pi/a}^{+\pi/a} \frac{d^4k}{(2\pi)^4} \mathrm{tr} \left\{ V_2^{\theta\chi}(k,k) S(k;\zeta,0,a) \right\}.
\end{align}\normalsize
Both integrals contain an overall factor~$ (-1) $ due to having \textit{one closed fermion loop}. The symmetry factor~$ 1/2 $ of the tadpole diagram is allotted to $ C_t^{ad} $. The SU(3)~structure is independent of the discretisation and collapses to $ C_2\, \delta^{a,d} $ using \mbox{eq.}~(\ref{eq: Casimir operator C_2}),
\begin{align}
  I_{b,c}^{ad} =& \sum\limits_{b,c,e,f=1}^{3}(T^a)^{fb}  \delta^{bc} (T^d)^{ce} \delta^{ef}= \mathrm{tr}(T^aT^d) = C_2 \delta^{ad}, \\
  I_{t,c}^{ad} =& \frac{1}{2} \sum\limits_{b,c}
   \delta^{bc}\left(\frac{1}{3}\delta^{bc}\delta^{ad}+\sum\limits_{b,c}d^{bce}(T^e)^{ad}\right)
 = \frac{1}{2} \delta^{ad} = C_2 \delta^{ad}.
\end{align}\normalsize
which is included in the integrals $ I_{b,t}^{\theta\chi}(p;\zeta,a) \equiv C_2 I_{b,t}^{0\,\theta\chi}(p;\zeta,a) $ as
\begin{align}
  I_b^{\theta\chi}(p;\zeta,a) &= -C_2 \int\limits_{-\pi/a}^{+\pi/a} \frac{d^4k}{(2\pi)^4} \mathrm{tr} \left\{ V_1^\theta(k,p+k) S(p+k;\zeta,0,a) V_1^\chi(p+k,k) S(k;\zeta,0,a) \right\},
  \label{eq: colourless vacuum polarisation bubble diagram} \\
  I_t^{\theta\chi}(p;\zeta,a) &=  -C_2 \int\limits_{-\pi/a}^{+\pi/a} \frac{d^4k}{(2\pi)^4} \mathrm{tr} \left\{ V_2^{\theta\chi}(k,k) S(k;\zeta,0,a) \right\} \equiv I_t^{\theta\chi}(\zeta,a).
  \label{eq: colourless vacuum polarisation tadpole diagram}
\end{align}\normalsize

\subsubsection{Tadpole diagram}
The tadpole diagram of \mbox{eq.}~(\ref{eq: colourless vacuum polarisation tadpole diagram}) is independent of the external four-momentum and yields a constant, which -- by virtue of its mass dimension -- must be a quadratic power divergence. Hence, the superficial degree of divergence of the tadpole diagram is negative, $ \mathrm{deg}\ I_t = -2$, IR~regularisation is not required and there is no problem with condition C3 of \mbox{eq.}~(\ref{eq: C3 condition}).
\noindent
For Karsten-Wilczek fermions, the tadpole diagram reads
\begin{equation}
  I_t^{\theta\chi}(\zeta,a) 
  = -\delta^{\theta\chi} a^2\frac{g_0^2 C_2}{4} \int \frac{d^4k}{(2\pi)^4} \mathrm{tr}\Big\{ 
  \left(\gamma^\theta \hat{s}^\theta_{2k} -\zeta \gamma^{\underline{\alpha}}\varrho^{\theta\underline{\alpha}}\hat{c}^\theta_{2k}\right)
  \frac{(\gamma \cdot s_k) +\frac{\zeta}{2}\gamma^{\underline{\alpha}}(\hat{s}_k)_\perp^2}{D^{KW}(k;\zeta,0,a)}\Big\}
,
\end{equation}\normalsize
which is simplified algebraically to
\begin{equation}
  I_t^{\theta\chi}(\zeta,a) 
  = -\delta^{\theta\chi} a^2 g_0^2 C_2 
  \int \frac{d^4k}{(2\pi)^4} \frac{
  2(s^\theta_k)^2 + \zeta^2 (\hat{s}_k)_\perp^2(c_k)_\perp
  +2\zeta s^{\underline{\alpha}}_k \big[\delta^{\theta\underline{\alpha}}-\varrho^{\theta\underline{\alpha}}(c_k)_\perp\big]}{D^{KW}(k;\zeta,0,a)}.
\end{equation}\normalsize
Because the odd power in $ \zeta s^{\underline{\alpha}}_k $ in the numerator must combine with the odd power of $ \zeta s^{\underline{\alpha}}_k $ in the denominator or be trivially integrated to zero, the integral is an even function of $ \zeta $. After setting $ \zeta=+1 $ and rescaling (\mbox{cf.} \mbox{eq.}~(\ref{eq: rescaling of integration momentum})), the integral is evaluated to
\begin{equation}
  I_t^{\theta\chi}(+1,a) 
  = \delta^{\theta\chi} \frac{g_0^2 C_2}{a^2} \left(-36.31464 \cdot \varrho^{\theta\underline{\alpha}}+7.12931 \cdot \delta^{\theta\underline{\alpha}} \right).
  \label{eq: KW vp tadpole}
\end{equation}\normalsize
\noindent
For Bori\c{c}i-Creutz fermions, the tadpole diagram reads
\begin{align}
  I_t^{\theta\chi}(\zeta,a) 
  =& -\delta^{\theta\chi} a^2\frac{g_0^2 C_2}{4} \int \frac{d^4k}{(2\pi)^4} \sum\limits_{\mu} \mathrm{tr}\Big\{ 
  \frac{\left(\gamma^\theta \hat{s}^\theta_{k+k}+\zeta \gamma^{\theta\,\prime}\hat{c}^\theta_{k+k}\right)
  \left(\gamma^\mu s^\mu_k -\zeta\gamma^{\mu\,\prime}\frac{1}{2}(\hat{s}^\mu_k)^2\right)}{D^{BC}(k;\zeta,0,a)}\Big\}
,
\end{align}\normalsize
which is simplified algebraically to
\begin{equation}
  I_t^{\theta\chi}(\zeta,a) 
  = -\delta^{\theta\chi} a^2 g_0^2 C_2 
  \int \frac{d^4k}{(2\pi)^4} \frac{
  2\big[(s^\theta_k)^2 + \zeta^2 \left((c_k)-(c_k)^2\right)\big]
  +6\zeta (s_k)}{D^{BC}(k;\zeta,0,a)}.
\end{equation}\normalsize
Since the odd power in $ \zeta (s_k) $ in the numerator must combine with the odd power of $ \zeta (s_k) $ in the denominator or be trivially integrated to zero, the integral is an even function of $ \zeta $. After setting $ \zeta=+1 $ and rescaling (\mbox{cf.}~\mbox{eq.}~(\ref{eq: rescaling of integration momentum})), the integral is evaluated to
\begin{equation}
  I_t^{\theta\chi}(+1,a) 
  = \delta^{\theta\chi} \frac{g_0^2 C_2}{a^2} \cdot \left(-73.71980 \right).
  \label{eq: BC vp tadpole}
\end{equation}\normalsize

\subsubsection{Bubble diagram}
The integral $ I_b^{\theta\chi}(p;\zeta,a) $ of the bubble diagram is split into two pieces, the lattice integral $ J^{\theta\chi}(p;\zeta,M,a) $, and the continuum integral $ I_b^{\theta\chi}(p;\zeta,a)-J^{\theta\chi}(p;\zeta,M,a) $. The external four-momentum $ p $ is treated as small and the integral $ J^{\theta\chi}(p;\zeta,M,a) $ is subjected to a Taylor expansion in $ p $ up to second order,
\small
\begin{align}
  J^{\theta\chi}(p;\zeta,a)&= J_2^{\theta\chi}(\zeta,a)+ \sum\limits_{\eta,\xi} p_\eta p_\xi J_{\eta\xi}^{\theta\chi}(\zeta,M,a)
  \label{eq: vp bubble int J} \\
  J_2^{\theta\chi} (\zeta,a) &= 
  -\int\!\!\mathrm{tr}
\left\{V_1^\theta(k,k) S(k;\zeta,0,a) V_1^\chi(k,k) S(k;\zeta,0,a) \right\}
  \label{eq: vp bubble int J2}, \\
  J_{\eta\xi}^{\theta\chi} (\zeta,M,a) &= -\frac{1}{2}
  \int\!\!\mathrm{tr}
  \left\{\frac{\partial^2 \left\{V_1^\theta(k,p+k) S(p+k;\zeta,M,a) V_1^\chi(p+k,k) S(k;\zeta,M,a)\right\}}{\partial p_\eta \partial p_\xi}\right\}_{p=0}
  \label{eq: vp bubble int J_etaxi},
\end{align}\normalsize
where each of the vacuum polarisation integrals in \mbox{eqs.} (\ref{eq: vp bubble int J2} and \ref{eq: vp bubble int J_etaxi} is integrated as
\begin{equation}
  \int\!\!\mathrm{tr} \left( \ldots \right) \equiv \int\limits_{-\pi/a}^{+\pi/a} \frac{d^4k}{(2\pi)^4} \mathrm{tr} \left( \ldots \right)
\end{equation}\normalsize
and where $ J_{\eta\xi}^{\theta\chi} (\zeta,M,a) $ is IR~regularised with a small mass term $ M^2 $ in every denominator. The indices $ \eta $ and $ \xi $ of the Taylor expansion in $ p $ do not necessarily have to match indices $ \theta $ and $ \chi $ of the vertices, since the anisotropy of the fermion action allows for the persistence of various combinations of indices in the continuum limit.

\subsubsection{Karsten-Wilczek fermions}

\noindent
For Karsten-Wilczek fermions, the power divergent integral $ J_2^{\theta\chi}(\zeta,a) $ reads
\small
\begin{align}
  J_2^{\theta\chi}(\zeta,a) =& 
  -a^2\frac{g_0^2 C_2}{4}
  \!\!\int\!\!\mathrm{tr}\ \Big\{
  \left(\gamma^\theta \hat{c}^\theta_{2k} +\zeta \gamma^{\underline{\alpha}}\varrho^{\theta\underline{\alpha}}\hat{s}^\theta_{2k}\right) 
  \frac{(\gamma \cdot s_k) + \frac{\zeta}{2} \gamma_{\underline{\alpha}} (\hat{s}_k)_\perp^2}{D^{KW}(k;\zeta,0,a)}
  \nonumber \\
  &\times
  \left(\gamma^\chi \hat{c}^\chi_{2k} +\zeta \gamma^{\underline{\alpha}}\varrho^{\theta\underline{\alpha}}\hat{s}^\chi_{2k}\right)
  \frac{(\gamma \cdot s_k) + \frac{\zeta}{2} \gamma_{\underline{\alpha}} (\hat{s}_k)_\perp^2}{D^{KW}(k;\zeta,0,a)}
  \Big\}.
  \label{eq: KW vp int J2}
\end{align}\normalsize
The part of the numerator, which is non-vanishing upon integration,
\small
\begin{align}
  N_0 = \delta^{\theta\chi}& \Big\{ 
  2(s^\theta_k)^2-4(s_k)^2(c^\theta_k)^2 
  -4\zeta s^{\underline{\alpha}}_k \big[(1 \!-\! 2\delta^{\theta\underline{\alpha}})(\hat{s}_k)_\perp^2(c^\theta_k)^2
  -\varrho^{\theta\underline{\alpha}}(s^\theta_k)^2 (4c^\theta_k+\zeta^2((\hat{s}_k)_\perp^2)^2)\big]
  \nonumber \\
  &- \zeta^2 \big[(1 \!-\! 2\delta^{\theta\underline{\alpha}})((\hat{s}_k)_\perp^2)^2(c^\theta_k)^2
  -
  \varrho^{\theta\underline{\alpha}}(s^\theta_k)^2 (4\left((s^{\underline{\alpha}}_k)^2-(s_k)_\perp^2\right)+\zeta^2((\hat{s}_k)_\perp^2)^2)
  \big]
  \Big\},
\end{align}\normalsize
has been simplified algebraically by taking into account that only $ k^{\underline{\alpha}} $ contributes as an odd power in the denominator. Therefore, the integral $ J_2(\zeta,a) $ is reduced to
\begin{equation}
  J_2^{\theta\chi}(\zeta,a) = 
  -\delta^{\theta\chi} a^2 g_0^2 C_2 \int \frac{d^4k}{(2\pi)^4} \frac{N_0}{(D^{KW}(k;\zeta,0,a))^2}.
\end{equation}\normalsize
Odd powers in $ \zeta s^{\underline{\alpha}}_k $ in the numerator must combine with odd powers of $ \zeta s^{\underline{\alpha}}_k $ in the denominator and the integral turns out to be an even function of $ \zeta $. 
The power divergent integral is evaluated after setting $ \zeta=+1 $ and rescaling (\mbox{cf.} \mbox{eq.} (\ref{eq: rescaling of integration momentum})) and exactly cancels the power divergent tadpole contribution of \mbox{eq.} (\ref{eq: KW vp tadpole}):
\begin{equation}
  J_2^{\theta\chi}(+1,a) 
  = \delta^{\theta\chi} \frac{g_0^2 C_2}{a^2} \left(36.31464 \cdot \varrho^{\theta\underline{\alpha}}-7.12931 \cdot \delta^{\theta\underline{\alpha}} \right).
  \label{eq: KW vp bubble pd}
\end{equation}\normalsize
\noindent
The integral $ J_{\eta\xi}^{\theta\chi} (\zeta,M,a) $ of \mbox{eq.} (\ref{eq: vp bubble int J_etaxi}) is far too cumbersome for explicit presentation. Due to the exclusive presence of two fermion propagators, it is IR~divergent at both poles and has to be regularised at both in order to satisfy condition C3 of \mbox{eq.} (\ref{eq: C3 condition}). A simple method to obtain both IR divergences is to calculate the standard IR divergence\footnote{
If the lattice divergence is automatically constructed from the lattice integral, it may pick up extra finite terms due to discretisation effects. These extra terms are due to higher derivatives of trigonometric function of external four-momenta. They are included in the regularised lattice integral (\mbox{cf.}~\mbox{eq.}~(\ref{eq: KW vp bubble reg int})).} 
from the continuum integral $ J_{\eta\xi}^{\theta\chi} (M,0) $, replace all loop momenta as $ k^\mu \to \hat{s}_k^\mu $ and finally add a second standard IR divergence that is shifted from the second pole to the first. The continuum integral $ J_{\eta\xi}^{\theta\chi} (\zeta,M,0) $ reads
\begin{align}
  J_{\eta\xi}^{\theta\chi} (M,0) =& \frac{g_0^2 C_2}{2}
  \int\limits_{-\infty}^{+\infty} \frac{d^dk}{(2\pi)^d}   \mathrm{tr}
  \left\{\gamma^\theta \left\{\frac{\partial^2 S(p+k;0,M,0)}{\partial p_\eta \partial p_\xi}\right\}_{p=0} \gamma^\chi S(k;0,M,0)\right\}
  \nonumber \\
  =& -g_0^2 C_2\, 4  \int\limits_{-\infty}^{+\infty} \frac{d^dk}{(2\pi)^d} \Big\{ 
 \frac{\Big[
 \delta^{\theta\chi}\delta^{\eta\xi}k^2
 -(\delta^{\eta\chi}\delta^{\theta\xi}+\delta^{\eta\theta}\delta^{\chi\xi})
 ((k^\theta)^2+(k^\chi)^2) \Big] }{(k^2+M^2)^3} 
    \nonumber \\& 
  -4 \frac{\Big[
  \delta^{\theta\chi}\delta^{\eta\xi}
  (k^\eta)^2(k^2-2(k^\theta)^2) 
  -2(\delta^{\eta\chi}\delta^{\theta\xi}+\delta^{\eta\theta}\delta^{\chi\xi})
  ((k^\theta)^2 (k^\chi)^2) \Big] }{(k^2+M^2)^4}\Big\}.
  \label{eq: vp continuum IR divergent int}
\end{align}\normalsize
It is transformed into a lattice integral for the first divergence without any obstacles,
\begin{align}
  \widetilde{J_1}_{\eta\xi}^{\theta\chi} (M,a)
  =& -4\,g_0^2 C_2 a^4 \int\limits_{-\pi/a}^{+\pi/a} \frac{d^dk}{(2\pi)^d} \Big\{ 
 \frac{\Big[
 \delta^{\theta\chi}\delta^{\eta\xi}(\hat{s}_k)^2
 -(\delta^{\eta\chi}\delta^{\theta\xi}+\delta^{\eta\theta}\delta^{\chi\xi})
 ((\hat{s}_k^\theta)^2+(\hat{s}_k^\chi)^2) \Big] }
 {((\hat{s}_k)^2+(aM)^2)^3} 
    \nonumber \\& 
  -4 \frac{\Big[
  \delta^{\theta\chi}\delta^{\eta\xi}
  (\hat{s}_k^\eta)^2((\hat{s}_k)^2-2(\hat{s}_k^\theta)^2) 
  -2(\delta^{\eta\chi}\delta^{\theta\xi}+\delta^{\eta\theta}\delta^{\chi\xi})
  ((\hat{s}_k^\theta)^2 (\hat{s}_k^\chi)^2) \Big] }
  {((\hat{s}_k)^2+(aM)^2)^4}\Big\}
  \nonumber \\
  =& 4\, g_0^2 C_2 \Big\{4\delta^{\theta\chi}\delta^{\eta\xi}
  \Big[\mathcal{B}(3,1)-(1-2\delta^{\eta\theta})\mathcal{B}(4,2)-(1+2\delta^{\eta\theta})\mathcal{B}(4,1,1)\Big]
  \nonumber \\&
  -2(\delta^{\eta\chi}\delta^{\theta\xi}+\delta^{\eta\theta}\delta^{\chi\xi})
  \Big[\mathcal{B}(3,1)-4\delta^{\theta\chi}\mathcal{B}(4,2)-4(1-\delta^{\theta\chi})\mathcal{B}(4,1,1)\Big]\Big\}.
  \label{eq: vp lattice first IR divergent int}
\end{align}\normalsize
The second divergence is obtained by a shift $ k^{\underline{\alpha}} \to k^{\underline{\alpha}} -\pi/a $, which transforms the trigonometric functions as
\begin{equation}
  (\hat{s}_k^\theta)^2 \to (\hat{t}_k^\theta)^2, \quad 
  (\hat{t}_k^\theta)^2 \equiv (\hat{s}_k^\theta)^2 + 2\delta^{\theta\underline{\alpha}}(2-(\hat{s}_k^{\underline{\alpha}})^2), \quad 
  (\hat{t}_k)^2 \equiv (\hat{s}_k)^2+2(2-(\hat{s}_k^{\underline{\alpha}})^2).
  \label{eq: KW vp divergence shift}
\end{equation}\normalsize
Thus, the second divergence reads
\begin{equation}
  \widetilde{J_2}_{\eta\xi}^{\theta\chi} (M,a) =
  \left.\widetilde{J_1}_{\eta\xi}^{\theta\chi} (M,a) \right|_{\hat{s}_k \to \hat{t}_k},
  \label{eq: vp lattice second IR divergent int}
\end{equation}\normalsize
which is numerically identical to \mbox{eq.} (\ref{eq: vp lattice first IR divergent int}) after integration. The tensor components\footnote{
Other tensor components of the divergences are integrated to $ 0 $ once symmetries between different Euclidean components of loop momenta in the numerator have been exploited.
} 
of each divergence $ \widetilde{J_j}_{\eta\xi}^{\theta\chi} (M,a)  $ are invariant under interchanges of $ \theta $ and $ \chi $ and interchanges of $ \eta $ and $ \xi $. Non-vanishing tensor components are evaluated to
\begin{equation}
  \left.\begin{array}{rll}
  \widetilde{J_j}_{\theta\chi}^{\theta\chi} (M,a) =& 
  +16 g_0^2 C_2 (\mathcal{B}(3,1)-4\mathcal{B}(4,1,1)) &= +4.13588 - \frac{4}{3} \log{(aM)^2}, \\
  \widetilde{J_j}_{\theta\theta}^{\eta\eta} (M,a) =& 
  -16 g_0^2 C_2 (\mathcal{B}(3,1)-\mathcal{B}(4,2)-\mathcal{B}(4,1,1)) &= -5.84941 + \frac{4}{3} \log{(aM)^2}.
  \end{array}\right.
  \label{eq: tilde Jj}
\end{equation}\normalsize
After subtraction of the divergences $ \widetilde{J_j}_{\eta\xi}^{\theta\chi} (M,a)  $, the lattice integral is automatically IR~finite by construction. The regulator is removed and loop momenta are rescaled (\mbox{cf.}~\mbox{eq.} (\ref{eq: rescaling of integration momentum})). For $ \zeta=+1 $, the integral evaluates to
\begin{align}
  J_{\eta\xi}^{\theta\chi} (+1,0,a)-&\sum\limits_{j=1}^2\widetilde{J_j}_{\eta\xi}^{\theta\chi}(0,a)
  = \delta^{\theta\chi}\delta^{\eta\xi}
  \Big\{ -2.51804-\frac{4}{3}\Big\}
  +(\delta^{\eta\chi}\delta^{\theta\xi}+\delta^{\eta\theta}\delta^{\chi\xi}) \Big\{ \frac{+7.2785}{2} \Big\}
  \nonumber \\
  &+ \Big[2\delta^{\theta\chi}\delta^{\eta\xi}(\delta^{\theta\underline{\alpha}}+\delta^{\eta\underline{\alpha}})-(\delta^{\eta\chi}\delta^{\theta\xi}+\delta^{\eta\theta}\delta^{\chi\xi})(\delta^{\theta\underline{\alpha}}+\delta^{\chi\underline{\alpha}})\Big]\frac{12.69766}{4}.
  \label{eq: KW vp bubble reg int}
\end{align}\normalsize
After adding the divergences $ \widetilde{J_j}_{\eta\xi}^{\theta\chi} (M,a) $ of \mbox{eq.} (\ref{eq: tilde Jj}) and including external momenta $ p_\eta $ and $ p_\xi $, the lattice integral finally yields
\begin{align}
  J^{\theta\chi} =& \left(p^\theta p^\chi -\delta^{\theta\chi}p^2\right)\frac{g_0^2 C_2}{16\pi^2} \left(-\frac{8}{3}\log{\left(aM\right)^2}+15.55024\right)
  \nonumber \\
  &-\left(p^\theta p^\chi
  (\delta^{\theta\underline{\alpha}}+\delta^{\chi\underline{\alpha}})
  -\delta^{\theta\chi}
  (p^2\delta^{\theta\underline{\alpha}}\delta^{\chi\underline{\alpha}}+(p^{\underline{\alpha}})^2) \right)\frac{g_0^2 C_2}{16\pi^2} \cdot 12.69766,
  \label{eq: KW vp bubble Pif}
\end{align}\normalsize
where it is noted explicitly that the term $ -8/3 \log{\left(aM\right)^2} $ is due to subtraction of IR~divergences for both poles. The Ward identity is satisfied despite the anisotropic contribution.

\subsubsection{Bori\c{c}i-Creutz fermions}

Bori\c{c}i-Creutz fermions mirror the strategy for Karsten-Wilczek fermions. The power divergent integral $ J_2^{\theta\chi}(\zeta,a) $ reads
\small
\begin{align}
  J_2^{\theta\chi}(\zeta,a) =& 
  -a^2\frac{g_0^2 C_2}{4}
  \!\!\int\!\!\mathrm{tr}\ \Big\{
  \left(\gamma^\theta \hat{c}^\theta_{2k} -\zeta \gamma^{\theta\,\prime}\hat{s}^\theta_{2k}\right)
  \frac{\left( \sum_\mu{\left(\gamma_\mu s^\mu_k 
  -\zeta \gamma_\mu^\prime (1-c^\mu_k)\right)} \right)}{D^{BC}(k;\zeta,0,a)}
  \nonumber \\
  &\times
  \left(\gamma^\chi \hat{c}^\chi_{2k} -\zeta \gamma^{\chi\,\prime}\hat{s}^\chi_{2k}\right)
  \frac{\left( \sum_\nu{\left(\gamma_\nu s^\nu_k 
  -\zeta \gamma_\nu^\prime (1-c^\nu_k)\right)} \right)}{D^{KBC}(k;\zeta,0,a)}
  \Big\}
  \label{eq: BC vp int J2}
\end{align}\normalsize
and is evaluated directly for $ \zeta=+1 $ after rescaling (\mbox{cf.}~\mbox{eq.}~(\ref{eq: rescaling of integration momentum})) as
\begin{equation}
  J_2^{\theta\chi}(\zeta,a) 
  =  \delta^{\theta\chi} \frac{g_0^2 C_2}{a^2} \cdot \left(73.71980 \right),
  \label{eq: BC vp bubble pd}
\end{equation}\normalsize
which precisely cancels the tadpole's power divergence of \mbox{eq.}~(\ref{eq: BC vp tadpole}). 
\noindent
The second divergence of $ J_{\eta\xi}^{\theta\chi} (\zeta,M,a) $ is obtained with a shift $ k^\mu \to k^\mu -\pi/(2a) \ \forall \mu $ between the two poles, which transforms the trigonometric functions as
\begin{equation}
  (\hat{s}_k^\theta)^2 \to (\hat{t}_k^\theta)^2, \quad 
  (\hat{t}_k^\theta)^2 \equiv 2(1-s_k^\theta), \quad 
  (\hat{t}_k)^2 \equiv 2(4-(s_k)).
  \label{eq: BC vp divergence shift}
\end{equation}\normalsize
After subtraction of the divergences $ \widetilde{J_j}_{\eta\xi}^{\theta\chi} (M,a) $, the lattice integral is automatically IR~finite by construction. The regulator is removed and loop momenta are rescaled (\mbox{cf.}~\mbox{eq.}~(\ref{eq: rescaling of integration momentum})). For $ \zeta=+1 $, the integral evaluates to
\begin{align}
  J_{\eta\xi}^{\theta\chi} (0,a)-\sum\limits_{j=1}^2\widetilde{J_j}_{\eta\xi}^{\theta\chi} (0,a)
  =& \delta^{\theta\chi}\delta^{\eta\xi}
  \Big\{ -4.38389-\frac{4}{3}\Big\}
  +(\delta^{\eta\chi}\delta^{\theta\xi}+\delta^{\eta\theta}\delta^{\chi\xi}) \Big\{ \frac{+9.1443}{2} \Big\}
  \nonumber \\
  &+ \Big[\delta^{\theta\chi}+\delta^{\eta\xi}-(\delta^{\eta\chi}+\delta^{\theta\xi}+\delta^{\eta\theta}+\delta^{\chi\xi})\Big] \cdot 0.9094.
  \label{eq: BC vp bubble reg int}
\end{align}\normalsize
After adding the divergences $ \widetilde{J_j}_{\eta\xi}^{\theta\chi} (M,a) $ of \mbox{eq.}~(\ref{eq: tilde Jj}) and including external momenta $ p_\eta $ and $ p_\xi $, the lattice integral finally yields
\begin{align}
  J^{\theta\chi} =& \left(p^\theta p^\chi -\delta^{\theta\chi}p^2\right)\frac{g_0^2 C_2}{16\pi^2} \left(-\frac{8}{3}\log{\left(aM\right)^2}+ 19.2349\right)
  \nonumber \\
  &-\left( (p^\theta+p^\chi) \left(\sum\limits_\lambda p^\lambda\right) - p^2 - \delta^{\theta\chi}\left(\sum\limits_\lambda p^\lambda\right)^2 \right)\frac{g_0^2 C_2}{16\pi^2} \cdot 0.9094,
  \label{eq: BC vp bubble Pif}
\end{align}\normalsize
where it is noted explicitly that the term $ -8/3 \log{\left(aM\right)^2} $ is due to subtraction of IR~divergences for both poles. The Ward identity is satisfied despite the anisotropic contribution.

\subsubsection{Continuum integral}
Evaluation of the continuum integral mirrors section \ref{sec: continuum integral for fermionic self-energy}. However, the limit $ a \to 0 $ for small $ p $ in $ I^{\theta\chi}(p;\zeta,0)-J^{\theta\chi}(p;\zeta,M,0) $ contains propagators with only one single fermion mode each and thus corresponds to a one-flavour theory. Hence, the continuum integral requires an extra factor~$ 2 $ for the two flavours,
\begin{equation}
  K^{\theta\chi}(p;M) = 2(I^{\theta\chi}(p;0,0)-\lim_{M^2\to 0} J^{\theta\chi}(p;0,M,0)).
\end{equation}\normalsize
$ K^{\theta\chi}(p;M) $ is evaluated using dimensional regularisation even though it is already UV~finite and IR~regularised. Cancellation of the presence of the IR~regulator of \mbox{eqs.}~(\ref{eq: KW vp bubble Pif}) and~(\ref{eq: BC vp bubble Pif}) would fail without the flavour factor~$ 2 $. The additional dimensional regulator cancels between $ I^{\theta\chi}(p;0,0) $ and $ J^{\theta\chi}(p;0,M,0) $ and the integral reads
\small
\begin{align}
  K^{\theta\chi}(p;M)
  =& +g_0^2 C_2 \Bigg(\int\limits_{-\infty}^{+\infty} \frac{d^dk}{(2\pi)^d} \mathcal{M}^{4-d}
  \mathrm{tr} 
  \left\{\gamma^\theta S(p+k;0,0,0) \gamma^\chi S(k;0,0,0)\right\}
  \nonumber \\
  &-\frac{1}{2}\sum\limits_{\eta,\xi} p^\eta p^\xi \mathrm{tr}
  \left\{\gamma^\theta \left\{\frac{\partial^2 S(p+k;0,M,0)}{\partial p_\eta \partial p_\xi}\right\}_{p=0} \gamma^\chi S(k;0,M,0)\right\}
  \Bigg).
\end{align}\normalsize
The power divergence of the zeroth order of the Taylor expansion vanishes in dimensional regularisation in the limit of vanishing IR~regulator~$ M $.
The integral evaluates to
\begin{align}
  K(p;M) =&
  \left(p^\theta p^\chi-\delta^{\theta\chi}p^2\right)\frac{g_0^2 C_2}{16\pi^2}\cdot 2\cdot\Bigg\{ -\frac{4}{3}\log{\frac{p^2}{M^2}} +\frac{20}{9}\Bigg\}.
  \label{eq: vp bubble continuum integral}
\end{align}\normalsize

\subsubsection{Full fermionic contribution to the vacuum polarisation}

The sum of \mbox{eqs.}~(\ref{eq: KW vp tadpole}),~(\ref{eq: KW vp bubble pd}),~(\ref{eq: KW vp bubble Pif}) and~(\ref{eq: vp bubble continuum integral}) completes the full fermionic contribution to the vacuum polarisation for Karsten-Wilczek fermions and reads
\begin{equation}
  I^{\theta\chi}(p) = \left(p^\theta p^\chi -\delta^{\theta\chi}p^2\right) \Pi(p^2)
  +A^{\theta\chi}(p) \widetilde{\Pi}(g_0),
  \label{eq: full vp}
\end{equation}\normalsize
where
\begin{align}
  \Pi(p^2) =& \frac{g_0^2 C_2}{16\pi^2} \left(-\frac{8}{3}\log{\left(ap\right)^2}+19.99468\right) 
  \label{eq: KW vp full Pi}\\
  A^{\theta\chi}(p) =& p^\theta p^\chi
  (\delta^{\theta\underline{\alpha}}+\delta^{\chi\underline{\alpha}})
  -\delta^{\theta\chi}
  (p^2\delta^{\theta\underline{\alpha}}\delta^{\chi\underline{\alpha}}+(p^{\underline{\alpha}})^2)
  \label{eq: KW anisotropic tensor structure}\\
  \widetilde{\Pi}(g_0) =& -\frac{g_0^2 C_2}{16\pi^2} \cdot 12.69766.
  \label{eq: KW vp full tildePi}
\end{align}\normalsize
Even before inclusion of counterterms, the Ward identity is satisfied:
\begin{equation}
  \sum\limits_\theta p_\theta I^{\theta\chi}(p) = 
  \sum\limits_\chi p_\chi I^{\theta\chi}(p) = 0.
  \label{eq: Ward identity for vp tensor}
\end{equation}\normalsize
Using $ \mathcal{Q} $ from \mbox{eq.} (\ref{eq: KW species rotation matrices}), the anisotropic term $ A^{\theta\chi}(p) $ is expressed as
\small
\begin{equation}
  A^{\theta\chi}(p) = \frac{1}{4} \left(
  \frac{1}{2}[\mathcal{Q},\slashed{p}]
  \left(\{\gamma^\theta,\slashed{p}\}[\gamma^\chi,\mathcal{Q}]+\{\gamma^\chi,\slashed{p}\}[\gamma^\theta,\mathcal{Q}]\right)
  -p^2 [\mathcal{Q},\gamma^\theta] [\gamma^\chi,\mathcal{Q}]
  - \delta^{\theta\chi}[\mathcal{Q},\slashed{p}] [\slashed{p},\mathcal{Q}] \right)
  \label{eq: anisotropic Dirac structure}
\end{equation}\normalsize
In the presence of the anisotropic term of \mbox{eq.} (\ref{eq: KW anisotropic tensor structure}), the tree-level tensor structure cannot be recoverd in the full one-loop gluon propagator. Once the anisotropic term has been removed by tuning the gluonic counterterm of \mbox{eq.} (\ref{eq: gluonic KW counterterm}) using \mbox{eq.} (\ref{eq: KW vp full tildePi}),
\begin{equation}
  d_P(g_0) = -\frac{g_0^2 C_2}{16\pi^2} \cdot 12.69766 + \mathcal{O}(g_0^4),
\end{equation}\normalsize
the one-loop gluon propagator takes the standard form
\begin{equation}
  \Gamma^{\mu\nu}(p) = Z_3\frac{\delta^{\mu\nu}-(1-\xi(1+\Pi(p^2)))\frac{p^\mu p^\nu}{p^2}}{p^2}
  \label{eq: 1-loop gluon propagator}
\end{equation}\normalsize
and the charge renormalisation factor $ Z_3 $ is obtained as
\begin{equation}
  Z_3 = (1-\Pi(0))^{-1}.
  \label{eq: Z_3}
\end{equation}\normalsize

\noindent
The sum of \mbox{eqs.}~(\ref{eq: BC vp tadpole}),~(\ref{eq: BC vp bubble pd}),~(\ref{eq: BC vp bubble Pif}) and~(\ref{eq: vp bubble continuum integral}) completes the full fermionic contribution to the vacuum polarisation for Bori\c{c}i-Creutz fermions \mbox{eq.}~(\ref{eq: full vp}) with
\begin{align}
  \Pi(p^2) =& \frac{g_0^2 C_2}{16\pi^2} \left(-\frac{8}{3}\log{\left(ap\right)^2}+23.6793\right) 
  \label{eq: BC vp full Pi}\\
  A^{\theta\chi}(p) =& (p^\theta+p^\chi) \left(\sum\limits_\lambda p^\lambda\right) - p^2 - \delta^{\theta\chi}\left(\sum\limits_\lambda p^\lambda\right)^2
  \label{eq: BC anisotropic tensor structure}\\
  \widetilde{\Pi}(g_0) =& -\frac{g_0^2 C_2}{16\pi^2} \cdot 0.9094
  \label{eq: BC vp full tildePi}
\end{align}\normalsize
Again, the Ward identity of \mbox{eq.}~(\ref{eq: Ward identity for vp tensor}) is satisfied before inclusion of counterterms and the anisotropic term is expressed through the Dirac structure of \mbox{eq.}~(\ref{eq: anisotropic Dirac structure}) using~$ \mathcal{Q} $ from \mbox{eq.}~(\ref{eq: BC species rotation matrices}). The anisotropy must be removed at one-loop level by tuning the gluonic counterterm of \mbox{eq.}~(\ref{eq: gluonic BC counterterm}) using \mbox{eq.}~(\ref{eq: BC vp full tildePi}),
\begin{equation}
  d_P(g_0) = -\frac{g_0^2 C_2}{16\pi^2} \cdot 0.9094 + \mathcal{O}(g_0^4)
\end{equation}\normalsize
in order to have the charge renormalisation factor of \mbox{eq.}~(\ref{eq: Z_3}) and the one-loop gluon propagator of \mbox{eq.}~(\ref{eq: 1-loop gluon propagator}) take their standard forms.\newline

\noindent
Whether there is any deeper meaning to the Dirac structure of \mbox{eq.} (\ref{eq: anisotropic Dirac structure}) is not fully understood at present. However, the presence of two matrices $ \mathcal{Q} $ seems to indicate that the anisotropic contribution is due to fermion loops which receive contributions from the propagation of different fermion species.

\sectionc{Boosted perturbation theory}{sec: Boosted perturbation theory}

\noindent
Boosted perturbation theory (BPT) has been suggested by  Lepage and Mackenzie~\cite{Lepage:1992xa} as a means to extend the validity of perturbative calculations into the non-perturbative regime. Non-perturbative effects are estimated with the average plaquette value,
\begin{equation}
  U_0 =\sqrt[4]{\frac{1}{|\Lambda|}\sum_{n\in\Lambda} \sum_{\mu<\nu}U^{\mu\nu}_n},
\end{equation}\normalsize
which is included in a boosted coupling constant (also called Parisi's coupling),
\begin{equation}
  g_P^2=g_0^2/U_0^4,
  \label{eq: Parisi's coupling}
\end{equation}\normalsize
which replaces $ g_0 $ in one-loop quantities. Predictions from BPT often serve as good starting points for non-perturbative studies. In the case of minimally doubled fermions, coefficients of the anisotropic counterterms have to be tuned non-perturbatively. Estimates from BPT are used to pin down a region of interest where anisotropic effects are expected to be mild compared with the untuned theory. BPT predictions are listed in table~\ref{tab: BPT predictions}.

\begin{table}[hbt]
\center
 \begin{tabular}{|c|c|c|c|c|c|c|c|}
  \hline
  \multicolumn{8}{|c|}{Karsten-Wilczek fermions} \\
  \hline
  $ \beta $ & $ U_{0}^4 $ & $ c_{1L} $  & $ c_{BPT} $ & $ d_{1L} $ & $ d_{BPT} $ & $ d_{P,\,1L} $ & $ d_{P,\,BPT} $ \\
  \hline
  $ 5.8 $ & $ 0.567 $ & $ -0.258 $ & $ -0.454 $ & $ -0.00110 $ & $ -0.00193 $ & $ -0.0924 $ & $ -0.163 $ \\
  \hline
  $ 6.0 $ & $ 0.594 $ & $ -0.249 $ & $ -0.420 $ & $ -0.00106 $ & $ -0.00179 $ & $ -0.0893 $ & $ -0.150 $ \\
  \hline
  $ 6.2 $ & $ 0.614 $ & $ -0.241 $ & $ -0.393 $ & $ -0.00103 $ & $ -0.00167 $ & $ -0.0865 $ & $ -0.141 $ \\
  \hline
  \hline
  \multicolumn{8}{|c|}{Bori\c{c}i-Creutz fermions} \\
  \hline
  $ \beta $ & $ U_{0}^4 $ & $ c_{1L} $  & $ c_{BPT} $ & $ d_{1L} $ & $ d_{BPT} $ & $ d_{P,\,1L} $ & $ d_{P,\,BPT} $ \\
  \hline
  $ 5.8 $ & $ 0.567 $ & $ +0.258 $ & $ +0.455 $ & $ +0.0133 $ & $ +0.0235 $ & $ -0.00298 $ & $ -0.00525 $ \\
  \hline
  $ 6.0 $ & $ 0.594 $ & $ +0.249 $ & $ +0.420 $ & $ +0.0129 $ & $ +0.0217 $ & $ -0.00288 $ & $ -0.00485 $ \\
  \hline
  $ 6.2 $ & $ 0.614 $ & $ +0.241 $ & $ +0.393 $ & $ +0.0125 $ & $ +0.0203 $ & $ -0.00279 $ & $ -0.00454 $ \\
  \hline
 \end{tabular}
 \caption{Boosted one-loop coefficients serve as starting point for non-perturbative renormalisation. Numerical values for $ U_0^4 $ are due to~\cite{Bali}.}
 \label{tab: BPT predictions}
\end{table}

\sectionc{Interim findings (I)}{sec: Interim findings (I)}

With the completion of perturbative studies at one-loop level, it is demonstrated that Karsten-Wilczek and Bori\c{c}i-Creutz actions are renormalisable quantum field theories. Due to their anisotropies, each action requires three counterterms. Even though the actions' anisotropies are different, their counterterms are exact analogues. \newline

\noindent
Both fermionic counterterms are required for recovery of isotropy of the fermion propagator. Once these counterterms are included with appropriate one-loop coefficients, the continuum limit of the fermion propagator takes the standard form of a one-flavour quark propagator. An exchange of the two doublers by a shift of the fermion's four-momentum from one pole to the other and application of the matrices $ \mathcal{Q} $ and $ \mathcal{Q}^\dagger $ (of \mbox{eqs.}~(\ref{eq: KW species rotation matrices}) or~(\ref{eq: BC species rotation matrices})) in the sense of the unitary transformation in \mbox{eq.}~(\ref{eq: naive species symmetry}) changes the sign of the Karsten-Wilczek (or Bori\c{c}i-Creutz) term. The coefficient $ c(g_0) $ of the dimension-three counterterm inherits this sign, whereas the other coeffcients of the self-energy -- $ Z_2 $, $ Z_m $ and $ d(g_0) $ -- are unchanged by this transformation. This verifies that the discretrised forms of the counterterm operators presented in section \ref{sec: Minimally doubled fermions} are appropriate choices for the actions over the full range of fermionic four-momenta. It is noteworthy that the coefficients of the relevant counterterms for both actions are almost equal in magnitude. \newline

\noindent
The fermionic contribution to the vacuum polarisation is anisotropic before inclusion of the gluonic counterterm, which recovers the gluon propagator's isotropy at one-loop level. Nevertheless, the Ward identity is satisfied even without counterterms and power divergences cancel between bubble and tadpole diagrams. 
Since all internal lines correspond to fermion propagators, both poles contribute on equal footing to the diagram and the logarithmic divergence indicates a two-flavour theory. 
The coefficient $ d_P(g_0) $ of the anisotropic term is finite and even under the analogue of \mbox{eq.}~(\ref{eq: naive species symmetry}).\newline

\noindent
Lastly, interacting minimally doubled fermions retain their chiral symmetry, which is reflected by the equality of renormalisation factors of local scalar and pseudoscalar densities as well as vector and axial vector currents, even though the latter undergo anisotropic renormalisation that necessitates additional counterterms. The tensor current is isotropic. 
\noindent
Minimally doubled fermions have two symmetry currents involving only fields at neighbouring sites. This vector current and axial current both satisfy chiral Ward identities. Such an ultralocal axial symmetry current is not available for the majority of fermion discretisations. The PCAC relation is satisfied at one-loop level only if the axial symmetry current is used. \newline

\noindent
For a theory of two quark flavours, the presence of a single, conserved axial symmetry current implies there is only one \mbox{(Pseudo-)}~Goldstone boson in the chiral limit. Thus, the spectrum must include two states which fail to be \mbox{(Pseudo-)}~Goldstone bosons at finite lattice spacing, but become degenerate with the \mbox{(Pseudo-)}~Goldstone boson in the continuum limit. Hence, a lattice artefacts contributes to their masses, which is the analogue of taste-breaking for staggered fermions~\cite{KlubergStern:1983dg, Blum:1996uf}.
\noindent
Of course, a third kind of pseudoscalar boson must have the role of the pseudoscalar singlet and retain a non-vanishing mass even in the continuum limit. 
Numerical simulations in chapter~\ref{sec: Numerical studies} aim at identifying at least one of the would-be \mbox{(Pseudo-)}~Goldstone bosons. 

\chapter{Structure and symmetry}\label{sec: Structure and symmetry}

\noindent
Perturbative studies of minimally doubled fermions reproduce standard behaviour of quarks in the continuum limit once three counterterms are included. At one loop, the propagator resembles the one-flavour expression, though the vacuum polarisation's logarithmic divergence implies a two-flavour theory. It seems as if minimally doubled fermions within purely fermionic loops resemble staggered fermions rather than Wilson fermions.
\noindent
Before embarking on an extensive numerical study, analytical methods are used to explore the limit $ a \to 0 $ of Karsten-Wilczek fermions by means of an expansion in the lattice spacing $ a $. The focus of section~\ref{sec: Decomposition into a pair of fields} is on a formal decomposition of Karsten-Wilczek fermions in terms of a pair of fields with different momentum support. This decomposition suggests additional oscillating contribution in some correlation functions -- as is known for the case of staggered fermions~(\mbox{e.g.}~\cite{Altmeyer:1992dd}) -- but with a frequency that depends on the coefficients of the counterterms.
\noindent
The focus of section \ref{sec: Remnant Theta symmetry and O(a) corrections} is on the interdependence of charge conjugation and Euclidean reflections for minimally doubled fermions, which is due to the $ CP\Theta $~symmetry of their actions. It is shown that observables in the quenched approximation with definite charge conjugation quantum number are automatically symmetric under the broken reflection symmetry
. A short summary of interim findings is presented in section~\ref{sec: Interim findings (II)}.

\sectionc{Decomposition into a pair of fields}{sec: Decomposition into a pair of fields}

\noindent
The aim of this section is to analyse Karsten-Wilczek fermions with an action that includes counterterms with arbitrary coefficients. This study in the na\"{i}ve continuum limit ($ a \to 0 $) reveals a dependence of correlation functions on mismatched counterterm coefficients, which is valuable for defining non-perturbative tuning schemes. 
\noindent
The relevant operator of \mbox{eq.}~(\ref{eq: ferm dim 4 KW counterterm operator}) diverges as $ a $ is taken to zero. Hence, its arbitrary coefficient has to be dealt with by a local field transformation before any expansion in $ a $ can be attempted. This is covered in section~\ref{sec: Absorption of a coefficient into a local field transformation}. A transformation along these lines was suggested for the first time by Pernici~\cite{Pernici:1994yj}. However, the coefficient of the divergent operator persistently modifies the boundary conditions. This was noted for the first time in~\cite{Bedaque:2008xs}. The new approach within this thesis is keeping the operator with a coefficient that is required for cancellation of power divergences while absorbing the \textit{mismatch} of its coefficient into a field transformation.
\noindent
Next, in section \ref{sec: Decomposition in the free theory}, the spinor fields of the free theory are related to a pair of fields, each having a different momentum support. The decomposition into this pair of \textit{components}, which must be defined with support on multiple lattice sites, follows ideas concerning a \textit{flavour interpretation}\footnote{It is not \textit{a priori} clear whether or not a flavour interpretation is eventually adequate for minimally doubled fermions. Hence, the pair of components is only a formal tool without flavour interpretation.} that has been applied to staggered fermions~\cite{vandenDoel:1983mf,Golterman:1984cy} for a long time and to minimally doubled fermions~\cite{Bedaque:2008xs,Kimura:2011ik,Tiburzi:2010bm} as well. The requirement that both components satisfy the same field equations in limit $ a \to 0 $ fixes their mixture and recovers the matrices of \mbox{eq.}~(\ref{eq: KW species rotation matrices}).
\noindent 
Later on, this definition of components is generalised to interacting fields in section~\ref{sec: Absorption of the coefficient into component fields}. Before support on different lattice sites can be defined, any mismatch of the divergent operator's coefficient must be shifted into local phase factors. It turns out that these phase factors for both components must be related by complex conjugation due to the components' non-trivial mixture that is observed in the free theory as well.
\noindent
Lastly, the decomposition is plugged into interpolating operators that are used in the construction of mesonic correlation functions in section~\ref{sec: Components in correlation functions}. These correlation functions contain sixteen different pieces due to $ 2^4 $ possible combinations of components, which can be sorted into five sets with different properties. Whereas two sets yield non-oscillating contributions with $ J^{PC} $ that is expected for the interpolating operator, two other sets yield oscillating contributions with different Dirac structure and different $ J^{PC} $. The oscillation's frequency depends on the mismatch of the divergent operator's coefficient. The last set with eight contributions must vanish due to symmetry violation.

\subsection{Absorption of a coefficient into a local field transformation}\label{sec: Absorption of a coefficient into a local field transformation}

\noindent
The full Karsten-Wilczek fermion action has been introduced in \mbox{eq.}~(\ref{eq: KW fermion action}) as
\begin{align}
  S^{f}[\psi,\bar\psi,U]  =&\ a^4\sum\limits_{n,m\in\Lambda} \bar\psi_n \left(D^N_{n,m}+D^{KW}_{n,m}+D^{3}_{n,m}+D^{4}_{n,m} + m_0\delta_{n,m}\right) \psi_m,
  \label{eq: KW fermion action on sites}
\end{align}\normalsize
which involves both fermionic counterterms. In the following, the case is considered where the counterterms' coefficients $ c $ and $ d $ do not match the tuned values $ c(g_0) $ and $ d(g_0) $ required for restoring the tree-level form to the propagator. This would be the case if either the counterterms were included in the free theory or if the coefficients were improperly tuned in the interacting theory. Hence, expressions for the coefficients read in these cases
\begin{align}
  c=c(g_0)+\delta c, \ d=d(g_0)+\delta d\,: & \quad c(g_0),d(g_0) =\mathcal{O}(g_0^2); \ \delta c, \delta d = \mathcal{O}(1).
  \label{eq: KW mismatched coefficients}
\end{align}
On the one hand, the mismatches $ \delta c $ and $ \delta d $ do not depend on the gauge coupling~$ g_0 $ and $ \delta c $~causes a uncancelled divergence of the theory in the limit $ a \to 0 $. On the other hand, the quantities $ c(g_0) $ and $ d(g_0) $ are necessary for a cancellation of interaction effects that would also diverge for $ a \to 0 $.
\noindent
Thus, the contribution proportional to $ \delta c $ from the relevant operator $ D^{3}_{n,m} $ of \mbox{eq.}~(\ref{eq: ferm dim 3 KW counterterm operator}) must be removed. The mismatch $ \delta c $ is absorbed into modified boundary conditions by a local field transformation of spinor fields
\begin{align}
  \psi_n = e^{-i\varphi n_{\underline{\alpha}}} \psi^c_n, &\qquad  
  \bar\psi_n = \bar\psi^c_n e^{+i\varphi n_{\underline{\alpha}}}.
  \label{eq: KW spinor transformation}
\end{align}\normalsize
The phase $ \varphi $ is fixed later in order to remove the mismatch $ \delta c $ in the divergent term. The gluon action is obviously invariant under the transformation of \mbox{eq.}~(\ref{eq: KW spinor transformation}). The fermion action is rewritten in terms of the transformed fields $ \psi^c $ and $ \bar\psi^c $ as
\begin{align}
  S^{f}[\psi^c,\bar\psi^c,U] 
  =&\  a^4\sum\limits_{n,m\in\Lambda} \bar\psi^c_n e^{+i\varphi n_{\underline{\alpha}}} \left(D^N_{n,m}+D^{KW}_{n,m}+D^{3}_{n,m}+D^{4}_{n,m} + m_0\delta_{n,m}\right) e^{-i\varphi m_{\underline{\alpha}}} \psi^c_m  \nonumber \\
  =&\  a^4\sum\limits_{n,m\in\Lambda} \bar\psi^c_n \Big(\sum\limits_{\mu\neq\underline{\alpha}} \gamma^\mu D^\mu_{n,m}[U]+D^{KW}_{n,m}+D^{3}_{n,m} + m_0\delta_{n,m}
  \nonumber \\
  &\hspace{66pt}+ e^{+i\varphi n_{\underline{\alpha}}}\left\{\gamma^{\underline{\alpha}} D^{\underline{\alpha}}_{n,m}[U]+D^{4}_{n,m}\right\} e^{-i\varphi m_{\underline{\alpha}}}\Big) \psi^c_m.
  \label{eq: KW fermion action with transformed fields}
\end{align}\normalsize
Both operators in the lower line of \mbox{eq.}~(\ref{eq: KW fermion action with transformed fields}) have the structure of the $ \hat{e}_{\underline{\alpha}} $~component of the lattice covariant derivative of \mbox{eq.}~(\ref{eq: lattice covariant derivative}). This term is abbreviated as
\begin{align}
    \{\mathcal{D}\psi\}_n =&\
  \tfrac{1+d}{2a}\left(U^{\underline{\alpha}}_n e^{-i\varphi (n_{\underline{\alpha}}+1)}\psi^c_{n+\hat{e}_{\underline{\alpha}}}-U^{\underline{\alpha}\dagger}_{n-\hat{e}_{\underline{\alpha}}}e^{-i\varphi (n_{\underline{\alpha}}-1)}\psi^c_{n-\hat{e}_{\underline{\alpha}}}\right)  
  \nonumber
\end{align}\normalsize
and satisfies a lattice product rule for sufficiently small values of $ a $ and $ \varphi $,
\begin{align}
  \{\mathcal{D}\psi\}_n
  =&\
  (1+d)e^{-i\varphi n_{\underline{\alpha}}}\left(D^{\underline{\alpha}}_{n,m}[U] \psi^c_m
  -i\tfrac{\varphi}{2a} \left\{U^{\underline{\alpha}}_n \psi^c_{n+\hat{e}_{\underline{\alpha}}}+U^{\underline{\alpha}\dagger}_{n-\hat{e}_{\underline{\alpha}}}\psi^c_{n-\hat{e}_{\underline{\alpha}}}\right\} +\mathcal{O}(\varphi^2) \right)
  \nonumber \\
  =&\
  (1+d)e^{-i\varphi n_{\underline{\alpha}}}\left(D^{\underline{\alpha}}_{n,m}[U] \psi^c_m
  -2i\tfrac{\varphi}{2a} \psi^c_{n} +\mathcal{O}(a,\varphi^2) \right)
  \nonumber \\
  =&\
   e^{-i\frac{\varphi}{a} a n_{\underline{\alpha}}} \left\{ (1+d) D^{\underline{\alpha}}_{n,m}[U] - i\frac{(1+d)\varphi}{a}\delta_{n,m} \right\} \psi^c_m
   +\mathcal{O}(a,\varphi^2).
  \label{eq: lattice product rule}
\end{align}\normalsize
When the lattice product rule of eq. (\ref{eq: lattice product rule}) is plugged into the fermion action of eq. (\ref{eq: KW fermion action with transformed fields}), the last remaining phase factors cancel and the action up to effects of $ \mathcal{O}(a,\varphi^2) $ reads
\begin{align}
  S^{f}[\psi^c,\bar\psi^c,U] 
  =&\  a^4\hspace{-5pt}\sum\limits_{n,m\in\Lambda}\hspace{-5pt} \bar\psi^c_n \left(D^N_{n,m} \!+\! D^{KW}_{n,m} \!+\! i\tfrac{c-(1+d)\varphi}{a}\gamma^{\underline{\alpha}}\delta_{n,m} \!+\! D^{4}_{n,m} \!+\! m_0\delta_{n,m}\right) \psi^c_m,
  \label{eq: KW fermion action with cancelled phase factors}
\end{align}\normalsize
where $ D^{3}_{n,m} $ and the second term of \mbox{eq.}~(\ref{eq: lattice product rule}) have been combined. After the phase $ \varphi $ is fixed to
\begin{equation}
  \varphi = \frac{\delta c}{1+d},
  \label{eq: phase parameter for spinor fields}
\end{equation}\normalsize
the mismatch $ \delta c $ of the relevant counterterm's coefficient appears exclusively in the combination of \mbox{eq.}~(\ref{eq: phase parameter for spinor fields}) in the boundary conditions of the fermion fields $ \psi^c $ and $ \bar\psi^c $ up to corrections of $ \mathcal{O}(a,\varphi^2) $. The remaining coefficient $ c(g_0) $ of the divergent operator exactly cancels power divergent interaction effects. Hence, fields with extended support on multiple sites can be constructed from $ \psi^c $ and $ \bar\psi^c $.

\subsection{Decomposition in the free theory}\label{sec: Decomposition in the free theory}

\noindent
Decomposition of spinor fields into components that differ in chirality and spin due to having different momentum support has been in use for various types of lattice fermions for a long time (\mbox{e.g.}~\cite{Karsten:1980wd,vandenDoel:1983mf,Golterman:1984cy}). Decomposition for minimally doubled fermions featured first in~\cite{Bedaque:2008xs,Tiburzi:2010bm} and is generalised in the following. Components $ \phi $ and $ \chi $ of spinors $ \psi $ for Karsten-Wilczek fermions are defined by a summation of the full lattice using \textit{decomposition kernels} $ g^\phi $ and $ g^\chi $ that restrict the components' momentum support as
\begin{align}
  \left.\begin{array}{rl}
  \psi_n =&\ \sum\limits_{k\in\Lambda} g^{\phi}_{n,k} \phi_k + (-1)^{n_{\underline{\alpha}}} \mathcal{Q} g^{\chi}_{n,k} \chi_k \\
  \bar\psi_n =&\ \sum\limits_{k\in\Lambda} \bar\phi_k (g^{\phi}_{n,k})^\dagger + (-1)^{n_{\underline{\alpha}}} \bar\chi_k \mathcal{\bar Q} (g^{\chi}_{n,k})^\dagger
  \end{array}\right..
  \label{eq: species components in the free theory}
\end{align}\normalsize
The symbols $ \mathcal{Q} $ and $ \mathcal{\bar Q} $ represent a priori unknown matrices that allow for mixtures of two components with different spin and chirality into the same original spinor fields. Because the matrices later turn out to be the same objects as in \mbox{eq.}~(\ref{eq: KW species rotation matrices}), the same symbols $ \mathcal{Q} $ and $ \mathcal{\bar Q} $ are used here as well.
\noindent
The alternating factor $ (-1)^{n_{\underline{\alpha}}} $ accounts for momentum support at different poles in the Brillouin zone according to \mbox{eq.}~(\ref{eq: KW doublers}).
\noindent
The only required properties of decomposition kernels here are a reasonably fast decrease for increasing distance $ |n-k| $ and a Kronecker symbol $ \delta_{n,k} $ in the limit $ a \to 0 $,
\begin{align}
  g^{\phi}_{n,k} = \delta_{n,k} + \mathcal{O}(a),&\qquad 
  g^{\chi}_{n,k} = \delta_{n,k} + \mathcal{O}(a).
  \label{eq: continuum limit of species kernels in the free theory}
\end{align}\normalsize
This decomposition is a generalisation of the decomposition in~\cite{Tiburzi:2010bm} with more general decomposition kernels. The field components $ \phi $ and $ \chi $ do not necessarily have a simple relation to the physical flavours of the theory (\mbox{cf.} the discussion in section~2 of~\cite{Golterman:1984cy}). Combinations of matrices and factors $ (-1)^{n_{\underline{\alpha}}} $ are wrapped up into a concise notation as
\begin{align}
  \mathcal{R}_n      =  (-1)^{n_{\underline{\alpha}}}\ \mathcal{Q}, &\qquad
  \mathcal{\bar R}_n =  (-1)^{n_{\underline{\alpha}}}\ \mathcal{\bar Q}.
  \label{eq: KW species conjugation matrices in the free theory}
\end{align}\normalsize
The decomposition of \mbox{eq.}~(\ref{eq: species components in the free theory}) is plugged into the Karsten-Wilczek fermion action of \mbox{eq.}~(\ref{eq: KW fermion action on sites}) neglecting counterterms and interactions initially. Hence, the action reads
\begin{align}
  S^{f}[\phi,\bar\phi,\chi,\bar\chi] 
  =&\ a^4\sum\limits_{k,l\in\Lambda} 
  \mathcal{L}^{\bar\phi\phi}[\phi_l,\bar\phi_k] + \mathcal{L}^{\bar\chi\chi}[\chi_l,\bar\chi_k] + \mathcal{L}^{\bar\phi\chi}[\chi_l,\bar\phi_k] + \mathcal{L}^{\bar\chi\phi}[\phi_l,\bar\chi_k] 
  \label{eq: KW species component Lagrangians in the free theory},
  \\
  \mathcal{L}^{\bar\phi\phi}[\phi_l,\bar\phi_k] =&\ \sum\limits_{n,m\in\Lambda}
  \bar\phi_k (g^{\phi}_{n,k})^\dagger \left(D^N_{n,m}+D^{KW}_{n,m}+ m_0\delta_{n,m}\right) g^{\phi}_{m,l} \phi_l,
  \label{eq: KW Lagrangian L[phi,phi] in the free theory} 
  \\
  \mathcal{L}^{\bar\chi\chi}[\chi_l,\bar\chi_k]=&\ \sum\limits_{n,m\in\Lambda}
  \bar\chi_k (g^{\chi}_{n,k})^\dagger \mathcal{\bar R}_{n} \left(D^N_{n,m}+D^{KW}_{n,m}+ m_0\delta_{n,m}\right) \mathcal{R}_{m} g^{\chi}_{m,l} \chi_l,
  \label{eq: KW Lagrangian L[chi,chi] in the free theory} 
  \\
  \mathcal{L}^{\bar\phi\chi}[\chi_l,\bar\phi_k]=&\ \sum\limits_{n,m\in\Lambda}
  \bar\phi_k (g^{\phi}_{n,k})^\dagger \left(D^N_{n,m}+D^{KW}_{n,m}+ m_0\delta_{n,m}\right) \mathcal{R}_{m} g^{\chi}_{m,l} \chi_l,
  \label{eq: KW Lagrangian L[phi,chi] in the free theory} 
  \\
  \mathcal{L}^{\bar\chi\phi}[\phi_l,\bar\chi_k]=&\ \sum\limits_{n,m\in\Lambda}
  \bar\chi_k (g^{\chi}_{n,k})^\dagger \mathcal{\bar R}_{n} \left(D^N_{n,m}+D^{KW}_{n,m}+ m_0\delta_{n,m}\right) g^{\phi}_{m,l} \phi_l.
  \label{eq: KW Lagrangian L[chi,phi] in the free theory}
\end{align}\normalsize
Next, decomposition kernels are treated as part of the Dirac kernels for field components. Thus, the Dirac kernel $ K^{\bar\chi,\chi}_{k,l} $ in the Lagrangian $ \mathcal{L}^{\bar\chi\chi}[\chi_l,\bar\chi_k] = \bar\chi_k K^{\bar\chi,\chi}_{k,l} \chi_l $ of \mbox{eq.}~(\ref{eq: KW Lagrangian L[chi,chi] in the free theory}) reads
\begin{align}
  K^{\bar\chi,\chi}_{k,l} 
  =  \sum\limits_{n\in\Lambda} (g^{\chi}_{n,k})^\dagger &\ \Big(
  \sum\limits_{\mu\neq\underline{\alpha}} \frac{1}{2a}\mathcal{\bar R}_n\gamma^{\mu}\mathcal{R}_n
   \left\{ g^{\chi}_{n+\hat{e}_{\mu},l} -g^{\chi}_{n-\hat{e}_{\mu},l} \right\} + m_0 \mathcal{\bar R}_n \mathcal{R}_n g^{\chi}_{n,l}
  \nonumber \\
  &
   +\frac{1}{2a}
   \left\{ \mathcal{\bar R}_n\gamma^{\underline{\alpha}}\mathcal{R}_{n+\hat{\underline{\alpha}}} g^{\chi}_{n+\hat{e}_{\underline{\alpha}},l} - \mathcal{\bar R}_n\gamma^{\underline{\alpha}}\mathcal{R}_{n-\hat{\underline{\alpha}}} g^{\chi}_{n-\hat{e}_{\underline{\alpha}},l} \right\}
  \nonumber \\  
  &+\frac{i\zeta}{2a}\mathcal{\bar R}_n\gamma^{\underline{\alpha}}\mathcal{R}_n \sum\limits_{\mu\neq\underline{\alpha}}
   \left\{ g^{\chi}_{n+\hat{e}_{\mu},l} +g^{\chi}_{n-\hat{e}_{\mu},l} - 2g^{\chi}_{n,l} \right\}\Big).
\end{align}\normalsize
The structure of the lattice derivatives in the $ \hat{e}_{\underline{\alpha}} $~direction simplifies, because
\begin{equation}
  \mathcal{\bar R}_n \mathcal{M} \mathcal{R}_{n\pm\hat{\underline{\alpha}}} = - \mathcal{\bar R}_n \mathcal{M} \mathcal{R}_{n} = -\mathcal{\bar Q} \mathcal{M} \mathcal{Q}
  \label{eq: species conjugation of matrix M in the free theory}
\end{equation}
is valid for arbitrary matrices $ \mathcal{M} $. Hence, the Dirac kernel $ K^{\bar\chi,\chi}_{k,l} $ reads
\begin{align}
  K^{\bar\chi,\chi}_{k,l} 
  =  \sum\limits_{n\in\Lambda} (g^{\chi}_{n,k})^\dagger &\ \Big(
   \sum\limits_{\mu\neq\underline{\alpha}} \frac{1}{2a} (\mathcal{\bar Q \gamma^{\mu} Q})
   \left\{ g^{\chi}_{n+\hat{e}_{\mu},l} -g^{\chi}_{n-\hat{e}_{\mu},l} \right\} + m_0 (\mathcal{\bar Q Q}) g^{\chi}_{n,l}
  \nonumber \\
  &
     -\frac{1}{2a} (\mathcal{\bar Q \gamma^{\underline{\alpha}} Q})
   \left\{ g^{\chi}_{n+\hat{e}_{\underline{\alpha}},l} - g^{\chi}_{n-\hat{e}_{\underline{\alpha}},l} \right\}
  \nonumber \\  
  &+\frac{i\zeta}{2a} (\mathcal{\bar Q \gamma^{\underline{\alpha}} Q}) \sum\limits_{\mu\neq\underline{\alpha}}
   \left\{ g^{\chi}_{n+\hat{e}_{\mu},l} +g^{\chi}_{n-\hat{e}_{\mu},l} - 2g^{\chi}_{n,l}  \right\}\Big).
\end{align}\normalsize
Components $ \phi $ and $ \chi $ satisfy the same field equations in the limit $ a \to 0 $ if and only if the Dirac kernels of Lagrangians $ \mathcal{L}^{\bar\phi\phi}[\phi_l,\bar\phi_k] $ and $ \mathcal{L}^{\bar\chi\chi}[\chi_l,\bar\chi_k] $ take the same form in the limit $ a \to 0 $. As the limit $ a \to 0 $ of the decomposition kernels is taken, this implies three conditions for the matrices $ \mathcal{Q} $ and $ \mathcal{\bar Q} $:
\begin{align}
  \left\{\begin{array}{rll}
    (\mathcal{\bar Q Q}) =& 1 & \text{equality of the mass term} \\
    (\mathcal{\bar Q \gamma^{\mu} Q}) =& \gamma^{\mu} & \text{equality of the perpendicular kinetic term} \\
    (\mathcal{\bar Q \gamma^{\underline{\alpha}} Q}) =& -\gamma^{\underline{\alpha}} & \text{equality of the parallel kinetic term}
  \end{array}\right\}.
  \label{eq: conditions on species rotation matrices}
\end{align}\normalsize
It is evident that the set of conditions requires that $ \mathcal{Q} $ and $ \mathcal{\bar Q} $ equal the matrices of \mbox{eq.}~(\ref{eq: KW species rotation matrices}) up to choices of the arbitrary phase $ \vartheta $. Because the Karsten-Wilczek term is the leading order correction to the na\"{i}ve continuum limit of the Karsten-Wilczek Dirac operator and has opposite sign in the kernels $ K^{\bar\phi,\phi}_{k,l} $ and $ K^{\bar\chi,\chi}_{k,l} $, the components satisfy field equations that differ at finite lattice spacing $ a $ by the sign of the Karsten-Wilczek term's contribution. This is the coordinate space analogue to \mbox{eq.}~(\ref{eq: KW relation between two species}). 
\noindent
Component-mixing Lagrangians $ \mathcal{L}^{\bar\phi\chi}[\chi_l,\bar\phi_k] $ and $ \mathcal{L}^{\bar\chi\phi}[\phi_l,\bar\chi_k] $ of \mbox{eqs.}~(\ref{eq: KW Lagrangian L[phi,chi] in the free theory}) and~(\ref{eq: KW Lagrangian L[chi,phi] in the free theory}) have Dirac kernels $ K^{\bar\phi,\chi}_{k,l} $ and $ K^{\bar\chi,\phi}_{k,l} $ which read
\begin{align}
  K^{\bar\phi,\chi}_{k,l}
  =  \sum\limits_{n\in\Lambda} e^{i(\frac{\pi}{2}+\vartheta)}(-1)^{n_{\underline{\alpha}}} (g^{\phi}_{n,k})^\dagger 
  &\ \Big(
    \sum\limits_{\mu,\nu,\lambda} \frac{1}{4a} ( -i\Sigma^{\lambda\nu} \epsilon^{\nu\lambda\mu\underline{\alpha}} )
   \left\{ g^{\chi}_{n+\hat{e}_{\mu},l} -g^{\chi}_{n-\hat{e}_{\mu},l} \right\}
  \nonumber \\
  &
   +\frac{1}{2a} \gamma^5 \left\{ g^{\chi}_{n+\hat{e}_{\underline{\alpha}},l} - g^{\chi}_{n-\hat{e}_{\underline{\alpha}},l} \right\} 
   + m_0 \gamma^{\underline{\alpha}}\gamma^5 g^{\chi}_{n,l}
  \nonumber \\  
  &
  +\frac{i\zeta}{2a} \gamma^5 \sum\limits_{\mu\neq\underline{\alpha}}
   \left\{ g^{\chi}_{n+\hat{e}_{\mu},l} +g^{\chi}_{n-\hat{e}_{\mu},l} - 2g^{\chi}_{n,l}  \right\}\Big),
  \label{eq: species non-conserving kernel K^phichi in the free theory} 
  \\
%
  K^{\bar\chi,\phi}_{k,l}
  =  \sum\limits_{n\in\Lambda} e^{i(\frac{\pi}{2}-\vartheta)}(-1)^{n_{\underline{\alpha}}} (g^{\chi}_{n,k})^\dagger 
  &\ \Big(
  \sum\limits_{\mu,\nu,\lambda} \frac{1}{4a} ( -i\Sigma^{\lambda\nu} \epsilon^{\nu\lambda\mu\underline{\alpha}} )
   \left\{ g^{\phi}_{n+\hat{e}_{\mu},l} -g^{\phi}_{n-\hat{e}_{\mu},l} \right\}
  \nonumber \\
  &
   +\frac{1}{2a} \gamma^5 \left\{ g^{\phi}_{n+\hat{e}_{\underline{\alpha}},l} - g^{\phi}_{n-\hat{e}_{\underline{\alpha}},l} \right\}
    + m_0 \gamma^{\underline{\alpha}}\gamma^5 g^{\phi}_{n,l}
  \nonumber \\  
  &
  -\frac{i\zeta}{2a} \gamma^5 \sum\limits_{\mu\neq\underline{\alpha}}
   \left\{ g^{\phi}_{n+\hat{e}_{\mu},l} +g^{\phi}_{n-\hat{e}_{\mu},l} - 2g^{\phi}_{n,l}  \right\}\Big).
  \label{eq: species non-conserving kernel K^chiphi in the free theory}
\end{align}\normalsize
The role of these terms in the limit $ a \to 0 $ is clarified by a Fourier transform of the component fields to momentum space. Hereby, the limit $ a \to 0 $ of decomposition kernels $ g^\phi $ and $ g^\chi $ is taken according to \mbox{eq.}~(\ref{eq: continuum limit of species kernels in the free theory}). The component-mixing piece of the action up to corrections of~$ \mathcal{O}(a) $ reads
\begin{align}
 S^\prime 
 =&\ e^{i(\tfrac{\pi}{2}+\vartheta)}\int \frac{d^4q}{(2\pi)^4} \bar \phi(q+\frac{\pi}{a}\hat{e}_{\underline{\alpha}}) \left\{\gamma^5 q^{\underline{\alpha}} + m_0 \gamma^{\underline{\alpha}}\gamma^5-\frac{i}{2}\Sigma^{\lambda\nu}\epsilon^{\nu\lambda\mu\underline{\alpha}}q^\mu\right\} \chi(q) 
 \nonumber \\
 +&\ e^{i(\tfrac{\pi}{2}-\vartheta)}\int \frac{d^4q}{(2\pi)^4} \bar \chi(q+\frac{\pi}{a}\hat{e}_{\underline{\alpha}}) \left\{\gamma^5 q^{\underline{\alpha}} + m_0 \gamma^{\underline{\alpha}}\gamma^5-\frac{i}{2}\Sigma^{\lambda\nu}\epsilon^{\nu\lambda\mu\underline{\alpha}}q^\mu\right\} \phi(q) 
 +\mathcal{O}(a).
 \label{eq: species non-conserving action in momentum space}
\end{align}\normalsize
If either component has small four-momentum $ q $ in the vicinity of the pole $ q^2=0 $, the four-momentum $ q+\pi/a\, \hat{e}_{\underline{\alpha}} $ of the other component in \mbox{eq.}~(\ref{eq: species non-conserving action in momentum space}) is inevitably near the pole at the cutoff. Thus, both components decouple in the limit $ a \to 0 $ of the free theory. This result has been discussed already by Pernici~\cite{Pernici:1994yj}.
\noindent
Moreover, decomposition kernels $ g^\phi $ and $ g^\chi $ can be defined on non-overlapping regions of the Brillouin zone enforcing $ |q_{\underline{\alpha}}| \leq \pi/(2a) $ for component momenta. This definition manifestly prohibits mixing of components without four-momentum exchange. Tiburzi applied this scheme in a calculation of the chiral anomaly for Karsten-Wilczek fermions~\cite{Tiburzi:2010bm}. 
\noindent
Putting everything together, the simplified free Lagrangians for the components read
\begin{align}
  S^{f}[\phi,\bar\phi,\chi,\bar\chi] 
  =&\ a^4\sum\limits_{k,l\in\Lambda} \mathcal{L}^{\bar\phi\phi}[\phi_l,\bar\phi_k] + \mathcal{L}^{\bar\chi\chi}[\chi_l,\bar\chi_k] 
  + \mathcal{L}^{\bar\phi\chi}[\chi_l,\bar\phi_k] + \mathcal{L}^{\bar\chi\phi}[\phi_l,\bar\chi_k] 
  \label{eq: KW species decomposed action in the free theory}
\end{align}\normalsize
\begin{align}
  \mathcal{L}^{\bar\phi\phi}[\phi_l,\bar\phi_k] =&\ \sum\limits_{n,m\in\Lambda}
  \bar\phi_k (g^{\phi}_{n,k})^\dagger \left(D^N_{n,m}+D^{KW}_{n,m} + m_0\delta_{n,m}\right) g^{\phi}_{m,l} \phi_l,
  \label{eq: KW simplified Lagrangian L[phi,phi] in the free theory} \\
  \mathcal{L}^{\bar\chi\chi}[\chi_l,\bar\chi_k]=&\ \sum\limits_{n,m\in\Lambda}
  \bar\chi_k (g^{\chi}_{n,k})^\dagger \left(D^N_{n,m}-D^{KW}_{n,m} + m_0\delta_{n,m}\right) g^{\chi}_{m,l} \chi_l,
  \label{eq: KW simplified Lagrangian L[chi,chi] in the free theory} \\
  \mathcal{L}^{\bar\phi\chi}[\chi_l,\bar\phi_k] =&\ \sum\limits_{n,m\in\Lambda}
  \bar\phi_k (-1)^{n_{\underline{\alpha}}}
  (g^{\phi}_{n,k})^\dagger \left(D^N_{n,m}+D^{KW}_{n,m} + m_0\delta_{n,m}\right)\mathcal{Q} g^{\chi}_{m,l} \chi_l,
  \label{eq: KW simplified Lagrangian L[chi,phi] in the free theory} \\
  \mathcal{L}^{\bar\chi\phi}[\phi_l,\bar\chi_k] =&\ \sum\limits_{n,m\in\Lambda}
  \bar\chi_k 
  (-1)^{n_{\underline{\alpha}}}
  (g^{\chi}_{n,k})^\dagger \left(D^N_{n,m}-D^{KW}_{n,m} + m_0\delta_{n,m}\right)\mathcal{\bar Q} g^{\phi}_{m,l} \phi_l.
  \label{eq: KW simplified Lagrangian L[phi,chi] in the free theory}
\end{align}\normalsize
It is made explicit how the Karsten-Wilczek term distinguishes between both components and how they decouple in the limit $ a \to 0 $ of the free theory. The component-mixing terms contain alternating factors $ (-1)^{n_{\underline{\alpha}}} $ that do not cancel because of contributions from different regions of the Brillouin zone. This phenomenon is well-known from staggered fermions~\cite{Golterman:1984cy}.

\subsection{Absorption of the coefficient into component fields}\label{sec: Absorption of the coefficient into component fields}

\noindent
The preceding derivation in section~\ref{sec: Decomposition in the free theory} defines field components for free Karsten-Wilczek fermions. In the limit $ a \to 0 $, they satisfy the same field equations and decouple. Nevertheless, the Karsten-Wilczek term contributes differently to their field equations at finite lattice spacing. It seems plausible that components can still be defined for sufficiently weak interactions. Such ideas have been extensively discussed for the case of staggered fermions in~\cite{Golterman:1984cy}. However, local gauge~invariance requires generalisation of the previous definition of components using decomposition kernels $ g^\phi[U] $ and $ g^\chi[U] $ as
\begin{align}
  \left.\begin{array}{rl}
  \psi_n =&\ \sum\limits_{k\in\Lambda} e^{-i \varphi^\phi n_{\underline{\alpha}}} g^{\phi}_{n,k}[U] \phi_k + (-e^{-i \varphi^\chi})^{n_{\underline{\alpha}}} \mathcal{Q} g^{\chi}_{n,k}[U] \chi_k \\
  \bar\psi_n =&\ \sum\limits_{k\in\Lambda} \bar\phi_k (g^{\phi}_{n,k}[U])^\dagger e^{i \varphi^\phi n_{\underline{\alpha}}}+ (-e^{i \varphi^\chi})^{n_{\underline{\alpha}}} \bar\chi_k \mathcal{\bar Q} (g^{\chi}_{n,k}[U])^\dagger
  \end{array}\right..
  \label{eq: species components in the interacting theory}
\end{align}\normalsize
Smooth transition from interacting to free theory without counterterms requires
\begin{equation}
  \lim\limits_{c,d,g_0 \to 0} \varphi^\phi = \lim\limits_{c,d,g_0 \to 0} \varphi^\chi = 0.
  \label{eq: phase parameters in weak coupling limit}
\end{equation}\normalsize
Gauge~invariance of component bilinears requires that the decomposition kernels $ g^\phi_{n,k}[U] $ and $ g^\chi_{n,k}[U] $ include parallel transport. This can be achieved by a definition as products of the free theory's decomposition kernels $ g^\phi_{n,k} $ and $ g^\chi_{n,k} $ and Wilson lines\footnote{In~\cite{Tiburzi:2010bm}, Tiburzi defines components of the free Karsten-Wilczek action ($ \underline{\alpha}=0 $) with extended support in the $ \hat{e}_{0} $~direction (``energy smearing''). In this case, parallel transport of the component fields can be achieved by the path-ordered products of gauge links $ U^{0} $ between $ n $ and $ k $ (the shortest possible Wilson line). 
There is no simple shortest Wilson line if the components' support is not restricted to a single axis. This is necessarily the case for Bori\c{c}i-Creutz fermions, where the components are spread out along the hypercube's diagonal.
In such a case, a scheme must either define a single path or the average of multiple paths for parallel transport~(\mbox{cf.} similar ideas in \cite{vonHippel:2013yfa}).
} 
between sites $ n $ and $ k $. The concise notation of \mbox{eq.}~(\ref{eq: KW species conjugation matrices in the free theory}) is generalised by including phase factors $ e^{\pm i\varphi^\chi} $ into matrices $ \mathcal{R}_n^{\varphi^\chi} $ and $ \mathcal{\bar R}_n^{\varphi^\chi} $ and phase factors $ e^{\pm i\varphi^\phi} $ into scalar factors $ r_n^{\varphi^\phi} $ and $ \bar r_n^{\varphi^\phi} $. These new abbreviations read
\begin{align}
  \left\{\begin{array}{rlrl}
  \mathcal{R}_n^{\varphi^\chi}      =&  (-e^{-i\varphi^\chi})^{n_{\underline{\alpha}}}\ \mathcal{Q}, &
  \mathcal{\bar R}_n^{\varphi^\chi} =&  (-e^{+i\varphi^\chi})^{n_{\underline{\alpha}}}\ \mathcal{\bar Q} \\
  r_n^{\varphi^\phi} =&  (+e^{-i\varphi^\phi})^{n_{\underline{\alpha}}}, &
  \bar r_n^{\varphi^\phi} =&  (+e^{+i\varphi^\phi})^{n_{\underline{\alpha}}}
  \end{array}\right\}.
  \label{eq: KW species conjugation matrices in the interacting theory}
\end{align}\normalsize

\noindent
The decomposition of \mbox{eq.}~(\ref{eq: species components in the interacting theory}) is plugged into the Karsten-Wilczek fermion action of \mbox{eq.}~(\ref{eq: KW fermion action on sites}) for the interacting theory as it is done for the free theory in \mbox{eq.}~(\ref{eq: KW species component Lagrangians in the free theory}). The fermion action is decomposed into the sum of four Lagrangians,
\small
\begin{align}
  S^{f}[\phi,\bar\phi,\chi,\bar\chi,U] 
  =&\ a^4 \hspace{-3pt}\sum\limits_{k,l\in\Lambda}\hspace{-3pt} 
  \mathcal{L}^{\bar\phi\phi}[\phi_l,\bar\phi_k,U] \!+\! \mathcal{L}^{\bar\chi\chi}[\chi_l,\bar\chi_k,U] \!+\! \mathcal{L}^{\bar\phi\chi}[\chi_l,\bar\phi_k,U] \!+\! \mathcal{L}^{\bar\chi\phi}[\phi_l,\bar\chi_k,U],
  \label{eq: KW species component Lagrangians in the interacting theory}
  \\
  \mathcal{L}^{\bar\phi\phi}[\phi_l,\bar\phi_k,U] =&\ \sum\limits_{n,m}
  \bar\phi_k (g^{\phi}_{n,k}[U])^\dagger \bar r_n^{\varphi^\phi} 
  \!\left(\! D^N_{n,m} \!+\! D^{KW}_{n,m} \!+\! D^{3}_{n,m} \!+\! D^{4}_{n,m} \!+\! m_0\delta_{n,m} \!\right)
  r_m^{\varphi^\phi} g^{\phi}_{m,l}[U] \phi_l,
  \label{eq: KW Lagrangian L[phi,phi] in the interacting theory} \\
  \mathcal{L}^{\bar\chi\chi}[\chi_l,\bar\chi_k,U]=&\ \sum\limits_{n,m}
  \bar\chi_k (g^{\chi}_{n,k}[U])^\dagger \mathcal{\bar R}_n^{\varphi^\chi} 
  \!\left(\! D^N_{n,m} \!+\! D^{KW}_{n,m} \!+\! D^{3}_{n,m} \!+\! D^{4}_{n,m} \!+\! m_0\delta_{n,m} \!\right)
  \mathcal{R}_m^{\varphi^\chi} g^{\chi}_{m,l}[U] \chi_l,
  \label{eq: KW Lagrangian L[chi,chi] in the interacting theory} \\
  \mathcal{L}^{\bar\phi\chi}[\chi_l,\bar\phi_k,U]=&\ \sum\limits_{n,m}
  \bar\phi_k (g^{\phi}_{n,k}[U])^\dagger \bar r_n^{\varphi^\phi} 
  \!\left(\! D^N_{n,m} \!+\! D^{KW}_{n,m} \!+\! D^{3}_{n,m} \!+\! D^{4}_{n,m} \!+\! m_0\delta_{n,m} \!\right)
  \mathcal{R}_{m}^{\varphi^\chi} g^{\chi}_{m,l}[U] \chi_l,
  \label{eq: KW Lagrangian L[phi,chi] in the interacting theory} \\
  \mathcal{L}^{\bar\chi\phi}[\phi_l,\bar\chi_k,U]=&\ \sum\limits_{n,m}
  \bar\chi_k (g^{\chi}_{n,k}[U])^\dagger \mathcal{\bar R}_n^{\varphi^\chi} 
  \!\left(\! D^N_{n,m} \!+\! D^{KW}_{n,m} \!+\! D^{3}_{n,m} \!+\! D^{4}_{n,m} \!+\! m_0\delta_{n,m} \!\right) 
  r_m^{\varphi^\phi} g^{\phi}_{m,l}[U] \phi_l.
  \label{eq: KW Lagrangian L[chi,phi] in the interacting theory}
\end{align}\normalsize
Dirac kernels (\mbox{e.g.}~$ \mathcal{K}^{\bar\chi,\chi}_{k,l} $) of the Lagrangians are defined analogously to the free theory in section~\ref{sec: Decomposition in the free theory}. The analogue of \mbox{eq.}~(\ref{eq: species conjugation of matrix M in the free theory}) in the interacting theory is
\begin{align}
  \mathcal{\bar R}_n^{\varphi^\chi}  \mathcal{M} \mathcal{R}_{n\pm\hat{\underline{\alpha}}}^{\varphi^\chi}  =  -e^{\mp i\varphi^\chi} \mathcal{\bar R}_n^{\varphi^\chi} \mathcal{M} \mathcal{R}_{n}^{\varphi^\chi}  = -e^{\mp i\varphi^\chi} \mathcal{\bar Q} \mathcal{M} \mathcal{Q}, \quad
  \bar r_n^{\varphi^\phi}  \mathcal{M} r_{n\pm\hat{\underline{\alpha}}}^{\varphi^\phi} = +e^{\mp i\varphi^\phi} \mathcal{M}.
  \label{eq: species conjugation of matrix M in the interacting theory}
\end{align}\normalsize
Hence, parallel kinetic terms of both Dirac kernels $ \mathcal{K}^{\bar\phi,\phi}_{k,l} $ and $ \mathcal{K}^{\bar\chi,\chi}_{k,l} $ include phase factors like those which feature in eq. (\ref{eq: KW fermion action with transformed fields}),
\small
\begin{align}
  K^{\bar\phi,\phi}_{k,l} 
  =  \sum\limits_{n\in\Lambda} (g^{\phi}_{n,k}[U])^\dagger &\ \Big(
  +\sum\limits_{\mu\neq\underline{\alpha}} \frac{1}{2a} \gamma^{\mu}
   \left\{ 
   U^{\mu}_n g^{\phi}_{n+\hat{e}_{\mu},l}[U] - 
   U^{\mu\dagger}_{n-\hat{e}_{\mu}}g^{\phi}_{n-\hat{e}_{\mu},l}[U] \right\} 
   + m_0 g^{\phi}_{n,l}
  \nonumber \\
  &
   +\gamma^{\underline{\alpha}} \left[ \tfrac{1+d}{2a} 
   \left\{ 
   e^{- i\varphi^\phi} U^{\underline{\alpha}}_n g^{\phi}_{n+\hat{e}_{\underline{\alpha}},l}[U] \!- 
   e^{+ i\varphi^\phi} U^{\underline{\alpha}\dagger}_{n-\hat{e}_{\underline{\alpha}}} g^{\phi}_{n-\hat{e}_{\underline{\alpha}},l}[U] \right\}
   \!+i \frac{c}{a} g^{\phi}_{n,l}[U] \right]
  \nonumber \\  
  &
  +\frac{i\zeta}{2a} \gamma^{\underline{\alpha}} \sum\limits_{\mu\neq\underline{\alpha}}
   \left\{ 
   U^{\mu}_n g^{\chi}_{n+\hat{e}_{\mu},l}[U] + 
   U^{\mu\dagger}_{n-\hat{e}_{\mu}} g^{\chi}_{n-\hat{e}_{\mu},l}[U] - 
   2g^{\chi}_{n,l}[U]  \right\}\Big),
  \nonumber \\
  K^{\bar\chi,\chi}_{k,l} 
  =  \sum\limits_{n\in\Lambda} (g^{\chi}_{n,k}[U])^\dagger &\ \Big(
  +\sum\limits_{\mu\neq\underline{\alpha}} \frac{1}{2a} \mathcal{\bar Q \gamma^{\mu} Q}
   \left\{ 
   U^{\mu}_n g^{\chi}_{n+\hat{e}_{\mu},l}[U] - 
   U^{\mu\dagger}_{n-\hat{e}_{\mu}} g^{\chi}_{n-\hat{e}_{\mu},l}[U] \right\} 
   + m_0 \mathcal{\bar Q Q} g^{\chi}_{n,l}[U]
  \nonumber \\
  &
   -\mathcal{\bar Q \gamma^{\underline{\alpha}} Q} \left[ \tfrac{1+d}{2a} 
   \left\{ 
   e^{- i\varphi^\chi} U^{\underline{\alpha}}_n g^{\chi}_{n+\hat{e}_{\underline{\alpha}},l}[U] \!-
   e^{+ i\varphi^\chi} U^{\underline{\alpha}\dagger}_{n-\hat{e}_{\underline{\alpha}}} g^{\chi}_{n-\hat{e}_{\underline{\alpha}},l}[U] \right\}
  \!-i \frac{c}{a} g^{\chi}_{n,l}[U] \right]
  \nonumber \\  
  &
  +\frac{i\zeta}{2a} \mathcal{\bar Q \gamma^{\underline{\alpha}} Q} \sum\limits_{\mu\neq\underline{\alpha}}
   \left\{ 
   U^{\mu}_n g^{\chi}_{n+\hat{e}_{\mu},l}[U] + 
   U^{\mu\dagger}_{n-\hat{e}_{\mu}} g^{\chi}_{n-\hat{e}_{\mu},l}[U] 
   - 2g^{\chi}_{n,l}[U]  \right\}\Big).
  \nonumber
\end{align}\normalsize
The lattice product rule of \mbox{eq.}~(\ref{eq: lattice product rule}) applies and yields up to~$ \mathcal{O}(a,\varphi^2) $
\begin{align}
  &\frac{1+d}{2a} 
   \left\{ 
   e^{- i\varphi^\phi} U^{\underline{\alpha}}_n g^{\phi}_{n+\hat{e}_{\underline{\alpha}},l}[U] - 
   e^{+ i\varphi^\phi} U^{\underline{\alpha}\dagger}_{n-\hat{e}_{\underline{\alpha}}} g^{\phi}_{n-\hat{e}_{\underline{\alpha}},l}[U] \right\}
   +i \tfrac{c}{a} g^{\phi}_{n,l}[U]  \nonumber \\ 
   &=
   \{1 \!+\! d\}
   D^{\underline{\alpha}}_{n,m} g^{\phi}_{m,l}[U]
   + i \frac{c-(1+d)\varphi^\phi}{a} g^{\phi}_{n,l}[U], 
   \label{eq: lattice product rule for 1st species phi} \\
  &\frac{1+d}{2a} 
   \left\{ 
   e^{- i\varphi^\chi} U^{\underline{\alpha}}_n g^{\chi}_{n+\hat{e}_{\underline{\alpha}},l}[U] - 
   e^{+ i\varphi^\chi} U^{\underline{\alpha}\dagger}_{n-\hat{e}_{\underline{\alpha}}} g^{\chi}_{n-\hat{e}_{\underline{\alpha}},l}[U] \right\}
   - i \frac{c}{a} g^{\chi}_{n,l}[U]  \nonumber \\ 
   &=
   \{1 \!+\! d\}
   D^{\underline{\alpha}}_{n,m} g^{\chi}_{m,l}[U]
   -i \tfrac{c+(1+d)\varphi^\chi}{a} g^{\chi}_{n,l}[U]. 
   \label{eq: lattice product rule for 2nd species chi}
\end{align}\normalsize
Unmatched divergent operators in \mbox{eqs.}~(\ref{eq: lattice product rule for 1st species phi}) and~(\ref{eq: lattice product rule for 2nd species chi}) must cancel before a expansion in $ a $ can be attempted. A divergence due to a mismatch of coefficients as in \mbox{eq.}~(\ref{eq: KW mismatched coefficients}) cancels if the phases~$ \varphi^\phi $ and~$ \varphi^\chi $ satisfy two different conditions,
\begin{equation}
  \varphi^\phi = \frac{\delta c}{1+d}, \qquad \varphi^\chi = -\frac{\delta c}{1+d}.
  \label{eq: phase parameter for species components}
\end{equation}\normalsize
This implies $ \varphi\equiv \varphi^\phi = -\varphi^\chi $.
Because $ \delta c \stackrel{c,d,g_0 \to0}{\to}0 $, these phases satisfy the requirement of \mbox{eq.}~(\ref{eq: phase parameters in weak coupling limit}) for a smooth transition to the free theory. The opposite sign of phases for both components has been already pointed out in~\cite{Pernici:1994yj}. The difference is not surprising in the light of section~\ref{sec: Fermionic self-energy}, since~$ c $ is an odd function of the Wilczek parameter~$ \zeta $. Hence, it affects the Dirac field $ \chi $ of the second component with momentum support near $ p=\pi/a\hat{e}_{\underline{\alpha}} $ as~$ -c $. The first component $ \phi $ with momentum support near $ p=0 $ is treated like the spinor field $ \psi^c $ that was discussed in section~\ref{sec: Absorption of a coefficient into a local field transformation}. The phase factors~$ \pm e^{\mp i\varphi} $ can be interpreted as plane wave terms of a Fourier transform that ensure that each pole corresponds to a component with vanishing four-momentum. If the conditions of \mbox{eq.}~(\ref{eq: phase parameter for species components}) are met, the mismatched divergent operator ($ \propto \delta c $) is removed from the Dirac kernels~$ \mathcal{K}^{\bar\phi,\phi}_{k,l} $ and~$ \mathcal{K}^{\bar\chi,\chi}_{k,l} $. The remainder of the counterterm operator ($ \propto c(g_0) $) exactly cancels the divergences due to interaction effects. The condition for the matrices $ \mathcal{Q} $ and $ \mathcal{\bar Q} $ in the interacting theory remains the same as in \mbox{eq.}~(\ref{eq: conditions on species rotation matrices}). \newline

\noindent
Though the mismatched coefficient $ \delta c $ of the divergent operator ceases to appear explicitly in the Lagrangians~$ \mathcal{L}^{\bar\phi\phi}[\phi_l,\bar\phi_k,U] $ and~$ \mathcal{L}^{\bar\chi\chi}[\chi_l,\bar\chi_k,U] $, it persists in the components' boundary conditions due to their definition in \mbox{eq.}~(\ref{eq: species components in the interacting theory}). Nevertheless, the component-mixing Lagrangians $ \mathcal{L}^{\bar\phi\chi}[\chi_l,\bar\phi_k,U] $ and $ \mathcal{L}^{\bar\chi\phi}[\phi_l,\bar\chi_k,U] $ of \mbox{eqs.}~(\ref{eq: KW Lagrangian L[phi,chi] in the interacting theory}) and~(\ref{eq: KW Lagrangian L[chi,phi] in the interacting theory}) retain its explicit appearance. Their Dirac kernels~$ K^{\bar\phi,\chi}_{k,l} $ and $ K^{\bar\chi,\phi}_{k,l} $ explictly depend on the mismatched coefficient $ \delta c $ through phases $ \varphi^\phi $ and $ \varphi^\chi $, which fail to cancel in
\small
\begin{align}
  K^{\bar\phi,\chi}_{k,l}
  =&\ \sum\limits_{n\in\Lambda} e^{i(\frac{\pi}{2}+\vartheta)}(-e^{+i(\varphi^\phi-\varphi^\chi)})^{n_{\underline{\alpha}}} (g^{\phi}_{n,k}[U])^\dagger \Big(
   +\frac{1+d}{2a} \gamma^5 \left\{ 
   U^{\underline{\alpha}}_{n} g^{\chi}_{n+\hat{e}_{\underline{\alpha}},l}[U] - 
   U^{\underline{\alpha}\dagger}_{n-\hat{e}_{\underline{\alpha}}} g^{\chi}_{n-\hat{e}_{\underline{\alpha}},l}[U] \right\}
  \nonumber \\  
  &+\sum\limits_{\mu,\nu,\lambda} \frac{1}{4a} ( -i\Sigma^{\lambda\nu} \epsilon^{\nu\lambda\mu\underline{\alpha}} )
  \left\{ 
  U^{\mu}_{n} g^{\chi}_{n+\hat{e}_{\mu},l}[U] -
  U^{\mu\dagger}_{n-\hat{e}_{\mu}} g^{\chi}_{n-\hat{e}_{\mu},l}[U] \right\}
  + m_0 \gamma^{\underline{\alpha}}\gamma^5 g^{\chi}_{n,l}[U]
  \nonumber \\
  &+\frac{i\zeta}{2a} \gamma^5 \sum\limits_{\mu\neq\underline{\alpha}}
  \left\{ 
  U^{\mu}_{n} g^{\chi}_{n+\hat{e}_{\mu},l}[U] + 
  U^{\mu\dagger}_{n-\hat{e}_{\mu}} g^{\chi}_{n-\hat{e}_{\mu},l}[U] - 2g^{\chi}_{n,l}[U]  \right\} 
  +i \gamma^5 \frac{c+(1+d)\varphi^\chi}{a} g^{\chi}_{n,l}[U]\Big),
  \label{eq: species non-conserving kernel K^phichi in the interacting theory} 
\end{align}\normalsize
\small
\begin{align}
  K^{\bar\chi,\phi}_{k,l}
  =&\  \sum\limits_{n\in\Lambda} e^{i(\frac{\pi}{2}-\vartheta)}(-e^{-i(\varphi^\phi-\varphi^\chi)})^{n_{\underline{\alpha}}} (g^{\chi}_{n,k}[U])^\dagger  \Big(
   +\frac{1+d}{2a}\gamma^5 \left\{ 
   U^{\underline{\alpha}}_{n} g^{\phi}_{n+\hat{e}_{\underline{\alpha}},l}[U] - 
   U^{\underline{\alpha}\dagger}_{n-\hat{e}_{\underline{\alpha}}} g^{\phi}_{n-\hat{e}_{\underline{\alpha}},l}[U] \right\}
  \nonumber \\  
  &+\sum\limits_{\mu,\nu,\lambda} \frac{1}{4a} ( -i\Sigma^{\lambda\nu} \epsilon^{\nu\lambda\mu\underline{\alpha}} )
  \left\{ 
  U^{\mu}_{n} g^{\phi}_{n+\hat{e}_{\mu},l}[U] - 
  U^{\mu\dagger}_{n-\hat{e}_{\mu}} g^{\phi}_{n-\hat{e}_{\mu},l}[U] \right\}
  + m_0 \gamma^{\underline{\alpha}}\gamma^5 g^{\phi}_{n,l}[U]
  \nonumber \\
  &-\frac{i\zeta}{2a} \gamma^5 \sum\limits_{\mu\neq\underline{\alpha}}
  \left\{ 
  U^{\mu}_{n} g^{\phi}_{n+\hat{e}_{\mu},l}[U] + 
  U^{\mu\dagger}_{n-\hat{e}_{\mu}} g^{\phi}_{n-\hat{e}_{\mu}[U],l} - 2g^{\phi}_{n,l}[U]  \right\}
  -i \gamma^5 \frac{c-(1+d)\varphi^\phi}{a} g^{\phi}_{n,l}[U]\Big).
  \label{eq: species non-conserving kernel K^chiphi in the interacting theory}
\end{align}\normalsize
The mismatched divergent operators ($ \propto \delta c $) in the last terms in each of \mbox{eqs.}~(\ref{eq: species non-conserving kernel K^phichi in the interacting theory}) and~(\ref{eq: species non-conserving kernel K^chiphi in the interacting theory}) cancel if the phases $ \varphi^\phi $ and $ \varphi^\chi $ are fixed according to \mbox{eq.}~(\ref{eq: phase parameter for species components}). However, there are remaining overall phase factors~$ \xi_n $ and~$ \xi_n^* $, which are defined as
\begin{equation}
  \xi_n \equiv (-e^{i(\varphi^\phi-\varphi^\chi)})^{n_{\underline{\alpha}}}=(-e^{i(2\varphi)})^{n_{\underline{\alpha}}}=(-e^{i(2\delta c)/(1+d)})^{n_{\underline{\alpha}}}.
  \label{eq: residual phase factors in species non-conserving terms}
\end{equation}
These phase factors bear witness that a mismatch $ \delta c $ affects the components of interacting fields not only at the boundary, but is presumably visible in component-mixing terms even in the bulk of the lattice. The signature of this mismatch shows in oscillating correlation functions that are considered in the next section~\ref{sec: Components in correlation functions}.
\noindent
This section is closed with a summary of the simplified Lagrangians for components of interacting fields:
\small
\begin{align}
  S^{f}[\phi,\bar\phi,\chi,\bar\chi] 
  =&\ a^4\sum\limits_{k,l} \mathcal{L}^{\bar\phi\phi}[\phi_l,\bar\phi_k,U] + \mathcal{L}^{\bar\chi\chi}[\chi_l,\bar\chi_k,U] + \mathcal{L}^{\bar\phi\chi}[\chi_l,\bar\phi_k,U] + \mathcal{L}^{\bar\chi\phi}[\phi_l,\bar\chi_k,U]
  \label{eq: KW species decomposed action in the interacting theory} ,
\\
  \mathcal{L}^{\bar\phi\phi}[\phi_l,\bar\phi_k,U] =& \sum\limits_{n,m}
  \bar\phi_k (g^{\phi}_{n,k}[U])^\dagger \left[D^N_{n,m}+\!D^{KW}_{n,m}+\!D^4_{n,m}+ (m_0+i\frac{c(g_0)}{a}\gamma^{\underline{\alpha}})\delta_{n,m}\right] g^{\phi}_{m,l}[U] \phi_l,
  \label{eq: KW simplified Lagrangian L[phi,phi] in the interacting theory} \\
  \mathcal{L}^{\bar\chi\chi}[\chi_l,\bar\chi_k,U]=& \sum\limits_{n,m}
  \bar\chi_k (g^{\chi}_{n,k}[U])^\dagger \left[D^N_{n,m}-\!D^{KW}_{n,m}+\!D^4_{n,m}+ (m_0-i\frac{c(g_0)}{a}\gamma^{\underline{\alpha}})\delta_{n,m}\right] g^{\chi}_{m,l}[U] \chi_l,
  \label{eq: KW simplified Lagrangian L[chi,chi] in the interacting theory} 
  \\
%
  \mathcal{L}^{\bar\phi\chi}[\chi_l,\bar\phi_k,U] =& \sum\limits_{n,m}
  \bar\phi_k
  (g^{\phi}_{n,k}[U])^\dagger 
   \xi_n
  \left[D^N_{n,m}+\!D^{KW}_{n,m}+\!D^4_{n,m}+ (m_0+i\frac{c(g_0)}{a}\gamma^{\underline{\alpha}})\delta_{n,m}\right]\!\mathcal{Q} g^{\chi}_{m,l}[U] \chi_l,
  \label{eq: KW simplified Lagrangian L[chi,phi] in the interacting theory} 
  \\
  \mathcal{L}^{\bar\chi\phi}[\phi_l,\bar\chi_k,U] =& \sum\limits_{n,m}
  \bar\chi_k 
  (g^{\chi}_{n,k}[U])^\dagger 
  \xi_n^*
  \left[D^N_{n,m}-\!D^{KW}_{n,m}+\!D^4_{n,m}+ (m_0-i\frac{c(g_0)}{a}\gamma^{\underline{\alpha}})\delta_{n,m}\right]\!\mathcal{\bar Q} g^{\phi}_{m,l}[U] \phi_l.
  \label{eq: KW simplified Lagrangian L[phi,chi] in the interacting theory}
\end{align}\normalsize
There is no explicit appearance of a mismatch $ \delta c $ in the relevant counterterm's coefficient up to~$ \mathcal{O}(a,\delta c^2) $ in the single-component Lagrangians of \mbox{eqs.}~(\ref{eq: KW simplified Lagrangian L[phi,phi] in the interacting theory}) and~ (\ref{eq: KW simplified Lagrangian L[chi,chi] in the interacting theory}), though it modifies their non-trivial boundary conditions. However, the component-mixing Lagrangians of \mbox{eqs.}~(\ref{eq: KW simplified Lagrangian L[chi,phi] in the interacting theory}) and~(\ref{eq: KW simplified Lagrangian L[phi,chi] in the interacting theory}) retain an explicit dependence on $ \delta c $  through phase factors $ \xi_n $ and $ \xi_n^* $ which manifest the mismatch even in the bulk of the lattice.

\newpage
\subsection{Components in correlation functions}\label{sec: Components in correlation functions}

\noindent
Next, decomposition into a pair of fields as defined in \mbox{eq.}~(\ref{eq: species components in the interacting theory}) is applied to hadronic correlation functions. A typical mesonic two-point correlation function is the expectation value (denoted by $ \big\langle \ldots \big\rangle_{U} $) of the product of two interpolating operators, which are fermionic bilinears with matrices $ \mathcal{M} $ or $ \mathcal{N} $. This mesonic correlation function
\begin{align}
  \mathcal{C}_{\mathcal{M},\mathcal{N}}(t) =&
  \big\langle \sum\limits_{n \in \Lambda_{t}^0} -\mathrm{tr} {\big( O_{\mathcal{N}}(t)  \bar O_{\mathcal{M}}(0) \big)}\big\rangle_{U} =
  \big\langle \sum\limits_{n \in \Lambda_{t}^0} -\mathrm{tr} {\big( \bar\psi_n \mathcal{N}^\dagger \psi_n \bar\psi_0 \mathcal{M} \psi_0 \big)}\big\rangle_{U}
  \label{eq: mesonic correlation function}
\end{align}\normalsize
has a connected and a quark-disconnected contribution, which are obtained from different Wick contractions of the fields. Only the former is discussed in detail in the following. As spinor fields are expressed through components fields, abbreviations from \mbox{eqs.}~(\ref{eq: KW species conjugation matrices in the interacting theory}) are employed extensively. The decomposed interpolating operators read
\small
\begin{align}
  \bar O_{\mathcal{M}}(0) =& 
  \left\{ \bar\phi_k (g^{\phi}_{0,k}[U])^\dagger \bar r_0^{\varphi} + \bar\chi_k (g^{\chi}_{0,k}[U])^\dagger \mathcal{\bar R}_0^{-\varphi} \right\} 
  \mathcal{M} 
  \left\{ r_0^{\varphi} (g^{\phi}_{0,l}[U]) \phi_l + \mathcal{ R}_0^{-\varphi} (g^{\chi}_{0,l}[U]) \chi_l  \right\}, 
 \\
  O_{\mathcal{N}}(t) =&
  \left\{ \bar\phi_v (g^{\phi}_{n,v}[U])^\dagger \bar r_n^{\varphi} + \bar\chi_v (g^{\chi}_{n,v}[U])^\dagger \mathcal{\bar R}_n^{-\varphi} \right\} 
  \mathcal{N}
  \left\{ r_n^{\varphi} (g^{\phi}_{n,w}[U]) \phi_w + \mathcal{ R}_n^{-\varphi} (g^{\chi}_{n,w}[U]) \chi_w  \right\}.
\end{align}\normalsize
The Lagrangians of \mbox{eqs.}~(\ref{eq: KW simplified Lagrangian L[phi,phi] in the interacting theory}),~\ldots,~(\ref{eq: KW simplified Lagrangian L[phi,chi] in the interacting theory}) are used in Wick contractions of the components. It is necessary to mention at this stage that transitions between the components are possible (at least through four-momentum exchanges in the interacting theory). It is understood that reference to a particular component on a fermion line is meaningful only at their endpoints. The correlation function is expressed through propagators of the components and split into sixteen pieces. Among these, twelve pieces require an odd number of transitions between different components on at least one of the fermion lines. They are arranged into three sets, which are denoted as the third, fourth and fifth set in the ensuing discussion. The remaining four pieces with an even number of transitions between components on each fermion line belong to two sets with fundamentally different properties. The two pieces of the first set, which have at each endpoint the same component on both fermion lines, exactly mirror two contributions from usual single-flavour theories (\mbox{e.g.}~Wilson fermions) and read
\begin{align}
  \mathcal{C}_{\mathcal{M},\mathcal{N}}^{\phi\bar\phi\phi\bar\phi}
  (t) = &
  \big\langle\sum\limits_{n \in \Lambda_{t}^0} 
  \mathrm{tr} \Big( 
  S^{\phi\bar\phi}_{w,k} (g^{\phi}_{0,k}[U])^\dagger \mathcal{M} (g^{\phi}_{0,l}[U]) S^{\phi\bar\phi}_{l,v} (g^{\phi}_{n,v}[U])^\dagger \mathcal{N}(g^{\phi}_{n,w}[U])
  \Big)\big\rangle_{U}, 
  \label{eq: species diagonal phi piece of correlation function} \\
  \mathcal{C}_{\mathcal{M},\mathcal{N}}^{\chi\bar\chi\chi\bar\chi}
  (t) = &
  \big\langle\sum\limits_{n \in \Lambda_{t}^0} 
  \mathrm{tr} \Big( 
  S^{\chi\bar\chi}_{w,k} (g^{\chi}_{0,k}[U])^\dagger \mathcal{\bar Q}\mathcal{M}\mathcal{Q} (g^{\chi}_{0,l}[U]) S^{\chi\bar\chi}_{l,v} (g^{\chi}_{n,v}[U])^\dagger \mathcal{\bar Q}\mathcal{N}\mathcal{Q} (g^{\chi}_{n,w}[U])  
  \Big)\big\rangle_{U}.
  \label{eq: species diagonal chi piece of correlation function}
\end{align}\normalsize
Dirac matrices for the second component can differ at most by a factor $ (-1) $, since
\begin{equation}
  \mathcal{\bar Q}\mathcal{M}\mathcal{Q} = \pm \mathcal{M},\quad \left\{\begin{array}{ccl}
  +: & \mathbf{1},i\gamma^{\underline{\alpha}}\gamma^5,\gamma^\mu,\Sigma^{\nu\lambda} & \forall\ \mu,\nu,\lambda \neq \underline{\alpha} \\
  -: & \gamma^5,\gamma^{\underline{\alpha}},i\gamma^\mu\gamma^5,\Sigma^{\underline{\alpha}\mu} & \forall\ \mu \neq \underline{\alpha} 
  \end{array}\right\}.
  \label{eq: Dirac matrices for 2nd species chi}
\end{equation}\normalsize
Thus, $ \mathcal{C}_{\mathcal{M},\mathcal{N}}^{\phi\bar\phi\phi\bar\phi}(t) $ and $ \mathcal{C}_{\mathcal{M},\mathcal{N}}^{\chi\bar\chi\chi\bar\chi}(t) $ are structurally identical. Quark-disconnected contributions to the processes of the first set are postponed until discussion of the third set. \newline
\noindent The two pieces of the second set have the same component at both endpoints of each fermion line, but different components on both fermion lines at each endpoint. They exhibit oscillating phase factors $ \xi_n $ and $ \xi_n^{*} $ that were defined in \mbox{eq.}~(\ref{eq: residual phase factors in species non-conserving terms}) and read
\begin{align}
  \mathcal{C}_{\mathcal{M},\mathcal{N}}^{\phi\bar\phi\chi\bar\chi}
  (t) = &
  \big\langle \sum\limits_{n \in \Lambda_{t}^0} 
  \xi_{n}^{*} \mathrm{tr} \Big( 
  S^{\phi\bar\phi}_{w,k} (g^{\phi}_{0,k}[U])^\dagger \mathcal{M}\mathcal{Q} (g^{\chi}_{0,l}[U]) S^{\chi\bar\chi}_{l,v} (g^{\phi}_{n,v}[U])^\dagger \mathcal{\bar Q}\mathcal{N}(g^{\phi}_{n,w}[U])
  \Big)\big\rangle_{U}, 
  \label{eq: species non-diagonal phibarphichibarchi piece of correlation function} \\
  \mathcal{C}_{\mathcal{M},\mathcal{N}}^{\chi\bar\chi\phi\bar\phi}
  (t) = &
  \big\langle \sum\limits_{n \in \Lambda_{t}^0} 
  \xi_{n} \mathrm{tr} \Big( 
  S^{\chi\bar\chi}_{w,k} (g^{\chi}_{0,k}[U])^\dagger \mathcal{\bar Q}\mathcal{M} (g^{\phi}_{0,l}[U]) S^{\phi\bar\phi}_{l,v} (g^{\phi}_{n,v}[U])^\dagger \mathcal{N}\mathcal{Q} (g^{\chi}_{n,w}[U])  
  \Big)\big\rangle_{U}.
  \label{eq: species non-diagonal chibarchiphibarphi piece of correlation function}
\end{align}\normalsize
These Dirac matrices are obviously different from those in \mbox{eqs.}~(\ref{eq: species diagonal phi piece of correlation function}) and~(\ref{eq: species diagonal chi piece of correlation function}). Therefore, the two sets overlap with $ q\bar q $ states that have different $ J^{PC} $. Matrix relations between matrices for both sets are listed in table \ref{tab: species non-diagonal matrices}. On the one hand, in the case $ \underline{\alpha}=0 $, the $ \hat{e}_{\underline{\alpha}} $~direction is parallel to the time direction. Thus, the two sets of component bilinears have different parity eigenvalues. Oscillating factors $ \xi_n $ and $ \xi_n^{*} $ are not summed over for $ \underline{\alpha}=0 $ and an oscillation along the time direction persists. On the other hand, in the case $ \underline{\alpha}\neq0 $, the $ \hat{e}_{\underline{\alpha}} $~direction is perpendicular to the time direction. Hence, the two sets of component field bilinears have different spin. Oscillating factors $ \xi_n $ and $ \xi_n^{*} $ are summed over for $ \underline{\alpha}\neq0 $ and strong cancellations that lead to suppression of component field bilinears from the second set are expected.

\begin{table}[hbt]
 \center
 \begin{tabular}{|c||c|c|c|c|}
  \hline
  $ \mathcal{M} $                 & $ \mathbf{1} $ & $ \gamma^5 $ & $ \gamma^\mu $ & $ \Sigma^{\underline{\alpha}\mu} $ \\
  $ \mathcal{M}\mathcal{Q} $      & $ i\gamma^{\underline{\alpha}}\gamma^5 $ & $ -i\gamma^{\underline{\alpha}} $ & $ \frac{1}{2}\Sigma^{\nu\lambda}\epsilon^{\lambda\nu\mu\underline{\alpha}} $ 
  & $ \gamma^\mu\gamma^5 $ \\
  $ \mathcal{\bar Q}\mathcal{M} $ & $ i\gamma^{\underline{\alpha}}\gamma^5 $ & $ +i\gamma^{\underline{\alpha}} $ & $ \frac{1}{2}\Sigma^{\nu\lambda}\epsilon^{\lambda\nu\mu\underline{\alpha}} $ 
  & $ -\gamma^\mu\gamma^5 $ \\
  \hline
 \end{tabular}
 \caption{Dirac matrices for both sets of components field bilinears with even numbers of transitions on each fermion line yield different quantum numbers for both sets.
 }
 \label{tab: species non-diagonal matrices}
 \vspace{-8pt}
\end{table}\normalsize

\begin{figure}[htb]
 \begin{flushleft}
 \begin{fmffile}{tnc_diagrams}
  \unitlength=1mm
  \hspace{10mm}
  \parbox{40mm}{
  \begin{fmfgraph*}(40,25)
   \fmfpen{thin}
   \fmfleft{i1} \fmfright{o1}
   \fmf{phantom}{i1,v1}
   \fmf{phantom}{v2,o1}
   \fmf{plain,right,tension=0.5,tag=1}{v1,v2}
   \fmf{plain,right,tension=0.5,tag=2}{v2,v1}
   \fmfblob{.10w}{v1,v2}
   \fmfv{label=$ \mathcal{\bar Q} \mathcal{M} \mathcal{Q} $,l.angle=-180,l.dist=+100}{v1}
   \fmfv{label=$ \mathcal{N} $,l.angle=00,l.dist=+100}{v2}
   \fmfposition

   \fmfipath{p[]}
   \fmfiset{p1}{vpath(__v1,__v2)}
   \fmfiset{p2}{vpath(__v2,__v1)}
   \fmfi{fermion,label=$ \chi $,l.side=left,l.dist=-130}{subpath (length(p2)/2,length(p2)) of p2}
   \fmfi{fermion,label=$ \chi $,l.side=right,l.dist=30}{subpath (0,length(p1)/2) of p1}
   \fmfi{fermion,label=$ \phi $,l.side=left,l.dist=-130}{subpath (length(p1)/2,length(p1)) of p1}
   \fmfi{fermion,label=$ \phi $,l.side=right,l.dist=30}{subpath (0,length(p2)/2) of p2}
   
   \fmfiv{decor.shape=cross}{point length(p1)/2 of p1}
   \fmfiv{decor.shape=cross}{point length(p2)/2 of p2}
  \end{fmfgraph*}
  }
  \hspace{25mm}
  \parbox{50mm}{
  \begin{fmfgraph*}(50,25)
   \fmfpen{thin}
   \fmfleft{i1} \fmfright{o1}
   \fmf{phantom}{i1,v1}
   \fmf{phantom}{v4,o1}
   \fmf{plain,right,tension=0.2,tag=1}{v1,v2}
   \fmf{plain,right,tension=0.2,tag=2}{v2,v1}
   \fmf{plain,right,tension=0.2,tag=1}{v3,v4}
   \fmf{plain,right,tension=0.2,tag=2}{v4,v3}
   \fmfblob{.10w}{v1,v4}
   \fmfv{label=$ \mathcal{\bar Q} \mathcal{M} \mathcal{Q} $,l.angle=-180,l.dist=+100}{v1}
   \fmfv{label=$ \mathcal{N} $,l.angle=0,l.dist=+100}{v4}
   \fmf{dbl_curly,straight, tension=0.5}{v2,v3} 
   \fmfposition

   \fmfipath{p[]}
   \fmfiset{p1}{vpath(__v1,__v2)}
   \fmfiset{p2}{vpath(__v2,__v1)}
   \fmfiset{p3}{vpath(__v3,__v4)}
   \fmfiset{p4}{vpath(__v4,__v3)}
   \fmfi{fermion,label=$ \chi $,l.side=left,l.dist=-130}{subpath (0,length(p1)) of p1}
   \fmfi{fermion,label=$ \chi $,l.side=right,l.dist=30}{subpath (0,length(p2)) of p2}
   \fmfi{fermion,label=$ \phi $,l.side=left,l.dist=-130}{subpath (0,length(p3)) of p3}
   \fmfi{fermion,label=$ \phi $,l.side=right,l.dist=30}{subpath (0,length(p4)) of p4}
   
   \fmfiv{decor.shape=cross}{point length(p1) of p1}
   \fmfiv{decor.shape=cross}{point 0 of p3}
  \end{fmfgraph*}
  }
 \end{fmffile}
 \unitlength=1pt
 \vspace{+5.5pt}
 \caption{Left: The third set are connected contributions to processes that involve different components at both endpoints. Right: The same structure occurs in quark-disconnected diagrams. The notation is explained in the text.
 }
 \label{fig: species non-conserving diagrams}
 \end{flushleft}
 \vspace{-8pt}
\end{figure}
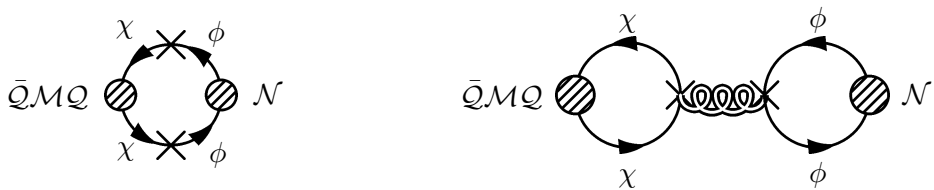

\noindent
In the end, the twelve remaining pieces with at least one fermion line with an odd number of transitions between the components fall into three different sets, which are denoted as third, fourth and fifth set in the following. Typical representatives of these sets are
\begin{align}
  \mathcal{C}_{\mathcal{M},\mathcal{N}}^{\phi\bar\chi\chi\bar\phi}
  (t) = &
  \big\langle \sum\limits_{n \in \Lambda_{t}^0} 
  \mathrm{tr} \Big( 
  S^{\phi\bar\chi}_{w,k} (g^{\chi}_{0,k}[U])^\dagger \mathcal{\bar Q}\mathcal{M}\mathcal{Q} (g^{\chi}_{0,l}[U]) S^{\chi\bar\phi}_{l,v} (g^{\phi}_{n,v}[U])^\dagger \mathcal{N}(g^{\phi}_{n,w}[U])
  \Big)\big\rangle_{U}, 
  \label{eq: species non-conserving phibarchichibarphi piece of correlation function} \\
  \mathcal{C}_{\mathcal{M},\mathcal{N}}^{\chi\bar\phi\chi\bar\phi}
  (t) = &
  \big\langle \sum\limits_{n \in \Lambda_{t}^0} 
  \xi_n \mathrm{tr} \Big( 
  S^{\chi\bar\phi}_{w,k} (g^{\phi}_{0,k}[U])^\dagger \mathcal{M}\mathcal{ Q}(g^{\chi}_{0,l}[U]) S^{\chi\bar\phi}_{l,v} (g^{\phi}_{n,v}[U])^\dagger \mathcal{N}\mathcal{Q} (g^{\chi}_{n,w}[U])  
  \Big)\big\rangle_{U},
  \label{eq: species non-conserving chibarphichibarphi piece of correlation function} \\
  \mathcal{C}_{\mathcal{M},\mathcal{N}}^{\chi\bar\chi\chi\bar\phi}
  (t) = &
  \big\langle \sum\limits_{n \in \Lambda_{t}^0} 
  \xi_n \mathrm{tr} \Big( 
  S^{\chi\bar\chi}_{w,k} (g^{\chi}_{0,k}[U])^\dagger \mathcal{\bar Q}\mathcal{M}\mathcal{ Q}(g^{\chi}_{0,l}[U]) S^{\chi\bar\phi}_{l,v} (g^{\phi}_{n,v}[U])^\dagger \mathcal{N}\mathcal{Q} (g^{\chi}_{n,w}[U])  
  \Big)\big\rangle_{U}.
  \label{eq: species non-conserving chibarchichibarphi piece of correlation function}
\end{align}\normalsize
The first representative of the third set is $ \mathcal{C}_{\mathcal{M},\mathcal{N}}^{\phi\bar\chi\chi\bar\phi}(t) $ of \mbox{eq.}~(\ref{eq: species non-conserving phibarchichibarphi piece of correlation function}). It is a connected contribution to a process, which has the same component on both fermion lines at each endpoint, but different components at each endpoint of each of the fermion lines. Thus, there must be an odd number of transitions between the components on each of the fermion lines, which sets the process apart from the first set. A pictorial representation of such a process is given by the diagram in the left part of figure~ \ref{fig: species non-conserving diagrams}. Crosses represent either two- or three-point insertions such as transitions between the components or interaction vertices with gauge fields and hatched circles represent the coupling of component fields to external meson fields. 
The second piece of the third set is $ \mathcal{C}_{\mathcal{M},\mathcal{N}}^{\chi\bar\phi\phi\bar\chi}(t) $ and requires exactly analogous treatment. 
\noindent Processes such as those described in \mbox{eq.}~(\ref{eq: species non-conserving phibarchichibarphi piece of correlation function}) are usually realised only in quark-disconnected contributions (right diagram in figure \ref{fig: species non-conserving diagrams}). The doubled curly line indicates interaction with non-perturbative gauge fields or exchange of at least three gluons, since such processes are OZI suppressed~\cite{Okubo:1963fa,Zweig:1964jf,Iizuka:1966fk}. The leading contribution is $ \mathcal{O}(a^2) $ if the components satisfy the same field equations in the continuum limit. Here, the quark-disconnected contributions to these processes involve only even numbers of transitions between the components on each fermion line and also exist for undoubled fermion fields (such as Wilson fermions). Moreover, quark-disconnected contributions to processes of the first set are analogous to quark-disconnected contributions to processes of the third set with the exception that the components on all fermion lines at all endpoints are the same. Hence, if oddness of the number of transitions between components on both fermions lines is used as the criterion for classification, these quark-disconnected contributions all belong to the first set. \newline

\begin{figure}[htb]
 \begin{flushleft}
 \begin{fmffile}{tc_diagrams}
  \unitlength=1mm
  \hspace{10mm}
  \parbox{40mm}{
  \begin{fmfgraph*}(40,25)
   \fmfpen{thin}
   \fmfleft{i1} \fmfright{o1}
   \fmf{phantom}{i1,v1}
   \fmf{phantom}{v2,o1}
   \fmf{plain,right,tension=0.5,tag=1}{v1,v2}
   \fmf{plain,right,tension=0.5,tag=2}{v2,v1}
   \fmfblob{.10w}{v1,v2}
   \fmfv{label=$ \mathcal{M} \mathcal{Q} $,l.angle=-180,l.dist=+100}{v1}
   \fmfv{label=$ \mathcal{N} \mathcal{Q} $,l.angle=0,l.dist=+100}{v2}
   \fmfposition

   \fmfipath{p[]}
   \fmfiset{p1}{vpath(__v1,__v2)}
   \fmfiset{p2}{vpath(__v2,__v1)}
   \fmfi{fermion,label=$ \phi $,l.side=left,l.dist=-130}{subpath (length(p2)/2,length(p2)) of p2}
   \fmfi{fermion,label=$ \chi $,l.side=right,l.dist=30}{subpath (0,length(p1)/2) of p1}
   \fmfi{fermion,label=$ \phi $,l.side=left,l.dist=-130}{subpath (length(p1)/2,length(p1)) of p1}
   \fmfi{fermion,label=$ \chi $,l.side=right,l.dist=30}{subpath (0,length(p2)/2) of p2}
   
   \fmfiv{decor.shape=cross}{point length(p1)/2 of p1}
   \fmfiv{decor.shape=cross}{point length(p2)/2 of p2}
  \end{fmfgraph*}
  }
  \hspace{25mm}
  \parbox{50mm}{
  \begin{fmfgraph*}(50,25)
   \fmfpen{thin}
   \fmfleft{i1} \fmfright{o1}
   \fmf{phantom}{i1,v1}
   \fmf{phantom}{v4,o1}
   \fmf{plain,right,tension=0.2,tag=1}{v1,v2}
   \fmf{plain,right,tension=0.2,tag=2}{v2,v1}
   \fmf{plain,right,tension=0.2,tag=1}{v3,v4}
   \fmf{plain,right,tension=0.2,tag=2}{v4,v3}
   \fmfblob{.10w}{v1,v4}
   \fmf{dbl_curly,straight, tension=0.5}{v2,v3} 
   \fmfv{label=$ \mathcal{M} \mathcal{Q} $,l.angle=-180,l.dist=+100}{v1}
   \fmfv{label=$ \mathcal{N} \mathcal{Q} $,l.angle=0,l.dist=+100}{v4}
   \fmfposition

   \fmfipath{p[]}
   \fmfiset{p1}{vpath(__v1,__v2)}
   \fmfiset{p2}{vpath(__v2,__v1)}
   \fmfiset{p3}{vpath(__v3,__v4)}
   \fmfiset{p4}{vpath(__v4,__v3)}
   \fmfi{fermion,label=$ \chi $,l.side=left,l.dist=-130}{subpath (0,length(p1)) of p1}
   \fmfi{fermion,label=$ \phi $,l.side=right,l.dist=30}{subpath (0,length(p2)) of p2}
   \fmfi{fermion,label=$ \phi $,l.side=left,l.dist=-130}{subpath (0,length(p3)) of p3}
   \fmfi{fermion,label=$ \chi $,l.side=right,l.dist=30}{subpath (0,length(p4)) of p4}
   
   \fmfiv{decor.shape=cross}{point length(p1) of p1}
   \fmfiv{decor.shape=cross}{point 0 of p3}
  \end{fmfgraph*}
  }
 \end{fmffile}
 \unitlength=1pt
 \vspace{+5.5pt}
 \caption{Left: The fourth set are connected contributions to processes that involve different components at both endpoints of each fermion line and on both fermion lines at each endpoints. Right: The same structure occurs in quark-disconnected diagrams as well. The notation is explained in the text.
 }
 \label{fig: species changing diagrams}
 \end{flushleft}
 \vspace{-8pt}
\end{figure}
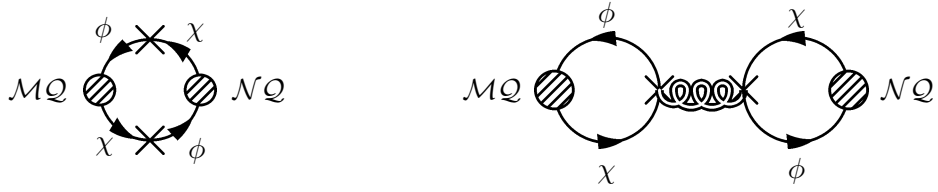

\begin{figure}[htb]
 \begin{flushleft}
 \begin{fmffile}{tv_diagrams}
  \unitlength=1mm
  \hspace{10mm}
  \parbox{40mm}{
  \begin{fmfgraph*}(40,25)
   \fmfpen{thin}
   \fmfleft{i1} \fmfright{o1}
   \fmf{phantom}{i1,v1}
   \fmf{phantom}{v2,o1}
   \fmf{plain,right,tension=0.5,tag=1}{v1,v2}
   \fmf{plain,right,tension=0.5,tag=2}{v2,v1}
   \fmfblob{.10w}{v1,v2}
   \fmfv{label=$ \mathcal{\bar Q} \mathcal{M} \mathcal{Q} $,l.angle=-180,l.dist=+100}{v1}
   \fmfv{label=$ \mathcal{N} \mathcal{Q} $,l.angle=0,l.dist=+100}{v2}
   \fmfposition

   \fmfipath{p[]}
   \fmfiset{p1}{vpath(__v1,__v2)}
   \fmfiset{p2}{vpath(__v2,__v1)}
   \fmfi{fermion,label=$ \chi $,l.side=left,l.dist=-130}{subpath (length(p2)/2,length(p2)) of p2}
   \fmfi{fermion,label=$ \chi $,l.side=right,l.dist=30}{subpath (0,length(p1)/2) of p1}
   \fmfi{fermion,label=$ \phi $,l.side=left,l.dist=-130}{subpath (length(p1)/2,length(p1)) of p1}
   \fmfi{fermion,label=$ \chi $,l.side=right,l.dist=30}{subpath (0,length(p2)/2) of p2}
   
   \fmfiv{decor.shape=cross}{point length(p1)/2 of p1}
   \fmfiv{decor.shape=cross}{point length(p2)/2 of p2}
  \end{fmfgraph*}
  }
  \hspace{25mm}
  \parbox{50mm}{
  \begin{fmfgraph*}(50,25)
   \fmfpen{thin}
   \fmfleft{i1} \fmfright{o1}
   \fmf{phantom}{i1,v1}
   \fmf{phantom}{v4,o1}
   \fmf{plain,right,tension=0.2,tag=1}{v1,v2}
   \fmf{plain,right,tension=0.2,tag=2}{v2,v1}
   \fmf{plain,right,tension=0.2,tag=1}{v3,v4}
   \fmf{plain,right,tension=0.2,tag=2}{v4,v3}
   \fmfblob{.10w}{v1,v4}
   \fmf{dbl_curly,straight, tension=0.5}{v2,v3} 
   \fmfv{label=$ \mathcal{\bar Q} \mathcal{M} \mathcal{Q} $,l.angle=-180,l.dist=+100}{v1}
   \fmfv{label=$ \mathcal{N} \mathcal{Q} $,l.angle=0,l.dist=+100}{v4}
   \fmfposition

   \fmfipath{p[]}
   \fmfiset{p1}{vpath(__v1,__v2)}
   \fmfiset{p2}{vpath(__v2,__v1)}
   \fmfiset{p3}{vpath(__v3,__v4)}
   \fmfiset{p4}{vpath(__v4,__v3)}
   \fmfi{fermion,label=$ \chi $,l.side=left,l.dist=-130}{subpath (0,length(p1)) of p1}
   \fmfi{fermion,label=$ \chi $,l.side=right,l.dist=30}{subpath (0,length(p2)) of p2}
   \fmfi{fermion,label=$ \phi $,l.side=left,l.dist=-130}{subpath (0,length(p3)) of p3}
   \fmfi{fermion,label=$ \chi $,l.side=right,l.dist=30}{subpath (0,length(p4)) of p4}
   
   \fmfiv{decor.shape=cross}{point length(p1) of p1}
   \fmfiv{decor.shape=cross}{point 0 of p3}
  \end{fmfgraph*}
  }
 \end{fmffile}
 \unitlength=1pt
 \vspace{+5.5pt}
 \caption{The fifth set requires an odd number of transitions between the components on fermion lines. Since symmetries are violated, the sum of all contributions must vanish.
 }
 \label{fig: species violating diagrams}
 \end{flushleft}
 \vspace{-8pt}
\end{figure}
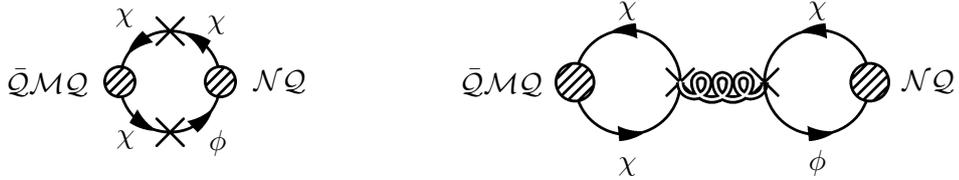

\noindent
It must be pointed out as a caveat that these diagrams serve only as a guide to the eyes and must not be understood as equivalent to a mathematical formula (in the sense of a Feynman diagram). This caveat holds for the entirety of diagrams in figures~\ref{fig: species non-conserving diagrams},~\ref{fig: species changing diagrams} and~\ref{fig: species violating diagrams}. First, assignment of any single component to a fermion line is meaningful only for asymptotically free fields at its endpoints, since there can be arbitrary numbers (though odd for the third set) of transitions between the components. Second, fermion lines for the components have decomposition kernels at their endpoints, which depend on local gauge fields appearing in parallel transports. Hence, perturbative treatment in the weak coupling regime must also include perturbative expansion of the decomposition kernels (which are not yet specified at this level of the discussion) before components can be considered as asymptotically free. \newline

\noindent
The first representative of the fourth set is  $ \mathcal{C}_{\mathcal{M},\mathcal{N}}^{\chi\bar\phi\chi\bar\phi}(t) $ of \mbox{eq.}~(\ref{eq: species non-conserving chibarphichibarphi piece of correlation function}). It is a connected contribution to a process, which has different components at each endpoint of each of the fermion lines and also different components on each fermion line at 
each endpoint. Hence, there must be an odd number of transitions between the components on each of the fermion lines, which sets the process apart from the second set. A pictorial representation of such a process is given by the diagram in the left part of figure~\ref{fig: species changing diagrams}, which may be understood in the sense of the aforementioned caveat as a guide to the eyes.
The same process also receives a quark-disconnected contribution, which is depicted in the right diagram of figure~\ref{fig: species changing diagrams}. An analogous piece $ \mathcal{C}_{\mathcal{M},\mathcal{N}}^{\phi\bar\chi\phi\bar\chi}(t) $ and a quark disconnected contribution to the same process exist as well. Furthermore, there are two quark-disconnected diagrams contributing to the processes of \mbox{eqs.}~(\ref{eq: species non-diagonal phibarphichibarchi piece of correlation function}) and~(\ref{eq: species non-diagonal chibarchiphibarphi piece of correlation function}). These contributions also involve an odd number of transitions between components on each fermion line. This is why these quark-disconnected contributions are considered as part of the fourth set. Neither of these diagrams is found in a theory of undoubled fermion fields. In order to obtain the correct continuum limit, the entirety of diagrams contributing to processes of the fourth set must either cancel mutually or vanish individually in the continuum limit. \newline

\noindent
The first representative of the fifth set is  $ \mathcal{C}_{\mathcal{M},\mathcal{N}}^{\chi\bar\chi\chi\bar\phi}(t) $ of \mbox{eq.}~(\ref{eq: species non-conserving chibarchichibarphi piece of correlation function}). It is a connected contribution to a process, which has equal components on both fermion lines at one endpoint and different components on both fermion lines at the other endpoint. Thus, there must be an odd number of transitions between the components on one of the fermion lines and an even number of transitions between the components on the other fermion line. The total number of transitions between the components is odd for the fifth set. This sets it apart from the other four sets, where the total number of transitions has always been even. A pictorial representation of such a process is given by the diagram in the left part of figure~\ref{fig: species violating diagrams}. Quark-disconnected contributions to the same kind of process exist as well and are depicted in the right diagram in figure \ref{fig: species violating diagrams}. In the sense of the aforementioned caveat, these diagrams must not be considered as more than a guide to the eyes. In total, there are eight processes in the fifth set. Since source and sink interpolating operators are multiplied by an odd number of matrices~$ \mathcal{Q} $ and~$ \mathcal{\bar Q} $, the component bilinears (assuming $ \mathcal{M}=\mathcal{N} $) accumulate an odd power of $ \gamma^5 $. Due to these contributions, either parity for $ \underline{\alpha}=0 $ or spin for $ \underline{\alpha}\neq0 $ would not be conserved. However, since either quantum number is conserved in the respective theory of minimally doubled fermions, the entirety of diagrams contributing to processes of the fifth set must cancel mutually or vanish individually on each gauge configuration. \newline

\noindent
In summary, decomposition of interpolating operators suggests five sets of contributions. The first set represents contributions that would also be observed for undoubled fermion actions. Moreover, the component bilinears have the same Dirac structure and $ J^{PC} $ as the original interpolating operators. Contributions from the first set do not include oscillating terms. The third set contributes only one kind of process that may interfere with processes of the first set. Nevertheless, there are no oscillating terms due to the third set and it matches the first set's Dirac structure. There are no additional processes that arise exclusively due to the third set. 
\noindent The second set represents contributions that would only be observed for undoubled fermion actions if both fermion lines would be assigned to different quark flavours. As an example, this kind of contribution is observed with Wilson fermions if one fermion line is considered as an \textit{up} quark and the other as a \textit{down} quark of a isospin symmetric doublet. Due to the possibility of transitions between the components, such an assignment is not possible for minimally doubled fermions. The component bilinears have different Dirac structure and thus different $ J^{PC} $ than the original interpolating operators. Moreover, contributions from the second set include oscillating terms with either $ \xi_n=(-e^{+i2\varphi})^{n_{\underline{\alpha}}} $ or  $ \xi_n^{*}=(-e^{-i2\varphi})^{n_{\underline{\alpha}}} $. The fact that these oscillating factors have frequencies that are not integer multiples of $ (2\pi)/(aN_{\underline{\alpha}}) $ seems unusual given that the eigenfrequencies of the lattice are restricted to integer multiples of $ (2\pi)/(aN_{\underline{\alpha}}) $. However, the frequencies can be realised as maxima of distributions of discrete eigenfrequencies of the lattice. A harmonic oscillator and a one-dimensional fermion with imaginary mass serve as toy models that exhibit similar behaviour. These toy models are discussed in the appendix~\ref{app: Oscillating lattice toy models} to illustrate this phenomenon. It seems clear that these contributions are very sensitive to a mismatch $ \delta c = (1+d)\varphi $ of the relevant counterterm's coefficient. It is expected that this yields oscillating contributions in the parallel case ($ \underline{\alpha}=0 $) and strong cancellations in perpendicular cases ($ \underline{\alpha}\neq0 $). Oscillating contributions of this kind (though without dependence on a mismatched counterterm's coeffcient) are a well-known observation in simulations with staggered fermions~\cite{Altmeyer:1992dd}. 
The contributions from the second set interfere with contributions from the fourth set. The same considerations about Dirac structure and oscillating terms are valid for the latter. However, the fourth set includes contributions that suggest non-conservation of the components. If the components have any physical meaning, such processes must not survive in the continuum limit.  
\noindent Lastly, contributions of the fifth set violate symmetries of the action. They must either cancel mutually or vanish individually on every gauge configuration even at finite lattice spacing. Hence, these terms do not contribute to any observable quantities and can be neglected in the interpetation. \newline

\noindent Since numerical simulations in section~\ref{sec: Numerical studies} are focused on only the connected contribution to two different mesonic channels ($ \mathcal{M}=\mathcal{N}=\gamma^5 $ and $ \mathcal{M}=\mathcal{N}=\gamma^0 $) with either $ \underline{\alpha}=0 $ or $ \underline{\alpha}=3 $, these mesonic correlations are constructed in detail. Using \mbox{eqs.}~(\ref{eq: residual phase factors in species non-conserving terms}) and (\ref{eq: species diagonal phi piece of correlation function})-(\ref{eq: species non-conserving chibarphichibarphi piece of correlation function}), $ \mathcal{C}_{\gamma^5,\gamma^5}(t) $ for $ \underline{\alpha}=0 $, $ \vartheta=0 $ and an infinite lattice can be written as
\small
\begin{align}
  \mathcal{C}_{\gamma^5,\gamma^5}(t) =& \mathcal{C}_{\gamma^5,\gamma^5}^{\text{std}}(t) +\mathcal{C}_{\gamma^0,\gamma^0}^{\text{osc}}(t) 
  \label{eq: decomposed gamma5 correlation function}, \\
  \mathcal{C}_{\gamma^5,\gamma^5}^{\text{std}}(t) =& \big\langle\sum\limits_{n \in \Lambda_{t}^0} 
  \mathrm{tr} \Big( 
  S^{\phi\bar\phi}_{w,k} (g^{\phi}_{0,k}[U])^\dagger \gamma^5 (g^{\phi}_{0,l}[U]) S^{\phi\bar\phi}_{l,v} (g^{\phi}_{n,v}[U])^\dagger \gamma^5 (g^{\phi}_{n,w}[U])
  \Big)\big\rangle_{U} \nonumber \\
  +& \big\langle\sum\limits_{n \in \Lambda_{t}^0} 
  \mathrm{tr} \Big( 
  S^{\chi\bar\chi}_{w,k} (g^{\chi}_{0,k}[U])^\dagger \gamma^5 (g^{\chi}_{0,l}[U]) S^{\chi\bar\chi}_{l,v} (g^{\chi}_{n,v}[U])^\dagger \gamma^5 (g^{\chi}_{n,w}[U])  
  \Big)\big\rangle_{U} \nonumber \\
  -& \big\langle \sum\limits_{n \in \Lambda_{t}^0} 
  \mathrm{tr} \Big( 
  S^{\phi\bar\chi}_{w,k} (g^{\chi}_{0,k}[U])^\dagger \gamma^5 (g^{\chi}_{0,l}[U]) S^{\chi\bar\phi}_{l,v} (g^{\phi}_{n,v}[U])^\dagger \gamma^5 (g^{\phi}_{n,w}[U])
  \Big)\big\rangle_{U} \nonumber \\
  -& \big\langle \sum\limits_{n \in \Lambda_{t}^0} 
  \mathrm{tr} \Big( 
  S^{\chi\bar\phi}_{w,k} (g^{\phi}_{0,k}[U])^\dagger \gamma^5 (g^{\phi}_{0,l}[U]) S^{\phi\bar\chi}_{l,v} (g^{\chi}_{n,v}[U])^\dagger \gamma^5 (g^{\chi}_{n,w}[U])
  \Big)\big\rangle_{U}
  \label{eq: standard piece of correlation function}, \\
  \mathcal{C}_{\gamma^0,\gamma^0}^{\text{osc}}(t) =&\big\langle \sum\limits_{n \in \Lambda_{t}^0} 
  e^{-i(\pi+2\varphi)n_0}\ \mathrm{tr} \Big( 
  S^{\phi\bar\phi}_{w,k} (g^{\phi}_{0,k}[U])^\dagger \gamma^0 (g^{\chi}_{0,l}[U]) S^{\chi\bar\chi}_{l,v} (g^{\chi}_{n,v}[U])^\dagger \gamma^0 (g^{\phi}_{n,w}[U])
  \Big)\big\rangle_{U} \nonumber \\
  +& \big\langle \sum\limits_{n \in \Lambda_{t}^0} 
  e^{+i(\pi+2\varphi)n_0}\ \mathrm{tr} \Big( 
  S^{\chi\bar\chi}_{w,k} (g^{\chi}_{0,k}[U])^\dagger \gamma^0 (g^{\phi}_{0,l}[U]) S^{\phi\bar\phi}_{l,v} (g^{\phi}_{n,v}[U])^\dagger \gamma^0 (g^{\chi}_{n,w}[U])  
  \Big)\big\rangle_{U} \nonumber \\
  -& \big\langle \sum\limits_{n \in \Lambda_{t}^0} 
  e^{+i(\pi+2\varphi)n_0}\  \mathrm{tr} \Big( 
  S^{\chi\bar\phi}_{w,k} (g^{\phi}_{0,k}[U])^\dagger \gamma^0 (g^{\chi}_{0,l}[U]) S^{\chi\bar\phi}_{l,v} (g^{\phi}_{n,v}[U])^\dagger \gamma^0 (g^{\chi}_{n,w}[U])  
  \Big)\big\rangle_{U} \nonumber \\
  -& \big\langle \sum\limits_{n \in \Lambda_{t}^0} 
  e^{-i(\pi+2\varphi)n_0}\ \mathrm{tr} \Big( 
  S^{\phi\bar\chi}_{w,k} (g^{\chi}_{0,k}[U])^\dagger \gamma^0 (g^{\phi}_{0,l}[U]) S^{\phi\bar\chi}_{l,v} (g^{\chi}_{n,v}[U])^\dagger \gamma^0 (g^{\phi}_{n,w}[U])  
  \Big)\big\rangle_{U}.
  \label{eq: oscillating piece of correlation function}
\end{align}\normalsize
The correlation function $ \mathcal{C}_{\gamma^0,\gamma^0}(t) $ is obtained by the trivial replacement $ \gamma^5 \leftrightarrow \gamma^0 $ throughout the expressions. Correlation functions for $ \underline{\alpha}=3 $ are obtained by the trivial replacement $ 0 \leftrightarrow 3 $ in $ n_0 $ and $ \gamma^0 $ throughout the expressions. Unless the decomposition kernels are sufficiently localised, this correlation function cannot be interpreted in terms of any states formed from the fields $ \phi $ and $ \chi $. However, if the decomposition kernels are sufficiently localised, each of the terms in $ \mathcal{C}_{\gamma^5,\gamma^5}^{\text{std/osc}}(t) $ represents states with $ J^{PC}=0^{-+} $ propagating from time~$ 0 $ to time~$ t $. Therefore, an oscillating contribution from $ 0^{-+} $ states should be be observed in the $ \gamma^0 $ channel. 
Each of the terms in $ \mathcal{C}_{\gamma^0,\gamma^0}^{\text{std/osc}}(t) $ represents states with $ J^{PC}=0^{+-} $ that are not part of the QCD spectrum and contribute supposedly only noise. Hence, the formal decomposition suggest existence of $ 0^{-+} $ states in $ \gamma^5 $ and $ \gamma^0 $ channels.  Numerical data indicating the presence of $ 0^{-+} $ states is presented in sections~\ref{sec: Oscillating correlation functions} and~\ref{sec: Chiral behaviour of the pseudoscalar ground state}.

\newpage
\sectionc{Remnant $ \Theta $ symmetry and $ \mathcal{O}(a) $ corrections}{sec: Remnant Theta symmetry and O(a) corrections}

\noindent The Karsten-Wilczek term breaks $ n_{\underline{\alpha}} $~reflection and charge conjugation symmetries, $ CP\Theta $ and both $ C\Theta_{\underline{\alpha}} $ and $ P_{\underline{\alpha}} $ are still good symmetries. Hence, generic observables that would have both symmetries if constructed using Wilson fermions may well have only $ C\Theta_{\underline{\alpha}} $ symmetry if constructed using Karsten-Wilczek fermions.
In particular, since correlation functions may well have only $ C\Theta_{\underline{\alpha}} $ symmetry, they potentially lack symmetry under time reflection in the parallel case ($ \underline{\alpha}=0 $). 
\noindent Nevertheless, both marginal anisotropic counterterms of \mbox{eqs.}~(\ref{eq: ferm dim 4 KW counterterm operator}) and~(\ref{eq: gluonic KW counterterm}) are symmetric under $ n_{\underline{\alpha}} $~reflection and charge conjugation. Hence, the perturbative vacuum at one-loop has the symmetries of the gluonic counterterm. If the perturbative vacuum has higher symmetry than generic observables, the question arises whether this may apply to other (perturbative or non-perturbative) observables as well. 
\noindent In section~\ref{sec: CTheta symmetry in the free theory}, invariance of the pseudoscalar correlation function under $ n_{\underline{\alpha}} $~reflection is derived analytically for free fields at first. Next, this invariance is deduced from $ C\Theta_{\underline{\alpha}} $ symmetry. 
\noindent In section \ref{sec: CTheta symmetry in the interacting theory}, general local correlation functions with charge conjugated gauge fields and an invariant vacuum are studied. It is shown that interpolating operators at source and sink with equal charge conjugation eigenvalue ensure symmetry under $ n_{\underline{\alpha}} $~reflection and suppress $ \mathcal{O}(a) $ corrections. This is a consequence of $ C\Theta_{\underline{\alpha}} $ symmetry.

\subsection{$ C\Theta $ symmetry in the free theory}\label{sec: CTheta symmetry in the free theory}

\subsubsection{The pseudoscalar correlation function in the free theory}\label{sec: The pseudoscalar correlation function in the free theory}

\noindent
The pseudoscalar correlation function is a special case of \mbox{eq.}~(\ref{eq: mesonic correlation function}) and reads
\begin{align}
  \mathcal{C}_{5,5}(t) \equiv \mathcal{C}_{\gamma^5,\gamma^5}(t) =&\
  \big\langle \sum\limits_{n \in \Lambda_{t}^0}-\mathrm{tr} {\left( \bar\psi_0 \gamma^5 \psi_0 \bar\psi_n \gamma^5 \psi_n \right)}\big\rangle_{1} .
  \label{eq: pseudoscalar correlation function}
\end{align}\normalsize
The spinor fields are Wick contracted and yield propagators as
\begin{align}
  \mathcal{C}_{5,5}(t) =& \big\langle \sum\limits_{n \in \Lambda_{t}^0}
  \mathrm{tr} {\left( S_{n,0} \gamma^5 S_{0,n} \gamma^5 \right)}
  -  \mathrm{tr} {\left( S_{0,0} \gamma^5 \right)} \mathrm{tr} {\left(S_{n,n} \gamma^5 \right)} \big\rangle_{1},
  \label{eq: Wick contracted pseudoscalar correlation function}
\end{align}\normalsize
where the first piece is the connected and the second piece is the quark-disconnected contribution. For a~$ \gamma^5 $~hermitian action, the connected contribution is written as
\begin{align}
  \mathcal{C}_{5,5}^{\mathrm{con.}}(t) =& \big\langle \sum\limits_{n \in \Lambda_{t}^0}
  \mathrm{tr} {\left( S_{n,0} \gamma^5 S_{0,n} \gamma^5 \right)}\big\rangle_{1}
  = \big\langle \sum\limits_{n \in \Lambda_{t}^0}
  \mathrm{tr} {\left( S_{n,0} S_{n,0}^\dagger \right)}\big\rangle_{1}.
  \label{eq: connected pseudoscalar correlation function}
\end{align}\normalsize
As gauge fields of the free theory are trivial ($ U=1 $), the expectation value can be omitted here. Furthermore, explicit indication as connected contribution is omitted as well. The free correlation function in the chiral limit is cast into momentum space as
\begin{align}
  \mathcal{C}_{5,5}^{}(t) =& \sum\limits_{n \in \Lambda_{t}^0} \int\limits \frac{d^4p\, d^4k}{(2\pi)^8} e^{in\cdot p}\ 
  \mathrm{tr} {\left( S(k+\frac{p}{2};\zeta,0,a) S^\dagger(k-\frac{p}{2};\zeta,0,a) \right)}
  \label{eq: pseudoscalar correlation function in momentum space}
\end{align}\normalsize
using the propagator of \mbox{eq.}~(\ref{eq: KW propagator in momentum space}). Summation of the sink slice introduces a delta function~$ \delta^{(3)}(\mathbf{p}) $ for external three-momenta. The trace acts upon the numerator $ N $ as
\begin{align}
  N=&\ \mathrm{tr} {\left\{\left( 
  -i(\gamma \cdot s_{k+\tfrac{p}{2}}) 
  -i\tfrac{\zeta}{2} \gamma_{\underline{\alpha}} (\hat{s}_{k+\tfrac{p}{2}})_\perp^2 \right)
  \left( +i(\gamma \cdot s_{k-\tfrac{p}{2}}) 
  +i\tfrac{\zeta}{2} \gamma_{\underline{\alpha}} (\hat{s}_{k-\tfrac{p}{2}})_\perp^2 \right)\right\}_{p=p_0\hat{e}_0}}.
  \nonumber
\end{align}\normalsize
The abbreviations are defined in \mbox{eqs.}~(\ref{eq: notation for LPT}) and~(\ref{eq: definition of varrho^munu}). The numerator is expanded as
\begin{align}
  N=&\ \big\{
  4(s_{k+\tfrac{p}{2}} \cdot s_{k-\tfrac{p}{2}})
  + \zeta^2 (\hat{s}_{k+\tfrac{p}{2}})_\perp^2(\hat{s}_{k-\tfrac{p}{2}})_\perp^2
  +2\zeta \left(
  s^{\underline{\alpha}}_{k+\tfrac{p}{2}}(\hat{s}_{k-\tfrac{p}{2}})_\perp^2+
  s^{\underline{\alpha}}_{k-\tfrac{p}{2}}(\hat{s}_{k+\tfrac{p}{2}})_\perp^2\right)
  \big\}_{p=p_0\hat{e}_0}
  \nonumber \\
  =&\ 
  \sum\limits_{\mu}(s^\mu_{k}\hat{c}^\mu_{p})^2-(c^0_{k}\hat{s}^0_{p})^2
  + \zeta 
  \Big\{\sum\limits_{\mu\neq\underline{\alpha}} \big(s^{\underline{\alpha}}_{k}\hat{c}^\mu_{p} (\hat{s}^\mu_{k}\hat{c}^\mu_{p/2})^2\big)+\varrho^{\underline{\alpha}0}s^{\underline{\alpha}}_{k}\hat{c}^0_{p} (\hat{c}^0_{k}\hat{s}^0_{p/2})^2 \Big\}
  \nonumber\\
  &\ 
  +\zeta^2 \Big(\frac{1}{16} \big\{\sum\limits_{\mu\neq\underline{\alpha}}(\hat{s}^\mu_{k}\hat{c}^\mu_{p/2})^2+\varrho^{\underline{\alpha}0}(\hat{c}^0_{k}\hat{s}^0_{p/2})^2\big\}^2
  - \varrho^{\underline{\alpha}0}\big(s^0_{k}\hat{s}^0_{p}\big)^2 \Big)
  ,
  \label{eq: simplified pseudoscalar numerator}
\end{align}\normalsize
until it finally simplifies to a manifestly even function of $ p_0 $. The denominator $ D $ reads
\begin{align}
  D=&\ \Big\{\big( 
  (s_{k+\tfrac{p}{2}})^2 +\frac{\zeta^2}{4} \left((\hat{s}_{k+\tfrac{p}{2}})_\perp^2\right)^2
  +\zeta s^{\underline{\alpha}}_{k+\tfrac{p}{2}}(\hat{s}_{k+\tfrac{p}{2}})_\perp^2
  \big) 
  \nonumber \\
  &\times\ 
  \big( (s_{k-\tfrac{p}{2}})^2 +\frac{\zeta^2}{4} \left((\hat{s}_{k-\tfrac{p}{2}})_\perp^2\right)^2
  +\zeta s^{\underline{\alpha}}_{k-\tfrac{p}{2}}(\hat{s}_{k-\tfrac{p}{2}})_\perp^2 
  \big)
  \Big\}_{p=p_0\hat{e}_0},
  \nonumber
\end{align}\normalsize
and is expanded as $ D=X+Y+Z $ with
\small
\begin{align}
  X=&\ 
  \big\{ (s_{k+\tfrac{p}{2}})^2 +\tfrac{\zeta^2}{4} \left((\hat{s}_{k+\tfrac{p}{2}})_\perp^2\right)^2 \big\} 
  \big\{ (s_{k-\tfrac{p}{2}})^2 +\tfrac{\zeta^2}{4} \left((\hat{s}_{k-\tfrac{p}{2}})_\perp^2\right)^2 \big\}
  = \sum\limits_{i=0}^2 \left(\frac{\zeta^2}{4}\right)^{i}X_{2i},
  \\
  Y=&\ \big\{
  \zeta s^{\underline{\alpha}}_{k+\tfrac{p}{2}}(\hat{s}_{k+\tfrac{p}{2}})_\perp^2
  \big\}
  \big\{
  \zeta s^{\underline{\alpha}}_{k-\tfrac{p}{2}}(\hat{s}_{k-\tfrac{p}{2}})_\perp^2 
  \big\}
  \\ 
 Z=&\ z(+p) + z(-p) = \zeta Z_1 + \zeta^3 Z_3,
  \\
  z(+p)=&\ 
  \big\{ (s_{k+\tfrac{p}{2}})^2 +\tfrac{\zeta^2}{4} \left((\hat{s}_{k+\tfrac{p}{2}})_\perp^2\right)^2 \big\}
  \big\{
  \zeta s^{\underline{\alpha}}_{k-\tfrac{p}{2}}(\hat{s}_{k-\tfrac{p}{2}})_\perp^2 
  \big\},
\end{align}\normalsize
where the restriction to $ p=p_0\hat{e}_0 $ is omitted. The individual terms read
\small
\begin{align}
  X_0=&\  
  \frac{1}{16}\big( \sum\limits_{\mu}\{(s^\mu_{k}\hat{c}^\mu_{p})^2+(c^\mu_{k}\hat{s}^\mu_{p})^2\}\big)^2 
  - \big(\sum\limits_{\mu}\{s^\mu_{k}c^\mu_{k}s^\mu_{p}\}\big)^2,
  \\
  X_2=&\ 
  \Bigg(
  \frac{1}{4} \sum\limits_{\mu}\{(s^\mu_{k}\hat{c}^\mu_{p})^2+(c^\mu_{k}\hat{s}^\mu_{p})^2\} \Big\{ 
  \frac{1}{16} \big( \sum\limits_{\nu\neq\underline{\alpha}} \{(\hat{s}^\nu_k \hat{c}^\nu_{p/2})^2+(\hat{c}^\nu_k \hat{s}^\nu_{p/2})^2\} \big)^2
  + \big( \sum\limits_{\nu\neq\underline{\alpha}} \{s^\nu_k \hat{s}^\nu_{p}\} \big)^2 \Big\}
  \nonumber \\
  &-\
  \big( \sum\limits_{\mu\neq\underline{\alpha}} \{(\hat{s}^\mu_k \hat{c}^\mu_{p/2})^2+(\hat{c}^\mu_k \hat{s}^\mu_{p/2})^2\} \big)
  \big(\sum\limits_{\nu}\{s^\nu_{k}c^\nu_{k}s^\nu_{p}\}\big) \big( \sum\limits_{\lambda\neq\underline{\alpha}} \{s^\lambda_k \hat{s}^\lambda_{p}\} \big)
  \Bigg),
  \\
  X_4=&\ 
  \Big\{ 
  \big( \sum\limits_{\mu\neq\underline{\alpha}} \{(\hat{s}^\mu_k \hat{c}^\mu_{p/2})^2+(\hat{c}^\mu_k \hat{s}^\mu_{p/2})^2\} \big)^2
  -\big( \sum\limits_{\mu\neq\underline{\alpha}} \{s^\mu_k \hat{s}^\mu_{p}\} \big)^2\Big\}^2,
  \\
  Y=&\ 
  \tfrac{\zeta^2}{4}
  \{(\hat{s}^{\underline{\alpha}}_k \hat{c}^{\underline{\alpha}}_{p/2})^2-(\hat{c}^{\underline{\alpha}}_k \hat{s}^{\underline{\alpha}}_{p/2})^2\} \Big\{ 
  \tfrac{1}{16} \big( \sum\limits_{\mu\neq\underline{\alpha}} \{(\hat{s}^\mu_k \hat{c}^\mu_{p/2})^2+(\hat{c}^\mu_k \hat{s}^\mu_{p/2})^2\} \big)^2
  - \big( \sum\limits_{\mu\neq\underline{\alpha}} \{s^\mu_k \hat{s}^\mu_{p}\} \big)^2 \Big\},
\end{align}\normalsize
\small
\begin{align}
  Z_1=&\ 
  \tfrac{\big( \sum\limits_{\mu} \{(s^{\mu}_k \hat{c}^{\mu}_{p})^2-(c^{\mu}_k \hat{s}^{\mu}_{p})^2\} \big) }{4}
  \Big\{ 
  \frac{\big(s^{\underline{\alpha}}_k \hat{c}^{\underline{\alpha}}_p\big)}{4}  \big( \sum\limits_{\nu\neq\underline{\alpha}} \{(\hat{s}^\nu_k \hat{c}^\nu_{p/2})^2+(\hat{c}^\nu_k \hat{s}^\nu_{p/2})^2\} \big)
  +\big(c^{\underline{\alpha}}_k \hat{s}^{\underline{\alpha}}_p\big)  \big( \sum\limits_{\nu\neq\underline{\alpha}} \{s^\nu_k \hat{s}^\nu_{p}\} \big) \Big\}
  \nonumber \\
  &-\
  \big( \sum\limits_{\mu}\{s^\mu_{k}c^\mu_{k}s^\mu_{p}\} \big)
  \Big\{ 
  \frac{\big(c^{\underline{\alpha}}_k \hat{s}^{\underline{\alpha}}_p\big)}{4} 
  \big( \sum\limits_{\nu\neq\underline{\alpha}} \{(\hat{s}^\nu_k \hat{c}^\nu_{p/2})^2+(\hat{c}^\nu_k \hat{s}^\nu_{p/2})^2\} \big)
  + \big(s^{\underline{\alpha}}_k \hat{c}^{\underline{\alpha}}_p\big) \big( \sum\limits_{\nu\neq\underline{\alpha}} \{s^\nu_k \hat{s}^\nu_{p}\} \big) 
  \Big\},
  \\
  Z_3=&\ 
  2 \Big\{ 
  \frac{\big(s^{\underline{\alpha}}_k \hat{c}^{\underline{\alpha}}_p\big)}{4} 
  \big( \sum\limits_{\nu\neq\underline{\alpha}} \{(\hat{s}^\nu_k \hat{c}^\nu_{p/2})^2+(\hat{c}^\nu_k \hat{s}^\nu_{p/2})^2\} \big)
  - \big(c^{\underline{\alpha}}_k \hat{s}^{\underline{\alpha}}_p\big) \big( \sum\limits_{\nu\neq\underline{\alpha}} \{s^\nu_k \hat{s}^\nu_{p}\} \big) 
  \Big\}
  \nonumber \\
  &\times\ \Big\{ 
  \frac{1}{16} \big( \sum\limits_{\mu\neq\underline{\alpha}} \{(\hat{s}^\mu_k \hat{c}^\mu_{p/2})^2+(\hat{c}^\mu_k \hat{s}^\mu_{p/2})^2\} \big)^2
  - \big( \sum\limits_{\mu\neq\underline{\alpha}} \{s^\mu_k \hat{s}^\mu_{p}\} \big)^2 \Big\}.
\end{align}\normalsize
Due to the delta function~$ \delta^{(3)}(\mathbf{p}) $ , these terms are simplified to 
\small
\begin{align}
  X_0=&\  
  \frac{1}{16}\big\{ \sum\limits_{\mu}(s^\mu_{k}\hat{c}^\mu_{p})^2+(c^0_{k}\hat{s}^0_{p})^2 \big\}^2 
  - \big\{ s^0_{k}c^0_{k}s^0_{p} \big\}^2,
  \\
  X_2=&\ 
  \Bigg(
  \frac{1}{4} \{\sum\limits_{\mu}(s^\mu_{k}\hat{c}^\mu_{p})^2+(c^0_{k}\hat{s}^0_{p})^2\} \Big( 
  \frac{1}{16} \big\{ \sum\limits_{\nu\neq\underline{\alpha}} (\hat{s}^\nu_k \hat{c}^\nu_{p/2})^2+\varrho^{\underline{\alpha}0}(\hat{c}^0_k \hat{s}^0_{p/2})^2 \big\}^2
  + \varrho^{\underline{\alpha}0}\{s^0_k \hat{s}^0_{p}\}^2 \Big)
  \nonumber \\
  &-\
  \varrho^{\underline{\alpha}0}\big\{ \sum\limits_{\mu\neq\underline{\alpha}} (\hat{s}^\mu_k \hat{c}^\mu_{p/2})^2+\varrho^{\underline{\alpha}0}(\hat{c}^0_k \hat{s}^0_{p/2})^2 \big\}
  \{s^0_{k}c^0_{k}s^0_{p}\} \{s^0_k \hat{s}^0_{p}\}
  \Bigg),
  \\
  X_4=&\ 
  \Big( 
  \big\{ \sum\limits_{\nu\neq\underline{\alpha}} (\hat{s}^\nu_k \hat{c}^\nu_{p/2})^2+\varrho^{\underline{\alpha}0}(\hat{c}^0_k \hat{s}^0_{p/2})^2 \big\}^2
  -\varrho^{\underline{\alpha}0}\{s^0_k \hat{s}^0_{p}\}^2 \Big)^2,
  \\
  Y=&\ 
  \tfrac{\zeta^2}{4}
  \Big\{\frac{(\hat{s}^{\underline{\alpha}}_k \hat{c}^{\underline{\alpha}}_{p/2})^2}{16} \Big( 
   \big\{ \sum\limits_{\nu\neq\underline{\alpha}} (\hat{s}^\nu_k \hat{c}^\nu_{p/2})^2+\varrho^{\underline{\alpha}0}(\hat{c}^0_k \hat{s}^0_{p/2})^2 \big\}^2
  + \varrho^{\underline{\alpha}0}\{s^0_k \hat{s}^0_{p}\}^2 \Big)
  \nonumber \\
  &\ 
  -\delta^{\underline{\alpha}0}(\hat{c}^{0}_k \hat{s}^{0}_{p/2})^2
  \big\{ \sum\limits_{\nu\neq\underline{\alpha}} (\hat{s}^\nu_k \hat{c}^\nu_{p/2})^2 \big\}^2
  \Big\},
  \\
  Z_1=&\ 
  \{ \sum\limits_{\mu} (s^{\mu}_k \hat{c}^{\mu}_{p})^2-(c^{0}_k \hat{s}^{0}_{p})^2\}
  \frac{\big(s^{\underline{\alpha}}_k \hat{c}^{\underline{\alpha}}_p\big)}{16}  \big( \sum\limits_{\nu\neq\underline{\alpha}} (\hat{s}^\nu_k \hat{c}^\nu_{p/2})^2+\varrho^{\underline{\alpha}0}(\hat{c}^0_k \hat{s}^0_{p/2})^2 \big)
  \nonumber \\
  &-\
  \{s^0_{k}c^0_{k}s^0_{p}\}
  \Big(
  \delta^{\underline{\alpha}0}\frac{\big(c^{0}_k \hat{s}^{0}_p\big)}{4} 
  \big\{ \sum\limits_{\mu\neq\underline{\alpha}} (\hat{s}^\mu_k \hat{c}^\mu_{p/2})^2 \big\}
  + \big(s^{\underline{\alpha}}_k \hat{c}^{\underline{\alpha}}_p\big) \varrho^{\underline{\alpha}0}\{s^0_k \hat{s}^0_{p}\} 
  \Big),
  \\
  Z_3=&\ 
  \tfrac{\big(s^{\underline{\alpha}}_k \hat{c}^{\underline{\alpha}}_p\big)}{2} 
  \big( \sum\limits_{\nu\neq\underline{\alpha}} (\hat{s}^\nu_k \hat{c}^\nu_{p/2})^2+\varrho^{\underline{\alpha}0}(\hat{c}^0_k \hat{s}^0_{p/2})^2 \big)
  \Big( 
  \tfrac{1}{16} \big\{ \sum\limits_{\nu\neq\underline{\alpha}} (\hat{s}^\nu_k \hat{c}^\nu_{p/2})^2+\varrho^{\underline{\alpha}0}(\hat{c}^0_k \hat{s}^0_{p/2})^2 \big\}^2
  -\varrho^{\underline{\alpha}0}\{s^0_k \hat{s}^0_{p}\}^2 \Big).
\end{align}\normalsize
Hence, all terms in the denominator are even functions of~$ p_0 $. Because both numerator and denominator are even functions of~$ p_0 $, the loop integral of the correlation function,
\begin{align}
  \widetilde{\mathcal{C}}(p_0) =&  \int\limits \frac{d^4k}{(2\pi)^4} 
  \mathrm{tr} {\left( S(k+\hat{e}_0\frac{p_0}{2};\zeta,0,a) S^\dagger(k-\hat{e}_0\frac{p_0}{2};\zeta,0,a) \right)} ,
  \label{eq: loop integral of the pseudoscalar correlation function in momentum space}
\end{align}\normalsize
is necessarily an even function of $ p_0 $. Thus, the exponential in \mbox{eq.}~(\ref{eq: pseudoscalar correlation function in momentum space}) can be replaced by a cosine and the connected part of the pseudoscalar correlation function is equal to
\begin{align}
  \mathcal{C}_{5,5}^{}(t) =& \int\limits \frac{d p_0}{(2\pi)} \cos{(t p_0)}\ \frac{\widetilde{\mathcal{C}}(p_0)+\widetilde{\mathcal{C}}(-p_0)}{2}. 
  \label{eq: time symmetric pseudoscalar correlation function in momentum space}
\end{align}\normalsize
Hence, even though the propagators explicitly break~$ n_{\underline{\alpha}} $~reflection symmetry, the symmetry is restored in the pseudoscalar correlation function of the free theory. The free pseudoscalar correlation function is manifestly invariant under time reflection and parity for any version~($ \underline{\alpha}=0 $ or~$ \underline{\alpha}\neq 0 $) of the Karsten-Wilczek action. Though it is not a priori clear whether this remarkable result can be reproduced with interacting fields, it certainly warrants further analytical and numerical studies. 

\subsubsection{$ C\Theta $, charge conjugation and reflection symmetry}\label{sec: CTheta, charge conjugation and reflection symmetry}

\noindent In order to understand the presence of~$ n_{\underline{\alpha}} $~reflection symmetry, it is related to charge conjugation symmetry using $ C\Theta_{\underline{\alpha}} $ symmetry of the theory. Using \mbox{eq.}~(\ref{eq: charge conjugation}), the pseudoscalar correlation function of \mbox{eq.}~(\ref{eq: pseudoscalar correlation function}) is charge conjugated and reads
\begin{align}
  \mathcal{C}_{5,5}(t) =&
  \big\langle \sum\limits_{n \in \Lambda_{t}^0} -\mathrm{tr} {\left( \bar\psi_0 \gamma^5 \psi_0 \bar\psi_n \gamma^5 \psi_n \right)} \big\rangle_{1}
  \stackrel{C}{\to} \nonumber \\
    \widehat{C}\mathcal{C}_{5,5}(t) \equiv&\ \mathcal{C}_{5,5}^C(t) =
  \big\langle \sum\limits_{n \in \Lambda_{t}^0} -\mathrm{tr} {\left( \bar\psi_0^C \gamma^5 \psi_0^C \bar\psi_n^C \gamma^5 \psi_n^C \right)}  \big\rangle_{1^C}
  \nonumber \\
  =&\ \big\langle \sum\limits_{n \in \Lambda_{t}^0} -\mathrm{tr} {\left( (-\psi_0^T) C \gamma^5 C (\bar\psi_0^T)  (-\psi_n^T) C \gamma^5 C (\bar\psi_n^T) \right)} \big\rangle_{1},
  \label{eq: charge conjugated pseudoscalar correlation function}
\end{align}\normalsize
where the charge conjugation matrix~$ C $ is defined in \mbox{eq.}~(\ref{eq: charge conjugation matrix}). The charge conjugated correlation function is simplified with basic Dirac and Grassmann algebra to
\begin{align}
  \mathcal{C}_{5,5}^C(t) =&
  \big\langle \sum\limits_{n \in \Lambda_{t}^0} -\mathrm{tr} {\left( (\psi_0^T \gamma^{5\,T} \bar\psi_0^T)  (\psi_n^T\gamma^{5\,T} \psi_n^T) \right)} \big\rangle_{1}
  = \big\langle  \sum\limits_{n \in \Lambda_{t}^0} -\mathrm{tr} {\left( (\bar\psi_0 \gamma^5 \psi_0)^T  (\bar\psi_n \gamma^5 \psi_n)^T \right)} \big\rangle_{1}.
\end{align}\normalsize
When the transposition of both bilinears in the trace is removed, manifest invariance of the correlation function under charge conjugation becomes evident as
\begin{align}
  \mathcal{C}_{5,5}^C(t) =&
   \big\langle \sum\limits_{n \in \Lambda_{t}^0} -\mathrm{tr} {\left( (\bar\psi_0 \gamma^5 \psi_0)  (\bar\psi_n \gamma^5 \psi_n) \right)} \big\rangle_{1}
  = \mathcal{C}_{5,5}(t).
\end{align}\normalsize
The relation to reflection symmetries becomes apparent, once the spinor fields are Wick contracted to propagators. The charge conjugated connected piece of \mbox{eq.}~(\ref{eq: connected pseudoscalar correlation function}) reads
\begin{align}
  \mathcal{C}_{5,5}^{C\,}(t) =& \big\langle \sum\limits_{n \in \Lambda_{t}^0}
  \mathrm{tr} {\left( S_{n,0}^C \gamma^5 S_{0,n}^C \gamma^5 \right)} \big\rangle_{1}
  \label{eq: charge conjugated connected pseudoscalar correlation function}.
\end{align}\normalsize
It is known from section~\ref{sec: KW fermions} that the Karsten-Wilczek action has a $ C\Theta_{\underline{\alpha}} $ symmetry, whose operator is the combination of charge conjugation and $ n_{\underline{\alpha}} $~reflection operators. The consequence of $ C\Theta_{\underline{\alpha}} $ symmetry is trivial for the perpendicular case~($ \underline{\alpha}\neq0 $), since the spatial directions are summed in \mbox{eq.}~(\ref{eq: charge conjugated connected pseudoscalar correlation function}). However, in the parallel case~($ \underline{\alpha}=0 $), $ C\Theta_{\underline{\alpha}} $ symmetry implies that
\begin{align}
  \mathcal{C}_{5,5}^{C\,}(t) =&\ \big\langle \sum\limits_{n \in \Lambda_{t}^0}
  \mathrm{tr} {\left( S_{n,0}^C \gamma^5 S_{0,n}^C \gamma^5 \right)} \big\rangle_{1}
= \big\langle \sum\limits_{n \in \Lambda_{t}^0}
  \mathrm{tr} {\left( (\widehat{C} S_{n,0} \widehat{C}^\dagger) \gamma^5 (\widehat{C} S_{0,n} \widehat{C}^\dagger) \gamma^5 \right)} \big\rangle_{1}
  \nonumber \\
=& \big\langle \sum\limits_{n \in \Lambda_{-t}^0}
  \mathrm{tr} {\left( (\widehat{\Theta} S_{n,0} \widehat{\Theta}^\dagger) \gamma^5 (\widehat{\Theta} S_{0,n} \widehat{\Theta}^\dagger) \gamma^5 \right)} \big\rangle_{1}
  = \big\langle \sum\limits_{n \in \Lambda_{-t}^0}
  \mathrm{tr} {\left( S_{n,0}^\Theta \gamma^5 S_{0,n}^\Theta \gamma^5 \right)} \big\rangle_{1}
  \label{eq: relation to time reflected connected pseudoscalar correlation function}.
\end{align}\normalsize
This expression is equal to the time-reflected correlation function defined as
\begin{align}
 \widehat{\Theta} \mathcal{C}_{5,5}(t) \equiv&\ \mathcal{C}_{5,5}^{\Theta\,}(-t)  
  = \big\langle \sum\limits_{n \in \Lambda_{-t}^0}
  \mathrm{tr} {\left( S_{n,0}^\Theta \gamma^5 S_{0,n}^\Theta \gamma^5 \right)} \big\rangle_{1}
  \label{eq: time reflected connected pseudoscalar correlation function}.
\end{align}\normalsize
Because charge-conjugated and time-reflected correlation functions are equal and since the charge conjugated correlation function is equal to the original correlation function, it follows that the pseudoscalar correlation function is manifestly invariant under time reflection. Restriction to the pseudoscalar correlation function is not necessary, since the proof can be easily extended to a larger group of observables (such as generic mesonic correlation functions with equal Dirac matrices at source and sink). However, a crucial ingredient of the proof is invariance of the vacuum under charge conjugation. In the quenched approximation, sea quarks are neglected and gauge configurations are pure Yang-Mills fields. Therefore, gauge configurations in the quenched approximation correspond to a vacuum that is invariant under charge conjugation. Since all numerical simulations in this thesis are restricted to the quenched approximation, the unquenched case with sea quarks is not considered here. Whether or not the unquenched vacuum is charge conjugation invariant or not is an open and non-trivial problem.

\subsection{$ C\Theta $~symmetry in the interacting theory}\label{sec: CTheta symmetry in the interacting theory}

Mesonic correlation functions consist of fermion loops with matrix insertions according to their interpolating operators. These operators may lead to non-trivial transformation behaviour under charge conjugation that is unrelated to the fermion action. After mesonic correlation functions with charge-conjugated gauge fields are discussed for invariant fermion actions (like na\"{i}ve or Wilson fermions), the pattern is compared to Karsten-Wilczek fermions. 
In order to understand the different behaviour of Karsten-Wilczek fermions, charge conjugation of general next-neighbour lattice Dirac operators is considered. Charge conjugation of Minkowski space-time gauge fields $ A^\mu $ is translated to gauge links $ U^\mu $ on a Euclidean space-time lattice, which transform as
\begin{align}
  U^\mu_n \stackrel{C}{\to} U^{C\,\mu}_n = (U^\mu_n)^* = (U^{\mu\dagger}_n)^T.
  \label{eq: cc of link U}
\end{align}\normalsize
Thus, a general interacting Dirac operator
\begin{align}
  D_{n,m}[U] =& \sum\limits_\mu 
  \mathcal{M}_{+}^\mu U^\mu_n \delta_{n+\hat{e}_\mu,m}
  -\mathcal{M}_{-}^\mu U^{\mu\dagger}_{n-\hat{e}_\mu,m} \delta_{n-\hat{e}_\mu}
  +\mathcal{M}_{m} \delta_{n,m}
  \label{eq: general Dirac operator}  
\end{align}\normalsize
with \textit{hopping term matrices} $ \mathcal{M}_{\pm}^\mu  $ and a \textit{mass matrix} $ \mathcal{M}_{m}  $ transforms as
\begin{align}
  D_{n,m}^C[U^C] 
   =&\ \sum\limits_\mu 
  C \mathcal{M}_{+}^\mu C  U^{\mu*}_n \delta_{n+\hat{e}_\mu,m}
  - C \mathcal{M}_{-}^\mu C  U^{\mu\dagger*}_{n-\hat{e}_\mu} \delta_{n-\hat{e}_\mu,m}
  + C \mathcal{M}_{m} C  \delta_{n,m}.
  \label{eq: charge conjugated general Dirac operator}
\end{align}\normalsize
Transposition in Dirac, colour and site space are labeled $ T_d $, $ T_c $ and $ T_s $. The definitions
\begin{align}
  (\mathcal{\widetilde{M}}_{\pm}^\mu)^{T_d} \equiv C \mathcal{M}_{\pm}^\mu C ,\quad &
  (\mathcal{\widetilde{M}}_{m})^{T_d}     \equiv C \mathcal{M}_{m} C.
  \label{eq: charge conjugated Dirac structure}
\end{align}\normalsize
allow for rewriting the charge conjugated Dirac operator with an overall transposition,
\begin{align}
  D_{n,m}^C[U^C] 
   =&\ \sum\limits_\mu 
  (\mathcal{\widetilde{M}}_{+}^\mu)^{T_d}  (U^{\mu\dagger}_{m-\hat{e}_\mu})^{T_c} \delta_{m-\hat{e}_\mu,n}^{T_s}
  - (\mathcal{\widetilde{M}}_{-}^\mu)^{T_d}  (U^{\mu}_{m})^{T_c} \delta_{m+\hat{e}_\mu,n}^{T_s}
  + (\mathcal{\widetilde{M}}_{m})^{T_d}  \delta_{m,n}^{T_s}
  \nonumber \\
  =&\ \sum\limits_\mu  \left(
  - \mathcal{\widetilde{M}}_{-}^\mu  U^{\mu}_{m}\delta_{m+\hat{e}_\mu,n}
  + \mathcal{\widetilde{M}}_{+}^\mu C  U^{\mu\dagger}_{m-\hat{e}_\mu} \delta_{m-\hat{e}_\mu,n}
  + \mathcal{\widetilde{M}}_{m}  \delta_{m,n} \right)^T
  = (\widetilde{D}_{m,n}[U])^T,
  \label{eq: individually transposed form of charge conjugated general Dirac operator}  
\end{align}\normalsize
where $ \widetilde{D} $ is to be understood in the sense of \mbox{eq.}~(\ref{eq: charge conjugated Dirac structure}). Except for the overall transposition, the original Dirac structure is recovered if
\begin{align}
   \mathcal{\widetilde{M}}_{\pm}^\mu = -\mathcal{M}_{\mp}^\mu,\quad &
   \mathcal{\widetilde{M}}_{m}     = \mathcal{M}_{m}.
  \label{eq: condition for charge conjugation invariance}
\end{align}\normalsize
In the case of the Karsten-Wilczek operator, the matrices read
\begin{align}
  {\mathcal{M}}_{\pm}^\mu = +(1+d\delta^{\underline{\alpha}\mu})\gamma^\mu \mp i\zeta\varrho^{\underline{\alpha}\mu}\gamma^{\underline{\alpha}},\quad&
  {\mathcal{M}}_{m}     = m_0 \mathbf{1} +i\frac{3\zeta+c}{a}\gamma^{\underline{\alpha}}  \label{eq: KW hopping term matrices M}, \\
  \widetilde{\mathcal{M}}_{\pm}^\mu = -(1+d\delta^{\underline{\alpha}\mu})\gamma^\mu \pm i\zeta\varrho^{\underline{\alpha}\mu}\gamma^{\underline{\alpha}},\quad&
  \widetilde{\mathcal{M}}_{m}     = m_0 \mathbf{1} -i\frac{3\zeta+c}{a}\gamma^{\underline{\alpha}}
    \label{eq: KW hopping term matrices widetilde M }.
\end{align}\normalsize
Thus, charge conjugation flips the sign of the Wilzcek parameter ($ c(-\zeta)=-c(+\zeta) $)
\begin{align}
   \mathcal{\widetilde{M}}_{\pm}^\mu(+\zeta) = -\mathcal{M}_{\mp}^\mu(-\zeta),\quad 
   \mathcal{\widetilde{M}}_{m}(+\zeta)     = \mathcal{M}_{m}(-\zeta),\quad &\widetilde{D^f}[U](+\zeta)     = D^f[U](-\zeta).
  \label{eq: relation for charge conjugation of KW fermions}
\end{align}\normalsize

\subsubsection{Charge conjugation invariant fermions}\label{sec: Charge conjugation invariant fermions}

\noindent
If \mbox{eq.}~(\ref{eq: condition for charge conjugation invariance}) is satisfied, the action is manifestly invariant under charge conjugation. Though na\"{i}ve and Wilson fermions satisfy \mbox{eq.}~(\ref{eq: condition for charge conjugation invariance}), Karsten-Wilczek fermions satisfy \mbox{eq.}~(\ref{eq: relation for charge conjugation of KW fermions}) instead. Because the Dirac operator of an invariant action transform as
\begin{align}
  D_{n,m}[U] \stackrel{C}{\to} C D_{n,m}[U^C]C =  D^C_{n,m}[U^C] = (D_{m,n}[U])^T,
  \label{eq: charge conjugation of C-invariant Dirac operator}
\end{align}\normalsize
its inverse, the fermion propagator, must transform as
\begin{align}
  S_{n,m}^C[U^C] = (S_{m,n}[U])^T, \quad & S_{n,m}[U^C] = C (S_{m,n}[U])^T C.
  \label{eq: charge conjugation of C-invariant fermion propagator}
\end{align}\normalsize
If a correlation function is computed from propagators that use charge conjugated gauge fields, it is related to correlation functions computed from propagators using the original gauge fields according to \mbox{eq.}~(\ref{eq: charge conjugation of C-invariant fermion propagator}) and reads
\begin{align}
  \mathcal{C}_{\mathcal{M},\mathcal{N}}(t)[U^C] =&\ 
  \big\langle \sum\limits_{n \in \Lambda_{t}^0}\mathrm{tr}{\left( 
  S_{n,0}[U^C] \mathcal{M} S_{0,n}[U^C] \mathcal{N}
  \right)} \big\rangle_{U^C}
  \nonumber \\
  =&\ 
  \big\langle\sum\limits_{n \in \Lambda_{t}^0}\mathrm{tr}{\left( 
  (S_{0,n}[U])^T C \mathcal{M} C (S_{n,0}[U])^T C \mathcal{N} C
  \right)} \big\rangle_{U}  \nonumber \\
  =&\ 
  \big\langle \sum\limits_{n \in \Lambda_{t}^0}\mathrm{tr}{\left( 
  S_{n,0}[U] \mathcal{\widetilde{M}} S_{n,0}[U] \mathcal{\widetilde{N}}
  \right)^T} \big\rangle_{U} = \mathcal{C}_{\mathcal{\widetilde{M}},\mathcal{\widetilde{N}}}(t)[U].
  \label{eq: correlation function with invariant action and charge conjugated gauge links}
\end{align}\normalsize
\begin{table}[hbt]
 \center
 \begin{tabular}{|c|ccccc|}
  \hline
  $ \mathcal{M},\mathcal{N} $ & $ \gamma^\nu $ & $ \Sigma^{\mu\nu} $ & $ \mathbf{1} $ & $ \gamma^5 $ & $ \gamma^5\gamma^\nu $ \\
  \hline
  $ \gamma^\theta $         & R & R & I & I & I \\
  $ \Sigma^{\chi\theta} $   & R & R & I & I & I \\
  $ \mathbf{1} $            & I & I & R & R & R \\
  $ \gamma^5 $              & I & I & R & R & R \\
  $ \gamma^5\gamma^\theta $ & I & I & R & R & R \\
  \hline
 \end{tabular}
 \caption{Mesonic correlation functions for Wilson fermions are either purely real or purely imaginary. The format is `R' for real and `I' for imaginary correlation functions. Real correlators preserve their sign and imaginary correlators change their sign under charge conjugation. Correlators with na\"{i}ve instead of Wilson fermions are always real and they are non-zero only if source and sink interpolators are symmetric ($ \mathcal{M}=\mathcal{N} $).
 }
 \label{tab: naive and Wilson fermion charge conjugation}
\end{table}\normalsize
\noindent The matrices~$ \mathcal{\widetilde{M}} $ and $ \mathcal{\widetilde{N}} $ must be understood in the sense of \mbox{eq.}~(\ref{eq: charge conjugated Dirac structure}). If both matrices transform with the same sign under left- and right-multiplication with the charge conjugation matrix~$ C $, the correlation function is independent of whether gauge fields are charge conjugated. Since all coefficients in Wilson and na\"{i}ve Dirac operators are real, charge conjugation of gauge fields can only change the sign of the imaginary part of the correlation function. Hence, correlation functions must be purely real if both matrices transform with the same sign and purely imaginary if both matrices transform with different signs. This behaviour is observed within errors in numerical simulations with Wilson and na\"{i}ve fermions and summarised in table~\ref{tab: naive and Wilson fermion charge conjugation}, where rows and columns represent matrices~$ \mathcal{M} $ and~$ \mathcal{N} $ of source and sink interpolating operators.

\subsubsection{Charge conjugation of Karsten-Wilczek fermions}\label{sec:Charge conjugation of Karsten-Wilczek fermions}

\begin{table}[hbt]
 \center
 \begin{tabular}{|c|cccccccc|}
  \hline
  $ \mathcal{M},\mathcal{N} $ & $ \gamma^{\underline{\alpha}} $ & $ \Sigma^{\underline{\alpha}\nu} $ & $ \gamma^\nu $ & $ \Sigma^{\mu\nu} $ & $ \mathbf{1} $ & $ \gamma^5\gamma^{\underline{\alpha}} $ & $ \gamma^5 $ & $ \gamma^5\gamma^\mu $ \\
  \hline
  $ \gamma^{\underline{\alpha}} $         & +R & +R & -R & -R & +I & +I & -I & -I \\
  $ \Sigma^{\underline{\alpha}\theta} $   & +R & +R & -R & -R & +I & +I & -I & -I  \\
  $ \gamma^{\theta} $                     & -R & -R & +R & +R & -I & -I & +I & +I  \\
  $ \Sigma^{\chi\theta} $                 & -R & -R & +R & +R & -I & -I & +I & +I  \\
  $ \mathbf{1} $                          & +I & +I & -I & -I & +R & +R & -R & -R  \\
  $ \gamma^5\gamma^{\underline{\alpha}} $ & +I & +I & -I & -I & +R & +R & -R & -R  \\
  $ \gamma^5 $                            & -I & -I & +I & +I & -R & -R & +R & +R  \\
  $ \gamma^5\gamma^{\theta} $             & -I & -I & +I & +I & -R & -R & +R & +R  \\
  \hline
 \end{tabular}
 \caption{Local mesonic correlation functions with Karsten-Wilczek fermions are either real or imaginary. Some change sign for charge conjugated gauge fields. The format is `+` for positive sign and `-' for negative sign under charge conjugation and `R' for real and `I' for imaginary correlation functions. $ \{\mu, \nu, \chi, \theta\} \neq\underline{\alpha} $.
 }
 \label{tab: KW fermion charge conjugation}
\end{table}\normalsize

\noindent For Karsten-Wilczek fermions, \mbox{eq.}~(\ref{eq: charge conjugation of C-invariant fermion propagator}) must be replaced due to \mbox{eq.}~(\ref{eq: relation for charge conjugation of KW fermions}) by
\begin{align}
  S_{n,m}^C[U^C] = (\widetilde{S}_{m,n}[U])^T, \quad & S_{n,m}[U^C] = C (\widetilde{S}_{m,n}[U])^T C.
  \label{eq: charge conjugation of KW fermion propagator}
\end{align}\normalsize
Hence, instead of \mbox{eq.}~(\ref{eq: correlation function with invariant action and charge conjugated gauge links}), correlation functions are described by
\begin{align}
  \mathcal{C}_{\mathcal{M},\mathcal{N}}(t)[U^C] =&\ 
  \big\langle \sum\limits_{n \in \Lambda_{t}^0}\mathrm{tr}{\left( 
  S_{n,0}[U^C] \mathcal{M} S_{0,n}[U^C] \mathcal{N}
  \right)} \big\rangle_{U^C}
  \nonumber \\
  =&\ 
  \big\langle\sum\limits_{n \in \Lambda_{t}^0}\mathrm{tr}{\left( 
  (\widetilde{S}_{0,n}[U])^T C \mathcal{M} C (\widetilde{S}_{n,0}[U])^T C \mathcal{N} C
  \right)} \big\rangle_{U}  \nonumber \\
  =&\ 
  \big\langle \sum\limits_{n \in \Lambda_{t}^0}\mathrm{tr}{\left( 
  \widetilde{S}_{n,0}[U] \mathcal{\widetilde{M}} \widetilde{S}_{n,0}[U] \mathcal{\widetilde{N}}
  \right)} \big\rangle_{U} \neq \mathcal{C}_{\mathcal{\widetilde{M}},\mathcal{\widetilde{N}}}(t)[U].
  \label{eq: correlation function with KW action and charge conjugated gauge links}
\end{align}\normalsize
The Wilczek parameter $ \zeta $ in the propagators $ S[U^C] $ is replaced by $ -\zeta $ in the propagators $ \widetilde{S}[U] $. Hence, \mbox{eq.}~(\ref{eq: correlation function with KW action and charge conjugated gauge links}) is rephrased with regard to the Wilczek parameter $ \zeta $ as
\begin{equation}
  \mathcal{C}_{\mathcal{M},\mathcal{N}}(t;+\zeta)[U^C] = \mathcal{C}_{\mathcal{\widetilde{M}},\mathcal{\widetilde{N}}}(t;-\zeta)[U]
  \label{eq: zeta dependence in correlation function with KW action and charge conjugated gauge links}
\end{equation}\normalsize
Given the Dirac matrices $ \mathcal{M} $ and $ \mathcal{N} $ of the interpolating operators, \mbox{eq.}~(\ref{eq: zeta dependence in correlation function with KW action and charge conjugated gauge links}) is an unambiguous statement about the $ \zeta $ depence of the correlation function if and only if the vacuum is invariant under charge conjugation such as the pure Yang-Mills vacuum of the quenched approximation. 
It is not clear at present whether or not this invariance is realised for full QCD with dynamical Karsten-Wilczek fermions. However, this issue is of no concern for predictions about numerical simulations in the quenched approximation that are considered in this thesis. 
The change of the Wilczek parameter can be compensated by left- and right-multiplication of the propagators by matrices $ \mathcal{R}_n = (-1)^{n_{\underline{\alpha}}} \mathcal{Q} $ and  $ \mathcal{R}_0 = \mathcal{Q} $ (\mbox{cf.}~paragraph below \mbox{eq.}~(\ref{eq: KW relation between two species}) or sections~\ref{sec: Interim findings (I)} and~\ref{sec: Decomposition in the free theory}). As $ 1=(\mathcal{R}_n)^2 $ is plugged into the correlation functions on both sides of each propagator, $ \mathcal{M} $ and~$ \mathcal{N} $ are also left-and right-multiplied by $ \mathcal{Q} $. Denoting total powers of $ \gamma^{\underline{\alpha}} $ and $ \gamma^5 $ in both interpolating operators as $ N_{\underline{\alpha}} $ and $ N_{5} $, they acquire factors~$ (-1)^{ N_{\underline{\alpha}}+ N_{5}} $ that must be combined with factors $ (\pm1) $ from left- and right-mutliplication with the charge conjugation matrix. 
Though mesonic correlation functions for Karsten-Wilczek fermions are purely real or purely imaginary according to the same pattern as na\"{i}ve and Wilson fermions. However, the pattern of signs in table~\ref{tab: naive and Wilson fermion charge conjugation} that real functions are positive and imaginary functions are negative under charge conjugation is broken up with extra factors~$ (-1)^{ N_{\underline{\alpha}}+ N_{5}} $. This behaviour is observed in numerical simulations and summarised in table~\ref{tab: KW fermion charge conjugation}, where rows and columns represent matrices~$ \mathcal{M} $ and~$ \mathcal{N} $ of source and sink interpolating operators. Because correlation functions for Karsten-Wilczek fermions satisfy \mbox{eq.}~(\ref{eq: zeta dependence in correlation function with KW action and charge conjugated gauge links}), those which are real and retain their sign under charge conjugation (indicated by +R in table~\ref{tab: KW fermion charge conjugation}) must be even functions of the Wilczek parameter $ \zeta $. Being even functions of $ \zeta $ implies that they are invariant under $ n_{\underline{\alpha}} $~reflection. This is a manifestation of the $ C\Theta_{\underline{\alpha}} $~symmetry in correlation functions.

\subsubsection{$ C\Theta $ symmetry and $ \mathcal{O}(a) $ corrections}\label{sec: CTheta symmetry and O(a) corrections}

\noindent Since invariance of correlation functions under $ n_{\underline{\alpha}} $~reflection is linked to invariance under charge conjugation by $ C\Theta_{\underline{\alpha}} $~symmetry, a condition for $ n_{\underline{\alpha}} $~reflection symmetry of correlation functions can be formulated in an operator language for more general states. The Euclidean correlation function of the state $ \mathcal{O}^\dagger|\Omega\rangle $ reads
\begin{align}
  \mathcal{C}_{\mathcal{O},\mathcal{O}^\dagger}(t) =&\
  \langle \Omega | \mathcal{\widehat{O}} e^{-\widehat{H}t} \mathcal{\widehat{O}}^\dagger | \Omega \rangle,
  \label{eq: correlation function in operator language}
\end{align}\normalsize
where $ |\Omega \rangle $ is the vacuum. Since numerical simulations of this thesis use the quenched approximation, the vacuum of full QCD is not considered here. The pure Yang-Mills vacuum of the quenched approximation is strictly invariant under both $ \widehat{C} $ and $ \widehat{\Theta}_{\underline{\alpha}} $ of \mbox{eqs.}~(\ref{eq: charge conjugation}) and~(\ref{eq: time reflection for Euclidean space-time}). The Hamiltonian $ \widehat{H}$ 
has the same $ C\Theta_{\underline{\alpha}} $~symmetry as the action,
\begin{align}
  \widehat{H} = (\widehat{\Theta_{\underline{\alpha}} C}) \widehat{H} (\widehat{\Theta_{\underline{\alpha}} C})^\dagger,
  \label{eq: Z_2-symmetry of the Hamiltonian}
\end{align}\normalsize
where $ \widehat{C} $ and $ \widehat{\Theta}_{\underline{\alpha}} $ are the charge conjugation and $ n_{\underline{\alpha}} $~reflection operators of \mbox{eqs.}~(\ref{eq: charge conjugation}) and~(\ref{eq: time reflection for Euclidean space-time}). 
Then the correlation function is equal to
\begin{align}
  \mathcal{C}_{\mathcal{O},\mathcal{O}^\dagger}(t) =&\
  \langle \Omega | \mathcal{\widehat{O}} (\widehat{\Theta_{\underline{\alpha}} C})^\dagger e^{-\widehat{H}(\widehat{\Theta}_{\underline{\alpha}}t)} (\widehat{\Theta_{\underline{\alpha}} C}) \mathcal{\widehat{O}}^\dagger | \Omega \rangle 
  = \langle \Omega | \mathcal{\widehat{O}}  \widehat{C}^\dagger \widehat{\Theta}_{\underline{\alpha}}^\dagger e^{-\widehat{H}(\widehat{\Theta}_{\underline{\alpha}}t)} \widehat{\Theta}_{\underline{\alpha}} \widehat{C} \mathcal{\widehat{O}}^\dagger | \Omega \rangle.
  \label{eq: correlation function with Z_2 operator}
\end{align}\normalsize
If the operator $ \mathcal{O}^\dagger $ generates an eigenstate of charge conjugation, $ \mathcal{O}^\dagger $ and $ \mathcal{O} $ satisfy
\begin{align}
 \widehat{\mathcal{O}}^\dagger = \pm \widehat{C} \widehat{\mathcal{O}}^\dagger \widehat{C}^\dagger,\  \widehat{\mathcal{O}} = \pm \widehat{C} \widehat{\mathcal{O}} \widehat{C}^\dagger,
  \label{eq: charge conjugation invariance of the operator O}
\end{align}\normalsize
and the charge conjugation operators may be moved past the operators $ \widehat{\mathcal{O}} $ and $ \widehat{\mathcal{O}}^\dagger $. 
\begin{align}
  \mathcal{C}_{\mathcal{O},\mathcal{O}^\dagger}(t) =&\
  \langle \Omega |\widehat{C}^\dagger \mathcal{\widehat{O}} \widehat{\Theta}_{\underline{\alpha}}^\dagger e^{-\widehat{H}(\widehat{\Theta}_{\underline{\alpha}}t)} \widehat{\Theta}_{\underline{\alpha}} \mathcal{\widehat{O}}^\dagger \widehat{C}| \Omega \rangle \nonumber\\
  =&\  \langle \Omega |(\widehat{C}^\dagger \widehat{\Theta}_{\underline{\alpha}}^\dagger) (\widehat{\Theta}_{\underline{\alpha}} \mathcal{\widehat{O}} \widehat{\Theta}_{\underline{\alpha}}^\dagger) e^{-\widehat{H}(\widehat{\Theta}_{\underline{\alpha}}t)} (\widehat{\Theta}_{\underline{\alpha}} \mathcal{\widehat{O}}^\dagger \widehat{\Theta}_{\underline{\alpha}}^\dagger) (\widehat{\Theta}_{\underline{\alpha}} \widehat{C}) | \Omega \rangle.
  \label{eq: correlation function with Theta_alpha operator}
\end{align}\normalsize
Since the vacuum satisfies $ (\widehat{\Theta}_{\underline{\alpha}}\widehat{C})|\Omega\rangle =|\Omega\rangle $, the correlation function is expressed through $ n_{\underline{\alpha}} $-reflected operators $ (\widehat{\Theta}_{\underline{\alpha}}\mathcal{O}\widehat{\Theta}_{\underline{\alpha}}^\dagger) $ as an overall $ n_{\underline{\alpha}} $-reflected correlation function,
\begin{align}
  \mathcal{C}_{\mathcal{O},\mathcal{O}^\dagger}(t) =&\
  = \mathcal{C}_{\widehat{\Theta}_{\underline{\alpha}}\mathcal{O}\widehat{\Theta}_{\underline{\alpha}}^\dagger,(\widehat{\Theta}_{\underline{\alpha}}\mathcal{O}\widehat{\Theta}_{\underline{\alpha}}^\dagger)^\dagger}(\widehat{\Theta}_{\underline{\alpha}}t)
  =\widehat{\Theta}_{\underline{\alpha}}\ \mathcal{C}_{\mathcal{O},\mathcal{O}^\dagger}(t).
  \label{eq: Theta_alpha transformed correlation function}
\end{align}\normalsize
In the parallel case of Karsten-Wilczek fermions ($ \underline{\alpha}=0 $) , \mbox{eq.}~(\ref{eq: Theta_alpha transformed correlation function}) implies that the correlation function in any channel with definite charge conjugation quantum number is invariant under time reflection. 
\noindent
The argument can be generalised to Bori\c{c}i-Creutz fermions at this stage. $ n_{\underline{\alpha}} $~reflection is along the $ \hat{f}_0 $-direction (diagonal of the hypercube, \mbox{cf.~eq.}~(\ref{eq: BC f-basis})) for Bori\c{c}i-Creutz fermions. Therefore, $ n_{\underline{\alpha}} $~reflection reflects all Euclidean directions at once. If the operators $ \widehat{\mathcal{O}}^\dagger $ and $ \widehat{\mathcal{O}} $ generate and destroy a state with definite parity and charge conjugation quantum numbers, the correlation function is invariant under any spatial reflection and under charge conjugation. Invariance under time reflection follows from the general $ CP\Theta $~invariance and both invariance under charge conjugation ($ \widehat{C} $) and under reflection of all spatial axes ($ \widehat{P} $) of the correlation function. Hence, correlation functions of operators with definite charge conjugation and parity quantum numbers are invariant under time reflection for Bori\c{c}i-Creutz fermions. \newline

\noindent Another consequence of $ n_{\underline{\alpha}} $~reflection symmetry concerns $ \mathcal{O}(a) $ corrections to the continuum limit. In both actions, any $ \mathcal{O}(a) $-suppressed operators break charge conjugation and $ n_{\underline{\alpha}} $~reflection symmetry to $ C\Theta_{\underline{\alpha}} $ symmetry, whereas the leading order terms respect both symmetries. Hence, because $ \mathcal{O}(a) $ corrections to any observable are generated by the $ \mathcal{O}(a) $-suppressed operators (in diagrammatical terms only one $ \mathcal{O}(a) $-suppressed insertion per diagram), they must also share their broken symmetry. If an observable has the unbroken symmetries, there cannot be any $ \mathcal{O}(a) $ corrections. Therefore, leading corrections to correlation functions which respect both symmetries are of $ \mathcal{O}(a^2) $. 
Of course, this statement is true only if the vacuum itself has the right symmetry and does not have $ \mathcal{O}(a) $ corrections on its own. This is the case for the pure Yang-Mills vacuum of the quenched approximation. Whether or not this is the case for the full QCD vacuum is a non-trivial question that is not considered here. 
This is not the same as automatical $ \mathcal{O}(a) $ improvement for Ginsparg-Wilson fermions, because the absence of $ \mathcal{O}(a) $ corrections is exclusive to observables with additional symmetry.

\sectionc{Interim findings (II)}{sec: Interim findings (II)}

\noindent
The preceding section contains analytical studies of Karsten-Wilczek fermions in terms of a decomposition of spinor fields $ \psi $ into a pair fields $ \phi $ and $ \chi $ with different four-momentum support and of higher symmetries of mesonic correlation functions. \newline

\noindent Decomposition into a pair of fields $ \phi $ and $ \chi $ with different four-momentum support requires use of decomposition kernels which extend support over multiple lattice sites. Definition of fields with extended support is not feasible in a theory with an unmatched power divergent operator. Hence, a mismatch $ \delta c $ of the relevant operator's coefficient is absorbed into two phase factors that modify the boundary conditions of the fields $ \phi $ and $ \chi $ differently. This must not be mistaken for a removal of the counterterm or its coefficient. 
Since the decomposition relies on linearisation both in $ \delta c $ and in the lattice spacing $ a $ for the lattice product rule of \mbox{eq.}~(\ref{eq: lattice product rule}), decomposition may well break down if either of $ \delta c $ or $ a $ is too large. 
\noindent
Decomposition of interpolating operators yields bilinear forms of the fields $ \phi $ and $ \chi $ that form two strikingly different terms in mesonic correlation functions. Whereas the first term with the same fields (\mbox{e.g.}~$ \bar\phi $ and $ \phi $) has the same Dirac structure and $ J^{PC} $ as the correlation function for undoubled fermions, the second term with different fields (\mbox{e.g.}~$ \bar\phi $ and $ \chi $) has different Dirac structure and $ J^{PC} $ and also oscillates in the $ \hat{e}_{\underline{\alpha}} $~direction. While such an oscillating term is well-known for staggered fermions~\cite{Altmeyer:1992dd}, its frequency depends on the mismatch $ \delta c $ for Karsten-Wilczek fermions. It is certainly worthwhile to investigate whether this dependence on $ \delta c $ is a viable non-perturbative tuning criterion. Moreover, the second term with different fields and different $ J^{PC} $ may contribute additional low-lying states or even the ground state. \newline

\noindent
The decomposition is based on linearising in the mismatch $ \delta c $ and the lattice spacing $ a $ and may or may not be legitimate for numerical simulations. Moreover, as it makes use of unspecified decomposition kernels that are functionals of the local gauge fields, a rigorous perturbative calculation would require a perturbative expansion of the decomposition kernels as well. Therefore, identification of components in any pictorial representation lacks the field-theoretical rigour of Feynman diagrams. This problem is inherent in the treatment of any doubled lattice fermion in position-space. However, the decomposition suggests that general concepts concerning doubled fermions should be also applicable to Karsten-Wilczek fermions. The presence of oscillating contributions with different $ J^{PC} $ due to fermion modes in different parts of the Brillouin zone in correlation functions is known for staggered fermions~\cite{Karsten:1980wd,Golterman:1984cy,Altmeyer:1992dd}. It seems as if similar findings apply to Karsten-Wilczek fermions as well. Whether or not the frequency of the oscillations depends on the mismatch $ \delta c $ is put to a test in the numerical simulations that make up section~\ref{sec: Oscillating correlation functions}. \newline

\noindent Correlation functions can have a higher symmetry than the fermion action, if gauge configurations correspond to a vacuum $ | \Omega \rangle $ that is invariant under $ (\widehat{\Theta}_{\underline{\alpha}}\widehat{C}) $. This is strictly satisfied for the pure Yang-Mills vacuum of the quenched approximation. Whether or not the full QCD vacuum has this symmetry is not within the scope of this thesis. For a symmetric vacuum, correlation functions with the same tree-level charge conjugation patterns at source and sink are also invariant under $ n_{\underline{\alpha}} $ reflection as a consequence of $ CP\Theta $~symmetry. 
This higher symmetry suggests that correlation functions with the same matrices in source and sink interpolating operators in the quenched approximation are symmetric under time reflection and have equal energies for forwards and backwards propagating states. Because $ \mathcal{O}(a) $ terms in minimally doubled fermion actions break both symmetries explicitly, $ \mathcal{O}(a) $ corrections should be absent in observables that retain the higher symmetry. Even if this behaviour were to hold true for the full QCD vacuum, this is not to be mistaken with $ \mathcal{O}(a) $ improvement because it requires an extra symmetry of the observables. Nevertheless, symmetric correlation functions are tested for this higher symmetry in the numerical simulations that make up section~\ref{sec: Anisotropy of hadronic quantities}. 

\chapter{Numerical studies}\label{sec: Numerical studies}

\noindent
The central objective of lattice QCD is to study QCD in the non-perturbative regime, which can be achieved with numerical simulations. In these simulations, the QCD path integral (the Euclidean analogue of \mbox{eq.}~(\ref{eq: QCD path integral, Minkowski space-time}) is evaluated on a limited number of field configurations, which are generated using an importance sampling method. A Markov process with a probabilistic acceptance condition is used to generate configurations. In order to satisfy detailed balance, the rejection probability is determined by the increase of the classical action. Field configurations are expressed in terms of gauge field variables that implicitly depend on sea quarks. Fermion fields are integrated out and yield the quark determinant, which is an additional contribution to an effective gauge field action.
\noindent
Once an ensemble of gauge configurations has been provided, observables are evaluated on each configuration as functions of fermionic and gluonic fields. Observables with valence fermions are calculated from a set of source fields on which the Dirac operator is inverted. These propagator components correspond to Wick contractions of fermion fields at different space-time points and are combined to form correlation functions of hadronic interpolating operators with appropriate spin and parity. Since interpolating operators have non-zero overlap with all physical states with the same quantum numbers, excited states as well as the ground state contribute in each channel. 
\noindent
In order to cleanly extract the ground state, correlation functions are studied for long Euclidean time separation. The ground state is isolated, since spectral weights of all states exponentially decrease with their energies. A typical Euclidean correlation function for a lattice with infinite extent in the time direction reads
\begin{equation}
  \mathcal{C}(t) = \sum\limits_{k=0}^\infty A_k e^{-E_k  t}
  =
  A_0 e^{-E_0 t} \sum\limits_{k=0}^\infty \frac{A_k}{A_0} e^{-(E_k-E_0) t} \stackrel{t \to \infty}{\to}  A_0 e^{-E_0 t} + \ldots,
  \label{eq: typical correlation function}
\end{equation}\normalsize
where $ A_k $ are spectral weights (at $ t=0 $) and $ E_k $ are the absolute energies of states. Whereas spectral weights depend on the peculiar details of the interpolating operators, the energies depend only on their quantum numbers. The set of spectral weights $ A_k $ can be manipulated in order to achieve dominance of the ground state for smaller $ t $ by applying smearing procedures, which alter geometrical shapes at source and sink. This is of particular importance for the study of anisotropies in section~\ref{sec: Anisotropy of hadronic quantities}. The energies of hadronic states satisfy dispersion relations $ E_k(\mathbf{p}) = \sqrt{m_k^2 +\mathbf{p}^2} $, where $ \mathbf{p} $ is the hadron's spatial momentum. Spatial momentum can be injected, if the local interpolating operator $ O_{\mathcal{N}}(t) $ at the sink site $ n=(t/a,\mathbf{n}) $ is multiplied by a plane wave $ e^{ia\mathbf{n}\cdot\mathbf{p}} $ before summation (\mbox{cf.~eq.}~(\ref{eq: mesonic correlation function})\,). The contribution of the ground state at smaller time separation is greatly enhanced with appropriate smearing techniques (\mbox{cf.}~section~\ref{sec: Setup of simulations}). Due to the $ W_4 $ symmetry of the lattice, any direction may serve as time direction for calculating correlation functions. Furthermore, \mbox{(anti-)~periodicity} of the finite space-time lattice is responsible for a superposition of hadronic states moving forward or backward in the time direction (\mbox{cf.}~section~\ref{sec: Setup of simulations}). Therefore, a generic correlation function on a lattice with periodic boundary conditions in the time direction ($ T=aN_t $, where $ N_t $ is the number of time slices) reads
\begin{equation}
  \mathcal{C}_{PB}(t) = \sum\limits_{k=0}^\infty A_k (e^{-E_k  t} + e^{-E_k  (T-t)})
  \label{eq: periodic lattice correlation function}
\end{equation}\normalsize 
\noindent
Correlation functions or ratios thereof on the full statistical ensemble are analysed with fit routines, which extract properties of hadronic states. The Jackknife method is applied for determination of statistical errors (\mbox{cf.}~appendix~\ref{app: Statistical analysis}). In the \textit{quenched approximation}, light quark masses close to the physical point lead to \textit{exceptional configurations} for Wilson fermions. The inversion of the Wilson Dirac operator $ D[U] $ breaks down for these exceptional configurations, which correspond to very small eigenvalues of the Wilson Dirac operator. Thus, Wilson fermions in the quenched approximation cannot reach pion masses below $ 300\,\mathrm{MeV} $. Because minimally doubled fermions have chiral symmetry, they are supposed to be protected against exceptional configurations. It is expected that simulations with pion masses below $ 300\,\mathrm{MeV} $ are feasible with Karsten-Wilczek fermions. Nevertheless, most of the simulations within this thesis use heavier quark masses, because they are numerically cheaper. Hence, it is necessary to perform a chiral extrapolation of hadronic quantities towards either the chiral limit or the physical point using chiral perturbation theory. 
Lastly, since hadronic observables inherit discretisation errors from the action, a continuum extrapolation on lines of constant physics is necessary for extraction of the continuum limit. 
\noindent
Technical apects of the numerical simulations in the quenched approximation are covered in section~\ref{sec: Setup of simulations}. Three different aspects of Karsten-Wilczek fermions are studied using data from numerical simulations. A study of the anisotropy of hadronic correlation functions~\cite{Weber:2013tfa} in section~\ref{sec: Anisotropy of hadronic quantities} and a study of oscillating correlation functions~\cite{proceeding4} in section~\ref{sec: Oscillating correlation functions} yield two independent tuning conditions for the relevant fermionic counterterm of \mbox{eq.}~(\ref{eq: ferm dim 3 KW counterterm operator}). Determination of a robust non-perturbative tuning condition for the marginal fermionic counterterm of \mbox{eq.}~(\ref{eq: ferm dim 4 KW counterterm operator}) is unsuccessful. Later on, in section~\ref{sec: Chiral behaviour of the pseudoscalar ground state}, actual physical information on QCD concerning the chiral behaviour of pseudoscalar mesons is extracted with simulations that reach into the chiral regime that is inaccessible with Wilson fermions in the quenched approximation. Lastly, interim findings from numerical studies are summarised in section~\ref{sec: Interim findings (III)}.

\newpage
\sectionc{Setup of simulations}{sec: Setup of simulations}

\subsection{Karsten-Wilczek fermions in the quenched approximation}\label{sec: Karsten-Wilczek fermions in the quenched approximation}

\noindent
Numerical simulations of Karsten-Wilczek fermions as one example of minimally doubled fermion actions are performed for the first time. Gauge configurations are calculated in the quenched approximation, where the full QCD path integral
\begin{align}
  \langle \mathcal{O} \rangle=&\ \frac{1}{Z} \int \mathcal{D}\bar\psi \mathcal{D}\psi \mathcal{D}U\ \mathcal{O}[\psi,\bar\psi,U]\ e^{-(S^{f}[\psi,\bar\psi,U]+S^{g}[U])}
  \nonumber \\
  =&\ \frac{1}{Z} \int \mathcal{D}U\ \det{(D[U])}\ \mathcal{O}[\psi,\bar\psi,U]\ e^{-S^{g}[U]}
  \label{eq: full LQCD path integral}
\end{align}\normalsize
is approximated by setting the fermion determinant $\det{(D[U])} $ to $ 1 $ instead of including it into an effective gauge action~$ S^{\mathrm{eff}}[U]=S^g[U]-\log\det{(D[U])} $ for full QCD. Herein, observables $ \mathcal{O}[\psi,\bar\psi,U] $ are expressed in terms of fermion propagators and gauge fields. This amounts to neglecting vacuum fermion loops due to sea quarks and the quenched vacuum is the pure Yang-Mills vacuum. The absence of fermion loops in the quenched approximation implies that the gauge fields are not affected by the anisotropy of the fermion sector. Hence, simulations in the quenched approximation can forgo the gluonic counterterm of \mbox{eq.}~(\ref{eq: gluonic KW counterterm}), though both fermionic counterterms of \mbox{eqs.}~(\ref{eq: ferm dim 3 KW counterterm operator}) and~(\ref{eq: ferm dim 4 KW counterterm operator}) must be included. 
\noindent
Due to the anisotropy of the fermion action, the QCD transfer matrix is anisotropic as well. Furthermore, the QCD transfer matrix lacks symmetry under $ n_{\underline{\alpha}} $~reflection and charge conjugation, since both are broken explicitly by the Karsten-Wilczek term of \mbox{eq.}~(\ref{eq: KW operator}). It was seen in section~\ref{sec: Fermionic self-energy} that requiring isotropy of the one-loop quark propagator is enough to fix the fermionic counterterms' coefficients perturbatively. It is worthwhile to investigate whether the requirement of isotropy for hadronic observables similarly provides a criterion for fixing the counterterms' coefficients. 
Even though `Lorentz scalars' like hadron masses or pseudoscalar densities might be na\"{i}vely anticipated as isotropic like in perturbation theory (\mbox{cf.}~section~\ref{sec: Local bilinears and symmetry currents}), this cannot be true in the non-perturbative context of a numerical simulation. Because they are calculated from directed correlation functions that require a choice of the Euclidean time direction (or \textit{direction of correlation}), even the most simple hadronic quantity, which is the pseudoscalar ground state mass, is necessarily sensitive to the lack of isotropy in the fermion action. The anisotropy is studied by calculating correlation functions with different orientations of the direction of correlation with respect to the alignment of the Karsten-Wilczek term in sections~\ref{sec: Anisotropy of hadronic quantities} and~\ref{sec: Oscillating correlation functions}. 
\noindent
Section~\ref{sec: CTheta symmetry in the interacting theory} suggests that correlation functions of states with definite charge conjugation quantum number on a pure Yang-Mills vacuum satisfy both $ n_{\underline{\alpha}} $~reflection and charge conjugation symmetries. These considerations are put to a numerical test in section~\ref{sec: Determination of the pseudoscalar mass} and methods for dealing with a potential lack of time-reflection symmetry are developed. 
\noindent

\subsection{Machine and code}\label{sec: Machine and code}

Numerical simulations are performed on the Wilson cluster of the Institut f\"{u}r Kernphysik at Johannes Gutenberg-Universit\"{a}t Mainz. The simulation code is based on the Kpipi Code~\cite{Giusti:2002sm,Giusti:2004yp,Giusti:2006mh}, which was originally designed for simulations of overlap fermions in the quenched approximation. The code is written in C, but makes frequent use of manual vectorisation using explicit inline assembly for SSE2 acceleration of particularly costly numerical operations. It was gradually adapted to numerical studies of minimally doubled fermion actions of Karsten-Wilczek and Bori\c{c}i-Creutz types. This section covers different stages of numerical simulations and summarises modifications to the code.

\subsection{Lattice geometry}\label{sec: Lattice geometry}

The preexisting Kpipi code sets the basic geometry constraints for numerical studies. The Euclidean direction $ N_0 $ is fixed as a free parameter $ T $ and the three other Euclidean directions $ N_1 $, $  N_2 $ and $ N_3 $ are fixed as another free parameter $ L $ at compile time. Both $ T $ and $ L $ must be integer multiples of $ 2 $ in the single processor version and of $ 8 $ in the parallelised version of the code. The parallelised version (\mbox{cf.}~section~\ref{sec: Parallelisation}) must have at least a $ 2\times2 $~grid of local sublattices in $ \hat{e}_0 $~and $ \hat{e}_1 $~directions, where local sublattices must have lengths of at least $ 4 $ in both $ \hat{e}_0 $~and $ \hat{e}_1 $~directions. Moreover, the sublattice length in the $ \hat{e}_1 $~direction must be an integer multiple of $ 2 $, $ 3 $ or $ 5 $.
\noindent
Lattice sites are grouped together in the memory in an alignment of blocks, which improved performance when the code was initially written (\mbox{ca.}~ 2004). Some parts of the code rely on the specific form of the geometry arrays. Since the effort of a geometry change for this preexisting code exceeds its benefits, the geometry was left unchanged. 
\noindent
Owing to the blocking structure, memory is not aligned in a pattern which allows for coherent even-odd ordering of sites, which is required for even-odd preconditioning of Dirac operators (\mbox{cf.}~section~\ref{sec: Even-odd preconditioning}). The problem was solved by introducing two additional geometry arrays, which provide a mapping between even-odd ordered and preexisting site-indices. Memory overhead due to two extra integer arrays is reasonably small and the loss of performance due to one additional intermediary hash table is acceptable.

\subsection{Gauge configurations and scale setting}\label{sec: Gauge configurations and scale setting}

\begin{table}[hbt]
\center
 \begin{tabular}{|c|c|c|c|c|}
  \hline
  $ \beta $ & $ U_{0}^4 $ & $ a [\mathrm{fm}]$ & $ r_0 $ & $ n_{JS} $   \\
  \hline
  $ 5.8 $ & $ 0.567 $ & $ 0.136 $ & $ 3.668 $ & $ 30 $ \\
  \hline
  $ 6.0 $ & $ 0.594 $ & $ 0.093 $ & $ 5.368 $ & $ 40 $  \\
  \hline
  $ 6.2 $ & $ 0.614 $ & $ 0.068 $ & $ 7.360 $ & $ 140 $  \\
  \hline
 \end{tabular}
 \caption{The scale $ a $ is set using \mbox{eq.}~(\ref{eq: scale setting}). Dimensionful quantities in lattice units are converted to physical quantities in units of $ 2\,\mathrm{fm}^{-1} $ by multiplication with $ r_0 $. $ n_{JS} $ is the iteration count of the Jacobi smearing algorithm (\mbox{cf.} section \ref{sec: Smearing}).}
 \label{tab: gauge coupling, scale setting and smearing}
\end{table}

\noindent
The code providing gauge configurations was taken over without modification. Gauge configurations are calculated with a Markov chain update procedure that satisfies a detailed balance condition. The update algorithm is a standard heatbath-overrelaxation algorithm. It consists of consecutive applications of the Cabibbo-Marinari heatbath algorithm~\cite{Creutz:1980zw,Cabibbo:1982zn} and microcanonical overrelaxation steps~\cite{Adler:1981sn, Whitmer:1984he}. The individual update procedures -- heatbath and overrelaxation -- are described in detail in the fourth chapter of~\cite{Gattringer:2010zz}. Gauge configurations throughout this thesis are always produced from a cold start with trivial gauge links~$ U^\mu_n=1 \ \forall\ \mu,n $ (free theory). Each update cycle consists of sequences of one heatbath and five overrelaxation  steps. The initial thermalisation uses $ 2000 $ iterations of the update cycle. Each configuration uses additional $ 100 $ iterations of the update cycle in order to reduce the correlation between configurations. The update algorithm provides gauge links in single precision. Double precision link variables are obtained as copies from the single precision gauge links, which are projected to SU(3) in double precision. Finally, double precision link variables are copied back to single precision links with an additional typecast. Otherwise, deviations between single and double precision fields occasionally cause exceptions in the code, which may halt execution of programs.
\noindent
The scale is set following~\cite{Guagnelli:1998ud} with the Sommer parameter~$ r_0 $ as reference scale~\cite{Sommer:1993ce}, which is defined by \mbox{eq.}~($ 1.1 $) of~\cite{Guagnelli:1998ud} as
\begin{equation}
  r_0 F(r_0) = 1.65,
  \label{eq: Sommer parameter}
\end{equation}\normalsize
where $ F(r) $ is the force between static charges in the fundamental representation~\cite{Sommer:1993ce}. The constant on the right hand side of \mbox{eq.}~(\ref{eq: Sommer parameter}) is chosen in order to fix the Sommer parameter at roughly $ 0.5\,\mathrm{fm} $. In the range $ 5.7 \leq \beta \leq 6.57 $, the inverse gauge coupling~$ \beta $ is related to the lattice spacing~$ a $ by \mbox{eq.}~($ 2.18 $) of~\cite{Guagnelli:1998ud}, which reads
\begin{equation}
  \log{\left(\frac{a}{r_0}\right)}= -1.6805-1.7139(\beta-6)+0.8155(\beta-6)^2-0.6667(\beta-6)^3.
  \label{eq: scale setting}
\end{equation}\normalsize

\subsection{Dirac operators}\label{sec: Dirac operators}

Dirac operators for Karsten-Wilczek fermions and Bori\c{c}i-Creutz fermions of \mbox{eqs.}~(\ref{eq: KW fermion action}) and~(\ref{eq: BC fermion action}) are implemented using Dirac matrices in the chiral representation of \mbox{eq.}~(\ref{eq: Euclidean Dirac matrices}). However, instead of a direct implementation of the Dirac operator~$ D $, which is $ \gamma^5 $~hermitian, the hermitian Dirac operator~$ Q=\gamma^5 D $ is used. Dirac operators~$ Q $ for implementation of Karsten-Wilczek and Bori\c{c}i-Creutz fermions are presented in appendix~\ref{app: Lattice Dirac operators}. The propagator~$ S_{m,\underline{n}}[U]\eta_{\underline{n}} $ is obtained by solving the equation
\begin{equation}
  \sum\limits_{m,\alpha,a} Q_{\underline{n},m}^{\underline{\beta}\alpha,\underline{b}a} \phi_m^{\alpha,a} = \gamma^5 \eta_{\underline{n}}^{\underline{\beta},\underline{b}}
  \label{eq: Dirac equation with source}
\end{equation}\normalsize
with numerical methods (\mbox{cf.}~sections~\ref{sec: Mixed precision CG inverter} and~\ref{sec: Even-odd preconditioning}). The source vector $ \eta_{\underline{n}} $ and the solution vector $ \phi_m $ are globally defined spinor fields. The solution $ \phi_m $, which is given by
\begin{equation}
  \phi_m^{\alpha,a}(\eta_{\underline{n}}^{\underline{\beta},\underline{b}}) 
  = (Q^{-1})_{m,\underline{n}}^{\alpha\underline{\beta},a\underline{b}} \gamma^5 \eta_{\underline{n}}^{\underline{\beta},\underline{b}} 
  = (D^{-1})_{m,\underline{n}}^{\alpha\underline{\beta},a\underline{b}} \eta_{\underline{n}}^{\underline{\beta},\underline{b}} 
  \equiv S_{m,\underline{n}}^{\alpha\underline{\beta},a\underline{b}}[U]\eta_{\underline{n}}^{\underline{\beta},\underline{b}},
  \label{eq: propagator components}
\end{equation}\normalsize
is the inverse Dirac operator applied to the source $ \eta_{\underline{n}} $ and describes the propagation of a fermion field from site $ \underline{n} $ to all other sites $ m $ in the background of the gauge field $ U $. The source is a local point source with twelve complex components $ \eta_{\underline{n}}^{\underline{\beta},\underline{b}} $ -- three colour components $ \underline{b} $ times four spin components $ \underline{\beta} $ -- at the site $ \underline{n} $, which is always located at $ \underline{n}=(0,0,0,0) $ for simulations of this thesis. Propagator components $ \phi_m^{\alpha,a}(\eta_{\underline{n}}^{\underline{\beta},\underline{b}})  $ are obtained from twelve inversions, where different source components are set to $ 1 $. Smeared sources have the gauge-invariant smearing operator of \mbox{eq.}~(\ref{eq: Jacobi smearing operator}) applied~$ n_{JS} $ times to local source components $ \eta_{\underline{n}}^{\underline{\beta},\underline{b}} $ before the Dirac operator is inverted.  

\subsection{Mixed precision CG inverter}\label{sec: Mixed precision CG inverter}

\begin{figure}[hbt]
 \flushleft
 \rule{\linewidth}{0.25mm}\\
 \textbf{Starting conditions:} $ k=0 $
 \begin{align*}
  x_k       =&\ 0                   &\qquad (A)\\
  r_k       =&\ b - A x_k           &\qquad (B)\\
  p_k       =&\ s_k = A r_k         &\qquad (C)
 \end{align*}\normalsize
 \textbf{Loop: iterate for} $ k=0,1,2,\ldots $
 \begin{align*}
  q_k       =&\ A p_k               & (a) \\
  \lambda_k =&\ |s_k|^2/|q_k|^2     & (b) \\
  r_{k+1}   =&\ r_k - \lambda_k q_k & (c) \\
  x_{k+1}   =&\ x_k +\lambda_k p_k  & (d) \\
  \nu_k     =&\ |s_k|^2             & (e) \\
  s_{k+1}   =&\ A r_{k+1}           & (f) \\
  \mu_k     =&\ |s_{k+1}|^2/\nu_k   & (g) \\
  p_{k+1}   =&\ s_{k+1}+ \mu_k p_k  & (h)
 \end{align*}\normalsize 
 \textbf{until convergence is achieved and } $ |r_{k+1}|^2 < \epsilon^2 $.
 \rule{\linewidth}{0.25mm}\\
 \caption{Pseudocode for the CG algorithm was taken from chapter thirteen of~\cite{Aoki:2005}.}
 \label{fig: CG pseudocode}
\end{figure}

Since inversion of the Dirac operator is the bottleneck of any simulation in the quenched approximation, the inversion algorithm must be optimised and accelerated to the utmost. Minimally doubled Dirac operators are inverted using a conjugate gradient (CG) algorithm~\cite{Geradin1971319,Kalkreuter:1995mm} that was taken from chapter thirteen of~\protect\cite{Aoki:2005} by solving \mbox{eq.}~(\ref{eq: Dirac equation with source}) as
\begin{equation}
  A x = b.
  \label{eq: CG initial equation}
\end{equation}\normalsize
Pseudocode of the CG algorithm is presented in figure~\ref{fig: CG pseudocode}. Convergence of a bi-conjugate gradient (Bi-CGstab) algorithm taken from chapter six of~\cite{Gattringer:2010zz} is unstable for minimally doubled fermions. \newline

\noindent
The inverter uses mixed precision as a runtime option, which conducts calculations $ (a) $, $ (f) $ and $ (h) $ in single and $ (c) $ and $ (d) $ in double precision. Thus, whereas residue $ r_k $ and solution vector $ x_k $ are double precision fields, search vectors $ p_k $, $ q_k $ and $ s_k $ use only single precision. 
\noindent
This implementation of the CG algorithm restricts costly matrix-times-vector operations of $ (a) $ and $ (f) $ to single precision, if $ (f) $ is performed after a typecast of $ r_{k+1} $ to single precision, which can be stored in $ q_k $. The mixed precision solver needs only one double precision spinor field $ r_k $ and three single precision spinor fields $ p_k $, $ q_k $ and $ s_k $ as well as three doubles\footnote{The double precision scalar $ \nu_k $ can be omitted since $ \lambda_k $ can be reused to store $ |s_k|^2 $ as well.} 
$ \lambda_k $, $ \nu_k $, $ \mu_k $ as workspace, which seems to be the most cost-efficient implementation of a straight CG algorithm without preconditioning for minimally doubled fermions.
\noindent 
In order to prevent the accumulation of numerical errors from single precision operations, convergence in double precision is ensured by recalculating the residue $ r_k $ from $ x_k $ using $ (B) $ in double precision after a fixed count ($ \sim10-30 $) of interations. In a final step after the precision goal $ |r_{k+1}|^2 < \epsilon^2 $ is reached, $ r_k $ is recalculated from $ x_k $ in double precision. In the rare case that the precision goal is missed, the solver runs through additional iterations at double precision\footnote{These final iterations require three additional double precision work space spinor fields, but ensure convergence in double precision. No more than two additional interations have been observed.}, until the precision goal is met.

\subsection{Even-odd preconditioning}\label{sec: Even-odd preconditioning}

In order to accelerate the inversion of Dirac operators for \mbox{eqs.}~(\ref{eq: KW fermion action}) and~(\ref{eq: BC fermion action}) with the CG algorithm, even-odd preconditioning~\cite{DeGrand:1990dk} is used to reduce the condition number of the operator. Lattice sites are separated into two sets of even or odd lattice sites, which are defined by \mbox{eq.}~(\ref{eq: even-odd site number}). Thus, the Dirac equation of \mbox{eq.}~(\ref{eq: Dirac equation with source}) is recast into
\begin{equation}
  \left.\begin{array}{rcl}
  Q_{\underline{n},m}^{ee} \phi_m^e + Q_{\underline{n},m}^{eo} \phi_m^o &=& \gamma^5 \eta_{\underline{n}}^e, \\
  Q_{\underline{n},m}^{oe} \phi_m^e + Q_{\underline{n},m}^{oo} \phi_m^o &=& \gamma^5 \eta_{\underline{n}}^o,
  \end{array}\right.
  \label{eq: even-odd Dirac equation with source}
\end{equation}\normalsize
where colour and spinor indices are omitted and indices~$ ^o $ and~$ ^e $ label fields restricted to odd or even sites. Matrix multiplications with gauge fields occur only in the hopping terms~$ Q_{n,m}^{eo} $ and~$ Q_{n,m}^{oe} $. $ Q_{n,m}^{ee} $ and $ Q_{n,m}^{oo} $ are constant matrices of type~$ Q_0  \delta_{n,m} $. Their colour structure is trivial and they are listed in table \ref{tab: eo pc constant Q_0}.
\begin{table}[hbt]
\center
 \begin{tabular}{|c|c|c|}
  \hline
  Action & $ Q_{0} $ & $ \sqrt{\det{Q_0}}$    \\
  \hline
  Wilson & $ (\frac{4r}{a}+m_0)\mathbf{1} $ & $ \left(\frac{4r+am_0}{a}\right)^2 $  \\
  \hline
  Karsten-Wilczek & $ i\frac{3\zeta+c}{a}\gamma^{\underline{\alpha}}+m_0\mathbf{1} $ & $ \left(\frac{3\zeta+c}{a}\right)^2+m_0^2 $ \\
  \hline
  Bori\c{c}i-Creutz & $ i\frac{c-2\zeta}{a}\Gamma+m_0\mathbf{1} $ & $ \left(\frac{c-2\zeta}{a}\right)^2+m_0^2 $ \\
  \hline
 \end{tabular}
 \caption{Constant matrices $ Q_0 $ have $ \sqrt{\det{Q_0}} > 1/a^2 $ and are easily inverted.}
 \label{tab: eo pc constant Q_0}
\end{table}

\noindent Following standard procedures, the upper row of \mbox{eq.}~(\ref{eq: even-odd Dirac equation with source}) is formally solved for~$ \phi_m^e $,
\begin{equation}
  \phi_m^e = (Q^{ee\,-1})_{m,\underline{n}}  \left( \gamma^5 \eta_{\underline{n}}^e - Q_{\underline{n},p}^{eo} \phi_p^o \right).
  \label{eq: eo pc complete}
\end{equation}\normalsize
Using \mbox{eq.}~(\ref{eq: eo pc complete}), $ \phi_m^e $ is eliminated from the lower row, which is reshuffled to 
\begin{equation}
 \left(Q_{\underline{n},m}^{oo} -  Q_{\underline{n},m_0}^{oe} (Q^{ee\,-1})_{m_0,m_1} Q_{m_1,m}^{eo}\right) \phi_m^o = 
  \gamma^5 \eta_{\underline{n}}^o - Q_{\underline{n},m_2}^{oe}(Q^{ee-1})_{m_2,\underline{n}} \gamma^5 \eta_{\underline{n}}^e.
  \label{eq: eo pc Qhat}
\end{equation}\normalsize
A new notation is introduced using a matrix $ \widehat{Q}_{n,m} $ and a preconditioned source $ \xi_{n}^o $,
\begin{align}
  \widehat{Q}_{n,m} =& \left(\delta_{n,m} - (Q^{oo\,-1})_{n,m_1} Q_{m_1,m_2}^{oe} (Q^{ee\,-1})_{m_2,m_3} Q_{m_3,m}^{eo}\right), 
  \label{eq: Qhat}\\
  \xi_{n}^o =& (Q^{oo\,-1})_{n,\underline{m}} \left(\gamma^5 \eta_{\underline{m}}^o - Q_{\underline{m},m_1}^{oe}(Q^{ee-1})_{m_1,\underline{m}} \gamma^5 \eta_{\underline{m}}^e \right).
  \label{eq: eo pc prepare}
\end{align}\normalsize
This notation allows for matching \mbox{eq.}~(\ref{eq: eo pc Qhat}) to the form of \mbox{eq.}~(\ref{eq: CG initial equation}) as
\begin{equation}
  \widehat{Q}_{n,m} \phi_m^o = \xi_{n}^o.
  \label{eq: eo pc Qhat phi}
\end{equation}\normalsize
Hence, the CG inverter can be applied to the system of \mbox{eq.}~(\ref{eq: eo pc Qhat phi}). After the solution~$ \phi_m^o = (\widehat{Q}^{-1})_{m,n} \xi_{n}^o $ is obtained, the even components $ \phi_m^e $ are calculated from \mbox{eq.}~(\ref{eq: eo pc complete}).
\noindent
The even-odd preconditioned operator $ \widehat{Q} $ requires the same number of matrix-times-vector operations like the original operator $ Q $, since the matrix multiplication is applied twice to only half the number of sites. However the condition number is reduced, since the gauge field dependent hopping term is suppressed against the constant on-site term by $ (\det{Q_0})^{1/2} $ instead of $ (\det{Q_0})^{1/4} $. Table~\ref{tab: eo pc constant Q_0} shows that suppression is best for Wilson fermions and worst for Bori\c{c}i-Creutz fermions. It should be noted here that even-odd preconditioning is computationally advantageous only if $ Q_{n,m}^{ee} $ and $ Q_{n,m}^{oo} $ can be inverted at marginal numerical cost.

\subsection{Contractions and interpolating operators}

Connected correlation functions are calculated according to \mbox{eq.}~(\ref{eq: mesonic correlation function}) from contractions of two fermion propagators at source and sink with a summation of the entire sink slice. Connected correlation functions depend on point-to-all propagators $ S_{n,0} $ and all-to-point propagators $ S_{0,n} $. The latter would require a prohibitive amount of inversions. However, the propagator inherits the $ \gamma^5 $~hermiticity from the Dirac operator and satisfies
\begin{equation}
  S_{0,n} = \gamma^5 (S^\dagger)_{n,0} \gamma^5.
  \label{eq: gamma^5 hermiticity for propagator}
\end{equation}\normalsize
Therefore, connected correlation functions require only point-to-all propagators,
\begin{equation}
  \mathcal{C}_{\mathcal{M},\mathcal{N}}(t)
  =  \sum\limits_{n \in \Lambda_{t}^0}\mathrm{tr}{\left( 
  S_{n,0} \mathcal{M} \gamma^5 (S^\dagger)_{n,0} \gamma^5\mathcal{N}
  \right)}.
  \label{eq: recast mesonic correlation function}
\end{equation}\normalsize
Contraction of all twelve propagator components completes the trace at the source. Source Dirac matrices are implemented through contraction of different propagator components and multiplication with phase factors ($ \pm 1 $ or $ \pm i $). 
The Dirac matrix at either source or sink must be transposed, since
\begin{equation}
  \mathcal{C}_{\mathcal{M},\mathcal{N}}(t)
  =  \sum\limits_{n \in \Lambda_{t}^0} \mathrm{tr}_c{\left( 
  S_{n,0}^{\alpha\beta} (S^*)_{n,0}^{\delta\gamma} ((\mathcal{M} \gamma^5)^T)^{\gamma\beta} (\gamma^5\mathcal{N})^{\delta\alpha}
  \right)}.
  \label{eq: mesonic correlation function for contractions}
\end{equation}\normalsize
Details of the procedure are listed in appendix~\ref{app: Contractions}.

\subsection{Smearing}\label{sec: Smearing}

\noindent A clean study of the anisotropy due to the Karsten-Wilczek action requires that other potential sources of anisotropy are avoided. Therefore, symmetric lattices are used. Compared to the usual choices for hadron spectroscopy, symmetric lattices have an unusually large spatial and unusually small temporal extent. These lattices have the disadvantage that the time interval in which the ground state is isolated becomes very short. It is therefore desirable to use smearing techniques to enhance the ground state's spectral weight. 
\noindent
Local and smeared interpolating operators for creation and annihilation of mesons are implemented in order to improve the signal of the pseudoscalar ground state. There is a heuristic rule that the overlap of an interpolating operator with a meson's creation operator is considerably improved if the average smearing radius of the spatial distribution of the smeared source operator is of similar magnitude than the physical radius of that meson. Thus, spectral weights of excited states are reduced and the ground state is isolated already for shorter Euclidean time separations (\mbox{cf.}~section~\ref{sec: Anisotropy of hadronic quantities}).
\noindent
Gauge-invariant smearing is achieved with (iterative) \textit{Jacobi smearing}\footnote{Jacobi smearing is related to Wuppertal smearing (also called Gaussian smearing), where the smearing procedure is repeated until convergence is reached.} algorithms~\cite{Gusken:1989ad,Allton:1993wc}. Jacobi smearing is an iterative procedure in which a gauge-invariant lattice Laplacian is added to the original spinor field with a fixed iteration count irrespective of convergence,
\begin{equation}
  O_{n,m}^J[U] = \frac{1}{1+6\kappa_{JS}}\left(\delta_{n,m}+ \frac{\kappa_{JS}}{a}\sum\limits_{\mu\neq0}\left(U^\mu_n\delta_{n+\hat{e}_\mu,m}+U^{\mu\dagger}_{n-\hat{e}_\mu}\delta_{n-\hat{e}_\mu,m}\right)\right),
  \label{eq: Jacobi smearing operator}
\end{equation}\normalsize
and includes a \textit{normalisation factor} $ 1/(1+6\kappa_{JS}) $, which prevents an unbounded growth of the smeared field. The smearing directions must be restricted to those perpendicular to the Euclidean time direction in order to avoid changes to the transfer matrix. In order to obtain a smooth spatial distribution, violent local fluctuations of spatial gauge fields have to be reduced. This is achieved with so-called \textit{fat links}. 
\noindent
\noindent Fat links $ V_n^\mu $ are calculated from \textit{thin links} $ U_n^\mu $ (the usual link variables) of the update procedure with link smearing algorithms that average gauge fields locally. As elementary step, thin links are replaced by the SU(3)~projected linear combinations of the thin links and their $ n_S $ pairs of \textit{staples},
\begin{equation}
  V_n^\mu = \text{Proj}_{SU(3)} \left[ (1-\xi) U_n^\mu + \frac{\xi}{2n_S} \sum_{\nu\neq\mu} U_n^{\nu} U_{n+\hat{e}_\nu}^\mu U_{n+\hat{e}_\mu}^{\mu\,\dagger} \right].
\end{equation}\normalsize 
In terms of the practical application, the smoothing properties of link smearing are similar to those of evolution with gradient flow~\cite{Bonati:2014tqa,Luscher:2010iy}. For the purpose of use in the numerical studies of this thesis,  APE smearing~\cite{Albanese:1987ds} and HYP smearing~\cite{Hasenfratz:2001hp} are implemented. The latter algorithm is applied with standard parameters ($ \alpha=0.75 $,~$ \beta=0.6 $,~$ \gamma=0.3 $). This amounts to using a single pair of opposite staples with parameter $ \xi=\gamma $ to get a first new set of links. Next, a second new set of links is constructed using the first set of new links for building two pairs of opposite staples with parameter $ \xi=\beta $. Finally, the fat links are constructed using the second new set of links for building three pairs of opposite staples with parameter $ \xi=\alpha $. Three full iterations of HYP smearing are used to provide smooth links for smeared interpolating operators at the source of correlation functions.

The Jacobi smearing parameter is fixed as $ \kappa_{JS}=4.0 $, which is a standard value used in many applications. The iteration number $ n_{JS} $ is adapted to the gauge coupling (\mbox{cf.}~table~\ref{tab: gauge coupling, scale setting and smearing}) in order to achieve smearing radii of approximately $ r \approx 0.5\, \mathrm{fm} $. This enhances the spectral weight of the pseudoscalar ground state. Lastly, it is noted that correlation functions which are smeared only at either the source or the sink may include states with negative spectral weights. In these cases, the correlation function is not necessarily a convex function and low-lying excited states may cancel partly with the ground state and reduce the quality of the signal.

\subsection{Parallelisation}\label{sec: Parallelisation}

\noindent Memory considerations as well as the general need to minimise the computational effort strongly motivate parallelisation of numerical simulations.
The full lattice~$ \Lambda $ is subdivided into equally-sized local sublattices~$ \lambda_j $, which must communicate the fields in the neighbourhood of their boundaries~$ \partial \lambda_j $ to their neighbouring subprocesses after every modification. Communication between subprocesses is handled by MPI routines. Data is communicated via the Infiniband network of the Wilson cluster. 
In parallelised code, extra memory is required for the boundaries~$ \partial \lambda_j $, which store copies of fields from neighbouring sublattices. Thus, a finer division into local sublattices incurs a larger memory overhead for boundaries. Moreover, actual communication of boundary fields must wait for termination of the previous subprocess on all local sublattices and communication itself is a bottleneck for application of the Dirac operator. \newline

\noindent
For Karsten-Wilczek fermions with $ \zeta=\pm1 $, kinetic terms of the Dirac operator in any direction other than the $ \hat{e}_{\underline{\alpha}} $~direction have only one linearly independent pair of one lower and one upper Dirac component (\mbox{cf.}~appendix~\ref{app: Lattice Dirac operators}). After this pair has been relayed to neighbouring sublattices, missing components are automatically reconstructed in communication routines. The present simulation code comprises Karsten-Wilczek fermions for two different directions of the Karsten-Wilczek term, either $ \underline{\alpha}= 0 $ or $ \underline{\alpha}=3 $, with arbitrary $ \zeta $ or $ \zeta=+1 $. Bori\c{c}i-Creutz ($ \zeta=+1 $) and Wilson fermions are also available. The Dirac structure of hopping terms for some of these Dirac operators (\mbox{e.g.}~$ \frac{1\pm\gamma^\mu}{2} $ for Wilson or $ \frac{\gamma^\mu\mp i\zeta\gamma^{\underline{\alpha}}}{2}$ for $ \mu\neq \underline{\alpha} $ for Karsten-Wilczek fermions) allows for reconstruction of some Dirac components from the others, which allows a reduction of the computational cost. This is due to a reduced number of SU(3)~multiplications and also due to a reduced number of components that have to be communicated across the boundaries. Each Dirac operator has its own optimised set of boundary field communication buffers and routines, which are initialised automatically in simulations.
\noindent

\newpage
\sectionc{Anisotropy of hadronic quantities}{sec: Anisotropy of hadronic quantities}

\begin{table}[hbt]
\center\footnotesize
 \begin{tabular}{|c|c|c|c|c|c|c|c|}
  \hline
  $ \beta $ & $ a\,[fm] $ & $ r_0 $ & $ L $ & $ n_{cfg} $ & $ am_0 $ & $ c $ & $ d $ \\
  \hline
  $ 6.0 $  & $ 0.093 $ & $ 5.368 $ & $ 32 $ & $ 100 $ & $ 0.02,0.03,0.04,0.05 $ &
  $ [-1.2,+0.3]  $ & $ 0.0 $ \\
  $ 6.0 $  & $ 0.093 $ & $ 5.368 $ & $ 32 $ & $ 100 $ & $ 0.01,0.02,0.03,0.04,0.05 $ &
  $ [-0.65,-0.20] $ & $ [-0.08,+0.02] $ \\
  $ 6.0 $  & $ 0.093 $ & $ 5.368 $ & $ 48 $ & $  40 $ & $ 0.02 $ &
  $ [-0.65,+0.0]  $ & $ 0.0 $ \\
  \hline
  $ 6.2 $  & $ 0.068 $ & $ 7.360 $ & $ 32 $ & $ 100 $ & $ 0.01,0.02,0.03,0.04,0.05 $ & 
  $ [-0.65,-0.20]  $ & $ [-0.08,+0.04] $ \\
  $ 6.2 $  & $ 0.068 $ & $ 7.360 $ & $ 48 $ & $  40 $ & $ 0.02 $ &
  $ [-0.65,+0.0]  $ & $ 0.0 $ \\
  \hline
  $ 5.8 $  & $ 0.136 $ & $ 3.668 $ & $ 32 $ & $ 100 $ & $ 0.02,0.04,0.05 $ & 
  $ [-0.65,+0.0]  $ & $ [0.0,+0.02] $ \\
  \hline
 \end{tabular}
 \caption{Symmetric lattices~($ T=L $) are used in studies of the anisotropy. The parameter $ c $ is varied in small steps close to the perturbative estimates (\mbox{cf.}~table~\ref{tab: BPT predictions}).}
 \label{tab: lattice parameters}
\end{table}\normalsize

\noindent
Numerical studies~\cite{Weber:2013tfa} of the anisotropy are performed with Karsten-Wilczek fermions of \mbox{eqs.}~(\ref{eq: KW fermion action}) and~(\ref{eq: KW gluon action}) in order to tune the coefficients $ c $ and $ d $ non-perturbatively. Boosted perturbation theory (BPT) provides an initial guess for the coefficients (\mbox{cf.}~table~\ref{tab: BPT predictions}), where the plaquette is used to estimate non-perturbative parameters from one-loop results (\mbox{cf.}~section~\ref{sec: Boosted perturbation theory}). Symmetric lattices with~$ T=L $ and periodic boundary conditions in all directions avoid anisotropic finite volume or boundary effects that might obscure the anisotropy due to the fermion action. Since simulations utilise the quenched approximation, the coefficient~$ d_p $ of the gluonic counterterm is set to zero.
\noindent
Correlation functions are calculated using Karsten-Wilczek fermion propagators with~$ \zeta=+1 $ and alignment~$ \underline{\alpha}=0 $ (\mbox{cf.}~section~\ref{sec: Dirac operators}), where the direction of correlation is either the~$ \hat{e}_0 $~or~$ \hat{e}_3 $~direction. With respect to the alignment of the Karsten-Wilczek operator, $ \hat{e}_0 $ yields \textit{parallel} and~$ \hat{e}_3 $ yields \textit{perpendicular correlation functions}. The statistical basis are $ 100 $ configurations (\mbox{cf.} section \ref{sec: Gauge configurations and scale setting}) of $ 32^4 $~lattices and $ 40 $~configurations of $ 48^4 $~lattices. Lattice parameters are listed in table~\ref{tab: lattice parameters}. Correlation functions use source-smeared interpolating operators (\mbox{cf.}~section~\ref{sec: Smearing}) with a smearing radius of $ r \approx 0.5\,\mathrm{fm} $, since the direction of correlation is quite short ($ T=32 $) and isolation of the ground state is not guaranteed otherwise. Scale setting using the Sommer parameter~\cite{Sommer:1993ce} is accomplished via \mbox{eq.}~(\ref{eq: scale setting}).

\subsection{Tuning with the mass anisotropy}\label{sec: Tuning with the mass anisotropy}

\noindent The gauge coupling is chosen as~$ \beta=6.0 $ for initial scans of parameter space. The bare fermion mass~$ am_0 $ is set to a relatively high value, which allows for inversions of the Dirac operator with moderate computational cost. The marginal counterterm's coefficient is fixed as~$ d=0 $ initially, because BPT suggests that its effects are mild (\mbox{cf.}~table~\ref{tab: BPT predictions}) compared to the relevant parameter~$ c $. $ c $ is varied in small steps near the BPT estimate~($ c\approx c_{BPT} $) and in wide steps in more outlying regions of parameter space~($ c\not\approx c_{BPT} $). The pseudoscalar~(PS) correlation function with vanishing hadron momentum (\mbox{cf.~eq.}~(\ref{eq: recast mesonic correlation function}))
\begin{equation}
    \mathcal{C}_{PS}^{\parallel,\perp}(t) \equiv \mathcal{C}_{\gamma^5,\gamma^5}(n_{\parallel,\perp}) \equiv \mathcal{C}_{5,5}(n_{\parallel,\perp})
  =  \sum\limits_{n \in \Lambda_{t}^{\parallel,\perp}} 
  \sum\limits_{a,b} \sum\limits_{\alpha,\beta}
  |S_{n,0}^{\alpha\beta,ab}|^2
  \label{eq: PS correlation function}
\end{equation}\normalsize
is computed in numerical simulations using Karsten-Wilczek fermion propagators with $ \underline{\alpha}=0 $. The direction of correlation is either $ \hat{e}_t=\hat{e}_0 $ for the \textit{parallel} (wrt the Karsten-Wilczek term) or $ \hat{e}_t=\hat{e}_3 $ for the \textit{perpendicular correlation function}. It is analysed by performing fits, which extract the mass~$ M_{PS}^{\parallel,\perp} $ of the pseudoscalar ground state. Statistical errors are determined with the Jackknife method (\mbox{cf.}~appendix~\ref{app: Statistical analysis}). Since the study of the anisotropy is restricted to the pseudoscalar channel, the index `$ PS $' is omitted in the remainder of section~\ref{sec: Anisotropy of hadronic quantities}. The mass~$ M_{\parallel} $, which is calculated from the parallel correlation function $ \mathcal{C}^{\parallel}(t) $ is called the \textit{parallel mass}, and the mass~$ M_{\perp} $, which is calculated from the perpendicular correlation function $ \mathcal{C}^{\perp}(t) $, is called the \textit{perpendicular mass}. The \textit{mass anisotropy} is the difference of squared pseudoscalar masses,
\begin{equation}
  \Delta(M_{PS}^2) \equiv \left(M_\parallel^2\right)-\left(M_\perp^2\right).
  \label{eq: PS mass anisotropy}
\end{equation}\normalsize
which is employed as a tuning criterion for~$ c $, while the other parameters~$ \beta $,~$ m_0 $ and~$ d $ are kept at fixed values. The mass anisotropy is interpolated as a function of~$ c $. The extremum of the mass anisotropy defines the non-perturbatively tuned coefficient~$ c_{M} $\footnote{$ c_M $ labels the value of $ c $ that is eventually obtained by tuning with the mass anisotropy. The label is used to allow a distinction to $ c $ from other methods of tuning~(\mbox{cf.}~sections~\ref{sec: Tuning with the frequency spectrum},\ref{sec: Interim findings (III)}).}. In the next step, the bare fermion mass~$ am_0 $ is lowered towards the chiral limit. An extrapolation of~$ c_{M} $ to the chiral limit is performed in order to compare with estimates from BPT. In the following step, sensitivity to variation of~$ d $ is studied. Lastly, the coupling~$ \beta $ is varied and the procedure is repeated for different lattice spacings.

\subsection{Determination of the pseudoscalar mass}\label{sec: Determination of the pseudoscalar mass}

\noindent
Because the fermion action of \mbox{eq.}~(\ref{eq: KW fermion action}) lacks a simple transformation behaviour under $ n_{\underline{\alpha}} $~reflection, backward (b) and forward (f) propagating states will in general be different in parallel correlation functions. Masses and spectral weights of backwards and forwards states are treated as independent. If both masses differ,~$ M_\parallel $ cannot be unambiguously defined. 
In line with section~\ref{sec: Components in correlation functions}, no oscillations are expected in the $ \gamma^5 $ channel. 
Hence, this correlation function can be described for vanishing hadronic momenta as
\begin{align}
  \mathcal{C}(t) = \sum\limits_{k=0}^\infty \Big(\frac{A_k^f}{2M_k^f} e^{-M_k^f \cdot t}+\frac{A_k^b}{2M_k^b} e^{-M_k^b \cdot (T-t)}\Big)
\end{align}\normalsize
and reaches asymptotic behaviour for large $ T $ and $ t \approx T/2 $ as
\begin{equation}
  \mathcal{C}(t) = \frac{A_0^f}{2M_0^f} e^{-M_0^f \cdot t} +  \frac{A_0^b}{2M_0^b} e^{-M_0^b \cdot (T-t)} + \ldots.
  \label{eq: generalised correlation function}
\end{equation}\normalsize
Using \mbox{eq.}~(\ref{eq: generalised correlation function}) as a four-parametric fit function with adequate fit ranges in clearly separated intervals of the Euclidean time direction simultaneously extracts masses of backwards and forwards states as independent parameters. A local effective mass of either backwards or forwards states is another tool for isolating their respective masses. 

\subsubsection{Local effective mass}\label{sec: Local effective mass}

\begin{figure}[hbt]
 \begin{picture}(360,140)
  \put(005  , 0.0){\includegraphics[bb=0 0 170 130, scale=0.85]{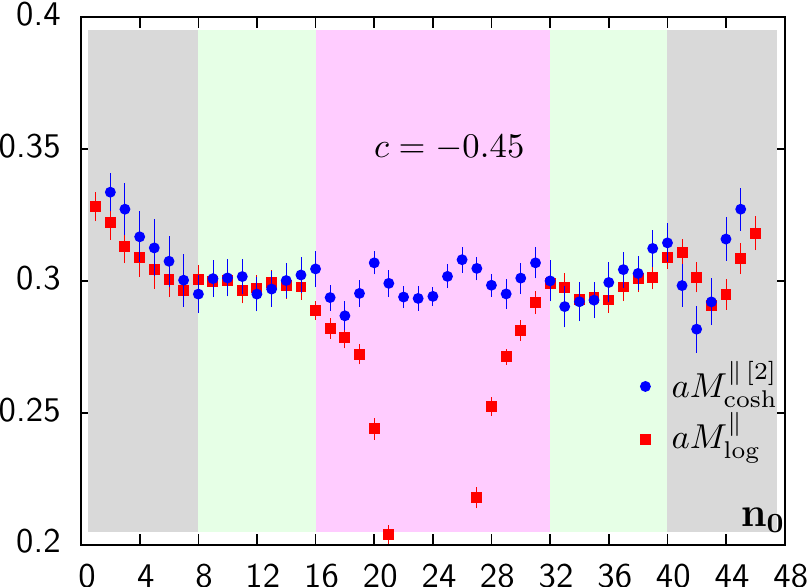}}
  \put(215.0, 0.0){\includegraphics[bb=0 0 170 130, scale=0.85]{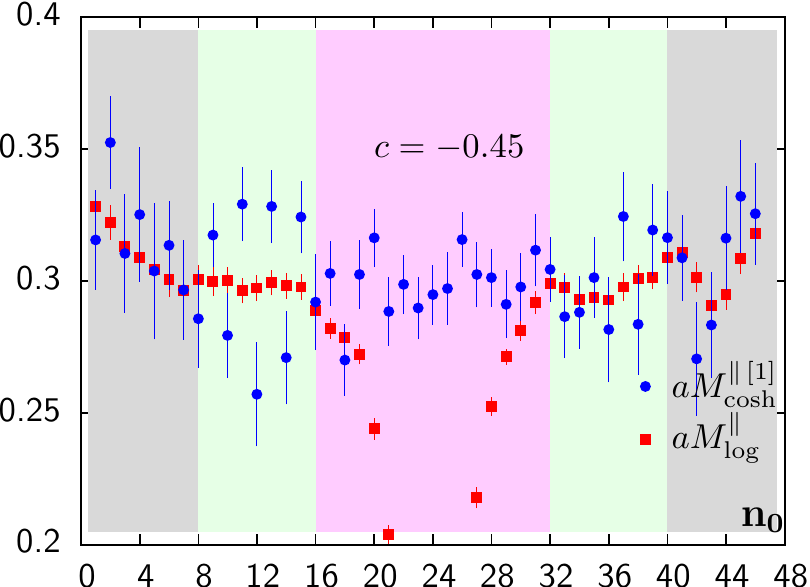}}
 \end{picture}
 \caption{Left: $ M_{\cosh}^{[2]}(t) $ and $ M_{\log}(t) $ agree in the plateau region within errors.
Right: Fluctuations around $ M_{\log}(t) $ beset $ M_{\cosh}^{[1]}(t) $ even in the plateau region. 
Symbols and data sets are explained in the text.
}
 \label{fig: parallel effective mass plot example}
\end{figure}

\noindent
Such a local effective mass, which is called the \textit{log mass}, is calculated as
\begin{equation}
  M_{\log}^{[s]}(t) = \frac{1}{s}\log{ \frac{|\mathcal{C}(t)|}{|\mathcal{C}(t+s)|}}
  \label{eq: log mass}
\end{equation}\normalsize
and reaches a plateau for large times if the oncoming state is negligible. The parameter $ s $ is the step size and is usually set to one. In an interval around the midpoint of a periodic lattice, the log mass is a poor quantity for deciding whether the ground state has been reached because of the oncoming state's contribution. Another definition of a local effective mass is the \textit{cosh mass}, which is calculated as
\begin{equation}
  M_{\cosh}^{[s]}(t) = \frac{1}{s}\log{\left(R+\sqrt{R^2-1}\right)},\ R=\frac{|\mathcal{C}(t+s)|+|\mathcal{C}(t-s)|}{2|\mathcal{C}(t)|}.
  \label{eq: cosh mass}
\end{equation}\normalsize

\begin{table}[hbt]
\center\footnotesize
 \begin{tabular}{|c|c|c|c|c|c|c|c|c|c|}
  \hline
  Eff.  & Fit range & \multicolumn{2}{|c|}{ Cor. fit } & \multicolumn{2}{|c|}{ Uncor. fit } & \multicolumn{2}{|c|}{ Cor. fit } & \multicolumn{2}{|c|}{ Uncor. fit } \\
   mass &  $ [\ldots] $ & $ M^{[1]} $ & $ \!\chi^2/\mathrm{dof}\! $ & $ M^{[1]} $ & $ \!\chi^2/\mathrm{dof}\! $ & $ M^{[2]} $ & $ \!\chi^2/\mathrm{dof}\! $ & $ M^{[2]} $ & $ \!\chi^2/\mathrm{dof}\! $ \\
  \hline
  $ {\cosh} $ & $ [8\!-\!16] $ &  
  $ \!0.305(3)\! $ & $ 0.950 $ & $ \!0.305(3)\! $ & $ 2.800 $ &
  $ \!0.299(3)\! $ & $ 0.692 $ & $ \!0.300(3)\! $ & $ 0.237 $ \\
  $ {\cosh} $ & $ [32\!-\!40] $ &  
  $ \!0.304(3)\! $ & $ 1.399 $ & $ \!0.301(3)\! $ & $ 0.915 $ &
  $ \!0.305(2)\! $ & $ 1.352 $ & $ \!0.301(3)\! $ & $ 1.406 $ \\
  $ {\cosh} $ & $ \!\![8\!-\!16,32\!-\!40]\!\! $ &  
  $ \!0.306(2)\! $ & $ 1.447 $ & $ \!0.303(3)\! $ & $ 1.769 $ &
  $ \!0.303(2)\! $ & $ 2.225 $ & $ \!0.300(2)\! $ & $ 0.786 $ \\
  $ {\cosh} $ & $ \!\![8\!-\!40]\!\! $ &  
  $ \!0.299(1)\! $ & $ 5.801 $ & $ \!0.301(1)\! $ & $ 1.349 $ &
  $ \!0.297(1)\! $ & $ 7.148 $ & $ \!0.299(1)\! $ & $ 1.126 $ \\
  \hline
  $ {\log} $ & $ [8\!-\!16] $ &  
  $ \!0.294(2)\! $ & $ 1.633 $ & $ \!0.297(3)\! $ & $ 0.833 $ &
  $ \!0.293(2)\! $ & $ 1.910 $ & $ \!0.296(3)\! $ & $ 0.886 $ \\
  $ {\log} $ & $ [32\!-\!40] $ &  
  $ \!0.301(2)\! $ & $ 1.461 $ & $ \!0.299(2)\! $ & $ 1.197 $ &
  $ \!0.299(1)\! $ & $ 2.303 $ & $ \!0.299(2)\! $ & $ 1.700 $ \\
  $ {\log} $ & $ \!\![8\!-\!16,32\!-\!40]\!\! $ &  
  $ \!0.299(1)\! $ & $ 2.100 $ & $ \!0.298(2)\! $ & $ 1.006 $ &
  $ \!0.302(1)\! $ & $ 3.916 $ & $ \!0.298(2)\! $ & $ 1.965 $ \\
  \hline
 \end{tabular}
 \caption{Fits to the effective mass mostly agree within $ 1\!-\!2\sigma $. 
 $ M_{\cosh}^{[2]} $ is more stable than $ M_{\cosh}^{[1]} $. 
 The data set is the same as in figure~\ref{fig: parallel effective mass plot example}.
 }
 \label{tab: effm fit mass example}
\end{table}\normalsize
\noindent
Effective mass plots in figure~\ref{fig: parallel effective mass plot example} have a lilac shaded band centered around midpoint \mbox{($ t=T/2 $)}, where the log mass (red filled squares) drops towards zero. 
Neither backwards nor forwards contributions are negligible in the central region. The grey-shaded bands near the temporal boundary of the lattice yield higher $ M_{\log}(t) $, since excited states cannot be neglected and the ground state is not isolated. In the green shaded plateau regions, $ M_{\log}(t) $ is constant within errors, since excited as well as oncoming states are negligible. Therefore, local effective masses in the green bands isolate the backward or forward propagating ground states repectively. The parallel pseudoscalar correlation function is obtained with~$ \beta=6.0 $ and~$ c=-0.45 $ on a $ 48^4 $~lattice from table~\ref{tab: lattice parameters}. The pseudoscalar mass is $ M_{PS}^\parallel \approx 630\,\mathrm{MeV} $.
\noindent
The cosh mass with~$ s=2 $ (blue bullets in the left plot of figure \ref{fig: parallel effective mass plot example}) agrees very well with $ M_{\log}(t) $ in the plateau region. Moreover,~$ M_{\cosh}^{[2]}(t) $ maintains the plateau value within errors within the central region around midpoint. Since the cosh mass with~$ s=1 $ (blue bullets in the right plot of figure~\ref{fig: parallel effective mass plot example}) strongly fluctuates around the plateau value of~$ M_{\log}(t) $,~$ M_{\cosh}^{[1]}(t) $ should not be used. The local effective mass is fitted as a constant within the plateau regions. Fit results in table~\ref{tab: effm fit mass example} provide no numerical evidence of different backwards and forwards masses. However, $ \chi^2/\mathrm{dof} $ is very large and its considerable variation between different fits is not understood at present.

\subsubsection{Parallel correlation functions}\label{sec: Parallel correlation functions}

\begin{table}[hbt]
\center\footnotesize
 \begin{tabular}{|c|c|c|c|c|c|c|c|}
  \hline
  Independent  & Fit range & \multicolumn{3}{|c|}{ Correlated fit } & \multicolumn{3}{|c|}{ Uncorrelated fit } \\
  parameters & $ [\ldots] $ &  $ M^f $ & $ M^b $ & $ \!\chi^2/\mathrm{dof}\! $ & $ M^f $ & $ M^b $ & $ \!\chi^2/\mathrm{dof}\! $ \\
  \hline
  $ M_0^f,\,A_0^f $ & $ [8\!-\!16] $ &  
  $ 0.301(3) $ &  & $ 1.261 $ & 
  $ 0.298(3) $ &  & $ 0.038 $ \\
  $ M_0^b,\,A_0^b $ & $ [32\!-\!40] $ &  
   & $ 0.300(2) $ & $ 0.690 $ & 
   & $ 0.297(3) $ & $ 0.059 $ \\
  \hline
  $ M_0^f\!=\!M_0^b,\,A_0^f\!=\!A_0^b $ & $ [8\!-\!16] $ &  
  $ 0.302(3) $ &  & $ 1.154 $ & 
  $ 0.299(3) $ &  & $ 0.035 $ \\
  $ M_0^f\!=\!M_0^b,\,A_0^f\!=\!A_0^b $ & $ [32\!-\!40] $ &  
   & $ 0.301(2) $ & $ 0.740 $ & 
   & $ 0.298(3) $ & $ 0.036 $ \\
  \hline
  $ M_0^f,\,M_0^b,\,A_0^f,\,A_0^b $ & $ [8\!-\!40] $ &  
  $ 0.300(1) $ & $ 0.299(1) $ & $ 3.923 $ & 
  $ 0.298(2) $ & $ 0.299(2) $ & $ 0.038 $ \\
  $ M_0^f,\,M_0^b,\,A_0^f,\,A_0^b $ & $ [10\!-\!38] $ &  
  $ 0.300(1) $ & $ 0.296(1) $ & $ 2.827 $ & 
  $ 0.298(2) $ & $ 0.299(2)  $ & $ 0.039 $ \\
  \hline
  $ M_0^f,\,M_0^b,\,A_0^f,\,A_0^b $ & $ \!\![8\!-\!16,32\!-\!40]\!\! $ &  
  $ 0.304(3) $ & $ 0.303(2) $ & $ 1.402 $ & 
  $ 0.299(3) $ & $ 0.298(3) $ & $ 0.036 $ \\
  $ M_0^f,\,M_0^b,\,A_0^f,\,A_0^b $ & $ \!\![10\!-\!16,32\!-\!38]\!\! $ &  
  $ 0.301(3) $ & $ 0.301(3) $ & $ 1.424 $ & 
  $ 0.299(4) $ & $ 0.296(3) $ & $ 0.033 $ \\
  $ M_0^f,\,M_0^b,\,A_0^f,\,A_0^b $ & $ \!\![8\!-\!14,34\!-\!40]\!\! $ &  
  $ 0.303(3) $ & $ 0.302(3) $ & $ 1.858 $ & 
  $ 0.299(4) $ & $ 0.298(3) $ & $ 0.042 $ \\
  \hline
 \end{tabular}
 \caption{Backwards and forwards fit masses agree within errors. 
 Fits with different functions or fit ranges are consistent but strongly deviate in their correlated $ \chi^2/\mathrm{dof} $. 
 The data set is the same as in figure~\ref{fig: parallel effective mass plot example}.
 }
 \label{tab: cfun fit mass example}
\end{table}\normalsize

\begin{table}[hbt]
\center\footnotesize
 \begin{tabular}{|c|c|c|c|c|c|c|}
  \hline
  Fit range & \multicolumn{3}{|c|}{ Correlated fit } & \multicolumn{3}{|c|}{ Uncorrelated fit }  \\
   $ [\ldots] $ & $ M $ & $ A\ [\times10^{-3}] $ & $ \!\chi^2/\mathrm{dof}\! $ & $ M $ & $ A\ [\times10^{-3}] $ & $ \!\chi^2/\mathrm{dof}\! $ \\
  \hline
  $ [10\!-\!38] $ & 
  $ 0.2983(5) $ & $ 6.02(10) $ & $ 3.215 $ & $ 0.298(1) $ & $ 6.2(2) $ & $ 0.566 $ \\
  $ [9\!-\!39] $ & 
  $ 0.2988(5) $ & $ 6.12(8) $ & $ 3.120 $ & $ 0.298(1) $ & $ 6.2(1) $ & $ 0.585 $ \\
  $ [8\!-\!40] $ & 
  $ 0.2996(4) $ & $ 6.24(8) $ & $ 3.839 $ & $ 0.298(1) $ & $ 6.2(1) $ & $ 0.612 $ \\
  $ \!\![8\!-\!14,34\!-\!40]\!\! $ & 
  $ 0.304(2) $ & $ 6.6(3) $ & $ 1.795 $ & $ 0.299(3) $ & $ 6.3(3) $ & $ 0.985 $ \\
  $ \!\![8\!-\!16,32\!-\!40]\!\! $ & 
  $ 0.304(2) $ & $ 6.6(2) $ & $ 1.432 $ & $ 0.299(3) $ & $ 6.3(3) $ & $ 0.969 $ \\
  $ \!\![10\!-\!16,32\!-\!38]\!\! $ & 
  $ 0.301(2) $ & $ 6.4(2) $ & $ 1.272 $ & $ 0.298(3) $ & $ 6.2(3) $ & $ 0.989 $ \\
  \hline
 \end{tabular}
 \caption{$ M_\parallel $ is determined with a symmetrised fit. 
 Though correlated $ \chi^2/\mathrm{dof} $ strongly depends on the fit range, results are stable and agree with uncorrelated fits within errors. 
 The data set is the same as in figure~\ref{fig: parallel effective mass plot example}.
 }
 \label{tab: sym fit mass example}
\end{table}\normalsize

\noindent
An alternative approach uses fits to the correlation function with an ansatz like \mbox{eq.}~(\ref{eq: generalised correlation function}). If the fit range is restricted to one of the plateau regions, backwards or forwards states are isolated for all fits of table~\ref{tab: cfun fit mass example}. Fits with independent parameters for forwards and backwards states simultaneously extract properties of backwards or forwards ground states. Examples of fit results are listed in table~\ref{tab: cfun fit mass example}. There is no numerical evidence for different backwards and forwards masses and consistency between both fit strategies is evident. Hereafter, $ n_{\underline{\alpha}} $~reflection symmetry is assumed and $ M_\parallel $ is defined as the fit mass obtained with a symmetrised fit ansatz. Local effective masses are calculated as cosh masses with step size $ s=2 $ due to better numerical stability. Fit results are stable (\mbox{cf.}~table~\ref{tab: sym fit mass example}), though correlated fits have large $ \chi^2/\mathrm{dof} $ if the interval around midpoint \mbox{($ t\approx T/2 $)} is included.

\subsubsection{Perpendicular correlation functions}\label{sec: Perpendicular correlation functions}

\begin{figure}[hbt]
 \begin{picture}(360,140)
  \put(015  , 0.0){\includegraphics[bb=0 0 170 130, scale=0.85]{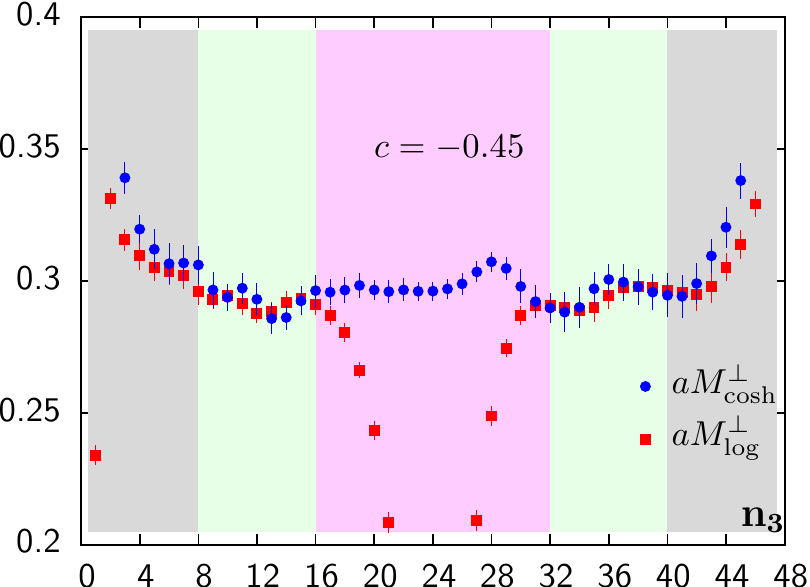}}
  \put(215.0, 0.0){\includegraphics[bb=0 0 170 130, scale=0.85]{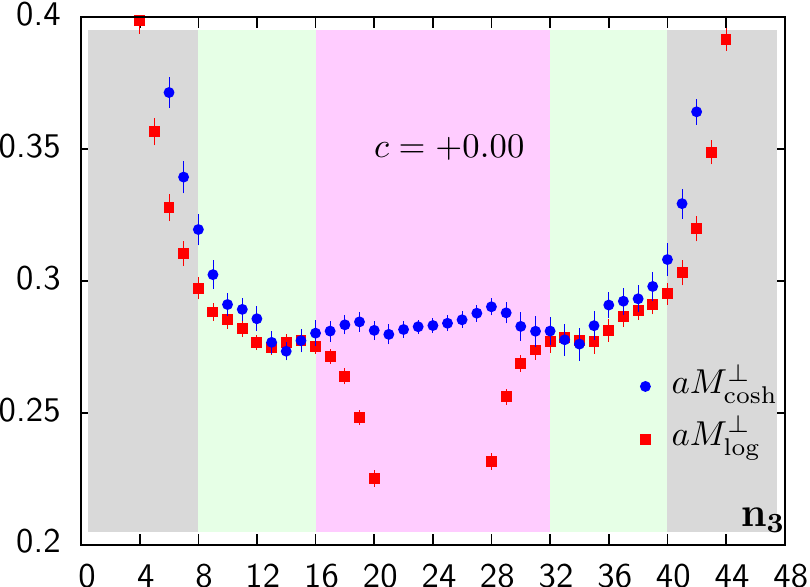}}
 \end{picture}
 \caption{Left: $ M_{\cosh}(t) $ and $ M_{\log}(t) $ agree in the plateau region within errors at $ c=-0.45 \approx c_{BPT} $.
Right: $ M_{\cosh}(t) $ and $ M_{\log}(t) $ agree only for $ t/a \in [13,16]\cup[32,35] $ at $ c=+0.00 $. 
Symbols and data sets are explained in the text.
}
 \label{fig: perpendicular effective mass plot example}
\end{figure}

\begin{table}[hbt]
\center\footnotesize
 \begin{tabular}{|c|c|c|c|c|c|c|
 c|c|
 }
  \hline
  & \multicolumn{4}{|c|}{$ c=-0.45 $ } & \multicolumn{4}{|c|}{$ c=+0.00 $ } \\
  Fit range & \multicolumn{2}{|c|}{ Cor. fit} & \multicolumn{2}{|c|}{ Unc. fit} & \multicolumn{2}{|c|}{ Cor. fit} & \multicolumn{2}{|c|}{ Unc. fit}
  \\
   $ [\ldots] $ &  $ M_{}^{[2]} $ & $ \!\chi^2/\mathrm{dof}\! $ &  $ M_{}^{[1]} $ & $ \!\chi^2/\mathrm{dof}\! $ & $ M_{}^{[2]} $ & $ \!\chi^2/\mathrm{dof}\! $ & $ M_{}^{[1]} $ & $ \!\chi^2/\mathrm{dof}\! $
  \\
  \hline
  $ [16\!-\!32] $ & 
  $ 0.297(1) $ & $ 0.862 $ & $ 0.299(2) $  & $ 0.061 $ & $ 0.283(1) $ & $ 1.583 $ & $ 0.284(1) $  & $ 0.069 $ 
  \\
  $ [14\!-\!34] $ & 
  $ 0.296(1) $ & $ 0.709 $ & $ 0.298(1) $  & $ 0.078 $ & $ 0.2829(9) $ & $ 1.363 $ & $ 0.284(1) $  & $ 0.101 $ 
  \\
  $ [12\!-\!36] $ & 
  $ 0.2953(8) $ & $ 1.640 $ & $ 0.297(1) $ & $ 0.121 $ & $ 0.2821(8) $ & $ 1.913 $ & $ 0.283(1) $  & $ 0.153 $ 
  \\
  $ [10\!-\!38] $ & 
  $ 0.2960(7) $ & $ 1.863 $ & $ 0.296(1) $ & $ 0.192 $ & $ 0.2840(6) $ & $ 2.114 $ & $ 0.283(1) $ & $ 0.211 $ 
  \\
  $ [8\!-\!40] $  & 
  $ 0.2953(4) $ & $ 1.729 $ & $ 0.296(1) $ & $ 0.266 $ & $ 0.2835(3) $ & $ 6.716 $ & $ 0.283(1) $ & $ 0.290 $ 
  \\
  $ \!\![8\!-\!14,34\!-\!40]\!\! $ & 
  $ 0.293(1) $ & $ 1.306 $ & $ 0.293(2) $ & $ 0.354 $ & $ 0.283(1) $ & $ 2.631 $ & $ 0.284(2) $ & $ 0.597 $ 
  \\
  $ \!\![8\!-\!16,32\!-\!40]\!\! $ & 
  $ 0.295(1) $ & $ 1.243 $ & $ 0.293(2) $ & $ 0.304 $ & $ 0.2834(9) $ & $ 2.076 $ & $ 0.283(2) $ & $ 0.510 $ 
  \\
  $ \!\![10\!-\!16,32\!-\!38]\!\! $ & 
  $ 0.295(1) $ & $ 0.912 $ & $ 0.292(2) $ & $ 0.208 $ & $ 0.282(1) $ & $ 1.633 $ & $ 0.281(2) $ & $ 0.374 $ 
  \\
  \hline
 \end{tabular}
 \caption{Fits to perpendicular functions hint at increasing mass for fit ranges closer to midpoint, but agree within errors. The data set is the same as in figure\ref{fig: perpendicular effective mass plot example}.}
 \label{tab: n3 fit mass example}
\end{table}\normalsize

\begin{table}[hbt]
\center\footnotesize
 \begin{tabular}{|c|c|c|c|c|c|c|}
  \hline
  Fit range & \multicolumn{3}{|c|}{ Correlated fit } & \multicolumn{3}{|c|}{ Uncorrelated fit }  \\
   $ [\ldots] $ & $ M $ & $ A\ [\times10^{-3}] $ & $ \!\chi^2/\mathrm{dof}\! $ & $ M $ & $ A\ [\times10^{-3}] $ & $ \!\chi^2/\mathrm{dof}\! $ \\
  \hline
  $ [10\!-\!38] $ & 
  $ 0.2960(7) $ & $ 3.05(5) $ & $ 1.863 $ & $ 0.296(1) $ & $ 3.10(8) $ & $ 0.192 $ \\
  $ [9\!-\!39] $ & 
  $ 0.2953(4) $ & $ 3.02(4) $ & $ 1.820 $ & $ 0.296(1) $ & $ 3.08(8) $ & $ 0.223 $ \\
  $ [8\!-\!40] $ & 
  $ 0.2953(4) $ & $ 3.01(3) $ & $ 1.729 $ & $ 0.296(1) $ & $ 3.06(8) $ & $ 0.266 $ \\
  $ \!\![8\!-\!14,34\!-\!40]\!\! $ & 
  $ 0.293(1) $ & $ 2.92(6) $ & $ 1.306 $ & $ 0.293(2) $ & $ 2.93(9) $ & $ 0.354 $ \\
  $ \!\![8\!-\!16,32\!-\!40]\!\! $ & 
  $ 0.295(1) $ & $ 2.96(5) $ & $ 1.243 $ & $ 0.293(2) $ & $ 2.93(9) $ & $ 0.304 $ \\
  $ \!\![10\!-\!16,32\!-\!38]\!\! $ & 
  $ 0.295(1) $ & $ 3.05(8) $ & $ 0.912 $ & $ 0.292(2) $ & $ 2.9(1) $ & $ 0.208 $ \\
  \hline
 \end{tabular}
 \caption{$ M_\perp $ is determined with a symmetrised fit ansatz. The data set is the same as in figure\ref{fig: perpendicular effective mass plot example} for $ c=-0.45 $.}
 \label{tab: sym n3 fit parameters}
\end{table}\normalsize

\noindent
Time reflection symmetry of perpendicular correlation functions is strictly satisfied and the asymptotic form of a correlation function is given by
\begin{equation}
  \mathcal{C}(t) = \frac{A_0}{2M_0} \left(e^{-M_0 \cdot t} + e^{-M_0 \cdot (T-t)}\right) + \ldots,
  \label{eq: symmetric correlation function}
\end{equation}\normalsize
which can be used as a two-parametric fit function. The local effective masses shown in figure~\ref{fig: perpendicular effective mass plot example} are obtained from perpendicular pseudoscalar correlation functions with~\mbox{$ \beta=6.0 $} on a $ 48^4 $~lattice from table~\ref{tab: lattice parameters}.
\noindent
The fit mass has a peculiar dependence on the fit range. Table~\ref{tab: n3 fit mass example} shows that the mass increases and~$ \chi^2/\mathrm{dof} $ decreases, if the fit range is contracted towards the central region around midpoint~($ t \approx T/2 $). Excited states contributing with negative spectral weights are a possible explanation for this counterintuitive behaviour. Such behaviour is possible as the correlation functions is smeared only at the source (\mbox{cf.~}section~\ref{sec: Smearing}). Fit results are listed in table~\ref{tab: sym n3 fit parameters} for comparison with parallel correlation functions (\mbox{cf.}~table~\ref{tab: sym fit mass example}). 

\FloatBarrier

\subsection{Minimisation of the mass anisotropy}\label{sec: Minimsation of the mass anisotropy}

\subsubsection{Interpolation of fit parameters}\label{sec: interpolation of fit parameters}

\begin{figure}[hbt]
 \begin{picture}(360,145)
   \put(090.0, 0.0){\includegraphics[bb=0 0 170 130, scale=0.90]{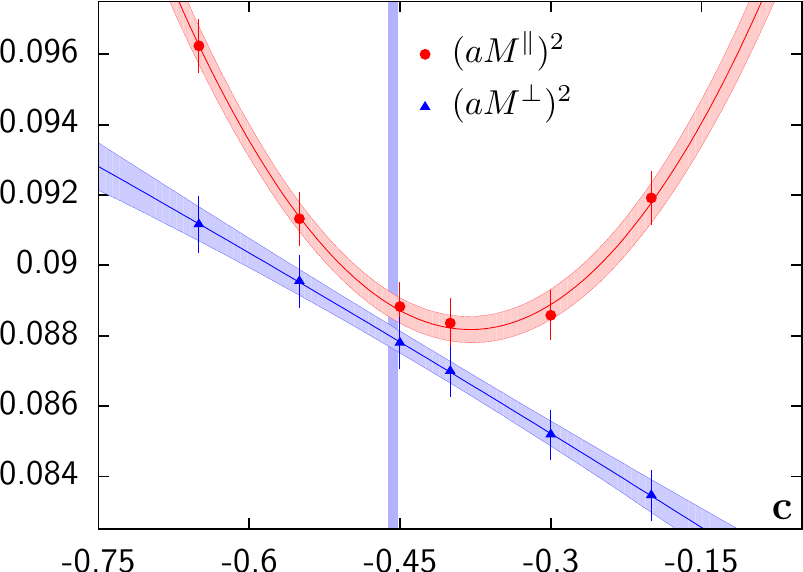}} 
 \end{picture}
 \caption{Interpolation of $ M_\parallel^2 $ and $ M_\perp^2 $ in $ c $: Both squared masses are interpolated within $ c \in [-0.65,-0.2]  $. Correlators have $ \beta=6.0 $ on a $ 48^4 $ lattice from table \ref{tab: lattice parameters}. The pseudoscalar mass is $ M_{PS} \approx 630\,\mathrm{MeV} $ for $ c = c_M $.
}
 \label{fig: interpolation example}
\end{figure}

\noindent
A fit to the pseudoscalar correlation function extracts the parameter~$ M $ for the ground state's mass. Residual mass anisotropy can persist at finite lattice spacing even after tuning the counterterms' coefficients. Calculating the mass anisotropy from squared pseudoscalar masses is beneficial for a chiral extrapolation that assumes Goldstone boson-like properties. An ansatz for the squared masses $ M_{\parallel,\perp}^2 $ as quadratic functions in $ c $,
\begin{equation}
  M_{\parallel,\perp}^2(c) = a_{0}^{\parallel,\perp} + a_{1}^{\parallel,\perp}\,c + a_{2}^{\parallel,\perp}\,c^2,
  \label{eq: interpolation ansatz}
\end{equation}\normalsize
is motivated by the expectation to achieve an minimum of the anisotropy after tuning. 
An example for the interpolation within~$ c \in [-0.65,-0.2] $ is shown in figure~\ref{fig: interpolation example}. All data within the range are matched within standard errors, but~$ M_{\parallel}^2 $ at~$ c=0.0 $ requires additional higher order terms. Fit masses~$ M_{\parallel,\perp}^2 $ do not match but retain a small difference, which is of the same order as statistical uncertainties. Therefore, the absolute mass anisotropy of \mbox{eq.}~(\ref{eq: PS mass anisotropy}) is not necessarily a good observable. The extremum of the mass anisotropy is calculated from the difference of interpolations as
\begin{equation}
  c_{M} = -\frac{a_{1}^{\parallel}-a_{1}^{\perp}}{2(a_{2}^{\parallel}-a_{2}^{\perp})}
  \label{eq: extremum of the parameter anisotropy}
\end{equation}\normalsize
and indicated as the blue vertical band in figure~\ref{fig: interpolation example}.
\begin{table}[hbt]
\center\footnotesize
 \begin{tabular}{|c|c|c|c|c|}
  \hline
  Fit range & \multicolumn{2}{|c|}{$ c_{M} $ for $ c \in [-0.65,-0.20] $} & \multicolumn{2}{|c|}{$ c_{M} $ for  $ c \in [-0.65,+0.00] $} \\
  $ [\ldots]\!\! $ & 
  { Cor. fit} & { Unc. fit} & { Cor. fit} & { Unc. fit}   \\
  \hline
  $ \!\![10\!-\!38]\!\! $ &
  $ \!-0.460(5)\! $ & $ \!-0.457(6)\! $ & 
  $ \!-0.457(4)\! $ & $ \!-0.456(4)\! $
  \\  
  $ \!\![9\!-\!39]\!\! $  &
  $ \!-0.519(9)\! $ & $ \!-0.457(5)\! $ & 
  $ \!-0.494(6)\! $ & $ \!-0.455(4)\! $
  \\
  $ \!\![8\!-\!40]\!\! $ &
  $ \!-0.508(9)\! $ & $ \!-0.457(5)\! $ & 
  $ \!-0.486(5)\! $ & $ \!-0.455(4)\! $
  \\
  \hline
 \end{tabular}
 \caption{
 Extrema of $ \Delta(M_{PS}^2) $ are stable within errors only for uncorrelated fits to the correlation function. The data set has $ \beta=6.0 $ on a $ 48^4 $ lattice from table~\ref{tab: lattice parameters}.
 }
 \label{tab: minimisation example}
\end{table}\normalsize

\noindent
Though the fit range dependence of $ c_{M} $ for interpolations of fit masses from correlated fits to the correlation function exceeds the statistical uncertainty, it is consistent with uncorrelated results for shorter fit intervals. This is shown in table~\ref{tab: minimisation example}.

\subsubsection{Smaller lattices}\label{sec: Smaller lattices}

\begin{figure}[hbt]
 \begin{picture}(360,260)
  \put(005  , 135.0){\includegraphics[bb=0 0 170 120, scale=0.80]{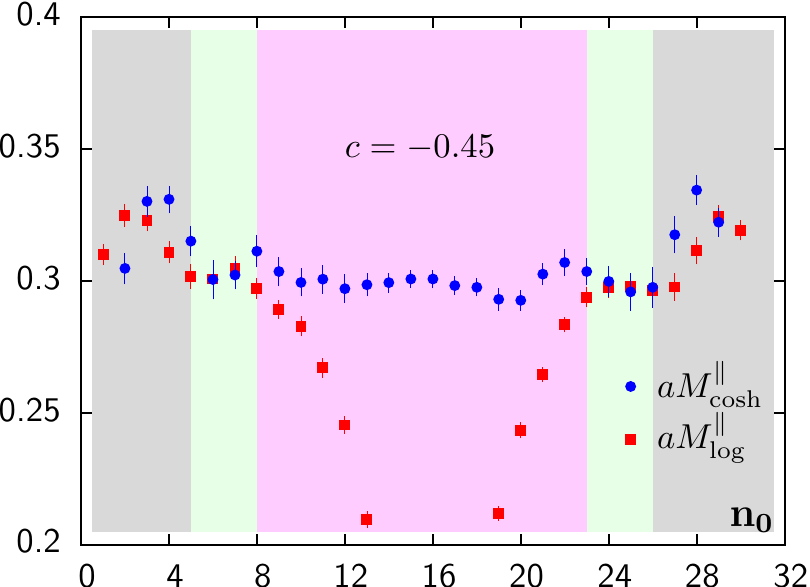}}
  \put(215.0, 135.0){\includegraphics[bb=0 0 170 120, scale=0.80]{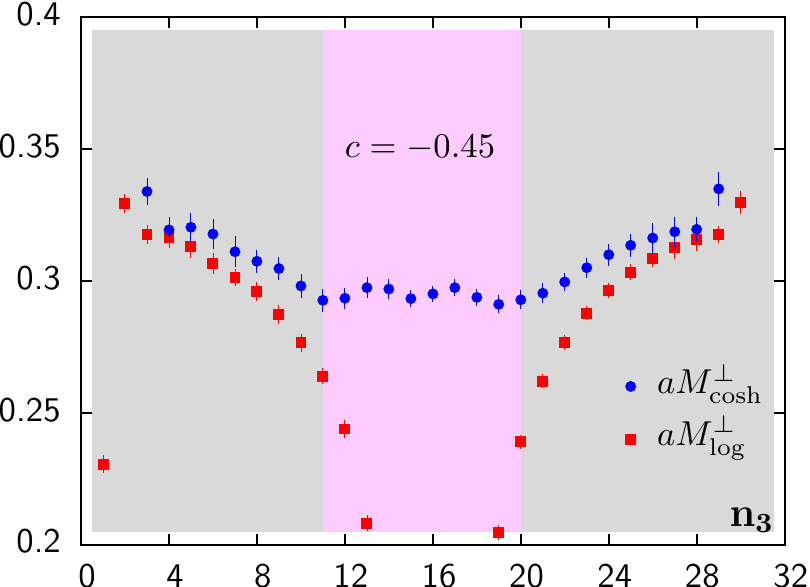}}
  \put(005  ,   0.0){\includegraphics[bb=0 0 170 120, scale=0.80]{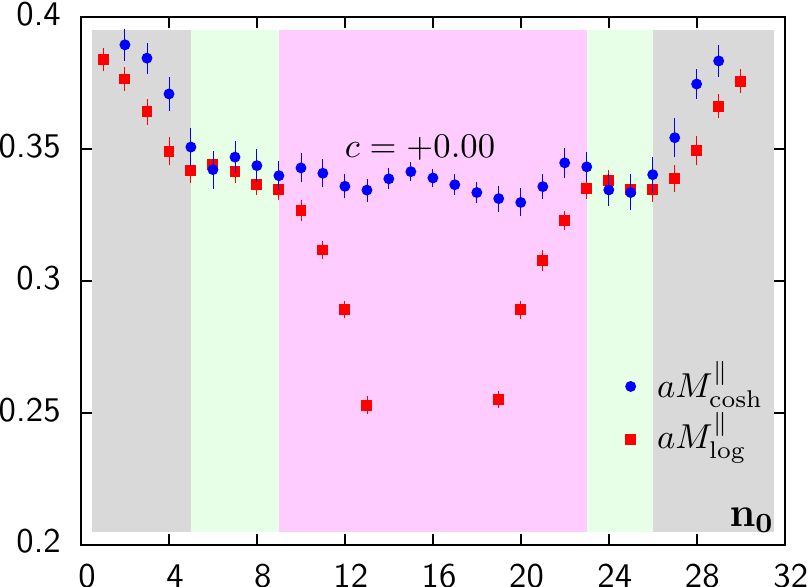}}
  \put(215.0,   0.0){\includegraphics[bb=0 0 170 120, scale=0.80]{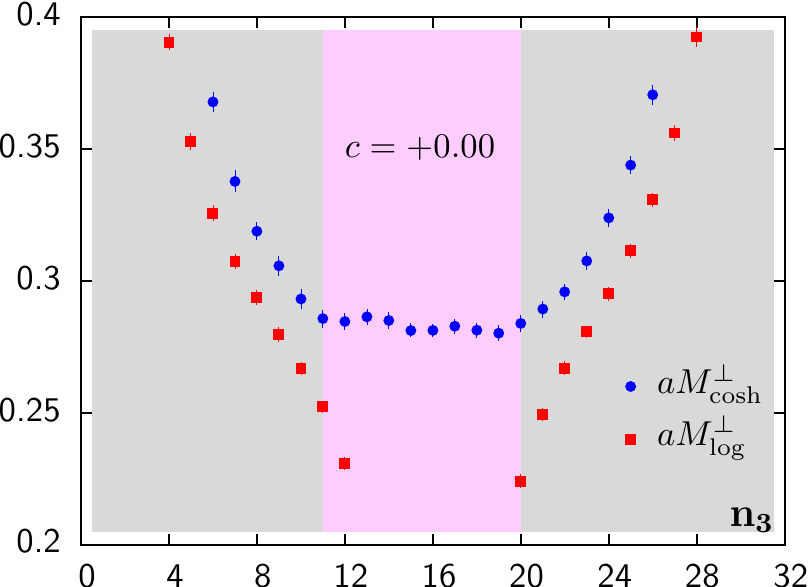}}
 \end{picture}
 \caption{Left: Plateaus of $ M_{\cosh}^\parallel(t) $ and $ M_{\log}^\parallel(t) $ are very short. Right: $ M_{\log}^\perp(t) $ lacks the plateau of $ M_{\cosh}^\perp(t) $.
 Symbols and data sets are explained in the text.
}
 \label{fig: 32^4 effective mass plot example}
\end{figure}

\begin{table}[hbt]
\center\footnotesize
 \begin{tabular}{|c|c|c|c|c|
 }
  \hline
  Fit range & \multicolumn{2}{|c|}{$ c_{M} $ for $ c \in [-0.65,-0.20] $} & \multicolumn{2}{|c|}{$ c_{M} $ for  $ c \in [-0.70,+0.00] $} \\
  $ [\ldots]\!\! $ & 
  { Cor. fit} & { Unc. fit} & { Cor. fit} & { Unc. fit}   \\
  \hline
  $ \!\![12\!-\!20]\!\! $ &
  $ \!-0.448(8)\! $ &
  $ \!-0.449(8)\! $ &
  $ \!-0.450(6)\! $ &
  $ \!-0.453(7)\! $ 
  \\
  $ \!\![10\!-\!22]\!\! $ &
  $ \!-0.444(6)\! $ & 
  $ \!-0.442(6)\! $ & 
  $ \!-0.446(5)\! $ &
  $ \!-0.447(5)\! $ 
  \\  
  $ \!\![9\!-\!23]\!\! $  &
  $ \!-0.446(6)\! $ & 
  $ \!-0.442(5)\! $ & 
  $ \!-0.447(5)\! $ & 
  $ \!-0.446(5)\! $ 
  \\
  $ \!\![8\!-\!24]\!\! $ &
  $ \!-0.440(6)\! $ & 
  $ \!-0.442(5)\! $ & 
  $ \!-0.442(5)\! $ & 
  $ \!-0.445(4)\! $ 
  \\
  \hline
 \end{tabular}
 \caption{$ c_M $ is slightly less negative on smaller volumes (\mbox{cf.}~table~\ref{tab: minimisation example}). The data set has $ \beta=6.0 $,~$ m_0=0.02 $ and~$ d=0.0 $ on a $ 32^4 $~lattice from table~\ref{tab: lattice parameters}.
 }
 \label{tab: 32^4 finite volume}
\end{table}\normalsize


\begin{figure}[hbt]
 \begin{picture}(360,140)
  \put(215.0, 0.0){\includegraphics[bb=0 0 170 130, scale=0.80]{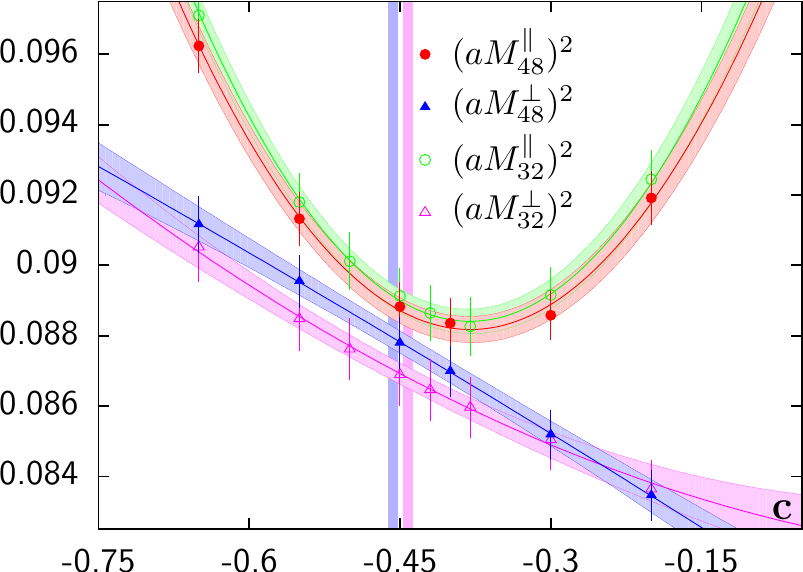}}
    \put(005.0, 0.0){\includegraphics[bb=0 0 170 130, scale=0.80]{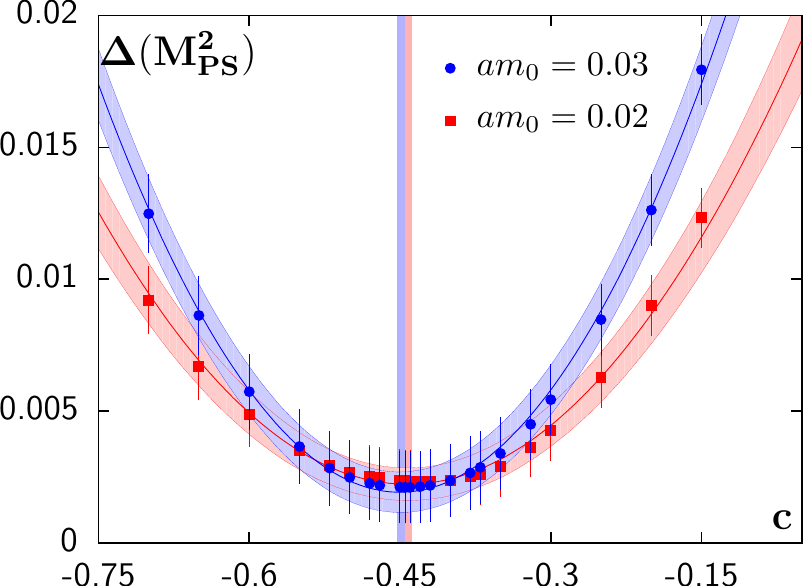}}
 \end{picture}
 \caption{Left: The extremum is shallower for lighter masses. Right: The mass anisotropy is consistent within errors. Symbols and data sets are explained in the text.
}
 \label{fig: 32^4 interpolation}
\end{figure}

\noindent
Smaller lattices~($ 32^4 $) of table~\ref{tab: lattice parameters} are used for scans of the parameter space in $ m_0 $ and $ d $ due to their smaller numerical cost. However, the time direction for symmetric lattices with $ T=L $ might be too short for a clean extraction of the ground state, which introduces systematical uncertainties. 
Scans of parameter space are required for clarification of the dependence of the mass anisotropy on $ d $ and for an extrapolation of $ c_M $ to the chiral limit. 
Local effective mass plateaus, which are invariably shorter due to the shorter time direction, are presented in figure~\ref{fig: 32^4 effective mass plot example}. The cosh mass with $ s=2 $ is represented by blue bullets and the log mass with $ s=1 $ by red squares. The cosh mass plateau of the perpendicular correlation function is very short. The pseudoscalar correlation functions have $ c=-0.45 $ in the upper plots and $ c=0.0 $ in the lower plots. Data are obtained with~\mbox{$ \beta=6.0 $},~$ m_0=0.02 $ and~$ d=0.0 $ on a $ 32^4 $~lattice from table~\ref{tab: lattice parameters}. The pseudoscalar mass for $ c=-0.45 $ is $ M_{PS} \approx 630\,\mathrm{MeV} $ and consistent with $ 48^4 $ within combined errors. \newline

\noindent
Direct subtraction of the interpolations yields~$ \Delta(M_{PS}^2) $ of \mbox{eq.}~(\ref{eq: PS mass anisotropy}). The left plot of figure~\ref{fig: 32^4 interpolation} displays the dependence of~$ \Delta(M_{PS}^2) $ on the fermion mass parameter in the neighbourhood of~$ c_M $, where the two overlapping vertical bands indicate $ c_M $ for the different masses. The lower mass $ am_0=0.02 $ (red squares) yields a shallower interpolation than the higher mass $ am_0=0.03 $ (blue bullets). 
\noindent
Interpolations of fit masses for different lattice sizes are plotted together in the right plot of figure~\ref{fig: 32^4 interpolation}. Masses on the smaller lattice are denoted by open symbols. Though the mass splitting is slightly wider on the smaller lattice as $ M^\parallel_{32} > M^\parallel_{48} $ and $ M^\perp_{32} < M^\perp_{48} $, the masses agree within errors. Minimisation results for $ \beta=6.0 $, $ m_0=0.02 $ and $ d=0.0 $ using different fit and interpolation ranges are stable within errors (\mbox{cf.}~table~\ref{tab: 32^4 finite volume}). Values for the larger ($ c_{M}^{(48)} $,~\mbox{cf.}~table~\ref{tab: minimisation example}) and smaller ($ c_{M}^{(32)} $,~\mbox{cf.}~table~\ref{tab: 32^4 finite volume}) lattices are indicated as vertical blue and lilac bands in the figure. Their difference $ \Delta^{\mathrm{fs}}_{c_M} \equiv c_M^{(48)}-c_M^{(32)} \approx -0.01 $ provides an estimate of finite size effects, which are considered as systematical errors. 

\subsubsection{Chiral extrapolation of the minimum of the mass anisotropy}\label{sec: Chiral extrapolation of the minimum of the mass anisotropy}

\begin{figure}[hbt]
 \begin{picture}(360,140)
  \put(005  , 0.0){\includegraphics[bb=0 0 170 130, scale=0.80]{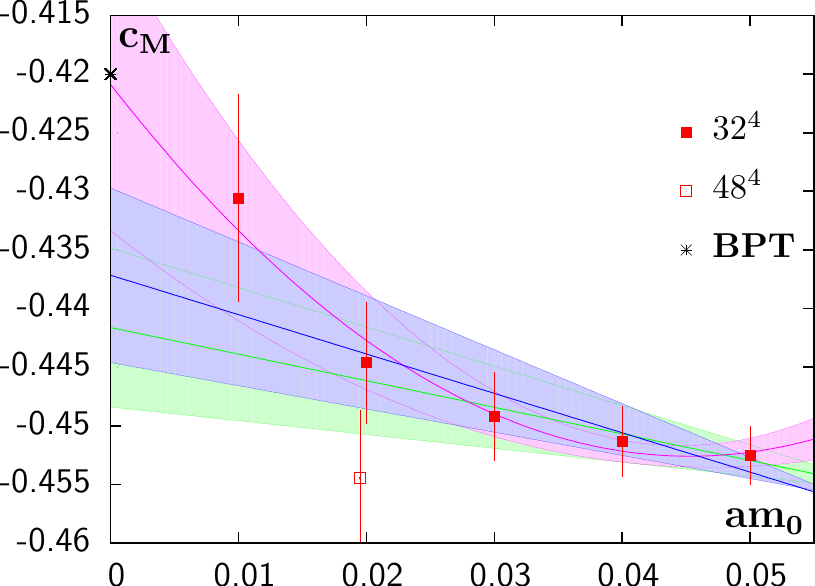}}
  \put(215.0, 0.0){\includegraphics[bb=0 0 170 130, scale=0.80]{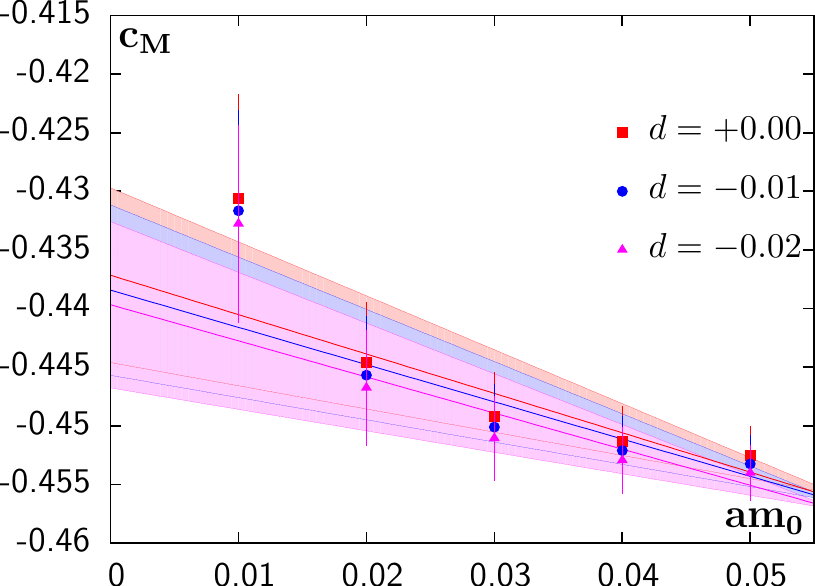}}
 \end{picture}
 \caption{
 Left: A chiral extrapolation of $ c_M $ can be compared to $ c_{BPT} $. Finite size effects seem to point to $ c_M $ being more negative by $ \Delta^{\mathrm{fs}}_{c_M} \approx -0.01 $. Right: Dependence on $ d $ is not resolved by $ c_M $. Symbols and data sets are explained in the text.
}
 \label{fig: c_M chiral extrapolation}
\end{figure}

\begin{table}[hbt]
\center\footnotesize
 \begin{tabular}{|c|c|c|c|c|
 }
  \hline
  Fit range & \multicolumn{2}{|c|}{$ c_{M} $, linear in $ m_0 $} & \multicolumn{2}{|c|}{$ c_{M} $, quadratic  in $ m_0 $} \\
   $ [\ldots]\!\! $ & \multicolumn{1}{|c|}{ Cor. fit}  & \multicolumn{1}{|c|}{ Unc. fit} & \multicolumn{1}{|c|}{ Cor. fit}  & \multicolumn{1}{|c|}{ Unc. fit}   \\
  \hline
  $ \!\![12\!-\!20]\!\! $ &
  $ \!-0.448(11)\! $ & $ \!-0.451(12)\! $ & 
  $ \!-0.429(22)\! $ & $ \!-0.436(21)\! $  \\
  $ \!\![10\!-\!22]\!\! $ &
  $ \!-0.439(8)\!  $ & $ \!-0.439(9)\! $ & 
  $ \!-0.425(14)\! $ & $ \!-0.422(14)\! $ \\  
  $ \!\![9\!-\!23]\!\! $  &
  $ \!-0.440(8)\!  $ & $ \!-0.437(13)\! $ & 
  $ \!-0.429(13)\! $ & $ \!-0.421(12)\! $ \\
  $ \!\![8\!-\!24]\!\! $ &
  $ \!-0.429(8)\!  $ & $ \!-0.435(7)\! $ & 
  $ \!-0.411(12)\! $ & $ \!-0.419(11)\! $ \\
  \hline
 \end{tabular}
 \caption{
  Quadratic instead of linear chiral extrapolation yields more negative $ c_M $ with larger errors. Correlators have $ \beta=6.0 $ on a $ 32^4 $ lattice from table \ref{tab: lattice parameters}.
 }
 \label{tab: 32^4 c_M chiral limit}
\end{table}\normalsize

\begin{figure}[hbt]
 \begin{picture}(360,140)
  \put(005.0, 0.0){\includegraphics[bb=0 0 170 130, scale=0.80]{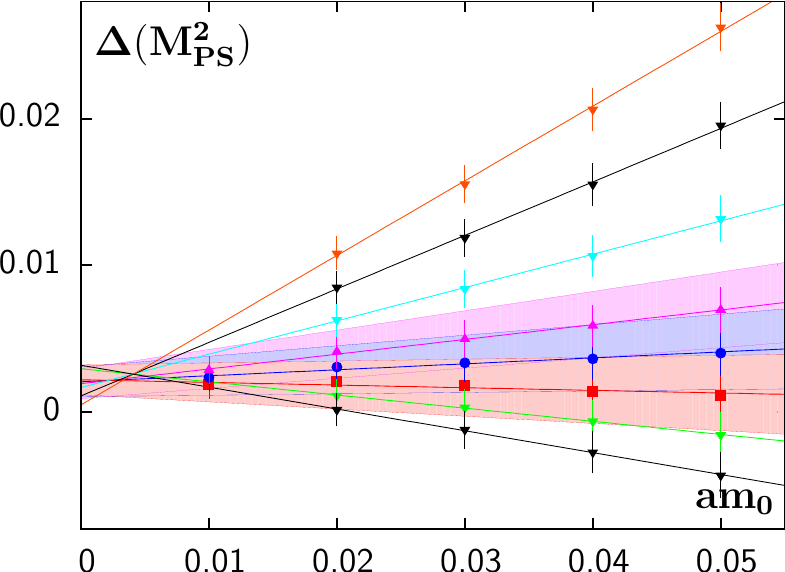}}
  \put(215.0, 0.0){\includegraphics[bb=0 0 170 130, scale=0.80]{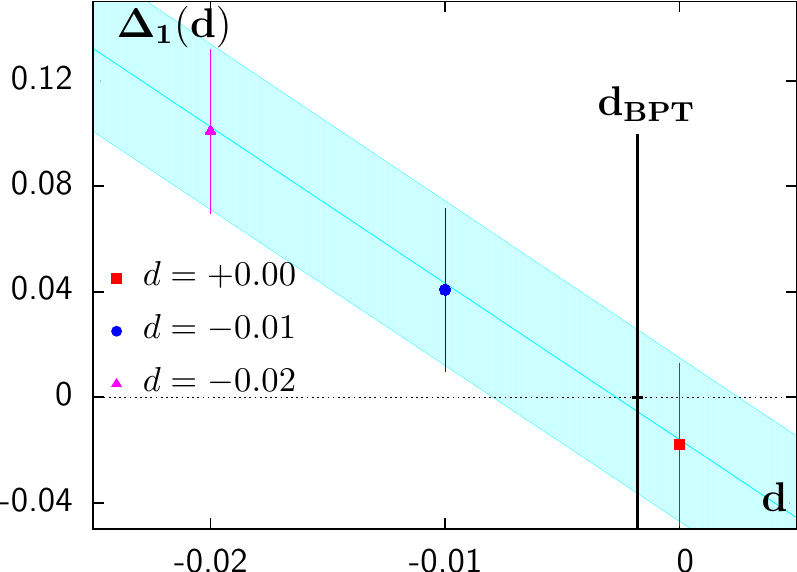}}
 \end{picture}
 \caption{Left: $ \Delta(M_{PS}^2) $ for $ c=c_M $ is linear in $ m_0 $. Right: The slope $ \Delta_1(d) $ is linear in $ d $ and vanishes close to $ d_{BPT} $. Symbols and data are explained in the text.
}
 \label{fig: residual mass anisotropy}
\end{figure}

\noindent
Extrema of $ \Delta(M_{PS}^2) $ on the $ 32^4 $~lattice with~$ \beta=6.0 $ and~$ d=0.0 $ are extrapolated towards the chiral limit. Linear and quadratic extrapolations of $ c_M $ are displayed in the left plot of figure~\ref{fig: c_M chiral extrapolation} and listed in table~\ref{tab: 32^4 c_M chiral limit}. The data set contains all masses and $ c \in [-0.65 , -0.2] $ from table~\ref{tab: lattice parameters} for fit ranges of $ t/a \in [9-23] $ for $ 32^4 $. The lightest mass ($ am_0=0.01 $) is not matched well in a linear extrapolation (blue line). However, curvatures of quadratic extrapolations (lilac line) seem too high as the extrapolation bends upwards at the highest mass ($ am_0=0.05 $). It is probably pure coincidence that its endpoint is very close to the perturbative estimate $ c_{BPT} $ (black burst). A linear extrapolation with the lightest mass left out yields slightly more negative $ c_M $, but still agrees with the extrapolation of the full sample within errors. The $ 48^4 $~lattice (open symbol at $ am_0=0.02 $, slightly shifted to the left for distinction of error bars) indicates more negative $ c_M $ (by $ \Delta^{\mathrm{fs}}_{c_M} \approx -0.01 $) than the $ 32^4 $~lattice (filled symbols). Thus, $ c_M $ for the lightest mass ($ am_0=0.01 $) agrees with the central value of the fit within combined statistical and estimated finite size errors. 
\noindent
Variation of $ d $ on the $ 32^4 $~lattice is shown in the right plot of figure \ref{fig: c_M chiral extrapolation}. Since results for $ d=0.0 $ (red squares), $ d=-0.01 $ (blue bullets) and $ d=-0.02 $ (lilac triangles) are consistent within fractions of the errors, the tuning condition $ c=c_M $ can be considered as independent of $ d $. \newline

\noindent
Once~$ c $ is tuned to~$ c_M $, the interpolating function for~$ \Delta(M_{PS}^2) $ yields the residual mass anisotropy, which is shown for a $ 32^4 $~lattice at~$ \beta=6.0 $ in the left plot of figure~\ref{fig: residual mass anisotropy}. The data set covers $ d \in [-0.08,+0.02] $ and is extrapolated linearly in the mass,
\begin{equation}
  \Delta(M_{PS}^2) = \Delta_0 + \Delta_1(d)\,m_0.
  \label{eq: residual mass anisotropy}
\end{equation}\normalsize
The crossing of lines at $ am_0>0 $ indicates that tuning does not remove the anisotropy completely. The slope parameter~$ \Delta_1(d) $ is linear in~$ d $ (\mbox{cf.}~right plot of figure~\ref{fig: residual mass anisotropy}) and is dominated by the $ d $~dependence of $ M_\parallel^2 $. The turquoise line is a regression line between $ d=0.0 $ and $ d=-0.02 $. The vertical line marks the BPT estimate ($ d_{BPT}=-0.00179 $).

\subsection{Dependence of the mass anisotropy on the gauge coupling}\label{sec: Dependence of the mass anisotropy on the gauge coupling}

\subsubsection{Finer lattices: $ \beta=6.2 $}\label{sec: Finer lattices}

\begin{figure}[hbt]
 \begin{picture}(360,140)
  \put(000.0, 0.0){\includegraphics[bb=0 0 170 130, scale=0.80]{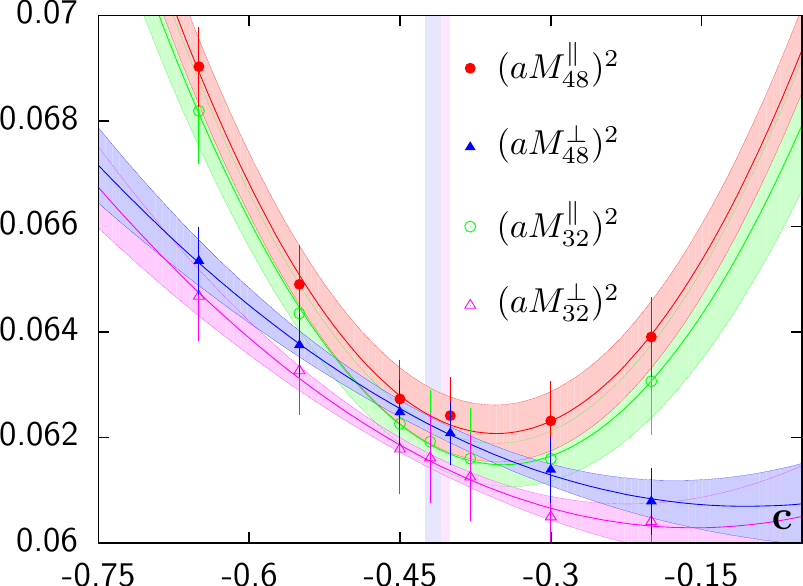}}
  \put(220.0, 0.0){\includegraphics[bb=0 0 170 130, scale=0.80]{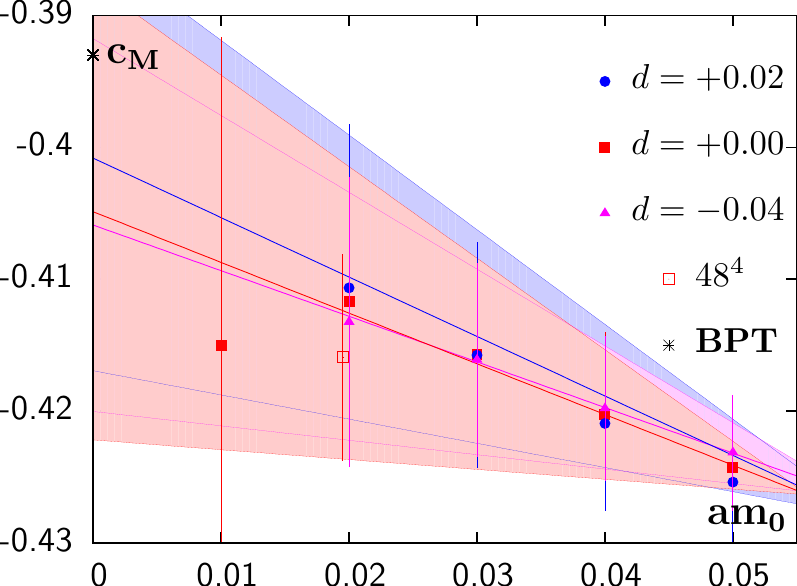}}
 \end{picture}
 \caption{
 Left: The mass anisotropy is consistent within errors. The pseudoscalar mass is $ M_{PS} \approx 725\,\mathrm{MeV} $ for $ c=c_M $. 
 Right: $ c_M $ is insensitive to $ d $ within errors.
 Symbols and data are explained in the text.
}
 \label{fig: beta=6.2 interpolation}
\end{figure}

\begin{figure}[hbt]
 \begin{picture}(360,140)
  \put(000.0, 0.0){\includegraphics[bb=0 0 170 130, scale=0.80]{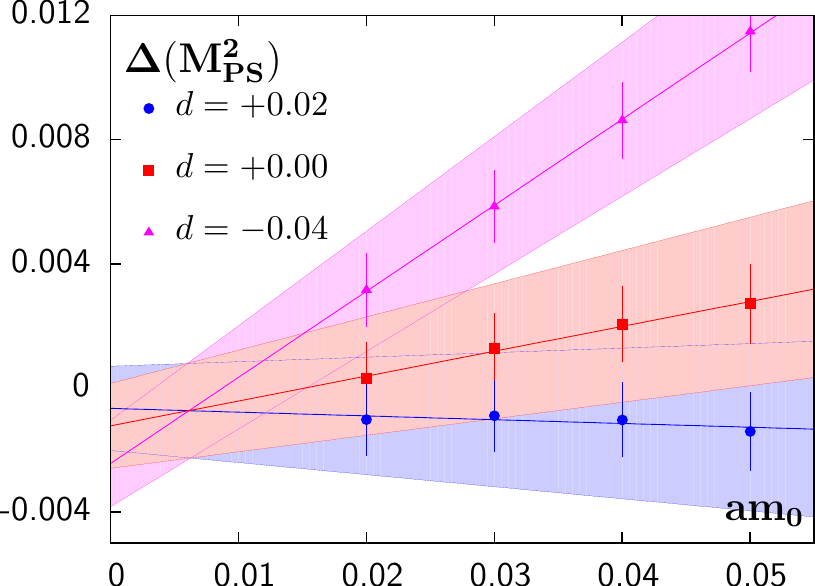}}
  \put(220.0, 0.0){\includegraphics[bb=0 0 170 130, scale=0.80]{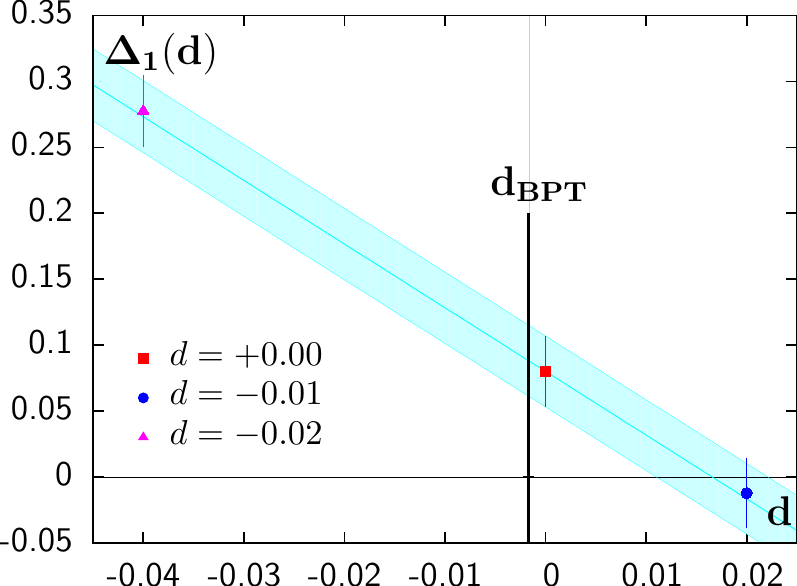}}
 \end{picture}
 \caption{Left: $ \Delta(M_{PS}^2) $ for $ c=c_M $ is linear in $ m_0 $. Right: The slope $ \Delta_1(d) $ is linear in $ d $ and vanishes close to $ d\approx+0.015 $. Symbols and data are explained in the text.
}
 \label{fig: beta=6.2 residual mass anisotropy}
\end{figure}

\noindent
The gauge coupling is changed to~$ \beta=6.2 $, which is closer to the continuum limit. The previous procedure for determination of the mass and interpolation using \mbox{eq.}~(\ref{eq: interpolation ansatz}) is repeated (\mbox{cf.}~interpolations for~$ am_0=0.02 $ in the left plot of figure~\ref{fig: beta=6.2 interpolation}). The smaller physical lattice size primarily affects perpendicular correlation functions and reasonable fit ranges are very short. Due to large variations of $ c_M $ for changes of the fit range, systematical uncertainties due to the fit range are estimated as $ \Delta_{c_M}^{\mathrm{fr}} \approx 0.02 $. A linear chiral extrapolation of~$ c_M $ is conducted on the $ 32^4 $~lattice with~$ \beta=6.2 $ as in section~\ref{sec: Chiral extrapolation of the minimum of the mass anisotropy}. Large statistical errors are observed and $ c_M $ is quite insensitive to~$ d $ (right plot of figure~\ref{fig: beta=6.2 interpolation}). The estimate $ c_{BPT} $ (black burst) lies within the error bands of $ c_M $. The $ 48^4 $~lattice (open symbol at $ am_0=0.02 $, slightly shifted to the left) suggests more negative values of $ c_M $ than the $ 32^4 $~lattice (filled symbols), which is also indicated by the two overlapping vertical bands (blue for $ 48^4$, lilac for $ 32^4 $) in the left plot.
Chiral behaviour of the residual mass anisotropy~$ \Delta(M_{PS}^2) $ for~$ c=c_M $ is described by \mbox{eq.}~(\ref{eq: residual mass anisotropy}) within errors. In the left plot of figure~\ref{fig: beta=6.2 residual mass anisotropy}, it is shown that $ \Delta_0 $ is consistent with zero but changes its sign with respect to~$ \beta=6.0 $. The right plot displays the linear regression of $ \Delta_1(d) $. The vertical line marks the BPT estimate.

\subsubsection{Coarser lattices: $ \beta=5.8 $}\label{sec: Coarser lattices}


\begin{figure}[hbt]
 \begin{picture}(360,140)
  \put(090.0, 0.0){\includegraphics[bb=0 0 170 130, scale=0.80]{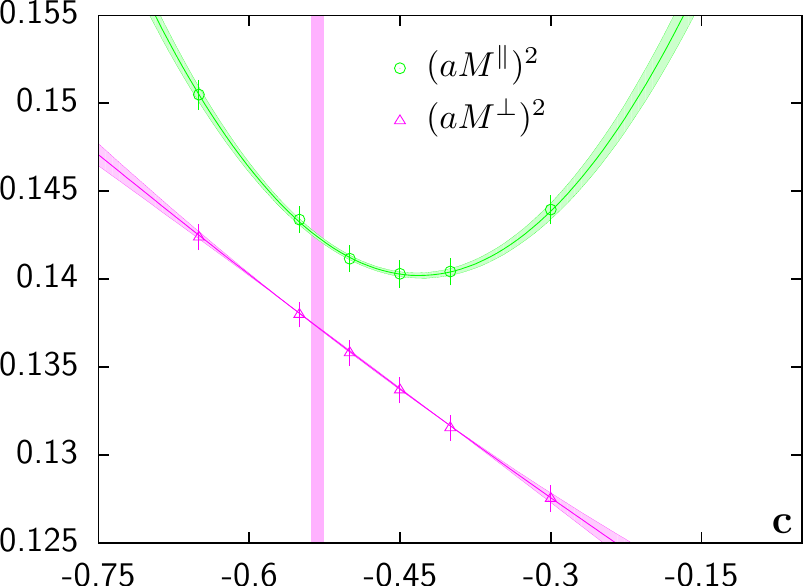}}
 \end{picture}
 \caption{
 The coarser lattice with $ \beta=5.8 $ is more anisotropic. Pseudoscalar masses for $ am_0=0.02 $ are $ M_{PS}^{\parallel} \approx 545\,\mathrm{MeV} $ and $ M_{PS}^{\perp} \approx 535\,\mathrm{MeV} $ for $ c=c_M $. Symbols and data sets are explained in the text below.
}
 \label{fig: beta=5.8 interpolation}
\end{figure}

\begin{figure}[hbt]
 \begin{picture}(360,140)
  \put(005.0, 0.0){\includegraphics[bb=0 0 170 130, scale=0.80]{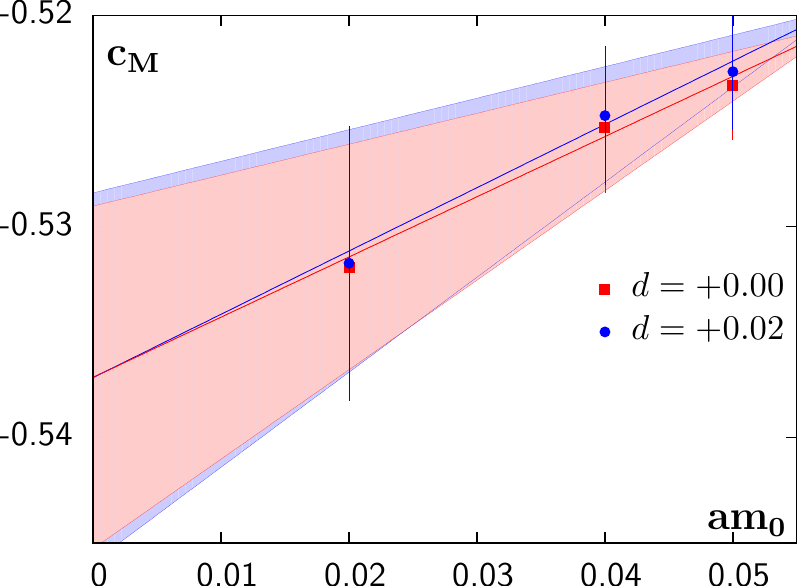}}
  \put(215.0, 0.0){\includegraphics[bb=0 0 170 130, scale=0.80]{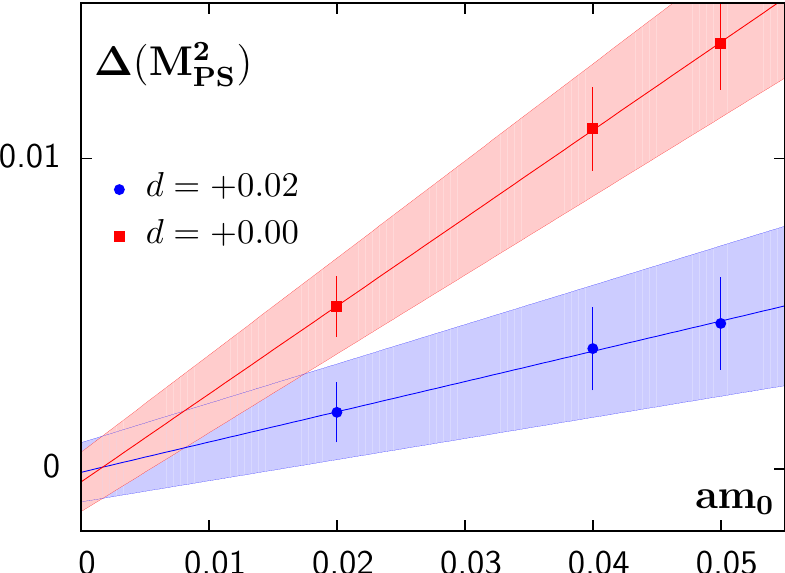}}
 \end{picture}
 \caption{
 Left: The chiral extrapolation of $ c_M $ has a positive slope. 
 Right: $ d_0\approx +0.03 $ would flat approach the chiral limit flatly. Symbols and data sets are explained below.
}
 \label{fig: beta=5.8 chiral extrapolation}
\end{figure}

\noindent
Figure~\ref{fig: beta=5.8 interpolation} shows that the mass anisotropy between parallel (green circles) and perpendicular (lilac triangles) masses is larger than for $ \beta=6.0 $. $ c_M $ is indicated by the lilac vertical band. Because data for larger lattices with $ \beta=5.8 $ is not available, uncertainties due to finite size of the lattice are estimated to be as large as for $ \beta=6.0 $ ($ \Delta^{\mathrm{fs}}_{c_M} \equiv c_M^{(48)}-c_M^{(32)} \approx -0.01 $ as defined in section~\ref{sec: Smaller lattices}). Significant dependendence on the fit range (for fits to the correlation function) is not observed.
\noindent
The left plot of figure \ref{fig: beta=5.8 chiral extrapolation} shows that $ c_M $ for $ d=+0.00 $ (red squares) and $ d=+0.02 $ (blue bullets) is independent of $ d $ within errors. The positive slope of the chiral extrapolation starkly differs from results for finer lattices. The estimate $ c_{BPT}=-0.454 $ is more than 15\% less negative than $ c_M $. The residual mass anisotropy $ \Delta(M_{PS}^2) $ suggests that the slope $ \Delta_1(d) $ of the approach to the chiral limit vanishes for $ d\approx+0.03 $. At present, no particular reason is known why trends in the chiral extrapolation for $ \beta=5.8 $ and $ \beta=6.0 $ or $ \beta=6.2 $ are opposite. However, data are compatible with $ c_M $ being independent of $ am_0 $.

\subsubsection{Qualitative aspects of gauge coupling dependence}\label{sec: Qualitative aspects of gauge coupling dependence}

\noindent
The non-perturbative value $ c_M $ is slightly more neagtive than the estimate $ c_{BPT} $. Unsurprisingly, this difference $ (c_{BPT}-c_M) $ increases for coarser lattices. $ c_M $ for all three gauge couplings is summarised in table~\ref{tab: c_M for all 3 couplings}. Dominant sources for systematical uncertainties are finite size effects, dependence on the fit range and the chiral extrapolation. Because the lightest mass is $ M_{PS} \approx 450\,\mathrm{MeV} $, the study of the anisotropy has not necessarily reached the chiral regime. 
Perpendicular correlation functions receive contributions from long-lived excited states. An increase of effective mass plateaus towards the midpoint of the lattice for source-smeared correlation functions suggests that these excited states may have negative spectral weights (\mbox{cf.}~section~\ref{sec: Perpendicular correlation functions}). Their contribution seems to become more important with detuning of $ c $. The associated systematical uncertainty is estimated to be similar in size to the other sources of systematical uncertainty.

\begin{table}[hbt]
 \center
 \begin{tabular}{|c|c|c|c|
  }
  \hline
  $ \beta $ & $ a\,[fm] $ & $ c_M $ & $ M_{PS} $ for $ c=c_M $
  \\
  \hline
  $ 5.8 $   & $ 0.136 $   & $ -0.537(8)(10) $  & $ 540\,\mathrm{MeV} $
  \\
  $ 6.0 $   & $ 0.093 $   & $ -0.437(7)(10) $  & $ 450\,\mathrm{MeV} $
  \\
  $ 6.2 $   & $ 0.068 $   & $ -0.405(17)(20) $ & $ 515\,\mathrm{MeV} $
  \\
  \hline
 \end{tabular}
 \caption{
 The format $ c_M(\delta)(\sigma) $ includes statistical and systematical errors. The last column indicates the lightest pseudoscalar mass in the chiral extrapolation of $ \Delta\left(M_{PS}^2\right) $.
 }
 \label{tab: c_M for all 3 couplings}
\end{table}\normalsize

\newpage
\sectionc{Oscillating correlation functions}{sec: Oscillating correlation functions}

\begin{table}[hbt]
\center\footnotesize
 \begin{tabular}{|c|c|c|c|c|c|c|c|}
  \hline
  $ \beta $ & $ a\,[\mathrm{fm}] $ & $ r_0 $ & $ T $ & $ n_{cfg} $ & $ am_0 $ & $ c $ & $ d $ \\
  \hline
  $ 6.0 $  & $ 0.093 $ & $ 5.368 $ & $ 48 $ & $ 20 $ & $ 0.02 $ &
  $ -0.45  $ & $ -0.001 $ \\
  \hline
  $ 6.2 $  & $ 0.068 $ & $ 7.360 $ & $ 128 $ & $ 10 $ & $ 0.00730 $ & 
  $ [-0.50,-0.30]  $ & $ -0.001 $ \\
  $ 6.0 $  & $ 0.093 $ & $ 5.368 $ & $ 128 $ & $ 10 $ & $ 0.01000 $ &
  $ [-0.55,-0.35] $ & $ -0.001 $ \\
  $ 5.8 $  & $ 0.136 $ & $ 3.668 $ & $ 128 $ & $ 10 $ & $ 0.01464 $ & 
  $ [-0.60,-0.40]  $ & $ -0.002 $ \\
  \hline
 \end{tabular}
 \caption{Oscillating correlation functions are studied with $ L=24 $. 
 }
 \label{tab: lattices for oscillation study}
\end{table}\normalsize

\noindent
Pseudoscalar correlation functions in the $ \gamma^5 $ channel do not exhibit visible oscillations. Nevertheless, oscillations are observed in many other channels of \textit{parallel correlation functions} for $ q\bar q $ states, while they are not observed in the same channels of \textit{perpendicular correlation functions} for $ q\bar q $ states. The direction of correlation is parallel to the Karsten-Wilczek term for parallel correlation functions and perpendicular to the Karsten-Wilczek term for perpendicular correlation functions.  Moreover, the frequency of the oscillations depends on the relevant counterterm's coefficient $ c $. Hence, the aim of this section is to clarify the origin and physical relevance of these oscillations without making explicit use of the decomposition that is defined in section~\ref{sec: Decomposition into a pair of fields} by answering the following questions:
\begin{enumerate}
  \item Are the states of the QCD spectrum observed in channels that contain oscillations?
  \item Are the differences between parallel and perpendicular correlation functions a result of residual anisotropies that can be removed with better tuning?
  \item Is it possibile to use the $ c $ dependence of the frequency for non-perturbative tuning?
\end{enumerate}
The first question is answered by comparing the ground state masses of the $ J=0 $ sector for Karsten-Wilczek fermions with tuned parameters to the ground state masses for Wilson fermions, which are known to correctly reproduce the spectrum of QCD. 
The second question is answered by a comparison to correlation functions for na\"{\i}ve fermions, which exhibit a similar pattern of oscillations without being anisotropic. Though na\"{\i}ve fermions are unphysical, they show a pattern of oscillating correlation functions that is similar to both tuned parallel Karsten-Wilczek fermions and staggered fermions~\cite{Altmeyer:1992dd}. This pattern is due to the presence of an additional pole of the quark propagators in the direction of correlation. Na\"{\i}ve fermions are realised with the routines for Karsten-Wilczek fermions by setting the Wilczek parameter~$ \zeta $ and all counterterms to zero and require no additional coding effort. 
\noindent
In section~\ref{sec: Na\"{i}ve and Wilson fermions}, the na\"{i}ve fermion action of \mbox{eq.}~(\ref{eq: naive fermion action}) is contrasted with the Wilson fermion action, which includes \mbox{eq.}~(\ref{eq: Wilson operator}). Next, approximately tuned \textit{parallel and perpendicular Karsten-Wilzcek fermions}\footnote{Parallel and perpendicular Karsten-Wilzcek fermions amount to calculation of parallel or perpendicular correlation functions (direction of correlation wrt alignment of the Karsten-Wilczek term).} are juxtaposed in section~\ref{sec: Oscillations of Karsten-Wilczek fermions}. The Wilczek parameter is set to $ \zeta=+1 $ and the counterterm coefficients are tuned to~$ c=-0.45 $ and~$ d=-0.001 $. Contrary to the approach in section~\ref{sec: Anisotropy of hadronic quantities}, perpendicular Karsten-Wilczek fermions are implemented by changing the direction of the Karsten-Wilczek term ($ \underline{\alpha}=3 $) while keeping the direction of correlation in the $ \hat{e}_t=\hat{e}_0 $ direction. All of these correlation functions are computed on~$ 20 $ configurations of a $ 48\times24^3 $~lattice at~$ \beta=6.0 $. This approach has the benefit of a longer time direction despite the reduction of the numerical cost due to smaller spatial volumes. This longer time direction is advantageous for isolating the ground state, in particular for perpendicular correlation functions. Moreover, local fluctuations of gauge links along the time direction are the same for both parallel and perpendicular correlation functions. The mass splitting between $ \gamma^5 $ channels for both parallel and perpendicular Karsten-Wilzcek fermions is consistent with zero and indicates that anisotropies discussed in section~\ref{sec: Anisotropy of hadronic quantities} have been removed even for lattices with $ T \neq L $.
The third question in the list is anwered in section~\ref{sec: Tuning with the frequency spectrum}, where a non-perturbative tuning condition is derived from a Fourier analysis of ratios of correlation functions. 

\subsection{Na\"{i}ve and Wilson fermions}\label{sec: Na\"{i}ve and Wilson fermions}

\noindent
Correlation functions for Wilson fermions use~$ am_0=-0.788 $. The bare mass parameter takes the critical mass~$ am_{cr}=-0.808 $ of Wilson fermions into account, which is determined as~$ am_0=1/(2\kappa)-4 $ from the critical hopping parameter~$ \kappa_{cr}=0.157131(9) $ for unimproved Wilson fermions at~$ \beta=6.0 $ that was taken from~\cite{Gupta:1996sa}. 
\noindent
Correlation functions for na\"{i}ve fermions use~$ am_0=0.02 $. Free na\"{i}ve fermions have 16 real fermion modes localised in the Brillouin zone at the na\"{i}ve doublers of \mbox{eq.}~(\ref{eq: naive doublers}) and satisfy a residual chiral symmetry. However, na\"{\i}ve fermions do not reproduce the spectrum of QCD due to their unphysically large number of degenerate fermions. Though the bare parameters of both actions correspond to a bare quark mass of $ am_q \approx 0.02 $, it must be kept in mind that use of the same bare mass parameter does not imply matching hadron masses for different fermion actions. 
The displayed absolute values of correlation functions are calculated with source-smeared interpolating operators. $ \mathcal{C}(n_0)\geq0 $ is indicated as red squares and $ \mathcal{C}(n_0)<0 $ as blue bullets in the following figures. The time direction in the following figures is always the $ \hat{e}_0 $~direction. Ground state masses of these correlation functions are summarised in table~\ref{tab: fit masses of the spin-0 sector} at the end. \newline

\begin{figure}[htb]
 \begin{picture}(360,250)
  \put(005.0,128.0){\includegraphics[bb=0 0 180 113, scale=0.75]{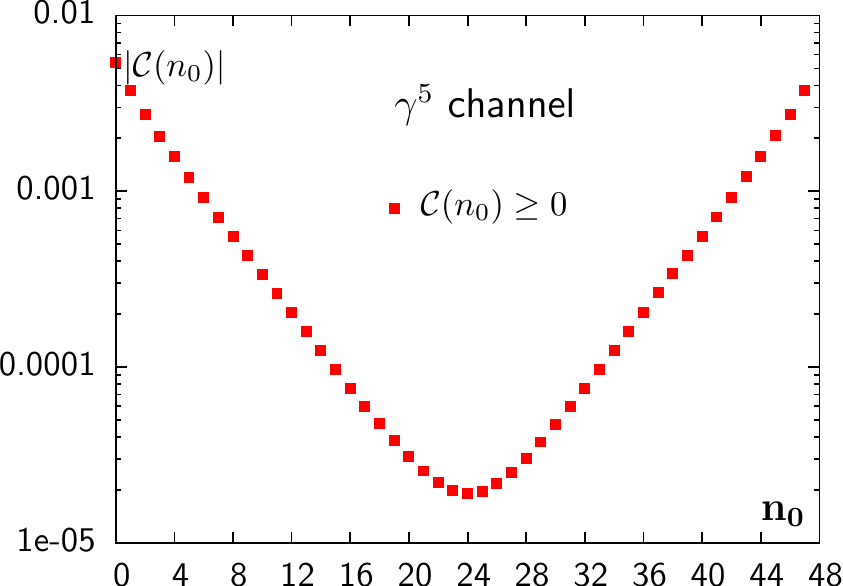}}
  \put(215.0,128.0){\includegraphics[bb=0 0 180 113, scale=0.75]{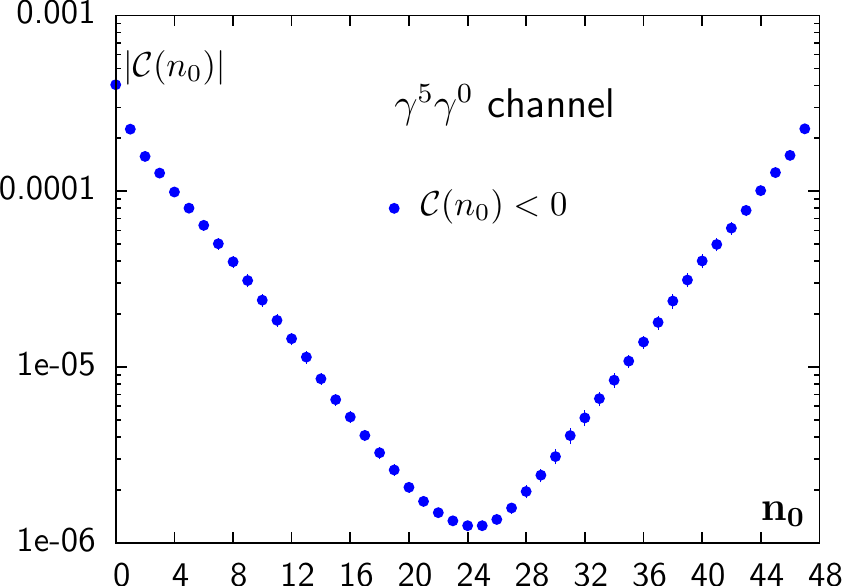}}
  \put(005.0,  0.0){\includegraphics[bb=0 0 180 113, scale=0.75]{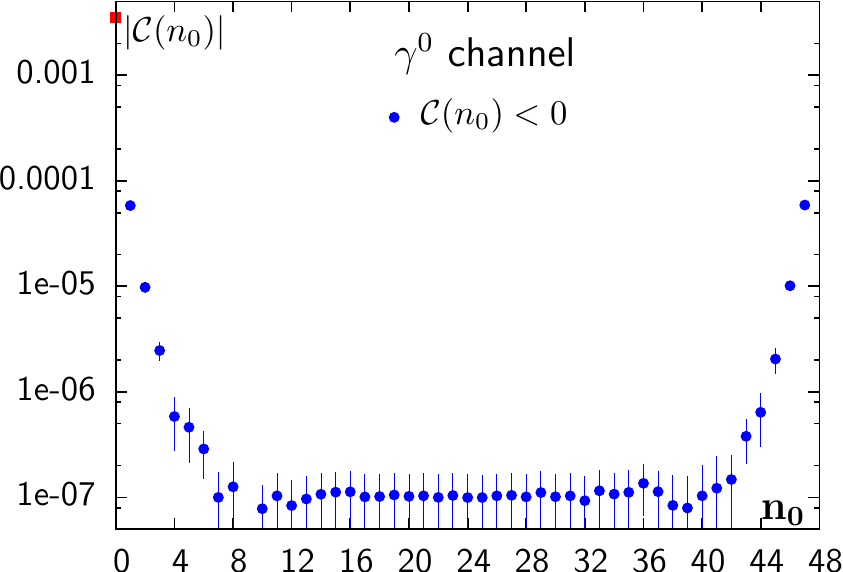}}
  \put(215.0,  0.0){\includegraphics[bb=0 0 180 113, scale=0.75]{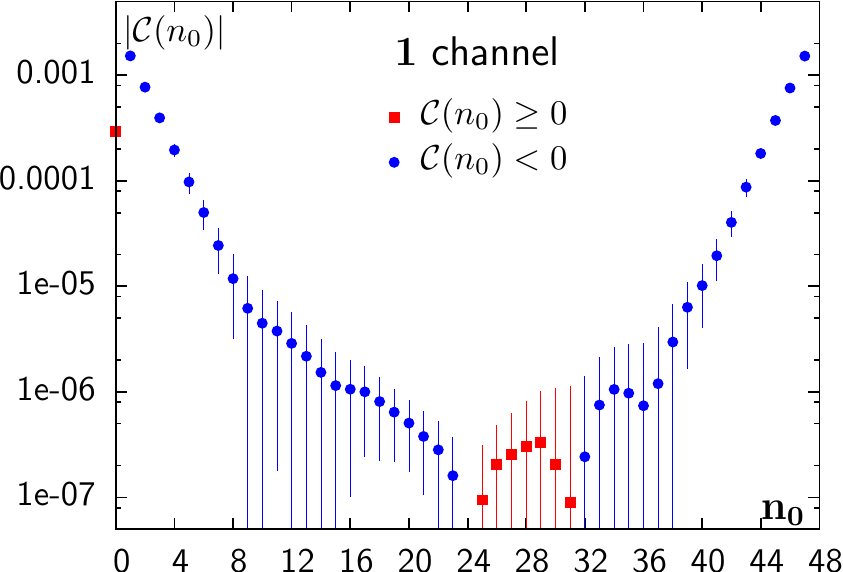}}
 \end{picture}
 \caption{Upper row: $ \gamma^5 $ (left) and $ \gamma^5\gamma^0 $~channels (right) are mass degenerate for Wilson fermions.
 Lower row: For Wilson fermions, the $ \gamma^0 $-channel (left) contains only noise, whereas the $ \mathbf{1} $-channel (right) contains low-lying scalar states.
  }
 \label{fig: Wilson correlators}
\end{figure}


\noindent
Correlation functions for Wilson fermions in the $ \gamma^5 $ and $ \gamma^5 \gamma^0 $~channels are displayed in the upper row of figure~\ref{fig: Wilson correlators}. 
Both quark bilinears $ \bar \psi \gamma^5 \psi $ and $ \bar\psi \gamma^5 \gamma^0 \psi $ have $ J^{PC}=0^{-+} $ and have the same ground state mass within statistical errors. Correlation functions for Wilson fermions in the $ \gamma^0 $ and $ \mathbf{1} $~channels are shown in the lower row of figure~\ref{fig: Wilson correlators}. Whereas the quark bilinear $ \bar \psi \mathbf{1} \psi $ has $ J^{PC}=0^{++} $, the other quark bilinear $ \bar \psi \gamma^0 \psi $ has $ J^{PC}=0^{+-} $. As there are no physical mesons with $ J^{PC}=0^{+-} $, no statistically significant signal is seen in the $ \gamma^0 $ channel and the correlation function rapidly plummets into statistical noise that is consistent with zero. The correlation function in the $ \mathbf{1} $ channel shows a clear signal of a state that is heavier than the ground state of the $ \gamma^5 $ channel. The change of sign for $ n_0 \in [25,31] $ is just a fluctuation that consistent with zero within errors. These observations reflect the meson spectrum of QCD. \newline

\begin{figure}[htb]
 \begin{picture}(360,250)
  \put(005.0,128.0){\includegraphics[bb=0 0 180 113, scale=0.75]{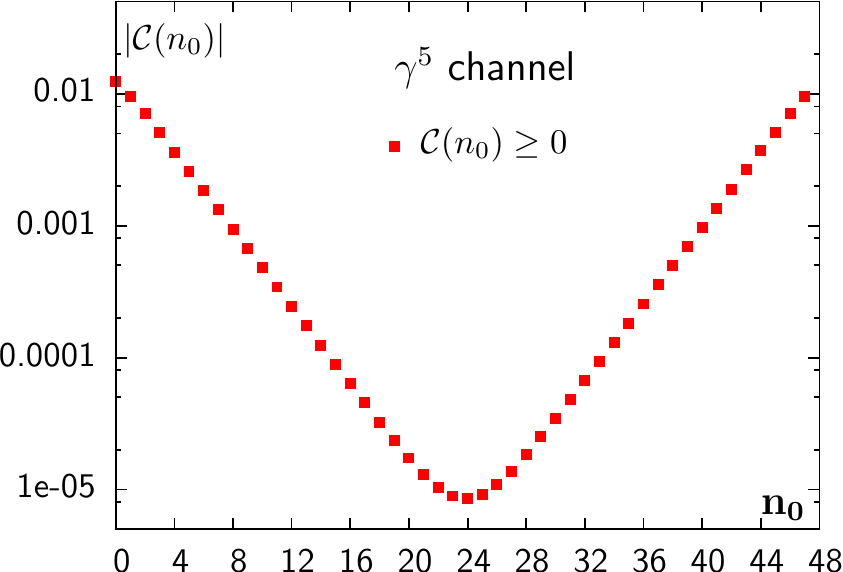}}
  \put(215.0,128.0){\includegraphics[bb=0 0 180 113, scale=0.75]{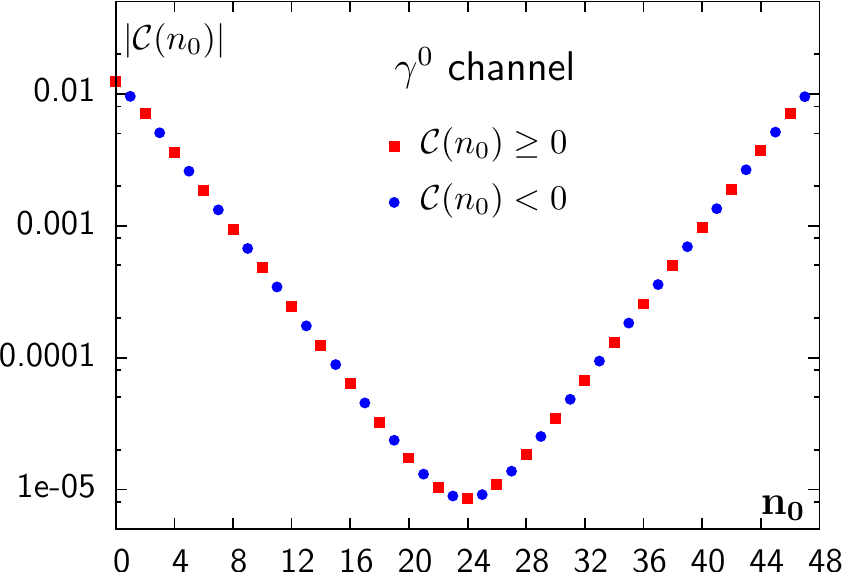}}
  \put(005.0,  0.0){\includegraphics[bb=0 0 180 113, scale=0.75]{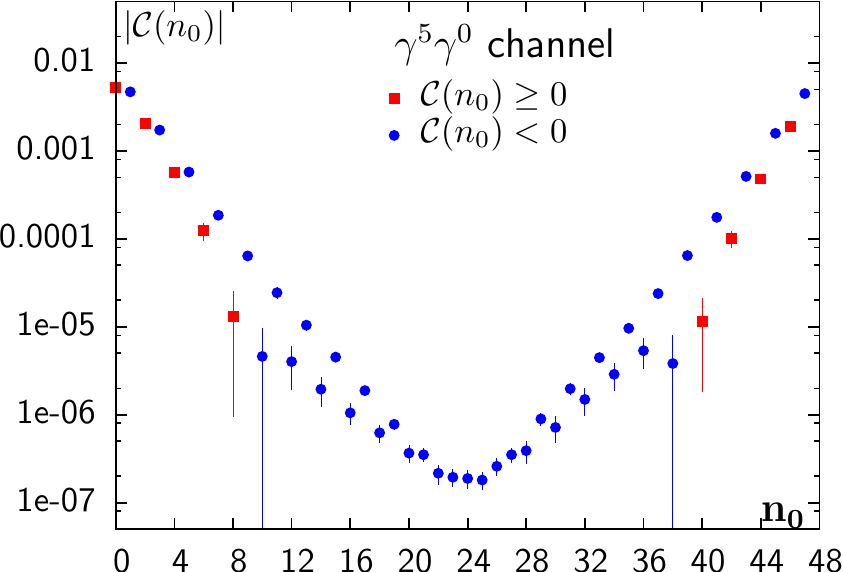}}
  \put(215.0,  0.0){\includegraphics[bb=0 0 180 113, scale=0.75]{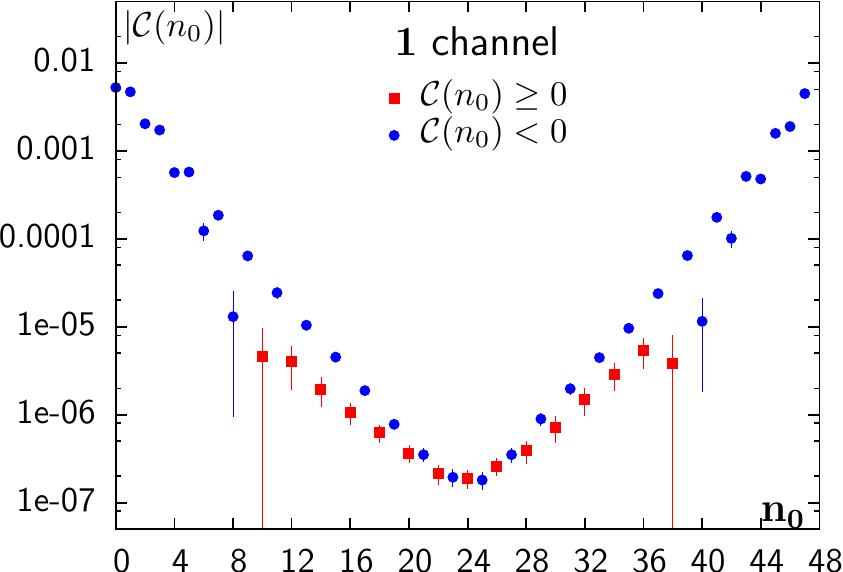}}
 \end{picture}
 \caption{Upper row: $ \gamma^5 $ (left) and $ \gamma^0 $ channels (right) have identical magnitude for na\"{i}ve fermions.
 Lower row: $ \gamma^5\gamma^0 $ (left) or $ \mathbf{1} $ channels contain superpositions of oscillating and non-oscillating terms for na\"{i}ve fermions. 
  }
 \label{fig: naive correlators}
\end{figure}


\noindent
Correlation functions for na\"{i}ve fermions in the $ \gamma^5 $ or $ \gamma^0 $~channels are displayed in the upper row of figure~\ref{fig: naive correlators}. As there are no physical mesons with $ J^{PC}=0^{+-} $, no statistically significant signal is expected for the $ \gamma^0 $ channel. However, the empirical observation is that the absolute value of both correlation functions agrees at machine precision and their ratio satisfies on all time slices (up to machine precision)
\begin{equation}
  R_{05}(n_0) \equiv R_{\gamma^0,\gamma^5}(n_0) = \frac{\mathcal{C}_{\gamma^0,\gamma^0}(n_0)}{\mathcal{C}_{\gamma^5,\gamma^5}(n_0)} = (-1)^{n_0}.
  \label{eq: R_05 correlator ratio}
\end{equation}\normalsize
Hence, there is an additional oscillating contribution in the $ \gamma^0 $~channel, which correponds to states that are mass-degenerate with states in the $ \gamma^5 $~channel (and have the same spectral weights). For na\"{\i}ve fermions, the spectra of the interpolating operators $ \bar \psi \gamma^5 \psi $ and $ \bar\psi \gamma^0 \psi $ are degenerate, but the non-oscillating contribution in the $ \gamma^5 $ channel is alternating in the other. \newline
\noindent
\noindent
Most other correlation functions are more complicated. An example is presented in the lower row of figure~\ref{fig: naive correlators}. Without any regard to a physical interpretation, the correlation functions have a form which appears to be a superposition of two different contributions on odd and even time slices or a superposition of a non-oscillating and an alternating contribution. The latter assessment is supported by the observation that the ratio of correlators in $ \gamma^5\gamma^0 $~and $ \mathbf{1} $~channels satisfies (up to machine precision) the same empirical relation up to a global factor $ (-1) $,
\begin{equation}
  R_{\mathbf{1},\gamma^5\gamma^0}(n_0) = \frac{\mathcal{C}_{\mathbf{1},\mathbf{1}}(n_0)}{\mathcal{C}_{\gamma^5\gamma^0,\gamma^5\gamma^0}(n_0)} = -(-1)^{n_0}.
\end{equation}\normalsize
Thus, for na\"{\i}ve fermions, the interpolating operators $ \bar \psi \gamma^5\gamma^0 \psi $ and $ \bar\psi \mathbf{1} \psi $ have degenerate spectra and the non-oscillating contributions in either of the channels are the alternating contributions in the other. \newline
\noindent 
It is known from Wilson fermions that both $ \bar \psi \gamma^5 \psi $ and $ \bar \psi \gamma^5\gamma^0 \psi $ generate $ 0^{-+} $ states. Therefore, if there were an exclusive relation between interpolating operators for na\"{\i}ve fermions and $ J^{PC} $, both interpolating operators would have to generate the same $ 0^{-+} $ states. Spectral weights would be different and even masses may differ due to lattice artefacts. 
However, it is evident from the correlation functions that the spectra of both interpolating operators generate different states in the oscillating contribution. 
Hence, an exclusive relation between interpolating operators for na\"{\i}ve fermions and $ J^{PC} $ cannot apply and their spectra may include different $ J^{PC} $. Thus, it is concluded that both $ \bar \psi \gamma^5\gamma^0 \psi $ and  $ \bar \psi \mathbf{1} \psi $ generate $ 0^{-+} $~and $ 0^{++} $~states. Since both $ J^{PC} $ are relevant for physical mesons, both the non-oscillating and the alternating contribution are observed in the correlation function.
This conclusion is consistent with the observation for the $ \gamma^5 $~and $ \gamma^0 $~channels (the $ 0^{+-} $ contribution is only noise) and can be generalised even further. For every channel of $ q\bar q $ states, an empirical correpondence to another channel is observed (up to machine precision):
\begin{align}
  R_{\mathcal{M}_\pm,\gamma^5\gamma^0\mathcal{M}_\pm}^{}(n_0)  =& \pm (-1)^{n_0}, &
  \left\{\begin{array}{lr}
  \mathcal{M}_+ \in& \left\{\gamma^5,\gamma^0,\gamma^5\gamma^i,\gamma^0\gamma^i \right\} \\ 
  \mathcal{M}_- \in& \left\{ \mathbf{1},\gamma^5\gamma^0,\gamma^i,\gamma^j\gamma^k \right\}
  \end{array}\right..
  \label{eq: naive conjugate correlator ratio}
\end{align}\normalsize

\subsection{Oscillations of Karsten-Wilczek fermions}\label{sec: Oscillations of Karsten-Wilczek fermions}

\begin{figure}[htb]
 \begin{picture}(360,250)
  \put(005.0,128.0){\includegraphics[bb=0 0 180 113, scale=0.75]{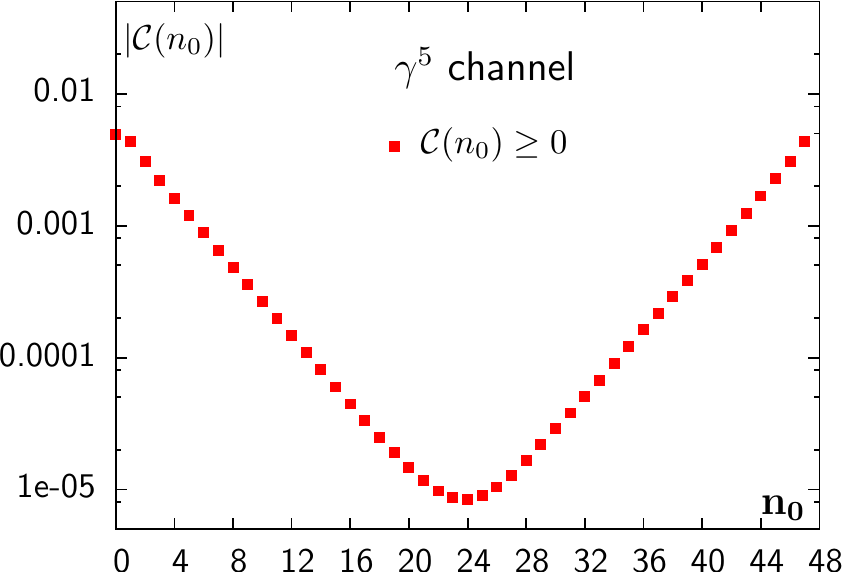}}
  \put(215.0,128.0){\includegraphics[bb=0 0 180 113, scale=0.75]{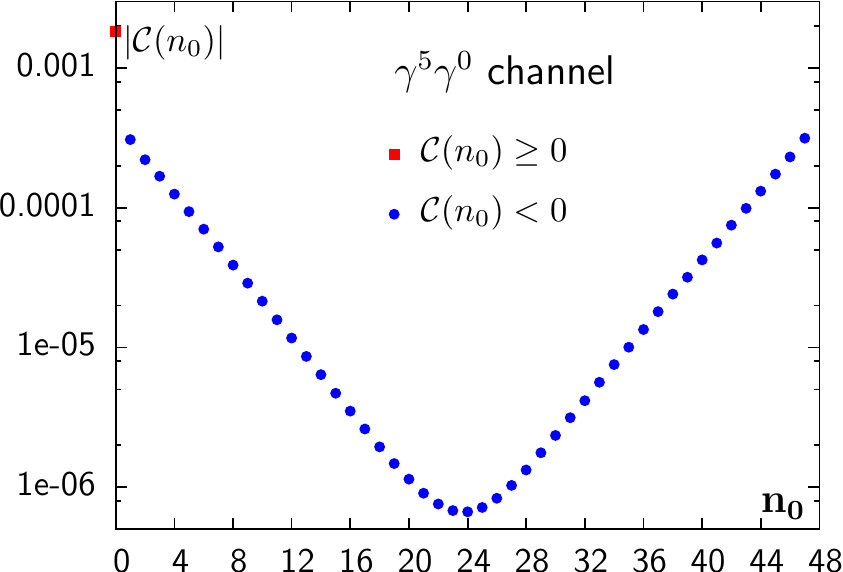}}
  \put(005.0,  0.0){\includegraphics[bb=0 0 180 113, scale=0.75]{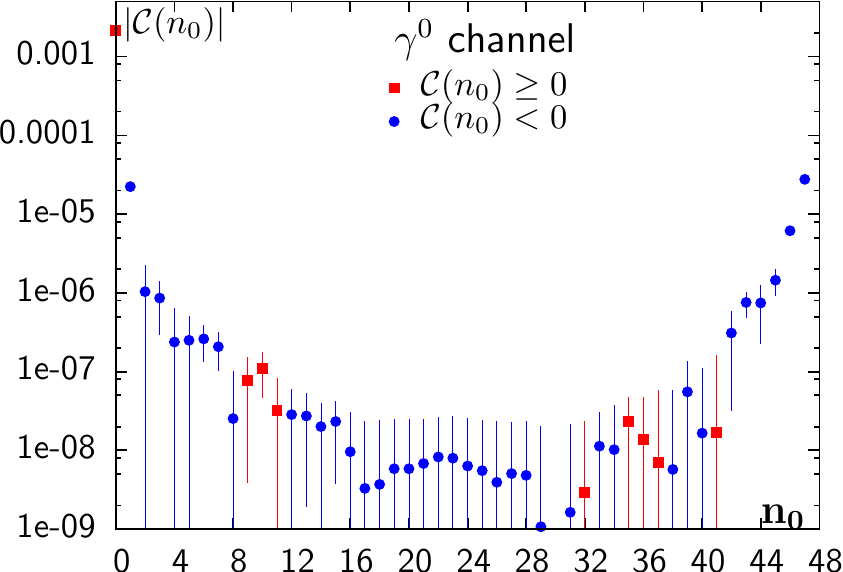}}
  \put(215.0,  0.0){\includegraphics[bb=0 0 180 113, scale=0.75]{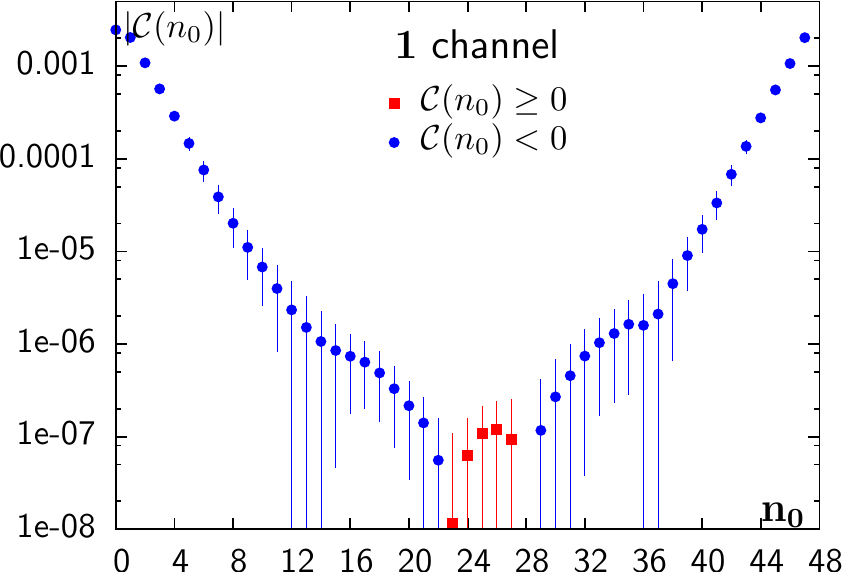}}
 \end{picture}
 \caption{Upper row: $ \gamma^5 $ (left) and $ \gamma^5\gamma^0 $ channels (right) are mass degenerate for perpendicular Karsten-Wilczek fermions.
 Lower row: The $ \gamma^0 $~channel (left) contains only noise and the $ \mathbf{1} $~channel (right) contains scalar states for perpendicular Karsten-Wilczek fermions.
  }
 \label{fig: KW perp correlators}
\end{figure}


\noindent
Perpendicular Karsten-Wilczek fermions yield mass degenerate ground states in $ \gamma^5 $~and $ \gamma^5\gamma^0 $~channels (\mbox{cf.}~upper row of figure~\ref{fig: KW perp correlators}). The $ \gamma^0 $~channel contains only statistical noise, while the $ \mathbf{1} $ channel exhibits a low-lying scalar state (\mbox{cf.}~lower row of figure~\ref{fig: KW perp correlators}). Thus, perpendicular Karsten-Wilczek fermions seem to closely resemble Wilson fermions. It appears appropriate to assign the same $ J^{PC} $. \newline

\begin{figure}[htb]
 \begin{picture}(360,250)
  \put(005.0,128.0){\includegraphics[bb=0 0 180 113, scale=0.75]{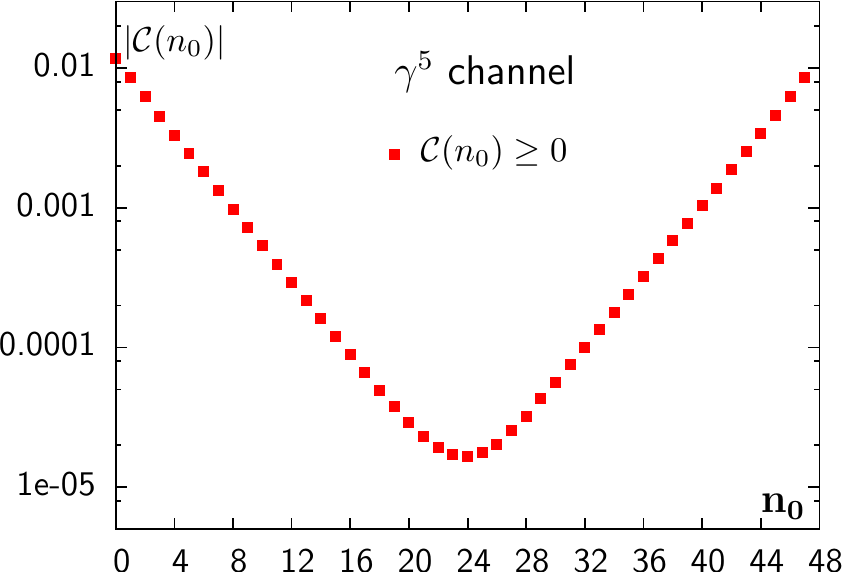}}
  \put(215.0,128.0){\includegraphics[bb=0 0 180 113, scale=0.75]{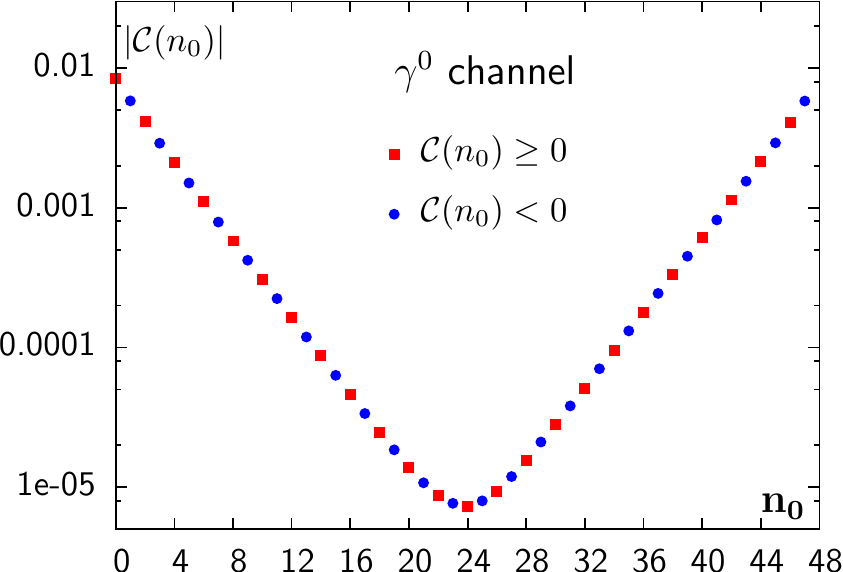}}
  \put(005.0,  0.0){\includegraphics[bb=0 0 180 113, scale=0.75]{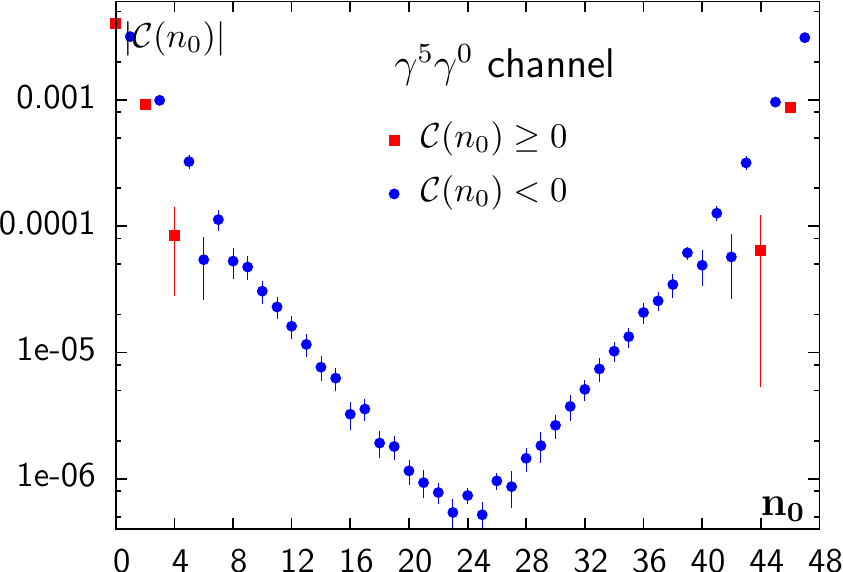}}
  \put(215.0,  0.0){\includegraphics[bb=0 0 180 113, scale=0.75]{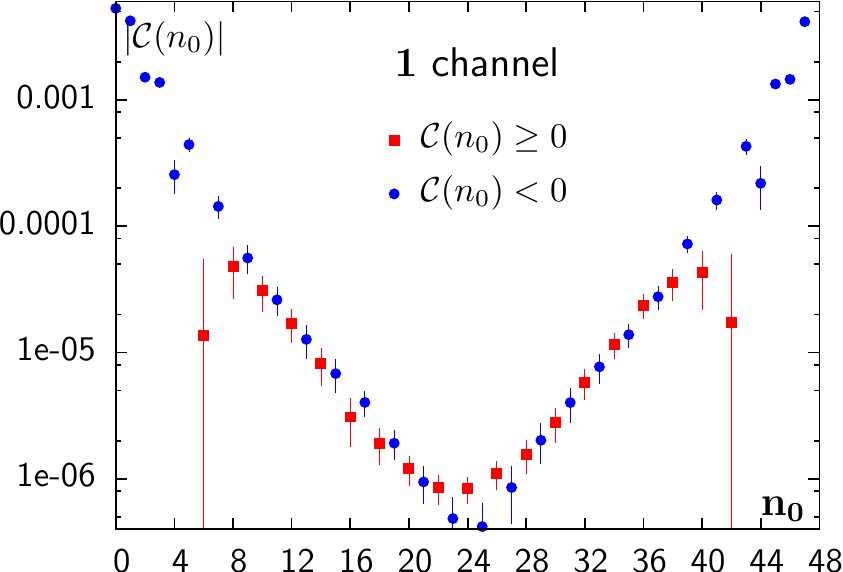}}
 \end{picture}
 \caption{Upper row: $ \gamma^5 $ (left) and $ \gamma^0 $~channels (right) have similar but different magnitude for parallel Karsten-Wilczek fermions. The masses are different.
 Lower row: $ \gamma^5\gamma^0 $ (left) or $ \mathbf{1} $~channels show both parity partners for parallel Karsten-Wilczek fermions.
 }
 \label{fig: KW parallel correlators}
\end{figure}


\noindent
Parallel Karsten-Wilczek correlation functions in~$ \gamma^5 $~and $ \gamma^0 $~channels are displayed in the upper row of figure~\ref{fig: KW parallel correlators} and exhibit striking similarity to na\"{i}ve correlation functions in figure~\ref{fig: naive correlators}. There are no visible oscillations in the $ \gamma^5 $~channel and only oscillating terms in the $ \gamma^0 $~channel. Nonetheless,~$ \gamma^5 $ and $ \gamma^0 $~channels for Karsten-Wilczek fermions do not have a common ground state mass. These observations would be expected if interpolating operators for Karsten-Wilczek fermions would overlap with states with the same $ J^{PC} $ as interpolating operators for na\"{\i}ve fermons. The ground state of the $ \gamma^0 $~channel would be considered to have $ J^{PC}=0^{-+} $ and the mass splitting would suggest that gauge fields discriminate between states with the same $ J^{PC} $ in different channels.
\noindent
Furthermore, the observed similarity of~$ \gamma^5\gamma^0 $ and $ \mathbf{1} $~channels for na\"{\i}ve fermions is preserved for parallel Karsten-Wilczek fermions. These channels are shown in the lower row of figure~\ref{fig: KW parallel correlators}. 
The oscillating terms in the $ \gamma^5\gamma^0 $~channel are observed only close to the boundaries ($ n_0\leq 8 $ or $ n_0\geq 40 $) and quickly fall to amplitudes close to noise level. On the contrary, the oscillations are long-lived in the $ \mathbf{1} $~channel. These observations lead to the conclusion that the ground state of the $ \mathbf{1} $~channel must belong to the oscillating contribution, while the ground state of the $ \gamma^5\gamma^0 $~channel must belong to the non-oscillating contribution. These observations would be expected if interpolating operators for Karsten-Wilczek fermions would overlap with states with the same $ J^{PC} $ as interpolating operators for na\"{\i}ve fermons. The ground state of both channels would have $ J^{PC}=0^{-+} $ and would belong to the non-oscillating contribution in the $ \gamma^5\gamma^0 $~channel and to the oscillating contribution in the $ \mathbf{1} $~channel. Because the other contributions in both channels would have $ J^{PC}=0^{++} $, their signal would be much weaker.

\begin{figure}[htb]
 \begin{picture}(360,120)
  \put(140.0, 0.0){\includegraphics[bb=0 0 180 113, scale=0.75]{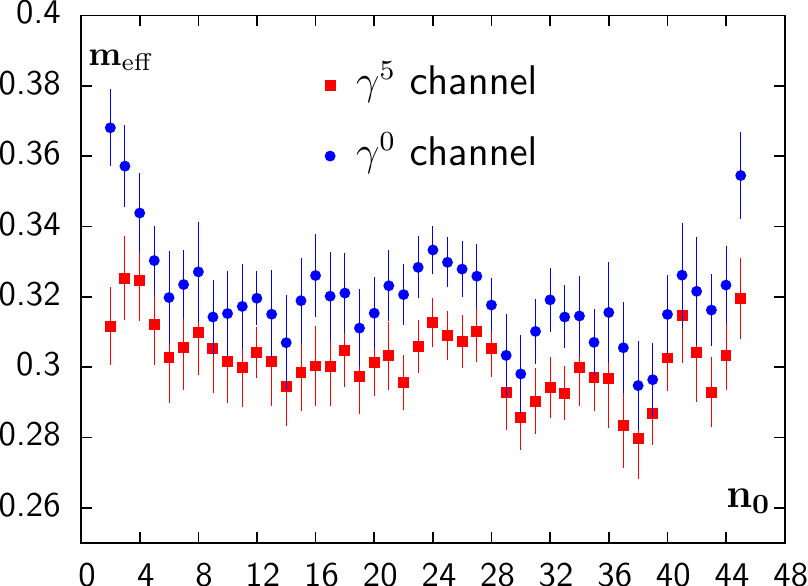}}
 \end{picture}
 \caption{There is a mass splitting between cosh masses of $ \gamma^0 $~ and $ \gamma^5 $~channels. Symbols and data sets are explained in the text.
  }
 \label{fig: KW parallel 55 and 00 effective mass}
\end{figure}

\noindent
Figure \ref{fig: KW parallel 55 and 00 effective mass} displays local effective masses for $ \gamma^0 $~ (blue bullets) and $ \gamma^5 $~channels (red squares). The former ($ aM_{00} $) exceeds the latter ($ aM_{55} $) consistently by approximately one or two standard errors over the full range of the effective mass plateau and precisely follows suit in any fluctuations of the plateau value. The mass splitting is
\begin{equation}
 aM_{00}-aM_{55} \approx 0.02
 \label{eq: M00-M55 mass spliting}
\end{equation}\normalsize
and amounts to ten standard errors on the level of the fit mass. This smallness of the mass splitting raises the question whether the splitting decreases or increases for finer lattices. A decreasing splitting would be interpreted as a lattice artefact and the $ \gamma^0 $~channel's ground state would have to be considered as a physical state that is mass-degenerate with the $ \gamma^5 $~channel's ground state in the continuum limit. An increasing splitting would have to be considered as an indication that the $ \gamma^0 $~channel's ground state were an unphysical state. This question is answered in a detailed study of the chiral behaviour and continuum extrapolation of both ground state masses in section~\ref{sec: Chiral behaviour of the pseudoscalar ground state}. 

\begin{table}[htb]
 \center
 \begin{tabular}{|c|c|c|c|c|}
  \hline
  Action           & $ aM_{55} $ & $ aM_{00} $ & $ aM_{50\,50} $  & $ aM_{\mathbf{1}\mathbf{1}} $
  \\
  \hline
  Na\"{i}ve        
  & $ 0.339(1) $ & $ 0.339(1) $ & $ 0.381(21) $ & $ 0.383(22) $
  \\
  KW $ \parallel $ 
  & $ 0.299(2) $ & $ 0.319(2) $ & $ 0.333(12) $ & $ 0.332(23) $
  \\
  KW $ \perp $     
  & $ 0.299(2) $ &              & $ 0.301(2) $  &              
  \\
  Wilson           
  & $ 0.254(2) $ &              & $ 0.261(5) $  &              
  \\
  \hline
 \end{tabular}
 \caption{Pseudoscalar fit masses $ aM_{55} $ are obtained using $ \gamma^5 $, $ aM_{50\,50} $ using $ \gamma^5\gamma^0 $, $ aM_{00} $ using $ \gamma^0 $ and $ aM_{\mathbf{1}\mathbf{1}} $ using $ \mathbf{1} $ in the interpolating operators. The data set of table~\ref{tab: lattices for oscillation study} is used and parameters are explained in the text. $ aM_{55} $ corresponds to $ M_{PS}\approx 630\,\mathrm{MeV} $ for Karsten-Wilczek fermions and $ M_{PS}\approx 540\,\mathrm{MeV} $ for Wilson fermions.
 }
 \label{tab: fit masses of the spin-0 sector}
\end{table}

\noindent
Fits to the correlation function for na\"{\i}ve and parallel Karsten-Wilczek fermions are performed with the ansatz
\begin{equation}
  \mathcal{C}(t) = \frac{A_0}{2M_0} \left(e^{-M_0 \cdot t} + e^{-M_0 \cdot (T-t)}\right) +\frac{\widetilde{A}_0}{2\widetilde{M}_0} (-1)^{t/a}\left(e^{-\widetilde{M}_0 \cdot t} + e^{-\widetilde{M}_0 \cdot (T-t)}\right),
  \label{eq: two state correlation function}
\end{equation}\normalsize
where $ A_0,M_0,\widetilde{A}_0,\widetilde{M}_0 $ are four independent parameters. For the $ \gamma^5 $~or $ \gamma^0 $~channels, where only either the non-oscillating or the oscillating contribution is observed, one term in the ansatz of \mbox{eq.}~(\ref{eq: two state correlation function}) is omitted in the fit function. Because they are considered as different $ J^{PC} $, both contributions each have their own ground state. 
Fit masses of pseudoscalar states for na\"{i}ve, parallel as well as perpendicular Karsten-Wilczek and Wilson fermions are summarised in table~\ref{tab: fit masses of the spin-0 sector}. It is noted again that the common bare quark mass parameter does not imply similar hadron masses because different fermion actions are used. The following empirical symmetry patterns are observed:
\begin{itemize}
  \item For Wilson and perpendicular Karsten-Wilczek fermions, the following are common observations: The $ \gamma^5 $~and $ \gamma^5\gamma^0 $~channels are mass degenerate (within errors) and both identified as $ 0^{-+} $. The $ \mathbf{1} $~channel is heavier than $ 0^{-+} $ and identified as $ 0^{++} $. The $ \gamma^0 $~channel contains only noise, because $ 0^{+-} $ is not realised in physical mesons.
  \item For na\"{\i}ve and parallel Karsten-Wilczek fermions, the following are common observations: Oscillating and non-oscillating contributions are observed in most channels. The non-oscillating contributions are interpreted as the same $ J^{PC} $ that is identified for Wilson fermions in the same channel. $ J^{PC} $ for the oscillating contributions is deduced by multiplying the Dirac matrix of the interpolating operators by $ \gamma^5\gamma^0 $. Thus, the $ \gamma^5 $~channel contains no visible oscillations and is interpreted as $ 0^{-+} $. The $ \gamma^0 $~channel contains only oscillating contributions that are also interpreted as $ 0^{-+} $. Contributions due to $ 0^{+-} $ in both channels may exist, but are indistinduishable from noise. The $ \gamma^5\gamma^0 $~and $ \mathbf{1} $~channels are interpreted as containing both $ 0^{-+} $ and $ 0^{++} $.
  For the $ \gamma^5\gamma^0 $~channel, non-oscillating contributions are interpreted as $ 0^{-+} $. For the $ \mathbf{1} $~channel, non-oscillating contributions are interpreted as $ 0^{++} $. The oscillating contribution is interpreted as the other $ J^{PC} $.
  \item For parallel and perpendicular Karsten-Wilczek fermions the following is observed: The $ \gamma^5 $~channels for both and the $ \gamma^5\gamma^0 $~channel for perpendicular Karsten-Wilczek fermions are mass degenerate (within errors). This indicates that the action is tuned correctly even for $ T \neq L $. The masses of $ \gamma^0 $, $ \gamma^5\gamma^0 $ and $ \mathbf{1} $~channels for parallel Karsten-Wilczek fermions are a few percent larger than the $ \gamma^5 $~channel's mass. The mass in the $ \gamma^0 $~channel is slightly lower than in $ \gamma^5\gamma^0 $ and $ \mathbf{1} $~channels, but all three masses are consistent within errors. The larger errors in the $ \gamma^5\gamma^0 $ and $ \mathbf{1} $~channels are a consequence of the oscillating contribution that generally enlarges the uncertainty of the fits. 
  \item For na\"{\i}ve fermions: Correlation functions satisfy the empirical relation of \mbox{eq.}~(\ref{eq: naive conjugate correlator ratio}) at machine precision. Therefore, both the pair of $ \gamma^5 $~and $ \gamma^0 $~channels and the pair of $ \gamma^5\gamma^0 $~and $ \mathbf{1} $~channels are mass-degenerate pairs. The mass splitting between the $ \gamma^5 $~and $ \gamma^5\gamma^0 $~channels is larger than for parallel Karsten-Wilczek fermions.
\end{itemize}
The observation of oscillations for na\"{\i}ve and parallel Karsten-Wilczek fermions in the same channels make it clear that oscillations are not due to anisotropy. As the anisotropy of the pseudoscalar mass is succesfully removed, incomplete tuning is not responsible for the oscillating terms. These oscillating terms are generated by the presence of the Dirac operator's extra pole in the direction of correlation and mass splittings between states with the same $ J^{PC} $ in different channels persist in the tuned theory.

\subsection{Tuning with the frequency spectrum}\label{sec: Tuning with the frequency spectrum}

\begin{figure}[htb]
 \begin{picture}(360,120)
  \put(005.0, 0.0){\includegraphics[bb=0 0 180 113, scale=0.75]{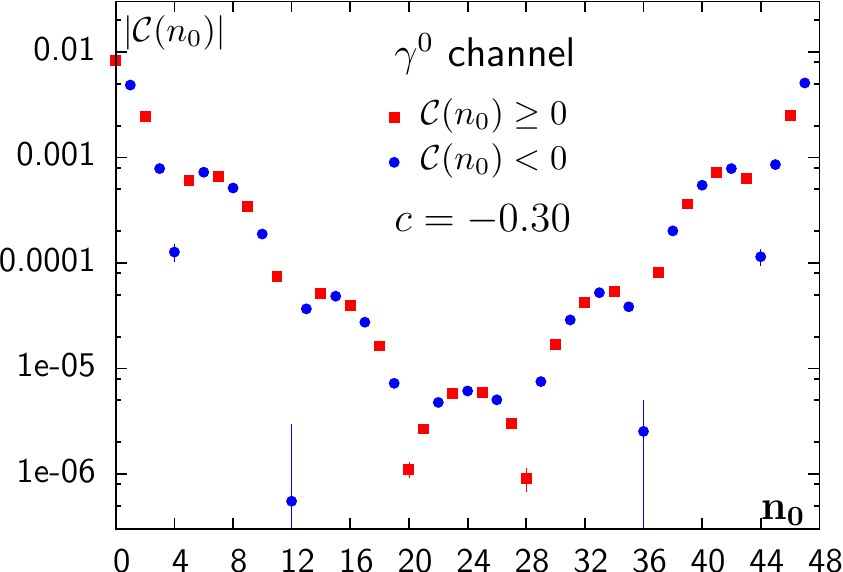}}
  \put(215.0, 0.0){\includegraphics[bb=0 0 180 113, scale=0.75]{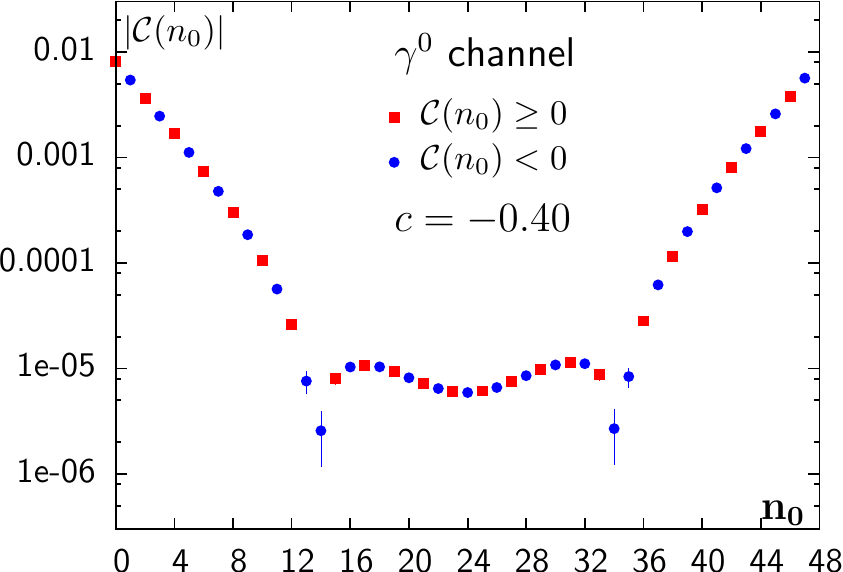}}
 \end{picture}
 \caption{
 The $ \gamma^0 $ channel exhibits oscillations whose frequency depends on the parameter $ c $. Data sets are explained in the text below.
  }
 \label{fig: KW 00 c dependence}
\end{figure}

\noindent
Correlation functions for approximately tuned parallel Karsten-Wilczek fermions (section~\ref{sec: Oscillations of Karsten-Wilczek fermions}) exhibit alternating terms in the same channels as correlation functions for na\"{i}ve fermions  (section~\ref{sec: Na\"{i}ve and Wilson fermions}). However, these patterns vary as the coefficients of the counterterms are changed, which is shown in figure~\ref{fig: KW 00 c dependence}. The depicted correlation functions in the $ \gamma^0 $~channel for $ c=-0.30 $ (left) and $ c=-0.40 $ (right) use $ \beta=6.0 $, $ am_0=0.02 $ and $ d=0.0 $ on a $ 48^4 $~lattice of table~\ref{tab: lattice parameters} and clearly show that the pattern of oscillations depends on the choice of the relevant counterterm's coefficient $ c $. These patterns have to be contrasted with approximately tuned Karsten-Wilczek fermions ($ c=-0.45 $ for $ \beta=6.0 $) that are shown in the upper right plot in figure~\ref{fig: KW parallel correlators}. The rapid oscillation of the parallel correlation function for approximately tuned Karsten-Wilczek fermions, which is described well by a factor $ (-1)^{n_0} $, is modified for detuned Karsten-Wilczek fermions. Thus, the question whether the change in oscillation patterns can serve as a criterion for tuning the relevant counterterm's coefficient.
\noindent
Figure~\ref{fig: KW 00 c dependence} indicates that the modification of the oscillation's pattern becomes more pronounced as $ c $ is detuned further. Hence, the oscillation's frequency $ \Omega $ can be parameterised as
\begin{equation}
  \Omega = \pi + \omega_c,\quad \partial \omega_c / \partial c \text{ finite},
\end{equation}\normalsize
where the \textit{frequency shift} $ \omega_c $ is a smooth function of $ c $ that vanishes in the tuned theory. Its variation may serve as an observable indicating a mismatch $ \delta c $. $ c_0 $ is defined as the value of $ c $ that restores the frequency spectrum to its tree level form (the \textit{tuned frequency spectrum}) so that the coefficient $ c $ can be written as $ c=c_0+\delta_c $. If $ \omega_c $ vanishes, the mismatch $ \delta c $ is zero and the coefficient is tuned correctly to $ c_0=c(g_0) $. Though there is no apparent link between a full restoration of hypercubic symmetry and the frequency spectrum, it is clear that neither restoration of hypercubic symmetry nor restoration of the frequency spectrum is possible unless $ c $ is tuned correctly. Instead of attempting a fit to an oscillating correlation function, where the frequency is an unknown parameter, methods of Fourier analysis are applied to a ratio of correlation functions. The frequency distribution of lattice eigenfrequencies is peaked at the oscillation frequencies $ \Omega $, even though $ \Omega $ does not have to be an eigenfrequency of the lattice. Two simple toy models in appendix~\ref{app: Oscillating lattice toy models} elucidate this concept. Of course, the resolution of a discrete Fourier transform is limited by the width of the frequency bins,
\begin{equation}
  \omega_b = \frac{2\pi}{N_0},
  \label{eq: spectral bin size}
\end{equation}\normalsize
which depends only on $ N_0 $, the temporal extent of the lattice. If $ |\omega_c| < \omega_b/2 $, the time direction is too short to provide sufficient resolution of the spectrum. Thus, $ N_0 $ should be as large as possible and lattices of table~\ref{tab: lattices for oscillation study} with very long time direction ($ N_0=128 $) are applied for this purpose.

\begin{figure}[htb]
 \begin{picture}(400,295)
  \put(005  ,200.0){\includegraphics[bb=0 0 180 90, scale=0.55]{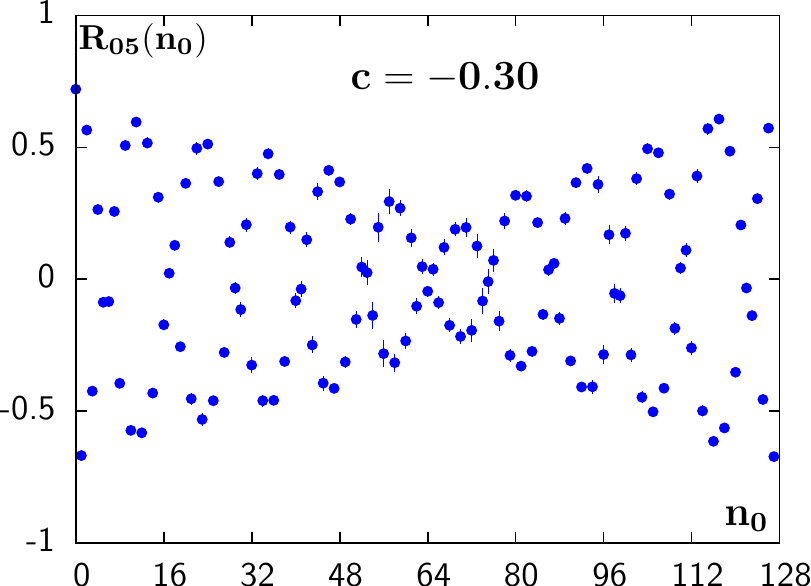}}
  \put(145.0,200.0){\includegraphics[bb=0 0 180 90, scale=0.55]{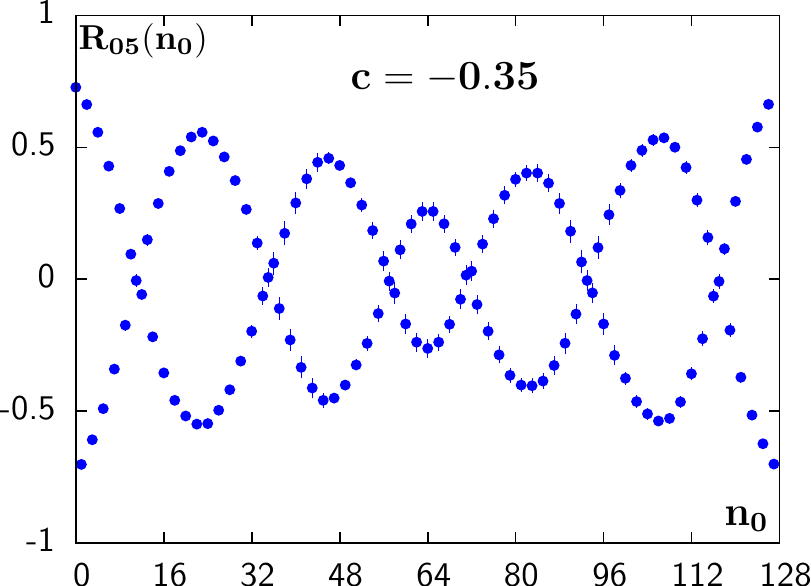}}
  \put(285  ,200.0){\includegraphics[bb=0 0 180 90, scale=0.55]{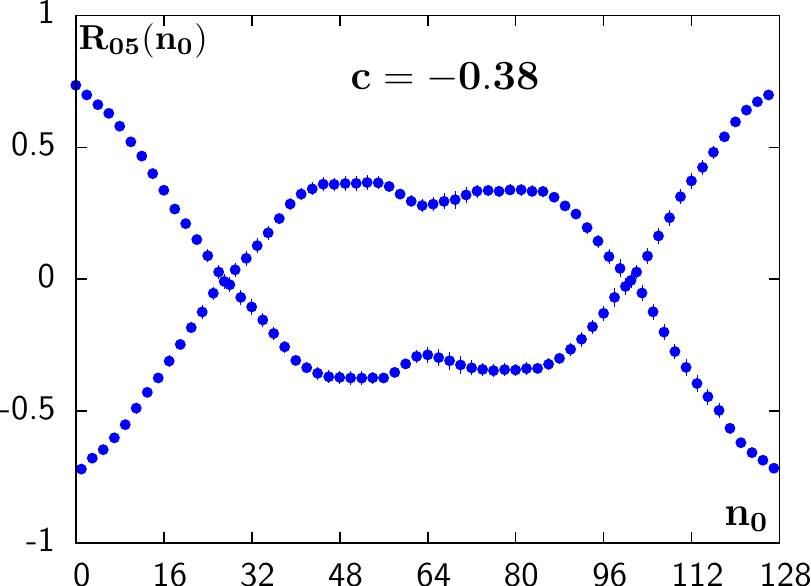}}
  \put(145.0,100.0){\includegraphics[bb=0 0 180 90, scale=0.55]{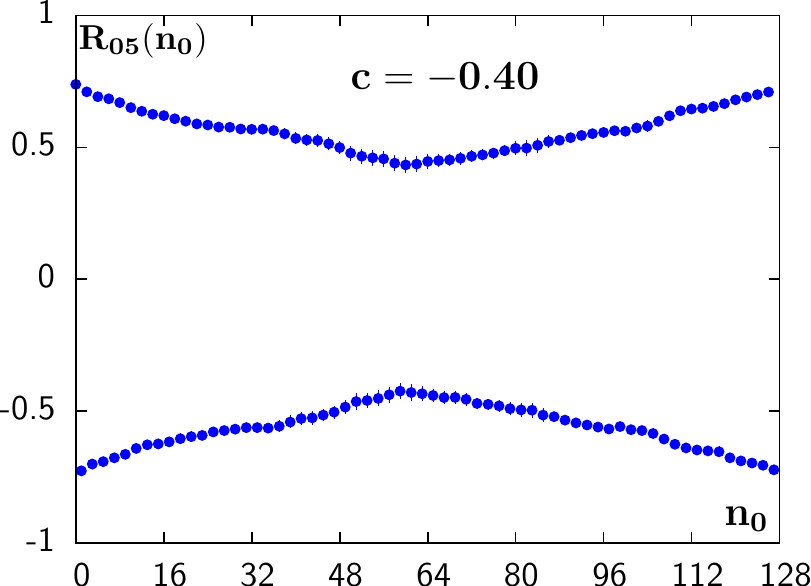}}
  \put(005  ,000.0){\includegraphics[bb=0 0 180 90, scale=0.55]{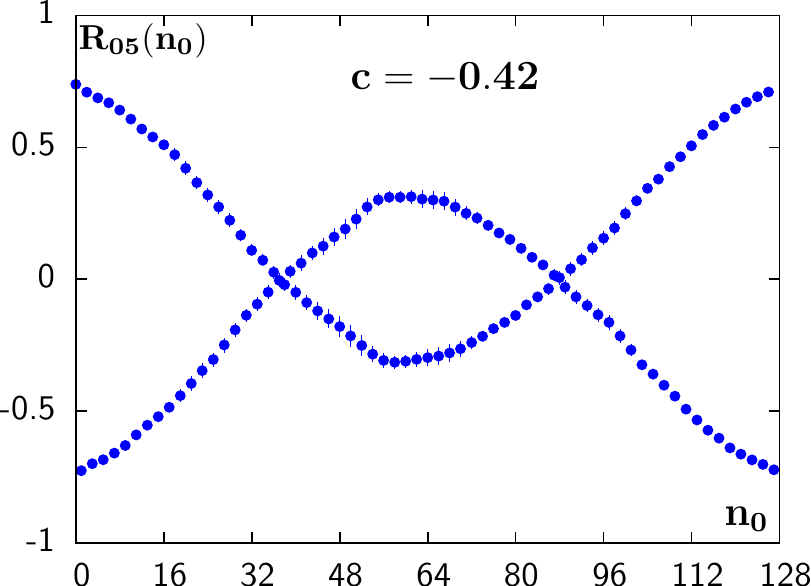}}
  \put(145.0,000.0){\includegraphics[bb=0 0 180 90, scale=0.55]{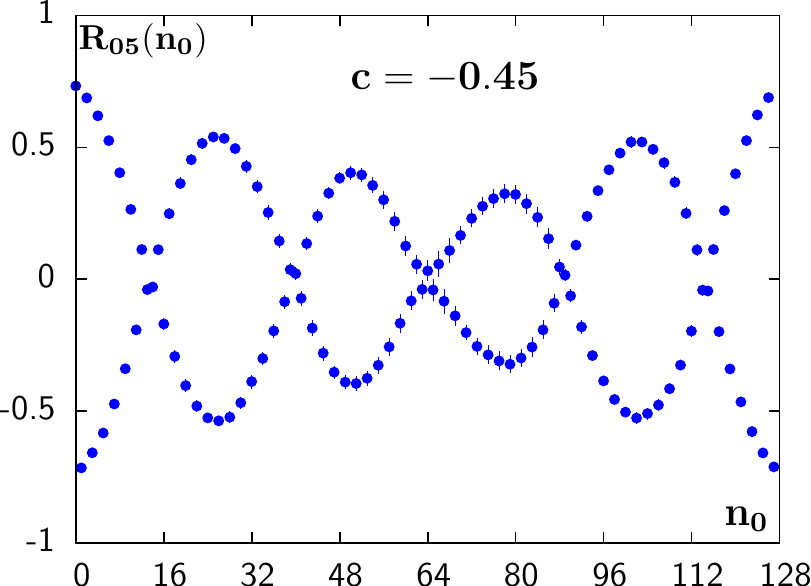}}
  \put(285.0,000.0){\includegraphics[bb=0 0 180 90, scale=0.55]{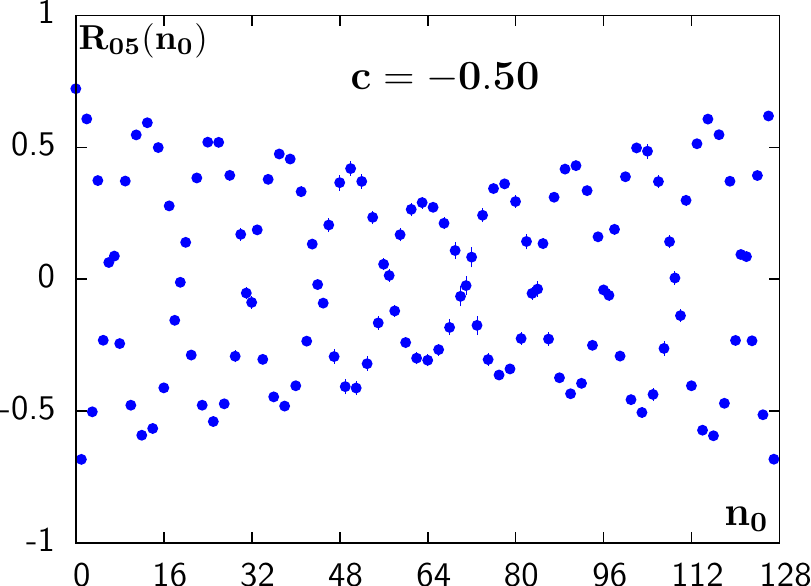}}
 \end{picture}
 \caption{Correlator ratios~$ R_{05}(n_0) $ on a $ 128\times24^3 $~lattice at~$ \beta=6.2 $ from table~\ref{tab: lattices for oscillation study} provide unmistakable evidence of the $ c $ dependence of the frequency shift $ \omega_c $.
  }
 \label{fig: ratio R_05 at beta 6.2}
\end{figure}
\noindent
It is observed in \mbox{eq.}~(\ref{eq: M00-M55 mass spliting} that the mass splitting between $ \gamma^0 $~and $ \gamma^5 $~channels is very small. Therefore, the leading order quark mass dependence cancels in the ratio~$ R_{05}(n_0) $ of these correlators, which is defined in \mbox{eq.}~(\ref{eq: R_05 correlator ratio}) for na\"{\i}ve fermions. Though \mbox{eq.}~(\ref{eq: R_05 correlator ratio}) is satified only approximately for tuned parallel Karsten-Wilczek fermions due to the small mass splitting, the residual decay of the correlator ratios (due to the mass splitting) is smaller for finer lattices. Because peaks in the frequency spectrum are narrower for a smaller residual decay, the following discussion is focused on $ \beta=6.2 $. However, it must be kept in mind that both correlation functions ($ \gamma^0 $~and $ \gamma^5 $~channels) have different excited states. Hence, even if the mass splitting between the ground states were zero, there would still be a residual decay due to the presence of different excited states. Seven ratios for~$ \beta=6.2 $ and $ c\in [-0.30,-0.50] $ are shown in figure~\ref{fig: ratio R_05 at beta 6.2}. $ c_M=-0.405(17)(20) $, the non-perturbative value that is obtained from the mass anisotropy in section~\ref{sec: Anisotropy of hadronic quantities}, agrees within errors with the parameter $ c=-0.40 $ of the central plot. Correlator ratios~$ R_{05}(n_0) $ for~$ \beta=6.0 $ and~$ \beta=5.8 $ are shown in figures~\ref{fig: ratio R_05 at beta 6.0} and~\ref{fig: ratio R_05 at beta 5.8} in appendix~\ref{app: Simulation parameters and data sets}. These ratios contain stronger exponential decays due to larger ground state mass differences. 

\begin{figure}[htb]
 \begin{picture}(360,280)
  \put(30.0, 0.0){\includegraphics[bb=0 0 220 113, scale=1.60]{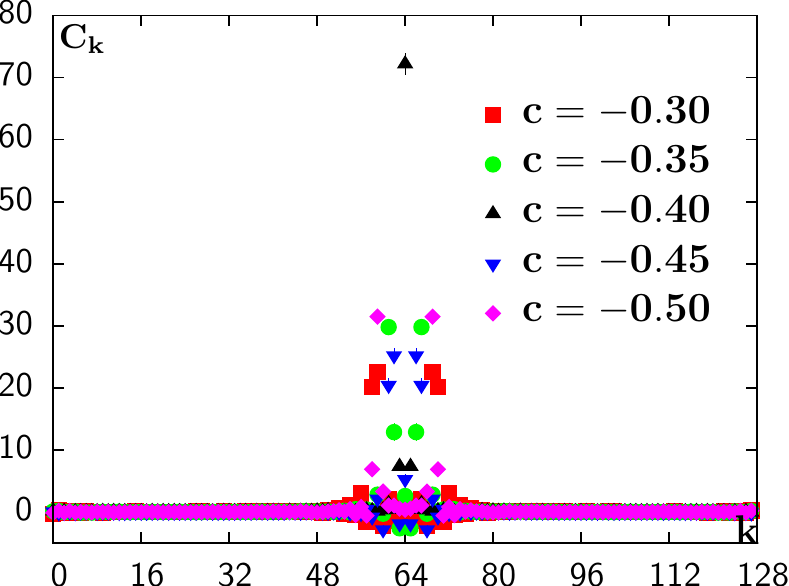}}
 \end{picture}
 \caption{
 The two peaks of the Fourier spectrum of the correlator ratio~$ \mathcal{R}_{05}(n_0) $ approach a single peak at $ k=N_0/2 $ if the coefficient $ c $ is tuned properly.
  }
 \label{fig: frequency spectra}
\end{figure}

\begin{figure}[htb]
 \begin{picture}(400,295)
  \put(005  ,200.0){\includegraphics[bb=0 0 180 90, scale=0.55]{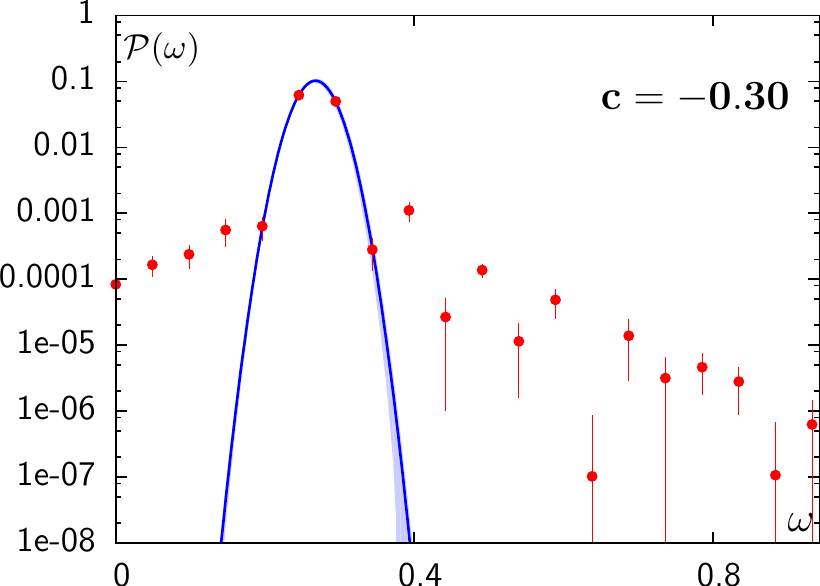}}
  \put(145.0,200.0){\includegraphics[bb=0 0 180 90, scale=0.55]{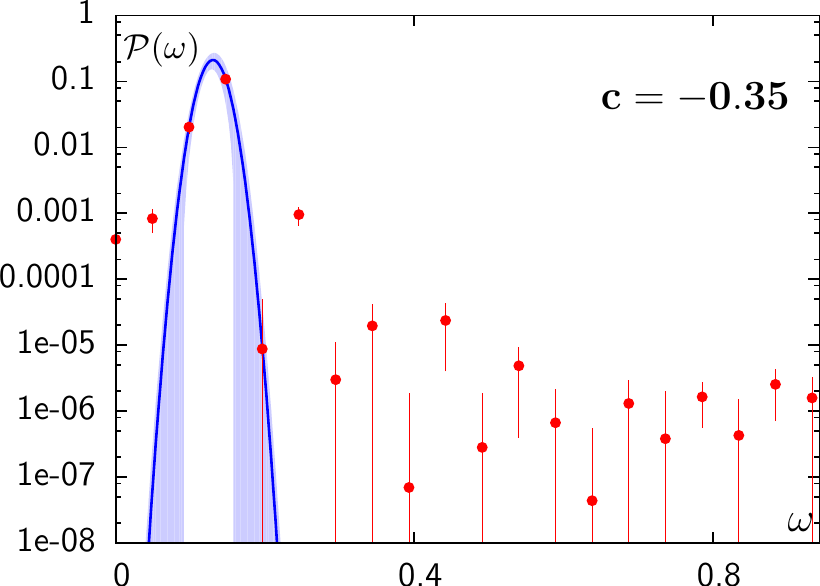}}
  \put(285  ,200.0){\includegraphics[bb=0 0 180 90, scale=0.55]{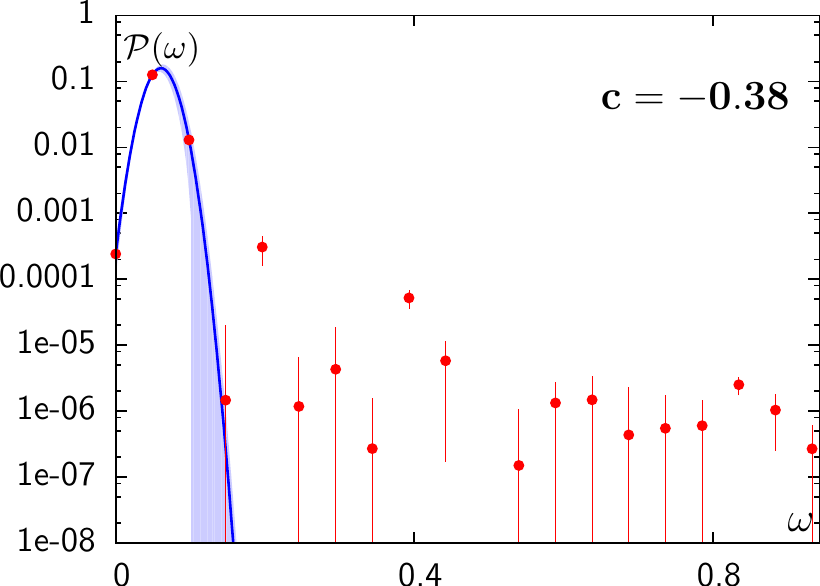}}
  \put(145.0,100.0){\includegraphics[bb=0 0 180 90, scale=0.55]{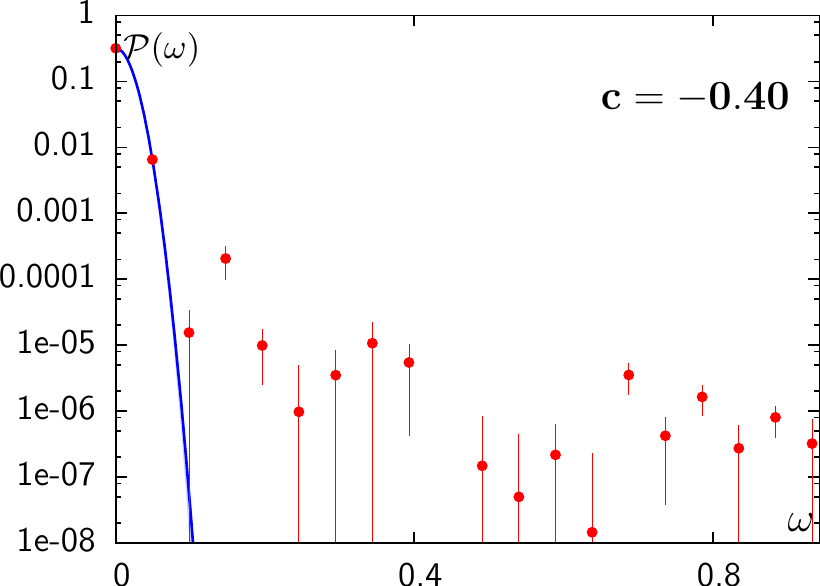}}
  \put(005  ,000.0){\includegraphics[bb=0 0 180 90, scale=0.55]{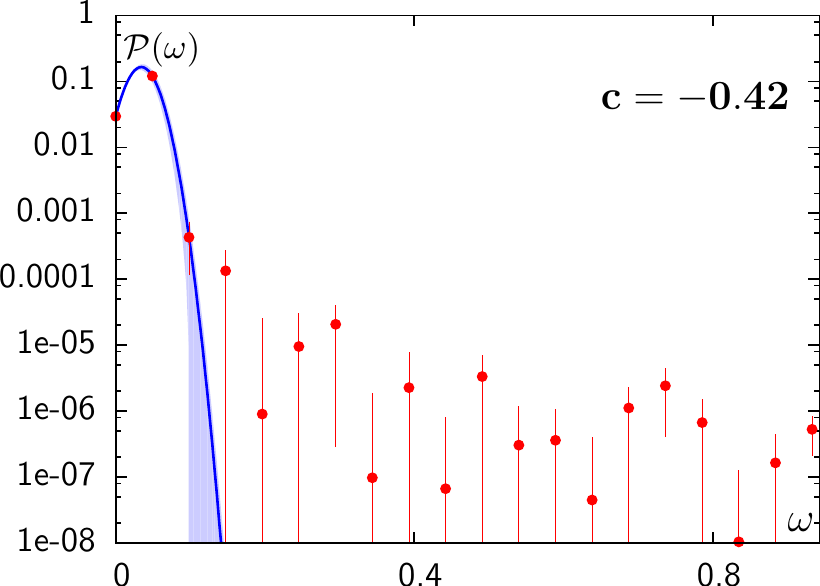}}
  \put(145.0,000.0){\includegraphics[bb=0 0 180 90, scale=0.55]{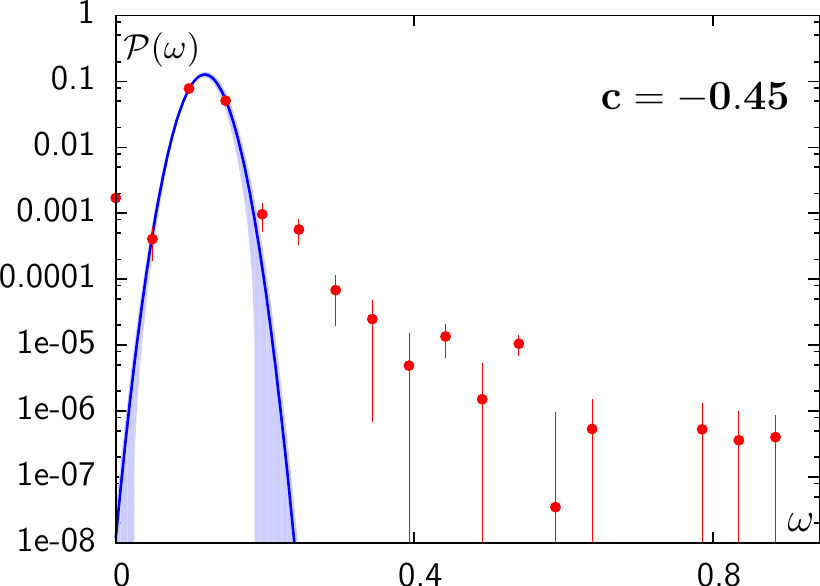}}
  \put(285.0,000.0){\includegraphics[bb=0 0 180 90, scale=0.55]{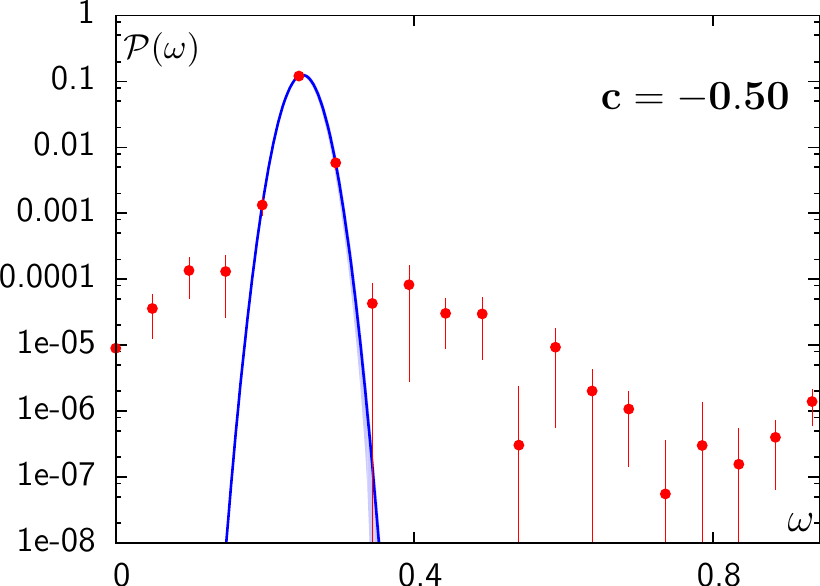}}
 \end{picture}
 \caption{Power spectral densities are displayed on a logarithmic scale. The curve is a gaussian function, which is used for estimating the maximum of the distribution.
  }
 \label{fig: psd P_05 at beta 6.2}
\end{figure}

\noindent
In order to obtain a frequency spectrum with two peaks near zero, an alternating factor~$ (-1)^{n_0} $ (due to~$ \Omega=\pi $ in the tuned frequency spectrum) is multiplied with the ratio before it is submitted to a discrete Fourier transform using the GSL FFT library~\cite{GSL:2009}. The Fourier coefficients $ C_k $ are computed as
\begin{equation}
  C_k = \sum\limits_{n_0=0}^{N_0-1} \mathcal{R}_{05}(n_0) (-1)^{n_0} e^{-2\pi i \frac{kn_0}{N_0}}.
  \label{eq: discrete FFT}
\end{equation}\normalsize
Because the FFT of the ratios is conducted independently on each sample of the full ensemble, no statistical information is lost in this process and statistical uncertainties of the spectra can be calculated with the Jackknife method. Fourier spectra that are defined without the alternating factor $ (-1)^{n_0} $ are shown in figure~\ref{fig: frequency spectra}. Two peaks are observed in the spectra that approach a single peak at $ k=N_0/2 $ for $ c=c_0 $. The spectra are symmetric (up to machine precision) about $ k=N_0/2 $. Distinction between $ c>c_0 $ and $ c<c_0 $ on the basis of the spectra in figure~\ref{fig: frequency spectra} is impossible. Outside of the peak region, the spectrum is almost consistent with zero. The power spectral density (PSD) takes advantage of the frequency spectrum's symmetry. Its definition is taken from~\cite{Press:2007},
\begin{equation}
  \left.\begin{array}{rl}
  \mathcal{P}(f_0) =& \frac{1}{N^2}|C_0|^2 \\
  \mathcal{P}(f_k) =& \frac{1}{N^2}\left(|C_k|^2+|C_{N-k}|^2\right)\qquad 
  k=1,2,\ldots,\left(\frac{N}{2}-1\right) \\
  \mathcal{P}(f_{N/2}) =& \frac{1}{N^2}|C_{N/2}|^2 
  \end{array}\right.,
  \label{eq: psd}
\end{equation}\normalsize
where the $ C_k $ are the Fourier coefficients defined in \mbox{eq.}~(\ref{eq: discrete FFT}). The two peaks of the frequency spectrum are mapped onto a single peak of the power spectral density. The power spectral densities within the frequency range~$ \omega \in[0,0.3\pi] $ are shown in figure~\ref{fig: psd P_05 at beta 6.2}. They are sharply peaked distributions with noise which is typically suppressed by $ 3\!-\!6 $ orders of magnitude. Due to the aforementioned residual exponential decays of the ground state mass difference and excited state contributions, which are not known in detail, the exact form of the frequency spectrum is unknown. 
%

\noindent
An estimate of the peak frequency and its systematical uncertainty are obtained as the maximum $ \omega_c $ and the width $ \sigma $ of a gaussian distribution,
\begin{equation}
  g(\omega) = \frac{P_0}{\sqrt{2\pi\sigma^2}}e^{-\frac{(\omega-\omega_c)^2}{2\sigma^2}},
  \label{eq: frequency peak gaussian estimator}
\end{equation}\normalsize
which is matched with a least squares fit to the power spectral density on each sample. 
\noindent
If the peak frequency $ \omega_c $ is smaller than $ \omega_b/2 $, all samples evaluate to $ \omega_c=0 $ and the statistical uncertainty vanishes. If $ \omega_c=0 $ with $ \delta_{\omega_c}=0 $ were used in an interpolation in $ c $, it would completely dominate the interpolation\footnote{If multiple coefficients $ c $ have $ \omega_c=0 $ and $ \delta_{\omega_c}=0 $, linear interpolation of $ \omega_c $ necessarily fails.
}. Hence, a balanced and sensitive approach to interpolation must combine statistical and systematical uncertainties. Systematic uncertainties are estimated as the root of the variance of the gaussian distribution. In the following, they are treated as independent errors and combined quadratically,
\begin{equation}
  \delta_{\omega_c}=\sqrt{(\delta_{\omega_c})^2+\sigma^2}.
  \label{eq: peak frequency error formula}
\end{equation}\normalsize

\begin{table}[htb]
   \begin{tabular}{|c||c|c|c|c|c|c|c|}
  \hline
  \multicolumn{8}{|l|}{$ \beta=5.8 $} \\
  \hline
  $ c $     & $ -0.40 $ & $ -0.45 $ & $ -0.48 $ & $ -0.50 $ & $ -0.53 $ & $ -0.55 $ & $ -0.60 $ \\
  $ \omega_c$ & $ 155(2)(37)\! $ & $ 74(1)(38)\! $ & $ 26(2)(38)\! $ & $ 11(2)(26)\! $ & $ 14(4)(35)\! $ & $ 40(2)(33)\! $ & $ 116(2)(35)\! $ \\
  \hline
  \multicolumn{8}{c}{ } \\
  \hline
  \multicolumn{8}{|l|}{$ \beta=6.0 $} \\
  \hline
  $ c $     & $ -0.35 $ & $ -0.40 $ & $ -0.42 $ & $ -0.45 $ & $ -0.47 $ & $ -0.50 $ & $ -0.55 $ \\
  $ \omega_c$ & $ 124(1)(12)\! $ & $ 54(1)(13)\! $ & $ 25(1)(13)\! $ & $ 0(0)(15)\! $ & $ 32(1)(13)\! $ & $ 74(1)(13)\! $ & $ 141(1)(13)\! $ \\
  \hline
  \multicolumn{8}{c}{ } \\
  \hline
  \multicolumn{8}{|l|}{$ \beta=6.2 $} \\
  \hline
  $ c $     & $ -0.30 $ & $ -0.45 $ & $ -0.38 $ & $ -0.40 $ & $ -0.42 $ & $ -0.45 $ & $ -0.50 $ \\
  $ \omega_c$ & $ 134(1)(11)\! $ & $ 65(1)(7)\! $ & $ 30(1)(8)\! $ & $ 0(0)(9)\! $ & $ 17(1)(9)\! $ & $ 60(1)(11)\! $ & $ 125(1)(9)\! $ \\
  \hline
 \end{tabular}
 \caption{The peak frequency $ \omega_c $ of the power spectral densities seems to be linear in $ |\delta_c| $. $ \omega_c $ in the table must be multiplied by $ 10^{-3} $. Statistical ($ \delta_{\omega_c}$) and systematical ($ \sigma $) uncertainties are in brackets. Ratios are calculated on $ 128\times24^3 $-lattices from table \ref{tab: lattices for oscillation study}.
 }
 \label{tab: psd maxima}
\end{table}

\begin{figure}[htb]
 \begin{picture}(360,200)
  \put(075.0,000.0){\includegraphics[bb=0 0 200 145, scale=1.20]{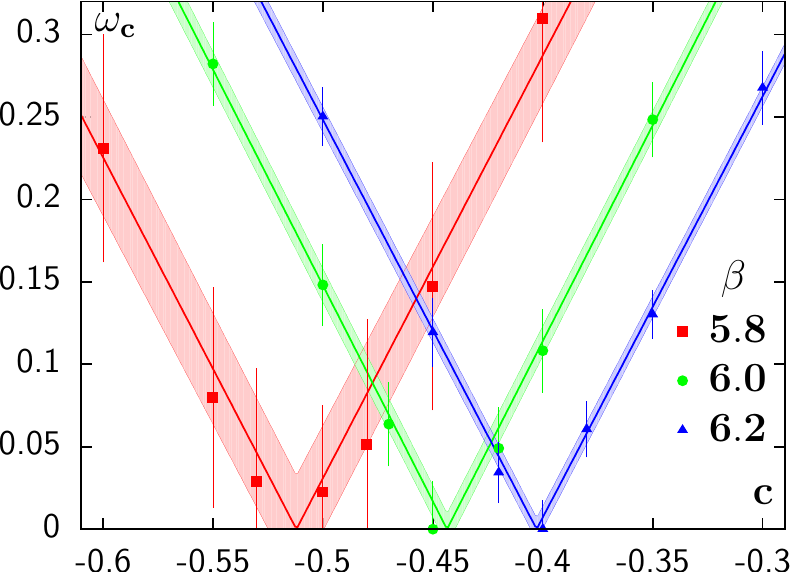}}
 \end{picture}
 \caption{
 Peak frequencies $ \omega_c $ for~$ \beta=5.8 $,~$ \beta=6.0 $ and~$ \beta=6.2 $ are interpolated in~$ c $ using \mbox{eq.}~(\ref{eq: peak frequency interpolation}). Error bands are dominated by systematical errors.
 }
 \label{fig: peak frequency interpolation}
\end{figure}

\noindent
Peak frequencies with statistical and systematical uncertainties in the format~$ \omega_c(\delta_{\omega_c})(\sigma) $ for~$ \beta=6.2 $, $ 6.0 $ and $ 5.8 $ are summarised in table~\ref{tab: psd maxima}. Power spectral densities of ratios for~$ \beta=6.0 $ and~$ \beta=5.8 $ are presented in figures~\ref{fig: psd P_05 at beta 6.0} and~\ref{fig: psd P_05 at beta 5.8} in appendix~\ref{app: Simulation parameters and data sets}. The width of the peak 
is larger on the coarse lattice~($ \beta=5.8 $), which is immediately clear from a comparison of the figures. This is the reason why the systematical errors for $ \beta=5.8 $ considerably exceed their counterparts on finer lattices~($ \beta=6.0 $,~$ \beta=6.2 $). The ratios contain stronger exponential decays due to larger ground state mass differences. 
\noindent
It is seen in figure~\ref{fig: frequency spectra} and in table~\ref{tab: psd maxima} that the frequency shift $ \omega_c $ does not allow for a distinction between $ \delta_c>0 $ and $ \delta_c<0 $. Two different approaches are applied. First, the peak frequencies~$ \omega_c $ are interpolated with the function
\begin{equation}
  \omega_c = A\cdot|c-c_0|,
  \label{eq: peak frequency interpolation}
\end{equation}\normalsize
which precludes from using a simple linear fit. Second, peak frequencies~$ \omega_c $ for~$ c < c_{\min} $ with~$ \omega_{c_{\min}} \leq \omega_c \ \forall\ c $ are multiplied by $ (-1) $ and a linear interpolation is conducted with
\begin{equation}
  \omega_c = A(c-c_0).
  \label{eq: linear peak frequency interpolation}
\end{equation}\normalsize
Both approaches are consistent. Interpolations of peak frequencies are displayed in figure~\ref{fig: peak frequency interpolation} and their zeros~$ c_0 $ are listed in table~\ref{tab: c_0 on T=128 lattices}. The frequency shift~$ \omega_c $ is described extraordinarily well over the full range of data (at $ \beta=6.2 $ and~$ \beta=6.0 $) by the ansatz of \mbox{eq.}~(\ref{eq: peak frequency interpolation}). Hence, even a shorter $ \hat{e}_0 $ direction (\mbox{e.g.}~$ N_0=48 $) yields sufficient resolution for linear interpolation. $ c_0 $ is consistent with $ c_M $ within combined errors.
\noindent
Since oscillations are exclusive to parallel correlation functions,~$ c_0 $ is obtained without use of perpendicular correlation functions (correlation functions, where the direction of correlation is perpendicular to alignment of the Karsten-Wilczek term). Therefore, restoration of the frequency spectrum to its tree-level form is a tuning condition that disregards the anisotropy. If an observable with sensitivity to the anisotropy, good statistical accuracy and strong sensitivity to~$ d $ were discovered, the residual anisotropy after setting~$ c=c_0 $ could be employed for tuning~$ d $ non-perturbatively.

\begin{table}[hbt]
 \center
 \begin{tabular}{|c|c|c|
  }
  \hline
  $ \beta $ & $ a\,[fm] $ & $ c_0 $ 
  \\
  \hline
  $ 5.8 $   & $ 0.136 $   & $ -0.5120(7)(100) $  
  \\
  $ 6.0 $   & $ 0.093 $   & $ -0.4435(8)(37) $   
  \\
  $ 6.2 $   & $ 0.068 $   & $ -0.4028(5)(27) $   
  \\
  \hline
 \end{tabular}
 \caption{
 The zero crossing~$ c_0 $ of \mbox{eq.}~(\ref{eq: linear peak frequency interpolation}) is obtained for ratios with $ N_0=128 $. The format $ c_0(\delta)(\sigma) $ includes statistical and systematical errors.
 }
 \label{tab: c_0 on T=128 lattices}
\end{table}\normalsize

\newpage
\sectionc{Chiral behaviour of the pseudoscalar ground state}{sec: Chiral behaviour of the pseudoscalar ground state}
%

\begin{table}[hbt]
\center\footnotesize
 \begin{tabular}{|c|c|c|c|c|c|c|c|c|}
  \hline
  $ \beta $ & $ am_0 $ & $ (r_0 m_0) $ & $ (r_0 M_{55}) $ & $ M_{55}\ [\mathrm{MeV}] $ & $ (r_0 M_{00}) $ & $ M_{00}\ [\mathrm{MeV}] $ & $ \mathcal{B}_{55} $  \\
  \hline
  $ 5.8 $ & $ 0.02928 $& $ 0.107 $ & $ 1.655(2) $ & $ 653(1) $ & $ 1.858(3) $ & $ 733(1) $ & $ 25.5(1) $ \\
  $ 5.8 $ & $ 0.02000 $   & $ 0.073 $ & $ 1.379(1) $ & $ 544(1) $ & $ 1.612(3) $ & $ 636(1) $ & $ 25.9(1) $ \\
  $ 5.8 $ & $ 0.01464 $& $ 0.054 $ & $ 1.188(1) $ & $ 469(1) $ & $ 1.450(4) $ & $ 572(2) $ & $ 26.2(1) $ \\
  $ 5.8 $ & $ 0.01000 $   & $ 0.037 $ & $ 0.989(1) $ & $ 390(1) $ & $ 1.294(5) $ & $ 511(2) $ & $ 26.6(1) $ \\
  $ 5.8 $ & $ 0.00732 $ & $ 0.027 $ & $ 0.851(1) $ & $ 336(1) $ & $ 1.194(6) $ & $ 471(2) $ & $ 26.9(1) $ \\
  $ 5.8 $ & $ 0.00534 $ & $ 0.020 $ & $ 0.731(1) $ & $ 288(1) $ & $ 1.115(8) $ & $ 440(3) $ & $ 27.2(1) $ \\
  \hline
  $ 6.0 $ & $ 0.02000 $   & $ 0.107 $ & $ 1.597(4) $ & $ 630(1) $ & $ 1.695(5) $ & $ 669(2) $ & $ 23.8(1) $ \\
  $ 6.0 $ & $ 0.01371 $& $ 0.074 $ & $ 1.333(4) $ & $ 526(1) $ & $ 1.444(5) $ & $ 570(2) $ & $ 24.2(1) $ \\
  $ 6.0 $ & $ 0.01000 $   & $ 0.054 $ & $ 1.149(4) $ & $ 453(2) $ & $ 1.274(5) $ & $ 503(2) $ & $ 24.6(2) $ \\
  $ 6.0 $ & $ 0.00685 $ & $ 0.037 $ & $ 0.963(4) $ & $ 380(2) $ & $ 1.109(6) $ & $ 438(2) $ & $ 25.3(2) $ \\
  $ 6.0 $ & $ 0.00500 $    & $ 0.027 $ & $ 0.832(4) $ & $ 328(2) $ & $ 0.997(7) $ & $ 393(3) $ & $ 25.9(3) $ \\
  $ 6.0 $ & $ 0.00365 $ & $ 0.020 $ & $ 0.719(5) $ & $ 284(2) $ & $ 0.905(8) $ & $ 357(3) $ & $ 26.5(3) $ \\
  \hline
  $ 6.2 $ & $ 0.01460 $ & $ 0.107 $ & $ 1.567(7) $ & $ 618(3) $ & $ 1.602(7) $ & $ 632(3) $ & $ 23.0(2) $ \\
  $ 6.2 $ & $ 0.01000   $ & $ 0.074 $ & $ 1.305(8) $ & $ 515(3) $ & $ 1.346(7) $ & $ 531(3) $ & $ 23.4(3) $ \\
  $ 6.2 $ & $ 0.00730 $ & $ 0.054 $ & $ 1.124(8) $ & $ 444(3) $ & $ 1.173(8) $ & $ 463(3) $ & $ 23.8(4) $ \\
  $ 6.2 $ & $ 0.00500   $ & $ 0.037 $ & $ 0.943(9) $ & $ 372(4) $ & $ 1.004(9) $ & $ 396(3) $ & $ 24.5(5) $ \\
  $ 6.2 $ & $ 0.00365 $ & $ 0.027 $ & $ 0.815(10) $ & $ 322(4) $ & $ 0.889(10) $ & $ 351(4) $ & $ 25.1(6) $ \\
  $ 6.2 $ & $ 0.00266 $ & $ 0.020 $ & $ 0.704(12) $ & $ 278(5) $ & $ 0.793(11) $ & $ 313(4) $ & $ 25.8(9) $ \\
  $ 6.2 $ & $ 0.00194 $ & $ 0.014 $ & $ 0.607(14) $ & $ 240(5) $ & $ 0.712(12) $ & $ 281(5) $ & $ 26.4(12) $ \\
  \hline
 \end{tabular}
 \caption{
 The tuned action is studied on lattices with $ T=48 $. $ L=24 $ for $ \beta=5.8,\, 6.0 $ and $ L=32 $ for $ \beta=6.2 $ are used.
}
 \label{tab: fermion mass parameters}
\end{table}

\begin{table}[hbt]
\center
 \begin{tabular}{|c|c|c||c|c|c|}
  \hline
  $ \beta $ & $ c  $ & $ d $ & $ T $ & $ L $ & $ n_{cfg} $ \\
  \hline
  $ 5.8 $ &  $ -0.51 $ & $ -0.002 $ & $ 48 $ &  $ 24 $ & $ 200 $ \\
  $ 6.0 $ &  $ -0.45 $ & $ -0.001 $ & $ 48 $ &  $ 24 $ & $ 200 $ \\
  $ 6.2 $ &  $ -0.40 $ & $ -0.001 $ & $ 48 $ &  $ 32 $ & $ 100 $ \\
  \hline
 \end{tabular}
 \caption{
 $ c $ is tuned non-perturbatively and $ \Delta(M_{PS}^2) $ as well as $ \omega_c $ are consistent with zero within errors. $ d $ is fixed perturbatively. The Wilczek parameter is set to $ \zeta=+1 $.
 }
 \label{tab: tuned parameter values}
\end{table}

\noindent
The preceding two sections concern non-perturbative tuning of Karsten-Wilczek fermions. The counterterm's coefficient~$ c $ of the Karsten-Wilczek action is tuned non-perturbatively by either making use of minimisation of the pseudoscalar mass anisotropy~(section \ref{sec: Anisotropy of hadronic quantities}) or restoration of the tree-level frequency spectrum~(section \ref{sec: Oscillating correlation functions}). The counterterm coefficient $ d $ is tuned using boosted perturbation theory. Simulation parameters are listed in table~\ref{tab: tuned parameter values}. Properties of QCD are studied using the Karsten-Wilczek fermion action. 
\noindent
The scale is set with the Sommer parameter $ r_0 $\cite{Sommer:1993ce} from table \ref{tab: gauge coupling, scale setting and smearing}. The only remaining simulation parameter is the bare fermion mass $ am_0 $, which is listed in table \ref{tab: fermion mass parameters} together with fit masses of the pseudoscalar ground states of the $ \gamma^5 $~channel~($ M_{55} $) and of the $ \gamma^{0} $~channel~($ M_{00} $). Chiral behaviour is studied by varying~$ am_0 $ within a factor of~$ 6\!-\!7 $. The 3rd column of table~\ref{tab: fermion mass parameters} is the bare mass~$ r_0 m_0 $ in physical units~($ 2\,\mathrm{fm}^{-1} $). 
Because mass renormalisation is required but not performed, different~$ r_0 M_{55} $ (in the 4th column) is obtained for the same $ r_0 m_0 $ at different $ \beta $.
Since~$ M_{55} $ and~$ M_{00} $ are quite similar and a pseudoscalar ground state is expected in both channels, both are tentatively treated as approximate Goldstone bosons with different discretisation effects. Local effective mass plots indicating the fit ranges are displayed in figure~\ref{fig: local effective mass of gamma5 and gamma0-channels} in the appendix.

\subsection{Chiral behaviour of the $ \gamma^5 $~channel}\label{sec: Chiral behaviour of the gamma^5 channel}

\noindent
In full QCD, chiral perturbation theory~(ChPT) at next-to-leading order~\cite{Gasser:1983yg,Gasser:1984gg} predicts that the pion mass scales with the average light quark mass~$ m_{ud} $ and the strange quark mass $ m_s $ as
\begin{align}
  M_{\pi,2}^2 =&\ 2B_0 m_{ud},\quad M_{\eta,2}^2 = \frac{2}{3}B_0 (m_{ud}+2m_s),
  \label{eq: chiral order 2 M_pi^2} \\
  M_{\pi,4}^2 =&\ M_{\pi,2}^2\left\{1 + X+\frac{M_{\pi,2}^2}{32\pi^2 F_0^2}\log{(\mu^{-2} M_{\pi,2}^2)}-\frac{M_{\eta,2}^2}{96\pi^2 F_0^2}\log{(\mu^{-2} M_{\eta,2}^2)}\right\}, 
  \label{eq: chiral order 4 M_pi^2}\\
  X =&\ \frac{16}{F_0^2} \left\{(2m_{ud}+m_s)B_0(2L_6^r-L_4^r)+(m_{ud}+m_s)B_0(2L_8^r-L_5^r)\right\}. \nonumber
\end{align}\normalsize
The notation follows~\cite{Scherer:2002tk}, where $ M_{\pi,2n}^2 $ and $ M_{\eta,2n}^2 $ indicate the pion and eta meson masses for degenerate up and down quarks at chiral order $ 2n $. $ 2B_0 $ and $ F_0 $ are the chiral condensate and the pion decay constant in the chiral limit. $ X $ is the sum of two contributions that depend on the quark masses, $ B_0 $, $ F_0 $ and four renormalised low-energy constants $ L_i^r $ from the chiral Lagrangian at chiral order four. All of these (quark masses, $ B_0 $, $ F_0 $ and $ L_i^r $) are undetermined parameters that must be obtained by matching ChPT calculations to calculations in QCD or to experimental data. 
The ratio~$ \mathcal{B}_{55} $ in the 8th column of table \ref{tab: fermion mass parameters} is defined as
\begin{equation}
  \mathcal{B}_{55} \equiv \frac{(r_0 M_{55})^2}{(r_0 m_0)}.
  \label{eq: ratio B_55}
\end{equation}\normalsize
Comparison of \mbox{eq.}~(\ref{eq: ratio B_55}) with \mbox{eqs.}~(\ref{eq: chiral order 2 M_pi^2})~and~(\ref{eq: chiral order 4 M_pi^2}) suggests that $ \mathcal{B}_{55} $ is related to $ 2B_0 $ in partially quenched QCD (the strange quark is considered as heavy) as
\begin{equation}
  \mathcal{B}_{55} \propto 2 B_0 \{1+X_0 m_{ud}+X_1 m_{ud}\log{(am_{ud})}\}.
  \label{eq: relation between B_55 and B_0}
\end{equation}\normalsize
As factors~$ X_0 $ and~$ X_1 $ are numerically positive, the prediction of partially quenched QCD is a monotonical increase of~$ \mathcal{B}_{55} $ with the quark mass. Instead, a decrease is observed at fixed~$ \beta $. There are two possible causes why this increase is not reflected in the data. \newline

\noindent
The first cause is that the quenched approximation neglects virtual quark loops, which cancel hairpin diagrams~\cite{Bernard:1992mk,Sharpe:1992ft} that contribute to mesonic correlation functions. Lack of this cancellation leads to additional logarithmic contributions to the pion mass~\cite{Bernard:1992mk} in the form of quenched chiral logarithms,
\begin{align}
  M_{\pi}^2 =&\ M_{\pi,2}^2\left\{(1-\delta) -\delta\log{(\mu^{-2} M_{\pi,2}^2)}\right\} + \mathcal{O}(M_{\pi,2}^4), 
  \label{eq: chiral order 4 M_pi^2 with quenched logs}
\end{align}\normalsize
where the new parameter $ \delta $ is related to the pseudoscalar flavour singlet mass $ M_{0}^2 $ by
\begin{equation}
  \delta=\frac{M_{0}^2}{N_f(4\pi F_0)^2}.
  \label{eq: quenched chiral log parameter delta}
\end{equation}\normalsize
The parameter~$ M_{0}^2 $ is related to the topological susceptibility $ \chi_t $ through the Witten-Veneziano formula~\cite{Witten1979269,Veneziano:1979ec}. A phenomenological estimate of~$ M_0^2 $ is obtained in terms of physical meson masses, 
\begin{equation}
  M_{0}^2=M_{\eta^\prime}^2+M_\eta^2-2M_K^2,
\end{equation}
which yields~$ \delta\approx0.18 $. Since~$ \delta $ is positive, quenched chiral logarithms cause a logarithmic divergence of~$ \mathcal{B}_{55} $ in the chiral limit.\newline

\noindent
The second cause for a rise of $ \mathcal{B}_{55} $ in the chiral limit are uncertainties in the choice of the coefficients. The anisotropy study in section~\ref{sec: Anisotropy of hadronic quantities} demonstrates that the mass anisotropy $ \Delta(M_{PS}^2 $) in the chiral limit is not removed perfectly at finite lattice spacing, which implies a mass shift in at least one of the correlation functions. As $ c $ is tuned only with a few percent accuracy and $ d $ only using perturbation theory, incomplete tuning must be considered another potential source of non-vanishing contributions to the ground state mass in the chiral limit. Nevertheless, a chiral extrapolation as
\begin{equation}
  r_0^2 M^2 = r_0^2 M_{\mathrm{res}}^2+ 2B_0 (r_0 m_0) \left\{(1-\delta) -\delta \log{(m_0/r_0)}\right\}
  \label{eq: quenched chiral extrapolation at leading order}
\end{equation}\normalsize
is suited for both $ \gamma^5 $~and $ \gamma^{0} $~channels. The small mass splitting of \mbox{eq.}~(\ref{eq: M00-M55 mass spliting}) between $ \gamma^5 $~and $ \gamma^0 $~channels indicates that the latter channel's ground state is a pseudoscalar which is affected by mild discretisation effects. 
Spontaneous chiral symmetry breaking would still affect all non-singlet pseudoscalars, if such discretisation effects could be treated like the quark mass as perturbations in a ChPT for Karsten-Wilczek fermions\footnote{ChPT for staggered fermions in~\cite{Lee:1999zxa} includes non-singlet discretisation effects in the chiral expansion. This sets the example that should be followed for Karsten-Wilczek fermions.}. \newline

\begin{figure}[htb]
 \begin{picture}(360,305)
  \put(000.0,153.0){\includegraphics[bb=0 0 170 130, scale=0.90]{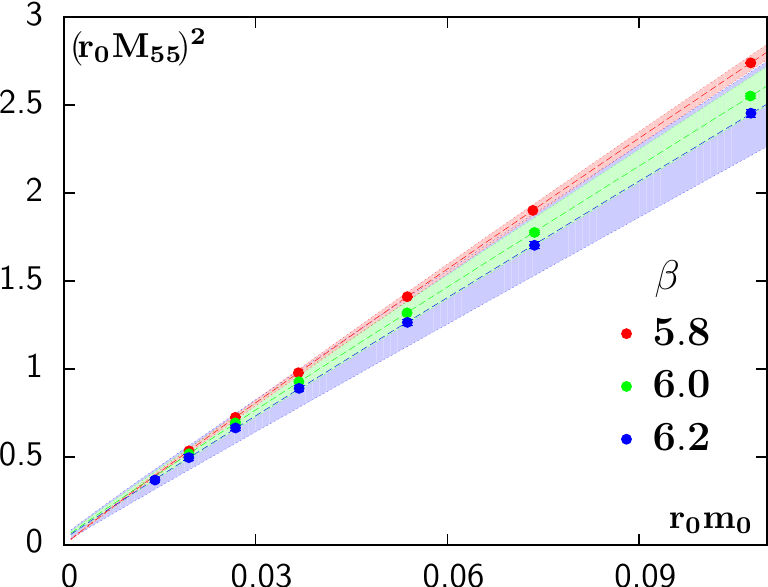}}
  \put(220.0,153.0){\includegraphics[bb=0 0 170 130, scale=0.90]{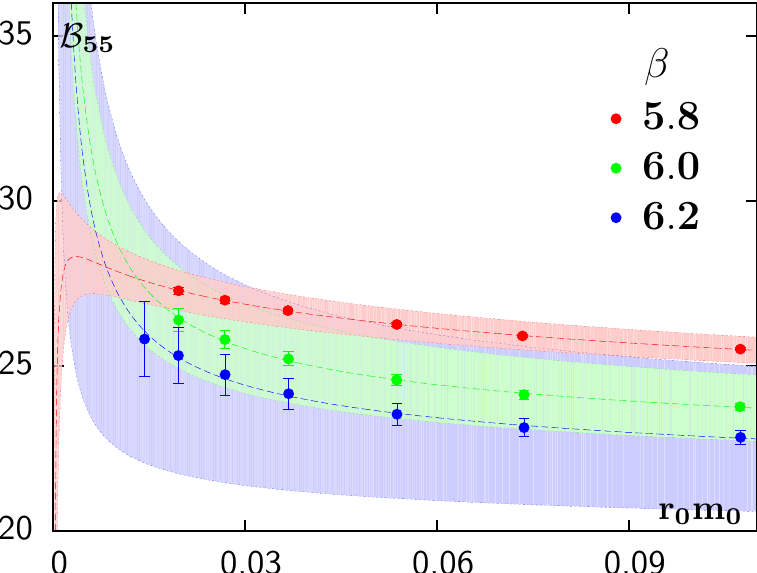}}
  \put(000.0,  0.0){\includegraphics[bb=0 0 170 130, scale=0.90]{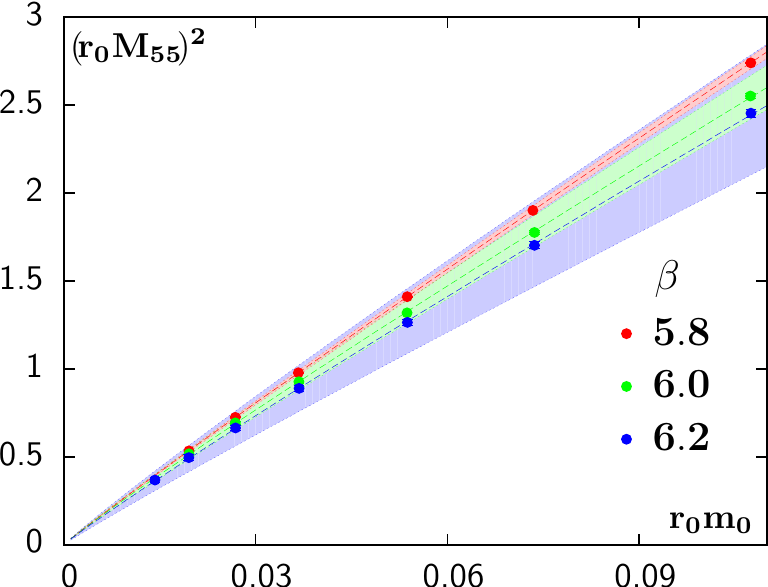}}
  \put(220.0,  0.0){\includegraphics[bb=0 0 170 130, scale=0.90]{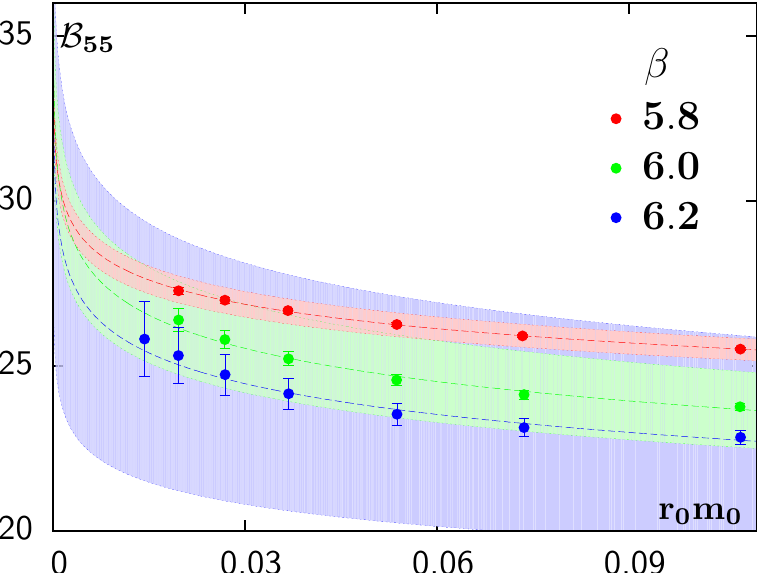}}
 \end{picture}
 \caption{Upper row: $ C $ is included as parameter in the chiral extrapolation of $ M_{55}^2 $. At $ \beta=5.8 $, it partly compensates for the rise of $ \mathcal{B}_{55} $ due to quenched chiral logarithms.
 Lower row: The offset parameter $ C $ in the chiral extrapolation of $ M_{55}^2 $ is fixed to $ 0 $.
  }
 \label{fig: chiral behaviour of M_55}
\end{figure}

\noindent
$ M_{55} $ is extracted by fitting a cosh to the correlation function for each fermion mass parameter~$ am_0 $. Fit masses are squared and subjected to a chiral extrapolation with
\begin{equation}
  M^2(am_0) = C+ (am_0) \cdot \left\{A+B\log{(am_0)}\right\}
  \label{eq: chiral behaviour fit function}
\end{equation}\normalsize
and~$ \delta = B/(B-A) $. The same chiral extrapolation is conducted in one case with three independent fit parameters (upper row of figure \ref{fig: chiral behaviour of M_55}) and in another case with the constraint~$ C=0 $~(lower row of figure \ref{fig: chiral behaviour of M_55}). Results of both chiral extrapolations are shown in table~\ref{tab: chiral behaviour of M_{55}^2}. 
\noindent
Extrapolation results with an offset~$ C $ agree well with each other and with the phenomenological estimate~($ \delta \approx 0.18 $). If the offset is fixed to~$ C=0 $,~$ \chi^2/\mathrm{dof} $ increases and~$ \delta $ varies within a factor of two for $ \beta=6.0 $ and $ \beta=6.2 $. Since $ \delta $ is only a phenomenological estimate and the spread of other lattice results for $ \delta $~\cite{Wittig:2002ux} is consistent with this result, strong conclusions must be avoided at this point. The variation might indicate incomplete tuning in the sense that the counterterms are tuned better for $ \beta=5.8 $ than for $ \beta=6.0 $ and $ \beta=6.2 $. A partial reexamination of the study with improved counterterm coefficients\footnote{The simulation parameter~$ d $ for~$ \beta=6.0 $ and~$ \beta=6.2 $ is closer to~$ d_{1L} $ than to~$ d_{BPT} $.} may partly solve this problem. Because~$ M_{55}^2 $ in the chiral limit at~$ \beta=5.8 $ is consistent with zero within less than~$ 2\sigma $, data support the claim that the ground state of the $ \gamma^5 $~channel behaves as a \mbox{(Pseudo-)}~Goldstone boson.

\begin{table}[hbt]
\center\footnotesize
 \begin{tabular}{|c||c|c|c|c|c||c|c|c|c|}
  \hline
  $ \beta $ & $ r_0 A $ & $ r_0 B $  & $ \delta $ & $ r_0^2 C $ & $ \!\chi^2/\mathrm{dof}\! $ & $ r_0 A $ & $ r_0 B $  & $ \delta $ & $ \!\chi^2/\mathrm{dof}\! $ \\
  \hline
  $ 5.8 $ & $ 21.5(2) $ & $ -4.2(2) $  & $ 0.16(1) $ & $ -0.004(3) $ & $ 0.05 $ & $ 21.8(2) $ & $ -3.9(2) $   & $ 0.15(1) $ & $ 0.04 $ \\
  \hline
  $ 6.0 $ & $ 20.6(4) $ & $ -3.9(7) $  & $ 0.15(1) $  & $ 0.036(8) $  & $ 0.10 $ & $ 18.0(6) $ & $ -7.8(8) $   & $ 0.30(2) $ & $ 0.59 $ \\
  \hline
  $ 6.2 $ & $ 19.7(9) $ & $ -5.0(19) $ & $ 0.20(6) $  & $ 0.031(21) $ & $ 0.05 $ & $ 17.1(16) $ & $ -9.9(28) $ & $ 0.37(7) $ & $ 0.16 $ \\
  \hline
 \end{tabular}
 \caption{The chiral extrapolation of $ M_{55}^2 $ with an offset in the chiral limit reproduces the phenomenological value of $ \delta\approx0.18 $ very well.
}
 \label{tab: chiral behaviour of M_{55}^2}
\end{table}

\subsection{Interpretation of the \mbox{$ \gamma^{0} $}~channel}\label{sec: Interpretation of the gamma^0 channel}

\noindent
For Wilson fermions, interpolating operators of the $ \gamma^0 $~channel are related to $ J^{PC}=0^{+-} $ and do not generate physical $ q\bar q $~states from the vacuum. Hence, the physical significance of the $ \gamma^0 $~channel's ground state for Karsten-Wilczek fermions is not immediately evident. However, the mass splitting between the ground states of $ \gamma^0 $~and $ \gamma^5 $~channels is about 3-15\% and decreases for finer lattices. The ground state would have to be considered as an unphysical state, if the mass splitting increased for finer lattices. Because the trend in data is opposite, the state must correspond to a physical meson in the continuum limit. Within this mass range, there are no states in the QCD spectrum other than isospin non-singlet pseudoscalars. Therefore, the continuum limit of the ground state must be a \mbox{(Pseudo-)}~Goldstone boson in the chiral limit. Assuming that the discretisation effects that eventually break the symmetries can be treated as perturbations within some parameter range (which is a priori unknown), its mass~$ M_{00} $ is tentatively squared and extrapolated using \mbox{eq.}~(\ref{eq: chiral behaviour fit function}). 
The same chiral extrapolation is conducted in one case with three independent fit parameters and in another case with~$ B=0 $ (upper and lower rows of figure~\ref{fig: chiral behaviour of M_00} in appendix~\ref{app: Addendum to chiral behaviour of the pseudoscalar ground state}). Results of both chiral extrapolations are in perfect agreement (\mbox{cf.} table~\ref{tab: chiral behaviour of M_{00}^2}) and provide no numerical evidence for quenched chiral logarithms in the $ \gamma^0 $~channel.

\begin{table}[hbt]
\center\footnotesize
 \begin{tabular}{|c||c|c|c|c||c|c|c|}
  \hline
  $ \beta $ & $ r_0 A $ & $ r_0 B $  & $ r_0^2 C $ & $ \!\chi^2/\mathrm{dof}\! $ & $ r_0 A $ & $ r_0^2 C $ & $ \!\chi^2/\mathrm{dof}\! $ \\
  \hline
  $ 5.8 $ & $ 25.2(2) $ & $ 0.0003(120) $  & $ 0.75(2) $ & $ 0.002 $ & $ 25.1(2) $ & $ 0.75(2) $   & $ 0.002 $ \\
  \hline
  $ 6.0 $ & $ 23.3(1) $ & $ -0.03(2) $  & $ 0.37(2) $  & $ 0.11 $ & $ 23.4(2) $ & $ 0.37(2) $   & $ 0.08 $ \\
  \hline
  $ 6.2 $ & $ 22.0(2) $ & $ 0.008(30) $ & $ 0.20(2) $ & $ 0.04 $ & $ 22.0(2) $ & $ 0.20(2) $ & $ 0.03 $ \\
  \hline
 \end{tabular}
 \caption{The chiral extrapolation of $ M_{00}^2 $ produces a large offset in the chiral limit. 
}
 \label{tab: chiral behaviour of M_{00}^2}
\end{table}

%

\noindent
In order to shed light on the physical interpretation of the $ \gamma^{0} $~channel, the quadratic mass difference $ \left( M_{00}^2-M_{55}^2 \right) $ is expressed in units of $ r_0^2 $,
\begin{equation}
  \Delta_{05} \equiv r_0^2\left( M_{00}^2-M_{55}^2 \right).
  \label{eq: mass difference Delta_{05}}
\end{equation}\normalsize
Because there are strong cancellations between statistical fluctuations in the difference of the squared masses, its statistical error decreases (compared to the individual masses) and turns out to be too small for covering the difference's variations over the full quark mass range. The difference in the chiral limit is obtained as the difference of the offset parameters~$ C $ (rescaled with~$ r_0^2 $) of the chiral extrapolations of the ground state masses of both channels and included as `ch.l.' in the table~\ref{tab: mass difference Delta_{05}}. $ \Delta_{05} $ is remarkably stable over the full range of quark masses and decreases towards the continuum limit. The scaling behaviour of its approach to the continuum limit is determined by the scaling behaviour of both correlation functions. Since the results of section~\ref{sec: Remnant Theta symmetry and O(a) corrections} indicate that mesonic correlation functions with definite charge conjugation quantum numbers on quenched cofigurations do not have $ \mathcal{O}(a) $~corrections, leading discretisation effects of~$ \Delta_{05} $ are~$ \mathcal{O}(a^2) $. Therefore, $ \Delta_{05} $ can be extrapolated to the continuum limit as
\begin{equation}
  \Delta_{05} = \Delta_0 + a^2 \Delta_2 +\mathcal{O}(a^3).
  \label{eq: continuum extrapolation of Delta_{05}}
\end{equation}\normalsize
Continuum extrapolation is conducted with different functions $ f_{02}(a) $ and $ f_{2}(a) $,
\begin{align}
  f_{02}(a^2)=&\ \Delta_0 + a^2 \Delta_2, 
  \label{eq: continuum extrapolation f_{02}} \\
  f_{2}(a^2) =&\ a^2 \Delta_2.
  \label{eq: continuum extrapolation f_{2}}
\end{align}\normalsize
\begin{table}[hbt]
\center
 \begin{tabular}{|c|c|c|c|c|}
  \hline
  $ \beta $ & $ (r_0 m_0) $ & $ (r_0 M_{55})^2 $ & $ (r_0 M_{00})^2 $ & $ \Delta_{05} $  \\
  \hline
  $ 5.8 $ & $ 0.107 $ & $ 2.740(5) $  & $ 3.454(11) $ & $ 0.714(8) $ \\
  $ 5.8 $ & $ 0.073 $ & $ 1.901(4) $  & $ 2.597(11) $ & $ 0.696(9) $ \\
  $ 5.8 $ & $ 0.054 $ & $ 1.410(3) $  & $ 2.103(12) $ & $ 0.692(10) $ \\
  $ 5.8 $ & $ 0.037 $ & $ 0.979(3) $  & $ 1.674(13) $ & $ 0.696(12) $ \\
  $ 5.8 $ & $ 0.027 $ & $ 0.725(2) $  & $ 1.427(15) $ & $ 0.702(14) $ \\
  $ 5.8 $ & $ 0.020 $ & $ 0.534(2) $  & $ 1.243(17) $ & $ 0.709(17) $ \\
  \hline
  $ 5.8 $ & ch.l. & $ -0.004(3) $ & $ 0.751(17) $   & $ 0.755(17) $ \\
  \hline
  $ 6.0 $ & $ 0.107 $ & $ 2.552(12) $ & $ 2.873(15) $ & $ 0.321(5) $ \\
  $ 6.0 $ & $ 0.074 $ & $ 1.776(10) $ & $ 2.085(14) $ & $ 0.309(6) $ \\
  $ 6.0 $ & $ 0.054 $ & $ 1.319(9) $  & $ 1.623(13) $ & $ 0.304(6) $ \\
  $ 6.0 $ & $ 0.037 $ & $ 0.927(8) $  & $ 1.229(13) $ & $ 0.302(7) $ \\
  $ 6.0 $ & $ 0.027 $ & $ 0.693(7) $  & $ 0.994(13) $ & $ 0.302(8) $ \\
  $ 6.0 $ & $ 0.020 $ & $ 0.517(7) $  & $ 0.818(14) $ & $ 0.301(10) $ \\
  \hline
  $ 6.0 $ & ch.l. & $ 0.036(8) $  & $ 0.366(15) $   & $ 0.330(13) $ \\
  \hline
  $ 6.2 $ & $ 0.107 $ & $ 2.454(22) $ & $ 2.566(22) $ & $ 0.112(5) $ \\
  $ 6.2 $ & $ 0.073 $ & $ 1.703(20) $ & $ 1.812(19) $ & $ 0.110(5) $ \\
  $ 6.2 $ & $ 0.054 $ & $ 1.264(18) $ & $ 1.377(18) $ & $ 0.112(5) $ \\
  $ 6.2 $ & $ 0.037 $ & $ 0.889(17) $ & $ 1.008(17) $ & $ 0.119(5) $ \\
  $ 6.2 $ & $ 0.027 $ & $ 0.665(17) $ & $ 0.791(17) $ & $ 0.127(5) $ \\
  $ 6.2 $ & $ 0.020 $ & $ 0.496(17) $ & $ 0.629(17) $ & $ 0.134(6) $ \\
  $ 6.2 $ & $ 0.014 $ & $ 0.369(16) $ & $ 0.507(17) $ & $ 0.139(7) $ \\
  \hline
  $ 6.2 $ & ch.l. & $ 0.031(21) $  & $ 0.196(18) $   & $ 0.165(13) $ \\
  \hline
 \end{tabular}
 \caption{
 The difference~$ \Delta_{05} $ of the squared ground state masses of $ \gamma^{0} $~and $ \gamma^5 $~channels has only a mild fermion mass dependence. The statistical errors of~$ \Delta_{05} $ are greatly reduced due to cancellations of fluctuations  between both channels.
}
 \label{tab: mass difference Delta_{05}}
\end{table}

\begin{figure}[htb]
 \begin{picture}(360,220)
  \put(50  , 0.0){\includegraphics[bb=0 0 220 160, scale=1.30]{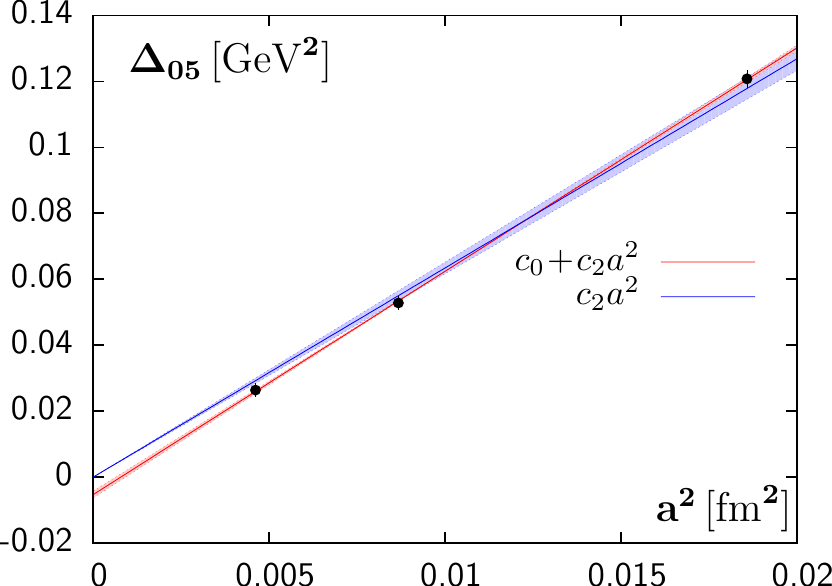}}
 \end{picture}
 \caption{The continuum extrapolation of~$ \Delta_{05} $ is conducted with two functions $ f_{02}(a^2) $~(red) and $ f_{2}(a^2) $~(blue). Data clearly supports $ f_{02}(a^2) $.
 }
 \label{fig: continuum scaling of Delta_{05}}
\end{figure}

\begin{table}[hbt]
\center
 \begin{tabular}{|c||c|c||c|}
  \hline
  extrapolation  & $ \Delta_0 $ & $ \Delta_2 $ & $ \!\chi^2/\mathrm{dof}\! $ \\
  \hline
  $ f_{02}(a) $  &  $ -0.033(6) $ & $ 42.3(6) $  & $ 0.15 $ \\
  $ f_{2}(a) $  &  $ 0 $ by def.  & $ 39.6(11) $  & $ 2.11 $ \\
  \hline
 \end{tabular}
 \caption{The continuum extrapolation of $ \Delta_{05} $ disfavours restriction to $ \mathcal{O}(a^2) $ corrections. This could be either due to incomplete tuning or due to higher orders.
 }
 \label{tab: continuum extrapolation of Delta_{05}}
\end{table}

\noindent
Both extrapolations are displayed in figure \ref{fig: continuum scaling of Delta_{05}} and their coefficients are listed in table \ref{tab: continuum extrapolation of Delta_{05}}. The extrapolation with $ f_{02}(a^2) $, which is non-zero in the continuum limit, agrees well with data and $ \chi^2/\mathrm{dof} $ is reasonably small, whereas $ f_2(a^2) $ seems to be disfavoured. This may have different reasons, among which incomplete tuning, fluctuations in the data points or unresolved higher order effects are the most obvious. Resolution of higher order discretisation effects is not deemed feasible with only three data points. Incomplete tuning is particularly suggestive because~$ C $ is non-vanishing in the chiral extrapolation of the $ \gamma^5 $~channel for both finer lattices ($ \beta=6.0 $ and~$ \beta=6.2 $).
It is pointed out with regard to \mbox{eq.}~(\ref{eq: mass difference Delta_{05}}) that the value of the expansion coefficient~$ \Delta_0 $ is numerically consistent with $ -C $ for~$ \beta=6.0 $ and~$ \beta=6.2 $ in table~\ref{tab: chiral behaviour of M_{55}^2}, which suggests that $ r_0^2 M_{00}^2 $ is consistent with zero in the combined chiral and continuum limit. Therefore, chiral behaviour of the ground state of the $ \gamma^0 $~channel is consistent with a \mbox{(Pseudo-)}~Goldstone boson in the continuum limit. As the fermion mass dependence of~$ \Delta_{05} $ is rather mild, this conclusion seems to be extendable to finite fermion mass. The difference at finite lattice spacing is evidence of discretisation effects that distinguish between both pseudoscalar states. This result is the analogue to taste-breaking effects for staggered fermions~\cite{KlubergStern:1983dg, Blum:1996uf}. It appears that (quenched) chiral perturbation theory is applicable and captures the quark mass dependence correctly even for \mbox{(Pseudo-)}~Goldstone bosons that are beset by these lattice artifacts. This remarkable result certainly warrants further dedicated studies.

\newpage
\sectionc{Interim findings (III)}{sec: Interim findings (III)}

\noindent
In an initial discussion, foreseeable difficulties in numerical studies with minimally doubled fermions are pointed out and technical details of the implementation are covered. Only some of these difficulties can be understood purely in terms of the broken discrete symmetries of the theory. The appearance of oscillating terms in certain channels can be interpreted in term of the decomposition of the spinor fields of section~\ref{sec: Decomposition into a pair of fields}. Two different approaches to non-perturbative tuning of Karsten-Wilczek fermions are developed. They make use of the anisotropy of the pseudoscalar mass or shifted frequency spectra of oscillating ratios of correlation functions. Lastly, a spectroscopic study of mesonic correlation functions in two different channels determines the chiral behaviour of their ground states and identifies them as pseudoscalars, which are degenerate in the continuum limit. \newline

\noindent
The first approach to non-perturbative renormalisation compares hadronic correlation functions, which use different directions of correlation -- parallel or perpendicular to the $ \hat{e}_{\underline{\alpha}} $~direction of the Karsten-Wilczek term. Without appropriately tuned counterterms, observables which are related to these correlators are anisotropic. The condition that the mass anisotropy $ \Delta(M_{PS}^2) $ of the pseudoscalar ground state has a mininum is used to tune the relevant counterterm's coefficient to a value~$ c_M $. This approach has three main weaknesses. First, because the mass anisotropy $ \Delta(M_{PS}^2) $ is a very shallow function of the relevant counterterm's coefficient, $ c_M $ has relatively large uncertainties. Second, since $ \Delta(M_{PS}^2) $ is not independent of the marginal counterterm's coefficient, disentangling the dependence on both coefficients is a non-trivial issue. Third, correlation functions with the direction of correlation perpendicular to the alignment of the Karsten-Wilczek term have short plateaus. It appears as if they include low-lying excited state contributions, which seem to have negative spectral weights in source-smeared correlation functions. These issues seem less prominent if~$ c \approx c_M $. 
\newline

\noindent
Two non-trivial aspects in the numerical evaluation in correlation functions with the direction of correlation parallel to the alignment of the Karsten-Wilczek term were studied in detail. First, signatures  of the broken time-reflection symmetry are not observed in numerical data for the pseudoscalar channel. 
Due to the considerations about the $ C\Theta_{\underline{\alpha}} $~symmetry in section~\ref{sec: Remnant Theta symmetry and O(a) corrections}, this is not surprising. Second, in all mesonic channels other than the $ \gamma^5 $~channel, an oscillating contribution can be observed directly. A possible explanation of the oscillatory pattern is offered by the formal decomposition in section~\ref{sec: Decomposition into a pair of fields}. In particular, a very small mass splitting between the ground states of $ \gamma^0 $~ and $ \gamma^5 $~channels is observed and their ratio $ R_{05}(n_0) $ for the finest lattice with $ \beta=6.2 $ is almost purely oscillating with only a very small exponential decay. If the relevant counterterm is not fully tuned, the ratio's frequency spectrum has two peaks that are aligned symmetrically around $ \pi/a $, while the peak is at $ \pi/a $ in the frequency spectra for na\"{i}ve or free Karsten-Wilczek fermions. The frequency shift's parameter dependence is empirically found to be $ \omega_c \propto |\delta c| $. The formal decomposition in section~\ref{sec: Decomposition into a pair of fields} also suggests oscillations with frequencies that are close to $ \pi/a $ but shifted linearly in $ \delta_c $. The condition that the frequency spectrum of $ R_{05}(n_0) $ is restored to its tree-level form is used to tune the relevant counterterm's coefficient to $ c_0 $ in this second approach. \newline

\begin{figure}[htb]
 \begin{picture}(360,240)
  \put(040.0, 0.0){\includegraphics[bb=0 0 360 240, scale=1.40]{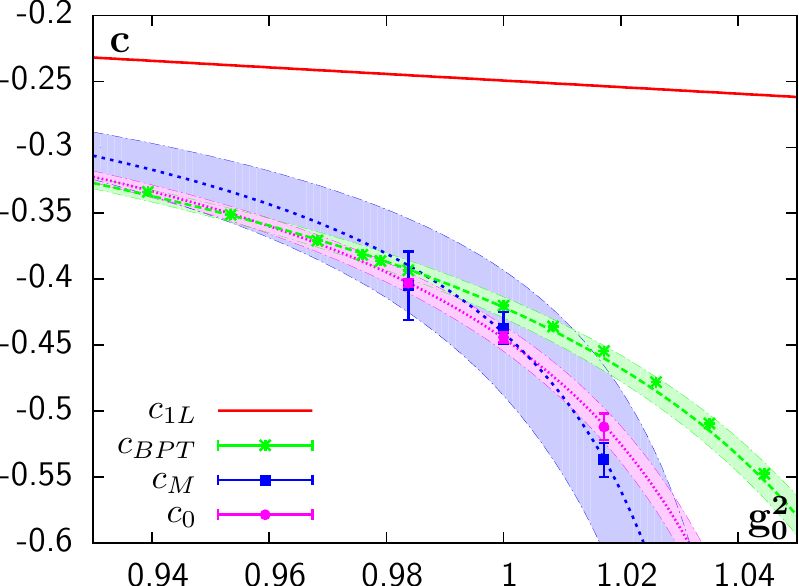}}
 \end{picture}
 \caption{Non-perturbative renormalisation schemes for~$ c(g_0^2) $ agree within errors. One-loop results are clearly too small, but BPT undershoots only by a few percent.
  }
 \label{fig: non-perturbative tuning of c}
\end{figure}

\begin{table}[hbt]
 \center
 \begin{tabular}{|c|c|c||c||c|c|}
  \hline
  $ \beta $ & $ a\,[fm] $ & $ c_{BPT} $ & $ \left|\frac{c_0-c_{BPT}}{c_0}\right| $ & $ c_M $   & $ c_0 $  \\
  \hline
  $ 5.8 $   & $ 0.136 $   & $ -0.454 $  & 11\% & $ -0.537(8)(10) $  & $ -0.5120(7)(100) $ \\
  $ 6.0 $   & $ 0.093 $   & $ -0.420 $  &  5\% & $ -0.437(7)(10) $  & $ -0.4435(8)(37) $  \\
  $ 6.2 $   & $ 0.068 $   & $ -0.393 $  &  2\% & $ -0.405(17)(20) $  & $ -0.4028(5)(27) $  \\
  \hline
 \end{tabular}
 \caption{
 Two different non-perturbative tuning schemes yield $ c_M $ and $ c_0 $ for the relevant counterterm. The first error is statistical, the second error is an estimate of systematical uncertainty. Results agree very well though $ c_0 $ yields smaller uncertainties.
 }
 \label{tab: non-perturbative tuning of c}
\end{table}\normalsize

\begin{table}[hbt]
 \center
 \begin{tabular}{|c|c|c|c|}
  \hline
  Scheme                  
  & $ c_M $         & $ c_0 $          \\
  \hline
  $ n_2 $                 
  & $ +0.2255(27) $ & $ +0.2163(8)  $  \\
  $ d_1 $                 
  & $ -0.9459(53) $ & $ -0.9256(18) $  \\
  $ \chi^2/\mathrm{dof} $ 
  & $ 0.458 $       & $ 0.081 $        \\
  \hline
 \end{tabular}
 \caption{The Padé approximation of $ c_0 $ using eq. (\ref{eq: Pade interpolation}) has the smallest uncertainties and agrees with $ c_{BPT} $ for fine lattices within $ 1\sigma $.
 }
 \label{tab: Pade interpolation}
\end{table}\normalsize

\noindent
Non-perturbative results for the relevant counterterm's coefficient $ c(g_0^2) $ are summarised in table~\ref{tab: non-perturbative tuning of c} and figure~\ref{fig: non-perturbative tuning of c}. $ c_M $~and~$ c_0 $ at each gauge coupling are in good agreement and are slightly (\mbox{cf.} fourth column in table~\ref{tab: non-perturbative tuning of c}) more negative than the estimate $ c_{BPT} $ from boosted perturbation theory. The systematical uncertainty of~$ c_M $ is due to finite size effects, the fit range dependence of correlation functions with direction of correlation perpendicular to the alignment of the Karsten-Wilczek term and uncertainties of the chiral extrapolation. The systematical uncertainty of $ c_0 $ is due to the width of the peak in the power spectral density, which is dominated by the ground state mass difference. Variation of~$ c_0 $ due to different~$ m_0 $ or~$ d $ or for different lattice sizes is within statistical uncertainties. The coefficient in each scheme is interpolated with a Padé approximant using~$ n_1 \equiv c_{1L}=-0.249351 $, which ensures that the interpolation reproduces the one-loop result for small couplings. The interpolation 
\begin{equation}
  c(g_0^2) = \frac{n_1 g_0^2 + n_2 g_0^4}{1 + d_1 g_0^2}
  \label{eq: Pade interpolation}
\end{equation}\normalsize
yields parameters that are listed in table~\ref{tab: Pade interpolation}. Finding a numerically robust approach to non-perturbative tuning of the marginal counterterm's coefficient $ d $ is still an open problem. Because estimates from BPT ($ d_{BPT} $ from table~\ref{tab: BPT predictions}) are very small, they may well be sufficient for non-perturbative tuning within uncertainties. \newline

\noindent
The spectroscopic study of the ground state of $ \gamma^5 $~and $ \gamma^{0} $~channels with an approximately tuned action verifies that~$ M_{55}^2 $ receives contributions from quenched chiral logarithms, which agree within uncertainties with phenomenological estimates for all studied couplings. Thus, data are consistent with an interpretation as a \mbox{(Pseudo-)}~Goldstone boson at finite lattice spacing. Consistency of the logarithms amongst different $ \beta $ and with phenomenological estimates improves if a finite offset is included in the chiral extrapolation of~$ M_{55}^2 $. This might indicate incomplete tuning of the counterterms. 
Data provide no numerical evidence for quenched chiral logarithms in the ground state mass~$ M_{00}^2 $ of the $ \gamma^{0} $~channel. Instead, it has a residual mass in the chiral limit that vanishes as $ \mathcal{O}(a^2) $ in the continuum limit. Deviation of the squared ground states mass  difference~$ \Delta_{05} $ from zero in the continuum limit is very small and consistent with the negative offset of~$ M_{55}^2 $ for the finer lattices. Results are inconclusive whether these deviations are due to systematical uncertainties (\mbox{e.g.}~incomplete tuning), due to neglecting higher order terms or simply due to statistical fluctuations. The ground states of $ \gamma^5 $~and $ \gamma^{0} $~channels for tuned parallel Karsten-Wilczek fermions are degenerate in the continuum limit within these uncertainties. Thus, the $ \gamma^{0} $~channel's  ground state is interpreted as a \mbox{(Pseudo-)}~Goldstone boson in the continuum limit. The non-vanishing residual mass term can be understood as a consequence of the explicitly broken non-singlet chiral symmetry at finite lattice spacing \mbox{cf.}~section~\ref{sec: Interim findings (I)}). Lastly, a spectroscopic study in the quenched approximation with pseudoscalar masses below $ 300\,\mathrm{MeV} $ would not have been feasible with Wilson fermions due to exceptional configurations. Clearly, Karsten-Wilczek fermions are protected from exceptional configurations by their residual chiral symmetry. 

\pagebreak
\chapter{Conclusions}\label{sec: Conclusions}

\noindent
For the first time, perturbative studies of minimally doubled fermions have been published in~\cite{Capitani:2009yn,Capitani:2010nn}. The first non-perturbative studies in the quenched approximation have been published in~\cite{Weber:2013tfa,proceeding4}. This thesis wraps up the results of these studies and presents new unpublished research about minimally doubled fermions in lattice QCD. The perturbative studies include two varieties of minimally doubled fermions -- Bori\c{c}i-Creutz and Karsten-Wilczek fermions -- whereas the numerical studies are restricted to Karsten-Wilczek fermions. 
Minimally doubled fermions are among the few types of lattice fermions with an exact local chiral symmetry at finite lattice spacing and an ultralocal Dirac operator that can be inverted with moderate numerical cost. Because minimally doubled fermions reproduce two mass-degenerate quarks, numerical results can be interpreted in terms of QCD in the isospin-symmetric limit. \newline

\noindent
Minimally doubled fermions realise an exact chiral symmetry at finite lattice spacing whilst explicitly breaking the hypercubic symmetry  and symmetry under both charge conjugation and reflections of one particular direction. This particular direction, along which the two poles of their Dirac operators are aligned, is different for Bori\c{c}i-Creutz and Karsten-Wilczek fermions and reflects the orientation of the dimension five operators in their actions. Nevertheless, the actions are invariant under the combination of charge conjugation and the broken reflection symmetry such that $ CP\Theta $~symmetry is always maintained. 
Due to the broken symmetries, renormalisation requires three counter\-terms, whose forms are determined by the broken symmetries and whose coefficients have been computed perturbatively in~\cite{Capitani:2009yn,Capitani:2010nn}. There are two fermionic counterterms -- a relevant counterterm with dimension three and a marginal counterterm with dimension four -- and a marginal gluonic counterterm with dimension four. The broken symmetries are manifest in the form of anisotropies unless the counterterms are tuned properly. Conservation of ultralocal vector and axial symmetry currents is verified on the one-loop level such that the PCAC relation is satisfied exactly. Estimates for the non-perturbative coefficients are obtained using boosted perturbation theory. \newline

\noindent
Different aspects of Karsten-Wilczek fermions are explored in numerical studies using the quenched approximation. In particular, the influence of the relevant counterterm's coefficient 
is studied extensively. 
Signatures of the broken reflection symmetry, which would be seen as a mass splitting between forward and backward propagating states in correlation functions parallel to the Karsten-Wilczek term's alignment, are consistent with zero. Interpretation as a consequence of $ CP\Theta $~symmetry of the action using pure Yang-Mills gauge configurations is suggestive. 
Anisotropies due to inaccurately tuned counterterms are observed in the mass determination for the pseudoscalar channel in directions which are either parallel or perpendicular to the alignment of the Karsten-Wilczek term. Minimisation of the anisotropy is used as a scheme for non-perturbative tuning of the relevant counterterm. Though the anisotropy depends on the marginal counterterm as well, the sensitivity is insufficient for the given statistical accuracy. Hence, use of its estimate from boosted perturbation theory seems to be numerically sufficient as it is very small. \newline

\noindent
Numerical studies reveal that the connection between interpolating operators for Karsten-Wilczek fermions and $ J^{PC} $ of hadrons is more complicated than for Wilson-type fermions, but correlation functions of Dirac bilinears can still be interpreted straightforwardly in terms of mesonic channels. These complications are indicated by the empirical observation of oscillations for most mesonic correlation functions in a direction that is parallel to the alignment of the Karsten-Wilczek term. These oscillations are most significant in the $ \gamma^0 $~and $ \mathbf{1} $~channels, where the ground state belongs to the oscillating contribution. Nevertheless, no oscillations are observed in correlation functions in a direction that is perpendicular to the Karsten-Wilczek term's alignment. The oscillating contributions are interpreted in the spirit of a formal decomposition of spinor fields and of their bilinear operators that are used in correlation functions. Conclusions from this decomposition about $ J^{PC} $ of the states in the oscillating contributions are consistent with their masses in numerical studies. 
Empirical observations indicate that the oscillation pattern changes upon variation of the relevant counterterm's coefficient. Frequency spectra for ratios of correlation functions in the direction that is parallel to the Karsten-Wilczek term's alignment are obtained with methods of discrete Fourier analysis. The numerical result that the frequency spectrum is restored to its tree-level form upon tuning of the relevant counterterm's coefficient serves as a second scheme for non-perturbative tuning. Conclusions from the formal decomposition about the frequency's dependence on the coefficient are consistent with observations. The results from the two different non-perturbative tuning schemes differ at the level of a few percent but are consistent within the overall uncertainties (\mbox{cf.}~table~\ref{tab: non-perturbative tuning of c}). \newline

\noindent
Correlation functions in the pseudoscalar sector are studied in the chiral regime for pseudoscalar masses well below $ 300\,\mathrm{MeV} $. This mass range is inaccessible with Wilson fermions in the quenched approximation due to exceptional configurations. The study shows that Karsten-Wilczek fermions are not affected by exceptional configurations. 
A chiral extrapolation of the pseudoscalar mass is performed in the spirit of chiral perturbation theory. Strictly speaking, the chiral limit does not exist due to pathologies of the quenched approximation, which manifest themselves in the form of quenched chiral logarithms. Their size as estimated from numerical data agrees with other determinations. The pseudoscalar mass is consistent with zero in the limit of vanishing quark mass. 
An approximate degeneracy between the ground state of the oscillating contribution in the $ \gamma^0 $~channel and the ground state of the pseudoscalar channel is observed empirically. This mass splitting is nearly independent of the quark mass and consistent with zero in the continuum limit. Since this state is interpreted as $ J^{PC}=0^{-+} $ in the spirit of the formal decomposition, both ground states are interpreted as pion states. \newline

\noindent
As a natural next step it is necessary to move on from the quenched approximation to simulations with \textit{dynamical minimally doubled fermions}. The best chances for success rests with an implementation of Karsten-Wilczek fermions within very general, preexisting frameworks such as USQCD~\cite{Holmgren:2005wf} or the MILC~\cite{milccode} code. In a simulation with dynamical fermions, the gluonic counterterm will be tuned by the prescription that the average plaquette value of \textit{parallel} and \textit{perpendicular plaquettes} (including or excluding the direction of the Karsten-Wilczek term's alignment) must be equal. Whether or not this prescription is sufficiently independent of the other counterterms' coefficients remains to be seen, but the idea seems very straightforward compared to the fermionic counterterms. 
\noindent
The method of stochastic sources can be applied to deal with the second systematical deficit of present simulations, as it may be applied in a calculation of \textit{quark-disconnected contributions}. These quark-disconnected contributions which discriminate between charged and neutral pions are indispensable for an unambiguous identification of the physical pion states that are seen in the different channels containing pseudoscalars. 
\noindent
The analytical and numerical studies within this thesis are the first proof that non-perturbative studies of QCD with minimally doubled fermions are feasible and yield a mass spectrum which exhibits the correct features as expected from QCD. Even though studies in the quenched approximation are in a way dissatisfying compared to the success of Lattice QCD with dynamical fermions using other discretisations, the present study will be the cornerstone upon which any future application with minimally doubled fermions will rest.

\chapter*{Acknowledgements}

\appendix
\chapter{Conventions}\label{app: Conventions}

\sectionc{Physical constants}{app: Physical constants}

\noindent
This section declares the unit conventions throughout this thesis. Natural units are applied throughout this thesis, i.e.
\begin{equation}
  c=1=\hbar.
\end{equation}
Therefore, physical quantities have the following length dimensions:
\begin{equation}
  [\text{length}] = [\text{time}] = [\text{energy}]^{-1} = [\text{mass}]^{-1}.
\end{equation}
For the purpose of conversion to other systems of units, the product of the quantum of action and the speed of light in vacuum as well as the speed of light in vacuum are given in SI units by
\begin{equation}
  \hbar c = 197.326968(17)\, \mathrm{MeV}\, \mathrm{fm},\quad
  c = 299792458\, \mathrm{m}\, \mathrm{s}^{-1}.
\end{equation}

\sectionc{Indices}{app: Indices}

\noindent
This section declares the index conventions throughout this thesis.

\subsubsection{Space-time indices}

\noindent
Space-time indices are labelled by Greek letters mostly from the middle of the Greek alphabet. If they are applied in Minkowski space-time $ \mathbb{M}^4 $, their range is $ \mu \in \{0,1,2,3\} $. If they are applied in Euclidean space-time $ \mathbb{R}^4 $, their range is $ \mu \in \{0,1,2,3\} $ as well. Space-time indices, which occur multiply are always considered to be summed over. Exceptions are explicitly highlighted. Fixed indices, which are never summed are marked by an underscore, \mbox{e.g.}~$ \underline{\mu} $.

\subsubsection{Spinor indices}

\noindent
Spinor indices are labelled by Greek letters mostly from the beginning of the Greek alphabet. Their range is $ \alpha \in \{1,2,3,4\} $. Spinor indices which occur twice are always considered to be summed over according to Einstein's summation convention.

\subsubsection{Colour indices}

\noindent
Colour indices are labelled by Latin letters mostly from the start of the alphabet. The range of colour indices is $ a \in \{1,2,3\}$ for the fundamental representation and $ a \in \{1,2,3,4,5,6,7,8\}$ for the adjoint representation. Colour indices which occur twice in either the fundamental or adjoint representation are always considered to be summed over according to Einstein's summation convention. Exceptions are explicitly highlighted. Fixed indices, which are never summed are marked by an underscore, \mbox{e.g.}~$ \underline{a} $.

\subsubsection{Flavour indices}

\noindent
Flavour indices in the fundamental representation are labelled by bracketed numbers. Their general range is $ f \in \{1,2,3,4,5,6\} $ and they are sorted in the order of ascending quark mass. The spinor's symbol $ \psi $ can be replaced by the one-letter label of the flavours, \mbox{e.g.}~$ \psi^{(1)}\equiv u $, $ \psi^{(2)}\equiv d $, $ \ldots $ and the quark mass parameters can be labeled as $ m^{(1)}=m_u $, $ m^{(2)}=m_d $, $ \ldots $. Summation of flavour indices is always declared explicitly.

\subsubsection{Space-time coordinates}

\noindent
Space-time coordinates $ x $ of field variables $ \psi $ defined on a space-time continuum are represented as arguments of the fields, \mbox{e.g.}~$ \psi(x) $. Space-time coordinates are labelled with latin letters from the end of the alphabet. In the contrary, site labels $ n $ on discretised space-time lattice are dimensionless numbers defined as $ n_\mu=x_\mu/a $. Site labels are represented as site indices $ \psi_n $. Sites labels are usually labelled by letters from the middle of the alphabet. Integration over space-time coordinates or summation over site labels are always declared explicitly. The summation range of site labels is usually omitted.

\subsubsection{Momentum space coordinates}

\noindent
The vector space of four-momenta is referred to as momentum space throughout this thesis. Its coordinates $ k_\mu $ are always represented as arguments of the fields, $ \psi(k) $. Integration over momentum space coordinates is always declared explicitly.

\sectionc{SU(N)~Matrices}{app: SU(N)-Matrices}
\noindent
This section declares the conventions concerning SU(N)~matrices.

\subsubsection{SU(2) -- Pauli matrices}
The Pauli matrices $ \sigma^{\mu} $ are defined by
\begin{equation}
  \sigma^1=\left(\begin{array}{cc} 0 & 1 \\ 1 & 0 \end{array}\right), \quad
  \sigma^2=\left(\begin{array}{cc} 0 & -i \\ i & 0 \end{array}\right), \quad
  \sigma^3=\left(\begin{array}{cc} 1 & 0 \\ 0 & -1 \end{array}\right).
\end{equation}
The $ 2\times2 $-unit matrix is included as
\begin{equation}
  \sigma^0=\left(\begin{array}{cc} 1 & 0 \\ 0 & 1 \end{array}\right).
\end{equation}
As generators of the symmetry group SU(2), they are defined with an extra factor $ 1/2 $,
\begin{equation}
  \tau_j=\frac{1}{2}\sigma^j
\end{equation}
and satisfy
\begin{equation}
  \tau^i \tau^j=\frac{1}{2}\left(\delta^{ij} +i\epsilon^{ijk}\tau^k\right).
\end{equation}

\subsubsection{SU(3) -- Gell-Mann matrices}
The Gell-Mann matrices $ \lambda^{j} $ are defined by
\small
\begin{align}
  \lambda^1=&\left(\begin{array}{ccc} 0 & 1 & 0 \\ 1 & 0 & 0 \\ 0 & 0 & 0 \end{array}\right), &&
  \lambda^2=&\left(\begin{array}{ccc} 0 & -i & 0 \\ i & 0 & 0 \\ 0 & 0 & 0 \end{array}\right), &&
  \lambda^3=&\left(\begin{array}{ccc} 1 & 0 & 0 \\ 0 & -1 & 0 \\ 0 & 0 & 0 \end{array}\right), \nonumber \\
  \lambda^4=&\left(\begin{array}{ccc} 0 & 0 & 1 \\ 0 & 0 & 0 \\ 1 & 0 & 0 \end{array}\right), &&
  \lambda^5=&\left(\begin{array}{ccc} 0 & 0 & -i \\ 0 & 0 & 0 \\ i & 0 & 0 \end{array}\right), &&
  \lambda^6=&\left(\begin{array}{ccc} 0 & 0 & 0 \\ 0 & 0 & 1 \\ 0 & 1 & 0 \end{array}\right), \nonumber \\
  \lambda^7=&\left(\begin{array}{ccc} 0 & 0 & 0 \\ 0 & 0 & -i \\ 0 & i & 0 \end{array}\right), &&
  \lambda^8=&\frac{1}{\sqrt{3}}\left(\begin{array}{ccc} 1 & 0 & 0 \\ 0 & 1 & 0 \\ 0 & 0 & -2 \end{array}\right). &&
\end{align}\normalsize
As generators of the symmetry group SU(3), they are defined with an extra factor $ 1/2 $,
\begin{equation}
  T_j=\frac{1}{2}\lambda^j
\end{equation}
and satisfy
\begin{equation}
  T^i T^j=\frac{1}{2}\left(\frac{1}{3}\delta^{ij}\mathbf{1}+(d^{ijk}+i f^{ijk})T_k\right).
  \label{eq: product of SU(3) generators}
\end{equation}

\subsubsection{Casimir operators}
The Casimir operators of the group SU(N) are given by
\begin{align}
  C(R)=&\ \frac{1}{2}
   = \mathrm{Tr}(T_{\underline{a}}T_{\underline{a}}) 
   \label{eq: Casimir operator C_2} \\ 
  C_2(R) 1_{N\times N}
  =&\ \sum\limits_{a}T_{a}T_{a}
   = \frac{N^2-1}{2N} 1_{N\times N},\quad C_2(R)\equiv C_F 
   \label{eq: Casimir operator C_F} \\ 
  C_2 \equiv C_2(G) 
  =&\ C(G) =f^{\underline{a}cd}f^{\underline{a}cd} = N,
\end{align}
where $ R $ is the fundamental and $ G $ is the adjoint representation.

\sectionc{Minkowski and Euclidean space-time}{app: Minkowski and Euclidean space-time}
\noindent
This section declares the conventions concerning Minkowski and Euclidean space-time throughout this thesis. Quantities defined on a Minkowski or Euclidean space-time are always labelled with upper indices $ M $ or $ E $ if there is any possibility of an ambiguity. If ambiguities can be excluded, these labels are omitted.

\subsection{Coordinates and four-vectors}
This thesis mostly uses Euclidean space-time. Four-vectors defined on Euclidean space-time are connected with four-vectors on Minkowski space-time by a Wick rotation. The Wick rotation of the $ x_0 $ coordinate is defined as
\begin{equation}
  x_0^E \equiv i x_0^M.
\end{equation}
The Minkowski space-time coordinate $ x_0^M $ is identical to the usual physical time $ t $. Four-momenta $ p_\mu $, derivatives $ \partial_\mu $ as well as four-vector fields $ A_\mu(x) $ have their temporal components transformed by the Wick-rotation with the conjugate factor $ (-i) $,
\begin{align}
  p_0^E =&\ -i p_0^M, \\
  \partial_0^E =\frac{\partial}{\partial x^{0\,E}} =&\ \frac{1}{i}\frac{\partial}{\partial x^{0\,M}} = -i \partial_0^M, 
  \label{eq: Wick rotation of derivatives} \\
  A_0^E(x^E) =&\ -i A_0^M(x^M) \label{eq: Wick rotation of vector fields}.
\end{align}
Four-vectors are defined in Minkowski and Euclidean space-time by
\begin{align}
  x^{M\,\mu} =&\ (x_0^M,\mathbf{x}^M), \\
  x_\mu^M=g_{\mu\nu}x^{M\,\mu} =&\ (x_0^M,-\mathbf{x}^M), \\
  x_\mu^E = x^{E\,\mu } =&\ (x_0^E,\mathbf{x}^E)
\end{align}
and scalar products are
\begin{align}
  x^M \cdot y^M =&\ x_\mu^M y^{M\,\mu} = (x_0^M y_0^M) - (\mathbf{x}^M \cdot \mathbf{y}^M), \\
  x^E \cdot y^E =&\ x_\mu^E y_\mu^E = (x_0^E y_0^E) + (\mathbf{x}^E \cdot \mathbf{y}^E).
\end{align}
Three-dimensional vectors are in bold print (\mbox{e.g.}~$ \mathbf{x} $). The scalar product of four-vectors and three-vectors are both abbreviated as dot products. The metric of Minkowski space-time is defined as
\begin{equation}
  g_{\mu\nu}=\mathrm{diag}\,(+1,-1,-1,-1).
\end{equation}
Due to the aforementioned transformation properties of derivatives in \mbox{eq.} (\ref{eq: Wick rotation of derivatives}) and vector fields in \mbox{eq.} (\ref{eq: Wick rotation of vector fields}), the field strength tensor $ F_{\mu\nu}(x) $ transforms as
\begin{equation}
  F^{E\,\mu\nu}(x^E) = F^{E}_{\mu\nu}(x^E) 
  = (-i)^{(\delta^{\mu0}+\delta^{\nu0})}\ F^{M}_{\mu\nu}(x^M) 
  = i^{(\delta^{\mu0}+\delta^{\nu0})}\ F^{M\,\mu\nu}(x^M).
\end{equation}
Thus, the product $ F^E_{\mu\nu}(x^E)F^E_{\mu\nu}(x^E)=F^M_{\mu\nu}(x^M)F^{M\,\mu\nu}(x^M) $ is not multiplied by any extra factors in the Wick rotation. \newline

\noindent
Spinor fields and adjoint spinor fields are treated as independent degrees of freedom in the Osterwalder-Schrader approach~\cite{Osterwalder:1973kn, Osterwalder:1973dx} to a Euclidean field theory, but are otherwise left unchanged:
\begin{equation}
  \psi^E(x^E) = \psi^M(x^M), \quad \bar\psi^E(x^E) = \bar\psi^M(x^M).
\end{equation}\normalsize
While this is not of particular importance in the path integral approach, which already treats $ \psi $ and $ \bar\psi $ as independent field in Minkowski space-time, the OS approach causes a loss of hermiticity for the Dirac action, which can be avoided by Waldron's elaborate scheme for the Wick rotation including spinor rotations~\cite{vanNieuwenhuizen:1996ip,vanNieuwenhuizen:1996tv,Waldron:1997re}. Throughout this thesis, the OS approach to Euclidean field theories is applied.

\subsection{Dirac matrices}\label{app: Dirac matrices}
Dirac matrices are defined as
\begin{align}
  \gamma^{M\,0}=&\ \left(\begin{array}{cc} 0 & \sigma^0 \\ \sigma^0 & 0 \end{array}\right), & 
  \gamma^{M\,j}=&\ \left(\begin{array}{cc} 0 & \sigma^j \\ -\sigma^j & 0 \end{array}\right), 
  \label{eq: Minkowski Dirac matrices} \\
  \gamma^{E\,0}=&\ \left(\begin{array}{cc} 0 & \sigma^0 \\ \sigma^0 & 0 \end{array}\right) = \gamma^{M\,0}, & 
  \gamma^{E\,j}=&\ \left(\begin{array}{cc} 0 & -i\sigma^j \\ i\sigma^j & 0 \end{array}\right) = -i\gamma^{M\,j},
  \label{eq: Euclidean Dirac matrices}
\end{align}\normalsize
which corresponds to the chiral representation in both Minkowski and Euclidean space-time. The Dirac matrix $ \gamma^0 $ is real, wheres the spatial Dirac matrices are anti-hermitian in Minkowski space-time and hermitian in Euclidean space-time. They satisfy the Clifford algebra,
\begin{equation}
  \left\{\gamma^{M\,\mu},\gamma^{M\,\nu}\right\} = 2g^{\mu\nu}, \quad
  \left\{\gamma^{E\,\mu},\gamma^{E\,\nu}\right\} = 2\delta^{\mu\nu}.
\end{equation}\normalsize
Contrary to standard conventions, Euclidean space-time Dirac matrices are usually written with upper indices throughout the thesis. 
The chirality matrix $ \gamma^5 $, which anticommutes with all Dirac matrices $ \gamma^\mu $, is defined by
\begin{equation}
  \gamma^{M\,5}= i\gamma^{M\,0}\gamma^{M\,1}\gamma^{M\,2}\gamma^{M\,3}, \quad
  \gamma^{E\,5}= \gamma^{E\,0}\gamma^{E\,1}\gamma^{E\,2}\gamma^{E\,3}
\end{equation}\normalsize
and is equal in Minkowski and Euclidean space-time:
\begin{equation}
  \gamma^5=\gamma^{M\,5}=\gamma^{E\,5}=\left(\begin{array}{cc} \sigma^0 & 0 \\ 0 & -\sigma^0 \end{array}\right).
\end{equation}
The chirality matrix is usually written with an upper index as $ \gamma^5 $ in this thesis. 
The chirality projectors are defined as
\begin{equation}
  P_R=\frac{1}{2}(1+\gamma^5),\quad P_L=\frac{1}{2}(1-\gamma^5)
  \label{eq: chirality projectors}
\end{equation}\normalsize
and parity projectors (for fermions) are defined as
\begin{equation}
  P_+=\frac{1}{2}(1+\gamma^0),\quad P_-=\frac{1}{2}(1-\gamma^0).
\end{equation}\normalsize
The charge conjugation matrix $ C $ is defined as
\begin{equation}
  C=i\gamma_0\gamma_2 = C^\dagger = C^{-1} =- C^T
  \label{eq: charge conjugation matrix}
\end{equation}\normalsize
both on Minkowski and Euclidean space-times and satisifies
\begin{equation}
  \gamma^\mu C = C^T\gamma^{\mu\,T}.
\end{equation}\normalsize

\subsection{Continuous space-time and discrete space-time lattices}\label{app: Continuous space-time and discrete space-time lattices}
Events in a four-dimensional continuous space-time $ V $ of infinite extent,
\begin{equation}
  V=\{x=(x_0,x_1,x_2,x_3)\ |\  -\infty<x_\mu<\infty\},
\end{equation}\normalsize
are labelled by four-component coordinate vectors $ x $, which have the dimension of a $ [\mathrm{length}] $. This definition can be restricted to any subspace of finite extent,
\begin{equation}
  V=\{x=(x_0,x_1,x_2,x_3)\ |\  0<x_\mu<L_\mu\},
\end{equation}\normalsize
where the size of the subspace introduces a length scale. Any finite subspace must be supplied with appropriate boundary conditions. Throughout this thesis, boundary conditions will be generally chosen as periodic or anti-periodic, e.g. any function satisfies
\begin{equation}
  f(x_\mu+L_\mu)= \pm f(x_\mu).
\end{equation}
In a discretisation, continuous space-time is replaced by a finite space-time lattice $ \Lambda $,
\begin{equation}
  \Lambda = \{ n=(n_0,n_1,n_2,n_3)\ |\ 0\leq n_\mu < N_\mu\}.
  \label{eq: Euclidean space-time lattice}
\end{equation}\normalsize
Events on this lattice are labelled by a four-component site vector $ n $, which is dimensionless. It is connected to a coordinate vector of dimension $ [\mathrm{length}] $ as 
\begin{equation}
  x_\mu= a n_\mu
  \label{eq: site vectors}
\end{equation}
by a lattice spacing $ a $, which defines the intrinsic length scale of the system (the intrinsic length scale of the continuous system is defined in terms of the lattice spacing and the number of sites: $ L_\mu = a N_\mu $). Sites $ n $ of a discretised space-time lattice can be sorted into even and odd sites, which are distinguished by whether the sum of components
\begin{equation}
  n_\Sigma=\sum\limits_{\mu=0}^{3}n_\mu
  \label{eq: even-odd site number}
\end{equation}
is an even or an odd number. The volume is defined as $ V=a^4\,|\Lambda| $, where $ |\Lambda| = \prod_\mu N_\mu $ takes the role of a dimensionless volume. Slices of the lattice are defined for any direction by
\begin{equation}
  \Lambda_m^{\underline{\mu}} = \{n\in \Lambda\ | n_{\underline{\mu}} = m = \mathrm{const} \}.
  \label{eq: slices}
\end{equation}\normalsize

\sectionc{Fourier transformations}{app: Fourier transformations}
\noindent
Fourier transformations connect space-time $ V $ to its Fourier space $ \widetilde{V} $, which is usually called momentum space throughout this thesis. 

\subsubsection{Infinite space-time}\label{app: Infinite continuous volume}
The Fourier space $ \widetilde{V} $ of an infinite space-time $ V $ is also infinite. The Fourier transform $ \widetilde{f}(p) $ of a real function $ f(x) $  is defined by
\begin{align}
  \widetilde{f}(p)=&\int\limits_{V} f(x) e^{-i x\cdot p} d^4x, \\
  f(x)=&\int\limits_{\widetilde{V}} \widetilde{f}(p) e^{i x\cdot p} \frac{d^4p}{(2\pi)^4}.
\end{align}\normalsize
Hence, complex functions $ \phi(x) $ are transformed to $ \widetilde{\phi}(p) $ according to
\begin{align}
  \widetilde{\phi}(p)=&\int\limits_{V} \phi(x) e^{-i x\cdot p} d^4x, \label{eq: complex funtion, Fourier transform, continuous space time, start} \\
  \phi(x)=&\int\limits_{\widetilde{V}} \widetilde{\phi}(p) e^{i x\cdot p} \frac{d^4p}{(2\pi)^4}, \\
  \widetilde{\phi^\dagger}(p)=&\int\limits_{V} \phi^\dagger(x) e^{i x\cdot p} d^4x, \\
  \phi^\dagger(x)=&\int\limits_{\widetilde{V}} \widetilde{\phi^\dagger}(p) e^{-i x\cdot p} \frac{d^4p}{(2\pi)^4}. \label{eq: complex funtion, Fourier transform, continuous space time, end} 
\end{align}\normalsize
Because it is always clear from the arguments, the Fourier transforms of the fields $ \psi $, $ \bar\psi $ and $ A $ are not marked with a tilde ($\,\widetilde{}\,$) explicitly. Delta functions, which naturally arise in Fourier transformations of local products of fields, are generated as
\begin{align}
  \delta(p-q)\equiv& \delta(p_0-q_0)\delta(p_1-q_1)\delta(p_2-q_2)\delta(p_3-q_3) = \int\limits_{V} e^{-i x\cdot (p-q)} d^4x, \label{eq: momentum delta function}\\
  \delta(x-y)\equiv& \delta(x_0-y_0)\delta(x_1-y_1)\delta(x_2-y_2)\delta(x_3-y_3) = \int\limits_{\widetilde{V}} e^{i (x-y)\cdot p} \frac{d^4p}{(2\pi)^4}. \label{eq: coordinate delta function}
\end{align}\normalsize

\subsubsection{Finite space-time lattice}
The discretised Fourier space is the Brillouin zone, $ \widetilde{V}=(2\pi/a)^4=|\Lambda|\,\prod_\mu(2\pi/(aN_\mu)) $. The Fourier transform $ f(k) $ of a real function $ f_n $ on a finite space-time lattice is defined by (complex functions are analogous to \mbox{eqs.}~(\ref{eq: complex funtion, Fourier transform, continuous space time, start})-(\ref{eq: complex funtion, Fourier transform, continuous space time, end}) )
\begin{align}
  \widetilde{f}(k)=&\frac{1}{\sqrt{|\Lambda|}}\sum\limits_{n \in \Lambda} f_n e^{-ia n\cdot k},   \label{eq: discrete Fourier transform}\\
  f_n=&\frac{1}{\sqrt{|\Lambda|}}\sum\limits_{k \in \widetilde{\Lambda}} \widetilde{f}(k) e^{ia n\cdot k}.
  \label{eq: inverse discrete Fourier transform}
\end{align}\normalsize
The analogues of the delta functions of \mbox{eqs.}~(\ref{eq: momentum delta function}) and~(\ref{eq: coordinate delta function}) are given by
\begin{align}
  \delta(k,l)\equiv \delta_{k_0 l_0}\delta_{k_1 l_1}\delta_{k_2 l_2}\delta_{k_3 l_3} &= \frac{1}{|\Lambda|}\sum\limits_{n \in \Lambda} e^{-ia n\cdot (k-l)}, \label{eq: lattice momentum delta function}\\
  \delta_{n,m}\equiv \delta_{n_0 m_0}\delta_{n_1 m_1}\delta_{n_2 m_2}\delta_{n_3 m_3} &= \frac{1}{|\Lambda|}\sum\limits_{k \in \widetilde{\Lambda}} e^{ia (n-m)\cdot k}. \label{eq: site delta function}
\end{align}\normalsize
Generalisation of the (inverse) discrete Fourier transform of real functions to complex functions is analogous to the treatment in the continuous case (\mbox{cf.} \ref{app: Infinite continuous volume}).

\chapter{Addendum to perturbative studies}\label{app: Addendum to perturbative studies}

\noindent
This appendix covers some particularly lengthy expression, which appear in perturbative calculations, in more detail.

\sectionc{Recursion relations for bosonic integrals}{sec: Recursion relations for bosonic integrals}

\noindent
This appendix summarises the recursion relations for basic bosonic integrals in LPT and clarifies the nomenclature. All recursion relations have been taken from~\cite{Capitani:2002mp}, chapter~18. Their practical application is done with a FORM~\cite{Vermaseren:2000nd, Kuipers:2012rf} program, which has been provided by Stefano Capitani. The basic integrals of \mbox{eq.}~(\ref{eq: general bosonic integral B}) satisfy recursion relations. Absent momentum components in the numerator are left out in the nomenclature, \mbox{e.g.}
\begin{equation}
  \mathcal{B}(p;n_0,n_3) \equiv \mathcal{B}(p;n_0,0,0,n_3),\quad 
  \mathcal{B}(p) \equiv \mathcal{B}(p;0,0,0,0).
\end{equation}\normalsize
\noindent
The first set of recursion relations removes a first power of a momentum component $ \hat{k}_\mu^2 $ in the numerator,
\begin{align}
  4\mathcal{B}(p;1) =& \mathcal{B}(p-1) - M^2\mathcal{B}(p) \\
  3\mathcal{B}(p;x,1) =& \mathcal{B}(p-1;x) - \mathcal{B}(p;x+1) - M^2\mathcal{B}(p;x) \\
  2\mathcal{B}(p;x,y,1) =& \mathcal{B}(p-1;x,y) - \mathcal{B}(p;x+1,y) - \mathcal{B}(p;x,y+1) - M^2\mathcal{B}(p;x,y) \\
  \mathcal{B}(p;x,y,z,1) =& \mathcal{B}(p-1;x,y,z) - \mathcal{B}(p;x+1,y,z) - \mathcal{B}(p;x,y+1,z) 
  \nonumber \\
  &- \mathcal{B}(p;x,y,z+1) - M^2\mathcal{B}(p;x,y,z).
\end{align}\normalsize
The second recursion relation lowers the power of a momentum component $ \hat{k}_\mu^{2r} $ in the numerator,
\small
\begin{equation}
  \mathcal{B}(p;\ldots,r) = \frac{r-1}{p-1}\mathcal{B}(p-1;\ldots,r-1) -\frac{4r-6}{p-1}\mathcal{B}(p-1;\ldots,r-2)+4 \mathcal{B}(p;\ldots,r-1).
\end{equation}\normalsize

\sectionc{Addendum to the self-energy calculation}{app: addendum to the self-energy calculation}

\subsection{Sunset diagram for Karsten-Wilczek fermions}\label{app: Sunset diagram for Karsten-Wilczek fermions}

\noindent
Various particular unwieldy expressions of the one-loop calculation of the fermionic self-energy are presented only in this addendum to the perturbative studies. The algebraically simplified numerators of $ J_3^\chi(\zeta,a) $ read
\small
\begin{align}
  N_0 =&\ s^\chi_k \big(\{2(\hat{c}^\chi_k)^2-(\hat{c}_k)^2\}
  +\delta^{\chi\underline{\alpha}}\zeta^2 (\hat{s}_k)_{\!\perp}^2\big)
  \nonumber \\&\
  +
  \zeta\big\{4 s^{\underline{\alpha}}_k s^\chi_k \varepsilon^{\underline{\alpha}\chi} +
   \delta^{\chi\underline{\alpha}}\big(4 (s_k)_{\!\perp}^2 + \frac{1}{2}(\hat{s}_k)_{\!\perp}^2 \{2(\hat{c}^{\underline{\alpha}}_k)^2-(\hat{c}_k)^2+\zeta^2(\hat{s}_k)_{\!\perp}^2\} \big)\big\}
 \label{eq: N_0 algebraically simplified}\\
 N_1 =&\ s^\chi_k \Big(\big\{4 (s_k)^2
  \!+\! \zeta^2 (1 \!+\! 2\delta^{\chi\underline{\alpha}})\{(\hat{s}_k)_\perp^2\}^2\big\}
  \!+\!
  \zeta(\hat{s}_k)_{\!\perp}^2 \big\{4 s^{\underline{\alpha}}_k s^\chi_k \!+\!
   \delta^{\chi\underline{\alpha}}\big(2 (s_k)^2 \!+\! \frac{\zeta^2}{2}\{(\hat{s}_k)_{\!\perp}^2\}^2 \big)\big\}\Big).
 \label{eq: N_1 algebraically simplified}
\end{align}\normalsize
The denominators of the sunset diagram that contribute to wavefunction renormalisation and the marginal counterterm's coefficient for Karsten-Wilczek fermions read
\begin{align}
  D_4 =&\ D^{KW}(k;\zeta,M,a)D^{g}(-k;M,a), 
  \label{eq: KW s-e sunset wf ren den D4} \\
  D_5 =&\ D^{KW}(k;\zeta,M,a)\left(D^{g}(-k;M,a)\right)^2, 
  \label{eq: KW s-e sunset wf ren den D5} \\
  D_6 =&\ D^{KW}(k;\zeta,M,a)\left(D^{g}(-k;M,a)\right)^3.
  \label{eq: KW s-e sunset wf ren den D6}
\end{align}\normalsize
Five numerators $ N_4 $ - $ N_8 $ contribute to wavefunction renormalisation and to the marginal counterterm's coefficient. The full numerators $ N_4 $ and $ N_5 $ of the part in Feynman gauge ($ J_4^{\chi\theta}|_{\xi=1} $) read
\footnotesize
\begin{align}
  N_{4} =&\ \sum\limits_{\mu,\nu,\lambda}\frac{\delta^{\mu\nu}}{8}\mathrm{tr}\Big\{
  \Big(\gamma^\chi \delta^{\mu\theta}
  \left(-\gamma^\mu \hat{s}^\mu_{k} +\zeta \gamma^{\underline{\alpha}}\varepsilon^{\underline{\alpha}\mu}\hat{c}^\mu_{k}\right)
  \left( \gamma^\lambda s^\lambda_k +\zeta \gamma^{\underline{\alpha}}\varepsilon^{\underline{\alpha}\lambda}\{1-c^\lambda_k\} \right) 
  \left(\gamma^\nu \hat{c}^\nu_{k} +\zeta \gamma^{\underline{\alpha}}\varepsilon^{\underline{\alpha}\nu}\hat{s}^\nu_{k}\right) 
  \Big)
  \nonumber \\
  &\ + \Big(\gamma^\chi \delta^{\nu\theta}
  \left(\gamma^\mu \hat{c}^\mu_{k} +\zeta \gamma^{\underline{\alpha}}\varepsilon^{\underline{\alpha}\mu}\hat{s}^\mu_{k}\right) 
  \left( \gamma^\lambda s^\lambda_k +\zeta \gamma^{\underline{\alpha}}\varepsilon^{\underline{\alpha}\lambda}\{1-c^\lambda_k\} \right) 
  \left(-\gamma^\nu \hat{s}^\nu_{k} +\zeta \gamma^{\underline{\alpha}}\varepsilon^{\underline{\alpha}\nu}\hat{c}^\nu_{k}\right)
  \Big)
  \Big\}, 
  \label{eq: KW s-e sunset wf ren full N4} \\
  N_{5} =&\ \sum\limits_{\mu,\nu,\lambda}\frac{\delta^{\mu\nu}}{4}\mathrm{tr}\Big\{
  \gamma^\chi 
  \left(\gamma^\mu \hat{c}^\mu_{k} +\zeta \gamma^{\underline{\alpha}}\varepsilon^{\underline{\alpha}\mu}\hat{s}^\mu_{k}\right) 
  \left( \gamma^\lambda s^\lambda_k +\zeta \gamma^{\underline{\alpha}}\varepsilon^{\underline{\alpha}\lambda}\{1-c^\lambda_k\} \right) 
  \left(\gamma^\nu \hat{c}^\nu_{k} +\zeta \gamma^{\underline{\alpha}}\varepsilon^{\underline{\alpha}\nu}\hat{s}^\nu_{k}\right) 
  \Big\}  \left(\hat{c}_{k}^\theta \hat{s}_{k}^\theta\right),
   \label{eq: KW s-e sunset wf ren full N5}
\end{align}\normalsize
and the numerators $ N_6 $, $ N_7 $ and $ N_8 $ of the gauge fixing part ($ -(1-\xi)\left(\partial J_4^{\chi\theta}/\partial \xi\right) $) read
\footnotesize
\begin{align}
  N_{6} =&\ \sum\limits_{\mu,\nu,\lambda}\frac{1}{8}\mathrm{tr}\Big\{
  \Big(\gamma^\chi \delta^{\mu\theta}
  \left(-\gamma^\mu \hat{s}^\mu_{k} +\zeta \gamma^{\underline{\alpha}}\varepsilon^{\underline{\alpha}\mu}\hat{c}^\mu_{k}\right)
  \left( \gamma^\lambda s^\lambda_k +\zeta \gamma^{\underline{\alpha}}\varepsilon^{\underline{\alpha}\lambda}\{1-c^\lambda_k\} \right) 
  \left(\gamma^\nu \hat{c}^\nu_{k} +\zeta \gamma^{\underline{\alpha}}\varepsilon^{\underline{\alpha}\nu}\hat{s}^\nu_{k}\right) 
  \Big)
  \nonumber \\
  &\ + \Big(\gamma^\chi \delta^{\nu\theta}
  \left(\gamma^\mu \hat{c}^\mu_{k} +\zeta \gamma^{\underline{\alpha}}\varepsilon^{\underline{\alpha}\mu}\hat{s}^\mu_{k}\right) 
  \left( \gamma^\lambda s^\lambda_k +\zeta \gamma^{\underline{\alpha}}\varepsilon^{\underline{\alpha}\lambda}\{1-c^\lambda_k\} \right) 
  \left(-\gamma^\nu \hat{s}^\nu_{k} +\zeta \gamma^{\underline{\alpha}}\varepsilon^{\underline{\alpha}\nu}\hat{c}^\nu_{k}\right)
  \Big)
  \Big\}\left(\hat{s}_{k}^\mu \hat{s}_{k}^\nu\right), 
  \label{eq: KW s-e sunset wf ren full N6} \\
  N_{7} =&\ \sum\limits_{\mu,\nu,\lambda} \!-\!\frac{\delta^{\mu\theta}\hat{c}_{k}^\mu \hat{s}_{k}^\nu \!+\! \delta^{\nu\theta}\hat{s}_{k}^\mu \hat{c}_{k}^\nu}{8}\mathrm{tr}\Big\{
  \gamma^\chi 
  \left(\gamma^\mu \hat{c}^\mu_{k} +\zeta \gamma^{\underline{\alpha}}\varepsilon^{\underline{\alpha}\mu}\hat{s}^\mu_{k}\right) 
  \left( \gamma^\lambda s^\lambda_k +\zeta \gamma^{\underline{\alpha}}\varepsilon^{\underline{\alpha}\lambda}\{1-c^\lambda_k\} \right) 
  \left(\gamma^\nu \hat{c}^\nu_{k} +\zeta \gamma^{\underline{\alpha}}\varepsilon^{\underline{\alpha}\nu}\hat{s}^\nu_{k}\right) 
  \Big\},
   \label{eq: KW s-e sunset wf ren full N7} \\
  N_{8} =&\ \sum\limits_{\mu,\nu,\lambda}-\frac{(\hat{c}_{k}^\theta \hat{s}_{k}^\theta)(\hat{s}_{k}^\mu \hat{s}_{k}^\nu)}{2}\mathrm{tr}\Big\{
  \gamma^\chi 
  \left(\gamma^\mu \hat{c}^\mu_{k} +\zeta \gamma^{\underline{\alpha}}\varepsilon^{\underline{\alpha}\mu}\hat{s}^\mu_{k}\right) 
  \left( \gamma^\lambda s^\lambda_k +\zeta \gamma^{\underline{\alpha}}\varepsilon^{\underline{\alpha}\lambda}\{1-c^\lambda_k\} \right) 
  \left(\gamma^\nu \hat{c}^\nu_{k} +\zeta \gamma^{\underline{\alpha}}\varepsilon^{\underline{\alpha}\nu}\hat{s}^\nu_{k}\right) 
  \Big\}.
   \label{eq: KW s-e sunset wf ren full N8}
\end{align}\normalsize
Summation of Euclidean indices and evaluation of the trace  simplifies the numerators to
\small
\begin{align}
  N_{4} =&\ s^\theta_k s^\chi_k \left\{2(1-2\delta^{\theta\chi}) + \zeta^2 \left(-1+2\delta^{\theta\underline{\alpha}}+4\delta^{\chi\underline{\alpha}}-\delta^{\theta\chi} \{4\delta^{\chi\underline{\alpha}}+(\hat{s_k})_\perp^2\varepsilon^{\chi\underline{\alpha}}\} \right)\right\}
  \nonumber \\
  &\ + \zeta\left\{s^{\underline{\alpha}}_k\delta^{\theta\chi}\left((\hat{c}^\theta_k)^2-\delta^{\theta\underline{\alpha}}\left\{(\hat{c}^\theta_k)^2+(\hat{s}_k)_\perp^2\right\}\right)
  + s^\theta_k \delta^{\chi\underline{\alpha}}\varepsilon^{\theta\underline{\alpha}}
  \left(2(\hat{c}^\theta_k)^2+(1+\zeta^2)(\hat{s}_k)_\perp^2\right)
  \right\}, 
  \label{eq: KW s-e sunset wf ren N4 algebraically simplified} \\
  N_{5} =&\ -2 s^\theta_k s^\chi_k \left\{
  (\hat{c}_k)^2-2(\hat{c}^\chi_k)^2
  -\zeta^2\delta^{\chi\underline{\alpha}}(\hat{s}_k)_\perp^2\right\}
  \nonumber \\
  &\ - 2\zeta s^\theta_k \Big\{
  4 s^{\underline{\alpha}}_k s^\chi_k +\delta^{\chi\underline{\alpha}}
  \big(4 \left\{(s_k)_\perp^2 - (s^{\chi}_k)^2\right\}
  +\frac{1}{2}(\hat{s}_k)_\perp^2 \left\{2(\hat{c}^\chi_k)^2-(\hat{c}_k)^2
  +\zeta^2 (\hat{s})_\perp^2 \right\}
  \big)
  \Big\}, 
  \label{eq: KW s-e sunset wf ren N5 algebraically simplified} \\
  N_{6} =&\ 
  -2 (\hat{s}^\theta_k)^2\delta^{\theta\chi} (s_k)^2 + \zeta^2\left\{2 s^\theta_k s^\chi_k \varepsilon^{\chi\underline{\alpha}}-\delta^{\theta\chi}(\hat{s}^\theta_k)^2((\hat{s}_k)_\perp^2)^2\right\}
  \nonumber \\
  &\ + \zeta (\hat{s}_k)_\perp^2\left\{
  (\hat{s}^\theta_k)^2 \left(s^\chi_k \delta^{\theta\underline{\alpha}}-s^\theta_k \delta^{\chi\underline{\alpha}}-3s^{\underline{\alpha}}_k \delta^{\theta\chi}\right)
  +\zeta^2 s^\theta_k \delta^{\chi\underline{\alpha}} (\hat{s}_k)_\perp^2
  \right\}, 
  \label{eq: KW s-e sunset wf ren N6 algebraically simplified} \\
  N_{7} =&\ 
  -2 \delta^{\theta\chi} (\hat{c}^\theta_k)^2 \left\{(s_k)^2 + \zeta^2 ((\hat{s}_k)_\perp^2)^2\right\}
  -2 \zeta^2 s^\theta_k s^\chi_k (\hat{s}_k)_\perp^2\left\{-2+\delta^{\theta\underline{\alpha}}+2\delta^{\chi\underline{\alpha}}\varepsilon^{\theta\chi}\right\}
  \nonumber \\
  &\ + \zeta \big(
  -2 s^{\underline{\alpha}}\delta^{\theta\chi} (\hat{c}^\theta_k)^2(\hat{s}_k)_\perp^2 
  -s^\theta_k \varepsilon^{\theta\underline{\alpha}}\left\{4\delta^{\chi\underline{\alpha}}(s_k)^2+\zeta^2((\hat{s}_k)_\perp^2)^2\right\}\big), 
  \label{eq: KW s-e sunset wf ren N7 algebraically simplified} \\
  N_{8} =&\ 
  -16 s^\theta_k s^\chi_k \big\{(s_k)^2 
  +
  \tfrac{\zeta^2}{4}\big(2\delta^{\chi\underline{\alpha}} + 1\big)\{(\hat{s}_k)_\perp^2\}^2\big\}
  - 8\zeta s^\theta_k (\hat{s}_k)_\perp^2\big\{
  2 s^\chi_k s^{\underline{\alpha}}_k 
  + 
   \delta^{\chi\underline{\alpha}} \big(1
  + 
  \tfrac{\zeta^2}{8} \{(\hat{s}_k)_\perp^2 \}^2\big)
  \big\}.
  \label{eq: KW s-e sunset wf ren N8 algebraically simplified}
\end{align}\normalsize
These numerators simplify further since the denominator includes odd powers of only $ k^{\underline{\alpha}} $. The next step of the calculation is shown in \mbox{eqs.}~(\ref{eq: KW s-e sunset wf ren N4}) - (\ref{eq: KW s-e sunset wf ren N8}) in the main part.

\subsection{Sunset diagram for Bori\c{c}i-Creutz fermions}\label{app: Sunset diagram for Bori\c{c}i-Creutz fermions}

\noindent
Bori\c{c}i-Creutz fermions involve one-loop calculations that are even more cumbersome than those for Karsten-Wilczek fermions. Since every Euclidean component $ k_\mu $ contributes as an odd power that multiplies $ \zeta $ in the denominator, simplification of the numerator using symmetry arguments is only possible in terms which are individually even in $ \zeta $ and loop momenta. However, any odd function of loop momenta in the numerators contributes to the integrals, since it combines to an even power with terms like $ \zeta\,(s_k) $ in the denominator. The integral of \mbox{eq.}~(\ref{eq: BC s-e int J3}) is split into a Feynman gauge part ($ \mathcal{J}_0=J_3^\chi|_{\xi=1} $) and a gauge fixing part ($ \mathcal{J}_1=(\xi-1) \left(\partial J_3^\chi/\partial \xi\right) $), which is the rest of \mbox{eq.}~(\ref{eq: BC s-e int J3}). Numerators $ N_0 $ of $ \mathcal{J}_0 $ and $ N_1 $ of $ \mathcal{J}_1 $ are simplified algebraically to
\small
\begin{align}
  N_0 =&\ s^\chi_k \{2(\hat{c}^\chi_k)^2-(\hat{c}_k)^2\}
  +\zeta^2 \big\{\{ 6(\hat{s}_k^\chi)^2+(\hat{s}_k)^2 \} s^\chi_k
  -\{2(\hat{s}_k^\chi)^2+\frac{1}{2}(\hat{s}_k)^2\}(s_k) -(s_k \hat{s}_k \hat{s}_k)\big\}
  \nonumber \\&\
  +
  \zeta\Big\{2 (s_k)^2 +4(s_k)s^\chi_k -8 (s^\chi_k)^2 + \frac{1}{4} \big\{2(\hat{c}^\chi_k)^2-(\hat{c}_k)^2\big\} \big\{2(\hat{s}^\chi_k)^2-(\hat{s}_k)^2\big\}
  \nonumber \\&\
  + \zeta^2 \big(
  \frac{1}{4} \big\{ (\hat{s}_k)^2 - 2(\hat{s}^\chi_k)^2 \big\}(\hat{s}_k)^2 
  + \frac{1}{2} \big\{ 2(\hat{s}^\chi_k)^4 - (\hat{s}_k)^4\big\}\big)
  \Big\},
 \label{eq: N_0 algebraically simplified, BC} \\
  N_1 =&\ s^\chi_k \big( 4(s_k)^2
  +\zeta^2 \{-(\hat{s}_k)^4+2((\hat{s}_k)^2)^2+2(s_k \hat{s}_k)^2\} \big)
  \nonumber \\&\
  +\zeta^2 \big( \{(\hat{s}_k^\chi)^2(\hat{s}_k)^2+\frac{1}{2}((\hat{s}_k)^2)^2\}(s_k) 
  +2\{(\hat{s}^\chi_k)^2 -(\hat{s}_k)^2\} (s_k \hat{s}_k \hat{s}_k) \big)
  \nonumber \\&\
  +
  \zeta\big( \{(\hat{s}_k)^2-4(\hat{s}^\chi_k)^2 \} (s_k)^2 + 2\{2(s_k \hat{s}_k \hat{s}_k)-(s_k)(\hat{s}_k)^2 \}s^\chi_k
  +\frac{\zeta^2}{2} \{(\hat{s}^\chi_k)^2-\frac{1}{2}(\hat{s}_k)^2\}(\hat{s}_k)^4 
  \big).
 \label{eq: N_1 algebraically simplified, BC}
\end{align}\normalsize
The terms in \mbox{eqs.}~(\ref{eq: N_0 algebraically simplified, BC}) and~(\ref{eq: N_1 algebraically simplified, BC}), are either even in $ \zeta $ and odd in $ k $ or odd in $ \zeta $ and even in $ k $. Since the denominator $ D^{BC}(k;\zeta,M,a) $ consists of terms that are either even or odd in both $ \zeta $ and $ k $, the full integral $ J_3^\chi(\zeta,a) $ is necessarily an odd function of $ \zeta $, since integrands which are odd in any loop momentum component $ k_\mu $ vanish upon integration. Further simplification of $ N_0 $ or $ N_1 $ due to symmetry arguments yields
\small
\begin{align}
  N_0 =&\ s^\chi_k \{2(\hat{c}^\chi_k)^2-(\hat{c}_k)^2\}
  +\zeta^2 \big\{\{ 6(\hat{s}_k^\chi)^2+(\hat{s}_k)^2 \} s^\chi_k
  -\{2(\hat{s}_k^\chi)^2+\frac{1}{2}(\hat{s}_k)^2\}(s_k) -(s_k \hat{s}_k \hat{s}_k)\big\}
  \nonumber \\&\
  +
  \zeta\Big\{2 (s_k)^2 -4 (s^\chi_k)^2 + \frac{1}{4} \big\{2(\hat{c}^\chi_k)^2-(\hat{c}_k)^2\big\} \big\{2(\hat{s}^\chi_k)^2-(\hat{s}_k)^2\big\}
  \nonumber \\&\
  + \zeta^2 \big(
  \frac{1}{4} \big\{ (\hat{s}_k)^2 - 2(\hat{s}^\chi_k)^2 \big\}(\hat{s}_k)^2 
  + \frac{1}{2} \big\{ 2(\hat{s}^\chi_k)^4 - (\hat{s}_k)^4\big\}\big)
  \Big\},
 \label{eq: N_0 simplified with symmetry, BC} \\
  N_1 =&\ s^\chi_k \big( 4(s_k)^2
  +\zeta^2 \{-(\hat{s}_k)^4+2((\hat{s}_k)^2)^2+2(s_k \hat{s}_k)^2\} \big)
  \nonumber \\&\
  +\zeta^2 \big( \{(\hat{s}_k^\chi)^2(\hat{s}_k)^2+\frac{1}{2}((\hat{s}_k)^2)^2\}(s_k) 
  +2\{(\hat{s}^\chi_k)^2 -(\hat{s}_k)^2\} (s_k \hat{s}_k \hat{s}_k) \big)
  \nonumber \\&\
  +
  \zeta\big( \{(\hat{s}_k)^2-4(\hat{s}^\chi_k)^2 \} (s_k)^2 + \{2(s^\chi_k)^2+\frac{\zeta^2}{4}(\hat{s}_k)^4 \}\{2(\hat{s}^\chi_k)^2-(\hat{s}_k)^2 \}
  \big).
 \label{eq: N_1 simplified with symmetry, BC}
\end{align}\normalsize

\noindent
The integral $ J_{m}(\zeta,M,a) $ of \mbox{eq.}~(\ref{eq: BC s-e int Jm}) contributes to mass renormalisation. It is split into a Feynman gauge part ($ \mathcal{J}_2 =J_m|_{\xi=1} $) and a gauge fixing part ($ \mathcal{J}_3 =(\xi-1)\left(\partial J_m/\partial \xi\right) $). Numerators $ N_2 $ of $ \mathcal{J}_2 $ and $ N_3 $ of $ \mathcal{J}_3 $ are simplified algebraically to
\begin{align}
  N_{2} &=\ (\hat{c}_k)^2 + \zeta^2 (\hat{s}_k)^2 + 2 \zeta (s_k),
 \label{eq: N_2 algebraically simplified, BC} \\
  N_{3} &=\ 4(s_k)^2 + \zeta^2 (\hat{s}_k)^4 + 2 \zeta (s_k \hat{s}_k \hat{s}_k).
 \label{eq: N_3 algebraically simplified, BC}
\end{align}\normalsize
The presence of an odd term in both $ \zeta $ and $ k $ is a remarkable difference to the case of Karsten-Wilczek fermions, where such terms had explicitly cancelled. Since it vanishes unless it is combined with the odd term in $ \zeta $ and $ k $ of the denominator, the overall integral $ J_{m}(\zeta,M,a) $ is an even function of $ \zeta $. The presence of these odd terms considerably increases the numerical effort of the numerical integration for Bori\c{c}i-Creutz fermions compared to Karsten-Wilczek fermions. \newline

\noindent
The most laboriuous part of the self-energy calculation is $ J_4 (p;\zeta,M,a) $, which contributes to the wavefunction renormalisation and the marginal counterterm.The denominators of the sunset diagram that contribute to wavefunction renormalisation and the marginal counterterm's coefficient for Bori\c{c}i-Creutz fermions read
\begin{align}
  D_4 =&\ D^{BC}(k;\zeta,M,a)D^{g}(-k;M,a), 
  \label{eq: BC s-e sunset wf ren den D4} \\
  D_5 =&\ D^{BC}(k;\zeta,M,a)\left(D^{g}(-k;M,a)\right)^2, 
  \label{eq: BC s-e sunset wf ren den D5} \\
  D_6 =&\ D^{BC}(k;\zeta,M,a)\left(D^{g}(-k;M,a)\right)^3.
  \label{eq: BC s-e sunset wf ren den D6}
\end{align}\normalsize
Five numerators $ N_4 $ - $ N_8 $ contribute to wavefunction renormalisation and to the marginal counterterm's coefficient. The full numerators $ N_4 $ and $ N_5 $ of the part in Feynman gauge ($ J_4^{\chi\theta}|_{\xi=1} $) read
\footnotesize
\begin{align}
  N_{4} =&\ \sum\limits_{\mu,\nu,\lambda}\frac{\delta^{\mu\nu}}{8}\mathrm{tr}\Big\{
  \Big(\gamma^\chi \delta^{\mu\theta}
  \left(-\gamma^\mu \hat{s}^\mu_{k} -\zeta \gamma^{\mu\prime}\hat{c}^\mu_{k}\right)
  \left( \gamma^\lambda s^\lambda_k -\zeta \gamma^{\lambda\prime}\{1-c^\lambda_k\} \right) 
  \left(\gamma^\nu \hat{c}^\nu_{k} -\zeta \gamma^{\nu\prime}\hat{s}^\nu_{k}\right) 
  \Big)
  \nonumber \\
  &\ + \Big(\gamma^\chi \delta^{\nu\theta}
  \left(\gamma^\mu \hat{c}^\mu_{k} -\zeta \gamma^{\mu\prime}\hat{s}^\mu_{k}\right) 
  \left( \gamma^\lambda s^\lambda_k -\zeta \gamma^{\lambda\prime}\{1-c^\lambda_k\} \right) 
  \left(-\gamma^\nu \hat{s}^\nu_{k} -\zeta \gamma^{\nu\prime}\hat{c}^\nu_{k}\right)
  \Big)
  \Big\}, 
  \label{eq: BC s-e sunset wf ren full N4} \\
  N_{5} =&\ \sum\limits_{\mu,\nu,\lambda}\frac{\delta^{\mu\nu}}{4}\mathrm{tr}\Big\{
  \gamma^\chi 
  \left(\gamma^\mu \hat{c}^\mu_{k} -\zeta \gamma^{\mu\prime}\hat{s}^\mu_{k}\right) 
  \left( \gamma^\lambda s^\lambda_k -\zeta \gamma^{\lambda\prime}\{1-c^\lambda_k\} \right) 
  \left(\gamma^\nu \hat{c}^\nu_{k} -\zeta \gamma^{\nu\prime}\hat{s}^\nu_{k}\right) 
  \Big\}  \left(\hat{c}_{k}^\theta \hat{s}_{k}^\theta\right),
   \label{eq: BC s-e sunset wf ren full N5}
\end{align}\normalsize
and the numerators $ N_6 $, $ N_7 $ and $ N_8 $ of the gauge fixing part ($ -(1-\xi)\left(\partial J_4^{\chi\theta}/\partial \xi\right) $) read
\footnotesize
\begin{align}
  N_{6} =&\ \sum\limits_{\mu,\nu,\lambda}\frac{1}{8}\mathrm{tr}\Big\{
  \Big(\gamma^\chi \delta^{\mu\theta}
  \left(-\gamma^\mu \hat{s}^\mu_{k} -\zeta \gamma^{\mu\prime}\hat{c}^\mu_{k}\right)
  \left( \gamma^\lambda s^\lambda_k -\zeta \gamma^{\lambda\prime}\{1-c^\lambda_k\} \right) 
  \left(\gamma^\nu \hat{c}^\nu_{k} -\zeta \gamma^{\nu\prime}\hat{s}^\nu_{k}\right) 
  \Big)
  \nonumber \\
  &\ + \Big(\gamma^\chi \delta^{\nu\theta}
  \left(\gamma^\mu \hat{c}^\mu_{k} -\zeta \gamma^{\mu\prime}\hat{s}^\mu_{k}\right) 
  \left( \gamma^\lambda s^\lambda_k -\zeta \gamma^{\lambda\prime}\{1-c^\lambda_k\} \right) 
  \left(-\gamma^\nu \hat{s}^\nu_{k} -\zeta \gamma^{\nu\prime}\hat{c}^\nu_{k}\right)
  \Big)
  \Big\}\left(\hat{s}_{k}^\mu \hat{s}_{k}^\nu\right), 
  \label{eq: BC s-e sunset wf ren full N6} \\
  N_{7} =&\ \sum\limits_{\mu,\nu,\lambda}-\frac{\delta^{\mu\theta}\hat{c}_{k}^\mu \hat{s}_{k}^\nu+\delta^{\nu\theta}\hat{s}_{k}^\mu \hat{c}_{k}^\nu}{8}\mathrm{tr}\Big\{
  \gamma^\chi 
  \left(\gamma^\mu \hat{c}^\mu_{k} -\zeta \gamma^{\mu\prime}\hat{s}^\mu_{k}\right) 
  \left( \gamma^\lambda s^\lambda_k -\zeta \gamma^{\lambda\prime}\{1-c^\lambda_k\} \right) 
  \left(\gamma^\nu \hat{c}^\nu_{k} -\zeta \gamma^{\nu\prime}\hat{s}^\nu_{k}\right) 
  \Big\}, 
   \label{eq: BC s-e sunset wf ren full N7} \\
  N_{8} =&\ \sum\limits_{\mu,\nu,\lambda}-\frac{(\hat{c}_{k}^\theta \hat{s}_{k}^\theta)(\hat{s}_{k}^\mu \hat{s}_{k}^\nu)}{2}\mathrm{tr}\Big\{
  \gamma^\chi 
  \left(\gamma^\mu \hat{c}^\mu_{k} -\zeta \gamma^{\mu\prime}\hat{s}^\mu_{k}\right) 
  \left( \gamma^\lambda s^\lambda_k -\zeta \gamma^{\lambda\prime}\{1-c^\lambda_k\} \right) 
  \left(\gamma^\nu \hat{c}^\nu_{k} -\zeta \gamma^{\nu\prime}\hat{s}^\nu_{k}\right) 
  \Big\}.
   \label{eq: BC s-e sunset wf ren full N8}
\end{align}\normalsize
Summation of Euclidean indices and evaluation of the trace  simplifies numerators to
\footnotesize
\begin{align}
  N_{4} =&\ s^\theta_k \{1-2\delta^{\theta\chi}\} \big\{ 2(1-\zeta^2) s^\chi_k +\zeta^2 (s_k)\big\}
  + \zeta^2 \Big( \{\delta^{\theta\chi}-\frac{1}{4}\} \big\{4(s^\theta_k)^2-(\hat{s}^\theta_k)^4\big\} 
  - c^\theta_k \big\{ (\hat{s}^\chi_k)^2+\varepsilon^{\theta\chi}(\hat{s}_k)^2\big\}\Big)
  \nonumber \\
  &\ + \zeta \big\{ s^\theta_k (1-2\delta^{\theta\chi}) 
  \big( \{1+\zeta^2\}\{(\hat{s}^\chi_k)^2-\frac{1}{2}(\hat{s}_k)^2\} - \zeta^2 (\hat{s}^\theta_k)^2 \big)
  +2 c^\theta_k \big( s^\theta_k+s^\chi_k+\delta^{\theta\chi}\{(s_k)-4s^\theta_k\}\big)
  \big\}, 
  \label{eq: BC s-e sunset wf ren N4 algebraically simplified} \\
  N_{5} =&\ +2 s^\theta_k \Big( s^\chi_k \{2(\hat{c}^\chi_k)^2-(\hat{c}_k)^2\}
  +\zeta^2 \big\{\{ 6(\hat{s}_k^\chi)^2+(\hat{s}_k)^2 \} s^\chi_k
  -\{2(\hat{s}_k^\chi)^2+\frac{1}{2}(\hat{s}_k)^2\}(s_k) -(s_k \hat{s}_k \hat{s}_k)\big\}
  \nonumber \\&\
  +
  \zeta\Big\{2 (s_k)^2 +4(s_k)s^\chi_k -8 (s^\chi_k)^2 + \frac{1}{4} \big\{2(\hat{c}^\chi_k)^2-(\hat{c}_k)^2\big\} \big\{2(\hat{s}^\chi_k)^2-(\hat{s}_k)^2\big\}
  \nonumber \\&\
  + \zeta^2 \big(
  \frac{1}{4} \big\{ (\hat{s}_k)^2 - 2(\hat{s}^\chi_k)^2 \big\}(\hat{s}_k)^2 
  + \frac{1}{2} \big\{ 2(\hat{s}^\chi_k)^4 - (\hat{s}_k)^4\big\}\big)
  \Big\} \Big), 
  \label{eq: BC s-e sunset wf ren N5 algebraically simplified} \\
  N_{6} =&\ 
  -2 (\hat{s}^\theta_k)^2\delta^{\theta\chi} (s_k)^2 
  + \zeta^2\Big\{s^\theta_k \big( \{4\delta^{\theta\chi}-1\} (s_k \hat{s}_k \hat{s}_k) + 2s^\theta_k (\hat{s}^\chi_k)^2 + 2s^\chi_k (\hat{s}_k)^2 + (s_k) \{(\hat{s}_k)^2-2(\hat{s}^\chi_k)^2\} \big)
  \nonumber \\&\ 
  - (\hat{s}^\theta_k)^2 \big( c^\theta_k \{2(\hat{s}^\chi_k)^2+(\hat{s}_k)^2\} + \frac{1}{4}(\hat{s}_k)^2
  +\frac{1}{2}\{\delta^{\theta\chi}(\{(\hat{s}_k)^2\}^2+(\hat{s}_k)^4)+(\hat{s}^\chi_k)^2(\hat{s}^\theta_k)^2\}
  \big)
  \Big\}
  \nonumber \\&\ 
  + \zeta \big\{
  s^\chi_k(\hat{s}^\theta_k)^2 \{ (\hat{s}_k)^2 + (\hat{s}^\theta_k)^2+c^\theta_k \} +
  s^\theta_k \big( \{2(s_k)^2-\frac{\zeta^2}{2} (\hat{s}_k)^4 \}\{1-2\delta^{\theta\chi}\}-4 s^\chi_k s^\theta_k - (\hat{s}^\theta_k)^2 (\hat{s}_k)^2 \big)
  \big\}, 
  \label{eq: BC s-e sunset wf ren N6 algebraically simplified} \\
  N_{7} =&\ 
  -\delta^{\theta\chi} (\hat{c}^\theta_k)^2 \big(2 (s_k)^2 + \frac{\zeta^2}{2} \big\{ ((\hat{s}_k)^2)^2-(\hat{s}_k)^4+(\hat{s}_k)^2(\hat{s}^\chi_k)^2 \big\}\big)
  \nonumber \\&\
  + \zeta^2 s^\theta_k \big( \{ 3 s^\theta_k - 2 s^\chi_k - (s_k) \} (\hat{s}_k)^2 + (1-4\delta^{\theta\chi}) (s_k \hat{s}_k \hat{s}_k) \big)
  \nonumber \\&\
  + \zeta \big( (\hat{c}^\theta_k)^2 \big\{ (\hat{s}_k)^2 \{s^\chi_k-s^\theta_k\}+2s^\theta_k(\hat{s}^\chi_k)^2 \big\}
  + s^\theta_k \big\{ 4 s^\theta_k s^\chi_k - \{ 2 (s_k)^2 - \frac{\zeta^2}{2} (\hat{s}_k)^4 \} (1-2\delta^{\theta\chi}) 
  \big\}\big), 
  \label{eq: BC s-e sunset wf ren N7 algebraically simplified} \\
  N_{8} =&\ 
  -4s^\theta_k \Big\{ s^\chi_k \big( 4(s_k)^2
  +\zeta^2 \{-(\hat{s}_k)^4+2((\hat{s}_k)^2)^2+2(s_k \hat{s}_k)^2\} \big)
  \nonumber \\&\
  +\zeta^2 \big( \{(\hat{s}_k^\chi)^2(\hat{s}_k)^2+\frac{1}{2}((\hat{s}_k)^2)^2\}(s_k) 
  +2\{(\hat{s}^\chi_k)^2 -(\hat{s}_k)^2\} (s_k \hat{s}_k \hat{s}_k) \big)
  \nonumber \\&\
  +
  \zeta\big( \{(\hat{s}_k)^2-4(\hat{s}^\chi_k)^2 \} (s_k)^2 + 2\{2(s_k \hat{s}_k \hat{s}_k)-(s_k)(\hat{s}_k)^2 \}s^\chi_k
  +\frac{\zeta^2}{2} \{(\hat{s}^\chi_k)^2-\frac{1}{2}(\hat{s}_k)^2\}(\hat{s}_k)^4 
  \big)\Big\}.
  \label{eq: BC s-e sunset wf ren N8 algebraically simplified}
\end{align}\normalsize
It is observed that each numerator consists of terms which are either even or odd in both $ \zeta $ and $ k $. Since these odd powers of $ k $ vanish unless combined with the denominator's term that is odd in $ \zeta $ and $ k $, both contributions to $ \Sigma_1 $ and $ d_{1L} $ are necessarily even functions of $ \zeta $. Further simplification is achieved by noting that odd powers of loop momentum components $ k_\mu $ in terms of the numerators which are even in powers $ k $ (with arbitrary Euclidean indices) are integrated to zero.
\footnotesize
\begin{align}
  N_{4} =&\ -(s^\theta_k)^2  \big\{ 2\delta^{\theta\chi} -\zeta^2\big\}
  + \zeta^2 \Big( \{\delta^{\theta\chi}-\frac{1}{4}\} \big\{4(s^\theta_k)^2-(\hat{s}^\theta_k)^4\big\} 
  - c^\theta_k \big\{ (\hat{s}^\chi_k)^2+\varepsilon^{\theta\chi}(\hat{s}_k)^2\big\}\Big)
  \nonumber \\
  &\ + \zeta \big\{ s^\theta_k (1-2\delta^{\theta\chi}) 
  \big( \{1+\zeta^2\}\{(\hat{s}^\chi_k)^2-\frac{1}{2}(\hat{s}_k)^2\} - \zeta^2 (\hat{s}^\theta_k)^2 \big)
  +2 c^\theta_k \big( s^\theta_k+s^\chi_k+\delta^{\theta\chi}\{(s_k)-4s^\theta_k\}\big)
  \big\}, 
  \label{eq: BC s-e sunset wf ren N4} \\
  N_{5} =&\ -2 s^\theta_k \Big( s^\theta_k \big( \delta^{\theta\chi} \{(\hat{c}_k)^2-2(\hat{c}^\theta_k)^2\}
  -\zeta^2 \big\{ (2\delta^{\theta\chi}-1) \{2(\hat{s}_k^\chi)^2 + (\hat{s}^\theta_k)^2 + \frac{1}{2}(\hat{s}_k)^2\} \big\} \big)
  \nonumber \\&\
  -
  \zeta\Big\{2 (s_k)^2 +4(s_k)s^\chi_k -8 (s^\chi_k)^2 + \frac{1}{4} \big\{2(\hat{c}^\chi_k)^2-(\hat{c}_k)^2\big\} \big\{2(\hat{s}^\chi_k)^2-(\hat{s}_k)^2\big\}
  \nonumber \\&\
  + \zeta^2 \big(
  \frac{1}{4} \big\{ (\hat{s}_k)^2 - 2(\hat{s}^\chi_k)^2 \big\}(\hat{s}_k)^2 
  + \frac{1}{2} \big\{ 2(\hat{s}^\chi_k)^4 - (\hat{s}_k)^4\big\}\big)
  \Big\} \Big), 
  \label{eq: BC s-e sunset wf ren N5} \\
  N_{6} =&\ 
  -2 (\hat{s}^\theta_k)^2\delta^{\theta\chi} (s_k)^2 
  + \zeta^2\Big\{(s^\theta_k)^2 \big( 2\delta^{\theta\chi} \{2(\hat{s}^\theta_k)^2+(\hat{s}_k)^2\}
  +(\hat{s}_k)^2 - (\hat{s}^\theta_k)^2 \big)
  \nonumber \\&\ 
  - (\hat{s}^\theta_k)^2 \big( c^\theta_k \{2(\hat{s}^\chi_k)^2+(\hat{s}_k)^2\} + \frac{1}{4}(\hat{s}_k)^2
  +\frac{1}{2}\{\delta^{\theta\chi}(\{(\hat{s}_k)^2\}^2+(\hat{s}_k)^4)+(\hat{s}^\chi_k)^2(\hat{s}^\theta_k)^2\}
  \big)
  \Big\}
  \nonumber \\&\ 
  + \zeta \big\{
  s^\chi_k(\hat{s}^\theta_k)^2 \{ (\hat{s}_k)^2 + (\hat{s}^\theta_k)^2+c^\theta_k \} +
  s^\theta_k \big( \{2(s_k)^2-\frac{\zeta^2}{2} (\hat{s}_k)^4 \}\{1-2\delta^{\theta\chi}\}-4 s^\chi_k s^\theta_k - (\hat{s}^\theta_k)^2 (\hat{s}_k)^2 \big)
  \big\}, 
  \label{eq: BC s-e sunset wf ren N6} \\
  N_{7} =&\ 
  -\delta^{\theta\chi} (\hat{c}^\theta_k)^2 \big(2 (s_k)^2 + \frac{\zeta^2}{2} \big\{ ((\hat{s}_k)^2)^2-(\hat{s}_k)^4+(\hat{s}_k)^2(\hat{s}^\chi_k)^2 \big\}\big)
  + \zeta^2 (s^\theta_k)^2 \big( 2\varepsilon^{\theta\chi} (\hat{s}_k)^2 + (1-4\delta^{\theta\chi}) (\hat{s}^\theta_k)^2 \big)
  \nonumber \\&\
  + \zeta \big( (\hat{c}^\theta_k)^2 \big\{ (\hat{s}_k)^2 \{s^\chi_k-s^\theta_k\}+2s^\theta_k(\hat{s}^\chi_k)^2 \big\}
  + s^\theta_k \big\{ 4 s^\theta_k s^\chi_k - \{ 2 (s_k)^2 - \frac{\zeta^2}{2} (\hat{s}_k)^4 \} (1-2\delta^{\theta\chi}) 
  \big\}\big), 
  \label{eq: BC s-e sunset wf ren N7} \\
  N_{8} =&\ 
  -4s^\theta_k \Big\{ s^\theta_k \big\{\delta^{\theta\chi} \big( 4(s_k)^2
  +\zeta^2 \{-(\hat{s}_k)^4+2((\hat{s}_k)^2)^2+2(s_k \hat{s}_k)^2\} \big)
  \nonumber \\&\
  +\zeta^2 \big( \{(\hat{s}_k^\chi)^2(\hat{s}_k)^2+\frac{1}{2}((\hat{s}_k)^2)^2\}
  +2\{(\hat{s}^\chi_k)^2 -(\hat{s}_k)^2\} (\hat{s}^\theta_k)^2 \big) \big\}
  \nonumber \\&\
  +
  \zeta\big( \{(\hat{s}_k)^2-4(\hat{s}^\chi_k)^2 \} (s_k)^2 + 2\{2(s_k \hat{s}_k \hat{s}_k)-(s_k)(\hat{s}_k)^2 \}s^\chi_k
  +\frac{\zeta^2}{2} \{(\hat{s}^\chi_k)^2-\frac{1}{2}(\hat{s}_k)^2\}(\hat{s}_k)^4 
  \big)\Big\}.
  \label{eq: BC s-e sunset wf ren N8}
\end{align}\normalsize
In particular, odd powers in $ \zeta $ in the numerators multiply various different combinations of odd powers of loop momentum components, which cannot be reduced because the denominator includes a sum of all loop momentum components in its odd term. This is the main reason why numerical integration is considerably more expensive for Bori\c{c}i-Creutz fermions than for Karsten-Wilczek fermions.

\chapter{Statistical analysis}\label{app: Statistical analysis}

\noindent
In the following, a brief overview of the statistical analysis within this thesis is given. This overview closely follows~\cite{Gattringer:2010zz}. The terminology of primary and secondary observables as well as correlated and uncorrelated fits are clarified. The subject of autocorrelations is touched and methods for estimating whether observables are sufficiently uncorrelated or not are discussed. Finally, the Jackknife method which is used in the statistical analysis of this thesis is discussed. \newline

\noindent
A primary observable $ X $ is calculated on $ N $ different configurations, where $ X_i $ denotes the value of $ X $ on the \textit{i}th sample. The expectation value of $ \langle X \rangle $ and the variance $ \sigma_{X}^2 $ are defined as
\begin{align}
  \langle X \rangle =&\ \langle X_i \rangle, 
  \label{eq: expectation value} \\
  \sigma_{X}^2 =&\ \langle (X_i- \langle X_i \rangle)^2 \rangle.
  \label{eq: variance} 
\end{align}\normalsize
Stochastic estimators $ \langle \widehat{X} \rangle $ and $ \widehat{\sigma}_{X}^2 $ are defined accordingly for $ \langle X \rangle $ and $ \sigma_{X}^2 $,
\begin{align}
  \widehat{X} =&\ \frac{1}{N} \sum\limits_{i=1}^{N} X_i, 
  \label{eq: estimator for expectation value} \\
  \widehat{\sigma}_{X}^2 =&\ \frac{1}{N-1} \sum\limits_{i=1}^{N} (X_i-X)^2.
  \label{eq: estimator for variance} 
\end{align}\normalsize
The expectation value of the estimator $ \langle \widehat{X} \rangle $ agrees with the mean as $ \langle \widehat{X} \rangle = \langle X \rangle $. If the $ X_i $ are uncorrelated, their expectation values factorise for $ i \neq j $ as
\begin{equation}
  \langle X_i X_j \rangle = \langle X_i \rangle \langle X_j \rangle = \langle X \rangle^2
  \label{eq: factorisation of expectation values for uncorrelated variables}
\end{equation}\normalsize
and the variance $ \sigma_{\widehat{X}}^2 $ of the estimator is given by
\begin{align}
  \sigma_{\widehat{X}}^2 =&\ \langle \left(\langle \widehat{X} \rangle - \langle X \rangle\right)^2 \rangle
  =\langle \left(\frac{1}{N} \sum\limits_{i=1}^{N} \left(X_i - \langle X \rangle\right) \right)^2 \rangle 
  \nonumber \\
  =&\ \frac{1}{N} \langle X^2 \rangle - \langle X \rangle^2 
  + \frac{1}{N^2} \sum\limits_{i\neq j=1}^N \langle X_i X_j \rangle.
\end{align}\normalsize
The last term yields $ \sum_{i\neq j=1}^N \langle X_i X_j \rangle = N(N-1)\langle X \rangle^2 $ for uncorrelated observables due to \mbox{eq.}~(\ref{eq: factorisation of expectation values for uncorrelated variables}) and the variance of the estimator is related to the variance of the observable by $ \sigma_{\widehat{X}}^2 = \sigma_{X}^2/N $. Using the estimator of the variance $ \widehat{\sigma}_X^2 $ instead of $ \sigma_{X}^2 $, the statistical error of the estimator $ \widehat{X} $ is given by
\begin{equation}
  \sigma = \frac{\widehat{\sigma}_X}{\sqrt{N}}.
\end{equation}\normalsize

\noindent
Since observables, which are computed from the same Markov chain are necessarily correlated to some extent, these correlations have to be accounted for. In principle, it would be necessary to calculate the autocorrelation function, 
\begin{equation}
  C_X(t) \equiv C_X(X_i,X_{i+t}) = (X_i -\langle X_i \rangle)  (X_{i+t} -\langle X_{i+t} \rangle),
  \label{eq: autocorrelation function}
\end{equation}\normalsize
and the integrated autocorrelation time, 
\begin{equation}
  \tau_{X,\mathrm{int}} = \frac{1}{2} + \sum\limits_{t=1}^{N}\Gamma_X(t), \quad \Gamma_X(t) = \frac{C_X(t)}{C_X(0)},
  \label{eq: integrated autocorrelation time}
\end{equation}\normalsize
where $ t $ is the computer time of the Markov chain update algorithm. The variance of the estimator which would yield the statistical error including autocorrelations reads
\begin{align}
  \sigma_{\widehat{X}}^2 =&\ \frac{1}{N^2}\sum\limits_{i,j} C_X(|i-j|) 
  = \frac{C_X(0)}{N} \sum\limits_{t=-N}^N \Gamma_X(|t|) \left(1-\frac{|t|}{N}\right)
  \approx \frac{\sigma_X^2}{N} 2\tau_{X,\mathrm{int}},
\end{align}\normalsize
where $ C_X(0) = \sigma_X^2 $ is used. However, calculation of the integrated autocorrelation time requires a cutoff of the summation for values of $ t $ where $ \Gamma_X(t) $ becomes unreliable. Since usually extremely large ensembles with at least $ 1000\,\tau_{X,\mathrm{int}} $ samples are required for a reliable estimate of $ \tau_{X,\mathrm{int}} $, this method is impractical for the small statistical ensembles of the numerical studies of chapter \ref{sec: Numerical studies}. \newline

\noindent
A typical representative of a secondary observable $ Y $ is a parameter obtained from a fit to a correlation function. The statistical error of $ Y $ is obtained as the square root of the diagonal elements of the covariance matrix
\begin{align}
  COV(Y,Z) =&\ \lim\limits_{N\to \infty} \frac{1}{N} \sum\limits_{i,j=1}^{N} \frac{\partial Y}{\partial X_i} \frac{\partial Z}{\partial X_j} \langle X_i X_j \rangle_N
  \label{eq: covariance matrix}
\end{align}\normalsize
Since the secondary observables $ Y $ are usually non-linear functions of the primary observables $ X $, the covariance matrix is not exactly known. The fit is performed as a minimisation of $ \chi^2 $ that is defined as
\begin{align}
  \chi^2 =&\ \sum_{s,s^\prime=n_{\min}}^{n_{\max}} \left( X(s)-f(Y,s) \right) w(s,s^\prime) \left( X(s^\prime)-f(Y,s^\prime) \right)
  \label{eq: chi2 for fits to primary observables}
\end{align}\normalsize
by varying the set of fit parameters $ Y $. The indices $ s,s^\prime $ denote the sampling range of the minimisation and typically represent indices of time slices. Ideally the inverse of the exact covariance matrix would be used as statistical weights $ w(s,s^\prime) = COV(s,s^\prime) $. However, since only a stochastic estimator $ COV_N(s,s^\prime) $ of the covariance matrix on $ N $ samples is known,
\begin{align}
  COV_N(s,s^\prime)  =&\ \frac{1}{N-1} \langle \left( X(s) - \langle X(s) \rangle_N \right) \left( X(s^\prime) - \langle X(s^\prime) \rangle_N \right) \rangle_N
  \label{eq: estimator for covariance matrix}
\end{align}\normalsize
is inverted numerically and used as a weight $ w(s,s^\prime)) = COV_N(s,s^\prime)^{-1} $ in the definition of $ \chi^2 $. In the case of truly uncorrelated data, the covariance matrix is diagonal and the weights take the form $ w(s,s^\prime) = \delta_{s,s^\prime}/\sigma_X^2(s) $. If the estimator of the covariance is badly determined due to statistical fluctuations, it may acquire accidental small eigenvalues that are detrimental to the stability of the fit. As a result, despite reasonably small values of $ \chi^2 $, the fit fails to agree with the primary observables. In such a case, the covariance matrix must either be smoothly approximated or the off-diagonal elements of the covariance matrix must be neglected. The latter case is the familiar form of an uncorrelated fit.\newline

\noindent
The numerical correlation of data is estimated in the statistical analysis with three different approaches. The first approach is data blocking. The ensemble is divided in $ N/K $ subsets, which are indicated by $ k \in [1,\ldots,N/K] $. The mean value $ \langle \widehat{X}_k \rangle $ and variance $ \sigma_{\widehat{X}_k}^2 $ are calulated on each subset $ k $ with $ K $ samples each. If the variance scales like $ 1/K $ upon variation of $ K $, the original observables can be considered sufficiently uncorrelated. The second approach omits samples of the total ensemble in a simplified version of the statistical bootstrap. This is done by either using only a subset of $ K $ samples which would be counted as the \textit{k}th subset in the data blocking approach or by using only every \textit{k}th sample and varying the offset $ i $ of the first sample within $ i \in [0,K-1] $. If the mean $ \langle X_k \rangle $ fluctuates between the subsets only within the statistical errors of any of the subsets, data can be considered sufficiently uncorrelated. \newline

\noindent
The third approach uses the Jackknife method. The Jackknife method computes sample averages of the original data, where the \textit{i}th sample is removed. Hence, the Jackknife bins read
\begin{equation}
  X_i^J = \frac{1}{N-1} \left( N\widehat{X} -X_i\right)
  \label{eq: Jackknife resampled bins}
\end{equation}\normalsize
and the variance is given by
\begin{equation}
  \sigma_{\widehat{X}}^2 = \frac{N-1}{N} \sum\limits_{i=1}^{N} \left( X_i^J - \langle \widehat{X} \rangle \right)^2
  \label{eq: Jackknife resampled variance}
\end{equation}\normalsize
The Jackknife method can be freely combined with the aforementioned blocking procedures. Its advantage is that secondary observables $ Y^J $ that are computed from primary observables $ X^J $ that have been processed with the Jackknife procedure are automatically stochastic estimators for $ \langle Y \rangle $. Their covariance matrix reads
\begin{align}
  COV(Y,Z) =&\ \lim\limits_{N\to \infty} \frac{N-1}{N} \sum\limits_{i=1}^{N} \left(Y_i^J - \langle Y \rangle \right) \left(Z_i^J - \langle Z \rangle \right).
  \label{eq: Jackknife resampled covariance matrix}
\end{align}\normalsize
Once the secondary observables $ Y^J $ are obtained with the Jackknife method, they can be used instead of primary observables in \mbox{eq.}~(\ref{eq: chi2 for fits to primary observables}), any secondary observables $ Y $ can be used to define a sampling range and the covariance matrix of the primary observables can be replaced by \mbox{eq.}~(\ref{eq: Jackknife resampled covariance matrix}). \newline

\noindent
Fits to secondary observables are performed for all extrapolations of fit parameters. The sampling range is defined in terms of simulation parameters of the primary simulations. The covariance matrix of fit parameters appears unstable using correlated fits for extrapolations due to having only small data sets. Thus, secondary observables from correlated fits are fitted with uncorrelated fits within this thesis. \newline

\noindent
Lastly, some values of the simulation parameter $ c $ in the analysis of power spectral densities in section \ref{sec: Tuning with the frequency spectrum} yield frequencies $ \omega_c \leq \omega_B/2 $, which are too close to zero for obtaining any statistical variation. In these cases, the analysis which determines the statistical error from Jackknife bins of secondary observables fails and yields zero as the statistical error. Thus, the statistical error is combined quadratically with an estimate of the systematical error and an uncorrelated fit using routines from the GSL library is conducted on any sample $ Y^J $. The combined statistical and systematical errors are used in the definition of the weights of the secondary observables. The statistical error is then calculated from the variance of ternary observabless and the systematical error is estimated from the variance-covariance matrix of ternary observables.

\chapter{Oscillating lattice toy models}\label{app: Oscillating lattice toy models}
\noindent Application of the decomposition of spinor fields in section~\ref{sec: Components in correlation functions} indicates that mesonic correlation functions receive oscillating contributions with a frequency that is a continuous function of the counterterms' coefficients. Thus, this frequency is not necessarily an eigenfrequency of a periodic lattice for arbitrary choices of the parameters. This appendix contains two toy models that show how oscillations with arbitrary frequencies are realised on a lattice with periodic boundary conditions.

\section{Harmonic oscillator as a toy model}\label{sec: Harmonic oscillator as a toy model}

\noindent The first toy model is a simple harmonic oscillator. Its equation of motion reads
\begin{equation}
  \left(\frac{d^2}{dt^2} +\omega_0^2\right)q(t)=0
\end{equation}\normalsize
and its solutions are
\begin{equation}
  q(t) = A \cos(\omega_0 t) +B \sin(\omega_0 t).
\end{equation}\normalsize
For a periodic system with $ q(t+T)=q(t) $, it is obvious that the only permissible homogenous solutions are those with $ \omega_0\stackrel{!}{=}\frac{2\pi n}{T} $. However, the situation is different for inhomogenous solutions. The Green's function of the harmonic oscillator is defined by
\begin{equation}
  \left(\frac{d^2}{dt^2} +\omega_0^2\right)G(t,t_0)=\delta(t-t_0),
\end{equation}\normalsize
and reads
\begin{equation}
  G(t,t_0) = \frac{1}{2\pi} \int_{-\infty}^{+\infty} dk \frac{e^{ik(t-t_0)}}{\omega_0^2-k^2+ i\epsilon}.
  \label{eq: continuum harmonic oscillator Green's function}
\end{equation}\normalsize
If the harmonic oscillator is set up with periodic boundary conditions $ q(t+T)=q(t) $, the Green's function becomes ($ T $ is the physical length of the periodic direction)
\begin{equation}
  G(t,t_0) = \frac{1}{T} \sum_{n_k=0}^{\infty} \frac{e^{i\frac{2\pi n_k}{T}(t-t_0)}}{\omega_0^2-\left(\frac{2\pi n_k}{T}\right)^2}
  \label{eq: periodic bc continuum harmonic oscillator Green's function}
\end{equation}\normalsize
due to its restricted frequency range ($ k=\frac{2\pi n_k}{T} $). The real part of the Green's function of \mbox{eq.}~(\ref{eq: periodic bc continuum harmonic oscillator Green's function})-- which is the same as the sum of Green's function and reflected Green's function -- is plotted as the points in figure~\ref{eq: periodic bc continuum harmonic oscillator Green's function}) for $ T=64 $ for three oscillator frequencies $ \frac{T\omega_0}{128\pi}=\left\{\frac{1}{11},\frac{2}{11},\frac{3}{11}\right\} $. The continuous curves are obtained by truncating the sum in \mbox{eq.}~(\ref{eq: periodic bc continuum harmonic oscillator Green's function}) at $ n_k=T-1 $. Effects of the truncation are visible near the inhomogenity. It is clearly visible that the Green's function oscillates with oscillator frequency $ \omega_0 $ even though it is not a permissible frequency $ k $ for homogenous solutions.
\begin{figure}[htb]
 \begin{picture}(360,145)
  \put(90.0, 0.0){\includegraphics[bb=0 0 300 136, scale=0.90]{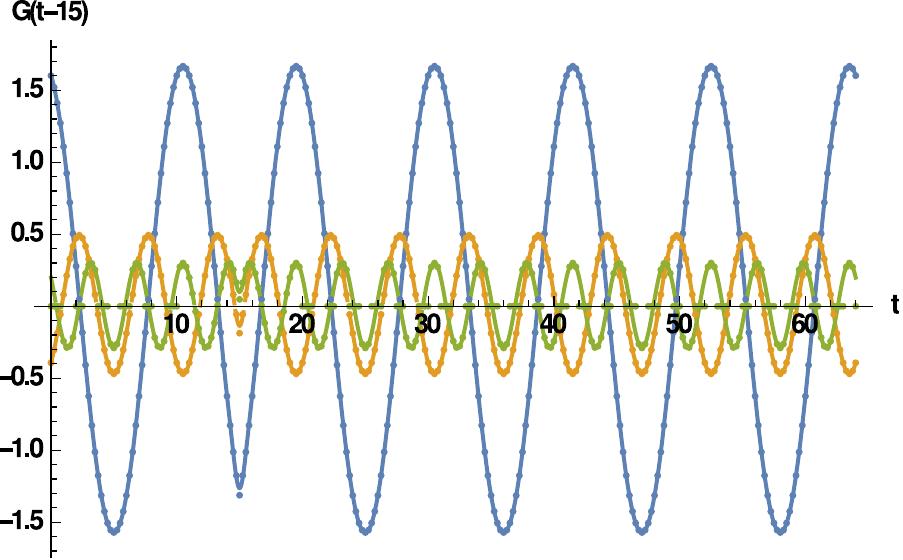}}
 \end{picture}
 \caption{Green's functions for harmonic oscillators of frequencies $ \frac{T\omega_0}{128\pi}=\left\{\frac{1}{11},\frac{2}{11},\frac{3}{11}\right\} $ (blue, yellow, green) oscillate with $ \omega_0 $.
  }
 \label{fig: toy model Green's function}
 \vspace{-8pt}
\end{figure}

\begin{figure}[htb]
 \begin{picture}(360,145)
  \put(90.0, 0.0){\includegraphics[bb=0 0 300 136, scale=0.90]{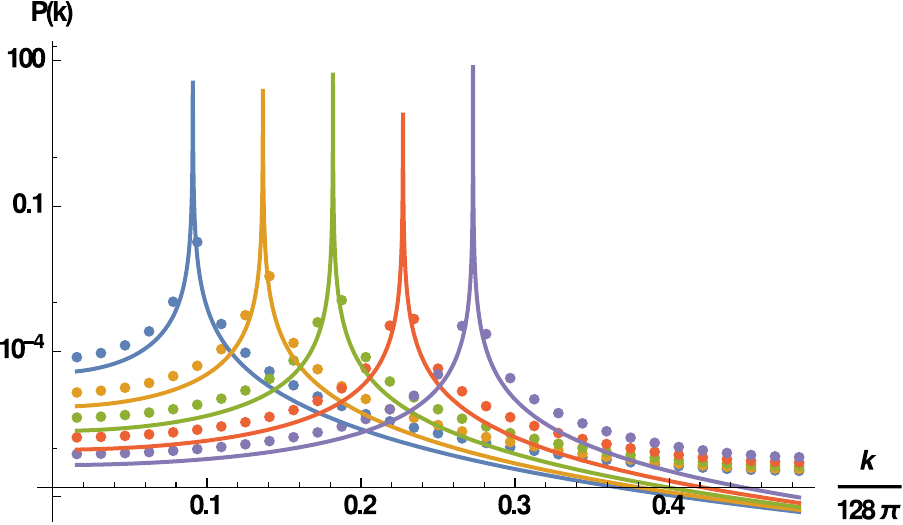}}
 \end{picture}
 \caption{The peaks of the power spectral densities mark the oscillators' frequencies $ \frac{T\omega_0}{64}=\frac{2\pi}{22}\times\{2,3,4,5,6\} $, which lie in between the lattice frequencies $ k $.
  }
 \label{fig: toy model continuum power spectral density}
 \vspace{-8pt}
\end{figure}

\noindent The power spectral density, which is defined in \mbox{eq.}~(\ref{eq: psd}), is computed from a discrete Fourier transform of the Green's function and plotted in figure~\ref{fig: toy model continuum power spectral density}. The curves are given by $ 1/[N^3(\omega_0^2-k^2)^2] $ and diverge at the peaks of the power spectral densities. The peak positions are not at frequencies $ k $ that are permissible in terms of the periodic boundary conditions, but at the eigenfrequencies $ \omega_0 $ of the harmonic oscillators.

\noindent Next, the harmonic oscillator is formulated on a lattice. This introduces an upper bound for the frequencies. One example of a discretised action for a harmonic oscillator on a lattice is given by
\begin{equation}
  S[q] = \frac{m}{2a} \sum_{n=0}^{N-1} q_n \left( ((a\omega)^2-2) q_n + q_{n-1} +q_{n+1} \right).
\end{equation}\normalsize
With a Fourier transform of the variable $ q_n $, the action is expressed in frequency space,
\begin{align}
  S[q] 
  =&\ \frac{m}{2aN} \frac{1}{\sqrt{N}^2}\sum_{n_k,n_\ell=0}^{N-1} q(2\pi \frac{n_k}{aN})q(2\pi \frac{n_\ell}{aN}) \sum_{n=0}^{N-1} e^{i(n_k+n_\ell)n\frac{2\pi}{N}} \left( (a\omega_0)^2-2+e^{+ia\ell}+e^{-ia\ell} \right) \nonumber \\
  =&\ \frac{m}{aN} \sum_{n_k=0}^{N-1} q\left(2\pi \frac{n_k}{aN}\right)q\left(2\pi \frac{N-n_k}{aN}\right) \left(\cos\left(2\pi \frac{n_k}{N}\right)-1+\frac{(a\omega)^2}{2} \right),
\end{align}\normalsize
where the different modes decouple. Hence, the equation of motion is
\begin{equation}
  \left(\cos\left(2\pi \frac{n_k}{N}\right)-1+\frac{(a\omega_0)^2}{2} \right)q\left(\frac{2\pi n_k}{aN}\right) = 0
\end{equation}\normalsize
and turns into a standard harmonic oscillator $ (-k^2+\omega_0^2)q(k) = 0 $ in its continuum limit ($ a \to 0 $ and $ k=\frac{2\pi}{a} \frac{n_k}{N} $). The homogenous solution of the lattice equation requires 
\begin{equation}
  \frac{2\pi}{a} \frac{n_k}{N}=\omega_0^{\mathrm{lat}} \equiv \frac{1}{a}\arccos\left(1-\frac{(a\omega_0)^2}{2}\right),
  \label{eq: discretised oscillator freuqency}
\end{equation}\normalsize
where $ n_k $ is an integer that labels one of the lattice eigenfrequencies. Due to the discretised second derivative, the lattice Green's function reads
\begin{equation}
  G(n,n_0) = \frac{a}{N} \sum_{n_k=0}^{N-1} \frac{e^{i2\pi\frac{n_k}{N}(n-n_0)}}{1-\frac{(a\omega_0)^2}{2}-\cos{(2\pi\frac{n_k}{N})}}.
  \label{eq: discretised harmonic oscillator Green's function}
\end{equation}\normalsize
\begin{figure}[htb]
 \begin{picture}(360,145)
  \put(90.0, 0.0){\includegraphics[bb=0 0 300 136, scale=0.90]{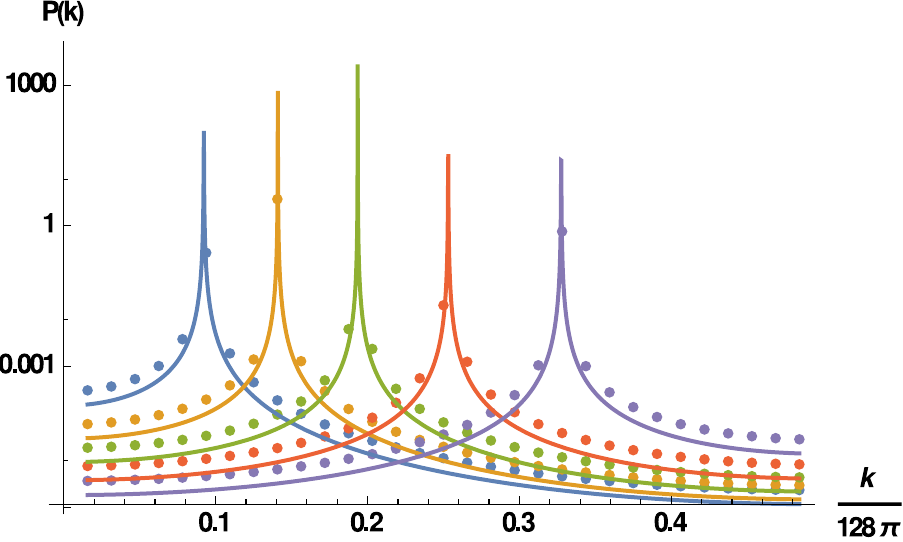}}
 \end{picture}
 \caption{The peaks of the power spectral densities mark the discretised oscillators' frequencies $ \omega_0^{\mathrm{lat}}(\omega_0),\ a\omega_0=\frac{2\pi}{22}\times\{2,3,4,5,6\} $.
  }
 \label{fig: toy model lattice power spectral density}
 \vspace{-8pt}
\end{figure}

\noindent The power spectral density is computed from a discrete Fourier transform of the lattice Green's function and plotted in figure~\ref{fig: toy model lattice power spectral density} together with curves that are given by
\begin{equation}
  C(n_k,\omega_0) = \frac{1}{N^3}\frac{1}{1-\frac{(a\omega_0)^2}{2}-\cos{(2\pi\frac{n_k}{N})}}.
\end{equation}\normalsize
The curves $ C(n_k,\omega_0) $ diverge at the peaks of the power spectral densities, which match the discretised oscillator's frequencies $ \omega_0^{\mathrm{lat}} $ instead of the lattice eigenfrequencies $ \frac{2\pi}{a}\frac{n_k}{N} $.

\section{One-dimensional, spinless lattice fermion}\label{sec: One-dimensional, spinless lattice fermion}

The second toy model, a one-dimensional colourless and spinless fermion field with imaginary mass $ i\omega_0 $ is closer to the thesis' main topic. The field's action is given by
\begin{equation}
  S[\psi,\psi^\dagger] = \sum_{n=0}^{N-1} \psi_n^\dagger \{ \frac{\psi_{n+1} -\psi_{n-1}}{2a} + i\omega_0 \psi_n \}.
\end{equation}\normalsize
The equation of motion reads
\begin{equation}
  \frac{\psi_{n+1} -\psi_{n-1}}{2a} + i\omega_0 \psi_n =0
\end{equation}\normalsize
and decouples in frequency space as
\begin{equation}
  \frac{i}{a}\left\{\sin\left(2\pi \frac{n_k}{N}\right) + a\omega_0\right\} \psi\left(2\pi \frac{n_k}{N}\right) =0.
\end{equation}\normalsize
\begin{figure}[htb]
 \begin{picture}(360,145)
  \put(90.0, 0.0){\includegraphics[bb=0 0 300 136, scale=0.90]{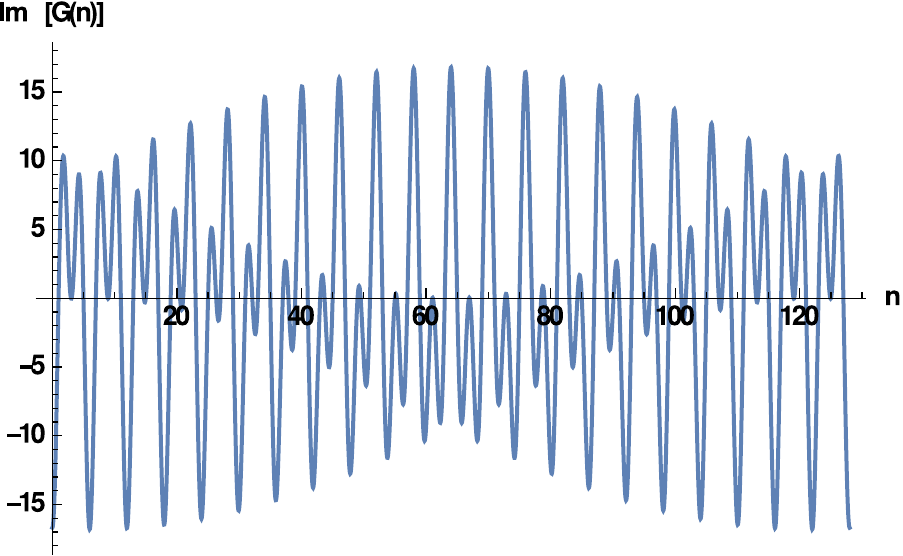}}
 \end{picture}
 \caption{The Green's function $ G(n,0) $ for $ \frac{a\omega_0}{2\pi}=-\frac{3}{22} $ exhibits interference patterns.
  }
 \label{fig: toy model KW Green's function}
 \vspace{-8pt}
\end{figure}

\begin{figure}[htb]
 \begin{picture}(360,145)
  \put(90.0, 0.0){\includegraphics[bb=0 0 300 136, scale=0.90]{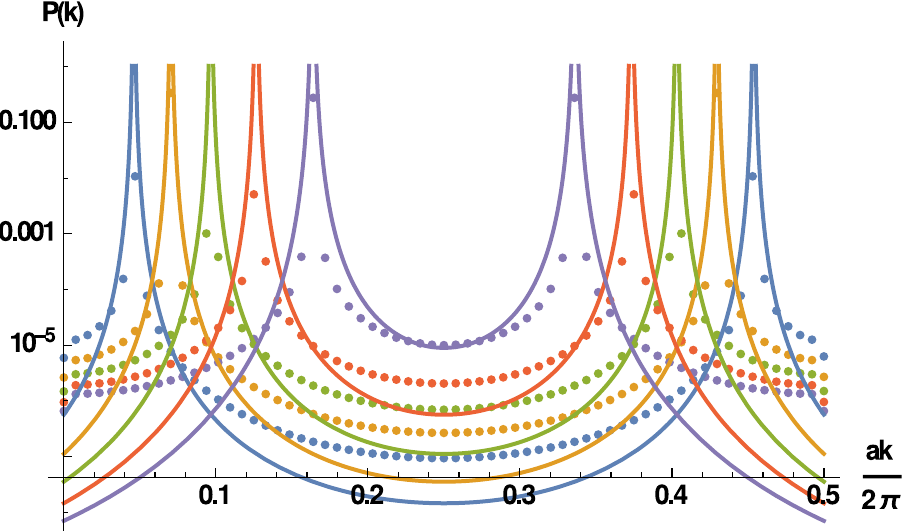}}
 \end{picture}
 \caption{The peaks of the power spectral densities mark both frequencies $ \Omega_1 $ and $ \Omega_2 $, which are defined as functions of $ \omega_0^{\mathrm{lat}} $, where $ a\omega_0=-\frac{2\pi}{22}\times\{2,3,4,5,6\} $.
  }
 \label{fig: toy model KW power spectral density}
 \vspace{-8pt}
\end{figure}

\noindent Similar to the discretised harmonic oscillator, homogenous solutions require
\begin{equation}
  \frac{2\pi}{a} \frac{n_k}{N}=\left\{\begin{array}{c}\Omega_1\\ \Omega_2 \end{array}\right. \equiv \left\{\begin{array}{c}-\omega_0^{\mathrm{lat}}\\ \frac{\pi}{a}+\omega_0^{\mathrm{lat}} \end{array}\right.,\quad \omega_0^{\mathrm{lat}} \equiv \frac{1}{a}\arcsin{(a\omega_0)},
  \label{eq: toy model KW eigenfrequencies}
\end{equation}\normalsize
where $ n_k $ is an integer that labels one of the lattice eigenfrequencies. Due to the discretised derivative, the lattice Green's function -- the one-dimensional fermion propagator -- reads
\begin{equation}
  G(n,n_0) = \frac{-i}{N} \sum_{n_k=0}^{N-1} \frac{e^{i2\pi\frac{n_k}{N}(n-n_0)}}{(a\omega_0) + \sin{(2\pi\frac{n_k}{N})}}.
  \label{eq: toy model KW Green's function}
\end{equation}\normalsize
\noindent The Green's function for $ N=128 $ is plotted in figure~\ref{fig: toy model KW Green's function} and exhibits interference patterns that are analysed in terms of its power spectral density, which is plotted in figure~\ref{fig: toy model KW power spectral density}. The continuous curves in the plot of the power spectral density are given by
\begin{equation}
  C(k,\omega_0) = \frac{1}{N^4}\frac{1}{(a\omega_0) + \sin{(2\pi\frac{n_k}{N})}}.
\end{equation}\normalsize
The power spectral density is peaked at the frequencies $ \Omega_1 $ and $ \Omega_2 $ that are defined in \mbox{eq.}~(\ref{eq: toy model KW eigenfrequencies}). Each lattice frequency $ \frac{2\pi}{a}\frac{n_k}{N} $ contributes with different weigths to each of the peaks for $ \Omega_1 $ and $ \Omega_2 $, whose interference pattern is visible in the Green's function.

\begin{figure}[htb]
 \begin{picture}(360,145)
  \put(90.0, 0.0){\includegraphics[bb=0 0 300 136, scale=0.90]{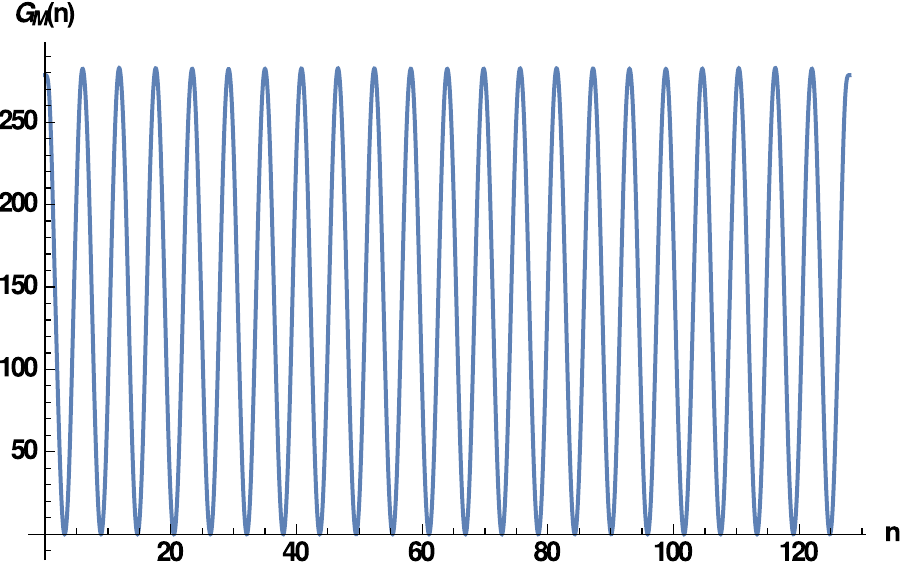}}
 \end{picture}
 \caption{The mesonic correlation function for $ \frac{a\omega_0}{2\pi}=-\frac{3}{22} $ has oscillating and non-oscillating contributions.
  }
 \label{fig: toy model KW mesonic correlation function}
 \vspace{-8pt}
\end{figure}

\noindent In the next step of the toy model, a mesonic correlation function is created from the contraction of two Green's functions as in \mbox{eqs.}~(\ref{eq: mesonic correlation function}) or~(\ref{eq: mesonic correlation function for contractions}),
\begin{equation}
  G_M(n,n_0) = |G(n,n_0)|^2.
  \label{eq: toy model KW mesonic correlation function}
\end{equation}\normalsize
This correlation function is manifestly positive and contains a constant and an oscillating term with equal weights as is depicted in figure~\ref{fig: toy model KW mesonic correlation function}. Hence, the correlation function can be written in terms of the oscillation frequency $ \Omega $ as
\begin{equation}
  G_M(n) \propto 1 + \frac{\cos{(a\Omega \frac{N}{2})}}{1+\cos{(a\Omega N)}}\cos{(a\Omega (n-\frac{N}{2}))}.
\end{equation}\normalsize
Thus, the toy model's correlation function formally resembles the decomposed correlation function in \mbox{eqs.}~(\ref{eq: decomposed gamma5 correlation function})-(\ref{eq: oscillating piece of correlation function}) that is derived from a spinor decomposition in section~\ref{sec: Decomposition into a pair of fields}.

\begin{figure}[htb]
 \begin{picture}(360,145)
  \put(90.0, 0.0){\includegraphics[bb=0 0 300 136, scale=0.90]{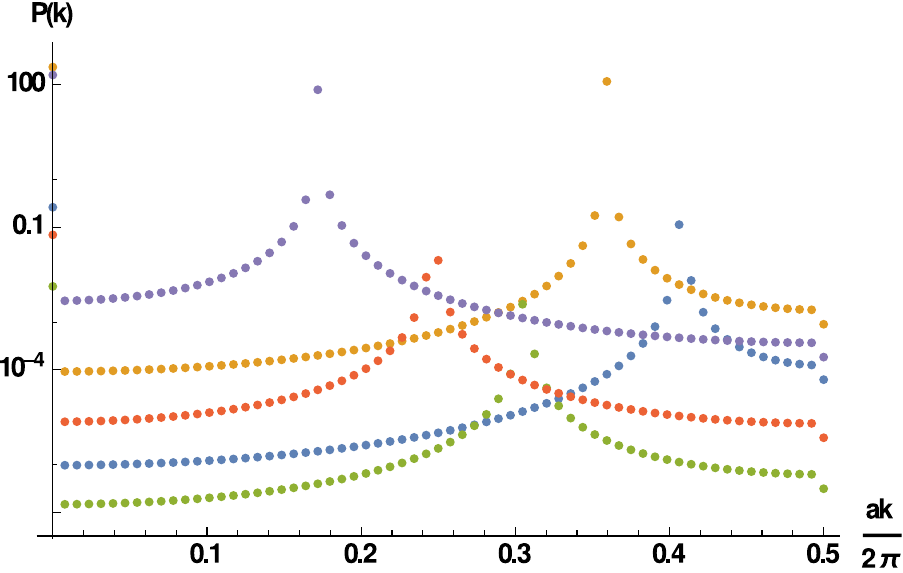}}
 \end{picture}
 \caption{The peak of the mesonic correlation function's power spectral densities marks the frequency difference $ \Omega=\Omega_2-\Omega_1 $. Both $ \Omega_1 $ and $ \Omega_2 $ are defined as functions of $ \omega_0^{\mathrm{lat}} $, where $ a\omega_0=-\frac{2\pi}{22}\times\{2,3,4,5,6\} $.
  }
 \label{fig: toy model KW mesonic power spectral density}
 \vspace{-8pt}
\end{figure}

\noindent Its frequency $ \Omega $ is analysed in terms of the mesonic correlation function's power spectral density, which is plotted in figure~\ref{fig: toy model KW mesonic power spectral density}. The first peak is exactly at the lattice's eigenfrequency $ \frac{2\pi}{a}\frac{n_k}{N}=0 $. The second peak has its maximum at
\begin{equation}
  \Omega \equiv \Omega_2-\Omega_1 = \frac{\pi}{a}+2\omega_0^{\mathrm{lat}},
\end{equation}\normalsize
which implies that the oscillation is due to the product of different fermionic modes (counted as $ \Omega_1 $ and $ \Omega_2 $) in both Green's functions. Frequencies other than the lattices's eigenfrequencies can be realised in terms of the peak position of the spectral distribution of the lattice's eigenfrequencies. Since this toy model is essentially a one-dimensional, non-interacting Karsten-Wilczek fermion, it is not suprising that similar features are reproduced in the numerical results of section~\ref{sec: Oscillating correlation functions}. 

\chapter{Simulation parameters and data sets}\label{app: Simulation parameters and data sets}

\FloatBarrier
\sectionc{Summary of data sets}{app: Summary of data sets}

\noindent
This appendix summarises all parameters sets of simulations for this thesis in tabular form. Karsten-Wilczek fermions are always simulated with $ \zeta=+1 $.

\begin{table}[hbt]
 \center\footnotesize
 \begin{tabular}{|c|c|c||p{6.6cm}|p{2.2cm}|p{1.6cm}|}
  \hline
  $ \beta $ & $  T $ & $  n_{cfg} $ & $ c $ & $ d $ & $ am_0 $  \\
  \hline
  $ 5.8 $   & $ 32 $ & $ 100 $ & $ 0.0 $, $ -0.3 $, $ -0.4 $, $ -0.45 $, $ -0.5 $, $ -0.55 $, $ -0.65 $ 
  & $ 0 $, $ 0.02 $ 
  & $ 0.02 $, $ 0.04 $, $ 0.05 $ \\
  \hline
  $ 6.0 $   & $ 32 $ & $ 100 $ & $ -0.2 $, $ -0.3 $, $ -0.38 $, $ -0.42 $, $ -0.45 $, $ -0.5 $, $ -0.55 $, $ -0.65 $ 
  & $ -0.02 $, $-0.01 $, $ 0 $ 
  & $ 0.01 $, $ 0.02 $, $ 0.03 $, $ 0.04 $, $ 0.05 $ \\
  \hline
  $ 6.0 $   & $ 32 $ & $ 100 $ & $ -0.3 $, $ -0.38 $, $ -0.42 $, $ -0.45 $, $ -0.5 $, $ -0.55 $ 
  & $ -0.08 $, $ -0.06 $, $-0.04 $, $ +0.01$, $ +0.02 $ 
  & $ 0.02 $, $ 0.03 $, $ 0.04 $,~~\,$ 0.05 $ \\
  \hline
  $ 6.0 $   & $ 32 $ & $ 100 $ & $ +0.3 $, $ +0.2 $, $ +0.1 $, $ 0.0 $, $ -0.1 $, $ -0.15 $, $ -0.2 $, $ -0.25 $, $ -0.3 $, $ -0.32 $, $ -0.35 $, $ -0.37 $, $ -0.38 $, $ -0.40 $, $ -0.42 $, $ -0.43 $, $ -0.44 $, $ -0.4445 $, $ -0.45 $, $ -0.47 $, $ -0.48 $, $ -0.5 $, $ -0.52 $, $ -0.55 $, $ -0.6 $, $ -0.65 $, $ -0.7 $, $ -0.8 $, $ -0.9 $, $ -1.0 $, $ -1.1 $, $ -1.2 $ 
  & $ 0 $ 
  & $ 0.02 $, $ 0.03 $, $ 0.04 $, $ 0.05 $ \\
  \hline
  $ 6.0 $   & $ 48 $ & $ 40 $ & $ 0.0 $, $ -0.2 $, $ -0.3 $, $ -0.4 $, $ -0.45 $, $ -0.55 $, $ -0.65 $  
  & $ 0 $ 
  & $ 0.02 $ \\
  \hline
  $ 6.2 $   & $ 32 $ & $ 100 $ & $ -0.2 $, $ -0.3 $, $ -0.38 $, $ -0.42 $, $ -0.45 $, $ -0.55 $, $ -0.65 $ 
  & $ 0 $ 
  & $ 0.01 $  \\
  \hline
  $ 6.2 $   & $ 32 $ & $ 100 $ & $ -0.2 $, $ -0.3 $, $ -0.38 $, $ -0.42 $, $ -0.45 $, $ -0.55 $, $ -0.65 $ 
  & $ -0.08 $, $-0.04 $, $+0.02 $ 
  & $ 0.02 $, $ 0.03 $, $ 0.04 $,~~\,$ 0.05 $   \\
  \hline
  $ 6.2 $   & $ 32 $ & $ 100 $ & $ -0.15 $, $ -0.2 $, $ -0.25 $, $ -0.27 $, $ -0.3 $, $ -0.32 $, $ -0.33 $, $ -0.35 $, $ -0.37 $, $ -0.38 $, $ -0.42 $, $ -0.43 $, $ -0.45 $, $ -0.47 $, $ -0.50 $, $ -0.55 $, $ -0.6 $, $ -0.65 $ 
  & $ 0 $ 
  & $ 0.02 $, $ 0.03 $, $ 0.04 $,~~\,$ 0.05 $   \\
  \hline
  $ 6.2 $   & $ 48 $ & $ 40 $ & $ 0.0 $, $ -0.2 $, $ -0.3 $, $ -0.4 $, $ -0.45 $, $ -0.55 $, $ -0.65 $  
  & $ 0 $ 
  & $ 0.02 $ \\
  \hline
 \end{tabular}
 \caption{The anisotropy study covers a wide range of parameter values. Parallel and perpendicular source-smeared correlators are available in $ \gamma^5 $, $ \mathbf{1} $ and all $ \gamma^\mu $~channels. 
 }
 \label{tab: anisotropy data sets}
\end{table}\normalsize

\begin{table}[hbt]
 \center\footnotesize
 \begin{tabular}{|c|c||p{7cm}|c|c|c|}
  \hline
  $ \beta $ & $ n_{cfg} $ & $ c $ & $ d $ & $ am_0 $ & $ r_0 m_0 $  \\
  \hline
  $ 5.8 $   & $ 10 $ & $ -0.4 $, $ -0.45 $, $ -0.48 $, $ -0.5 $, $ -0.53 $, $ -0.55 $, $ -0.6 $ & $ -0.002 $ & $ 0.01464 $ & $ 0.054 $ \\
  \hline
  $ 6.0 $   & $ 10 $ & $ -0.35 $, $ -0.40 $, $ -0.42 $, $ -0.45 $, $ -0.47 $, $ -0.50 $, $ -0.55 $ & $ -0.001 $ & $ 0.01 $ & $ 0.054 $ \\
  \hline
  $ 6.2 $   & $ 10 $ & $ -0.3 $, $ -0.35 $, $ -0.38 $, $ -0.40 $, $ -0.42 $, $ -0.45 $, $ -0.50 $ & $ -0.001 $ & $ 0.0073 $ & $ 0.054 $ \\
  \hline
 \end{tabular}
 \caption{All datasets of the frequency study use the same bare mass $ (r_0m_0) $ in physical units. Mesonic correlators are source smeared and available in all $ \mathcal{M} $,$ \mathcal{N} $~channels.
 }
 \label{tab: frequency data sets}
\end{table}\normalsize

\begin{table}[hbt]
 \center\footnotesize
 \begin{tabular}{|c|c|c||c|c|c||c|c|}
  \hline
  Action & $ \beta $ & $ n_{cfg} $ & $ c $ & $ d $ & $ m_{\mathrm{cr}} $ & $ m_0 $ & $ (r_0 M_{55})\ [\mathrm{MeV}] $  \\
  \hline
  Karsten-Wilczek, $ \parallel $ & $ 6.0 $ & $ 20 $ & $ -0.45 $ & $ -0.001 $ & n.a       & $ 0.02 $   & $ 642(4) $ \\
  Karsten-Wilczek, $ \perp $     & $ 6.0 $ & $ 20 $ & $ -0.45 $ & $ -0.001 $ & n.a       & $ 0.02 $   & $ 642(4) $ \\
  Na\"{i}ve                      & $ 6.0 $ & $ 20 $ & n.a.      & n.a        & n.a       & $ 0.02 $   & $ 728(2) $ \\
  Wilson                         & $ 6.0 $ & $ 20 $ & n.a.      & n.a        & $ -0.808 $ & $ -0.788 $ & $ 545(4) $ \\
  \hline
 \end{tabular}
 \caption{Different values of $ r_0 M_{55} $ for equal bare masses $ m_0 $ (or $ m_0-m_{cr} $ for Wilson fermions) are due to use of different fermion actions.
 }
 \label{tab: comparison data sets}
\end{table}\normalsize

\begin{table}[hbt]
\center\footnotesize
 \begin{tabular}{|c|c|c|c|c||c|c|c|c|c|}
  \hline
  $ \beta $ & $ L $ & $ n_{cfg} $ & $ c $ & $ d $ & $ am_0 $ & $ (r_0 m_0) $ & $ (r_0 M_{55}) $ & $ M_{55}\ [\mathrm{MeV}] $ & $ \mathcal{B}_{55} $  \\
  \hline
  $ 5.8 $ & $ 24 $ & $ 200 $ & $ -0.51 $ & $ -0.002 $ & $ 0.02928 $ & $ 0.107 $ & $ 1.655(2) $ & $ 653(1) $ & $ 25.5(1) $ \\
  $ 5.8 $ & $ 24 $ & $ 200 $ & $ -0.51 $ & $ -0.002 $ & $ 0.02000 $ & $ 0.073 $ & $ 1.379(1) $ & $ 544(1) $ & $ 25.9(1) $ \\
  $ 5.8 $ & $ 24 $ & $ 200 $ & $ -0.51 $ & $ -0.002 $ & $ 0.01464 $ & $ 0.054 $ & $ 1.188(1) $ & $ 469(1) $ & $ 26.3(1) $ \\
  $ 5.8 $ & $ 24 $ & $ 200 $ & $ -0.51 $ & $ -0.002 $ & $ 0.01000 $ & $ 0.037 $ & $ 0.990(1) $ & $ 391(1) $ & $ 26.7(1) $ \\
  $ 5.8 $ & $ 24 $ & $ 200 $ & $ -0.51 $ & $ -0.002 $ & $ 0.07320 $ & $ 0.027 $ & $ 0.852(1) $ & $ 336(1) $ & $ 27.0(1) $ \\
  $ 5.8 $ & $ 24 $ & $ 200 $ & $ -0.51 $ & $ -0.002 $ & $ 0.05340 $ & $ 0.020 $ & $ 0.731(1) $ & $ 288(1) $ & $ 27.3(1) $ \\
  \hline
  $ 6.0 $ & $ 24 $ & $ 200 $ & $ -0.45 $ & $ -0.001 $ & $ 0.02000 $ & $ 0.107 $ & $ 1.598(3) $ & $ 631(1) $ & $ 23.8(1) $ \\
  $ 6.0 $ & $ 24 $ & $ 200 $ & $ -0.45 $ & $ -0.001 $ & $ 0.01371 $ & $ 0.074 $ & $ 1.333(3) $ & $ 526(1) $ & $ 24.2(1) $ \\
  $ 6.0 $ & $ 24 $ & $ 200 $ & $ -0.45 $ & $ -0.001 $ & $ 0.01000 $ & $ 0.054 $ & $ 1.149(3) $ & $ 453(1) $ & $ 24.6(1) $ \\
  $ 6.0 $ & $ 24 $ & $ 200 $ & $ -0.45 $ & $ -0.001 $ & $ 0.06850 $ & $ 0.037 $ & $ 0.964(4) $ & $ 380(1) $ & $ 25.3(2) $ \\
  $ 6.0 $ & $ 24 $ & $ 200 $ & $ -0.45 $ & $ -0.001 $ & $ 0.00500 $ & $ 0.027 $ & $ 0.833(4) $ & $ 329(2) $ & $ 25.9(3) $ \\
  $ 6.0 $ & $ 24 $ & $ 200 $ & $ -0.45 $ & $ -0.001 $ & $ 0.03650 $ & $ 0.020 $ & $ 0.719(4) $ & $ 284(2) $ & $ 26.4(3) $ \\
  \hline
  $ 6.2 $ & $ 32 $ & $ 100 $ & $ -0.40 $ & $ -0.001 $ & $ 0.01460 $ & $ 0.107 $ & $ 1.575(5) $ & $ 621(2) $ & $ 23.1(2) $ \\
  $ 6.2 $ & $ 32 $ & $ 100 $ & $ -0.40 $ & $ -0.001 $ & $ 0.01000 $ & $ 0.074 $ & $ 1.312(6) $ & $ 518(2) $ & $ 23.4(2) $ \\
  $ 6.2 $ & $ 32 $ & $ 100 $ & $ -0.40 $ & $ -0.001 $ & $ 0.07300 $ & $ 0.054 $ & $ 1.131(6) $ & $ 446(2) $ & $ 23.8(3) $ \\
  $ 6.2 $ & $ 32 $ & $ 100 $ & $ -0.40 $ & $ -0.001 $ & $ 0.00500 $ & $ 0.037 $ & $ 0.947(7) $ & $ 374(3) $ & $ 24.4(3) $ \\
  $ 6.2 $ & $ 32 $ & $ 100 $ & $ -0.40 $ & $ -0.001 $ & $ 0.03650 $ & $ 0.027 $ & $ 0.819(7) $ & $ 323(3) $ & $ 25.0(4) $ \\
  $ 6.2 $ & $ 32 $ & $ 100 $ & $ -0.40 $ & $ -0.001 $ & $ 0.02660 $ & $ 0.020 $ & $ 0.709(8) $ & $ 280(3) $ & $ 25.7(6) $ \\
  $ 6.2 $ & $ 32 $ & $ 100 $ & $ -0.40 $ & $ -0.001 $ & $ 0.01940 $ & $ 0.014 $ & $ 0.615(8) $ & $ 243(3) $ & $ 26.5(7) $ \\
  \hline
 \end{tabular}
 \caption{
 Chiral behaviour is studied on lattices with $ T=48 $. Different $ r_0 M_{55} $ for equal $ r_0 m_0 $ at different $ \beta $ indicates lack of mass renormalisation.
}
 \label{tab: chiral study data sets}
 \vspace{-8pt}
\end{table}

\FloatBarrier
\sectionc{Addendum to tuning with the frequency spectrum}{app: Addendum to tuning with the frequency spectrum}

\noindent
Ratios $ R_{05}(n_0) $ of correlation functions for the coarser lattices~($ \beta=6.0 $,~$ \beta=5.8 $) are displayed in the following. Since the ratio at the source $ R_{05}(0) $ varies only within one or two standard errors over the range of~$ c $ for all lattices, the ratio of renormalisation factors of the~$ \gamma^5 $ and $ \gamma^0 $~bilinears seems to have only a weak dependence on~$ c $. \newline 

\begin{figure}[htb]
 \begin{picture}(400,142)
  \put(000  ,080.0){\includegraphics[bb=0 0 180 90, scale=0.45]{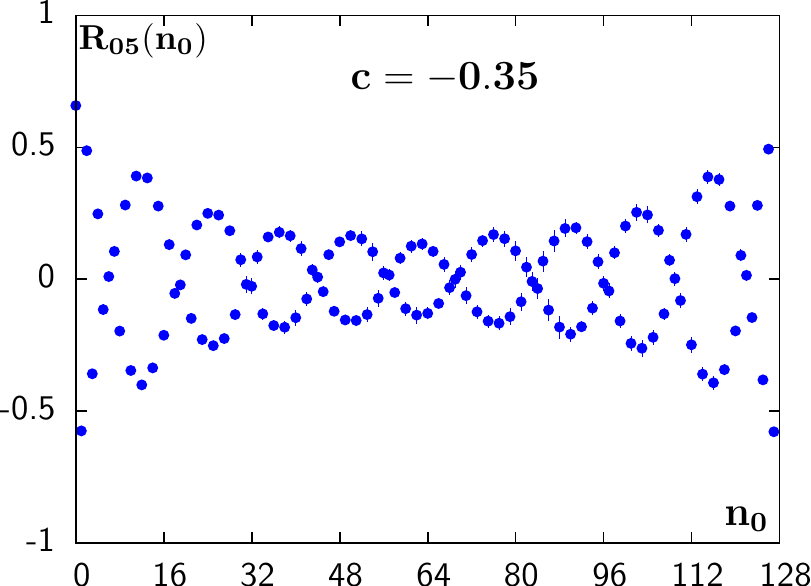}}
  \put(105.0,080.0){\includegraphics[bb=0 0 180 90, scale=0.45]{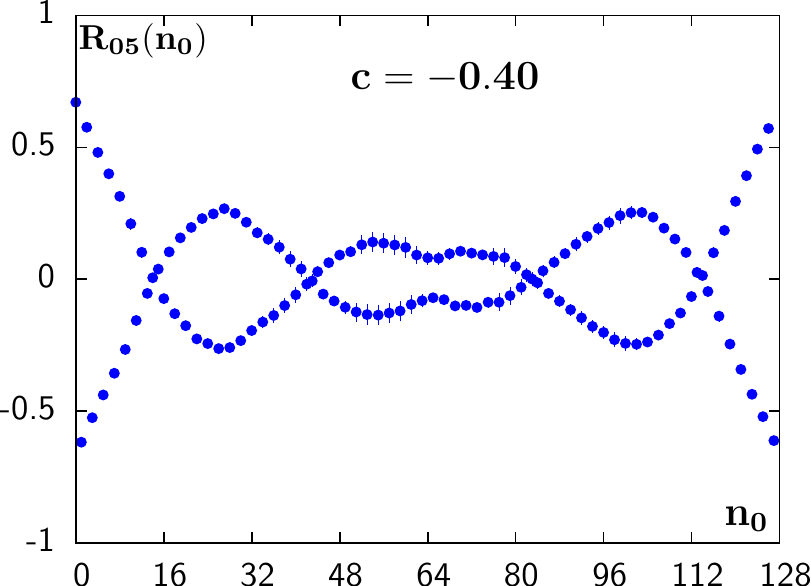}}
  \put(210  ,080.0){\includegraphics[bb=0 0 180 90, scale=0.45]{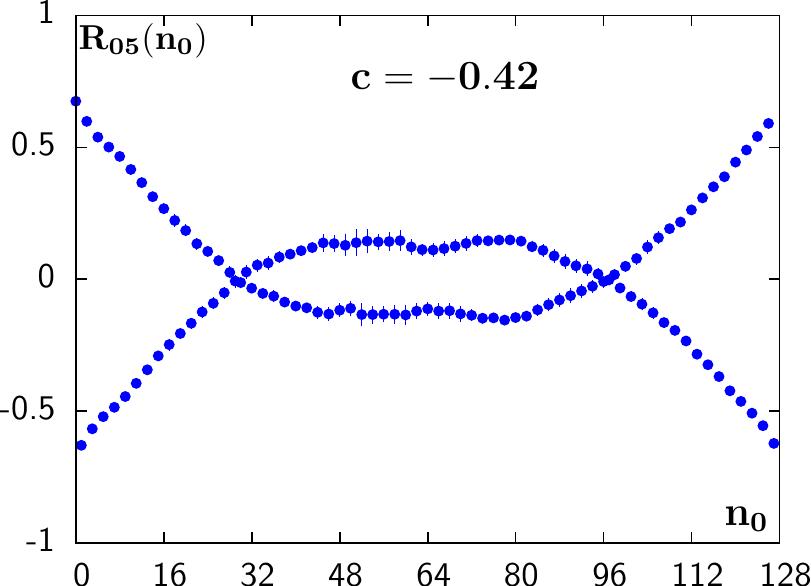}}
  \put(315.0,080.0){\includegraphics[bb=0 0 180 90, scale=0.45]{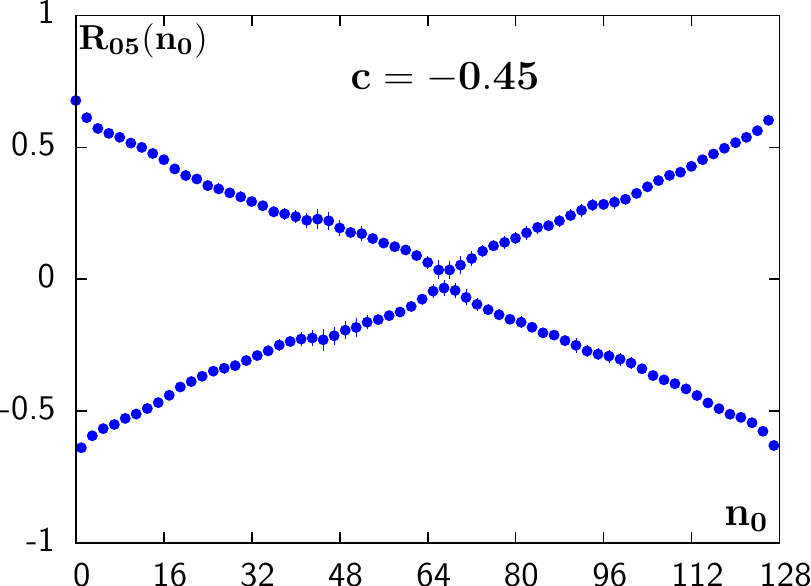}}
  \put(000  ,000.0){\includegraphics[bb=0 0 180 90, scale=0.45]{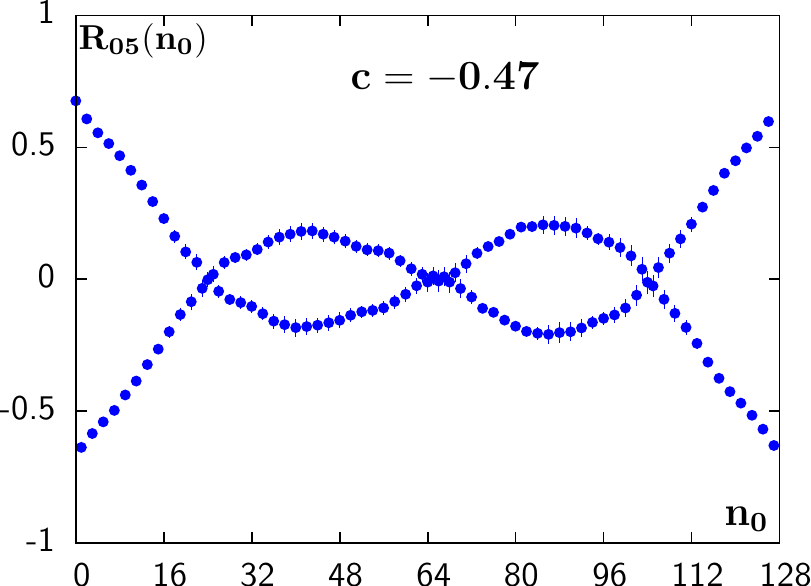}}
  \put(105.0,000.0){\includegraphics[bb=0 0 180 90, scale=0.45]{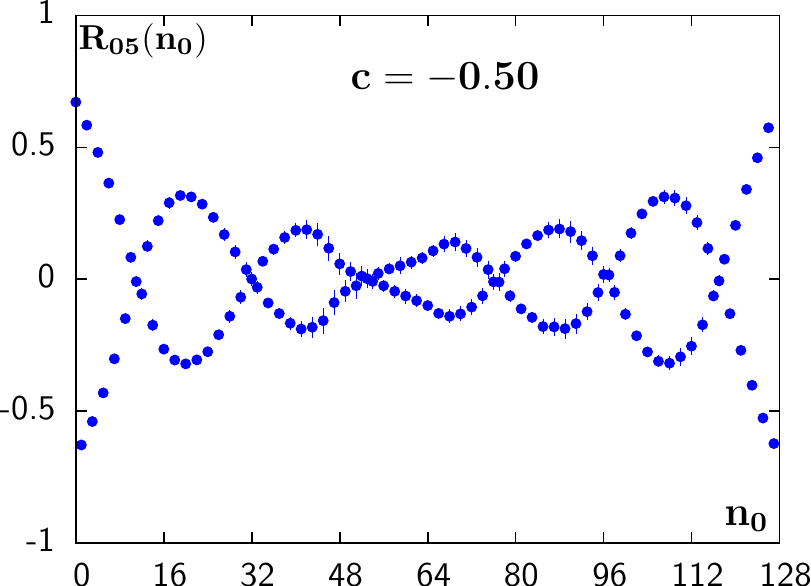}}
  \put(210.0,000.0){\includegraphics[bb=0 0 180 90, scale=0.45]{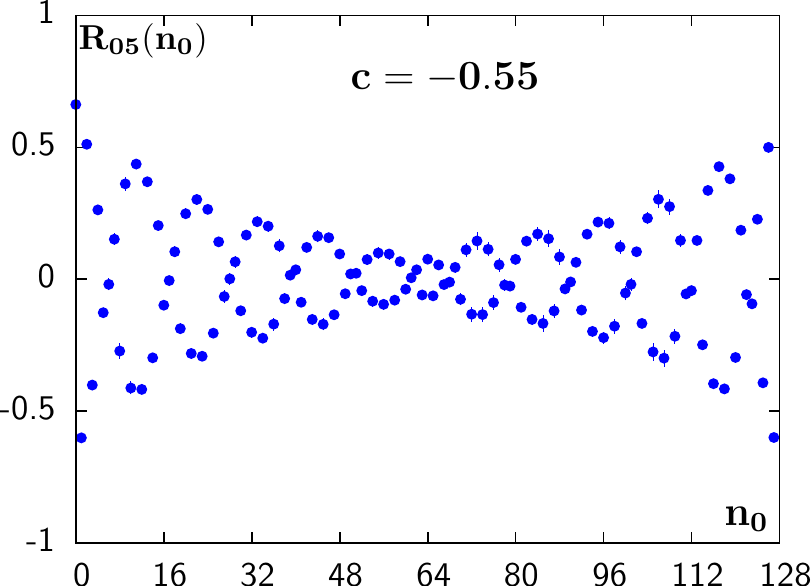}}
 \end{picture}
 \caption{Ratios $ R_{05}(n_0) $ on a $ 128\times24^3 $~lattice at~$ \beta=6.0 $ from table~\ref{tab: lattices for oscillation study} are unmistakable evidence of the~$ c $ dependence of the frequency shift.
  }
 \label{fig: ratio R_05 at beta 6.0}
\end{figure}

\begin{figure}[htb]
 \begin{picture}(400,142)
  \put(000  ,080.0){\includegraphics[bb=0 0 180 90, scale=0.45]{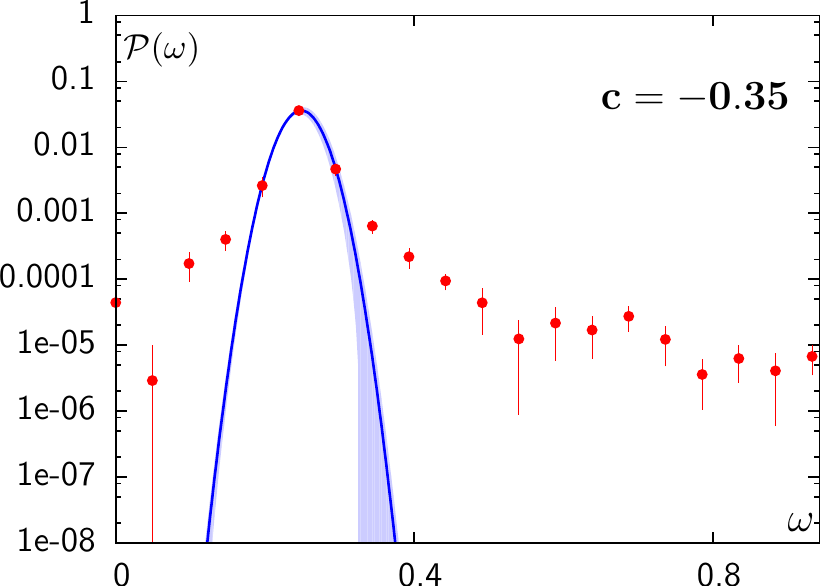}}
  \put(105.0,080.0){\includegraphics[bb=0 0 180 90, scale=0.45]{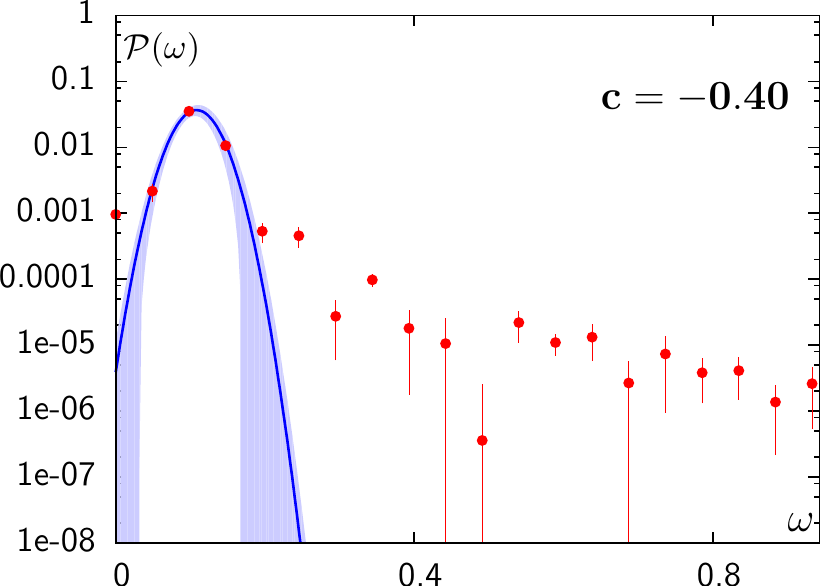}}
  \put(210  ,080.0){\includegraphics[bb=0 0 180 90, scale=0.45]{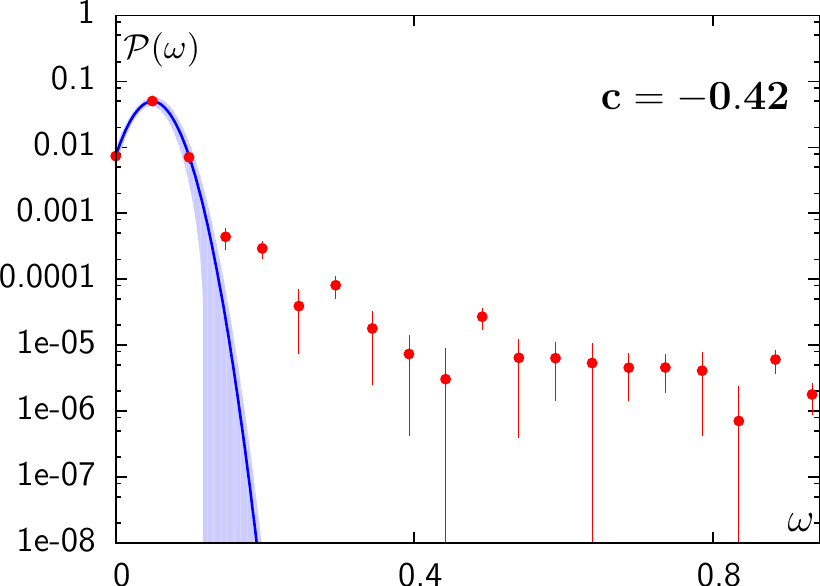}}
  \put(315.0,080.0){\includegraphics[bb=0 0 180 90, scale=0.45]{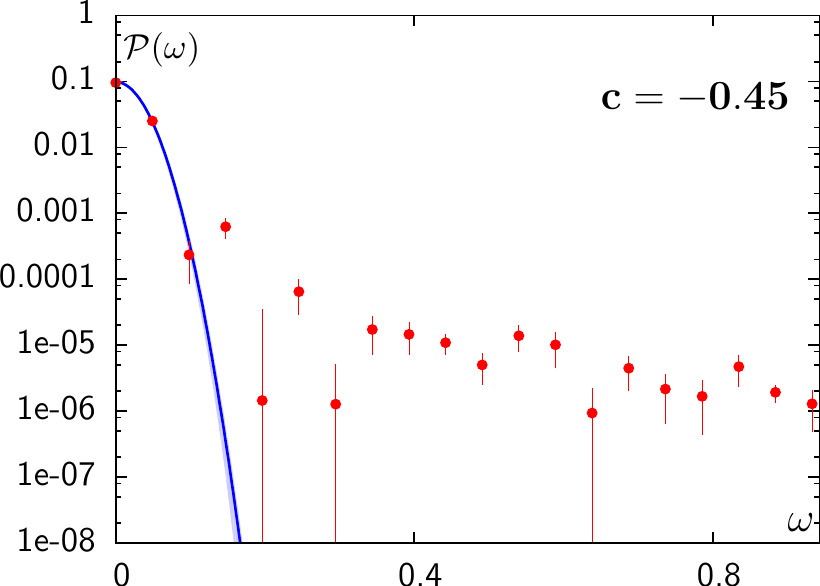}}
  \put(000  ,000.0){\includegraphics[bb=0 0 180 90, scale=0.45]{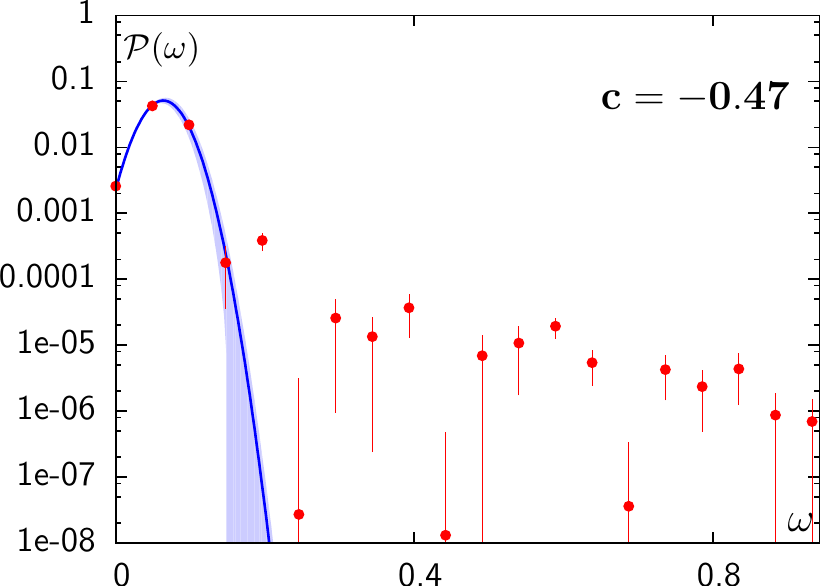}}
  \put(105.0,000.0){\includegraphics[bb=0 0 180 90, scale=0.45]{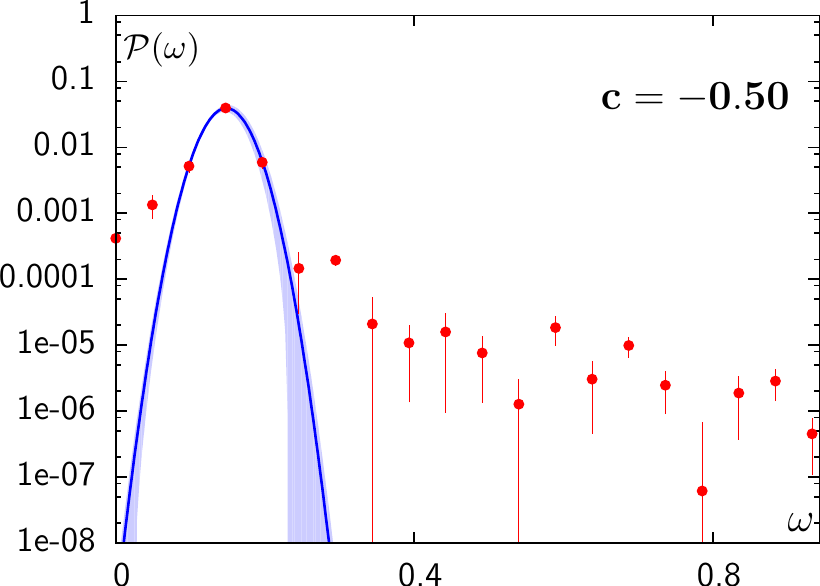}}
  \put(210.0,000.0){\includegraphics[bb=0 0 180 90, scale=0.45]{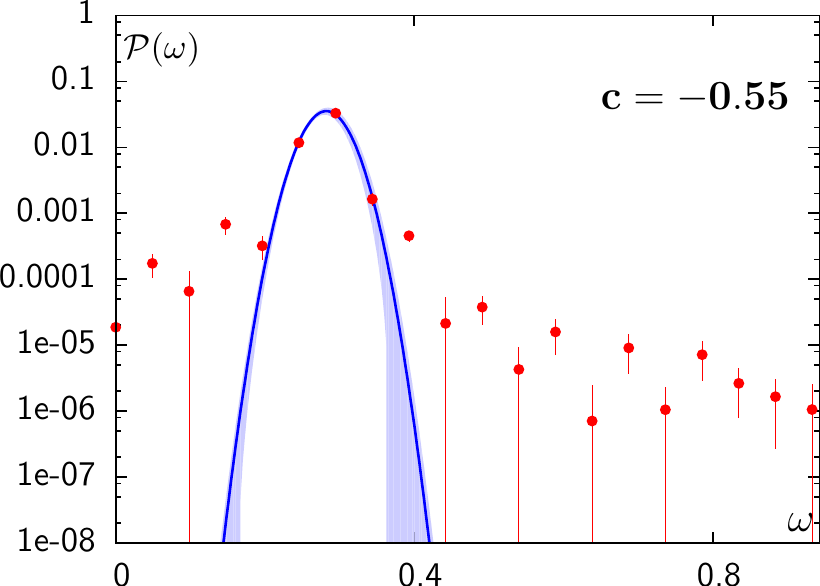}}
 \end{picture} \caption{
 Power spectral densities ($ \beta=6.0 $) are displayed on a logarithmic scale. The curve is a gaussian function, which is used to estimate the maximum of the distribution.
  }
 \label{fig: psd P_05 at beta 6.0}
\end{figure}

\noindent
The residual exponential decay at $ \beta=6.0 $ can be approximated linearly (\mbox{cf.} $ c=-0.45 $ in figure \ref{fig: ratio R_05 at beta 6.0}). Hence, peak broadening in the power spectral density is still relatively mild and the peaks consist of $ 3\!-\!4 $ data points. The width of the peak does not increase significantly above the mininum that is defined by half of the bin size,
\begin{equation}
   \sigma_{\min}=\omega_b/2.
   \label{eq: minimal width}
\end{equation}\normalsize

\begin{figure}[htb]
 \begin{picture}(400,142)
  \put(000  ,080.0){\includegraphics[bb=0 0 180 90, scale=0.45]{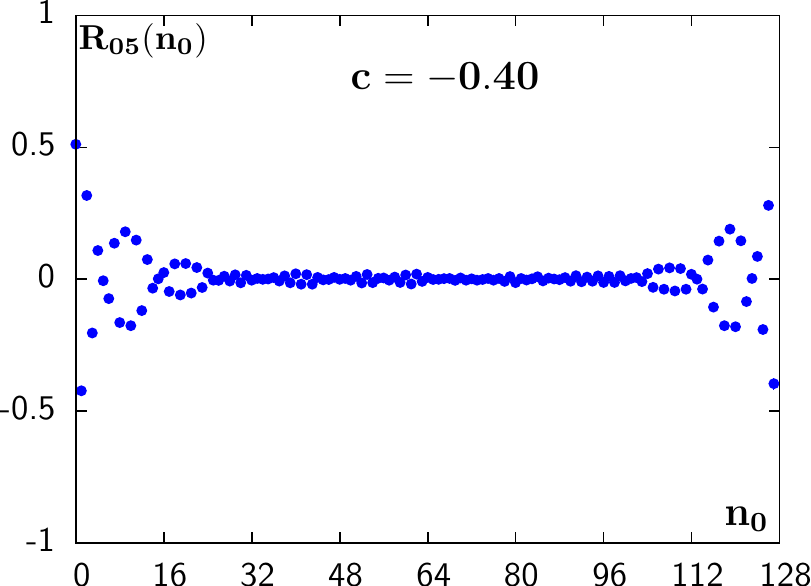}}
  \put(105.0,080.0){\includegraphics[bb=0 0 180 90, scale=0.45]{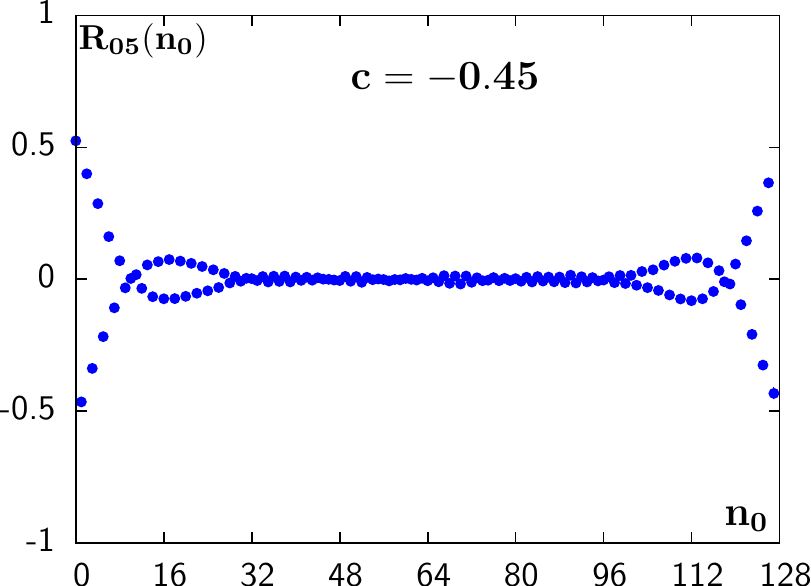}}
  \put(210  ,080.0){\includegraphics[bb=0 0 180 90, scale=0.45]{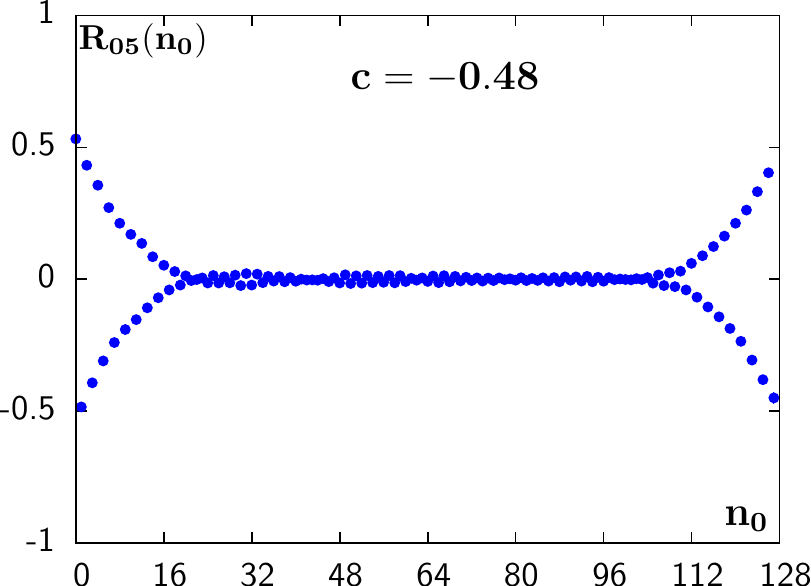}}
  \put(315.0,080.0){\includegraphics[bb=0 0 180 90, scale=0.45]{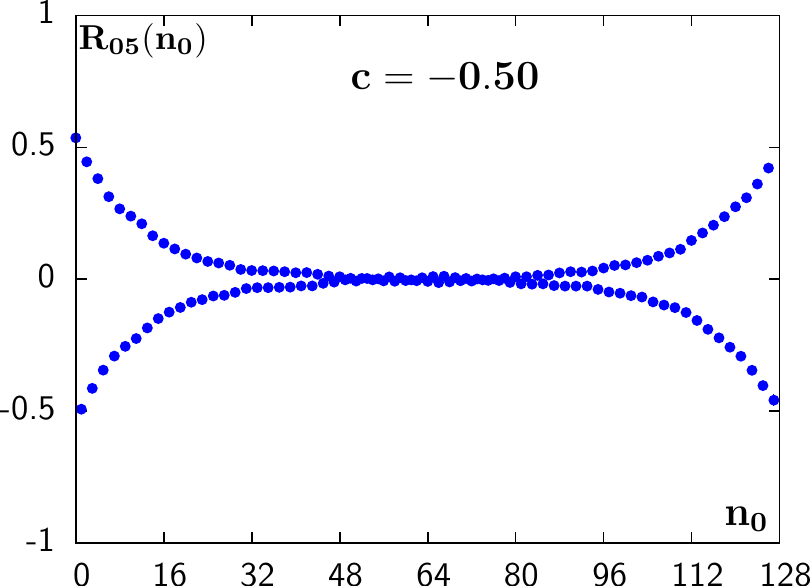}}
  \put(000  ,000.0){\includegraphics[bb=0 0 180 90, scale=0.45]{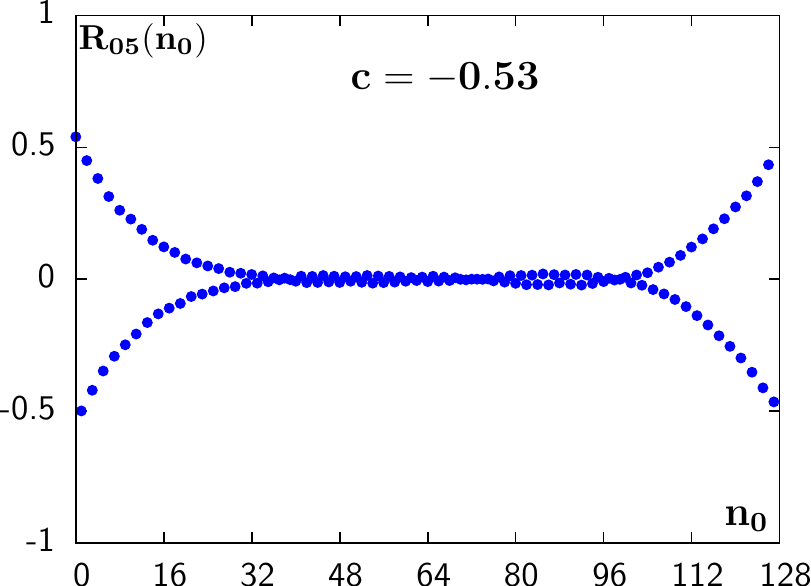}}
  \put(105.0,000.0){\includegraphics[bb=0 0 180 90, scale=0.45]{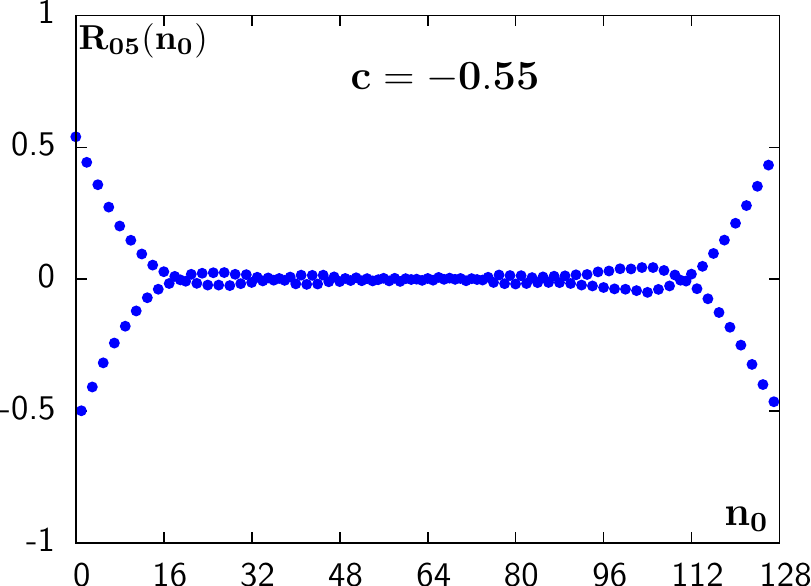}}
  \put(210.0,000.0){\includegraphics[bb=0 0 180 90, scale=0.45]{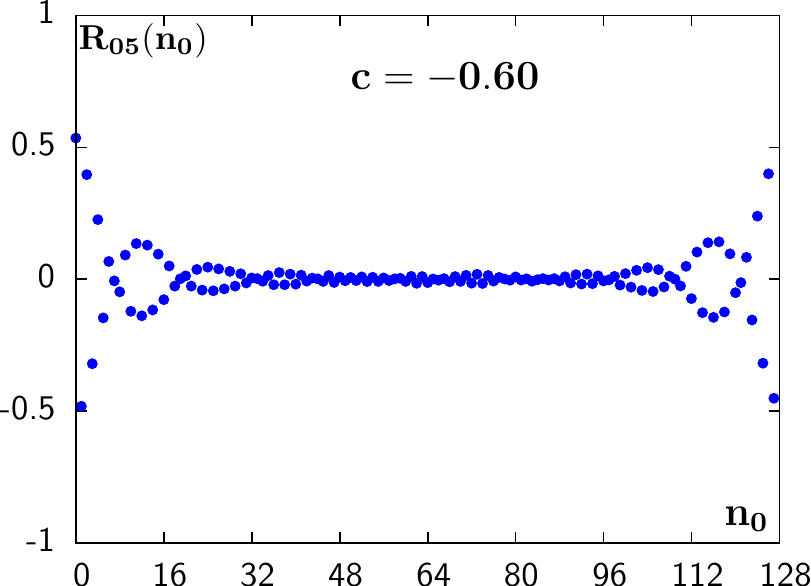}}
 \end{picture}
 \caption{Ratios~$ R_{05}(n_0) $ on a $ 128\times24^3 $~lattice for~$ \beta=5.8 $ from table~\ref{tab: lattices for oscillation study} still feature large exponential decays.
  }
 \label{fig: ratio R_05 at beta 5.8}
\end{figure}

\FloatBarrier
\pagebreak
\noindent
The residual exponential decay for~$ \beta=5.8 $ cannot be approximated as a linear funciton~(\mbox{cf.}~$ c=-0.50 $ in figure~\ref{fig: ratio R_05 at beta 5.8}) and oscillations can hardly be resolved visually. Presumably, the oscillation can be isolated cleanly if the ratio is divided by a hyperbolic cosine of the ground state mass difference for~$ c\approx c_0 $. \newline

\begin{figure}[htb]
 \begin{picture}(400,142)
  \put(000  ,080.0){\includegraphics[bb=0 0 180 90, scale=0.45]{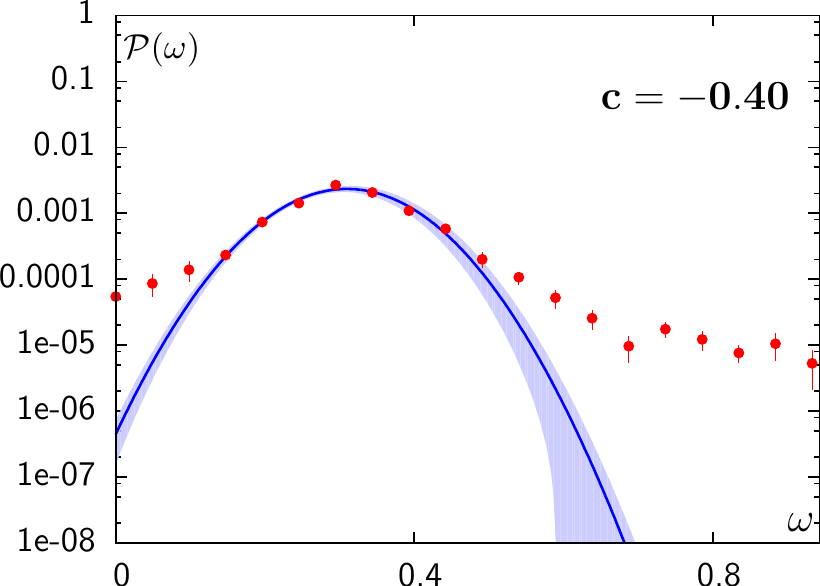}}
  \put(105.0,080.0){\includegraphics[bb=0 0 180 90, scale=0.45]{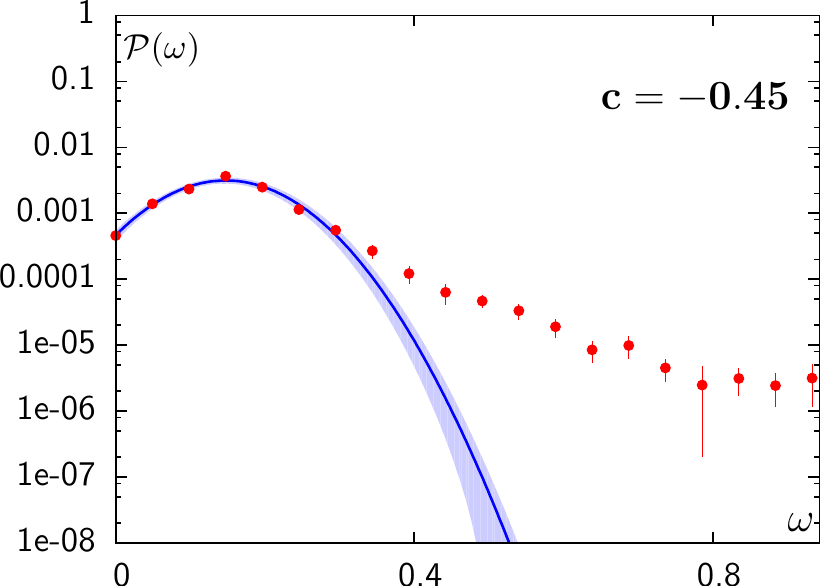}}
  \put(210  ,080.0){\includegraphics[bb=0 0 180 90, scale=0.45]{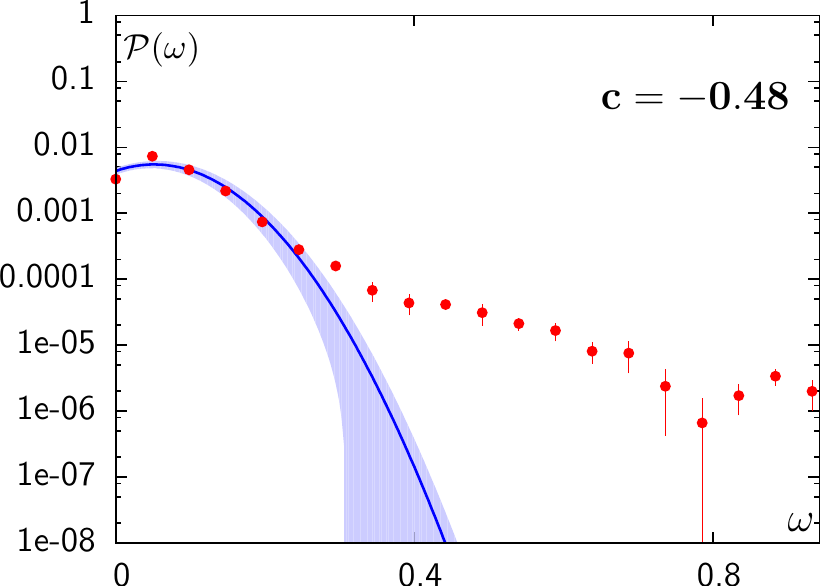}}
  \put(315.0,080.0){\includegraphics[bb=0 0 180 90, scale=0.45]{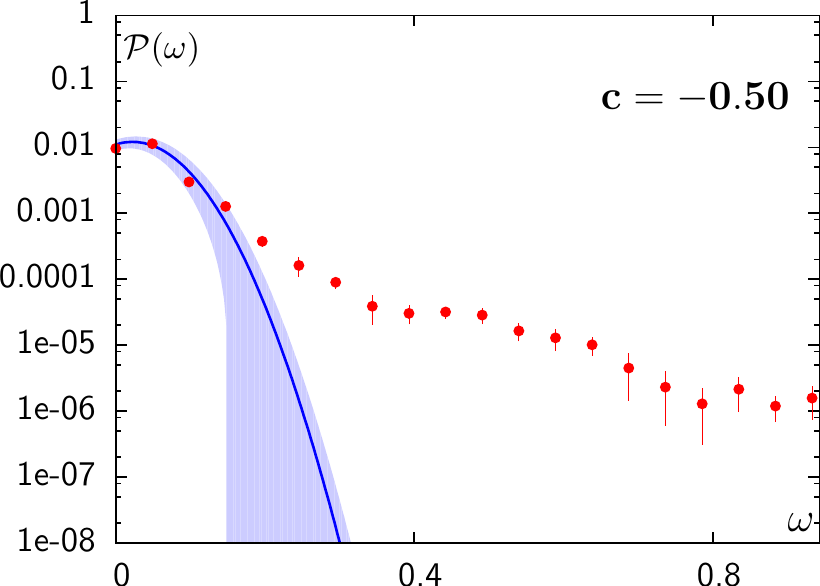}}
  \put(000  ,000.0){\includegraphics[bb=0 0 180 90, scale=0.45]{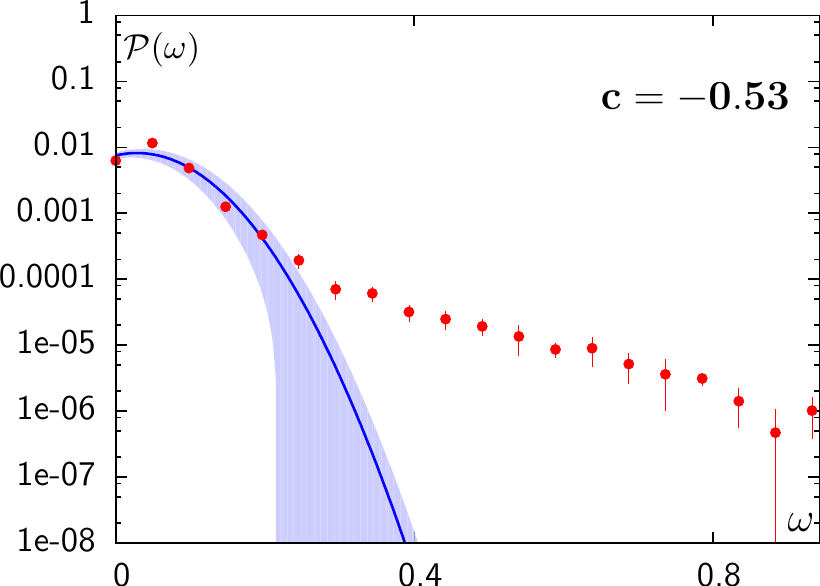}}
  \put(105.0,000.0){\includegraphics[bb=0 0 180 90, scale=0.45]{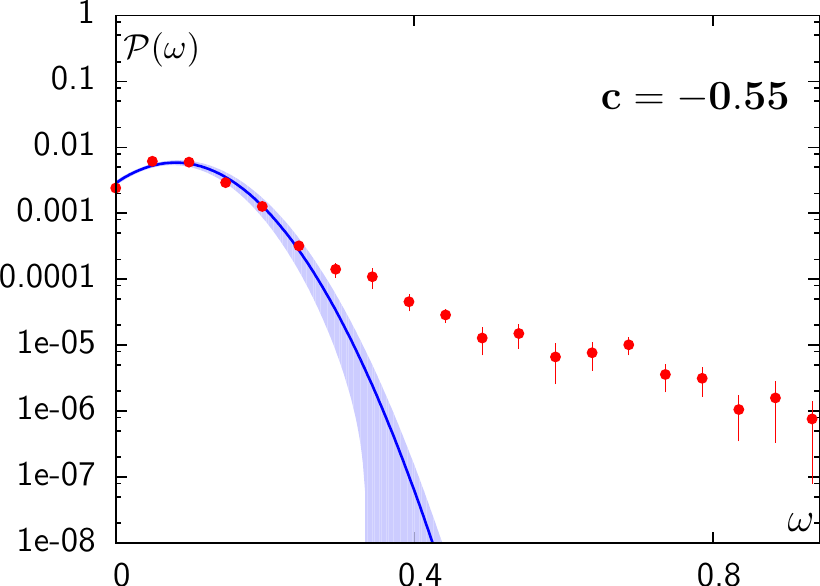}}
  \put(210.0,000.0){\includegraphics[bb=0 0 180 90, scale=0.45]{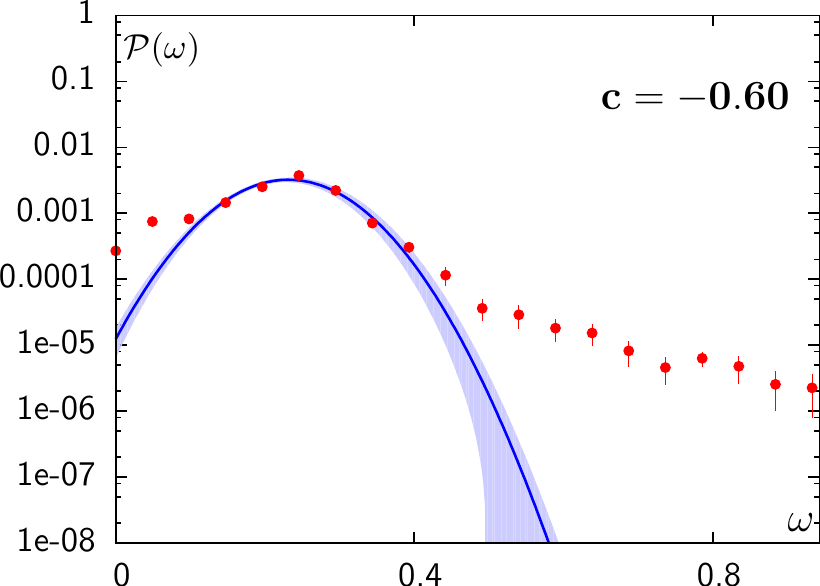}}
 \end{picture}
 \caption{
 Power spectral densities~($ \beta=5.8 $) are displayed on a logarithmic scale. The curve is a gaussian function, which is used to estimate the maximum of the distribution.
  }
 \label{fig: psd P_05 at beta 5.8}
\end{figure}

\noindent
Peak broadening in the power spectral density is quite severe and each peak consists of~$ 7\!-\!10 $ data points~(\mbox{cf.}~figure~\ref{fig: psd P_05 at beta 5.8}). Its width is therefore approximately three times as large as the minimal width of \mbox{eq.}~(\ref{eq: minimal width}). Thus, the systematical error can be presumably reduced by a factor~$ 2\!-\!4 $, if the residual decay is removed. \newline

\newpage
\sectionc{Addendum to chiral behaviour of the pseudoscalar ground state}{app: Addendum to chiral behaviour of the pseudoscalar ground state}

\begin{figure}[htb]
 \begin{picture}(360,270)
  \put(000.0,138.0){\includegraphics[bb=0 0 170 130, scale=0.75]{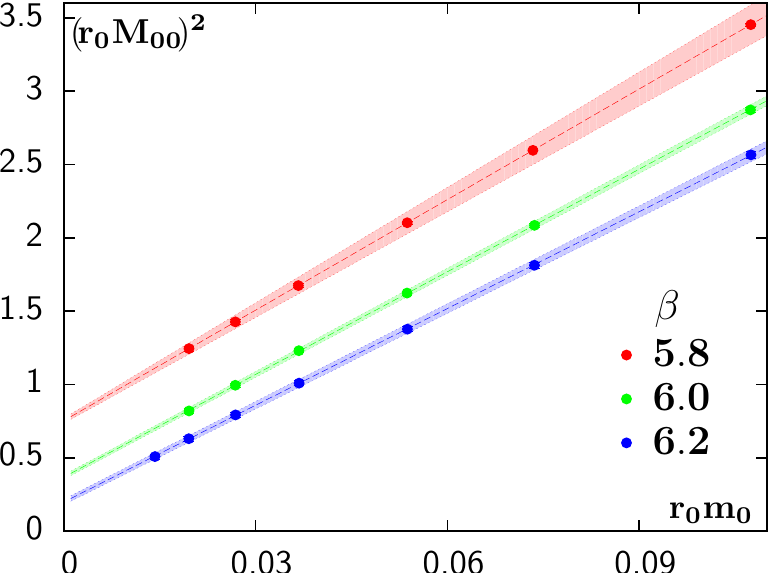}}
  \put(220.0,138.0){\includegraphics[bb=0 0 170 130, scale=0.75]{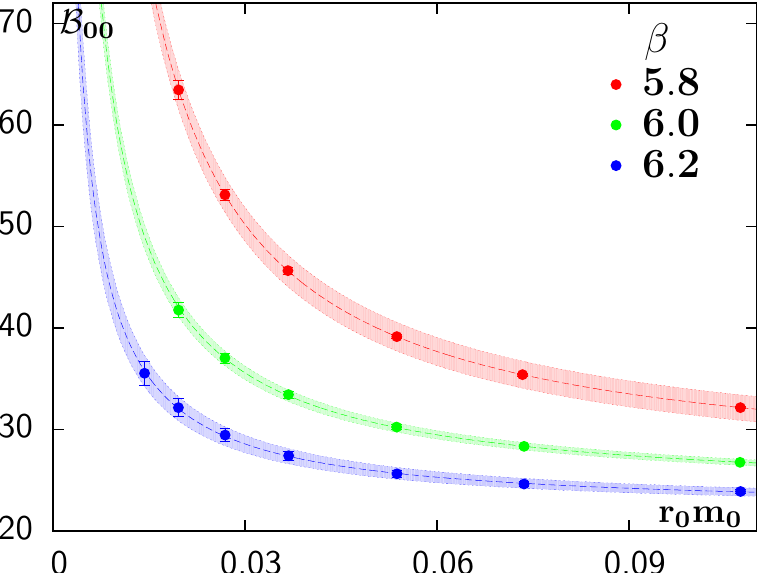}}
  \put(000.0,  0.0){\includegraphics[bb=0 0 170 130, scale=0.75]{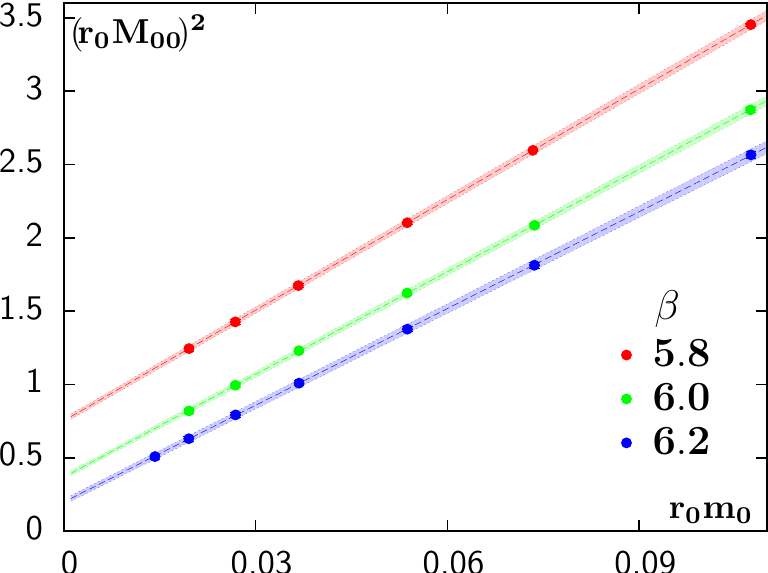}}
  \put(220.0,  0.0){\includegraphics[bb=0 0 170 130, scale=0.75]{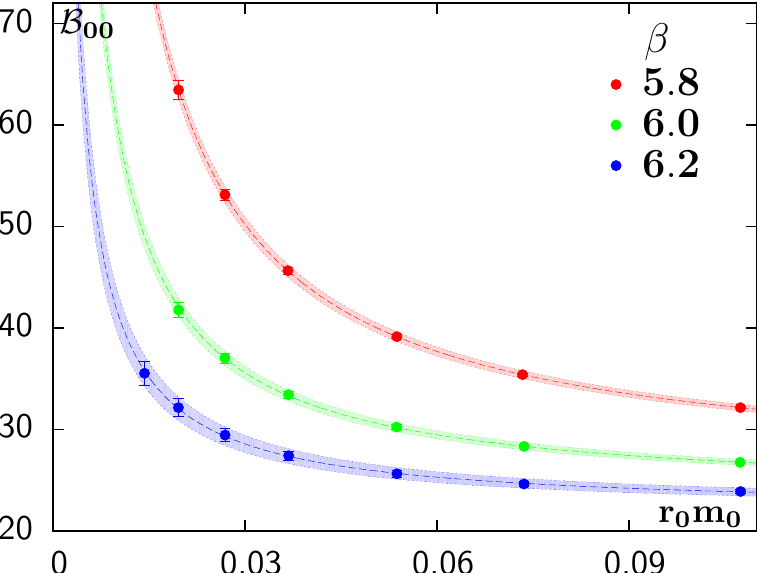}}
 \end{picture}
 \caption{Upper row: The chiral extrapolation of~$ M_{00}^2 $ including quenched chiral logarithms obtains the residual mass~$ M_{\mathrm{res}}^2 $ with small errors.
 Lower row: Without logarithms, the extrapolation of~$ M_{00}^2 $ has narrower error bands.
  }
 \label{fig: chiral behaviour of M_00}
\end{figure}

The ground state mass $ M_{00} $ is extracted with a fit using \mbox{eq.}~(\ref{eq: chiral behaviour fit function}). The left column of figure~\ref{fig: chiral behaviour of M_00} shows \mbox{$ \mathcal{B}_{00}=(r_0 M_{00})^2/(r_0 m_0) $}, the analogue of $ \mathcal{B}_{55} $ defined in \mbox{eq.}~(\ref{eq: ratio B_55}).

\noindent
The study of the chiral behaviour of the pseudoscalar ground states of $ \gamma^5 $~and $ \gamma^0 $~channels involves fits to correlation functions. Local effective (cosh) masses are displayed in the following together with the fit masses with $ 1\sigma $~bands within the fit ranges. The grey-shaded bands are considered as potentially affected by excited states. Fit parameters are stable within $ 1\sigma $ upon variation of the fit range within the central region. The error bands of the $ \gamma^0 $~channel are wider by a factor $ 1.5 \sim 2 $. Fluctuations as well as errors increase with a decrease of the lattice spacing as well as with a decrease of the quark mass. Since fluctuations of both channels are very similar, there are large cancellations in the ratio of correlation functions, which explain the decrease of the error of the mass difference in the chiral limit.

\begin{figure}[htb]
 \begin{picture}(360,480)
  \put( 05  ,320.0){\includegraphics[bb=0 0 180 90, scale=0.90]{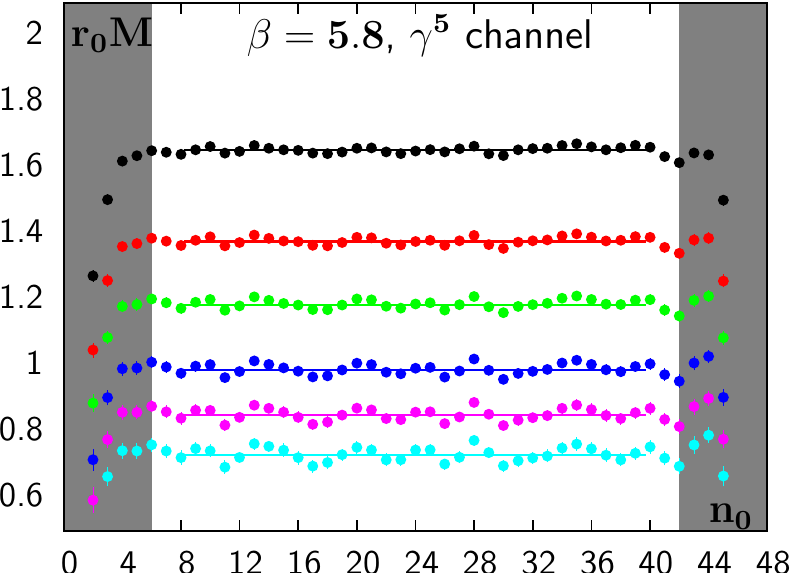}}
  \put( 05  ,160.0){\includegraphics[bb=0 0 180 90, scale=0.90]{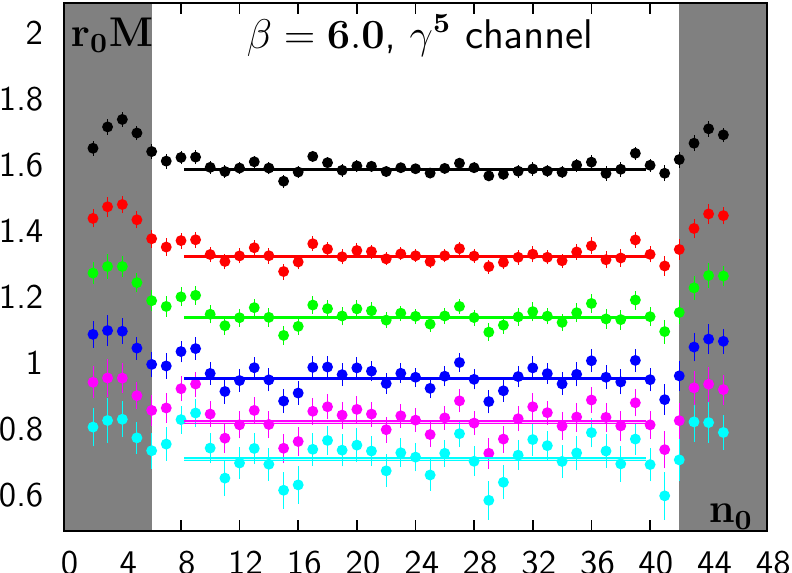}}
  \put( 05  ,000.0){\includegraphics[bb=0 0 180 90, scale=0.90]{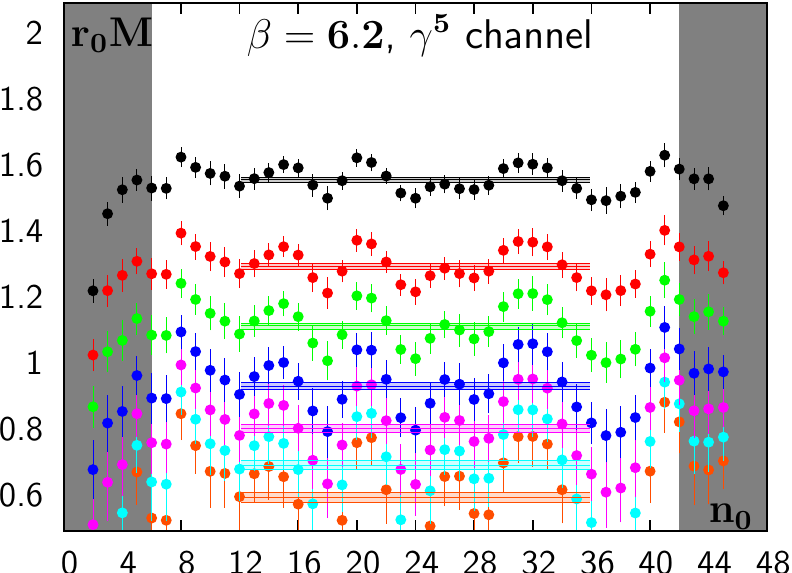}}
  \put( 210 ,320.0){\includegraphics[bb=0 0 180 90, scale=0.90]{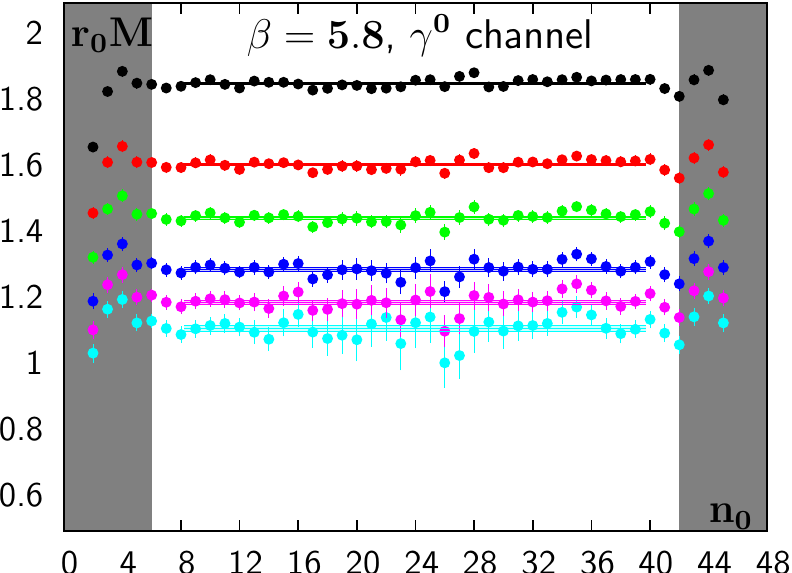}}
  \put( 210 ,160.0){\includegraphics[bb=0 0 180 90, scale=0.90]{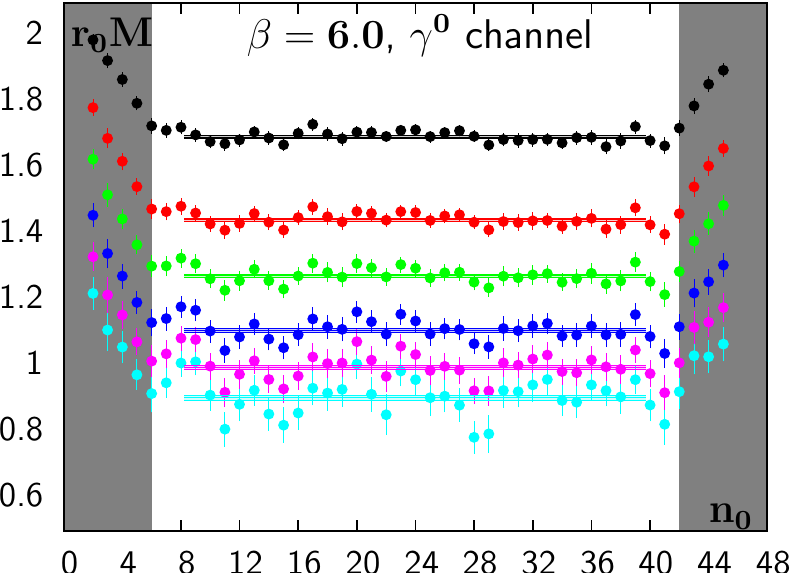}}
  \put( 210 ,000.0){\includegraphics[bb=0 0 180 90, scale=0.90]{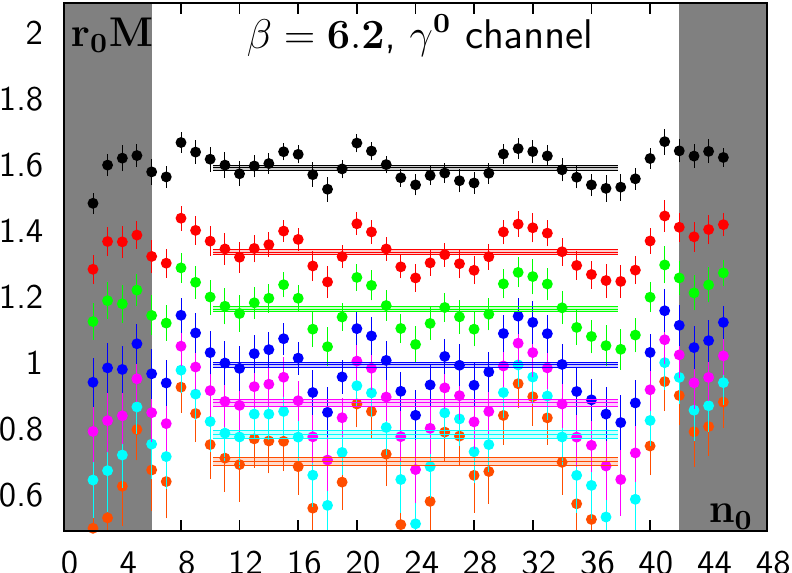}}
 \end{picture}
 \caption{Local effective mass plots of the $ \gamma^5 $ and $ \gamma^0 $ channels exhibit almost the same fluctuations for all masses from table \ref{tab: fermion mass parameters}.
  }
 \label{fig: local effective mass of gamma5 and gamma0-channels}
\end{figure}

\chapter{Numerical implementation}\label{app: Numerical implementation}

\sectionc{Lattice Dirac operators}{app: Lattice Dirac operators}

\noindent
This appendix covers the implementation of the Karsten-Wilczek and the Bori\c{c}i-Creutz Dirac operators in numerical studies of minimally doubled fermions. The hermitian Dirac operators $ Q= \gamma^5 D $ are applied in numerical simulations. The application of $ Q $ is implemented with an input field $ \phi $ and an output field $ \psi $ as
\begin{equation}
  \psi = Q \phi = \delta_m \psi +\frac{1}{2} \sum\limits_{\mu}{\left(\delta^\mu_+ \psi+ \delta^\mu_- \psi\right)}.
  \label{eq: numerical Dirac operator application}
\end{equation}\normalsize
The on-site term $ \delta_m \psi $ and the hopping terms $ \delta^\mu_\pm \psi $ are defined for Karsten-Wilczek and Bori\c{c}i-Creutz operators in the following.

\subsection{Karsten-Wilczek Dirac operator}

The Karsten-Wilczek Dirac operator with $ \underline{\alpha}=0 $ is taken from \mbox{eq.}~(\ref{eq: KW fermion action}). The hopping terms $ \delta^\mu_\pm \psi $ of \mbox{eq.}~(\ref{eq: numerical Dirac operator application}) add $ 1/2 $ times the following contributions:
\begin{align}
  \left\{\begin{array}{rcl} 
  \delta^0_- \psi^1_n &=& U_n^{0\dagger} (-\left[1+d\right]\phi^3_{n-\hat{e}_0}) \\
  \delta^0_- \psi^2_n &=& U_n^{0\dagger} (-\left[1+d\right]\phi^4_{n-\hat{e}_0}) \\
  \delta^0_- \psi^3_n &=& U_n^{0\dagger} (+\left[1+d\right]\phi^1_{n-\hat{e}_0}) \\
  \delta^0_- \psi^4_n &=& U_n^{0\dagger} (+\left[1+d\right]\phi^2_{n-\hat{e}_0})
  \end{array}\right\} &
  \left\{\begin{array}{rcl} 
  \delta^0_+ \psi^1_n &=& U_n^0\ (+\left[1+d\right]\phi^3_{n+\hat{e}_0}) \\
  \delta^0_+ \psi^2_n &=& U_n^0\ (+\left[1+d\right]\phi^4_{n+\hat{e}_0}) \\
  \delta^0_+ \psi^3_n &=& U_n^0\ (-\left[1+d\right]\phi^1_{n+\hat{e}_0}) \\
  \delta^0_+ \psi^4_n &=& U_n^0\ (-\left[1+d\right]\phi^2_{n+\hat{e}_0})
  \end{array}\right\} 
  \label{eq: KW 0 delta^0 psi}   \\
  \left\{\begin{array}{rcl} 
  \delta^3_- \psi^1_n &=& U_n^{3\dagger} (-i\left[1+\zeta\right]\phi^3_{n-\hat{e}_3}) \\
  \delta^3_- \psi^2_n &=& U_n^{3\dagger} (+i\left[1-\zeta\right]\phi^4_{n-\hat{e}_3}) \\
  \delta^3_- \psi^3_n &=& U_n^{3\dagger} (-i\left[1-\zeta\right]\phi^1_{n-\hat{e}_3}) \\
  \delta^3_- \psi^4_n &=& U_n^{3\dagger} (+i\left[1+\zeta\right]\phi^2_{n-\hat{e}_3})
  \end{array}\right\} &
  \left\{\begin{array}{rcl} 
  \delta^3_+ \psi^1_n &=& U_n^{3}\ (+i\left[1-\zeta\right]\phi^3_{n+\hat{e}_3}) \\
  \delta^3_+ \psi^2_n &=& U_n^{3}\ (-i\left[1+\zeta\right]\phi^4_{n+\hat{e}_3}) \\
  \delta^3_+ \psi^3_n &=& U_n^{3}\ (+i\left[1+\zeta\right]\phi^1_{n+\hat{e}_3}) \\
  \delta^3_+ \psi^4_n &=& U_n^{3}\ (-i\left[1-\zeta\right]\phi^2_{n+\hat{e}_3})
  \end{array}\right\}
  \label{eq: KW 0 delta^3 psi}
\end{align}\normalsize
\begin{align}
  \left\{\begin{array}{rcl} 
  \delta^1_- \psi^1_n &=& U_n^{1\dagger} (+i\left[\phi^4_{n-\hat{e}_1}-\zeta \phi^3_{n-\hat{e}_1}\right]) \\
  \delta^1_- \psi^2_n &=& U_n^{1\dagger} (+i\left[\phi^3_{n-\hat{e}_1}-\zeta \phi^4_{n-\hat{e}_1}\right]) \\
  \delta^1_- \psi^3_n &=& U_n^{1\dagger} (+i\left[\phi^2_{n-\hat{e}_1}+\zeta \phi^1_{n-\hat{e}_1}\right]) \\
  \delta^1_- \psi^4_n &=& U_n^{1\dagger} (+i\left[\phi^1_{n-\hat{e}_1}+\zeta \phi^2_{n-\hat{e}_1}\right])
  \end{array}\right\} &
  \left\{\begin{array}{rcl} 
  \delta^1_+ \psi^1_n &=& U_n^{1}\ (-i\left[\phi^4_{n-\hat{e}_1}+\zeta \phi^3_{n-\hat{e}_1}\right]) \\
  \delta^1_+ \psi^2_n &=& U_n^{1}\ (-i\left[\phi^3_{n-\hat{e}_1}+\zeta \phi^4_{n-\hat{e}_1}\right]) \\
  \delta^1_+ \psi^3_n &=& U_n^{1}\ (-i\left[\phi^2_{n-\hat{e}_1}-\zeta \phi^1_{n-\hat{e}_1}\right]) \\
  \delta^1_+ \psi^4_n &=& U_n^{1}\ (-i\left[\phi^1_{n-\hat{e}_1}-\zeta \phi^2_{n-\hat{e}_1}\right])
  \end{array}\right\} 
  \label{eq: KW 0 delta^1 psi} \\
  \left\{\begin{array}{rcl} 
  \delta^2_- \psi^1_n &=& U_n^{2\dagger} (+\left[\phi^4_{n-\hat{e}_2}-i\zeta \phi^3_{n-\hat{e}_2}\right]) \\
  \delta^2_- \psi^2_n &=& U_n^{2\dagger} (-\left[\phi^3_{n-\hat{e}_2}+i\zeta \phi^4_{n-\hat{e}_2}\right]) \\
  \delta^2_- \psi^3_n &=& U_n^{2\dagger} (+\left[\phi^2_{n-\hat{e}_2}+i\zeta \phi^1_{n-\hat{e}_2}\right]) \\
  \delta^2_- \psi^4_n &=& U_n^{2\dagger} (-\left[\phi^1_{n-\hat{e}_2}-i\zeta \phi^2_{n-\hat{e}_2}\right])
  \end{array}\right\} &
  \left\{\begin{array}{rcl} 
  \delta^2_+ \psi^1_n &=& U_n^{2}\ (-\left[\phi^4_{n-\hat{e}_2}+i\zeta \phi^3_{n-\hat{e}_2}\right]) \\
  \delta^2_+ \psi^2_n &=& U_n^{2}\ (+\left[\phi^3_{n-\hat{e}_2}-i\zeta \phi^4_{n-\hat{e}_2}\right]) \\
  \delta^2_+ \psi^3_n &=& U_n^{2}\ (-\left[\phi^2_{n-\hat{e}_2}-i\zeta \phi^1_{n-\hat{e}_2}\right]) \\
  \delta^2_+ \psi^4_n &=& U_n^{2}\ (+\left[\phi^1_{n-\hat{e}_2}+i\zeta \phi^2_{n-\hat{e}_2}\right])
  \end{array}\right\},
  \label{eq: KW 0 delta^2 psi}
\end{align}\normalsize
and the on-site term $ \delta_m \psi $ of eq. (\ref{eq: numerical Dirac operator application}) contributes
\begin{align}
  \left\{\begin{array}{rcl} 
  \delta_m \psi^1_n &=& (+m_0 \phi^1_{n} +i\frac{3\zeta+c}{a}\phi^3_{n}) \\
  \delta_m \psi^2_n &=& (+m_0 \phi^2_{n} +i\frac{3\zeta+c}{a}\phi^4_{n}) \\
  \delta_m \psi^3_n &=& (-m_0 \phi^3_{n} -i\frac{3\zeta+c}{a}\phi^1_{n}) \\
  \delta_m \psi^4_n &=& (-m_0 \phi^4_{n} -i\frac{3\zeta+c}{a}\phi^2_{n})
  \end{array}\right\}.
  \label{eq: KW 0 delta_m psi}
\end{align}\normalsize
Some spatial hopping terms $ \delta^j_\pm \psi $ for $ \underline{\alpha}=0 $ and $ \zeta=\pm1 $ are related (\mbox{cf.} table \ref{tab: KW n_0 spatial components' linear dependence}). Components which are linearly related are reconstructed with reduced numerical effort.
\begin{table}[hbt]
\center
 \begin{tabular}{|cccc|}
  \hline
  $ \delta^1_- \psi^1_n = \mp  \zeta \delta^1_- \psi^2_n $, & $ \delta^1_- \psi^3_n = \pm  \zeta \delta^1_- \psi^4_n $, & 
  $ \delta^1_+ \psi^1_n = \pm  \zeta \delta^1_+ \psi^2_n $, & $ \delta^1_+ \psi^3_n = \mp  \zeta \delta^1_+ \psi^4_n $, \\
  $ \delta^2_- \psi^1_n = \pm i\zeta \delta^2_- \psi^2_n $, & $ \delta^2_- \psi^3_n = \mp i\zeta \delta^2_- \psi^4_n $, & 
  $ \delta^2_+ \psi^1_n = \mp i\zeta \delta^2_+ \psi^2_n $, & $ \delta^2_+ \psi^3_n = \pm i\zeta \delta^2_+ \psi^4_n $, \\
  \hline
  \hline
  \multicolumn{2}{|r}{$ \zeta=+1 $:} &$ \delta^3_- \psi^2_n = \delta^3_- \psi^3_n =0 $, & $ \delta^3_+ \psi^1_n = \delta^3_+ \psi^4_n =0 $, \\
  \multicolumn{2}{|r}{$ \zeta=-1 $:} &$ \delta^3_- \psi^1_n = \delta^3_- \psi^4_n =0 $, & $ \delta^3_+ \psi^2_n = \delta^3_+ \psi^3_n =0 $. \\
  \hline
 \end{tabular}
 \caption{For $ \underline{\alpha}=0 $ and $ \zeta=\pm1 $, the Dirac components of $ \delta^1_\pm \psi $ and $ \delta^1_\pm \psi $ are linearly dependent and half of the $ \delta^3_\pm \psi $ trivially vanish. Moreover, the communications for spatial boundaries are halved in each spatial direction.}
 \label{tab: KW n_0 spatial components' linear dependence}
\end{table}

\noindent
The Karsten-Wilczek Dirac operator with $ \underline{\alpha}=3 $ is used for studies of the anisotropy. The hopping terms $ \delta^\mu_\pm \psi $ of \mbox{eq.}~(\ref{eq: numerical Dirac operator application}) add $ 1/2 $ times the following contributions:
\begin{align}
  \left\{\begin{array}{rcl} 
  \delta^0_- \psi^1_n &=& U_n^{0\dagger} (-\left[1-\zeta\right]\phi^3_{n-\hat{e}_0}) \\
  \delta^0_- \psi^2_n &=& U_n^{0\dagger} (-\left[1+\zeta\right]\phi^4_{n-\hat{e}_0}) \\
  \delta^0_- \psi^3_n &=& U_n^{0\dagger} (+\left[1-\zeta\right]\phi^1_{n-\hat{e}_0}) \\
  \delta^0_- \psi^4_n &=& U_n^{0\dagger} (+\left[1+\zeta\right]\phi^2_{n-\hat{e}_0})
  \end{array}\right\} &
  \left\{\begin{array}{rcl} 
  \delta^0_+ \psi^1_n &=& U_n^0\ (+\left[1+\zeta\right]\phi^3_{n+\hat{e}_0}) \\
  \delta^0_+ \psi^2_n &=& U_n^0\ (+\left[1-\zeta\right]\phi^4_{n+\hat{e}_0}) \\
  \delta^0_+ \psi^3_n &=& U_n^0\ (-\left[1+\zeta\right]\phi^1_{n+\hat{e}_0}) \\
  \delta^0_+ \psi^4_n &=& U_n^0\ (-\left[1-\zeta\right]\phi^2_{n+\hat{e}_0})
  \end{array}\right\} 
  \label{eq: KW 3 delta^0 psi} \\
  \left\{\begin{array}{rcl} 
  \delta^3_- \psi^1_n &=& U_n^{3\dagger} (-i\left[1+d\right]\phi^3_{n-\hat{e}_3}) \\
  \delta^3_- \psi^2_n &=& U_n^{3\dagger} (+i\left[1+d\right]\phi^4_{n-\hat{e}_3}) \\
  \delta^3_- \psi^3_n &=& U_n^{3\dagger} (-i\left[1+d\right]\phi^1_{n-\hat{e}_3}) \\
  \delta^3_- \psi^4_n &=& U_n^{3\dagger} (+i\left[1+d\right]\phi^2_{n-\hat{e}_3})
  \end{array}\right\} &
  \left\{\begin{array}{rcl} 
  \delta^3_+ \psi^1_n &=& U_n^{3}\ (+i\left[1+d\right]\phi^3_{n+\hat{e}_3}) \\
  \delta^3_+ \psi^2_n &=& U_n^{3}\ (-i\left[1+d\right]\phi^4_{n+\hat{e}_3}) \\
  \delta^3_+ \psi^3_n &=& U_n^{3}\ (+i\left[1+d\right]\phi^1_{n+\hat{e}_3}) \\
  \delta^3_+ \psi^4_n &=& U_n^{3}\ (-i\left[1+d\right]\phi^2_{n+\hat{e}_3})
  \end{array}\right\} 
  \label{eq: KW 3 delta^3 psi}
\end{align}\normalsize
\begin{align}
  \left\{\begin{array}{rcl} 
  \delta^1_- \psi^1_n &=& U_n^{1\dagger} (+i\left[\phi^4_{n-\hat{e}_1}-i\zeta \phi^3_{n-\hat{e}_1}\right]) \\
  \delta^1_- \psi^2_n &=& U_n^{1\dagger} (+i\left[\phi^3_{n-\hat{e}_1}+i\zeta \phi^4_{n-\hat{e}_1}\right]) \\
  \delta^1_- \psi^3_n &=& U_n^{1\dagger} (+i\left[\phi^2_{n-\hat{e}_1}+i\zeta \phi^1_{n-\hat{e}_1}\right]) \\
  \delta^1_- \psi^4_n &=& U_n^{1\dagger} (+i\left[\phi^1_{n-\hat{e}_1}-i\zeta \phi^2_{n-\hat{e}_1}\right])
  \end{array}\right\} &
  \left\{\begin{array}{rcl} 
  \delta^1_+ \psi^1_n &=& U_n^{1}\ (-i\left[\phi^4_{n-\hat{e}_1}+i\zeta \phi^3_{n-\hat{e}_1}\right]) \\
  \delta^1_+ \psi^2_n &=& U_n^{1}\ (-i\left[\phi^3_{n-\hat{e}_1}-i\zeta \phi^4_{n-\hat{e}_1}\right]) \\
  \delta^1_+ \psi^3_n &=& U_n^{1}\ (-i\left[\phi^2_{n-\hat{e}_1}-i\zeta \phi^1_{n-\hat{e}_1}\right]) \\
  \delta^1_+ \psi^4_n &=& U_n^{1}\ (-i\left[\phi^1_{n-\hat{e}_1}+i\zeta \phi^2_{n-\hat{e}_1}\right])
  \end{array}\right\} 
  \label{eq: KW 3 delta^1 psi} \\
  \left\{\begin{array}{rcl} 
  \delta^2_- \psi^1_n &=& U_n^{2\dagger} (+\left[\phi^4_{n-\hat{e}_2}+\zeta \phi^3_{n-\hat{e}_2}\right]) \\
  \delta^2_- \psi^2_n &=& U_n^{2\dagger} (-\left[\phi^3_{n-\hat{e}_2}+\zeta \phi^4_{n-\hat{e}_2}\right]) \\
  \delta^2_- \psi^3_n &=& U_n^{2\dagger} (+\left[\phi^2_{n-\hat{e}_2}-\zeta \phi^1_{n-\hat{e}_2}\right]) \\
  \delta^2_- \psi^4_n &=& U_n^{2\dagger} (-\left[\phi^1_{n-\hat{e}_2}-\zeta \phi^2_{n-\hat{e}_2}\right])
  \end{array}\right\} &
  \left\{\begin{array}{rcl} 
  \delta^2_+ \psi^1_n &=& U_n^{2}\ (-\left[\phi^4_{n-\hat{e}_2}-\zeta \phi^3_{n-\hat{e}_2}\right]) \\
  \delta^2_+ \psi^2_n &=& U_n^{2}\ (+\left[\phi^3_{n-\hat{e}_2}-\zeta \phi^4_{n-\hat{e}_2}\right]) \\
  \delta^2_+ \psi^3_n &=& U_n^{2}\ (-\left[\phi^2_{n-\hat{e}_2}+\zeta \phi^1_{n-\hat{e}_2}\right]) \\
  \delta^2_+ \psi^4_n &=& U_n^{2}\ (+\left[\phi^1_{n-\hat{e}_2}+\zeta \phi^2_{n-\hat{e}_2}\right])
  \end{array}\right\}, 
  \label{eq: KW 3 delta^2 psi} 
\end{align}\normalsize
and the on-site term $ \delta_m \psi $ of eq. (\ref{eq: numerical Dirac operator application}) contributes
\begin{align}
  \left\{\begin{array}{rcl} 
  \delta_m \psi^1_n &=& (+m_0 \phi^1_{n} +\frac{3\zeta+c}{a}\phi^3_{n}) \\
  \delta_m \psi^2_n &=& (+m_0 \phi^2_{n} -\frac{3\zeta+c}{a}\phi^4_{n}) \\
  \delta_m \psi^3_n &=& (-m_0 \phi^3_{n} -\frac{3\zeta+c}{a}\phi^1_{n}) \\
  \delta_m \psi^4_n &=& (-m_0 \phi^4_{n} +\frac{3\zeta+c}{a}\phi^2_{n})
  \end{array}\right\}.
  \label{eq: KW 3 delta_m psi}
\end{align}\normalsize
Perpendicular hopping terms ($ \delta^0_\pm \psi $, $ \delta^1_\pm \psi $, $ \delta^2_\pm \psi $) for $ \underline{\alpha}=3 $ and $ \zeta=\pm1 $, are related (\mbox{cf.} table \ref{tab: KW n_3 spatial components' linear dependence}). Components which are linearly related are reconstructed with reduced numerical effort.
\begin{table}[hbt]
\center
 \begin{tabular}{|cccc|}
  \hline
  $ \delta^1_- \psi^1_n = \mp i\zeta \delta^1_- \psi^2_n $, & $ \delta^1_- \psi^3_n = \pm i\zeta \delta^1_- \psi^4_n $, & 
  $ \delta^1_+ \psi^1_n = \pm i\zeta \delta^1_+ \psi^2_n $, & $ \delta^1_+ \psi^3_n = \mp i\zeta \delta^1_+ \psi^4_n $, \\
  $ \delta^2_- \psi^1_n = \mp  \zeta \delta^2_- \psi^2_n $, & $ \delta^2_- \psi^3_n = \pm  \zeta \delta^2_- \psi^4_n $, & 
  $ \delta^2_+ \psi^1_n = \pm  \zeta \delta^2_+ \psi^2_n $, & $ \delta^2_+ \psi^3_n = \mp  \zeta \delta^2_+ \psi^4_n $, \\
  \hline
  \hline
  \multicolumn{2}{|r}{$ \zeta=+1 $:} &$ \delta^0_- \psi^1_n = \delta^0_- \psi^3_n =0 $, & $ \delta^0_+ \psi^2_n = \delta^0_+ \psi^4_n =0 $, \\
  \multicolumn{2}{|r}{$ \zeta=-1 $:} &$ \delta^0_- \psi^2_n = \delta^0_- \psi^4_n =0 $, & $ \delta^0_+ \psi^1_n = \delta^0_+ \psi^3_n =0 $. \\
  \hline
 \end{tabular}
 \caption{For $ \underline{\alpha}=3 $ and $ \zeta=\pm1 $, the Dirac components of $ \delta^1_\pm \psi $ and $ \delta^1_\pm \psi $ are linearly dependent and half of the $ \delta^0_\pm \psi $ trivially vanish. Moreover, the communications for spatial boundaries are halved in each perpendicular direction.}
 \label{tab: KW n_3 spatial components' linear dependence}
\end{table}

\noindent
Independent of the particular choice of $ \underline{\alpha} $, the tally of floating point operations, which is dominated by the SU(3) matrix-times-vector operations in the $ \delta^\mu_\pm \psi $ , is reduced to approximately $ 5/8 $ for $ \zeta=\pm1 $, if the symmetries of $ \delta^\mu_\pm \psi $ are incorporated into the code. It is reasonable to provide extra versions of the Karsten-Wilczek Dirac operator for $ \zeta=\pm1 $ due to their superior numerical efficiency.

\subsection{Bori\c{c}i-Creutz Dirac operator}

The Bori\c{c}i-Creutz Dirac operator is taken from \mbox{eq.}~(\ref{eq: BC fermion action}). The hopping terms $ \delta^\mu_\pm \psi $ of \mbox{eq.}~(\ref{eq: numerical Dirac operator application}) add $ 1/2 $ times the following contributions:
\begin{align}
  &\left\{\begin{array}{rcl} 
  \delta^0_- \psi^1_n &=& U_n^{0\dagger} ((-1-i\zeta)\phi^3_{n-\hat{e}_0}
  +(i\zeta+d/2)\left[(-1+i)\phi^3_{n-\hat{e}_0}+(+1+i)\phi^4_{n-\hat{e}_0}\right]) \\
  \delta^0_- \psi^2_n &=& U_n^{0\dagger} ((-1-i\zeta)\phi^4_{n-\hat{e}_0}
  +(i\zeta+d/2)\left[(-1+i)\phi^3_{n-\hat{e}_0}+(-1-i)\phi^4_{n-\hat{e}_0}\right]) \\
  \delta^0_- \psi^3_n &=& U_n^{0\dagger} ((+1+i\zeta)\phi^1_{n-\hat{e}_0}
  +(i\zeta+d/2)\left[(+1+i)\phi^1_{n-\hat{e}_0}+(+1+i)\phi^2_{n-\hat{e}_0}\right]) \\
  \delta^0_- \psi^4_n &=& U_n^{0\dagger} ((+1+i\zeta)\phi^2_{n-\hat{e}_0}
  +(i\zeta+d/2)\left[(-1+i)\phi^1_{n-\hat{e}_0}+(+1-i)\phi^2_{n-\hat{e}_0}\right])
  \end{array}\right\} 
  \label{eq: BC delta^0_- psi} \\
  &\left\{\begin{array}{rcl} 
  \delta^0_+ \psi^1_n &=& U_n^0\ ((+1-i\zeta)\phi^3_{n+\hat{e}_0}
  +(i\zeta-d/2)\left[(+1-i)\phi^3_{n-\hat{e}_0}+(-1-i)\phi^4_{n-\hat{e}_0}\right]) \\
  \delta^0_+ \psi^2_n &=& U_n^0\ ((+1-i\zeta)\phi^4_{n+\hat{e}_0}
  +(i\zeta-d/2)\left[(+1-i)\phi^3_{n-\hat{e}_0}+(+1+i)\phi^4_{n-\hat{e}_0}\right]) \\
  \delta^0_+ \psi^3_n &=& U_n^0\ ((-1+i\zeta)\phi^1_{n+\hat{e}_0}
  +(i\zeta-d/2)\left[(-1-i)\phi^1_{n-\hat{e}_0}+(-1-i)\phi^2_{n-\hat{e}_0}\right]) \\
  \delta^0_+ \psi^4_n &=& U_n^0\ ((-1+i\zeta)\phi^2_{n+\hat{e}_0}
  +(i\zeta-d/2)\left[(+1-i)\phi^1_{n-\hat{e}_0}+(-1+i)\phi^2_{n-\hat{e}_0}\right])
  \end{array}\right\} 
  \label{eq: BC delta^0_+ psi} \\
  &\left\{\begin{array}{rcl} 
  \delta^1_- \psi^1_n &=& U_n^{1\dagger} ((+i-\zeta)\phi^3_{n-\hat{e}_1}
  +(i\zeta+d/2)\left[(-1+i)\phi^3_{n-\hat{e}_1}+(+1+i)\phi^4_{n-\hat{e}_1}\right]) \\
  \delta^1_- \psi^2_n &=& U_n^{1\dagger} ((+i-\zeta)\phi^4_{n-\hat{e}_1}
  +(i\zeta+d/2)\left[(-1+i)\phi^3_{n-\hat{e}_1}+(-1-i)\phi^4_{n-\hat{e}_1}\right]) \\
  \delta^1_- \psi^3_n &=& U_n^{1\dagger} ((+i-\zeta)\phi^1_{n-\hat{e}_1}
  +(i\zeta+d/2)\left[(+1+i)\phi^1_{n-\hat{e}_1}+(+1+i)\phi^2_{n-\hat{e}_1}\right]) \\
  \delta^1_- \psi^4_n &=& U_n^{1\dagger} ((+i-\zeta)\phi^2_{n-\hat{e}_1}
  +(i\zeta+d/2)\left[(-1+i)\phi^1_{n-\hat{e}_1}+(+1-i)\phi^2_{n-\hat{e}_1}\right])
  \end{array}\right\} 
  \label{eq: BC delta^1_- psi} \\
  &\left\{\begin{array}{rcl} 
  \delta^1_+ \psi^1_n &=& U_n^1\ ((-i-\zeta)\phi^3_{n+\hat{e}_1}
  +(i\zeta-d/2)\left[(+1-i)\phi^3_{n-\hat{e}_1}+(-1-i)\phi^4_{n-\hat{e}_1}\right]) \\
  \delta^1_+ \psi^2_n &=& U_n^1\ ((-i-\zeta)\phi^4_{n+\hat{e}_1}
  +(i\zeta-d/2)\left[(+1-i)\phi^3_{n-\hat{e}_1}+(+1+i)\phi^4_{n-\hat{e}_1}\right]) \\
  \delta^1_+ \psi^3_n &=& U_n^1\ ((-i-\zeta)\phi^1_{n+\hat{e}_1}
  +(i\zeta-d/2)\left[(-1-i)\phi^1_{n-\hat{e}_1}+(-1-i)\phi^2_{n-\hat{e}_1}\right]) \\
  \delta^1_+ \psi^4_n &=& U_n^1\ ((-i-\zeta)\phi^2_{n+\hat{e}_1}
  +(i\zeta-d/2)\left[(+1-i)\phi^1_{n-\hat{e}_1}+(-1+i)\phi^2_{n-\hat{e}_1}\right])
  \end{array}\right\} 
  \label{eq: BC delta^1_+ psi} \\
  &\left\{\begin{array}{rcl} 
  \delta^2_- \psi^1_n &=& U_n^{2\dagger} ((+1+i\zeta)\phi^3_{n-\hat{e}_2}
  +(i\zeta+d/2)\left[(-1+i)\phi^3_{n-\hat{e}_2}+(+1+i)\phi^4_{n-\hat{e}_2}\right]) \\
  \delta^2_- \psi^2_n &=& U_n^{2\dagger} ((-1-i\zeta)\phi^4_{n-\hat{e}_2}
  +(i\zeta+d/2)\left[(-1+i)\phi^3_{n-\hat{e}_2}+(-1-i)\phi^4_{n-\hat{e}_2}\right]) \\
  \delta^2_- \psi^3_n &=& U_n^{2\dagger} ((+1+i\zeta)\phi^1_{n-\hat{e}_2}
  +(i\zeta+d/2)\left[(+1+i)\phi^1_{n-\hat{e}_2}+(+1+i)\phi^2_{n-\hat{e}_2}\right]) \\
  \delta^2_- \psi^4_n &=& U_n^{2\dagger} ((-1-i\zeta)\phi^2_{n-\hat{e}_2}
  +(i\zeta+d/2)\left[(-1+i)\phi^1_{n-\hat{e}_2}+(+1-i)\phi^2_{n-\hat{e}_2}\right])
  \end{array}\right\} 
  \label{eq: BC delta^2_- psi} \\
  &\left\{\begin{array}{rcl} 
  \delta^2_+ \psi^1_n &=& U_n^2\ ((-1+i\zeta)\phi^3_{n+\hat{e}_2}
  +(i\zeta-d/2)\left[(+1-i)\phi^3_{n-\hat{e}_2}+(-1-i)\phi^4_{n-\hat{e}_2}\right]) \\
  \delta^2_+ \psi^2_n &=& U_n^2\ ((+1-i\zeta)\phi^4_{n+\hat{e}_2}
  +(i\zeta-d/2)\left[(+1-i)\phi^3_{n-\hat{e}_2}+(+1+i)\phi^4_{n-\hat{e}_2}\right]) \\
  \delta^2_+ \psi^3_n &=& U_n^2\ ((-1+i\zeta)\phi^1_{n+\hat{e}_2}
  +(i\zeta-d/2)\left[(-1-i)\phi^1_{n-\hat{e}_2}+(-1-i)\phi^2_{n-\hat{e}_2}\right]) \\
  \delta^2_+ \psi^4_n &=& U_n^2\ ((+1-i\zeta)\phi^2_{n+\hat{e}_2}
  +(i\zeta-d/2)\left[(+1-i)\phi^1_{n-\hat{e}_2}+(-1+i)\phi^2_{n-\hat{e}_2}\right])
  \end{array}\right\} 
  \label{eq: BC delta^2_+ psi}
\end{align}\normalsize
\begin{align}
  &\left\{\begin{array}{rcl} 
  \delta^3_- \psi^1_n &=& U_n^{3\dagger} ((-i+\zeta)\phi^3_{n-\hat{e}_3}
  +(i\zeta+d/2)\left[(-1+i)\phi^3_{n-\hat{e}_3}+(+1+i)\phi^4_{n-\hat{e}_3}\right]) \\
  \delta^3_- \psi^2_n &=& U_n^{3\dagger} ((+i-\zeta)\phi^4_{n-\hat{e}_3}
  +(i\zeta+d/2)\left[(-1+i)\phi^3_{n-\hat{e}_3}+(-1-i)\phi^4_{n-\hat{e}_3}\right]) \\
  \delta^3_- \psi^3_n &=& U_n^{3\dagger} ((-i+\zeta)\phi^1_{n-\hat{e}_3}
  +(i\zeta+d/2)\left[(+1+i)\phi^1_{n-\hat{e}_3}+(+1+i)\phi^2_{n-\hat{e}_3}\right]) \\
  \delta^3_- \psi^4_n &=& U_n^{3\dagger} ((+i-\zeta)\phi^2_{n-\hat{e}_3}
  +(i\zeta+d/2)\left[(-1+i)\phi^1_{n-\hat{e}_3}+(+1-i)\phi^2_{n-\hat{e}_3}\right])
  \end{array}\right\} 
  \label{eq: BC delta^3_- psi} \\
  &\left\{\begin{array}{rcl} 
  \delta^3_+ \psi^1_n &=& U_n^3\ ((+i+\zeta)\phi^3_{n+\hat{e}_3}
  +(i\zeta-d/2)\left[(+1-i)\phi^3_{n-\hat{e}_3}+(-1-i)\phi^4_{n-\hat{e}_3}\right]) \\
  \delta^3_+ \psi^2_n &=& U_n^3\ ((-i-\zeta)\phi^4_{n+\hat{e}_3}
  +(i\zeta-d/2)\left[(+1-i)\phi^3_{n-\hat{e}_3}+(+1+i)\phi^4_{n-\hat{e}_3}\right]) \\
  \delta^3_+ \psi^3_n &=& U_n^3\ ((+i+\zeta)\phi^1_{n+\hat{e}_3}
  +(i\zeta-d/2)\left[(-1-i)\phi^1_{n-\hat{e}_3}+(-1-i)\phi^2_{n-\hat{e}_3}\right]) \\
  \delta^3_+ \psi^4_n &=& U_n^3\ ((-i-\zeta)\phi^2_{n+\hat{e}_3}
  +(i\zeta-d/2)\left[(+1-i)\phi^1_{n-\hat{e}_3}+(-1+i)\phi^2_{n-\hat{e}_3}\right])
  \end{array}\right\},
  \label{eq: BC delta^3_+ psi}
\end{align}\normalsize
and the on-site term $ \delta_m \psi $ of \mbox{eq.}~(\ref{eq: numerical Dirac operator application}) is
\begin{align}
  \left\{\begin{array}{rcl} 
  \delta_m \psi^1_n &=& (+m_0 \phi^1_{n} 
  +\frac{2\zeta+c}{2}\left[(+1+i)\phi^3_{n}+(+1-i)\phi^4_{n}\right]) \\
  \delta_m \psi^2_n &=& (+m_0 \phi^2_{n} 
  +\frac{2\zeta+c}{2}\left[(+1+i)\phi^3_{n}+(-1+i)\phi^4_{n}\right]) \\
  \delta_m \psi^3_n &=& (-m_0 \phi^3_{n} 
  +\frac{2\zeta+c}{2}\left[(+1-i)\phi^1_{n}+(+1-i)\phi^2_{n}\right]) \\
  \delta_m \psi^4_n &=& (-m_0 \phi^4_{n} 
  +\frac{2\zeta+c}{2}\left[(+1+i)\phi^1_{n}+(-1-i)\phi^2_{n}\right])
  \end{array}\right\}.
  \label{eq: BC delta_m psi}
\end{align}\normalsize
For $ \zeta=+1 $, all coefficients of Bori\c{c}i-Creutz hopping terms $ \delta^\mu_\pm \psi $ in the linear combinations on the right hand sides of \mbox{eqs.}~(\ref{eq: BC delta^0_- psi}), \ldots, (\ref{eq: BC delta^3_+ psi}) are covered by three real constants,
\begin{equation}
  \left.\begin{array}{rl} 
  h_{1+} =& 1+d/2 \\
  h_{1-} =& 1-d/2 \\
  h_{3+} =& 3+d/2
  \end{array}\right.,
  \label{eq: comprehensive coefficients for BC hopping terms}
\end{equation}\normalsize
which can be used to minimise the tally of complex-times-vector operations. This is the most efficient implementation of the Bori\c{c}i-Creutz Dirac operator at present knowledge.

\sectionc{Contractions}{app: Contractions}

\noindent
Contractions of two fermionic propagators into mesonic correlation functions rely on \mbox{eq.}~(\ref{eq: mesonic correlation function for contractions}), which reads
\begin{equation}
  \mathcal{C}_{\mathcal{M},\mathcal{N}}(t)
  =  \sum\limits_{n \in \Lambda_{t}^0} \mathrm{tr}_c{\left( 
  S_{n,0}^{\alpha\beta} (S^*)_{n,0}^{\delta\gamma} ((\mathcal{M} \gamma^5)^T)^{\gamma\beta} (\gamma^5\mathcal{N})^{\delta\alpha}
  \right)}.
  \label{eq: contraction of mesonic correlation function}
\end{equation}\normalsize
Propagator components  (hereafter: propagators) are stored in the memory as spinor fields `$ \psi $' which are defined on the full lattice in a memory alignment of 24 doubles following
\begin{equation}
  \mathrm{Re} \psi_1^1\ \mathrm{Im} \psi_1^1\ \mathrm{Re} \psi_1^2\ \ldots\ \mathrm{Im} \psi_1^3\quad 
  \mathrm{Re} \psi_2^1\ \ldots\ \mathrm{Im} \psi_2^3\quad 
  \mathrm{Re} \psi_3^1\ \ldots\ \mathrm{Im} \psi_3^3\quad 
  \mathrm{Re} \psi_4^1\ \ldots\ \mathrm{Im} \psi_4^3.
\end{equation}\normalsize

\noindent
In the following, the indication of Dirac matrices within the simulation program for interpolating operators at source and sink is outlined in table \ref{tab: indication of Dirac matrices}. The simulation uses Dirac matrices in the chiral representation that are defined in \mbox{eq.}~(\ref{eq: Euclidean Dirac matrices}). Instead of using the original matrices $ \mathcal{M} $ and $ \mathcal{N} $ in interpolating operators, contractions directly use the transposed matrix product $ (\mathcal{M} \gamma^5)^T $ for the source and the matrix product $ (\gamma^5\mathcal{N}) $ for the sink:
\begin{equation}
  \mathcal{M} \to (\mathcal{M} \gamma^5)^T, \qquad \mathcal{N} \to (\gamma^5\mathcal{N}).
  \label{eq: matrix products in interpolating operators}
\end{equation}\normalsize

\begin{table}[hbt]
 \center
 \begin{tabular}{c|cccccc}
  Dirac matrix $ \mathcal{M},\mathcal{N} $ & $ \gamma^\mu $ & $ \mathbf{1} $ & $ \gamma^5 $ & 
  $ \gamma^5\gamma^\mu $ & $ \gamma^0\gamma^i $ & $ \gamma^i\gamma^j $, $ i<j $  \\
  \hline
  Index $ \mu $                & $ \mu $        & $ 4 $          & $ 5 $        & 
  $ 6+\mu $                    & $ 9+i $        & $ 10+i+j $  \\
 \end{tabular}
 \caption{$ 16 $ Dirac matrices that are used in interpolating operators are labeled by only one index $ \mu $. Care has to be taken since only $ \gamma^\mu $, $ \mathbf{1} $ and $ \gamma^5 $ are hermitian.
 }
 \label{tab: indication of Dirac matrices}
\end{table}\normalsize

\noindent
Propagators are calculated one by one in sequence, where the colour component ranges between $ i_c \in[1,2,3] $ in an outer loop and the spin component (or Dirac component) ranges between $ i_d \in [1,2,3,4] $ in an inner loop. Each Dirac component is directly used in the calculation of $ 2 $-point functions with either $ \mathbf{1} $ or $ \gamma^5 $ at the source and added up to the total correlation function. Once the second ($ i_d=2 $) Dirac component (or fourth ($ i_d=4 $) Dirac component) is available, $ 2 $ contributions to correlation functions with source interpolators with tensor structure ($ \gamma^\mu\gamma^\nu $) are calculated. The propagator is contracted with the first ($ j_d=1 $) Dirac component (or third ($ j_d=3 $) Dirac component) for $ \mu = \{10,11,14,15\} $ and it is contracted with itself for $ \mu=\{12,13\} $. Next, contractions of $ j_d $ with $ i_d $ are performed for $ \mu = \{10,11,14,15\} $ and of $ j_d $ with $ j_d $ for $ \mu=\{12,13\} $ added to the total correlation functions. Once the third ($ i_c=3 $) Dirac component is available, $ 2 $ contributions to correlation functions with source interpolators with vector ($ \gamma^\mu $) or axial vector ($ \gamma^5\gamma^\mu $) structure are calculated. The propagator is contracted with the first ($ j_d^0=1 $) Dirac component for $ \mu = \{0,3,6,9\} $ and it is contracted with the second ($ j_d^1=2 $) for $ \mu=\{1,2,7,8\} $. Next, contractions of $ j_d^{0,1} $ with $ i_d $ are performed and added to the total correlation functions. Once the fourth ($ i_d=3 $) Dirac component is available, $ 2 $ contributions to correlation functions with source interpolators with vector ($ \gamma^\mu $) or axial vector ($ \gamma^5\gamma^\mu $) structure are calculated. The propagator is contracted with the first ($ j_d^1=1 $) Dirac component for $ \mu = \{1,2,7,8\} $ and it is contracted with the second ($ j_d^0=2 $) for $ \mu=\{0,3,6,9\} $ and added to the total correlation function. Next, contractions of $ j_d^{0,1} $ with $ i_d $ are performed and added to the total correlation functions. Finally, the last $ 2 $ contributions to correlation functions with source interpolators with tensor structure ($ \gamma^\mu\gamma^\nu $) are calculated as described above. \newline

\noindent
The procedure is repeated for all colour components. Propagators are temporarily dumped to the hard disk for later use as $ j_d $~component in contractions. For the contractions, these dumped propagators are restored and kept in work space that is reserved for the CG solver in the rest of the program. After processing the fourth $ i_d=4 $ components, the dumped propagators are deleted. \newline

\noindent
The subroutine 
\begin{tabbing}
\textbf{void}\ add\_to\_M\!M\_2pt\ \= (arguments) \kill 
\textbf{void}\ \textit{add\_to\_MM\_2pt}\ \> (\textbf{int}\  \textit{p1},\ \textbf{int}\  \textit{p2},\ \textbf{int}\  \textit{l},\ \textbf{int}\  \textit{mu},\ \textbf{int}\  \textit{nu},\ \textbf{int}\  \textit{dir}, \\ 
  \> \ \textbf{int}\  \textit{*vect\_ps},\ \textbf{complex\_dble}\  \textit{*corr}),
\end{tabbing}
which performs the contractions, receives the locations \textit{p1}, \textit{p2} in memory for $ 2 $ possibly different propagators that are computed by the inverter for different Dirac operators or for different smearing options at source or sink. The index \mbox{$ \mathit{l}=3(i_d-1)+(i_c-1) $} denotes which propagator is treated and is factored into a macro that determines whether the contribution is added or subtracted from the correlator. The indices \textit{mu} and \textit{nu} label the Dirac matrices of interpolating operators at sink (\textit{mu}) and source (\textit{nu}). The index \textit{nu} determines the Dirac components that are accessed in the contractions of $ j_d^0 $, $ j_d^{1} $ or $ j_d $ with $ i_d $. Either index $ j_d^0 $, $ j_d^{1} $ or $ j_d $ is calculated only once for every $ i_d $ and replaces \textit{l} within the subroutine. The index \textit{dir} incidates the Euclidean time direction and may be either set to $ 0 $ or $ 3 $. The contractions are performed for all lattice sites individually and all sites in a slice that is perpendicular to this time direction are summed up in the aftermath. The pointer \textit{*vect\_ps} indicates a four-component array, which holds integers that determine an external momentum insertion at the sink. The momentum is given by $ p_\mu = 2\pi/a \cdot $(\textit{vect\_ps}$ [\mu $]/N$ \mu $) . Only three components have a physical role as spatial hadron momenta. However, due to the flexibility in terms of defining the time direction, external momenta are handled most naturally as a four-component field. Finally, the pointer \textit{corr} indicates an array, which stores the value of the correlation function. \newline
\begin{table}[hbt]
 \center
 \begin{tabular}{c|ccc}
  \textit{mu}\       & \textit{amu} & \textit{tmu} & \textit{imu} \\
  \hline
  $ 0\leq mu < 5  $  & $ 0 $ & $ 0 $ & $ mu $    \\
  $ mu = 5 $         & $ 1 $ & $ 0 $ & $ 4 $     \\
  $ 6\leq mu < 10 $  & $ 1 $ & $ 0 $ & $ mu-6 $  \\
  $ 10\leq mu < 15 $ & $ 0 $ & $ 1 $ & $ mu-10 $ \\
 \end{tabular}
 \caption{Flags for contractions are used for determination of signs with simple integer operations. Flags \textit{anu}, \textit{tnu} and \textit{inu} are defined accordingly.
 }
 \label{tab: flags for contractions}
\end{table}

\noindent
In each call of the subroutine, eight integers (\textit{amu}, \textit{tmu}, \textit{imu}, \textit{xmu}, \textit{anu}, \textit{tnu}, \textit{inu}, \textit{xnu}) are defined according to table \ref{tab: flags for contractions}. \textit{amu} is set to one, if the sink Dirac structure contains $ \gamma^5 $ and to zero otherwise. \textit{tmu} is set to one if the sink Dirac structure has two Euclidean indices and to zero otherwise. \textit{xmu} is defined as one if the sink Dirac structure is imaginary and zero otherwise,
\begin{equation}
  xmu = (((imu\&1)\&\&(!tmu))||(((imu)\,\widehat{\ }\,1)\&\&((imu)\,\widehat{\ }\,4)\&\&(tmu))).
  \label{eq: xmu}
\end{equation}\normalsize
\textit{anu}, \textit{tnu}, \textit{xnu} are defined accordingly. \textit{imu} is an index running within 0 and 5, which is used within the macros \textbf{k(imu,\,tmu,\,j)}, \textbf{f(imu,\,amu,\,tmu,\,j)}, \textbf{g(inu,\,anu,\,tnu,\,l)}. 
The 24 real components of the propagators are looped with an index $ j $ using steps of two. The spinor component of the second propagator is determined by the macro \textbf{k(imu,\,tmu,\,j)}, which is not affected by the replacement rule of \mbox{eq.}~(\ref{eq: matrix products in interpolating operators}). Variation of the sign due to different components of the Dirac matrices at source and sink are determined by the macros \textbf{g(inu,\,anu,\,tnu,\,l)} and \textbf{f(imu,\,amu,\,tmu,\,j)}, which take the matrix products of \mbox{eq.}~(\ref{eq: matrix products in interpolating operators}) into account. Then the full contraction for one sink site is given by
\begin{align}
  A = \sum\limits_{j^\prime=0}^{j^\prime<12} \delta_{j,2j^\prime}\ \mathbf{f}\cdot \mathbf{g}\cdot &\ ((\mathbf{double}*)(\psi_0^0[p1]+j  )*(\mathbf{double}*)(\psi_0^0[p2]+\mathbf{k}) 
    \nonumber \\
                                   &\ +(\mathbf{double}*)(\psi_0^0[p1]+j+1)*(\mathbf{double}*)(\psi_0^0[p2]+\mathbf{k}+1)) 
    \label{eq: contraction A} \\
  B = \sum\limits_{j^\prime=0}^{j^\prime<12} \delta_{j,2j^\prime}\ \mathbf{f}\cdot \mathbf{g}\cdot &\ ((\mathbf{double}*)(\psi_0^0[p1]+j  )*(\mathbf{double}*)(\psi_0^0[p2]+\mathbf{k}+1) 
    \nonumber \\
                                   &\ -(\mathbf{double}*)(\psi_0^0[p1]+j+1)*(\mathbf{double}*)(\psi_0^0[p2]+\mathbf{k})), 
    \label{eq: contraction B}
\end{align}
where $ k $, $ f $ and $ g $ are determined with the macros for each $ j $. After the summation, \textbf{Re(xmu,\,xnu,\,A,\,B)} is added to the real part and \textbf{Im(xmu,\,xnu,\,A,\,B)} is added to the imaginary part of the correlator for this sink site. The calculation is repeated for all sink sites. The role of the index $ j_d $ for the contraction at the source is mirrored by the index k for the contraction at the sink.

\begin{figure}[hbt]
 \flushleft
 \rule{\linewidth}{0.25mm}\\
 \normalsize\textit{Macro} k=\textbf{k(imu,\,tmu,\,j)}:
 \footnotesize
 \begin{tabbing}
   xx \= xxxx \= x \= xxxxxxx  \= x \= xxxxxxx \= x \= \hspace{1cm} \= \kill
   k= \> (tmu \> ? \> (imu\&2) \> ? \> (j) \\
      \>      \>   \>          \> : \> (1-((((j)/6)\&1)<< 1))*6+(j) \\
      \>      \> : \> (imu\&4) \> ? \> (j) \\
      \>      \>   \>          \> : \>(imu\%3) \> ? \> ((j)\%6+6*(3-((j)/6))) \\
      \>      \>   \>          \>   \>         \> : \> (((j)+12)\%24))
 \end{tabbing}
 \normalsize\textit{Macro} f=\textbf{f(imu,\,amu,\,tmu,\,j)}:
 \footnotesize
 \begin{tabbing}
   xx \= xxxx \= x \= xxxxxxxxxx                   \= x \= xxxxxxxxxxxxx             \= x \=  xxxxxxxxxxxxxxxx       \= x \= xxxxxxxxx    \= x \= xxxxxxxxx  \kill
   f= \> (tmu \> ? \> (imu\&2)                    \> ? \> (((j)/6)\%imu)             \> ? \> (imu\&1)                \> ? \> ($ -1 $) : ($ +1 $) \\
      \>      \>   \>                             \>   \>                            \> : \> (imu\&1)                \> ? \> ($ +1 $) : ($ -1 $) \\
      \>      \>   \>                             \> : \> (imu\&4)                   \> ? \> (imu\&1)                \> ? \> (((j)/6)>>1) \> ? \> ($ -1 $) : ($ +1 $) \\
      \>      \>   \>                             \>   \>                            \>   \>                         \> : \> (((j)/6)\%3) \> ? \> ($ +1 $) : ($ -1 $) \\
      \>      \>   \>                             \>   \>                            \> : \> (imu\&1)                \> ? \> (((j)/6)\%2) \> ? \> ($ -1 $) : ($ +1 $) \\
      \>      \>   \>                             \>   \>                            \>   \>                         \> : \> ($ -1 $) \\
      \>      \> : \> ((imu\&(\textasciitilde 1)) \> ? \> (imu\&(\textasciitilde 3)) \> ? \> (amu)                   \> ? \> ($ +1 $) \\
      \>      \>   \>                             \>   \>                            \>   \>                         \> : \> (((j)/6)>>1) \> ? \> ($ -1 $) : ($ +1 $) \\
      \>      \>   \>                             \>   \>                            \> : \> (((j)/6)\%(2\^{}amu))   \> ? \> ($ -1 $) : ($ +1 $) \\
      \>      \>   \>                             \> : \> ((imu\&1)\^{}amu)          \> ? \> (amu)                   \> ? \> ($ -1 $) : ($ +1 $) \\
      \>      \>   \>                             \>   \>                            \> : \> ((((j)/6)>>1)\^{}(amu)) \> ? \> ($ +1 $) : ($ -1 $)))))
 \end{tabbing}
 \normalsize\textit{Macro} g=\textbf{g(inu,\,anu,\,tnu,\,l)}:
 \footnotesize
 \begin{tabbing}
   xx \= xxxx \= x \= xxxxxxxxxx                   \= x \= xxxxxxxxxxxxx             \= x \=  xxxxxxxxxxxxxxxx         \= x \= xxxxxxxxxxxx      \= x \= xxxxxxxxx  \kill
   g= \> (tnu \> ? \> (inu\&2)                    \> ? \> (((l)/3)\%inu)             \> ? \> (inu\&1)                  \> ? \> ($ -1 $) : ($ +1 $) \\
      \>      \>   \>                             \>   \>                            \> : \> (inu\&1)                  \> ? \> ($ +1 $) : ($ -1 $) \\
      \>      \>   \>                             \> : \> (inu\&4)                   \> ? \> (inu\&1)                  \> ? \> (((l)/3)>>1)      \> ? \> ($ -1 $) : ($ +1 $) \\
      \>      \>   \>                             \>   \>                            \>   \>                           \> : \> (((l)/3)\%3)      \> ? \> ($ -1 $) : ($ +1 $) \\
      \>      \>   \>                             \>   \>                            \> : \> (inu\&1)                  \> ? \> (((l)/3)\%2)      \> ? \> ($ +1 $) : ($ -1 $) \\
      \>      \>   \>                             \>   \>                            \>   \>                           \> : \> ($ -1 $) \\
      \>      \> : \> ((inu\&(\textasciitilde 1)) \> ? \> (inu\&(\textasciitilde 3)) \> ? \> (anu)                     \> ? \> ($ +1 $) \\
      \>      \>   \>                             \>   \>                            \>   \>                           \> : \> (((l)/3)>>1) \> ? \> ($ -1 $) : ($ +1 $) \\
      \>      \>   \>                             \>   \>                            \> : \> (((l)/3)\%(2\^{}anu))     \> ? \> ((inu\&1)\^{}anu) \> ? \>  ($ +1 $) : ($ -1 $) \\
      \>      \>   \>                             \> : \> ((inu\&1)\^{}anu)          \>   \>                           \> : \> ((inu\&1)\^{}anu) \> ? \>  ($ -1 $) : ($ +1 $) \\
      \>      \>   \>                             \> : \> ((inu\&1)\^{}anu)          \> ? \> (anu)                     \> ? \> ($ +1 $) : ($ -1 $) \\
      \>      \>   \>                             \>   \>                            \> : \> ((((l)/3)>>1)\^{}(anu)) \> ? \> ($ +1 $) : ($ -1 $)))))
 \end{tabbing}
 \normalsize\textit{Macros} Re=\textbf{Re(xmu,\,xnu,\,A,\,B)}, Im=\textbf{Im(xmu,\,xnu,\,A,\,B)}:
 \footnotesize
 \begin{tabbing}
   xxxx \= xxxxxxxxxxxxx     \= x \= xxxxxxxxxxxxxx       \= x \= xxxxxxx \kill
   Re=  \> (((xmu)\^{}(xnu)) \> ? \> (-B) \\
        \>            \> : \>  ((xmu)\&\&(xnu)) \> ? \> (-A) : (+A)) \\
   Im=  \> (((xmu)\^{}(xnu)) \> ? \> (A) \\
        \>            \> : \>  ((xmu)\&\&(xnu)) \> ? \> (-B) : (+B)) \\
 \end{tabbing}
 \rule{\linewidth}{0.25mm}\\
 \caption{Macros for contractions spread out factors $ \pm1 $, factors $ i $ and selection of components due to Dirac structure of interpolators to different parts of the program.}
 \label{fig: macros}
 \vspace{-8pt}
\end{figure}

\addcontentsline{toc}{chapter}{Bibliography}  
\bibliography{library}{}

\begin{thebibliography}{100}

\bibitem{Aad:2012tfa}
Georges Aad et~al.
\newblock {Observation of a new particle in the search for the Standard Model
  Higgs boson with the ATLAS detector at the LHC}.
\newblock {\em Phys.Lett.}, B716:1--29, 2012.

\bibitem{Adler:1981sn}
Stephen~L. Adler.
\newblock {An Overrelaxation Method for the Monte Carlo Evaluation of the
  Partition Function for Multiquadratic Actions}.
\newblock {\em Phys.Rev.}, D23:2901, 1981.

\bibitem{Albanese:1987ds}
M.~Albanese et~al.
\newblock {Glueball Masses and String Tension in Lattice QCD}.
\newblock {\em Phys.Lett.}, B192:163--169, 1987.

\bibitem{Allton:1993wc}
C.R. Allton et~al.
\newblock {Gauge invariant smearing and matrix correlators using Wilson
  fermions at Beta = 6.2}.
\newblock {\em Phys.Rev.}, D47:5128--5137, 1993.

\bibitem{Altmeyer:1992dd}
R.~Altmeyer et~al.
\newblock {The Hadron spectrum in QCD with dynamical staggered fermions}.
\newblock {\em Nucl.Phys.}, B389:445--512, 1993.

\bibitem{1933PhRv...43..491A}
C.~D. {Anderson}.
\newblock {The Positive Electron}.
\newblock {\em Physical Review}, 43:491--494, March 1933.

\bibitem{PhysRev.50.263}
Carl~D. Anderson and Seth~H. Neddermeyer.
\newblock Cloud chamber observations of cosmic rays at 4300 meters elevation
  and near sea-level.
\newblock {\em Phys. Rev.}, 50:263--271, Aug 1936.

\bibitem{PhysRev.91.155}
H.~L. Anderson, E.~Fermi, R.~Martin, and D.~E. Nagle.
\newblock Angular distribution of pions scattered by hydrogen.
\newblock {\em Phys. Rev.}, 91:155--168, Jul 1953.

\bibitem{PhysRev.86.793.2}
H.~L. Anderson, E.~Fermi, D.~E. Nagle, and G.~B. Yodh.
\newblock Angular distribution of pions scattered by hydrogen.
\newblock {\em Phys. Rev.}, 86:793--794, Jun 1952.

\bibitem{Anderson:1952nw}
H.L. Anderson, E.~Fermi, E.A. Long, and D.E. Nagle.
\newblock {Total Cross-sections of Positive Pions in Hydrogen}.
\newblock {\em Phys.Rev.}, 85:936, 1952.

\bibitem{1963PhRv..130..439A}
P.~W. {Anderson}.
\newblock {Plasmons, Gauge Invariance, and Mass}.
\newblock {\em Physical Review}, 130:439--442, April 1963.

\bibitem{Aoki:2005}
Sinya Aoki.
\newblock {\em \japanesetext{格子上の場の理論}}.
\newblock Springer-Verlag Tokyo, 2005.
\newblock (KOUSHIJOUNOBANORIRON, literally: ``Field theory on a lattice'').

\bibitem{Aoki:2013ldr}
Sinya Aoki, Yasumichi Aoki, Claude Bernard, Tom Blum, Gilberto Colangelo,
  et~al.
\newblock {Review of lattice results concerning low energy particle physics}.
\newblock 2013.

\bibitem{Aubert:2005rm}
Bernard Aubert et~al.
\newblock {Observation of a broad structure in the $\pi^+ \pi^- \to J/\psi$
  mass spectrum around 4.26-GeV/c$^2$}.
\newblock {\em Phys.Rev.Lett.}, 95:142001, 2005.

\bibitem{Baaquie:1977hz}
Belal~E. Baaquie.
\newblock {Gauge Fixing and Mass Renormalization in the Lattice Gauge Theory}.
\newblock {\em Phys.Rev.}, D16:2612, 1977.

\bibitem{Bali}
G.~Bali.
\newblock private communication.

\bibitem{PhysRevLett.12.204}
V.~E. Barnes, P.~L. Connolly, D.~J. Crennell, B.~B. Culwick, W.~C. Delaney,
  W.~B. Fowler, P.~E. Hagerty, E.~L. Hart, N.~Horwitz, P.~V.~C. Hough, J.~E.
  Jensen, J.~K. Kopp, K.~W. Lai, J.~Leitner, J.~L. Lloyd, G.~W. London, T.~W.
  Morris, Y.~Oren, R.~B. Palmer, A.~G. Prodell, D.~Radoji\ifmmode
  \check{c}\else \v{c}\fi{}i\ifmmode~\acute{c}\else \'{c}\fi{}, D.~C. Rahm,
  C.~R. Richardson, N.~P. Samios, J.~R. Sanford, R.~P. Shutt, J.~R. Smith,
  D.~L. Stonehill, R.~C. Strand, A.~M. Thorndike, M.~S. Webster, W.~J. Willis,
  and S.~S. Yamamoto.
\newblock Observation of a hyperon with strangeness minus three.
\newblock {\em Phys. Rev. Lett.}, 12:204--206, Feb 1964.

\bibitem{Bedaque:2008xs}
Paulo~F. Bedaque, Michael~I. Buchoff, Brian~C. Tiburzi, and Andre Walker-Loud.
\newblock {Broken Symmetries from Minimally Doubled Fermions}.
\newblock {\em Phys.Lett.}, B662:449--455, 2008.

\bibitem{Bedaque:2008jm}
Paulo~F. Bedaque, Michael~I. Buchoff, Brian~C. Tiburzi, and Andre Walker-Loud.
\newblock {Search for Fermion Actions on Hyperdiamond Lattices}.
\newblock {\em Phys.Rev.}, D78:017502, 2008.

\bibitem{PDG:2012}
J.~Beringer and others (Particle Data~Group).
\newblock The review of particle physics.
\newblock {\em Phys. Rev.}, D86:010001, 2012.

\bibitem{Bernard:1992mk}
Claude~W. Bernard and Maarten~F.L. Golterman.
\newblock {Chiral perturbation theory for the quenched approximation of QCD}.
\newblock {\em Phys.Rev.}, D46:853--857, 1992.

\bibitem{PhysRev.72.339}
H.~A. Bethe.
\newblock The electromagnetic shift of energy levels.
\newblock {\em Phys. Rev.}, 72:339--341, Aug 1947.

\bibitem{Bjorken:1968dy}
J.D. Bjorken.
\newblock {Asymptotic Sum Rules at Infinite Momentum}.
\newblock {\em Phys.Rev.}, 179:1547--1553, 1969.

\bibitem{Blum:1996uf}
Tom Blum, Carleton~E. Detar, Steven~A. Gottlieb, Kari Rummukainen, Urs~M.
  Heller, et~al.
\newblock {Improving flavor symmetry in the Kogut-Susskind hadron spectrum}.
\newblock {\em Phys.Rev.}, D55:1133--1137, 1997.

\bibitem{Bonati:2014tqa}
Claudio Bonati and Massimo D'Elia.
\newblock {Comparison of the gradient flow with cooling in $SU(3)$ pure gauge
  theory}.
\newblock {\em Phys.Rev.}, D89:105005, 2014.

\bibitem{Borici:2007kz}
Artan Borici.
\newblock {Creutz fermions on an orthogonal lattice}.
\newblock {\em Phys.Rev.}, D78:074504, 2008.

\bibitem{Borici:2008ym}
Artan Borici.
\newblock {Minimally Doubled Fermion Revival}.
\newblock {\em PoS}, LATTICE2008:231, 2008.

\bibitem{PhysRev.86.106}
Keith~A. Brueckner.
\newblock Meson-nucleon scattering and nucleon isobars.
\newblock {\em Phys. Rev.}, 86:106--109, Apr 1952.

\bibitem{Cabibbo:1982zn}
N.~Cabibbo and E.~Marinari.
\newblock {A New Method for Updating SU(N) Matrices in Computer Simulations of
  Gauge Theories}.
\newblock {\em Phys.Lett.}, B119:387--390, 1982.

\bibitem{Cabibbo:1963yz}
Nicola Cabibbo.
\newblock {Unitary Symmetry and Leptonic Decays}.
\newblock {\em Phys.Rev.Lett.}, 10:531--533, 1963.

\bibitem{Capitani:2002mp}
Stefano Capitani.
\newblock {Lattice perturbation theory}.
\newblock {\em Phys.Rept.}, 382:113--302, 2003.

\bibitem{Capitani:2013fda}
Stefano Capitani.
\newblock {New actions for minimally doubled fermions and their counterterms}.
\newblock 2013.

\bibitem{Capitani:2013iha}
Stefano Capitani.
\newblock {New chiral lattice actions of the Borici-Creutz type}.
\newblock 2013.

\bibitem{Capitani:2013zta}
Stefano Capitani.
\newblock {Reducing the number of counterterms with new minimally doubled
  actions}.
\newblock {\em Phys.Rev.}, D89:014501, 2014.

\bibitem{Capitani:2010ht}
Stefano Capitani, Michael Creutz, Johannes Weber, and Hartmut Wittig.
\newblock {Minimally doubled fermions and their renormalization}.
\newblock {\em PoS}, LATTICE2010:093, 2010.

\bibitem{Capitani:2010nn}
Stefano Capitani, Michael Creutz, Johannes Weber, and Hartmut Wittig.
\newblock {Renormalization of minimally doubled fermions}.
\newblock {\em JHEP}, 1009:027, 2010.

\bibitem{Capitani:2009yn}
Stefano Capitani, Johannes Weber, and Hartmut Wittig.
\newblock {Minimally doubled fermions at one loop}.
\newblock {\em Phys.Lett.}, B681:105--112, 2009.

\bibitem{Capitani:2009ty}
Stefano Capitani, Johannes Weber, and Hartmut Wittig.
\newblock {Minimally doubled fermions at one-loop level}.
\newblock {\em PoS}, LAT2009:075, 2009.

\bibitem{Caracciolo:1991cp}
Sergio Caracciolo, Pietro Menotti, and Andrea Pelissetto.
\newblock {One loop analytic computation of the energy momentum tensor for
  lattice gauge theories}.
\newblock {\em Nucl.Phys.}, B375:195--242, 1992.

\bibitem{Chadwick:1932ma}
J.~Chadwick.
\newblock {Possible Existence of a Neutron}.
\newblock {\em Nature}, 129:312, 1932.

\bibitem{Choi:2003ue}
S.K. Choi et~al.
\newblock {Observation of a narrow charmonium - like state in exclusive $ B^\pm
  \to K^\pm \pi^+ \pi^- J/\psi $ decays}.
\newblock {\em Phys.Rev.Lett.}, 91:262001, 2003.

\bibitem{Cichy:2008gk}
K.~Cichy, J.~Gonzalez~Lopez, K.~Jansen, A.~Kujawa, and A.~Shindler.
\newblock {Twisted Mass, Overlap and Creutz Fermions: Cut-off Effects at
  Tree-level of Perturbation Theory}.
\newblock {\em Nucl.Phys.}, B800:94--108, 2008.

\bibitem{Creutz:1980zw}
M.~Creutz.
\newblock {Monte Carlo Study of Quantized SU(2) Gauge Theory}.
\newblock {\em Phys.Rev.}, D21:2308--2315, 1980.

\bibitem{Creutz:2008nk}
Michael Creutz.
\newblock {Comments on staggered fermions: Panel discussion}.
\newblock {\em PoS}, CONFINEMENT8:016, 2008.

\bibitem{Creutz:2007af}
Michael Creutz.
\newblock {Four-dimensional graphene and chiral fermions}.
\newblock {\em JHEP}, 0804:017, 2008.

\bibitem{Creutz:2008sr}
Michael Creutz.
\newblock {Local chiral fermions}.
\newblock {\em PoS}, LATTICE2008:080, 2008.

\bibitem{Creutz:2010qm}
Michael Creutz.
\newblock {Minimal doubling and point splitting}.
\newblock {\em PoS}, LATTICE2010:078, 2010.

\bibitem{Creutz:2010bm}
Michael Creutz, Taro Kimura, and Tatsuhiro Misumi.
\newblock {Index Theorem and Overlap Formalism with Naive and Minimally Doubled
  Fermions}.
\newblock {\em JHEP}, 1012:041, 2010.

\bibitem{DeGrand:1990dk}
Thomas~A. DeGrand and Pietro Rossi.
\newblock {Conditioning Techniques for Dynamical Fermions}.
\newblock {\em Comput.Phys.Commun.}, 60:211--214, 1990.

\bibitem{1931RSPSA.133...60D}
P.~A.~M. {Dirac}.
\newblock {Quantised Singularities in the Electromagnetic Field}.
\newblock {\em Royal Society of London Proceedings Series A}, 133:60--72,
  September 1931.

\bibitem{PhysRev.88.1053}
S.~D. Drell and E.~M. Henley.
\newblock Pseudoscalar mesons with applications to meson-nucleon scattering and
  photoproduction.
\newblock {\em Phys. Rev.}, 88:1053--1064, Dec 1952.

\bibitem{Durr:2008zz}
S.~Durr, Z.~Fodor, J.~Frison, C.~Hoelbling, R.~Hoffmann, et~al.
\newblock {Ab-Initio Determination of Light Hadron Masses}.
\newblock {\em Science}, 322:1224--1227, 2008.

\bibitem{PhysRev.75.486}
F.~J. Dyson.
\newblock The radiation theories of tomonaga, schwinger, and feynman.
\newblock {\em Phys. Rev.}, 75:486--502, Feb 1949.

\bibitem{PhysRev.75.1736}
F.~J. Dyson.
\newblock The $s$ matrix in quantum electrodynamics.
\newblock {\em Phys. Rev.}, 75:1736--1755, Jun 1949.

\bibitem{1964PhRvL..13..321E}
F.~{Englert} and R.~{Brout}.
\newblock {Broken Symmetry and the Mass of Gauge Vector Mesons}.
\newblock {\em Physical Review Letters}, 13:321--323, August 1964.

\bibitem{Faddeev:1967fc}
L.D. Faddeev and V.N. Popov.
\newblock {Feynman Diagrams for the Yang-Mills Field}.
\newblock {\em Phys.Lett.}, B25:29--30, 1967.

\bibitem{RevModPhys.20.367}
R.~P. Feynman.
\newblock Space-time approach to non-relativistic quantum mechanics.
\newblock {\em Rev. Mod. Phys.}, 20:367--387, Apr 1948.

\bibitem{PhysRev.76.769}
R.~P. Feynman.
\newblock Space-time approach to quantum electrodynamics.
\newblock {\em Phys. Rev.}, 76:769--789, Sep 1949.

\bibitem{PhysRev.76.749}
R.~P. Feynman.
\newblock The theory of positrons.
\newblock {\em Phys. Rev.}, 76:749--759, Sep 1949.

\bibitem{PhysRev.80.440}
R.~P. Feynman.
\newblock Mathematical formulation of the quantum theory of electromagnetic
  interaction.
\newblock {\em Phys. Rev.}, 80:440--457, Nov 1950.

\bibitem{Fritzsch:1973pi}
H.~Fritzsch, Murray Gell-Mann, and H.~Leutwyler.
\newblock {Advantages of the Color Octet Gluon Picture}.
\newblock {\em Phys.Lett.}, B47:365--368, 1973.

\bibitem{GSL:2009}
M.~et~al. Galassi.
\newblock {GNU Scientific Library Reference Manual}.
\newblock 2009.

\bibitem{Gasser:1983yg}
J.~Gasser and H.~Leutwyler.
\newblock {Chiral Perturbation Theory to One Loop}.
\newblock {\em Annals Phys.}, 158:142, 1984.

\bibitem{Gasser:1984gg}
J.~Gasser and H.~Leutwyler.
\newblock {Chiral Perturbation Theory: Expansions in the Mass of the Strange
  Quark}.
\newblock {\em Nucl.Phys.}, B250:465, 1985.

\bibitem{Gattringer:2010zz}
Christof Gattringer and Christian~B. Lang.
\newblock {Quantum chromodynamics on the lattice}.
\newblock {\em Lect.Notes Phys.}, 788:1--343, 2010.

\bibitem{GellMann:1961ky}
Murray Gell-Mann.
\newblock {The Eightfold Way: A Theory of strong interaction symmetry}.
\newblock 1961.

\bibitem{PhysRev.125.1067}
Murray Gell-Mann.
\newblock Symmetries of baryons and mesons.
\newblock {\em Phys. Rev.}, 125:1067--1084, Feb 1962.

\bibitem{GellMann:1964nj}
Murray Gell-Mann.
\newblock {A Schematic Model of Baryons and Mesons}.
\newblock {\em Phys.Lett.}, 8:214--215, 1964.

\bibitem{GellMann:1964tf}
Murray Gell-Mann.
\newblock {The Symmetry group of vector and axial vector currents}.
\newblock {\em Physics}, 1:63--75, 1964.

\bibitem{GellMann:1964xy}
Murray Gell-Mann and Yuval Neemam.
\newblock {The Eightfold way: a review with a collection of reprints}.
\newblock 1964.

\bibitem{Geradin1971319}
M.~Geradin.
\newblock The computational efficiency of a new minimization algorithm for
  eigenvalue analysis.
\newblock {\em Journal of Sound and Vibration}, 19(3):319 -- 331, 1971.

\bibitem{Ginsparg:1981bj}
Paul~H. Ginsparg and Kenneth~G. Wilson.
\newblock {A Remnant of Chiral Symmetry on the Lattice}.
\newblock {\em Phys.Rev.}, D25:2649, 1982.

\bibitem{Giusti:2006mh}
Leonardo Giusti, P.~Hernandez, M.~Laine, C.~Pena, J.~Wennekers, et~al.
\newblock {On K $ \to $ pi pi amplitudes with a light charm quark}.
\newblock {\em Phys.Rev.Lett.}, 98:082003, 2007.

\bibitem{Giusti:2004yp}
Leonardo Giusti, P.~Hernandez, M.~Laine, P.~Weisz, and H.~Wittig.
\newblock {Low-energy couplings of QCD from current correlators near the chiral
  limit}.
\newblock {\em JHEP}, 0404:013, 2004.

\bibitem{Giusti:2002sm}
Leonardo Giusti, C.~Hoelbling, M.~Luscher, and H.~Wittig.
\newblock {Numerical techniques for lattice QCD in the epsilon regime}.
\newblock {\em Comput.Phys.Commun.}, 153:31--51, 2003.

\bibitem{Glashow:1961tr}
S.L. Glashow.
\newblock {Partial Symmetries of Weak Interactions}.
\newblock {\em Nucl.Phys.}, 22:579--588, 1961.

\bibitem{Goldberger:1958tr}
M.L. Goldberger and S.B. Treiman.
\newblock {Decay of the pi meson}.
\newblock {\em Phys.Rev.}, 110:1178--1184, 1958.

\bibitem{Goldberger:1958vp}
M.L. Goldberger and S.B. Treiman.
\newblock {Form-factors in Beta decay and muon capture}.
\newblock {\em Phys.Rev.}, 111:354--361, 1958.

\bibitem{PhysRev.127.965}
Jeffrey Goldstone, Abdus Salam, and Steven Weinberg.
\newblock Broken symmetries.
\newblock {\em Phys. Rev.}, 127:965--970, Aug 1962.

\bibitem{Golterman:1984cy}
Maarten~F.L. Golterman and Jan Smit.
\newblock {Self-energy and Flavor Interpretation of Staggered Fermions}.
\newblock {\em Nucl.Phys.}, B245:61, 1984.

\bibitem{Gross:1973id}
David~J. Gross and Frank Wilczek.
\newblock {Ultraviolet Behavior of Nonabelian Gauge Theories}.
\newblock {\em Phys.Rev.Lett.}, 30:1343--1346, 1973.

\bibitem{Gross:1973ju}
D.J. Gross and Frank Wilczek.
\newblock {Asymptotically Free Gauge Theories. 1}.
\newblock {\em Phys.Rev.}, D8:3633--3652, 1973.

\bibitem{Gross:1974cs}
D.J. Gross and Frank Wilczek.
\newblock {Asymptotically free gauge theories. 2.}
\newblock {\em Phys.Rev.}, D9:980--993, 1974.

\bibitem{Guagnelli:1998ud}
Marco Guagnelli, Rainer Sommer, and Hartmut Wittig.
\newblock {Precision computation of a low-energy reference scale in quenched
  lattice QCD}.
\newblock {\em Nucl.Phys.}, B535:389--402, 1998.

\bibitem{Gupta:1996sa}
Rajan Gupta and Tanmoy Bhattacharya.
\newblock {Light quark masses from lattice QCD}.
\newblock {\em Phys.Rev.}, D55:7203--7217, 1997.

\bibitem{1964PhRvL..13..585G}
G.~S. {Guralnik}, C.~R. {Hagen}, and T.~W. {Kibble}.
\newblock {Global Conservation Laws and Massless Particles}.
\newblock {\em Physical Review Letters}, 13:585--587, November 1964.

\bibitem{Gusken:1989ad}
S.~Gusken, U.~Low, K.H. Mutter, R.~Sommer, A.~Patel, et~al.
\newblock {Nonsinglet Axial Vector Couplings of the Baryon Octet in Lattice
  {QCD}}.
\newblock {\em Phys.Lett.}, B227:266, 1989.

\bibitem{Hasenfratz:2001hp}
Anna Hasenfratz and Francesco Knechtli.
\newblock {Flavor symmetry and the static potential with hypercubic blocking}.
\newblock {\em Phys.Rev.}, D64:034504, 2001.

\bibitem{Hasenfratz:1993sp}
P.~Hasenfratz and F.~Niedermayer.
\newblock {Perfect lattice action for asymptotically free theories}.
\newblock {\em Nucl.Phys.}, B414:785--814, 1994.

\bibitem{Heisenberg:1932dw}
W.~Heisenberg.
\newblock {On the structure of atomic nuclei}.
\newblock {\em Z.Phys.}, 77:1--11, 1932.

\bibitem{Hernandez:1998et}
Pilar Hernandez, Karl Jansen, and Martin Luscher.
\newblock {Locality properties of Neuberger's lattice Dirac operator}.
\newblock {\em Nucl.Phys.}, B552:363--378, 1999.

\bibitem{1964PhRvL..13..508H}
P.~W. {Higgs}.
\newblock {Broken Symmetries and the Masses of Gauge Bosons}.
\newblock {\em Physical Review Letters}, 13:508--509, October 1964.

\bibitem{Holmgren:2005wf}
D.~Holmgren.
\newblock {U.S. lattice clusters and the USQCD project}.
\newblock {\em PoS}, LAT2005:105, 2006.

\bibitem{Iizuka:1966fk}
Jugoro Iizuka.
\newblock {Systematics and phenomenology of meson family}.
\newblock {\em Prog.Theor.Phys.Suppl.}, 37:21--34, 1966.

\bibitem{Geim:2004}
{K. S. Novoselov, A. K. Geim, S. V. Morozov, D. Jiang, Y. Zhang, S. V. Dubonos,
  I. V. Grigorieva, and A. A. Firsov}.
\newblock {Electric Field Effect in Atomically Thin Carbon Films}.
\newblock {\em Science}, 306:666--669, October 2004.

\bibitem{Kalkreuter:1995mm}
Thomas Kalkreuter and Hubert Simma.
\newblock {An Accelerated conjugate gradient algorithm to compute low lying
  eigenvalues: A Study for the Dirac operator in SU(2) lattice QCD}.
\newblock {\em Comput.Phys.Commun.}, 93:33--47, 1996.

\bibitem{Kaplan:1992bt}
David~B. Kaplan.
\newblock {A Method for simulating chiral fermions on the lattice}.
\newblock {\em Phys.Lett.}, B288:342--347, 1992.

\bibitem{Karsten:1981gd}
Luuk~H. Karsten.
\newblock {Lattice Fermions in Euclidean Space-time}.
\newblock {\em Phys.Lett.}, B104:315, 1981.

\bibitem{Karsten:1980wd}
Luuk~H. Karsten and Jan Smit.
\newblock {Lattice Fermions: Species Doubling, Chiral Invariance, and the
  Triangle Anomaly}.
\newblock {\em Nucl.Phys.}, B183:103, 1981.

\bibitem{Kawai:1980ja}
Hikaru Kawai, Ryuichi Nakayama, and Koichi Seo.
\newblock {Comparison of the Lattice Lambda Parameter with the Continuum Lambda
  Parameter in Massless QCD}.
\newblock {\em Nucl.Phys.}, B189:40, 1981.

\bibitem{Kimura:2011ik}
Taro Kimura, Shota Komatsu, Tatsuhiro Misumi, Toshifumi Noumi, Shingo Torii,
  et~al.
\newblock {Revisiting symmetries of lattice fermions via spin-flavor
  representation}.
\newblock {\em JHEP}, 1201:048, 2012.

\bibitem{Kimura:2009qe}
Taro Kimura and Tatsuhiro Misumi.
\newblock {Characters of Lattice Fermions Based on the Hyperdiamond Lattice}.
\newblock {\em Prog.Theor.Phys.}, 124:415--432, 2010.

\bibitem{Kimura:2009di}
Taro Kimura and Tatsuhiro Misumi.
\newblock {Lattice Fermions Based on Higher-Dimensional Hyperdiamond Lattices}.
\newblock {\em Prog.Theor.Phys.}, 123:63--78, 2010.

\bibitem{KlubergStern:1983dg}
H.~Kluberg-Stern, A.~Morel, O.~Napoly, and B.~Petersson.
\newblock {Flavors of Lagrangian Susskind Fermions}.
\newblock {\em Nucl.Phys.}, B220:447, 1983.

\bibitem{Kobayashi:1973fv}
Makoto Kobayashi and Toshihide Maskawa.
\newblock {CP Violation in the Renormalizable Theory of Weak Interaction}.
\newblock {\em Prog.Theor.Phys.}, 49:652--657, 1973.

\bibitem{Kuipers:2012rf}
J.~Kuipers, T.~Ueda, J.A.M. Vermaseren, and J.~Vollinga.
\newblock {FORM version 4.0}.
\newblock {\em Comput.Phys.Commun.}, 184:1453--1467, 2013.

\bibitem{PhysRev.72.241}
Willis~E. Lamb and Robert~C. Retherford.
\newblock Fine structure of the hydrogen atom by a microwave method.
\newblock {\em Phys. Rev.}, 72:241--243, Aug 1947.

\bibitem{Lattes:1947mw}
C.M.G. Lattes, H.~Muirhead, G.P.S. Occhialini, and C.F. Powell.
\newblock {Processes involving charged mesons}.
\newblock {\em Nature}, 159:694--697, 1947.

\bibitem{Lee:1999zxa}
Weon-Jong Lee and Stephen~R. Sharpe.
\newblock {Partial flavor symmetry restoration for chiral staggered fermions}.
\newblock {\em Phys.Rev.}, D60:114503, 1999.

\bibitem{Lepage:1992xa}
G.~Peter Lepage and Paul~B. Mackenzie.
\newblock {On the viability of lattice perturbation theory}.
\newblock {\em Phys.Rev.}, D48:2250--2264, 1993.

\bibitem{Luscher:1985wf}
M.~Luscher and P.~Weisz.
\newblock {Efficient Numerical Techniques for Perturbative Lattice Gauge Theory
  Computations}.
\newblock {\em Nucl.Phys.}, B266:309, 1986.

\bibitem{Luscher:2010iy}
Martin Luscher.
\newblock {Properties and uses of the Wilson flow in lattice QCD}.
\newblock {\em JHEP}, 1008:071, 2010.

\bibitem{milccode}
\mbox{MILC collaboration}.
\newblock \mbox{http://www.physics.utah.edu/~detar/milc/milc\_qcd.html}.

\bibitem{Misumi:2012eh}
Tatsuhiro Misumi.
\newblock {New fermion discretizations and their applications}.
\newblock {\em PoS}, LATTICE2012:005, 2012.

\bibitem{Misumi:2012uu}
Tatsuhiro Misumi.
\newblock {Phase structure for lattice fermions with flavored chemical
  potential terms}.
\newblock {\em JHEP}, 1208:068, 2012.

\bibitem{Misumi:2010ea}
Tatsuhiro Misumi, Michael Creutz, and Taro Kimura.
\newblock {Classification and Generalization of Minimal-doubling actions}.
\newblock {\em PoS}, LATTICE2010:260, 2010.

\bibitem{PhysRev.122.345}
Y.~Nambu and G.~Jona-Lasinio.
\newblock Dynamical model of elementary particles based on an analogy with
  superconductivity. i.
\newblock {\em Phys. Rev.}, 122:345--358, Apr 1961.

\bibitem{Ne'eman:1961cd}
Yuval Ne'eman.
\newblock {Derivation of strong interactions from a gauge invariance}.
\newblock {\em Nucl.Phys.}, 26:222--229, 1961.

\bibitem{Neuberger:1997fp}
Herbert Neuberger.
\newblock {Exactly massless quarks on the lattice}.
\newblock {\em Phys.Lett.}, B417:141--144, 1998.

\bibitem{Niedermayer:1998bi}
Ferenc Niedermayer.
\newblock {Exact chiral symmetry, topological charge and related topics}.
\newblock {\em Nucl.Phys.Proc.Suppl.}, 73:105--119, 1999.

\bibitem{Nielsen:1981hk}
Holger~Bech Nielsen and M.~Ninomiya.
\newblock {No Go Theorem for Regularizing Chiral Fermions}.
\newblock {\em Phys.Lett.}, B105:219, 1981.

\bibitem{Novoselov:2005kj}
K.S. Novoselov, A.K. Geim, S.V. Morozov, D.~Jiang, M.I. Katsnelson, et~al.
\newblock {Two-dimensional gas of massless Dirac fermions in graphene}.
\newblock {\em Nature}, 438:197, 2005.

\bibitem{Okubo:1963fa}
S.~Okubo.
\newblock {Phi meson and unitary symmetry model}.
\newblock {\em Phys.Lett.}, 5:165--168, 1963.

\bibitem{Osterwalder:1973kn}
K.~Osterwalder and R.~Schrader.
\newblock {Feynman-kac formula for euclidean fermi and bose fields}.
\newblock {\em Phys.Rev.Lett.}, 29:1423--1425, 1972.

\bibitem{Osterwalder:1977pc}
K.~Osterwalder and E.~Seiler.
\newblock {Gauge Field Theories on the Lattice}.
\newblock {\em Annals Phys.}, 110:440, 1978.

\bibitem{Osterwalder:1973dx}
Konrad Osterwalder and Robert Schrader.
\newblock {Axioms Foe Euclidean Green's functions}.
\newblock {\em Commun.Math.Phys.}, 31:83--112, 1973.

\bibitem{PhysRev.86.663}
A.~Pais.
\newblock Some remarks on the $v$-particles.
\newblock {\em Phys. Rev.}, 86:663--672, Jun 1952.

\bibitem{Pauli:1930pc}
W.~Pauli.
\newblock {Dear radioactive ladies and gentlemen}.
\newblock {\em Phys.Today}, 31N9:27, 1978.

\bibitem{Pernici:1994yj}
M.~Pernici.
\newblock {Chiral invariance and lattice fermions with minimal doubling}.
\newblock {\em Phys.Lett.}, B346:99--105, 1995.

\bibitem{Peskin:1995ev}
Michael~E. Peskin and Daniel~V. Schroeder.
\newblock {An Introduction to quantum field theory}.
\newblock 1995.

\bibitem{Politzer:1974fr}
H.~David Politzer.
\newblock {Asymptotic Freedom: An Approach to Strong Interactions}.
\newblock {\em Phys.Rept.}, 14:129--180, 1974.

\bibitem{Press:2007}
William H. et~al. Press.
\newblock {Numerical Recipes: The Art of Scientific Computing}.
\newblock 2007.

\bibitem{Reisz:1987pw}
T.~Reisz.
\newblock {A Convergence Theorem for Lattice Feynman Integrals With Massless
  Propagators}.
\newblock {\em Commun.Math.Phys.}, 116:573, 1988.

\bibitem{Reisz:1987px}
T.~Reisz.
\newblock {Renormalization of Feynman Integrals on the Lattice}.
\newblock {\em Commun.Math.Phys.}, 117:79, 1988.

\bibitem{Reisz:1987hx}
T.~Reisz.
\newblock {Renormalization of Lattice Feynman Integrals With Massless
  Propagators}.
\newblock {\em Commun.Math.Phys.}, 117:639, 1988.

\bibitem{Reisz:1987da}
Thomas Reisz.
\newblock {A Power Counting Theorem for Feynman Integrals on the Lattice}.
\newblock {\em Commun.Math.Phys.}, 116:81, 1988.

\bibitem{Rochester:1947mi}
G.D. Rochester and C.C. Butler.
\newblock {Evidence for the Existence of New Unstable Elementary Particles}.
\newblock {\em Nature}, 160:855--857, 1947.

\bibitem{Salam:1968rm}
Abdus Salam.
\newblock {Weak and Electromagnetic Interactions}.
\newblock {\em Conf.Proc.}, C680519:367--377, 1968.

\bibitem{Scherer:2002tk}
Stefan Scherer.
\newblock {Introduction to chiral perturbation theory}.
\newblock {\em Adv.Nucl.Phys.}, 27:277, 2003.

\bibitem{PhysRev.73.416}
Julian Schwinger.
\newblock On quantum-electrodynamics and the magnetic moment of the electron.
\newblock {\em Phys. Rev.}, 73:416--417, Feb 1948.

\bibitem{PhysRev.74.1439}
Julian Schwinger.
\newblock Quantum electrodynamics. i. a covariant formulation.
\newblock {\em Phys. Rev.}, 74:1439--1461, Nov 1948.

\bibitem{Sharpe:1992ft}
Stephen~R. Sharpe.
\newblock {Quenched chiral logarithms}.
\newblock {\em Phys.Rev.}, D46:3146--3168, 1992.

\bibitem{Sommer:1993ce}
R.~Sommer.
\newblock {A New way to set the energy scale in lattice gauge theories and its
  applications to the static force and alpha-s in SU(2) Yang-Mills theory}.
\newblock {\em Nucl.Phys.}, B411:839--854, 1994.

\bibitem{Susskind:1976jm}
Leonard Susskind.
\newblock {Lattice Fermions}.
\newblock {\em Phys.Rev.}, D16:3031--3039, 1977.

\bibitem{Tiburzi:2010bm}
Brian~C. Tiburzi.
\newblock {Chiral Lattice Fermions, Minimal Doubling, and the Axial Anomaly}.
\newblock {\em Phys.Rev.}, D82:034511, 2010.

\bibitem{Tomonaga:1946zz}
S.~Tomonaga.
\newblock {On a relativistically invariant formulation of the quantum theory of
  wave fields}.
\newblock {\em Prog.Theor.Phys.}, 1:27--42, 1946.

\bibitem{vandenDoel:1983mf}
Cees van~den Doel and Jan Smit.
\newblock {Dynamical Symmetry Breaking in Two Flavor SU($N$) and SO($N$)
  Lattice Gauge Theories}.
\newblock {\em Nucl.Phys.}, B228:122, 1983.

\bibitem{vanNieuwenhuizen:1996ip}
Peter van Nieuwenhuizen and Andrew Waldron.
\newblock {A Continuous Wick rotation for spinor fields and supersymmetry in
  Euclidean space}.
\newblock pages 394--403, 1996.

\bibitem{vanNieuwenhuizen:1996tv}
Peter van Nieuwenhuizen and Andrew Waldron.
\newblock {On Euclidean spinors and Wick rotations}.
\newblock {\em Phys.Lett.}, B389:29--36, 1996.

\bibitem{Veneziano:1979ec}
G.~Veneziano.
\newblock {U(1) Without Instantons}.
\newblock {\em Nucl.Phys.}, B159:213--224, 1979.

\bibitem{Vermaseren:2000nd}
J.A.M. Vermaseren.
\newblock {New features of FORM}.
\newblock 2000.

\bibitem{vonHippel:2013yfa}
Georg~M. von Hippel, Benjamin Jäger, Thomas~D. Rae, and Hartmut Wittig.
\newblock {The Shape of Covariantly Smeared Sources in Lattice QCD}.
\newblock {\em JHEP}, 1309:014, 2013.

\bibitem{Waldron:1997re}
Andrew Waldron.
\newblock {A Wick rotation for spinor fields: The Canonical approach}.
\newblock {\em Phys.Lett.}, B433:369--376, 1998.

\bibitem{proceeding4}
Johannes Weber.
\newblock {Correlation functions with Karsten-Wilczek fermions}.
\newblock {\em PoS}, LAT2014:071, 2014.

\bibitem{Weber:2013tfa}
Johannes~Heinrich Weber, Stefano Capitani, and Hartmut Wittig.
\newblock {Numerical studies of Minimally Doubled Fermions}.
\newblock 2013.

\bibitem{PhysRevLett.18.188}
Steven Weinberg.
\newblock Dynamical approach to current algebra.
\newblock {\em Phys. Rev. Lett.}, 18:188--191, Jan 1967.

\bibitem{PhysRevLett.19.1264}
Steven Weinberg.
\newblock A model of leptons.
\newblock {\em Phys. Rev. Lett.}, 19:1264--1266, Nov 1967.

\bibitem{Whitmer:1984he}
C.~Whitmer.
\newblock {Overrelaxation methods for Monte Carlo simulations of quadratic and
  multiquadratic actions}.
\newblock {\em Phys.Rev.}, D29:306--311, 1984.

\bibitem{Wilczek:1987kw}
Frank Wilczek.
\newblock {On lattice fermions}.
\newblock {\em Phys.Rev.Lett.}, 59:2397, 1987.

\bibitem{Wilson:1974sk}
Kenneth~G. Wilson.
\newblock {Confinement of Quarks}.
\newblock {\em Phys.Rev.}, D10:2445--2459, 1974.

\bibitem{Witten1979269}
E.~Witten.
\newblock Current algebra theorems for the u(1) “goldstone boson”.
\newblock {\em Nuclear Physics B}, 156(2):269 -- 283, 1979.

\bibitem{Wittig:2002ux}
Hartmut Wittig.
\newblock {Chiral effective Lagrangian and quark masses}.
\newblock {\em Nucl.Phys.Proc.Suppl.}, 119:59--70, 2003.

\bibitem{Yang:1954ek}
Chen-Ning Yang and Robert~L. Mills.
\newblock {Conservation of Isotopic Spin and Isotopic Gauge Invariance}.
\newblock {\em Phys.Rev.}, 96:191--195, 1954.

\bibitem{Yukawa:1935xg}
Hideki Yukawa.
\newblock {On the interaction of elementary particles}.
\newblock {\em Proc.Phys.Math.Soc.Jap.}, 17:48--57, 1935.

\bibitem{Zweig:1964jf}
G.~Zweig.
\newblock {An SU(3) model for strong interaction symmetry and its breaking.
  Version 2}.
\newblock pages 22--101, 1964.

\end{thebibliography}
\bibliographystyle{plain}

\end{document}